\documentclass[twocolumn,tighten]{aastex631}

\usepackage{graphicx}
\usepackage{bm}
\usepackage{url}
\usepackage{threeparttable}
\usepackage{amsmath,amssymb}
\usepackage{xspace}
\usepackage{ulem}
\usepackage{subfigure}

\newcommand{\ergs}{erg s$^{-1}$}

\shorttitle{Hard X-ray Radiative AGN Feedback}
\shortauthors{Kawamuro et al.}

\begin{document}
\title{
    Hard X-ray Irradiation Potentially Drives Negative AGN Feedback \\ 
    by Altering Molecular Gas Properties
} 

\correspondingauthor{Taiki Kawamuro}
\email{taiki.kawamuro@mail.udp.cl}

\author{Taiki Kawamuro}
\altaffiliation{FONDECYT postdoctoral fellow}
\affil{Nu\'{c}leo de Astronom\'{i}a de la Facultad de Ingenier\'{i}a, Universidad Diego Portales, Av. Ej\'{e}ercito Libertador 441, Santiago, Chile}
\affil{National Astronomical Observatory of Japan, Osawa, Mitaka, Tokyo 181-8588, Japan}

\author{Claudio Ricci} 
\affil{Nu\'{c}leo de Astronom\'{i}a de la Facultad de Ingenier\'{i}a, Universidad Diego Portales, Av. Ej\'{e}ercito Libertador 441, Santiago, Chile}
\affil{Kavli Institute for Astronomy and Astrophysics, Peking University, Beijing 100871, People’s Republic of China}
\affil{George Mason University, Department of Physics \& Astronomy, MS 3F3, 4400 University Drive, Fairfax, VA 22030, USA}

\author{Takuma Izumi}
\altaffiliation{NAOJ fellow} 
\affil{National Astronomical Observatory of Japan, Osawa, Mitaka, Tokyo 181-8588, Japan}
\affil{Department of Astronomy, School of Science, Graduate University for Advanced Studies (SOKENDAI), 2-21-1 Osawa, Mitaka, Tokyo 181-8588, Japan}

\author{Masatoshi Imanishi} 
\affil{National Astronomical Observatory of Japan, Osawa, Mitaka, Tokyo 181-8588, Japan}
\affil{Department of Astronomy, School of Science, Graduate University for Advanced Studies (SOKENDAI), 2-21-1 Osawa, Mitaka, Tokyo 181-8588, Japan}

\author{Shunsuke Baba} 
\altaffiliation{JSPS fellow (PD)}
\affil{National Astronomical Observatory of Japan, Osawa, Mitaka, Tokyo 181-8588, Japan}

\author{Dieu D. Nguyen}
\affil{National Astronomical Observatory of Japan, Osawa, Mitaka, Tokyo 181-8588, Japan}

\author{Kyoko Onishi}
\affil{Department of Earth and Space Sciences, Chalmers University of Technology, Onsala Space Observatory, 439 94 Onsala, Sweden}

\begin{abstract}

To investigate the role of active galactic nucleus (AGN) X-ray irradiation on the interstellar medium (ISM), we systematically analyzed Chandra and ALMA CO($J$=2--1) data for 26 ultra-hard X-ray ($>$ 10\,keV) selected AGNs at redshifts below 0.05. While Chandra unveils the distribution of X-ray-irradiated gas via Fe-K$\alpha$ emission, the CO($J$=2--1) observations reveal that of cold molecular gas. At high resolutions $\lesssim$ 1\arcsec, we derive Fe-K$\alpha$ and CO($J$=2--1) maps for the nuclear 2\arcsec\ region, and for the external annular region of 2\arcsec--4\arcsec, where 2\arcsec\ is $\sim$ 100--600 pc for most of our AGNs. First, focusing on the external regions, we find the Fe-K$\alpha$ emission for six AGNs above 2$\sigma$. Their large equivalent widths ($\gtrsim$ 1\,keV) suggest a fluorescent process as their origin. Moreover, by comparing 6--7\,keV/3--6\,keV ratio, as a proxy of Fe-K$\alpha$, and CO($J$=2--1) images for three AGNs with the highest significant Fe-K$\alpha$ detections, we find a possible spatial separation. These suggest the presence of X-ray-irradiated ISM and the change in the ISM properties. Next, examining the nuclear regions, we find that (1) The 20--50\,keV luminosity increases with the CO($J$=2--1) luminosity. 
(2) The ratio of CO($J$=2--1)-to-HCN($J$=1--0) luminosities increases with 20--50\,keV luminosity, suggesting a decrease in the dense gas fraction with X-ray luminosity. (3) The Fe-K$\alpha$-to-X-ray continuum luminosity ratio decreases with the molecular gas mass. This may be explained by a negative AGN feedback scenario: the mass accretion rate increases with gas mass, and simultaneously, the AGN evaporates a portion of the gas, which possibly affects star formation.

\end{abstract}


\section{Introduction}

Since the discovery of a potential link between the spheroidal components of galaxies  and their central supermassive black holes \citep[SMBHs; e.g.,][]{Mag98,Geb00,Mar03,Gul09,Kor13}, 
the co-evolution of galaxies and SMBHs has been one of the most debated topics in astronomy. 
Generally, galaxies have grown by converting gas into stars, whereas the SMBHs are expected to have obtained a significant fraction of their mass by gas accretion \citep[e.g.,][]{Sol82,Mar04,Ued14}. 
Thus, in any scenario related to the growth of galaxies and SMBHs \citep[e.g., ][]{DiM05,Cro06}, 
it is fundamental to understand 
detailed conversion process of gas into key components (e.g., stars and SMBHs).

The properties of the ISM around an accreting SMBH, or an active galactic nucleus (AGN), likely differ from those in pure star-forming regions because 
the X-ray emission of an AGN is much stronger than that of star-forming regions  \cite[e.g.,][]{Elv94,Min12a,Hic18}. 
Given that hard X-ray ($\gtrsim$ 1\,keV) emission has high penetrating power, it can deeply pierce through the ISM. Accordingly, the physical and chemical properties of the ISM are expected to be widely altered \citep[e.g.,][]{Kro84,Mal96,Mei05,Mei07}. 
Regions where the X-ray emission is particularly strong are usually referred to as X-ray dominated regions (XDRs).
Following several theoretical studies \citep[e.g.,][]{Mei05,Mei07}, observational evidence supporting a strong impact of X-ray radiation on the ISM has been reported in several studies \citep[e.g., higher kinetic temperature and molecular gas destruction;][]{Gar10,Vit14,Izu20b}.
Given the crucial role of molecular gas in the creation of stars, 
the likelihood of X-ray emission affecting the star-formation process was inferred. 
Supportive results were presented based on numerical studies \cite[e.g.,][]{Hoc10,Hoc11}. 
By considering gas clouds exposed to 
X-ray emission from an AGN, \cite{Hoc11} found that strong X-ray emission can compress a part of the gas to form stars but eventually reduce the total mass of formed stars by evaporating a major fraction of the gas. This could further lead to a change in the initial mass function of stars \citep[][]{Hoc10,Hoc11}.

Since the advent of Chandra \citep{Gar03}, 
extended hard X-ray emission around AGNs has been
confirmed by several studies owing to its high spatial resolution (down to $\sim$ 0\arcsec.1), which is currently the highest in the X-ray band \citep[e.g.,][]{You01,Wan09,Mar12,Mar13,Fab17,Mar17,Gom17,Kaw19b,Kaw20,Fer20,Ma20,Nak21}. 
While this extended X-ray emission has been found in different energy bands, 
understanding the spatial extension of the Fe-K$\alpha$ line ($\simeq 6.4$\,keV) is extremely important for constraining the X-ray irradiation of the ISM.
Photons with $E>7.1$\,keV can deeply penetrate the ISM and produce Fe-K$\alpha$ photons, which retain a high penetrating power, and can indicate whether gas is irradiated by the X-ray emission \citep[e.g.,][]{Fab77}.
Although the Fe-K$\alpha$ fluorescent emission is produced preferentially in dense gas with a column density of $\sim 10^{23-24}$ cm$^{-2}$, the use of the line is valuable in discussing the association between X-ray irradiation and dense gas, which is expected to be associated with star-forming regions.
Another advantage of the K$\alpha$ line is that, unlike the emission in the soft band ($<$ 2~keV), emission at $\sim$ 6.4\,keV is little contaminated by star-formation activity, which is usually faint in the hard X-ray band \citep[$>$ 2\,keV; e.g., ][]{LaM12,LaM17,Kaw13}.
Therefore, the observation of this line enables 
the effect of the AGN on the ISM to be studied. 

While extended X-ray emission has been studied in detail over the past few decades, as described above, it has been difficult to investigate the physical and chemical properties of  X-ray-irradiated gas. However, the advent of the 
Atacama Large Millimeter/submillimeter Array (ALMA), with a  very high angular resolution and sensitivity,  
has allowed the properties of the ISM around the AGN 
to be studied \citep[][]{Gar14,Kaw19a,Kaw20,Fer20}. 
Over the past few years, several systematic surveys using ALMA have been performed to reveal the molecular and atomic gas distribution around nearby AGNs \citep[e.g.,][]{Miy18,Izu18,Com19,Ram19}. 
Among the various molecular and atomic emission lines observed, 
the CO($J$=2--1) line at the rest frequency of 230.538 GHz has frequently been targeted. 
Compared with the lower CO($J$=1--0) emission, 
the flux density of CO($J$=2--1) can be higher as long as the CO molecules are optically thick and thermalized.
In addition, the critical density of $\sim 3\times10^3$ cm$^{-3}$, which is lower than those of CO molecules at higher rotational energies, allows detailed mapping of the global gas distribution. 

In this paper, we address one important question, 
\textit{what impact does the hard X-ray irradiation have on the ISM and on star formation?}. 
To answer this question, we performed a systematic analysis of Chandra X-ray and ALMA CO($J$=2--1) data taken for nearby AGNs from the Swift/BAT ultra-hard X-ray 70-month catalog \citep{Bau13}. 
From the results obtained in our study, we suggest two points: (1) The physical and chemical conditions of the ISM irradiated by X-ray emission are altered and, accordingly, such gas cannot be traced by CO($J$=2--1), or cold gas tracers. (2) Strong X-ray radiation may have an impact on star formation by evaporating a fraction of gas from nuclear regions.

\begin{figure*}
    \centering
    \includegraphics[width=13.5cm]{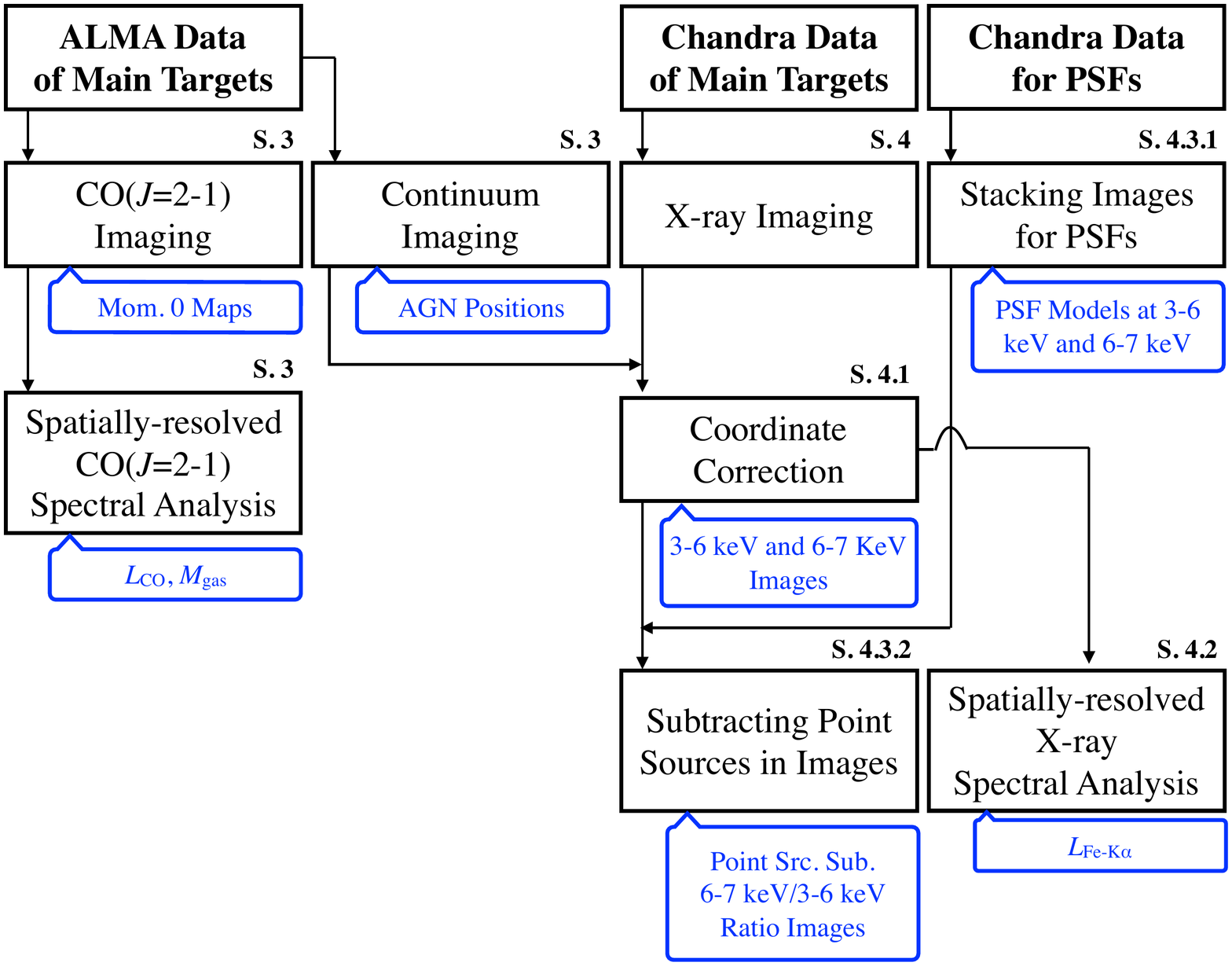}
    \caption{Flowchart for our main analyses. 
    The numbers on the right shoulders of the black boxes indicate the sections where the corresponding analyses are explained. 
    Additionally, the main products are indicated in the blue boxes. 
    }
    \label{fig:flow_chart}
\end{figure*}

The remainder of this paper is structured as follows.
In Section~\ref{sec:sample}, we introduce our sample selection,
as well as the ALMA and Chandra data.
The reduction and analysis of the ALMA and Chandra data are described in Sections~\ref{sec:alma_data} and ~\ref{sec:cxo_data}, respectively. 
As a guide to these data analyses, we present a flowchart in Figure~\ref{fig:flow_chart}.
The results are discussed in Section~\ref{sec:dis}. 
Finally, Section~\ref{sec:sum} summarizes our results and conclusions.
The Appendix provides figures produced throughout this paper: spatially resolved X-ray spectra, radial profiles of X-ray surface brightness, X-ray images, spectra around CO($J$=2--1) emission, and CO($J$=2--1) moment 0 maps.

Throughout this paper, we adopt standard cosmological parameters ($H_0$ = 70 km s$^{-1}$ Mpc$^{-1}$, $\Omega_{\rm m}$ = 0.3, $\Omega_\Lambda$ = 0.7). Unless otherwise stated, the  uncertainties at the 1$\sigma$ confidence level are quoted. 

\floattable
\renewcommand{\arraystretch}{1.}
\begin{deluxetable}{ccccccccccc}
\tabletypesize{\footnotesize}
  \tablecaption{Target information}
  \tablehead{
      \colhead{(1)} & \colhead{(2)} & \colhead{(3)} & \colhead{(4)} & \colhead{(5)} & \colhead{(6)}  & \colhead{(7)} & \colhead{(8)} 
      & \colhead{(9)}  & \colhead{(10)} \\ 
      \colhead{Number} & 
       \colhead{Target name} & \colhead{R.A.} & \colhead{Decl.} &  \colhead{$z$} & \colhead{$D$} 
       & \colhead{pc/1\arcsec}
       & \colhead{$\Gamma$} 
       & \colhead{$\log(N_{\rm H}$/cm$^{-2})$}
       & \colhead{$\log[L_{\rm 20-50~keV}$/(erg/s)]} 
     }
    \startdata
\thispagestyle{empty} 
01 & NGC 424      & 17.865170 & -38.083467  & 0.0118 & 51.0 & 247 & 2.11 & 24.25 & 42.98 \\
02 & NGC 526A     & 20.976574 & -35.065481  & 0.0191 & 83.0 & 402 & 1.43 & 22.01 & 43.21 \\
03 & NGC 612      &  23.490399 & -36.493187 & 0.0305 & 133.7 & 648 & 1.71$^\dagger$ & 23.97$^\dagger$ & 43.80$^\dagger$ \\
04 & NGC 788      & 30.276933 & -6.815881  & 0.0136 & 58.9 & 286 & 1.64 & 23.82 & 43.22 \\
05 & NGC 1052     & 40.269992 & -8.255763  & 0.005 & 19.5 & 95 & 1.44 & 22.95 & 41.73 \\
06 & NGC 1068     & 40.669625 & -0.013323  & 0.0038 & 12.7 & 62 & 1.9 & 25.0 & 42.39 \\
07 & NGC 1125     & 42.918575 & -16.650651  & 0.011 & 47.5 & 230 & 1.9 & 24.21 & 42.49 \\
08 & NGC 1365     & 53.401529 & -36.140430  & 0.0055 & 18.2 & 88 & 1.76 & 22.21 & 42.23 \\
09 & 2MASX J0521-2521 & 80.255840 & -25.362585 & 0.0426 & 188.4 & 913 & 2.6 & 22.92 & 43.23 \\
10 & NGC 2110     & 88.047406 & -7.456249  & 0.0078 & 35.6 & 173 & 1.71 & 22.94 & 43.14 \\
11 & ESO 005-G004 & 91.422034 & -86.631555  & 0.0062 & 22.4 & 109 & 1.64 & 24.18 & 42.23 \\
12 & Mrk 1210     & 121.024420 & 5.113847  & 0.0135 & 58.4 & 283  & 2.07 & 23.4 & 42.98 \\
13 & NGC 3081     &  149.873106 & -22.826329 & 0.008 & 26.5 & 128 & 1.88 & 23.91 & 42.82 \\
14 & NGC 3393     & 162.097528 & -25.162230  & 0.0125 & 54.0 & 262  & 2.22 & 24.4 & 42.57 \\
15 & NGC 4507     & 188.902654 & -39.909359  & 0.0118 & 51.0 & 247 & 1.71 & 23.95 & 43.54 \\
16 & NGC 4945     & 196.364514 & -49.468156  & 0.0019 & 8.1 & 39 & 1.38 & 24.8 & 42.62 \\
17 & ESO 323-077  & 196.608839 & -40.414612  & 0.015 & 65.0 & 315 & 2.0 & 22.81 & 42.83 \\
18 & Mrk 463      & 209.012026 & 18.371872  & 0.0504 & 224.1 & 1086 & 1.64 & 23.81 & 43.35 \\
19 & NGC 5506     & 213.311996 & -3.207678  & 0.0062 & 23.8 & 115  & 1.8 & 22.44 & 42.91 \\
20 & NGC 5643     & 218.169579 & -44.174419  & 0.004 & 16.9 & 82 & 1.46 & 25.0 & 42.49 \\
21 & NGC 5728     & 220.599479 & -17.253054  & 0.0093 & 30.6 & 148 & 1.5 & 24.14 & 42.87 \\
22 & NGC 6240     & 253.245380 & 2.400932  & 0.0245 & 106.9 & 518  & 1.95 & 24.25 & 44.25 \\
23 & Fairall 49   & 279.242649 & -59.402303  & 0.0202 & 87.9 & 426  & 2.16 & 22.03 & 42.74 \\
24 & IC 5063      & 313.009841 & -57.068772  & 0.0114 & 49.3 & 239 & 1.9 & 23.56 & 42.96 \\
25 & NGC 7130     & 327.081343 & -34.951313  & 0.0162 & 70.2  & 340 & 2.03 & 24.22 & 42.52 \\
26 & NGC 7582     & 349.598514 & -42.370433  & 0.0052 & 20.9  & 101 & 1.89 & 24.15 & 42.86 \\ 
\enddata
\thispagestyle{empty}
\tablenotetext{}{
Notes. (1) Source number. (2) Target name. (3), (4) Right ascension and declination (ICRS) in units of degrees, determined from ALMA data. 
NGC 6240 and Mrk 463 are in mergers, and their AGN positions listed here are for the X-ray brighter nucleus (see detail in Section~\ref{sec:alma_data}).
For ESO 005-G004, we adopted an optical AGN position from NED, as no significant emission was detected from the ALMA data. 
(5) Redshift.
(6) Distance in units of Mpc. 
(7) Physical resolution in units of pc achievable at 1\arcsec. 
(8) Photon index. 
(9) Hydrogen column density on logarithmic scale.
(10) Absorption-corrected hard X-ray 20--50\,keV luminosity on logarithmic scale.
The X-ray data of (8),(9), and (10) are from \cite{Ric17c}, 
except for those of NGC 612, which are from  \cite{Urs18}$^\dagger$. 
}\label{tab:sample}
\end{deluxetable}
\setlength{\topmargin}{0in}

\section{Sample selection}\label{sec:sample}

\begin{figure*}
    \centering
    \includegraphics[width=5.5cm]{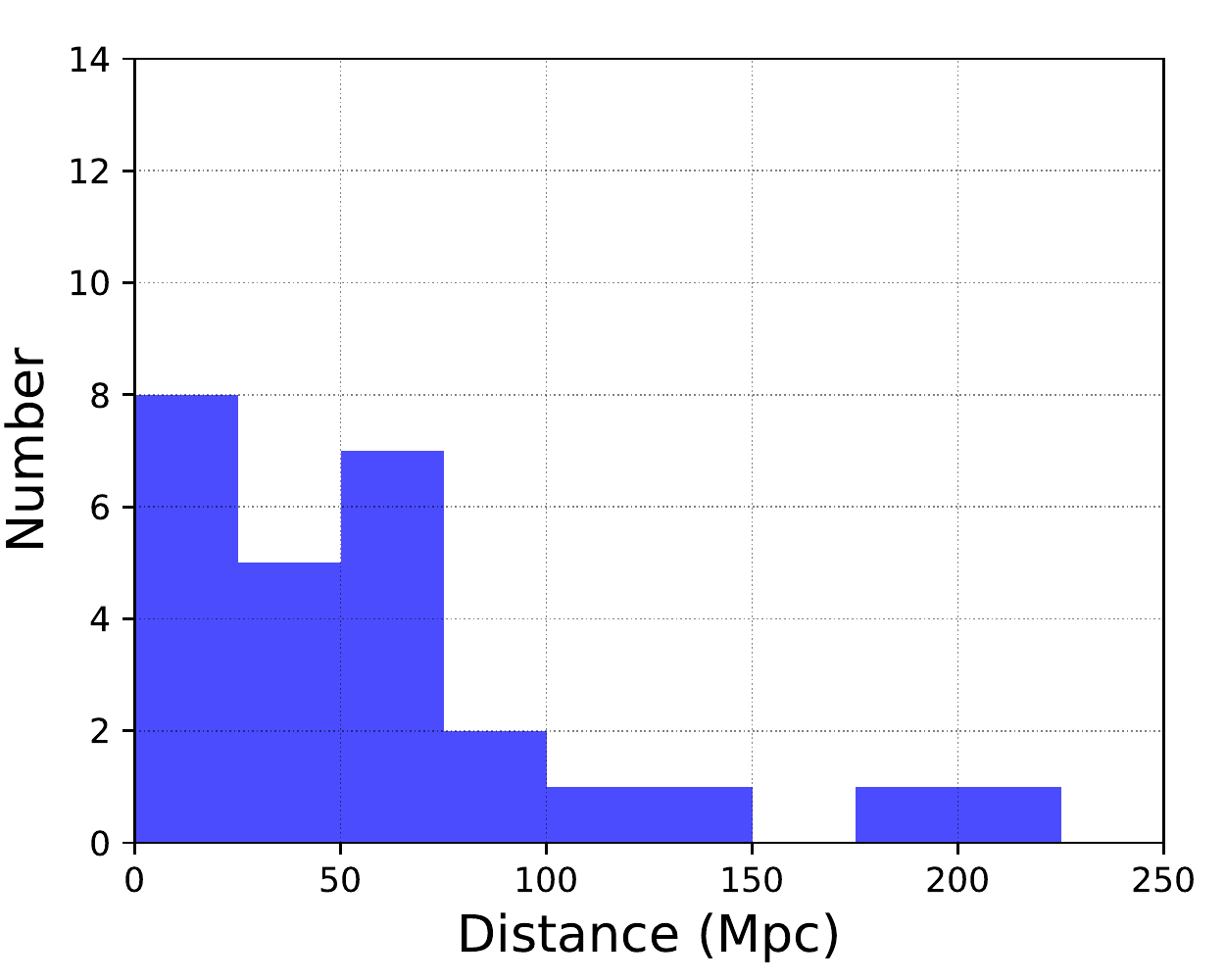}
    \includegraphics[width=5.5cm]{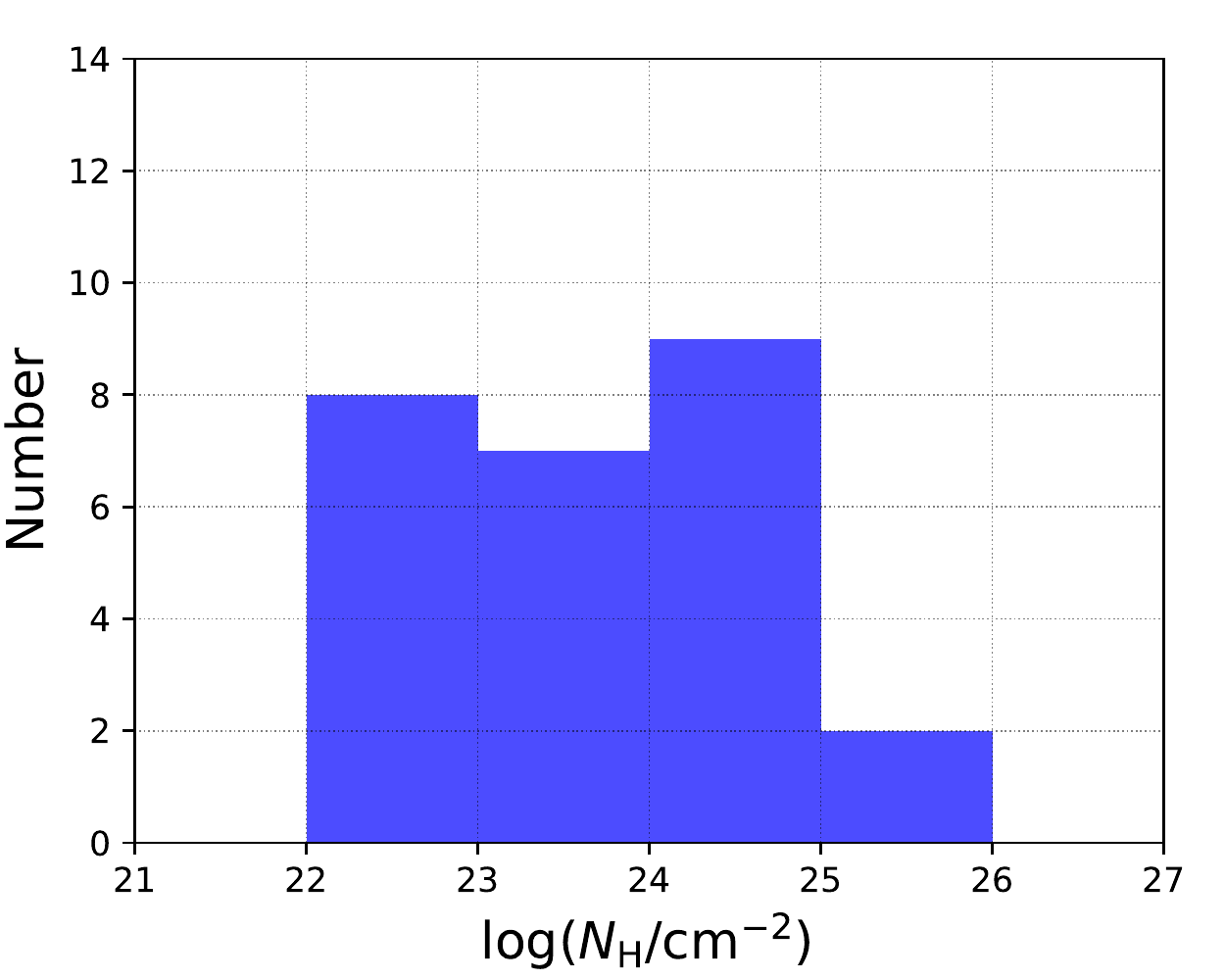}
    \includegraphics[width=5.5cm]{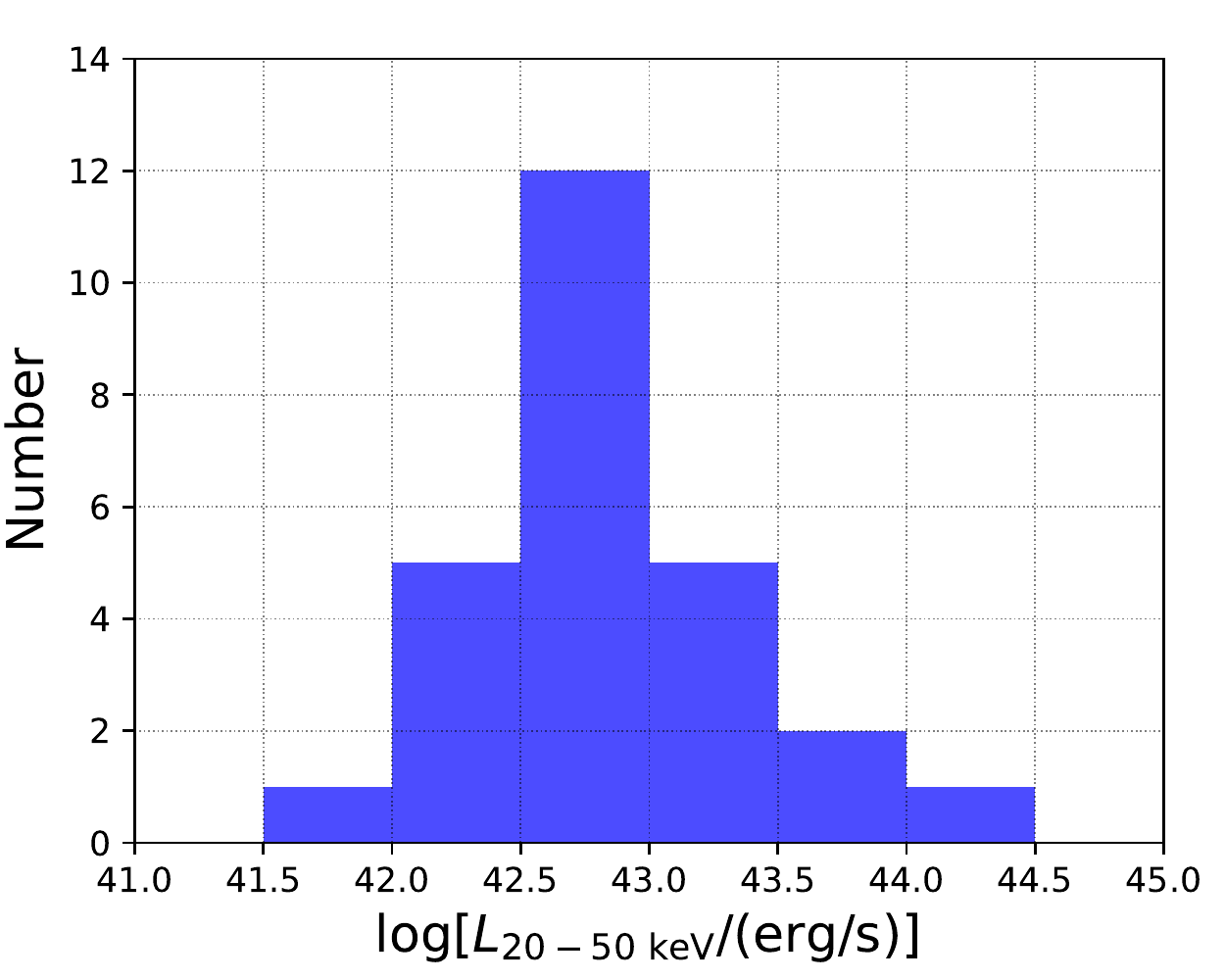}
    \caption{Distributions of the distances, the line-of-sight hydrogen column densities, and the absorption-corrected 20--50\,keV luminosities of our sample. 
    }
    \label{fig:sample_hists}
\end{figure*}

To address our main question based on a sample that is not strongly biased by obscuration, we considered hard X-ray selected AGNs from the Swift/BAT AGN Spectroscopic Survey (BASS) project \citep{Kos17,Ric17c}\footnote{https://www.bass-survey.com/}. This sample is based on the 70-month BAT catalog \citep{Bau13}, and provides not only precise nuclear X-ray properties \citep{Ric17c}, but also information obtained by follow-up observations at various wavelengths \citep[e.g.,][]{Lamperti17,Bae19,Smi20,Kos21}. 

From the BASS Data Release 1 catalog, we first selected obscured AGNs with $\log (N_{\rm H}/{\rm cm}^{-2}) > 22$ at redshifts $\lesssim0.05$, corresponding to $\approx$ 200~Mpc, 
suited to revealing extended X-ray emission. 
Line-of-sight obscuration makes it easier to detect extended faint emission. In addition, their proximity enables us to achieve high physical resolution. Then, we searched for their CO($J$=2--1) ALMA data, which were obtained with 12 m telescope arrays and were publicly available as of 2020 August.
Next, we searched the ChaSeR web interface for Chandra archived ACIS data with a search radius of 1\arcmin. The radius was set to avoid poorer angular resolution due to the large offset between the aim-point and the targeted object. Here, we retained AGNs observed by either imaging or grating observations, but excluded objects if any of their observations were obtained with 
exposures less than 10 ks. To avoid confusion, we kept the objects that have at least one observation taken with an exposure larger than 10 ks, and considered all their available data even if some data were taken with exposures less than 10 ks.
Finally, we checked the data quality, and, for example, we excluded Cen A, whose Chandra data are quite heavily affected by photon pileup around the nucleus.
We then defined a sample consisting of 26 AGNs.
The basic properties are summarized in Table~\ref{tab:sample}, and some of them are presented as histograms in Figure~\ref{fig:sample_hists}. 

We assessed the quality of all publicly available Chandra data as of 2020 August and considered only those that were not severely affected by either the photon pileup or particle events. 
We selected the highest resolution ALMA data for each target.
However, for NGC 1068, we used the 2nd highest resolution data as the maximum recoverable scale (MRS) of the highest resolution less than 1\arcsec\ cannot be used in conjunction with Chandra to discuss the properties of the circumnuclear gas, as is done for the other targets. 
To the contrary, the selected data have MRSs larger than 1\arcsec.3 and thus, retain good sensitivities to the angular scale of interest (e.g., $\sim$ 1\arcsec). Tables~\ref{tab:cxo_data} and \ref{tab:alma_data} summarize the  Chandra and ALMA data adopted in this study.

As described in detail in Section~\ref{sec:cxo_data_red}, we  define and separately discuss two types of regions: 
a central 2\arcsec\ nuclear region and an external annular region of 2\arcsec--4\arcsec. 
The former and latter regions 
correspond to $\lesssim$ 100--600 pc and $\sim$ 100--1300 pc, respectively, for most of our AGNs (i.e., 19/26 or 73\%). 
Hereafter, we refer to the two regions as the ``nuclear'' and ``external'' regions, respectively. 
Because of this definition, the physical scales of the regions do not match between nearby and distance objects. 
For example, the external regions of the nearest objects in our sample (e.g., $\sim$ 300--600 pc for a object at $D \sim$ 30 Mpc) can be within the nuclear regions of distant objects 
(e.g., $\lesssim$ 600 pc for a object at $D \sim$ 60 Mpc). 
However, our discussions are made while considering this fact as much as possible.  
Specifically, we discuss the relevance between 
Fe-K$\alpha$ and CO($J$=2--1) emission while extracting them from the same regions in Section~\ref{sec:fe_vs_co21}, therefore, 
being unaffected by differences in object distance. 
Also, we limit our sample only to the nearest AGNs at distances $\lesssim$ 30 Mpc in Section~\ref{sec:acc_vs_gasmass}.

\renewcommand{\arraystretch}{0.9}
\begin{table*} 
\caption{Chandra Data List}
\begin{minipage}{0.5\textwidth}
\begin{tabular}{ccccc}
\toprule
(1)   & (2) & (3) & (4) & (5) \\ 
Number & Target name & ObsID & Grating & Exp. \\ \hline
  01 &      NGC 424 & 3146 & NONE & 9.18 \\
  01 &      NGC 424 &21417 & NONE &15.28 \\
  02 &     NGC 526A & 4376 & HETG &28.68 \\
  02 &     NGC 526A & 4437 & HETG &29.07 \\
  03 &      NGC 612 &16099 & NONE &10.78 \\
  03 &      NGC 612 &17577 & NONE &24.73 \\
  04 &      NGC 788 &11680 & NONE &13.61 \\
  05 &     NGC 1052 &  385 & NONE & 2.34 \\
  05 &     NGC 1052 & 5910 & NONE &59.20 \\
  06 &     NGC 1068 & 9148 & HETG &79.57 \\
  06 &     NGC 1068 & 9149 & HETG &88.73 \\
  06 &     NGC 1068 & 9150 & HETG &41.08 \\
  06 &     NGC 1068 &10815 & HETG &19.07 \\
  06 &     NGC 1068 &10816 & HETG &16.16 \\
  06 &     NGC 1068 &10817 & HETG &32.65 \\
  06 &     NGC 1068 &10823 & HETG &34.54 \\
  06 &     NGC 1068 &10829 & HETG &38.44 \\
  06 &     NGC 1068 &10830 & HETG &42.90 \\
  07 &     NGC 1125 &21418 & NONE &53.15 \\
  08 &     NGC 1365 & 3554 & NONE &14.61 \\
  08 &     NGC 1365 & 6868 & NONE &14.61 \\
  08 &     NGC 1365 & 6869 & NONE &15.54 \\
  08 &     NGC 1365 & 6870 & NONE &14.55 \\
  08 &     NGC 1365 & 6871 & NONE &13.36 \\
  08 &     NGC 1365 & 6872 & NONE &14.62 \\
  08 &     NGC 1365 & 6873 & NONE &14.64 \\
  08 &     NGC 1365 &13920 & HETG &88.53 \\
  08 &     NGC 1365 &13921 & HETG &      108.20 \\
  09 &    2MASX J0521-2521 & 3432 & NONE &14.86 \\
  10 &     NGC 2110 &  883 & NONE &45.75 \\
  10 &     NGC 2110 & 3143 & HETG &25.05 \\
  10 &     NGC 2110 & 3417 & HETG &24.47 \\
  10 &     NGC 2110 & 3418 & HETG &56.07 \\
  10 &     NGC 2110 & 4377 & HETG &94.88 \\
  11 & ESO 005-G004 &21421 & NONE &21.94 \\
  12 &     Mrk 1210 & 4875 & NONE &10.41 \\
  12 &     Mrk 1210 & 9264 & NONE & 9.82 \\
  12 &     Mrk 1210 & 9265 & NONE & 9.41 \\
  12 &     Mrk 1210 & 9266 & NONE & 9.42 \\
  12 &     Mrk 1210 & 9267 & NONE & 9.80 \\
  12 &     Mrk 1210 & 9268 & NONE & 9.79 \\
  13 &     NGC 3081 &20622 & NONE &29.39 \\
\hline
 \hline
\end{tabular}
\vspace{-0.cm}
\end{minipage} \hfill
\begin{minipage}{0.5\textwidth}
\begin{tabular}{cccccccc}
\toprule
(1)   & (2) & (3) & (4) & (5) \\ 
Number & Target name & ObsID & Grating & Exp. \\ \hline
  14 &     NGC 3393 & 4868 & NONE &29.33 \\
  14 &     NGC 3393 &12290 & NONE &69.16 \\
  14 &     NGC 3393 &13967 & HETG &      176.78 \\
  14 &     NGC 3393 &13968 & HETG &28.06 \\
  14 &     NGC 3393 &14403 & HETG &77.72 \\
  14 &     NGC 3393 &14404 & HETG &56.68 \\
  15 &     NGC 4507 & 2150 & HETG &      138.20 \\
  15 &     NGC 4507 &12292 & NONE &39.60 \\
  16 &     NGC 4945 & 4899 & HETG &77.37 \\
  16 &     NGC 4945 & 4900 & HETG &95.60 \\
  16 &     NGC 4945 &14984 & NONE &      128.76 \\
  16 &     NGC 4945 &14985 & NONE &68.73 \\
  17 &  ESO 323-077 &11848 & HETG &45.90 \\
  17 &  ESO 323-077 &11849 & HETG &      118.14 \\
  17 &  ESO 323-077 &12139 & HETG &59.55 \\
  17 &  ESO 323-077 &12204 & HETG &67.34 \\
  18 &      Mrk 463 & 4913 & NONE &49.33 \\
  18 &      Mrk 463 &18194 & NONE & 9.57 \\
  19 &     NGC 5506 &  357 & NONE & 0.71 \\
  19 &     NGC 5506 & 1598 & HETG &88.90 \\
  20 &     NGC 5643 & 5636 & NONE & 7.63 \\
  20 &     NGC 5643 &17031 & NONE &72.12 \\
  20 &     NGC 5643 &17664 & NONE &41.53 \\
  21 &     NGC 5728 & 4077 & NONE &18.73 \\
  22 &     NGC 6240 & 1590 & NONE &36.69 \\
  22 &     NGC 6240 & 6908 & HETG &      157.00 \\
  22 &     NGC 6240 & 6909 & HETG &      141.23 \\
  22 &     NGC 6240 &12713 & NONE &      145.36 \\
  23 &   Fairall 49 & 3148 & HETG &55.98 \\
  23 &   Fairall 49 & 3452 & HETG &50.26 \\
  24 &      IC 5063 & 7878 & NONE &34.10 \\
  24 &      IC 5063 &21466 & NONE &87.67 \\
  24 &      IC 5063 &21467 & NONE &26.93 \\
  24 &      IC 5063 &21999 & NONE &34.13 \\
  24 &      IC 5063 &22000 & NONE &15.64 \\
  24 &      IC 5063 &22001 & NONE &29.27 \\
  24 &      IC 5063 &22002 & NONE &43.87 \\
  25 &     NGC 7130 & 2188 & NONE &38.64 \\
  26 &     NGC 7582 &  436 & NONE &13.44 \\
  26 &     NGC 7582 & 2319 & NONE & 5.87 \\
  26 &     NGC 7582 &16079 & HETG &      173.39 \\
  26 &     NGC 7582 &16080 & HETG &19.71 \\ \hline \hline 
\end{tabular}
\end{minipage}\vspace{0.45cm}
\tablenotetext{}{
Notes.--- (1) Source number. 
(2) Target name. 
(3) Chandra observed data ID.
(4) Tag for imaging (NONE) and grating (HETG) observations.
(5) Observation exposure time in units of ks, taken from ChaSeR.
}\label{tab:cxo_data}
\end{table*}
\renewcommand{\arraystretch}{1.0}

\setlength{\topmargin}{3.0cm}
\floattable
\renewcommand{\arraystretch}{1.1}
\begin{deluxetable}{ccccrccc}
\thispagestyle{empty}
  \tablecaption{ALMA Data}
  \tablehead{
      \colhead{(1)} & \colhead{(2)} & \colhead{(3)} & \colhead{(4)} & \colhead{(5)} & \colhead{(6)} & \colhead{(7)}  & \colhead{(8)} \\ 
      \colhead{Number} & 
       \colhead{Target name} & \colhead{Project code} & \colhead{Beam size} &  \colhead{P.A.} & \colhead{MRS}
       & \colhead{Mosaic} & \colhead{rms} 
     }
    \startdata
\thispagestyle{empty} 
01 & NGC 424          & 2017.1.00236.S & 0\arcsec.11$\times$0\arcsec.09 & 1$^\circ$.4 
   & 1\arcsec.9  & no & 0.025 \\ 
02 & NGC 526A         & 2018.1.00538.S & 0\arcsec.34$\times$0\arcsec.31 & 68$^\circ$.4 
   & 3\arcsec.7  &  no & 0.037 \\ 
03 & NGC 612          & 2015.1.01572.S & 0\arcsec.33$\times$0\arcsec.28 & $-$75$^\circ$.4
   & 3\arcsec.5 &  no & 0.084 \\
04 & NGC 788          & 2018.1.00538.S & 0\arcsec.41$\times$0\arcsec.30 & 65$^\circ$.3
   & 4\arcsec.0 & no & 0.099 \\
05 & NGC 1052         & 2013.1.01225.S & 0\arcsec.25$\times$0\arcsec.19 & 1$^\circ$.1
   & 1\arcsec.9 &  no & 0.048 \\
06 & NGC 1068         & 2016.1.00232.S & 0\arcsec.34$\times$0\arcsec.28  & 1$^\circ$.4
   & 2\arcsec.8  &  no & 0.046 \\
07 & NGC 1125         & 2017.1.00236.S & 0\arcsec.17$\times$0\arcsec.12 & $-$69$^\circ$.7 
   & 2\arcsec.5 &  no & 0.028 \\
08 & NGC 1365         & 2013.1.01161.S & 0\arcsec.92$\times$0\arcsec.69 & 80$^\circ$.3
   & 1\arcsec.9 & yes & 0.618 \\
09 & 2MASX J0521-2521 & 2018.1.00699.S & 0\arcsec.58$\times$0\arcsec.49 & $-$74$^\circ$.4
   & 5\arcsec.9 & no & 0.048 \\
10 & NGC 2110         & 2012.1.00474.S & 0\arcsec.87$\times$0\arcsec.54 & $-$1$^\circ$.3
   & 1\arcsec.3 & no & 0.060 \\
11 & ESO 005-G004     & 2013.1.00623.S & 2\arcsec.02$\times$1\arcsec.20   & 65$^\circ$.9
   & 9\arcsec.3 & no & 0.218 \\
12 & Mrk 1210         & 2017.1.01439.S & 0\arcsec.45$\times$0\arcsec.40 & $-$88$^\circ$.7
   & 4\arcsec.3 & no & 0.074 \\
13 & NGC 3081         & 2015.1.00086.S & 0\arcsec.60$\times$0\arcsec.52 & $-$1$^\circ$.4
   & 5\arcsec.2 & no & 0.055 \\
14 & NGC 3393         & 2016.1.01553.S & 0\arcsec.43$\times$0\arcsec.32 & 56$^\circ$.3
   & 4\arcsec.6 & no & 0.064 \\ 
15 & NGC 4507         & 2018.1.00538.S & 0\arcsec.34$\times$0\arcsec.33 & 63$^\circ$.2
   & 3\arcsec.7 & no & 0.089 \\
16 & NGC 4945         & 2016.1.01279.S & 0\arcsec.50$\times$0\arcsec.48 & $-$3$^\circ$.5
   & 4\arcsec.5 & yes & 0.071 \\
17 & ESO 323-077      & 2018.1.00538.S & 0\arcsec.35$\times$0\arcsec.31 & $-$73$^\circ$.7
   & 3\arcsec.9 & no & 0.051 \\
18 & Mrk 463          & 2013.1.00525.S & 0\arcsec.41$\times$0\arcsec.21 & $-$0$^\circ$.9
   & 1\arcsec.8 & no & 0.080 \\
19 & NGC 5506         & 2017.1.00236.S & 0\arcsec.32$\times$0\arcsec.26 & 1$^\circ$.0
   & 3\arcsec.6 & yes & 0.135 \\
20 & NGC 5643         & 2016.1.00254.S & 0\arcsec.31$\times$0\arcsec.24 & $-$50$^\circ$.0
   & 1\arcsec.7 & no & 0.047 \\
21 & NGC 5728         & 2015.1.00086.S & 0\arcsec.59$\times$0\arcsec.51 & $-$1$^\circ$.3
   & 5\arcsec.0 & no & 0.070 \\
22 & NGC 6240         & 2015.1.00370.S & 0\arcsec.80$\times$0\arcsec.69 & 0$^\circ$.9
   & 5\arcsec.0 & no & 0.247 \\
23 & Fairall 49       & 2017.1.00904.S & 0\arcsec.19$\times$0\arcsec.13 & $-$5$^\circ$.9 
   & 4\arcsec.1 & no & 0.070 \\
24 & IC 5063          & 2016.1.01279.S & 0\arcsec.48$\times$0\arcsec.31 & 54$^\circ$.8
   & 4\arcsec.0 & yes & 0.063 \\
25 & NGC 7130         & 2017.1.00255.S & 0\arcsec.46$\times$0\arcsec.34 & 1$^\circ$.2
   & 3\arcsec.7 & no & 0.029 \\
26 & NGC 7582         & 2016.1.00254.S & 0\arcsec.16$\times$0\arcsec.12 & $-$12$^\circ$.3
   & 1\arcsec.3 & no & 0.028 \\
    \enddata
    \thispagestyle{empty}
\tablenotetext{}{
Notes. (1) Source number. (2) Target name. (3) Project code. (4) Beam size in units of arcsec$\times$arcsec. 
(5) Position angle in degrees. 
(6) Maximum recoverable scale in arcsec. 
(7) Mosaic observation.
(8) Root-mean-square in units of Jy beam$^{-1}$ km s$^{-1}$, derived from a line-free region. 
}\label{tab:alma_data}
\end{deluxetable}
\setlength{\topmargin}{0in}

\section{ALMA data and analysis}\label{sec:alma_data}

The ALMA data were analyzed to produce continuum and CO($J$=2--1) images.
The positions of our AGNs were determined based on the continuum images and were then used to mitigate systematic offsets of coordinates between ALMA and Chandra images by matching Chandra- and ALMA-determined AGN positions (Section~\ref{sec:cxo_data_red}).
This is crucial to reliably compare ALMA and Chandra images.
Additionally, the CO($J$=2--1) images were used to estimate spatially resolved gas masses.

We analyzed the ALMA data using the Common Astronomy Software Applications package \citep[CASA; ][]{McM07}. 
For data reduction and calibration, we adopted the same CASA versions as those used in the quality verification by the ALMA Regional Center. After completing the above tasks, we uniformly used  CASA ver. 5.6.1. for image reconstruction and to derive the physical values. 

To produce images of the continuum emission, we first identified spectral channels free of any emission lines via the \texttt{plotms} task. 
For each target, a single image at a frequency is produced from the defined channels by 
\texttt{tclean} in the multi-frequency synthesis mode, with the Briggs weighting (\texttt{robust} = 0.5). 
For data obtained in a mosaic mode, we set \texttt{gridder} to mosaic. 
Alternatively, to produce a continuum-subtracted CO($J$=2--1) cube, we first determined the continuum level by fitting line-free channels with zeroth- or first-order functions, and subtracted it via the task \texttt{uvcontsub}. The product was then deconvolved using \texttt{tclean} with the same weighting as that used for continuum emission. Considering that velocity resolutions were different among the targets, we combined native bins to achieve resolutions of $\approx$ 10 km s$^{-1}$ for all, except for ESO 005-G009, becauase this source was originally observed at a larger native resolution of $\approx$ 20 km s$^{-1}$. 
Notably, some images were produced by adopting different parameters in the \texttt{tclean} task, such as that for \texttt{robust}, and/or by applying a mask for possible CO($J$=2--1) emitting regions. This is because the default setting yielded severe systematic residuals.
Basic information of the ALMA data is summarized in Table~\ref{tab:alma_data}. 

Significant nuclear continuum emission was detected for all targets except for ESO 005-G004. While we were not able to 
detect nuclear continuum emission of NGC 1365 by 
combining all three data available through  project 2013.1.01161.S, we successfully detected it by considering only the highest-resolution data with a beam size of 0\arcsec.19$\times$0\arcsec.21.
For those with significant detections, we estimated the AGN positions by applying the \texttt{imfit} CASA task to probable AGN continuum components with reference to the literature, NED\footnote{https://ned.ipac.caltech.edu/} and SIMBAD\footnote{http://simbad.u-strasbg.fr/simbad/}. 
For two merger systems, NGC 6240 and Mrk 463, 
we found two bright nuclei for each. 
NGC 6240 showed south and north nuclei, and their positions were close to those of X-ray nuclei \citep[e.g., ][]{Kom03,Pac16,Tre20}. As the southern one was apparently brighter, we adopted it as the main AGN.
While Mrk 463 showed east and west nuclei, known to have intrinsically similar X-ray luminosities \citep[e.g.,][]{Tre18,Yam18}, the eastern nucleus was apparently brighter in both the X-ray and millimeter bands. Thus, the eastern nucleus was adopted as the main AGN. 
Because no significant emission was observed from ESO 005-G004, we adopted an optical AGN position from the NED.
The determined and adopted AGN positions are summarized in Table~\ref{tab:sample}. 

Regarding the nondetection of ESO 005-G004, that may be partly due to its low X-ray luminosity ($\sim$ $10^{42}$ \ergs). If a correlation between X-ray and millimeter luminosities proposed in some studies \citep[e.g., ][]{Beh15,Beh18} is valid for this object, its 100 GHz luminosity may be as low as $\approx 3\times10^{38}$ erg s$^{-1}$. By considering the millimeter spectral shapes of nearby AGNs unveiled by \cite{Ino18}, an expected luminosity at $\sim$ 200 GHz is $\sim 10^{38}$ erg s$^{-1}$. As the achieved sensitivity for continuum emission is $\sim$ $2\times10^{-4}$ Jy beam$^{-1}$ and a luminosity of $10^{38}$ erg s$^{-1}$ corresponds to $\sim$ 5$\sigma$, the nondetection is thus unsurprising.

\begin{figure}
    \centering
    \hspace{-0.7cm}
    \includegraphics[width=7.8cm]{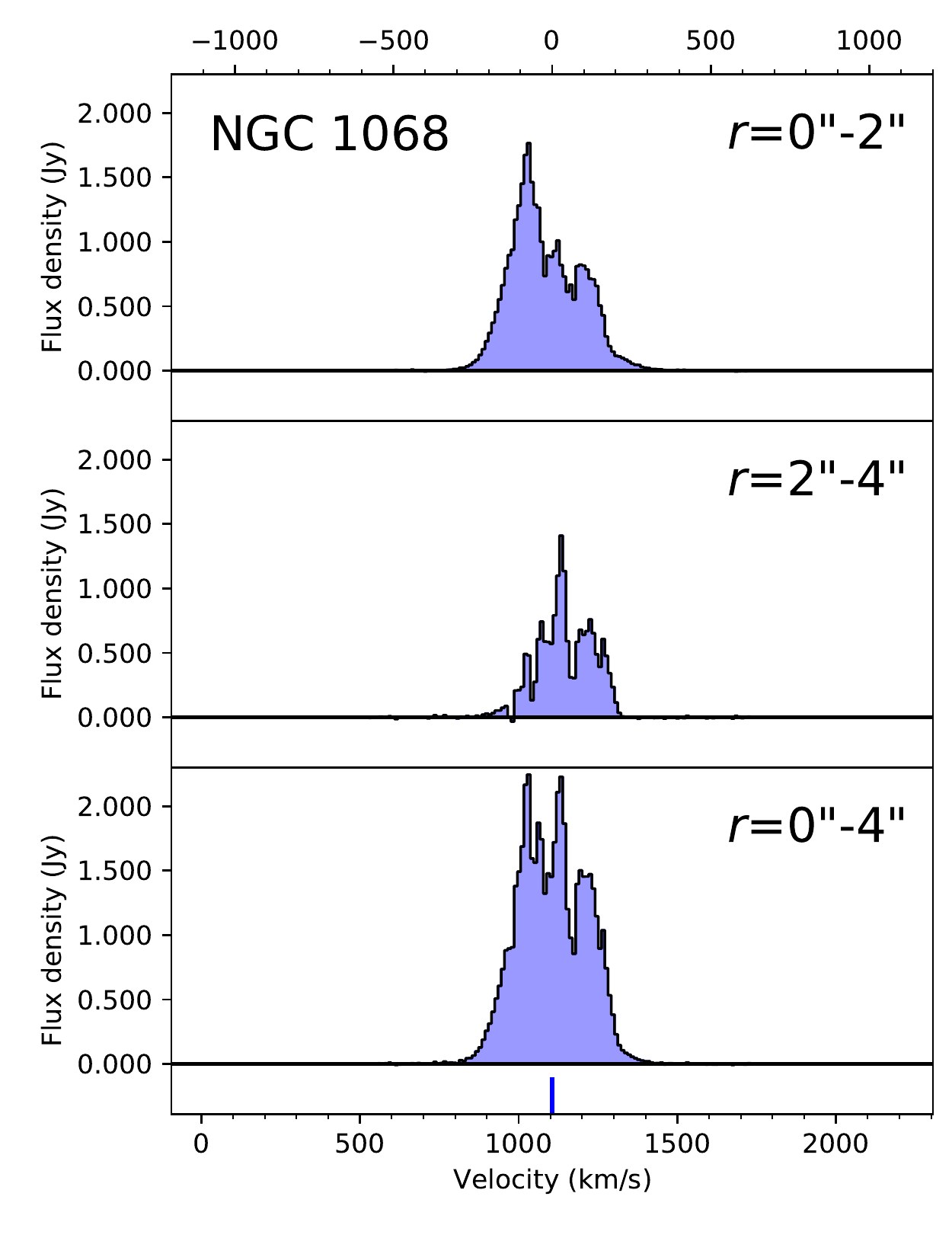}\vspace{-0.5cm}
    \caption{Spectra around CO($J$=2--1) emission taken for NGC 1068. 
    From top to bottom, they were extracted from 
    a circle of radius 2\arcsec, an annulus of radius  2\arcsec--4\arcsec, and a circle of radius 4\arcsec.
    Based on the spectrum from the 4\arcsec\ circle, 
    a channel range to calculate the flux was defined (i.e., blue shaded regions).
    The centroid of the emission was derived based on the range and is indicated by the blue vertical line in the bottom panel. 
    The top and bottom scales denote 
    the relative velocity to the line centroid and 
    the velocity in the optical convention, respectively.
    The latter velocity at the line centroid is thus a CO($J$=2--1)-estimated systematic velocity of the galaxy.
    The spectra of all targets can be seen in Appendix~\ref{app:co_spec}.}
    \label{fig:co_spec} 
\end{figure}

We extracted CO($J$=2--1) spectra from the nuclear ($< $ 2\arcsec) and external ($r = $2\arcsec--4\arcsec) regions centered at the AGN positions.
Significant CO($J$=2--1) emission was found for 22 targets in both regions. As an example, Figure~\ref{fig:co_spec} shows the spectra of NGC\,1068 
for three different extraction regions, and all the spectra are shown in Appendix~\ref{app:co_spec}. 
Throughout this paper, velocities are denoted in the optical convention (i.e., speed of light $\times$ redshift). 
Uniquely, the nuclear spectrum of NGC 1052 was dominated by absorption. Specifically, \cite{Kam20} confirmed that the absorption was concentrated within $\approx$ 0.\arcsec5 of the center, which is consistent with our finding.
The remaining three AGNs (NGC 424, NGC 526A, and Mrk 463) showed no significant emission in either the nuclear or external regions.

We adopted a nonparametric method to derive CO($J$=2--1) fluxes. 
This is because for a non-negligible number of targets, the spectral shapes of the CO($J$=2--1) emission were jagged and could not be reproduced by a few Gaussian functions (Figure~\ref{fig:co_spec} and Appendix~\ref{app:co_spec}). 
We first extracted a spectrum from the 4\arcsec-radius circular region centered at each AGN position, and then defined contiguous channels 
around the systemic velocity where the emission was significant above 1$\sigma$. 
For example, a defined channel range is shown by
blue shaded areas in Figure~\ref{fig:co_spec}. 
Such areas are also denoted in Appendix~\ref{app:co_spec}. 
Finally, we integrated flux densities over the defined channel range. 
Note that NGC 424 was exceptionally treated by using its 2\arcsec-radius circular region, as its CO($J$=2--1) emission was slightly smeared out 
in the 4\arcsec-radius spectrum. 

Table~\ref{tab:co_data} summarizes the CO($J$=2--1) luminosities, calculated according to \cite{Sol05}, and their errors. 
The errors consider the statistical error and conventional 10\% systematic error, originating from uncertainty in the absolute flux calibration. As the nuclear spectrum of NGC 1052 was dominated by the absorption, we listed the sum of the 1-$\sigma$ uncertainty and the absorption strength as the upper limit of the CO($J$=2--1) luminosity. 
In addition to NGC 1052, other AGNs whose CO($J$=2--1) emission is significantly absorbed can be expected. However, the effect would be negligible (i.e., $\sim$ 0.1 dex). To confirm this, we performed a detailed analysis of the spectrum taken from the center of NGC 4945, the details of which are presented in Appendix~\ref{app:co_analysis}. This AGN was selected because 
its nuclear 2\arcsec\ spectrum can be most easily affected by absorption owing to its proximity ($\sim 8$ Mpc), except for NGC 1052. 
This is also suggested from previous reports on clear absorption lines due to HCN($J$=1--0) and HCO$^{+}$($J$=1--0)  \citep[e.g., ][]{Hen18}.

Table~\ref{tab:co_data} lists the molecular gas masses derived from the CO($J$=2--1) luminosities. 
We adopted a CO($J$=1--0)-to-CO($J$=2--1) brightness temperature ratio 
of 1.4 and an $\alpha_{\rm CO}$ conversion factor of 0.8 $M_\odot$(K km s$^{-1}$ pc$^{2}$)$^{-1}$ \citep{San13}. Throughout this paper, we do not discuss the absolute values of the gas masses. Therefore, a different choice of ratio or conversion factor does not have any effect on our discussion and conclusion. 

As the final analysis of the ALMA data, we produced the moment 0 (intensity) maps of the CO($J$=2--1) emission. For each data cube,  
zeroth moments were calculated over a 1000 km s$^{-1}$ range centered 
at the centroid of the emission. 
The centroid was computed based on the spectrum obtained from the 4\arcsec-radius region. 
As NGC 6240 showed very broad CO($J$=2--1) emission, we  considered a 2000 km s$^{-1}$ width. 
All the moment 0 images are shown in Appendix~\ref{app:co_image}.

\begin{figure*}
    \centering \vspace{-0.2cm}
    \includegraphics[width=8.cm]{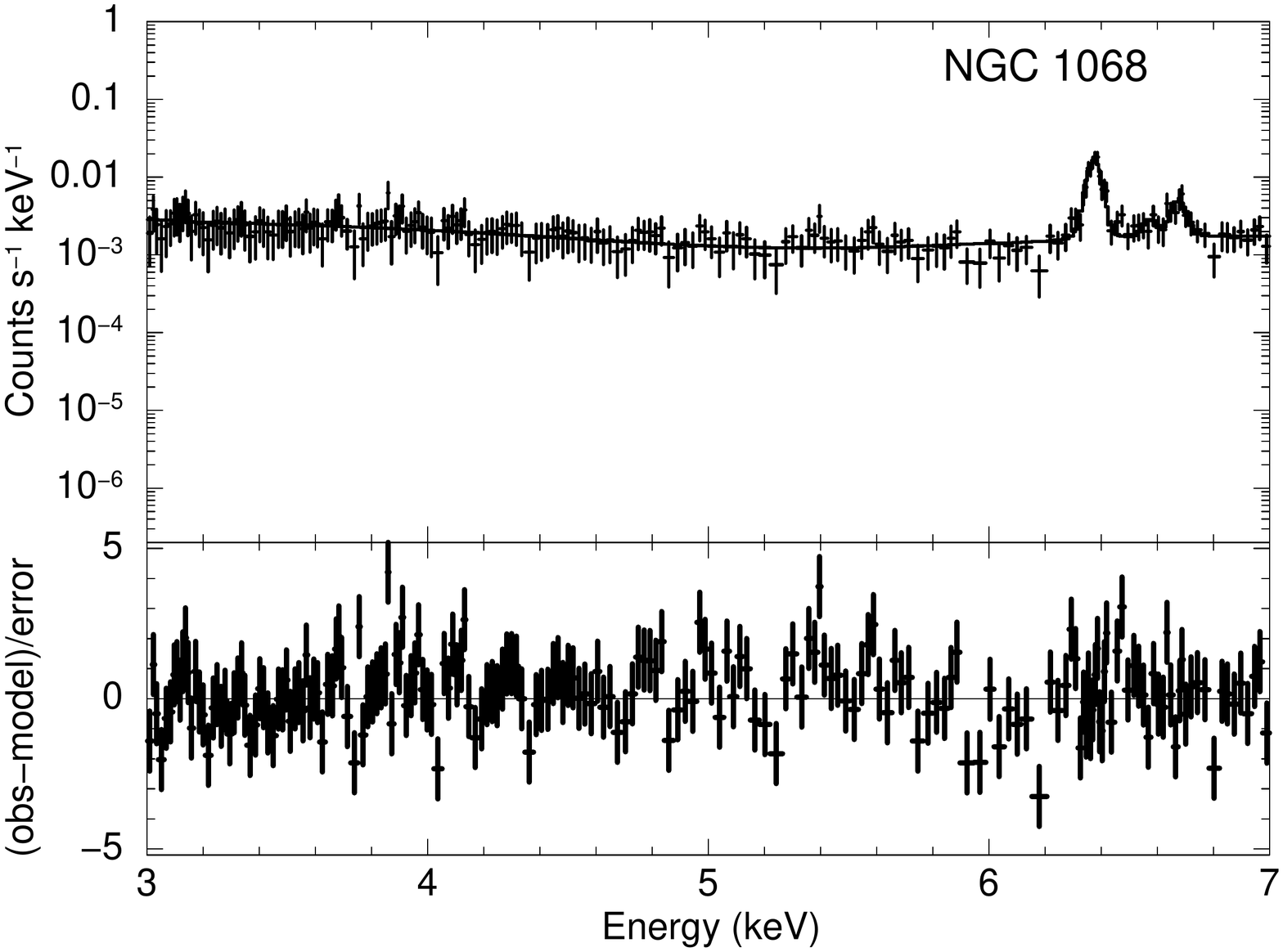} %
    \includegraphics[width=8.cm]{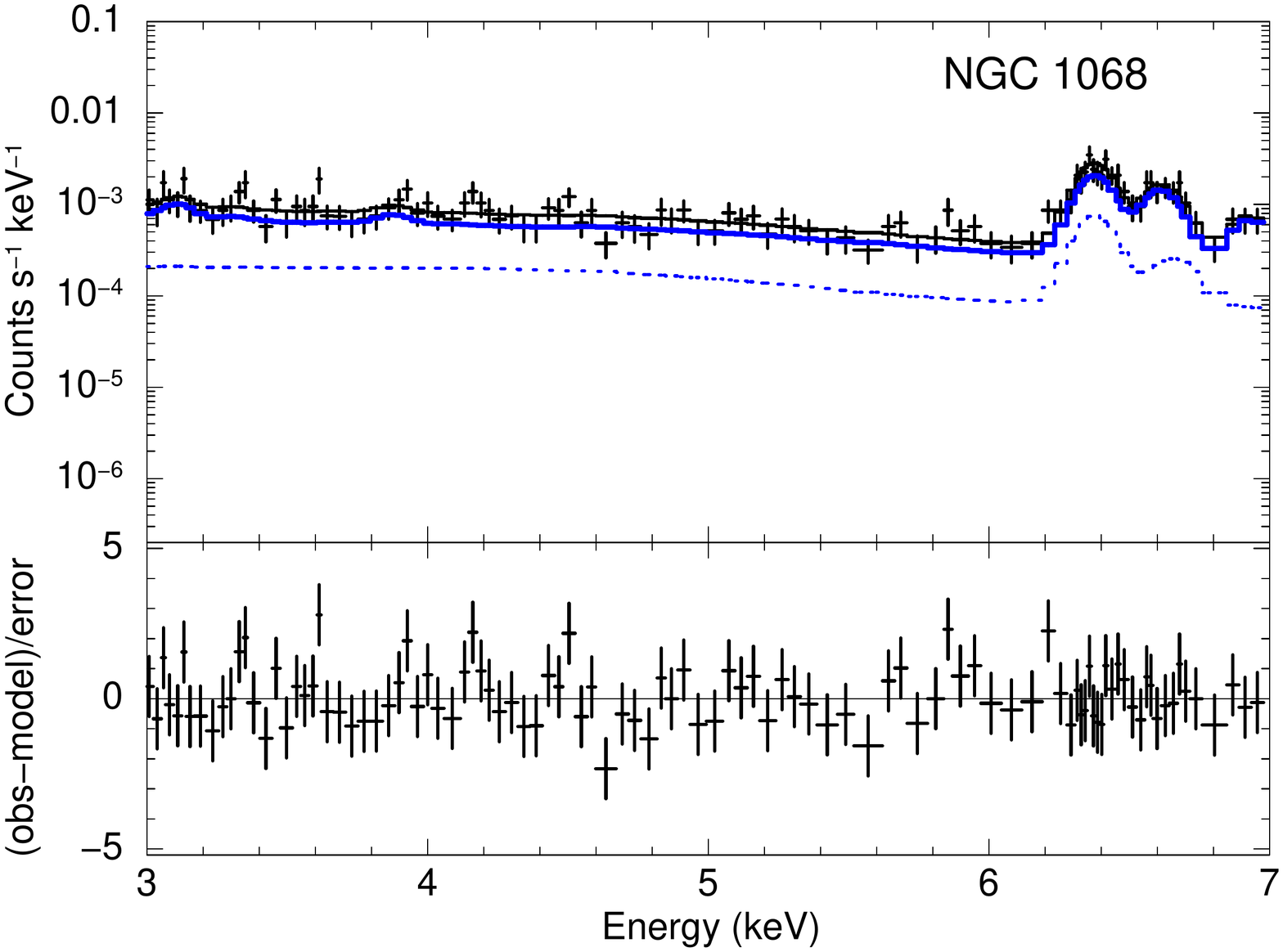} \vspace{-0.3cm} 
    \caption{
    (Left) X-ray spectrum taken from the circle of radius 2\arcsec\ centered at the AGN position of NGC 1068. The upper window shows the observed data (black crosses) and the best-fit model (black solid line), and the lower window shows the residuals. 
    (Right) X-ray spectrum taken from the central annulus of radius 2\arcsec--4\arcsec\ for NGC 1068. 
    The spectrum is reproduced by considering 
    essentially two components: one is from the extracted 2\arcsec--4\arcsec\ region (blue solid line), and the other is from the central point source extended by the PSF (blue dotted line). 
    }
    \label{fig:xspec_ngc_1068} 
\end{figure*}

\setlength{\topmargin}{3.0cm}
\floattable
\renewcommand{\arraystretch}{1.1}
\begin{deluxetable}{cccccccccc}
\thispagestyle{empty}
  \tablecaption{CO($J$=2--1) Line Luminosities and Derived Gas Mass}
  \tablehead{
      \colhead{(1)} & \colhead{(2)} & \colhead{(3)} & \colhead{(4)} & \colhead{(5)} & \colhead{(6)} & \colhead{(7)}  & \colhead{(8)} 
      & \colhead{(9)}  & \colhead{(10)} \\ 
      \colhead{Number} & 
       \colhead{Target name} 
       & \colhead{$L^{\rm nuc}_{\rm CO}$}
       & \colhead{$\Delta L^{\rm nuc}_{\rm CO}$}
       & \colhead{$M^{\rm nuc}_{\rm gas}$}
       & \colhead{$\Delta M^{\rm nuc}_{\rm gas}$}       
       & \colhead{$L^{\rm ext}_{\rm CO}$}
       & \colhead{$\Delta L^{\rm ext}_{\rm CO}$}
       & \colhead{$M^{\rm ext}_{\rm gas}$}
       & \colhead{$\Delta M^{\rm ext}_{\rm gas}$}   
     }
    \startdata
\thispagestyle{empty} 
 01 &       NGC 424 &  6.78 &  0.78 &  7.59 &  0.87 &  $< 0.76$ & ... &  $< 0.85$ &  ... \\ 
 02 &      NGC 526A &  $< 0.43$ & ... &  $< 0.48$ & ... &  4.92 &  0.99 &  5.51 &  1.11 \\ 
 03 &       NGC 612 & 498 & 52 & 557 & 58 & 684 & 73 & 766 & 81 \\ 
 04 &       NGC 788 & 16.8 &  2.1 & 18.8 &  2.3 &  8.3 &  2.1 &  9.3 &  2.4 \\ 
 05 &      NGC 1052 &  $< 0.17$ & ... &  $< 0.19$ & ... &  $< 0.01$ & ... &  $< 0.01$ & ... \\ 
 06 &      NGC 1068 & 33.6 &  3.4 & 37.7 &  3.8 & 17.9 &  1.8 & 20.0 &  2.0 \\ 
 07 &      NGC 1125 & 22.7 &  2.3 & 25.4 &  2.6 &  4.63 &  0.79 &  5.18 &  0.88 \\ 
 08 &      NGC 1365 & 11.3 &  1.2 & 12.7 &  1.3 & 63.0 &  6.3 & 70.6 &  7.1 \\ 
 09 & 2MASX J05210136-2521450 & 2150 & 220 & 2400 & 240 & 234 & 24 & 262 & 27 \\ 
 10 &      NGC 2110 & 11.0 &  1.1 & 12.3 &  1.2 &  9.73 &  0.99 & 10.9 &  1.1 \\ 
 11 &  ESO 005-G004 &  1.94 &  0.20 &  2.17 &  0.22 &  4.03 &  0.41 &  4.51 &  0.46 \\ 
 12 &      Mrk 1210 & 15.6 &  1.6 & 17.4 &  1.8 & 29.4 &  3.0 & 32.9 &  3.4 \\ 
 13 &      NGC 3081 &  9.31 &  0.93 & 10.4 &  1.0 &  2.96 &  0.31 &  3.31 &  0.35 \\ 
 14 &      NGC 3393 & 14.8 &  1.6 & 16.5 &  1.7 & 24.5 &  2.6 & 27.4 &  2.9 \\ 
 15 &      NGC 4507 & 70.4 &  7.1 & 78.8 &  7.9 & 43.1 &  4.6 & 48.2 &  5.1 \\ 
 16 &      NGC 4945 & 56.6 &  5.7 & 63.4 &  6.3 & 69.4 &  6.9 & 77.8 &  7.8 \\ 
 17 &   ESO 323-077 & 238 & 24 & 266 & 27 & 332 & 33 & 372 & 37 \\ 
 18 &       Mrk 463 & 466 & 50 & 522 & 56 &  $< 28$ & ... &  $< 31$ & ... \\ 
 19 &      NGC 5506 & 38.9 &  3.9 & 43.6 &  4.4 & 22.8 &  2.4 & 25.5 &  2.7 \\ 
 20 &      NGC 5643 & 40.2 &  4.0 & 45.0 &  4.5 & 36.8 &  3.7 & 41.2 &  4.1 \\ 
 21 &      NGC 5728 & 36.7 &  3.7 & 41.1 &  4.1 & 35.6 &  3.6 & 39.9 &  4.0 \\ 
 22 &      NGC 6240 & 6540 & 650 & 7320 & 730 & 2510 & 250 & 2810 & 280 \\ 
 23 &    Fairall 49 & 386 & 39 & 432 & 44 & 151 & 17 & 169 & 19 \\ 
 24 &       IC 5063 & 34.0 &  3.4 & 38.1 &  3.8 & 17.6 &  1.9 & 19.7 &  2.1 \\ 
 25 &      NGC 7130 & 936 & 94 & 1050 & 110 & 341 & 34 & 382 & 38 \\ 
 26 &      NGC 7582 & 120 & 12 & 134 & 13 & 60.4 &  6.0 & 67.6 &  6.8 \\ 
  \hline
     \enddata
    \thispagestyle{empty}
\tablenotetext{}{
Notes.--- (1) Source number. (2) Target name. 
(3,4) CO($J$=2--1) luminosity in units of 10$^6$ 
K~km~s$^{-1}$~pc$^{-2}$ derived from the nuclear region ($r <$ 2\arcsec) 
and its 1$\sigma$ error. 
(5,6) Gas mass in units of $10^6~M_\odot$ derived from (3) and its 1$\sigma$ error based on (4,5). 
(7,8,9,10) Same as (3,4,5,6) but for the external region ($r =$ 2\arcsec--4\arcsec).
For those with only upper limits, we list the limits at the 1$\sigma$ level. 
}\label{tab:co_data}
\end{deluxetable}
\setlength{\topmargin}{0in}

\section{Chandra data and analysis}\label{sec:cxo_data}

\subsection{Data Reduction}\label{sec:cxo_data_red}

We analyzed Chandra data with two aims: (i) to measure spatially resolved Fe-K$\alpha$ fluxes and (ii) to map the Fe-K$\alpha$ emission. Throughout the analyses of the Chandra data, the CIAO 4.12 package was used along with CALDB version 4.9.1. 
Raw data obtained from ChaSeR were reprocessed in a standard manner using \texttt{chandra\_repro}. Then, 
we filtered events in a period when a flare of background counts, likely due to cosmic rays, was observed. Some of our targets were bright enough to cause pileup, making it difficult to recover the intrinsic flux.
Thus, we discarded data where the pileup with a fraction of 5\% extended over 2\arcsec\ around a nuclear X-ray peak. The fraction was estimated from the observed counts per read time using the CIAO task \texttt{pileup\_map}. 
The 2\arcsec\ radius was selected so that we could extract the nuclear spectra that were unaffected by the photon pileup. 

As first mentioned in Section~\ref{sec:sample}, 
we defined the nuclear region of 2\arcsec\ and the external region of 2\arcsec--4\arcsec. 
The 2\arcsec\ outer radius of the nuclear region was selected as 
a significant fraction of the Chandra data were affected by the photon pileup within $\sim$ 1\arcsec, but the effect did not frequently extend over 2\arcsec. 
Thus, 2\arcsec\ was a reasonable optimization that allowed us to extract nuclear spectra from either a circular 2\arcsec\ region or an annular 1\arcsec--2\arcsec\ region, depending on the degree of the pileup effect. 
With this determination, we measured the nuclear and external CO($J$=2--1) fluxes discussed in Section~\ref{sec:alma_data}. 

As the final data reprocessing step, we corrected the systematic offsets between the coordinates of the Chandra and ALMA images. 
In each Chandra image, we first calculated the centroid of the nuclear X-ray photons in the 3--7\,keV band, where nuclear emission 
was likely to be dominant. 
Then, we modified the coordinates of the Chandra image such that the X-ray centroid matched the AGN position determined from the ALMA continuum image. Notably, the positional offsets before the adjustment were generally less than 0\arcsec.5. 

\subsection{Spatially resolved X-ray Spectroscopy}\label{sec:cxo_fe}

We derived Fe-K$\alpha$ fluxes in both the nuclear ($r <$ 2\arcsec) and external ($r$ = 2\arcsec--4\arcsec) regions. 
For the nongrating data, we estimated the fluxes simply based on the spectra extracted from the two regions. If the pileup effect was severe within 1\arcsec, we adopted an annulus of 1\arcsec--2\arcsec\ instead of the circular region. 
If there were more than one observation, we combined the spectra into a single spectrum. Response files for the nuclear and external spectra were produced by assuming a point source and an extended source, respectively.
For the grating data, we used 1st-order HEG spectra to 
measure nuclear Fe-K$\alpha$ fluxes because they are not affected significantly by the pile-up. Notably, the contamination from extended or surrounding sources is expected to be not severe given their creation process\footnote{Diffraction distance depends on the photon energy, and thus, for the position of a target, the absolute coordinates of diffracted photons on the CCDs are determined. Accordingly, in each CCD region, the photons from the target can be selectively obtained, whereas those from surrounding sources with different energies are rejected. Then, a spectrum is obtained by compiling the target photons.}. 
For the external components, we extracted spectra from annuli of 2\arcsec--4\arcsec\, based on the 0th-order image. 
If multiple spectra were available for each target, they were merged into one. 
All the extracted spectra were binned to have at least one count per bin. 
In almost the same manner as adopted for the imaging observations, their response files were created. 

As an example, Figure~\ref{fig:xspec_ngc_1068} shows the spectra of NGC 1068 taken from its nuclear and external regions together with the best-fit models, detailed in the following paragraphs. Appendix~\ref{app:fig:xspec} 
summarizes the corresponding figures for all our targets. 

The nuclear spectra in the 3--7\,keV band were fitted with C-statistics. 
This energy range was selected because we were not interested in soft X-ray components, such as thermal emission due to stellar activity, a soft-X-ray excess linked to the accretion disk, scattered nuclear light, and emission from hot extended cluster gas.
When both imaging and grating data were available for a single object, we fitted the two spectra jointly without merging them.
As our fiducial model, we employed a phenomenological model of \texttt{constant*(zTBabs*zpowerlw + zgauss + zgauss)} in XSPEC terminology. 
When imaging and grating data were available, the \texttt{constant} model was allowed to vary by 10\% to account for any calibration uncertainty in the normalization of the different spectra\footnote{https://cxc.harvard.edu/proposer/POG/html/chap8.html}.
Then, we considered an absorbed transmitted power-law emission from the central X-ray corona as \texttt{zTBabs*zpowerlw}, and two Gaussian functions (i.e., \texttt{zgauss + zgauss}) to reproduce the Fe-K$\alpha$ and Fe-K$\beta$ lines at 6.40 and 7.06\,keV. 
Given the low photon statistics of the majority of the spectra, we fixed the power-law photon indices at those determined based on broadband X-ray spectra ($\approx$ 0.5--200\,keV) by \cite{Ric17c}, except for NGC 612. 
The photon index of NGC 612 was estimated to be 1.07 in the hard X-ray catalog \citep{Ric17c}, which is particularly low and could be the result of heavy obscuration and low signal-to-noise ratio. Thus, we adopted a more likely value of $\Gamma =$ 1.71, which was derived based on a broadband spectrum in \cite{Urs18}.
As long as we had a grating spectrum, we retained the linewidth of Fe-K$\alpha$ as a free parameter and tied it to the width of Fe-K$\beta$. 
The other remaining free parameters were 
the absorbing hydrogen column density and 
the normalizations of the power-law and emission lines. 
The atomic abundance was set to the solar value determined by \cite{Wil00}. 
If non-negligible residuals were found after fitting the fiducial model, according to their shape, we considered additional components, such as line emission from highly ionized iron atoms (e.g., Fe~{\sc xxv} and Fe~{\sc xxvi}), an unabsorbed power-law component representing scattered emission, and high-temperature ($\sim$ 1\,keV) thermal emission. 
If any of these components improved the fit at the 90\% confidence level (i.e., $\Delta C >$ 2.71), we incorporated it into the model. 

The external spectra in the 3--7\,keV band were fitted by adopting C-statistics. As our fiducial model for the emission from  external regions, we considered \texttt{constant*(pexrav + zgauss)}. The first \texttt{constant} was employed to absorb the calibration uncertainty between the grating and imaging spectra found in the fits to the nuclear spectra and was accordingly fixed. If either the grating or imaging spectrum was available, it was removed. The last two terms account for a reflection continuum \citep{Mag95} and Fe-K$\alpha$ emission. 
In addition to these components, we considered the contributions from the nuclear sources determined previously. These contributions are usually not negligible. 
As with the analyses for the nuclear spectra, we added certain components, provided that they improved the fits. 

The derived Fe-K$\alpha$ luminosities in both the nuclear and external regions ($L^{\rm nuc}_{\rm Fe-K\alpha}$ and $L^{\rm ext}_{\rm Fe-K\alpha}$) are summarized in Table~\ref{tab:fe_data}. 
As shown in the table, we detected Fe-K$\alpha$ emission above 1$\sigma$ 
from all but two (Fairall 49 and 2MASX J0521-2521) on the nuclear scale ($r <$ 2\arcsec) and nine objects (i.e., NGC 526A, NGC1068, NGC 2110, NGC 3393, NGC 4945, NGC 5643, NGC 6240, NGC 7130, and NGC 7582) on the external scale ($r =$ 2\arcsec--4\arcsec). 
Using 20--50\,keV luminosities ($L_{\rm 20-50~keV}$) determined based on BAT spectra averaged over 70 months 
\citep{Ric17c,Urs18}, we calculated the ratios of $L^{\rm nuc}_{\rm Fe-K\alpha}/L_{\rm 20-50~keV}$ and $L^{\rm ext}_{\rm Fe-K\alpha}/L_{\rm 20-50~keV}$. 
These can be used as a proxy of the solid angle of 
gas irradiated by hard X-ray emission. 
Compared with the Fe-K$\alpha$ equivalent width (EW), the ratios are less affected by the absorption of underlying continuum emission. 
Additionally, because we adopt luminosities derived from the long-term-averaged spectra, they are less subject to short-time variability of continuum emission that is not associated with the narrow Fe-K$\alpha$ emission line.
Here, we note two clear merging systems of NGC 6240 and Mrk 463 in the X-ray band \citep[e.g., ][]{Kom03,Bia08}. 
The central 2\arcsec\ region of NGC 6240 encompasses bright double nuclei. Thus, the Fe-K$\alpha$ luminosity includes both contributions. As the 20--50~keV luminosity was also calculated as the sum of the nuclei \citep{Ric17c}, the line-to-continuum luminosity ratio was calculated consistently. 
Regarding Mrk 463, the second nucleus is located at $\sim$ 4\arcsec\ from the primary nucleus. Thus, the Fe-K$\alpha$ 
luminosity on the external scale was derived from the spectrum between 2\arcsec\ and 4\arcsec, from which a 1\arcsec\ circular region around the second nucleus was excluded.

\floattable
\renewcommand{\arraystretch}{1.1}
\begin{deluxetable}{ccccccccc}
\thispagestyle{empty}
  \tablecaption{Fe-K$\alpha$ Luminosities} 
  \tablehead{
      \colhead{(1)} & \colhead{(2)} & \colhead{(3)} & \colhead{(4)} & \colhead{(5)} & \colhead{(6)} & \colhead{(7)}  & \colhead{(8)} \\ 
      \colhead{Number} & 
       \colhead{Target name} 
       & \colhead{$L^{\rm nuc}_{\rm Fe-K\alpha}$} 
       & \colhead{$\Delta L^{\rm nuc}_{\rm Fe,low}$}
       & \colhead{$\Delta L^{\rm nuc}_{\rm Fe,upp}$}
       & \colhead{$L^{\rm ext}_{\rm Fe-K\alpha}$}
       & \colhead{$\Delta L^{\rm ext}_{\rm Fe,low}$}
       & \colhead{$\Delta L^{\rm ext}_{\rm Fe,upp}$}
     }
    \startdata
\thispagestyle{empty} 
 01 &       NGC 424 &    4.74 &    0.78 &    0.87 &  $<7.0$ &  ... &  ... \\ 
 02 &      NGC 526A &    19.1 &     7.8 &     8.2 &      51 &   45 &   56 \\ 
 03 &       NGC 612 &    13.3 &     2.7 &     2.9 &   $<34$ &  ... &  ... \\ 
 04 &       NGC 788 &     3.1 &     1.5 &     1.6 &   $<62$ &  ... &  ... \\ 
 05 &      NGC 1052 &    0.55 &    0.18 &    0.19 & $<0.48$ &  ... &  ... \\ 
 06 &      NGC 1068 &    0.86 &    0.05 &    0.05 &    5.89 & 0.71 & 0.77 \\ 
 07 &      NGC 1125 &    1.47 &    0.26 &    0.28 &  $<2.0$ &  ... &  ... \\ 
 08 &      NGC 1365 &    0.50 &    0.09 &    0.10 & $<0.15$ &  ... &  ... \\ 
 09 & 2MASX J05210136-2521450 & $<28$ & ... & ... &   $<94$ &  ... &  ... \\ 
 10 &      NGC 2110 &    8.82 &    1.84 &    0.93 &    46.0 &  8.9 &  8.3 \\ 
 11 &  ESO 005-G004 &    0.96 &    0.14 &    0.16 & $<3.70$ &  ... &  ... \\ 
 12 &      Mrk 1210 &    14.9 &     4.5 &     5.0 &  $<5.2$ &  ... &  ... \\ 
 13 &      NGC 3081 &     6.0 &     1.7 &     1.9 &  $<1.2$ &  ... &  ... \\ 
 14 &      NGC 3393 &    1.61 &    0.22 &    0.23 &    12.8 &  3.9 &  4.5 \\ 
 15 &      NGC 4507 &    18.9 &     1.3 &     1.3 & $<10.5$ &  ... &  ... \\ 
 16 &      NGC 4945 &    0.17 &    0.01 &    0.01 &    2.95 & 0.29 & 0.29 \\ 
 17 &   ESO 323-077 &     5.7 &     1.1 &     1.2 &   $<19$ &  ... &  ... \\ 
 18 &       Mrk 463 &     8.8 &     3.2 &     3.6 &  $<100$ &  ... &  ... \\ 
 19 &      NGC 5506 &    4.74 &    1.15 &    0.79 &  $<2.2$ &  ... &  ... \\ 
 20 &      NGC 5643 &    0.52 &    0.03 &    0.04 &    1.57 & 0.76 & 0.88 \\ 
 21 &      NGC 5728 &    2.21 &    0.43 &    0.47 &  $<3.8$ &  ... & ...  \\ 
 22 &      NGC 6240 &    18.2 &     1.1 &     1.3 &      98 &   22 &   24 \\ 
 23 &    Fairall 49 & $<4.4$ &      ... &     ... &   $<26$ &  ... &  ... \\ 
 24 &       IC 5063 &     9.1 &     1.7 &     1.7 &  $<1.8$ &  ... &  ... \\ 
 25 &      NGC 7130 &    1.71 &    0.43 &    0.50 &      12 &   11 &   18 \\ 
 26 &      NGC 7582 &    1.02 &    0.24 &    0.17 &     1.6 &  1.4 &  1.7 \\ 
     \enddata
    \thispagestyle{empty}
\tablenotetext{}{
Notes.--- (1) Source number. (2) Target name. 
(3),(4), and (5) Fe-K$\alpha$ luminosities and its lower and upper 1$\sigma$ errors in units of $10^{40}$ erg s$^{-1}$, measured from the nuclear spectra.
(6),(7), and (8) Fe-K$\alpha$ luminosities and its lower and upper 1$\sigma$ errors
in units of $10^{38}$ erg s$^{-1}$, measured from the external spectra. 
For those with only upper limits, we list the limits at the 1$\sigma$ level. 
}\label{tab:fe_data}
\end{deluxetable}
\setlength{\topmargin}{0in}

\begin{figure}
    \centering
    \includegraphics[width=8.5cm]{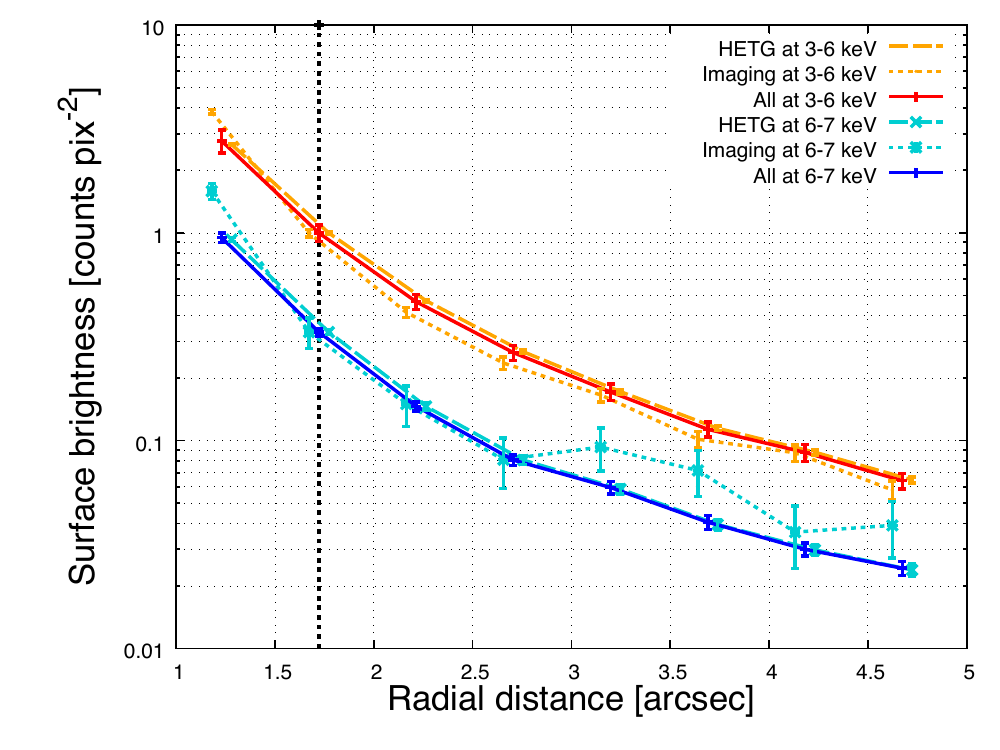}
    \caption{Radial profiles of surface brightness (counts pix$^{-2}$) as a function of radial distance (arcsec) for different energy bands (3--6 and 6--7\,keV) and types of data (imaging and grating data). The profiles at 6--7\,keV are shifted downward by 0.5 dex for clarity. 
    In addtion, those solely from the HEG and imaging data are slightly shifted around their averaged profiles in the horizontal direction. Note that the left-most points correspond to those at 3.5 pixel from the central pixels.}
    \label{fig:cxo_psfs}
\end{figure}

\subsection{X-ray Imaging}\label{sec:xray_image}

The Chandra data were further used to reveal
the distributions of extended Fe-K$\alpha$ emission at $\approx$ 0\arcsec.5, which are comparable to those of the ALMA data. 
For this analysis, we first empirically constructed point spread functions (PSFs) in the 3--6 and 6--7 \,keV bands (Section~\ref{sec:cxo_psfs}). Then, 
we modeled point source images based on the PSFs and subtracted them from the observed ones (Section~\ref{sec:xray_sub_image}). 

We note that although one popular method 
to obtain a PSF is to perform a Monte Carlo simulation with MARX, we found that for bright objects, there seems to be a significant difference between the simulated and observed profiles. This discrepancy cannot be ignored in bright objects. Therefore, we finally decided to adopt an empirical approach.

\subsubsection{Construction of PSF models of Chandra}\label{sec:cxo_psfs}

To construct the PSFs, we first compiled Chandra data of bright galactic X-ray sources and then stacked their observed images in the 3--6 and 6--7\,keV bands.
We searched ChaSeR for ACIS data of galactic X-ray sources listed 
in all-sky catalogs created by Monitor of All-sky X-ray Image \citep[MAXI;][]{Kaw18,Hor18}. 
The galactic X-ray sources included those classified as ALGOL Star, CV, Dwarf Nova, high-mass X-ray binary, low-mass X-ray binary, Nova, Pulsar, RS CVn Star, Rotationally variable Star, Spectroscopic binary, Star, Symbiotic Star, Variable Star, and X-ray binary. 
The catalogs were suitable because the energy band of 4--10\,keV covered the band of interest and provided one of the largest samples of bright X-ray sources. 
At this point, we excluded objects known to have bright X-ray extended emission, such as an object with a highly extended jet and a system consisting of double X-ray nuclei (e.g., Circinus X-1 and CH Cyg).  
In addition, the effect of X-ray halos produced by dust scattering \citep[e.g., ][]{Pre95} is likely negligible, as discussed in the last paragraph of this section.
Even after these selections and considerations, there may have remained X-ray sources that showed emission resolvable by Chandra but were difficult to identify by visual inspection. 
Although such objects expand our empirical PSFs, the expanded PSFs avoid the overestimation of extended emission, enabling conservative discussion of the presence of extended emission. 

For PSF construction, only ACIS-S imaging and ACIS-S/HETG data were selected to match the data of our original targets. 
We excluded data affected by photon pileup. 
Furthermore, to avoid noisy and nonmeaningful data, we only used 10 data with the highest net-source counts at 3--7\,keV within 5\arcsec\ around the targeted X-ray sources. The 5\arcsec\ radius was selected to cover the outer 4\arcsec\ radii of the external regions. 
To verify the agreement between the PSFs of the grating and nongrating data, we eventually analyzed 10 data for each observation mode, that is, 20 data in total. 
Table~\ref{tab:psf_data} summarizes the information of the 20 data.

The selected data were reprocessed in the same manner as those used for our main targets. Then, for each data, we defined multiple concentric annuli with 0\arcsec.49 width (i.e., the ACIS-S native resolution) centered at the AGN position, producing the radial profiles of surface brightness in units of counts pix$^{-2}$ at 3--6 and 6--7\,keV. 
In each dataset of the imaging and the grating, we calculated 
weighted means of the profiles of the two bands. Figure~\ref{fig:cxo_psfs} shows 
the obtained profiles of the surface brightness normalized to 1 at 3.5 pixels from the centers. 
At pixels below 1\arcsec.5, the grating profiles show slightly lower surface brightness than the imaging profiles. This is because of the stronger photon pileup in the grating observations. The grating observation is less affected by the pileup, but the stronger pileup was due to the observations of the very bright sources.
On the contrary, at larger radii beyond $\sim$ 1\arcsec.5, the profiles 
obtained from the imaging and grating observations were in agreement in both energy bands. 
Although we can see a bump for the imaging 6--7\,keV data in the range of 3\arcsec--5\arcsec, which can be ascribed to poor count statistics, it is compatible with the grating data within 2$\sigma$.
Finally, we derived a single PSF model for each energy band by taking the average of the imaging and grating profiles. 
In the following analysis, however, each PSF within a radius of 1\arcsec.5 was ignored, given that the pileup likely disturbed the PSF. A similar empirical method was adopted by Andonie et al. (in prep.) to produce PSF models.

\begin{figure*}
    \centering
    \includegraphics[width=7.9cm]{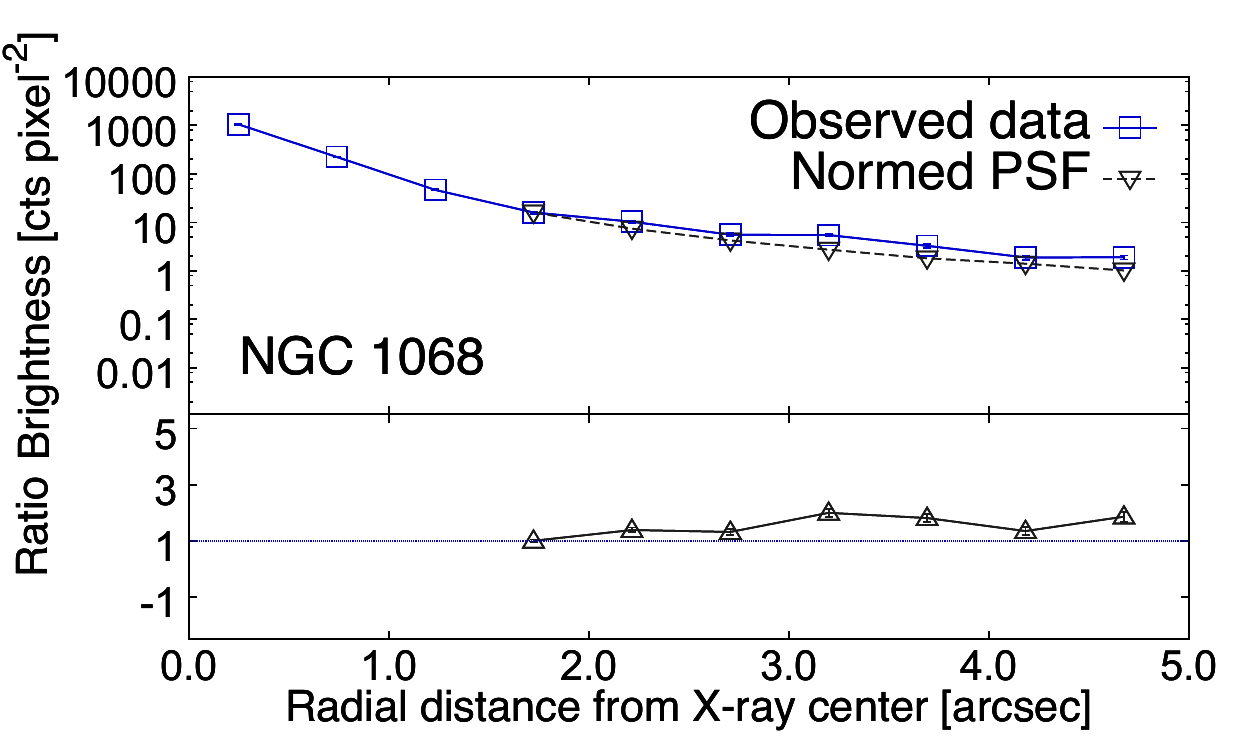}
    \includegraphics[width=7.9cm]{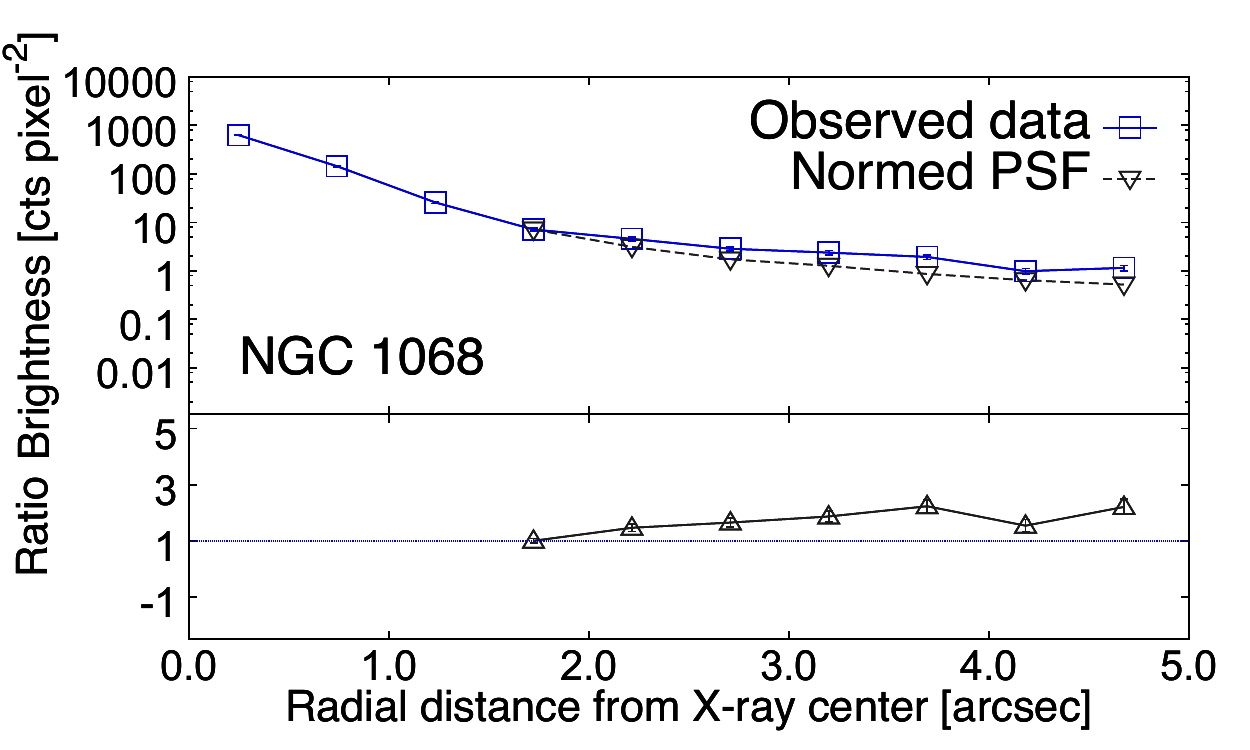}\\
    \includegraphics[width=7.9cm]{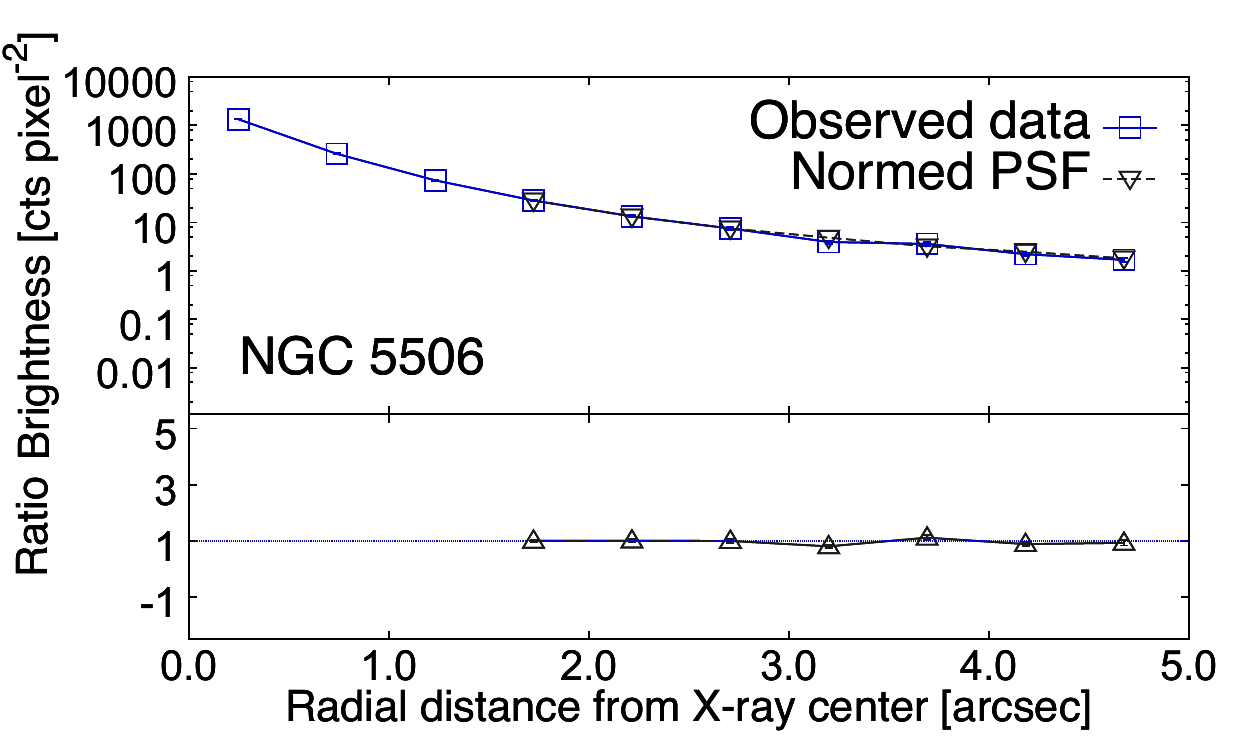}
    \includegraphics[width=7.9cm]{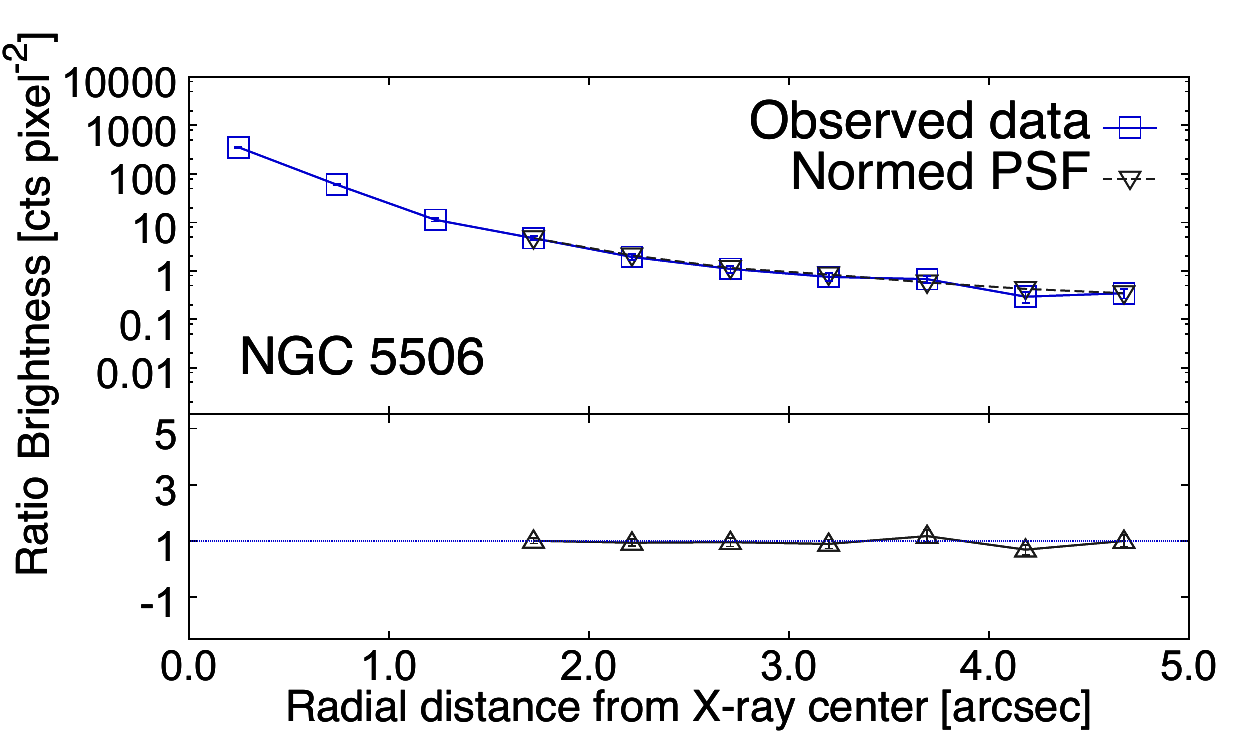}
    \caption{
    Top: Comparisons of observed profiles 
    and those modeled by assuming a point source at 3--6\,keV (left) and 6--7\,keV (right) for NGC 1068. The ratios are shown in the lower panels. 
    Bottom: Same figures for NGC 5506, one of the brightest AGNs in our sample.
    }
    \label{fig:cxo_profiles} 
\end{figure*}

\begin{figure*}
    \centering
    \includegraphics[width=5.5cm]{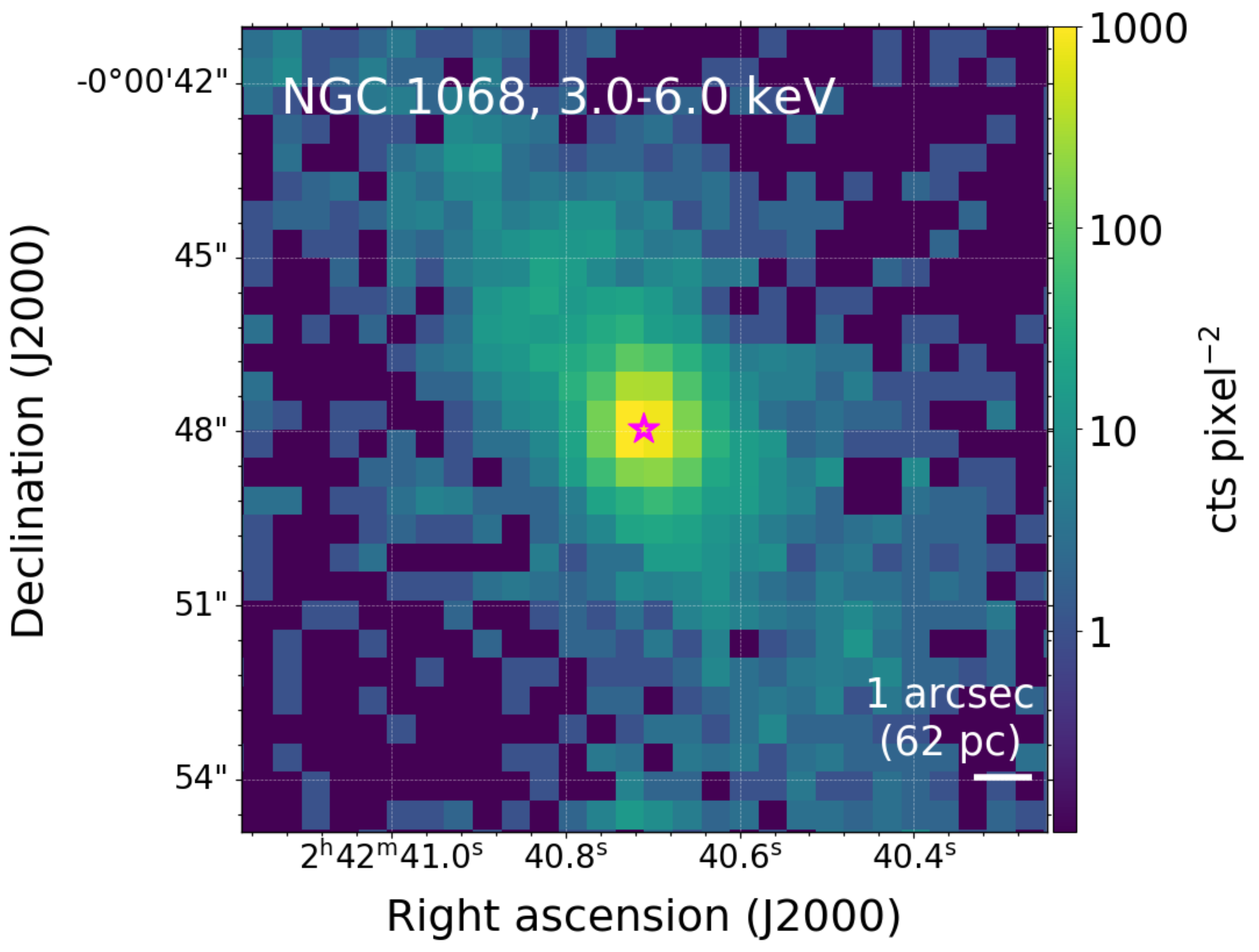}    
    \includegraphics[width=5.5cm]{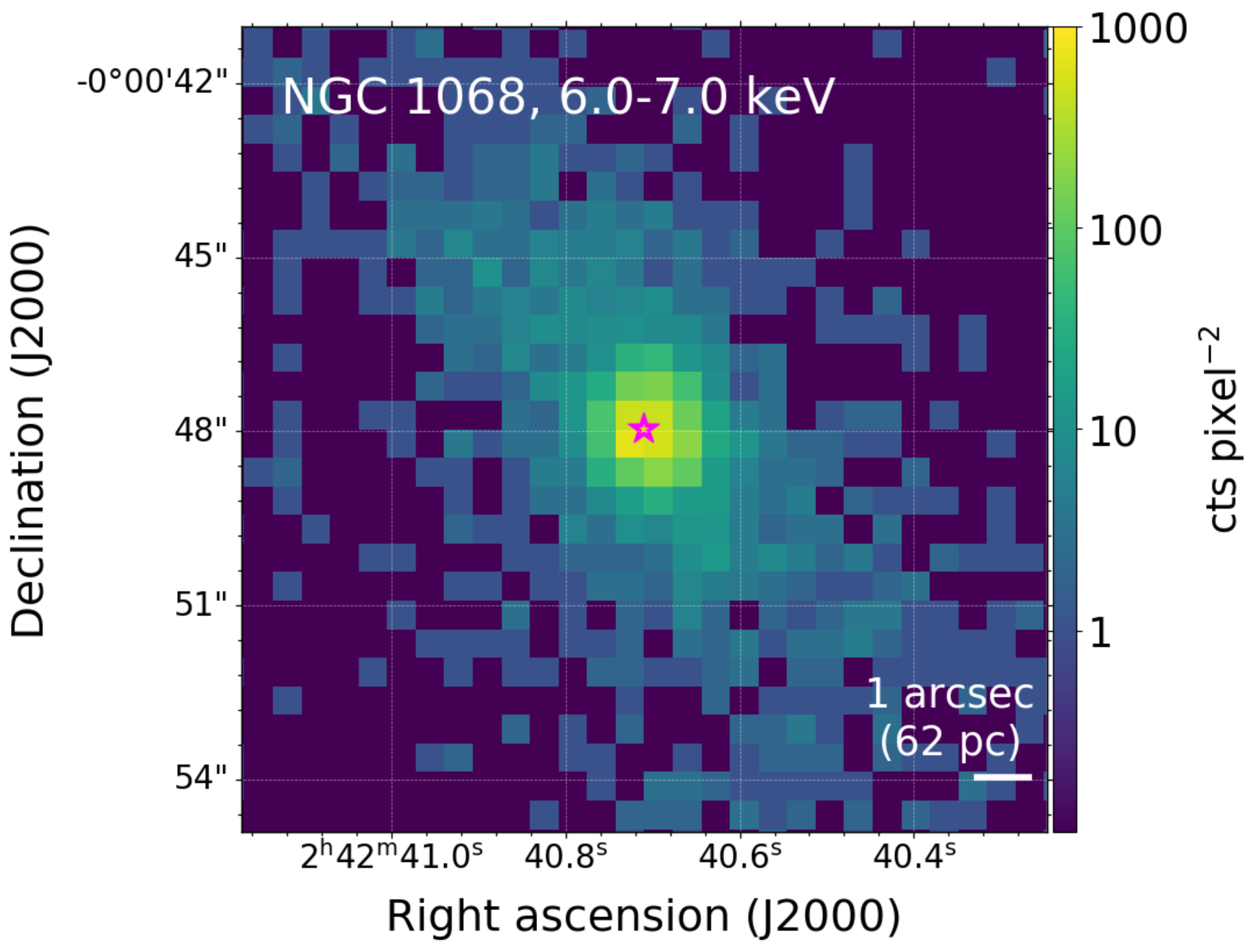}
    \includegraphics[width=5.35cm]{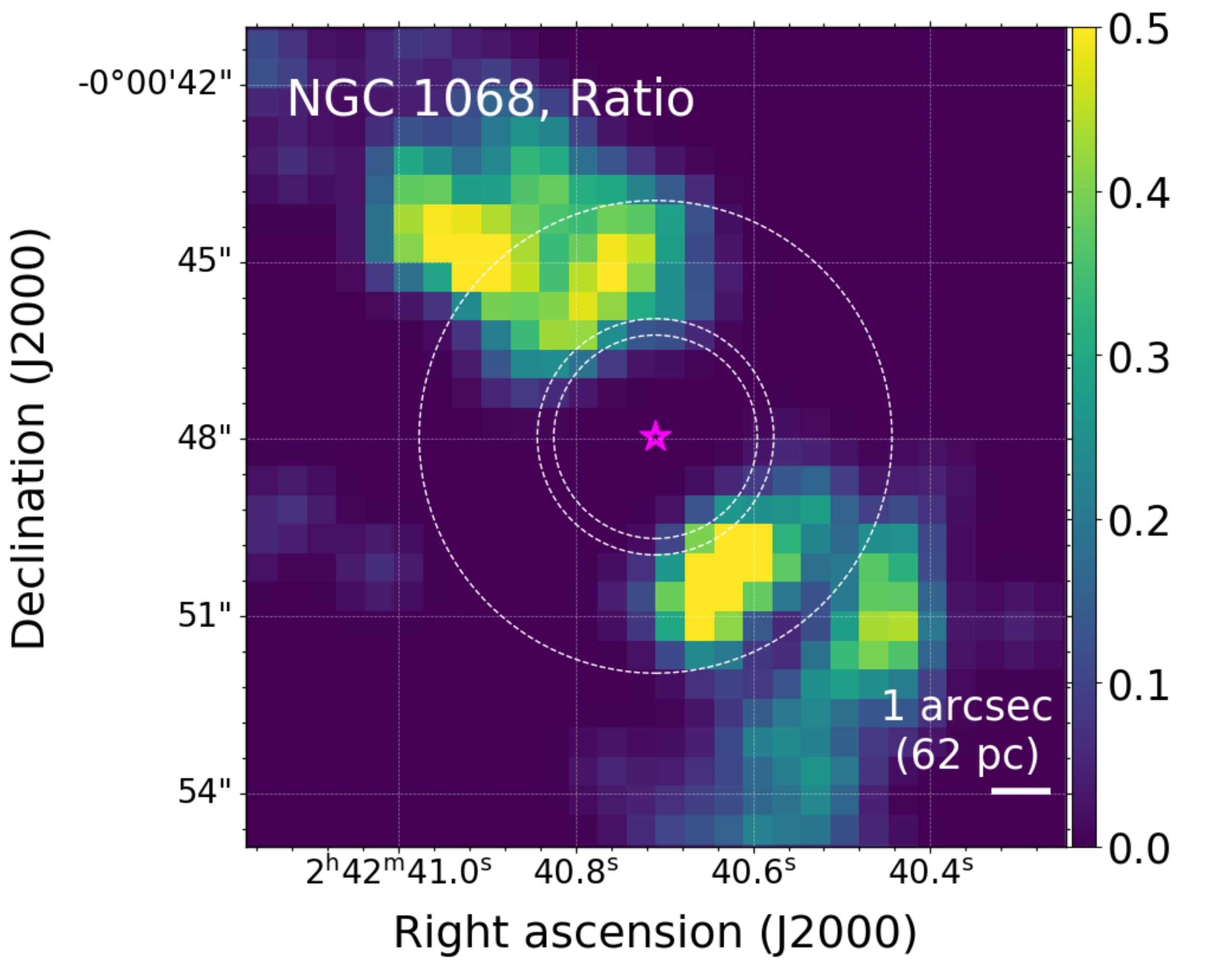}\\
    \includegraphics[width=5.5cm]{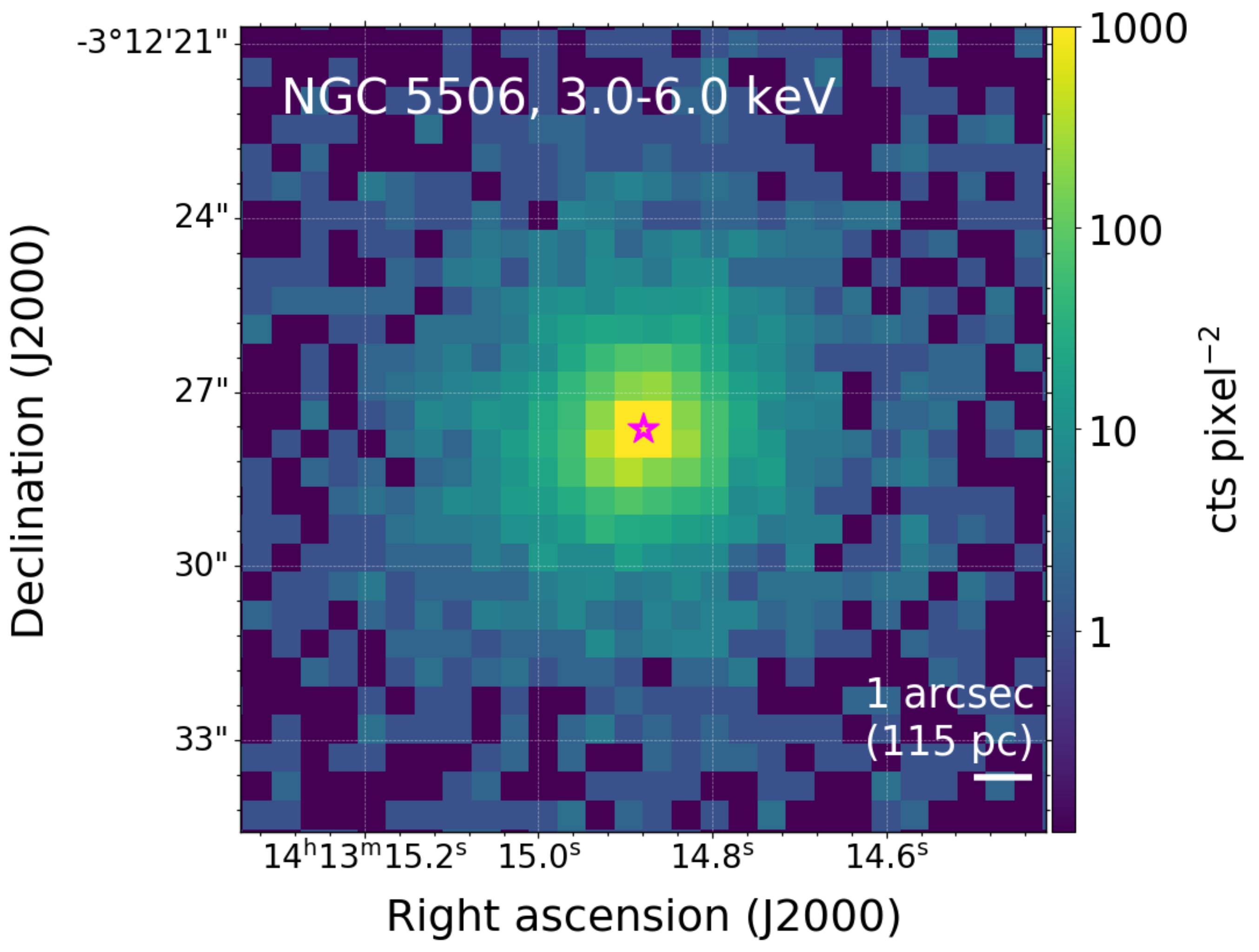}    
    \includegraphics[width=5.5cm]{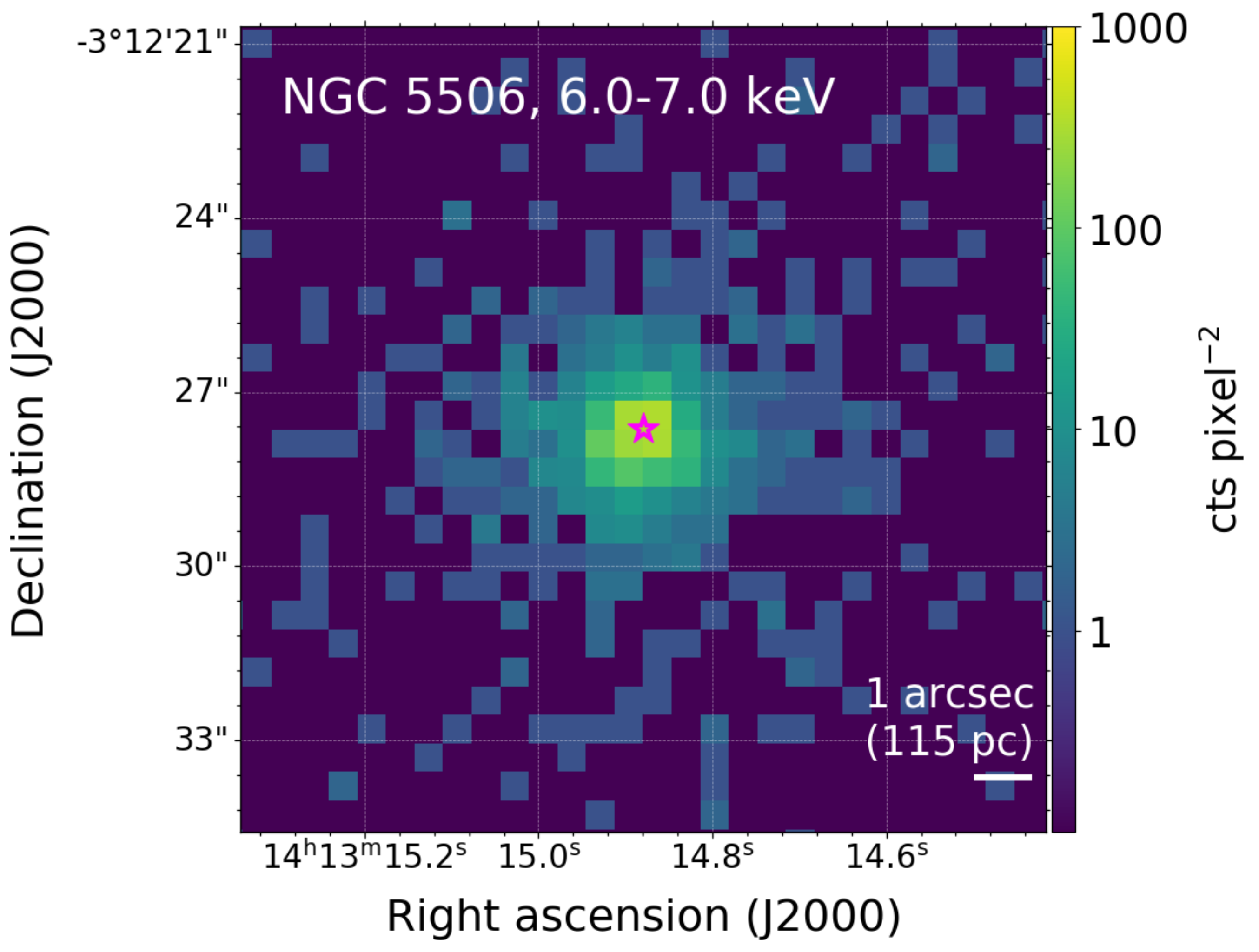}
    \includegraphics[width=5.35cm]{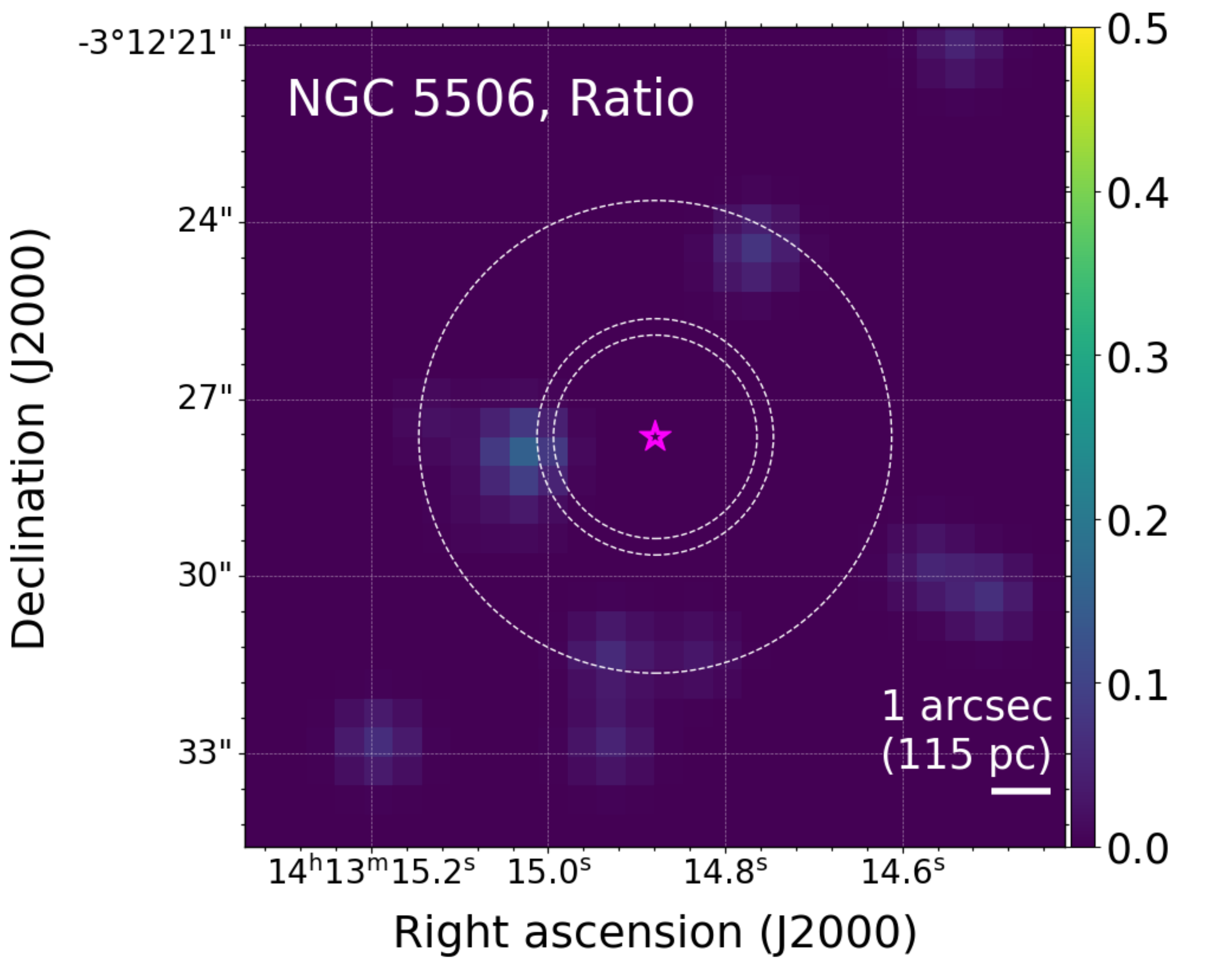}
    \caption{From left to right, 3--6 and 6--7\,keV images and the ratios of 6--7 and 3--6\,keV images from which nuclear point source emission is subtracted. 
    The color bars in the 1st and 2nd columns indicate 
    counts per pixel$^2$, and those in the 3rd one indicate 
    count ratios.
    North is up and east is left.
    The upper and lower panels show images of NGC 1068 and NGC 5506, respectively. The central magenta stars indicate the AGN positions. 
    Circles from the inner to outer ones in the right panels denote the 3.5 pixel at which the PSF is normalized to model a nuclear point source, a 2\arcsec\ radius, and a 4\arcsec\ radius. Note that no ratios in the central $\sim$ 2\arcsec\ are shown because the modeled emission from the central point sources are 
    subtracted. 
    } 
    \label{fig:cxo_images}
\end{figure*}

It can be expected that the PSF depends on the energy, but the effect is negligible. This is inferred from the fact that the 3--6\,keV profiles, dominated by softer photons, are close to those at 6--7\,keV. 
A Chandra team also reported a supportive result that showed almost the same profiles between the 4.5 and 6.4\,keV bands\footnote{
https://cxc.cfa.harvard.edu/proposer/POG/html/
HRMA.html\#tth\_sEc4.2.3}. 

Finally, we suggest that the impact of dust-scattering X-ray halos on the PSF model is insignificant. 
For this suggestion, we compared PSF models created 
for two subsamples consisting of objects with low column densities below $10^{21}$  cm$^{-2}$ and those with high column densities above $10^{21}$ cm$^{-2}$ (Table~\ref{tab:psf_data}). We then obtained a supportive result that the PSF models agreed within 1$\sigma$.

\subsubsection{Point-Source-Subtracted X-ray Images}\label{sec:xray_sub_image}

Using the PSF models obtained in Section~\ref{sec:cxo_psfs}, we produced  point-source-subtracted images at 3--6 and 6--7\,keV. 
From the reprocessed data of the main targets, we produced 
images in the two bands at an ACIS-S native resolution of 0\arcsec.49. If multiple data were available for a single object, the images were stacked into one. 
Then, for each image, we derived photon counts at 3.5 pixels from the AGN position and modeled a point source image by renormalizing the PSF model in an appropriate energy band to match the photon counts. 

In Figure~\ref{fig:cxo_profiles}, 
we representatively show the observed radial profiles and those of the modeled point source images for NGC 1068 and NGC 5506. 
The former is adopted to show a case where clear extended emission is present, and the latter is an example showing good agreement between the observed and modeled profiles. 
All the figures are shown in Appendix~\ref{app:xrad}, and it can be seen that the observed profiles are generally not overestimated by those of the modeled point source images, suggesting the success of our modeling. 
Then, we subtracted the models from the observed images and calculated the ratios between the 6--7 and 3--6\,keV images to indicate the sites with bright Fe-K$\alpha$ emission. Note that in some objects, emission from highly ionized iron atoms also contribute to the ratio (e.g., NGC 1068, NGC 4945, and NGC 6240, as inferred from the spectra in Appendix~\ref{app:fig:xspec}). 
Here, to avoid extreme ratios due to very low denominator, or low counts in the 3.0--6.0\,keV image, we excluded pixels with counts smaller than 2. This criterion corresponds approximately to the 1$\sigma$ level for a mean value of zero counts \citep{Geh86}. The ratio images were then smoothed with a Gaussian kernel with a full width at half maximum of 1\arcsec.0. 
Figure~\ref{fig:cxo_images} shows the ratio images created for NGC 1068 and NGC 5506. 

Notably, NGC 6240 hosts two bright nuclei that are quite close ($<$ 1\arcsec), and it was difficult to estimate their PSF normalizations accurately. However, the 3--6\,keV, 6--7\,keV, and ratio images were created for completeness in the same manner as adopted for the others.
Therefore, in this study, we do not discuss the obtained images in detail. Detailed imaging analyses can be found in \cite{Fab20}. 

\begin{table}
\begin{center}
\caption{Data for Empirical PSF Construction}\label{tab:psf_data}
\begin{tabular}{ccccc}
\toprule
(1)   & (2) & (3) & (4) \\ 
Target name & ObsID & Grating & Exp. \\ \hline
                  HD 110432 &  9947 &     HETG &   139.49 \\
                   AO PSC$^\dagger$ &  1898 &     HETG &    98.50 \\
               4U 1626$-$67$^\dagger$ &  3504 &     HETG &    94.81 \\
              AM HERCULIS$^\dagger$ &  3769 &     HETG &    91.46 \\
              4U 1323-619 & 13721 &     HETG &    79.00 \\
            XTE J1710-281 & 12468 &     HETG &    73.31 \\
            XTE J1710-281 & 12469 &     HETG &    73.30 \\
                 NGC 1851$^\dagger$ & 18759 &     HETG &    62.57 \\
              4U 1908+075 &  5477 &     HETG &    49.00 \\
               4U 1907+09 & 21123 &     HETG &    41.28 \\ \hline 
        SAX J2103.5+4545 & 15780 &     NONE &    45.41 \\
                 NGC 6652$^\dagger$ & 12461 &     NONE &    45.32 \\
           MAXI J0556-332$^\dagger$ & 14227 &     NONE &    18.20 \\
           MAXI J1659-152 & 12441 &     NONE &    18.15 \\
           MAXI J0556-332$^\dagger$ & 14429 &     NONE &    13.67 \\
                 NGC 6652$^\dagger$ & 18987 &     NONE &    10.04 \\
    1RXS J045707.4+452751 &  3878 &     NONE &     4.93 \\
             GRO J1008-57 & 14639 &     NONE &     4.60 \\
              KS 1947+300 & 14647 &     NONE &     4.60 \\
               4U 0728-25 & 14636 &     NONE &     4.60 \\ \hline 
\end{tabular}
\end{center}
\tablenotetext{}{
Notes.--- 
(1) Target name. 
(2) Chandra data ID.
(3) Tag to identify imaging (NONE) or grating (HETG) observations.
(4) Observation exposure time in units of ks, taken from ChaSeR.
$^\dagger$ Object with a line-of-sight column density from an HI map \citep{Kal05} less than 10$^{21}$ cm$^{-2}$. 
}
\end{table}

\section{Discussion}\label{sec:dis}

Based on the results obtained in Sections~\ref{sec:alma_data} and \ref{sec:cxo_data}, we address the main question, \textit{what impact does the hard X-ray irradiation have on the ISM and on star formation?}. 
In the first half of this section (Sections~\ref{sec:fe_origin} and \ref{sec:ext_fe_co}), we focus on the external regions. Then, in the second half (Section~\ref{sec:dis_nuc_scale}), we focus on the nuclear regions.

\subsection{Origin of extended Fe-K$\alpha$ emission}\label{sec:fe_origin}

\begin{figure}[h!]
    \centering
    \includegraphics[width=8.5cm]{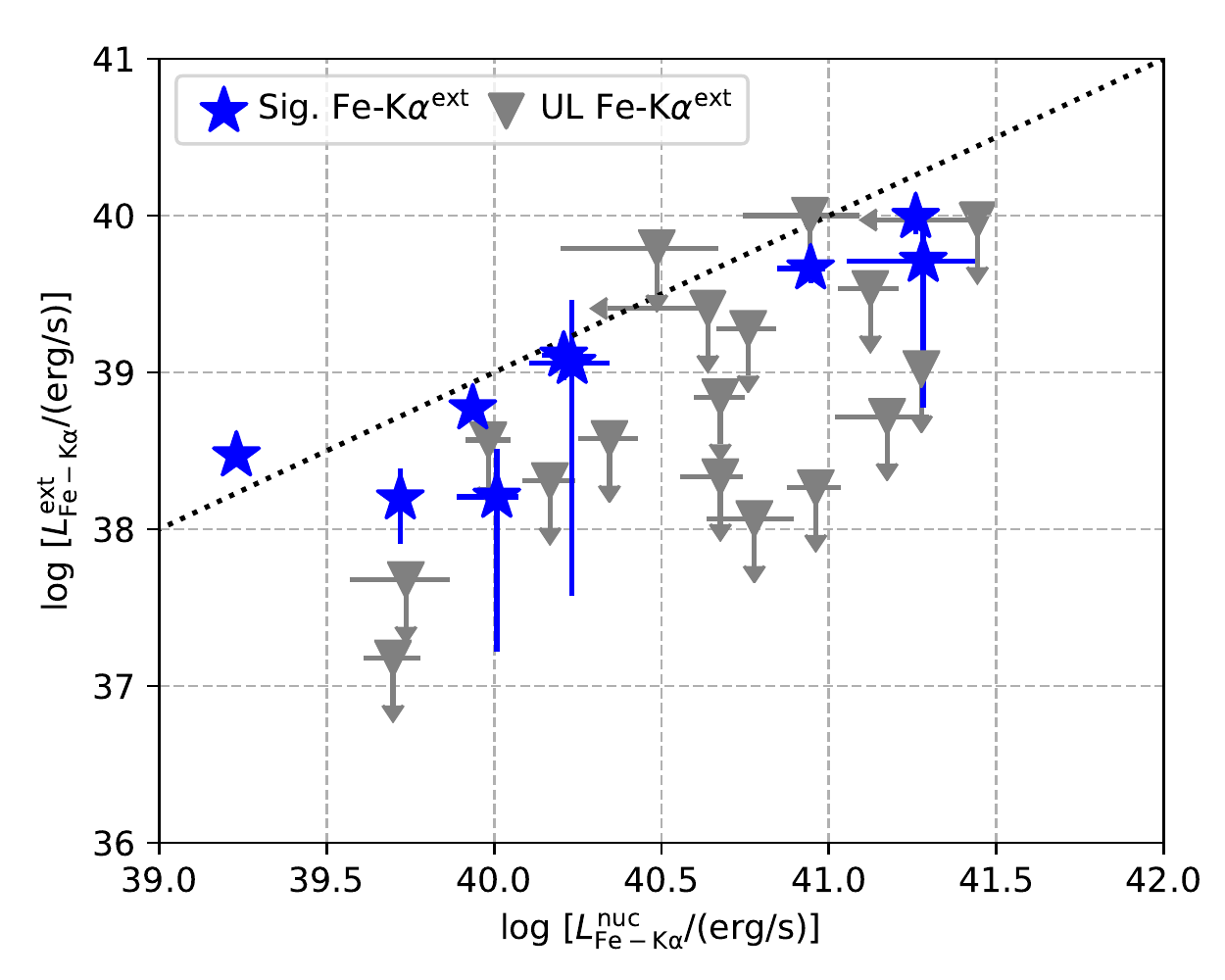}
    \caption{Scatter plot of Fe-K$\alpha$ luminosities derived from the nuclear and external regions. 
    The dotted line is drawn where the ratio is 0.1. 
    AGNs with significant detection and upper limits (UL) for Fe-K$\alpha$ emission 
    in their external regions are indicated by blue stars and gray triangles, respectively.   
    }\label{fig:nuc_to_ext_fe}
\end{figure}

As reported in Section~\ref{sec:cxo_fe}, Fe-K$\alpha$ emission was detected in the external regions of nine AGNs above 1$\sigma$. More reliably, six of them showed Fe-K$\alpha$ emission above 2$\sigma$ (i.e., NGC 1068, NGC 2110, NGC 3393, NGC 4945, NGC 5643, and NGC 6240). 
The luminosity ratios between the extended and nuclear Fe-K$\alpha$ emission are at most 0.1 (Figure~\ref{fig:nuc_to_ext_fe}). 
Thus, a large fraction of the Fe-K$\alpha$ emission originates
from regions within a radius of $\sim$ 2\arcsec, corresponding to 100--600 pc for the majority of our targets (i.e., 19/26). 

Although the extended Fe-K$\alpha$ emission is faint, its detailed study suggests the potential effect of hard X-ray irradiation on the ISM. 
We thus first discuss the following four conceivable mechanisms as the origin of the spatially extended Fe-K$\alpha$ emission: 
(1) This is the result of the Thomson scattering of Fe-K$\alpha$ photons produced by dense nuclear material (e.g., the putative torus; \citealt{Fab17}).
(2) This is caused by the collisional ionization of Fe atoms by cosmic rays including electrons and ions \citep[e.g., ][]{Val00,Yus02,Yus07}. 
(3) The fluorescent X-ray emission following photoionization by high-energy X-ray photons above 7.1\,keV produced by the AGN \citep[e.g., ][]{Kat07,Nob10,Pon10}. 
(4) The sum of contributions from multiple high-mass X-ray binaries \citep[e.g., ][]{Tor10}.

In Section~\ref{sec:thomson}, we test the first scenario. Then, we discuss the next two scenarios in Section~\ref{sec:ionization}. Finally, the fourth idea is examined in Section~\ref{sec:hxrb4fe}. Following these discussions, we suggest that the detected Fe-K$\alpha$ emission can be explained based on the third scenario (i.e., AGN photoionization).

\subsubsection{Thomson Scattering of Nuclear Fe-K$\alpha$ Photons?}\label{sec:thomson}

We first test the Thomson scattering scenario. 
The fraction of Thomson scattered light compared to direct emission (scattering fraction or $f_{\rm scat}$) has been well estimated for obscured AGNs using broadband X-ray spectroscopy \cite[e.g.,][]{Bia07,Nog10,Kaw16b,Kaw16c,Ric17c,Gup21}. 
If the external Fe-K$\alpha$ emission is also the result of scattering, its luminosity with respect to primary X-ray emission ($L^{\rm ext}_{\rm Fe-K\alpha}/L_{\rm X}$), a proxy for the solid angle of gas seen from the nucleus, should correlate with the scattering fraction. 
To assess the correlation, we took the scattered fractions of our sample from \cite{Ric17c} \citep[see also][]{Gup21}, except for NGC 612, for which we referred to \cite{Urs18}.

\begin{figure}
    \begin{center}
    \includegraphics[width=8.5cm]{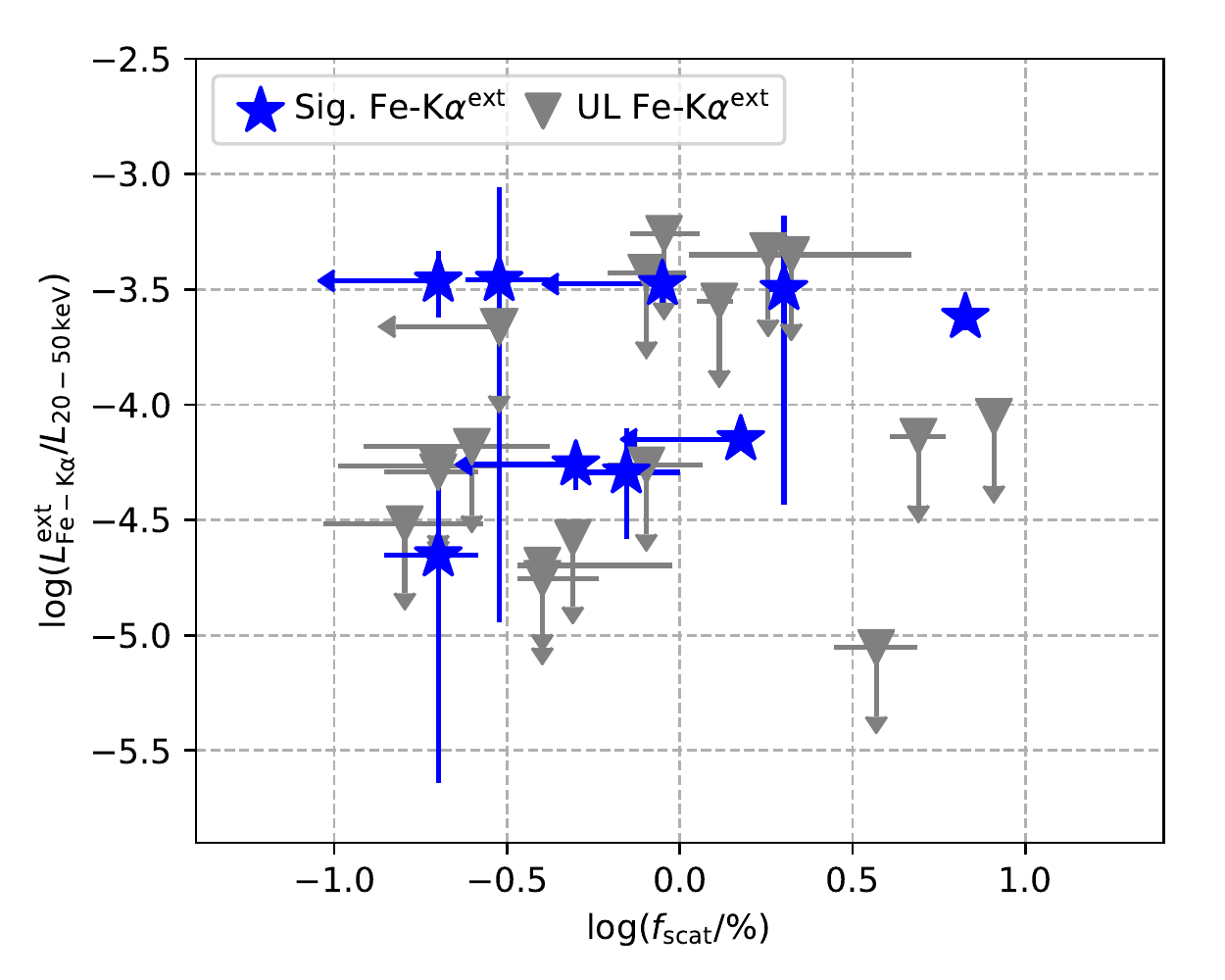}
    \caption{Scatter plot of the 
    fraction of scattered X-ray light
    ($f_{\rm scat}$) and the Fe-K$\alpha$-to-continuum luminosity ratio ($L^{\rm ext}_{\rm Fe-K\alpha}$/$L_{\rm 20-50~keV}$). Sources with detected Fe-K$\alpha$ emission in their external regions are indicated by blue stars. In contrast, those with upper limits (UL) are shown as gray triangles. 
    No significant correlation is found for these two values. 
    }\label{fig:fscat_vs_lfelx}
    \end{center}
\end{figure}

Figure~\ref{fig:fscat_vs_lfelx} shows a scatter plot of $\log(f_{\rm scat}/\%)$ and $\log(L^{\rm ext}_{\rm Fe-K\alpha}/L_{\rm 20-50~keV})$, and no clear correlation is seen between these two parameters. 
To confirm this, we derive the $p$-value ($P$) and the Spearman rank coefficient ($\rho_{\rm S}$) using a bootstrap method. 
Following previous works \cite[e.g.,][]{Ric14b,Gup21}, we draw a number of datasets based on the 26 real data, considering their uncertainties. 
For data with upper and lower errors, we randomly draw values from a Gaussian distribution whose mean 
and standard deviation are set to the best value and 
its 1$\sigma$ error, respectively.
If the upper and lower 1$\sigma$ errors are asymmetric, we simply use the mean of the two. 
For the data that only have an upper limit, we use a uniform distribution between zero and the upper limit. 
Any drawn datasets including unphysical negative values were discarded.
For each dataset, we compute the $p$-value and
the Spearman rank coefficient using a Python program of \texttt{spearmnr}. 
We then fit the data using a relation of the form $\log Y = a + b\times \log X$, using the ordinary least-squares 
bisector regression fitting algorithm \citep{Iso90}. 
Finally, we consider the 16th and 84th percentiles of the distributions of the parameters ($P$, $\rho_{\rm S}$, $a$, and $b$) as their lower and upper limits, respectively. 
For $X$ = $(f_{\rm scat}/\%)$ and $Y$ = $(L^{\rm ext}_{\rm Fe-K\alpha}/L_{\rm 20-50~keV})$,
we draw $\approx$ 10000 datasets and find $P = 0.65^{+0.25}_{-0.29}$, indicating no correlation. 
Finally, we could not obtain evidence supporting the Thomson scattering scenario as the general origin of the external Fe-K$\alpha$ emission. However, it should be noted that there may be objects such as ESO 428-G014 \citep[e.g.,][]{Fab17,Fab18a,Fab18b}, in which the extended Fe-K$\alpha$ is due to Thomson scattering.

\subsubsection{Photoionized Gas or Collisionally-ionized Gas 
as the Origin of Fe-K$\alpha$ Emission}\label{sec:ionization}

\begin{table}
\begin{center}
\caption{Fe-K$\alpha$ EWs in the external regions}\label{tab:ews}
\begin{tabular}{ccc}
\toprule
(1) & (2) & (3) \\ 
Number & Target name & EW(Fe-K$\alpha^{\rm ext}$) \\ \hline
16 & NGC 4945 & 620$^{+60}_{-70}$ \\
22 & NGC 6240 & 700$^{+100}_{-260}$ \\ \hline 
06 & NGC 1068 & 1400$^{+260}_{-170}$\\
10 & NGC 2110 & 1500$^{+2100}_{-1000}$/1600$^{+2600}_{-1100}$\,$^\dagger$ \\ 
14 & NGC 3393 & 1800$^{+800}_{-650}$ \\
20 & NGC 5643 & 1500$^{+880}_{-730}$ \\ \hline
\end{tabular}
\end{center}
\tablenotetext{}{
Notes. (1) Source number. (2) Target name. (3) EW of the Fe-K$\alpha$ emission  from the external region in units of eV. 
This lists only six sources with significant Fe-K$\alpha$ emission in their external regions above 2$\sigma$.
In the upper rows, we list AGNs with EWs smaller than 1\,keV. 
\,$^\dagger$The two EWs were taken from \cite{Kaw20}, who performed a spatially resolved spectral analysis in more detail and derived them in the northwest and southeast regions with bright Fe-K$\alpha$ emission within 2\arcsec--5\arcsec.}
\end{table}

Based on the equivalent width of the Fe-K$\alpha$ emission [EW(Fe-K$\alpha^{\rm ext}$)], we discuss whether the external Fe-K$\alpha$ emission is from excitation following the collision of low-energy cosmic rays \citep[e.g., electrons and ions;][]{Yus07} or photoionized Fe atoms \citep[e.g., ][]{Nob10}. This kind of discussion has often been considered to interpret the Fe-K$\alpha$ emission distribution in our galaxy \citep[e.g., ][]{Koy96,Mur00,Yus07,Tat12}. 
Theoretically, it is expected that if the collisional ionization by cosmic rays is significant, Fe-K$\alpha$ photons are accompanied by strong bremsstrahlung continuum emission, thereby resulting in EWs with $\sim$ 0.3--0.8\,keV for the typical solar metal abundance \citep[e.g., ][]{Nob10,Tat12}. 
By contrast, in a photoionization model, continuum emission 
arises from the Compton scattering of X-rays from an external source.
Because the scattered continuum emission is weak, a higher EW above $\sim$ 1\,keV is expected \citep[e.g.,][]{Nob10}. 
Such high EWs were found in the X-ray spectra of Compton-thick AGNs
and have been explained by the fluorescent process \cite[e.g.,][]{Mat04,Mar12,Tan16,Tan18,Yam20}. 
Thus, the EW can be used to discriminate these ionization models. 

\begin{figure}
    \centering
    \includegraphics[width=8.3cm]{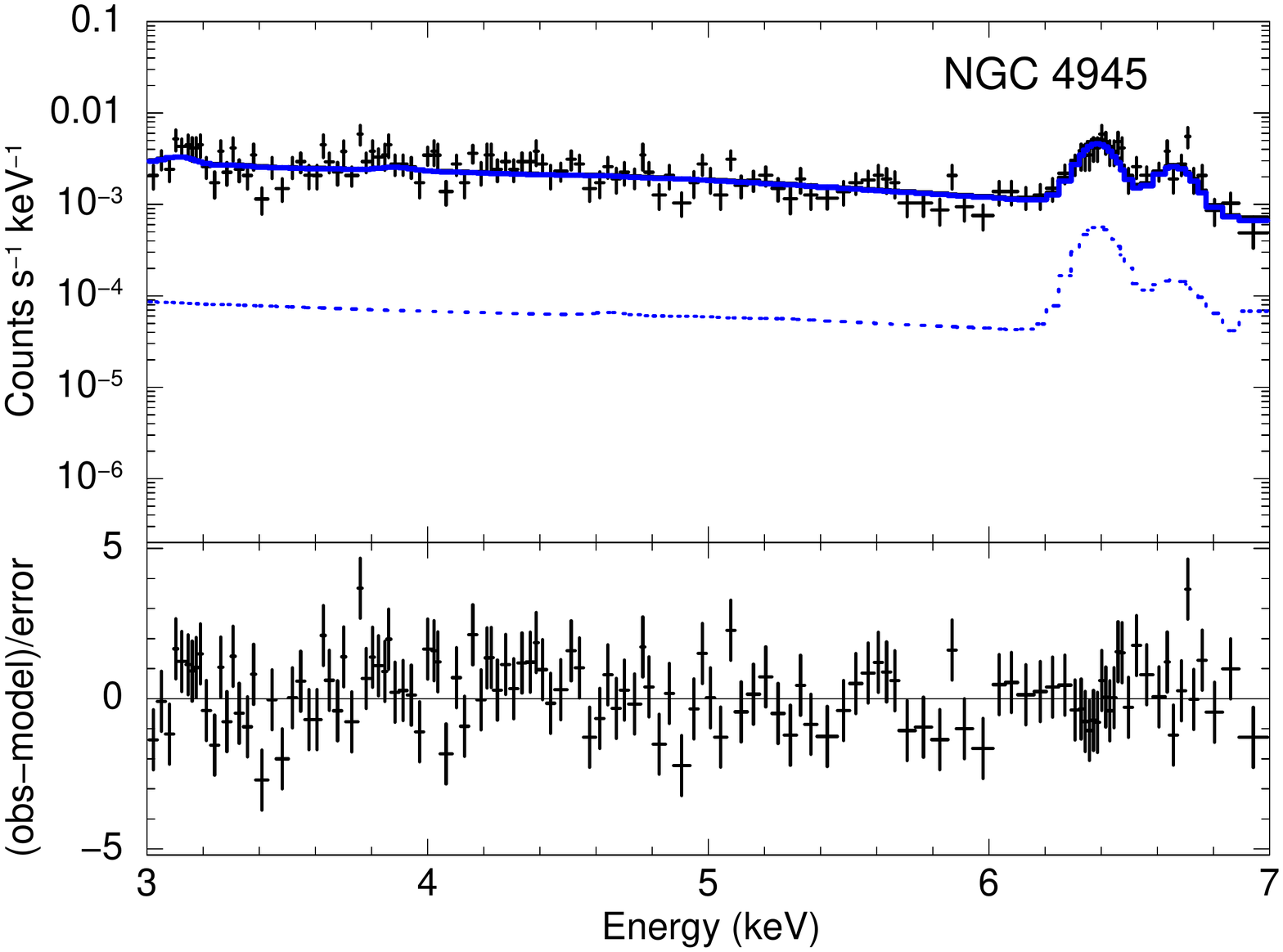}\vspace{-0.8cm}
    \includegraphics[width=8.3cm]{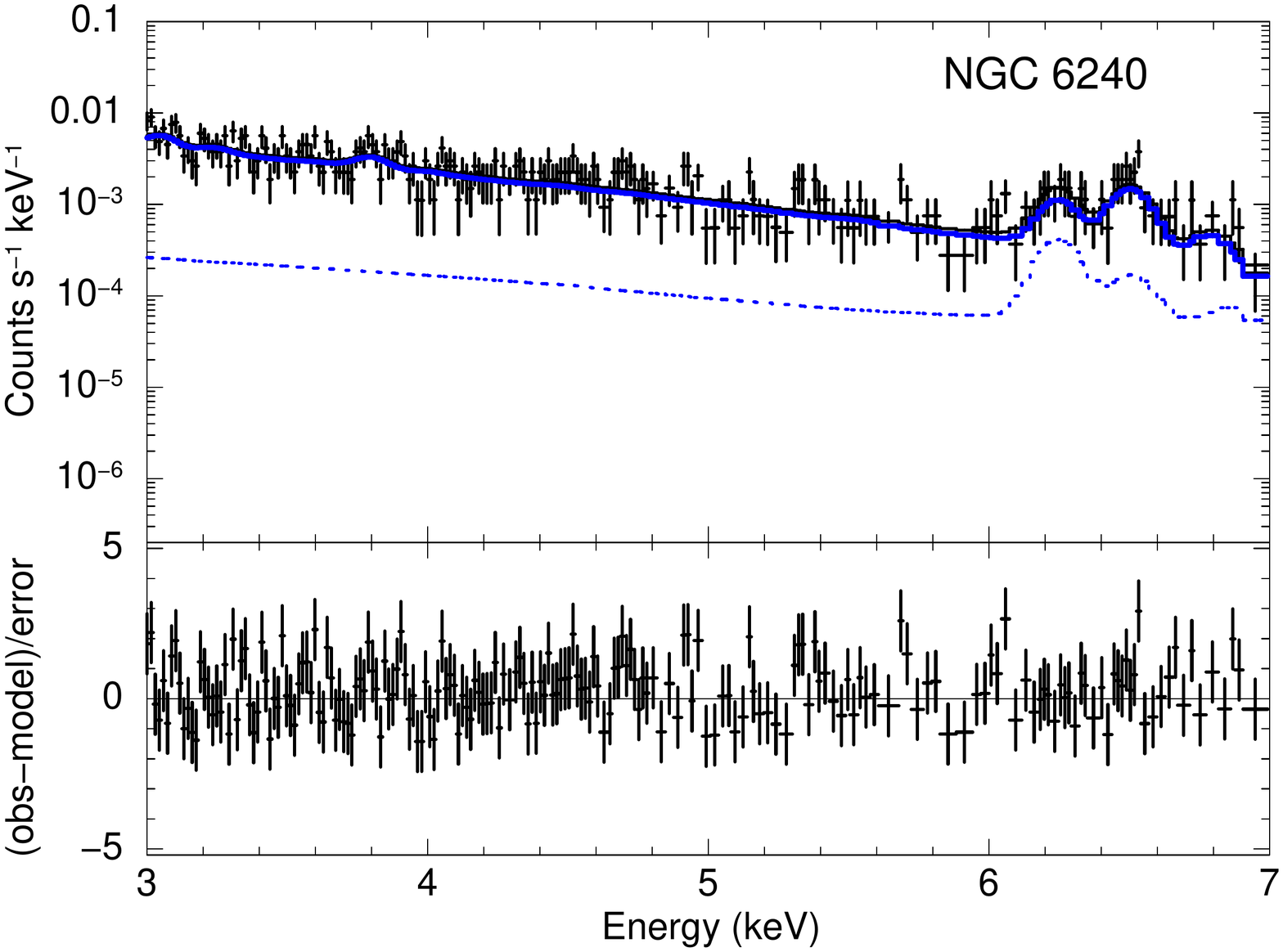}\vspace{-0.3cm}
    \caption{X-ray spectra of NGC 4945 and NGC 6240 taken from the external regions. 
    The upper windows show the observed data (black crosses) and the best-fit models (black solid line), and the lower ones show residuals.
    The spectra are reproduced by considering the sum of two components: one is from the extracted 2\arcsec--4\arcsec\ region (blue solid line), and the other is from the central point source extended by the PSF (blue dotted line).
    }
    \label{fig:xspec_ngc_4945_ngc_6240}
\end{figure}

We compute the EWs with respect to the underlying continuum emission for the six AGNs with Fe-K$\alpha$ emission detected above 2$\sigma$. 
Here, the continuum emission includes 
the Compton reflection component (\texttt{pexrav}) and/or the scattered emission modeled with a power-law function (\texttt{powerlaw}). 
Table~\ref{tab:ews} lists the derived EWs. 
Notably, four of them have EW(Fe-K$\alpha^{\rm ext}$) larger than $\sim$ 1\,keV, supporting  the dominance of the Fe-K$\alpha$ emission produced by the fluorescent process. 
For the other two AGNs (NGC 4945 and NGC 6240), 
we find lower values of $\sim$ 600--700\,eV. 
Their spectra are shown in 
Figure~\ref{fig:xspec_ngc_4945_ngc_6240}.
We discuss the two objects in the following paragraphs, remarking that the Fe-K$\alpha$ emission of the two objects may originate from the photoionization of cold Fe atoms.

NGC 4945: as inferred from the spectrum in the upper panel of Figure~\ref{fig:xspec_ngc_4945_ngc_6240}, NGC 4945 shows 
a significant emission line at 6.67$\pm0.02$\,keV in the rest frame. 
The line energy suggests that the forbidden (6.637 keV) and intercombination Fe~{\sc xxv} lines (6.682 and 6.668\,keV) are stronger than the resonance line at 6.700 keV. \cite{Mar17} also confirmed the line using Chandra data and suggested that this was favored by a photoionization model. 
As an additional test to determine whether the line does not originate from optically thin thermal plasma, we performed a fit by replacing the line model with a plasma model of \texttt{apec}. 
The resultant fit was worse, as indicated by $\Delta C =$ 19 for the same degree of freedom. Thus, the line suggests that a fraction of the Fe atoms should be photoionized.

We further mention the results of \cite{Mar17}. 
At higher resolutions ($\sim$ 0\arcsec.25), they identified clumps 
with Fe-K$\alpha$ EWs of 0.45--0.75 keV. These EWs apparently favor 
a collisional model. However, 
while the EWs were reproduced by adjusting the column density and inclination angle of a reflecting gas model \citep[MYTORUS;][]{Mur09}, 
they also suggested that the EWs were due to significant electron-scattered radiation with little Fe-K$\alpha$ emission. 
Their argument was based on the significant Fe~{\sc xxv} emission from the clumps. In highly ionized gas associated with such line emission, 
the photoelectric absorption by Fe does not occur readily  \citep[e.g., ][]{Kro87,Iwa97}, and accordingly, electron-scattered continuum emission with little Fe-K$\alpha$ emission is expected. 
Finally, such a component can decrease the apparent Fe-K$\alpha$ EW. 
In summary, the ionized line supports that photoionization takes place, and the Fe-K$\alpha$ line may originate from less-ionized iron atoms irradiated by nuclear X-ray emission.

NGC 6240: the spectrum of NGC 6240 was reproduced by 
\texttt{pexrav + apec + zgauss + zgauss} where, in addition to the 6.4 keV emission, another line with a rest-frame energy of 6.95$^{+0.06}_{-0.05}$\,keV was included (lower panel of Figure~\ref{fig:xspec_ngc_4945_ngc_6240}). 
This is consistent with the emission from H-like Fe atoms (Fe~{\sc xxvi}). By analyzing the XMM-Newton data of NGC 6240, \cite{Bol03} suggested that ionized Fe lines (Fe~{\sc xxv} and Fe~{\sc xxvi}) could be produced in 
optically thin thermalized plasma with a high temperature of $\approx 5.5$\,keV.
Accordingly, we replaced the line component at 6.95\,keV with the plasma model \texttt{apec}. 
Thus, the adopted model was \texttt{pexrav + apec + apec + zgauss} and fitted the spectrum well at a temperature of 6.0$^{+1.0}_{-1.1}$\,keV. This is consistent with the result obtained by \cite{Bol03}. We computed the EW(Fe-K$\alpha^{\rm ext}$) with respect to the Compton reflection continuum \texttt{pexrav} and found it to be 
2.5$^{+1.1}_{-2.0}$\,keV. Thus, cold Fe gas irradiated by X-ray emission may be embedded in optically thin hot gas. 

In summary, by analyzing the spectra more carefully, we found that the observed EWs of the 6.4\,keV emission may be suppressed by unrelated continuum emission (i.e., electron-scattered emission or optically thin thermalized plasma emission). 
Thus, the Fe-K$\alpha$ emission may still originate from the gas ionized by the AGN X-ray emission. For more detailed analyses of these two AGNs, we refer to previous studies \citep[e.g.,][]{Bol03,Mar17,Fab20}.

\subsubsection{Spatially distributed HMXBs for the external Fe-K$\alpha$ emission?}\label{sec:hxrb4fe}

\begin{figure*}
    \centering
    \includegraphics[width=8.5cm]{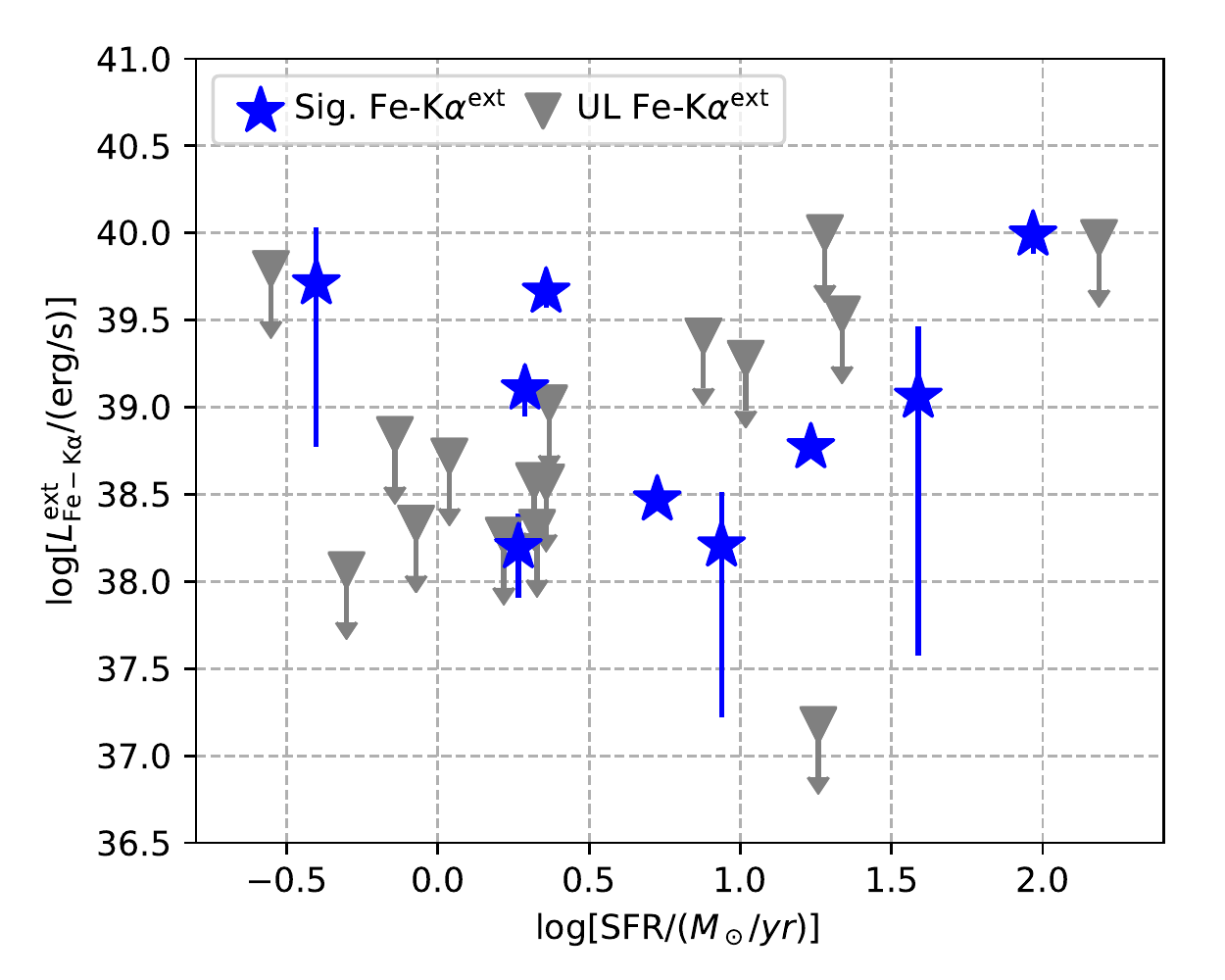}
    \includegraphics[width=8.5cm]{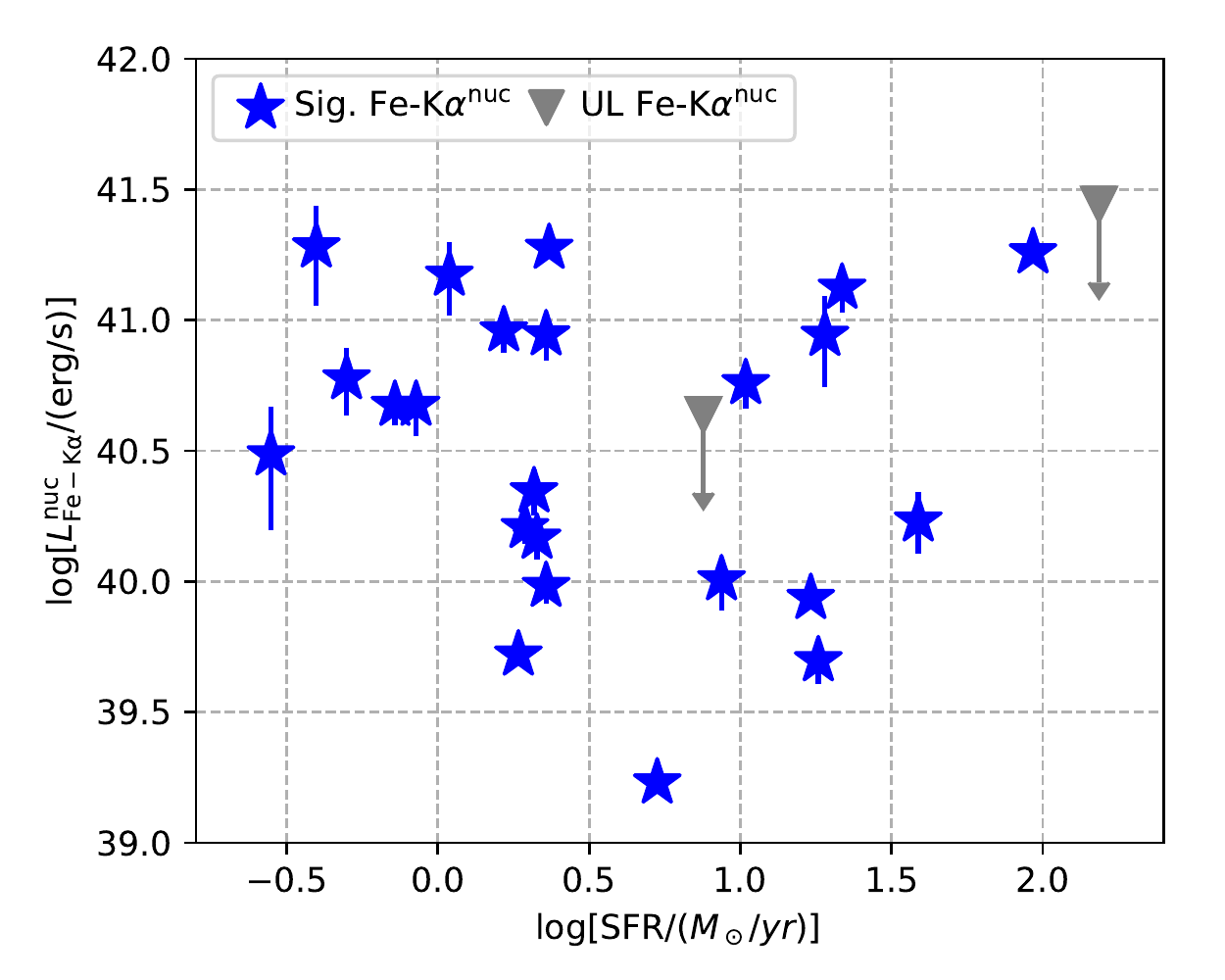} 
    \caption{Scatter plots of the SFR and Fe-K$\alpha$ luminosities on the external (left) and nuclear (right) scales. AGNs with significant Fe-K$\alpha$ emission 
    ($L^{\rm ext}_{\rm Fe-K\alpha}/\Delta L^{\rm ext}_{\rm Fe-K\alpha} > 1$ or $L^{\rm nuc}_{\rm Fe-K\alpha}/\Delta L^{\rm nuc}_{\rm Fe-K\alpha} > 1$)
    are indicated by blue stars.
    In contrast, those with upper limits (UL) are shown as gray triangles. 
    No significant correlation is found in either case. 
    } 
    \label{fig:sfr_vs_lfe}
\end{figure*}

Finally, we discard a scenario in which X-ray binaries (XRBs) strongly contribute to the Fe-K$\alpha$ emission. 
Among XRBs, high-mass XRBs (HMXBs)  
contribute more than low-mass XRBs (LMXBs), given the higher detection rate of the Fe-K$\alpha$ emission in HXMBs than in LMXBs \citep{Tor10}. 
Because a good correlation was found between the total X-ray luminosity of HMXBs in a galaxy and the galaxy-scale star-formation rate (SFR) in a wide SFR range of 0.1--1000 $M_\odot$ yr$^{-1}$ \citep{Min12a}, we examine the relation between the Fe-K$\alpha$ luminosity ($L^{\rm ext}_{\rm Fe-K\alpha}$) and the SFR, as shown in the left panel of Figure~\ref{fig:sfr_vs_lfe}. 
The SFRs were taken from \cite{Shi17} and \cite{Ich19}, who derived the SFRs by decomposing spectral energy distributions (SEDs) into AGN and star-formation components. 
Notably, the far-infrared (FIR) photometry data of the SEDs where star-formation emission is usually significant were obtained at an angular resolution of $\approx$ 6\arcsec\ of Herschel/PACS at best. 
However, the FIR source sizes of hard X-ray selected AGN host galaxies at redshifts below 0.05, as with our targets, were found to be at most $\sim$ 3\arcsec\ at 70 $\mu$m \citep{Mus14}. 
Therefore, it is reasonable to compare the SFRs we adopted with  the Fe-K$\alpha$ luminosities because they were derived at similar angular scales (2\arcsec--4\arcsec). 
As illustrated in Figure~\ref{fig:sfr_vs_lfe}, the SFR and the Fe-K$\alpha$ luminosity are not correlated ($P$ = $0.19^{+0.10}_{-0.08}$), suggesting that HMXBs do not provide a strong contribution to the observed Fe-K$\alpha$ flux. 

As a supplemental check, we examine a relation 
of the SFR versus the nuclear-scale Fe-K$\alpha$ luminosity (right panel of Figure~\ref{fig:sfr_vs_lfe}). No significant correlation is found
($P = 0.85^{+0.11}_{-0.24}$). 

\begin{figure*}
    \centering
    \includegraphics[width=8.5cm]{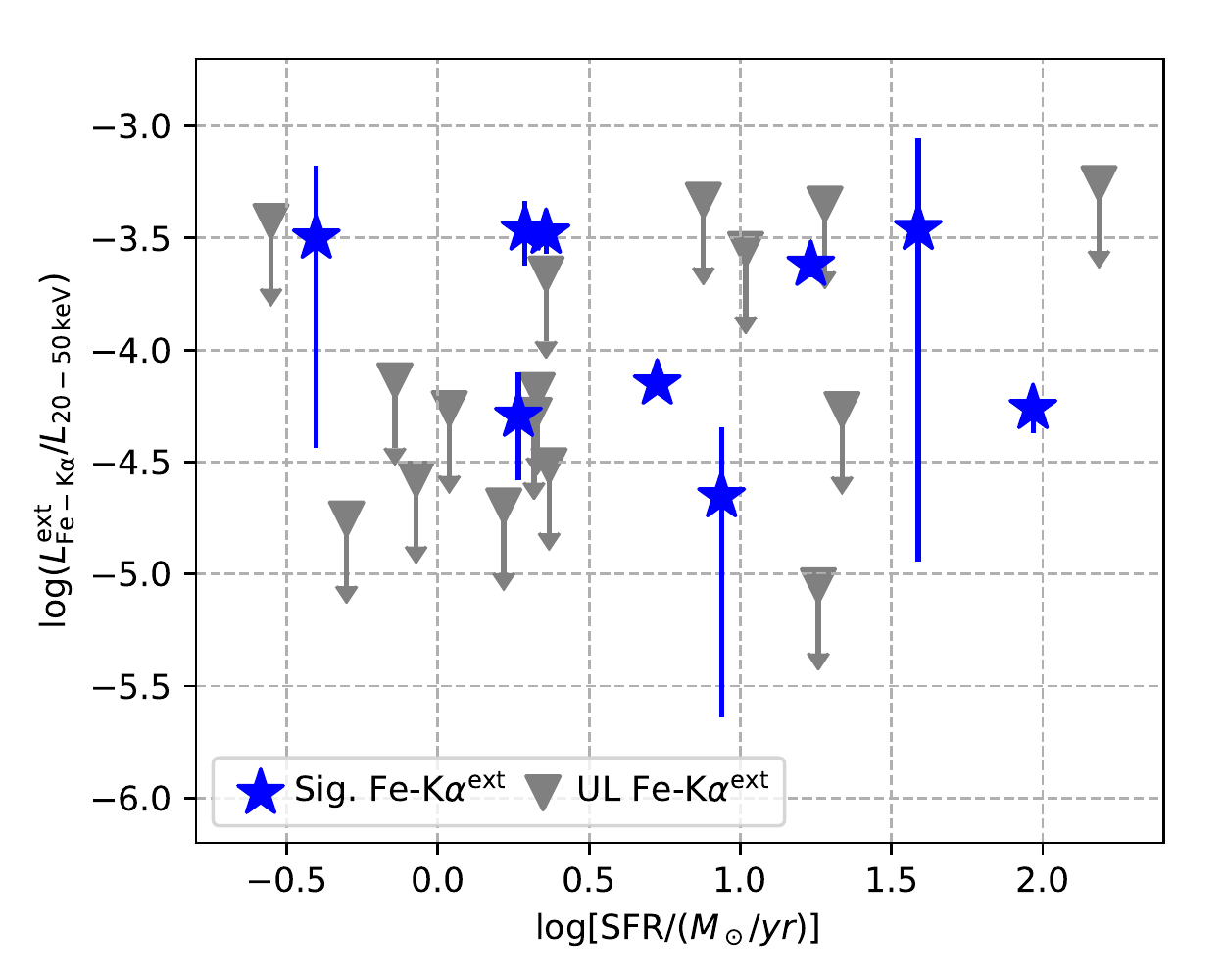}
    \includegraphics[width=8.5cm]{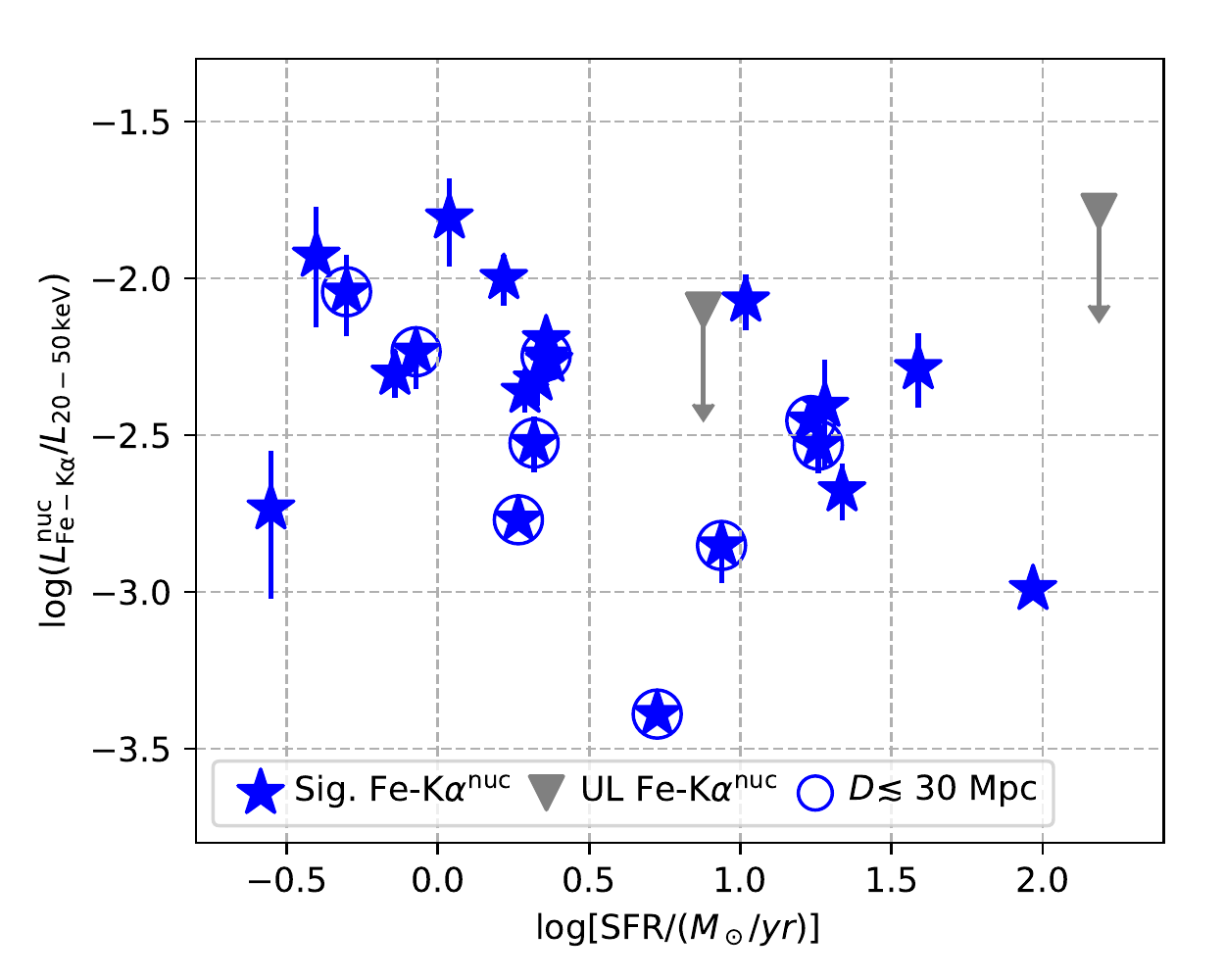}
    \caption{Scatter plots of the SFR and 
    the Fe-K$\alpha$-to-X-ray luminosity ratio on the external
    (left) and nuclear (right) scales. 
    AGNs with significant Fe-K$\alpha$ emission
    ($L^{\rm ext}_{\rm Fe-K\alpha}/\Delta L^{\rm ext}_{\rm Fe-K\alpha} > 1$
    or 
    $L^{\rm nuc}_{\rm Fe-K\alpha}/\Delta L^{\rm nuc}_{\rm Fe-K\alpha} > 1$)
    are indicated by blue stars.
    In contrast, those with upper limits (UL) are shown by gray triangles.  
    Only in the right panel, open circles are over-plotted for AGNs at distances $\lesssim$ 30 Mpc. 
    No significant correlation is found in either case. 
    } 
    \label{fig:sfr_vs_lfelx}
\end{figure*}

Finally, we assess a correlation between SFR and 
$L^{\rm ext}_{\rm Fe-K\alpha}/L_{\rm 20-50~keV}$, motivated by 
a recent study by \cite{Yan21}. They suggested a connection between the SFR on a galaxy scale and the EW of the Fe-K$\alpha$ line in a redshift range of 0.5--2. Contrary to this suggestion, no strong link was found between the two physical parameters, as shown in
the left panel of Figure~\ref{fig:sfr_vs_lfelx}. 
However, the SFRs of most of our targets are below 17 $M_\sun$ yr$^{-1}$, which is the threshold adopted by \cite{Yan21} to observe a difference in the Fe-K$\alpha$ EW between active and less-active star-forming galaxies. Thus, further studies focusing on rapidly star-forming galaxies are needed to assess the relation between the extended Fe-K$\alpha$ emission and SFR.

\begin{figure*}[t]
    \centering
    \includegraphics[width=5.7cm]{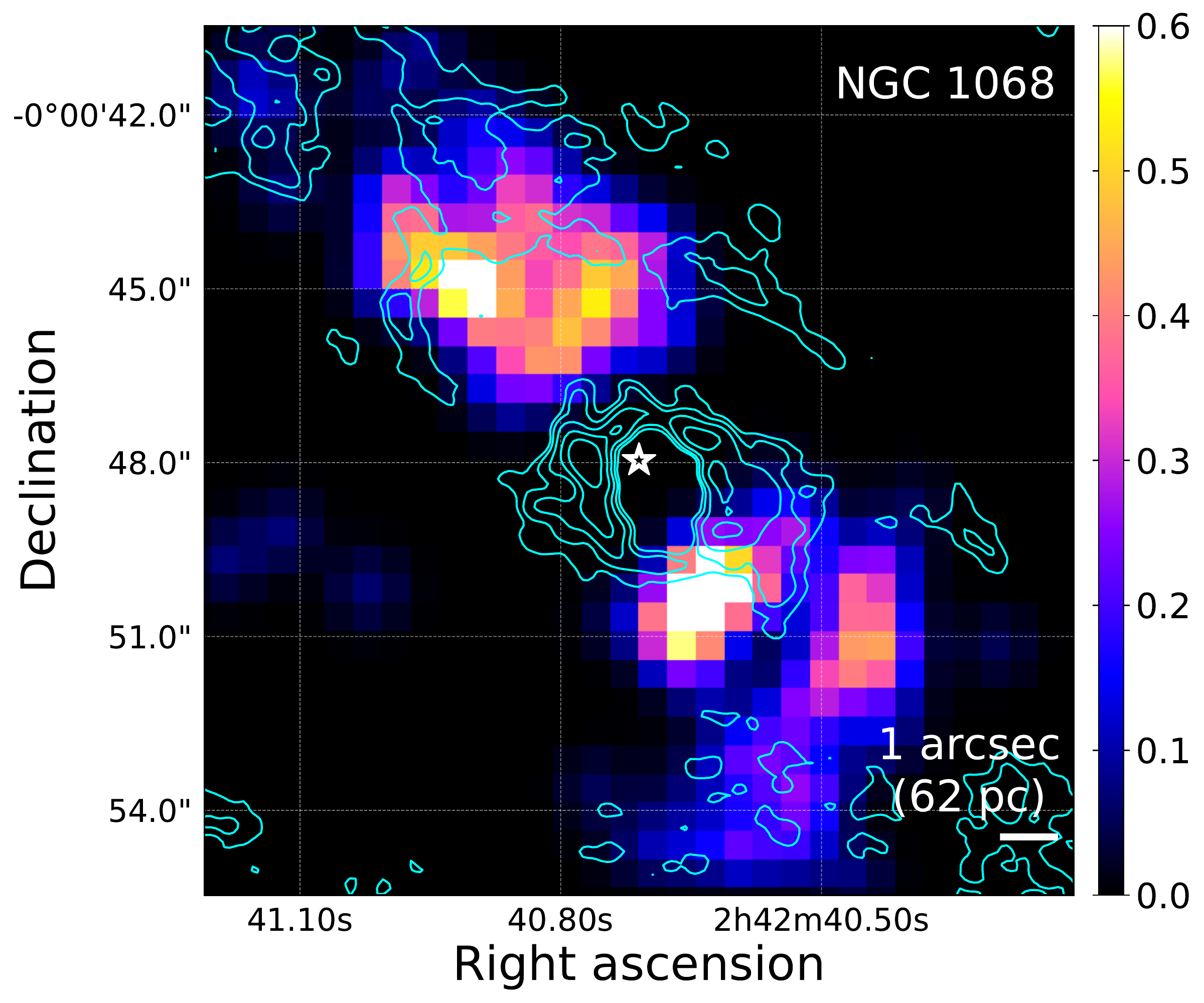}
    \includegraphics[width=5.7cm]{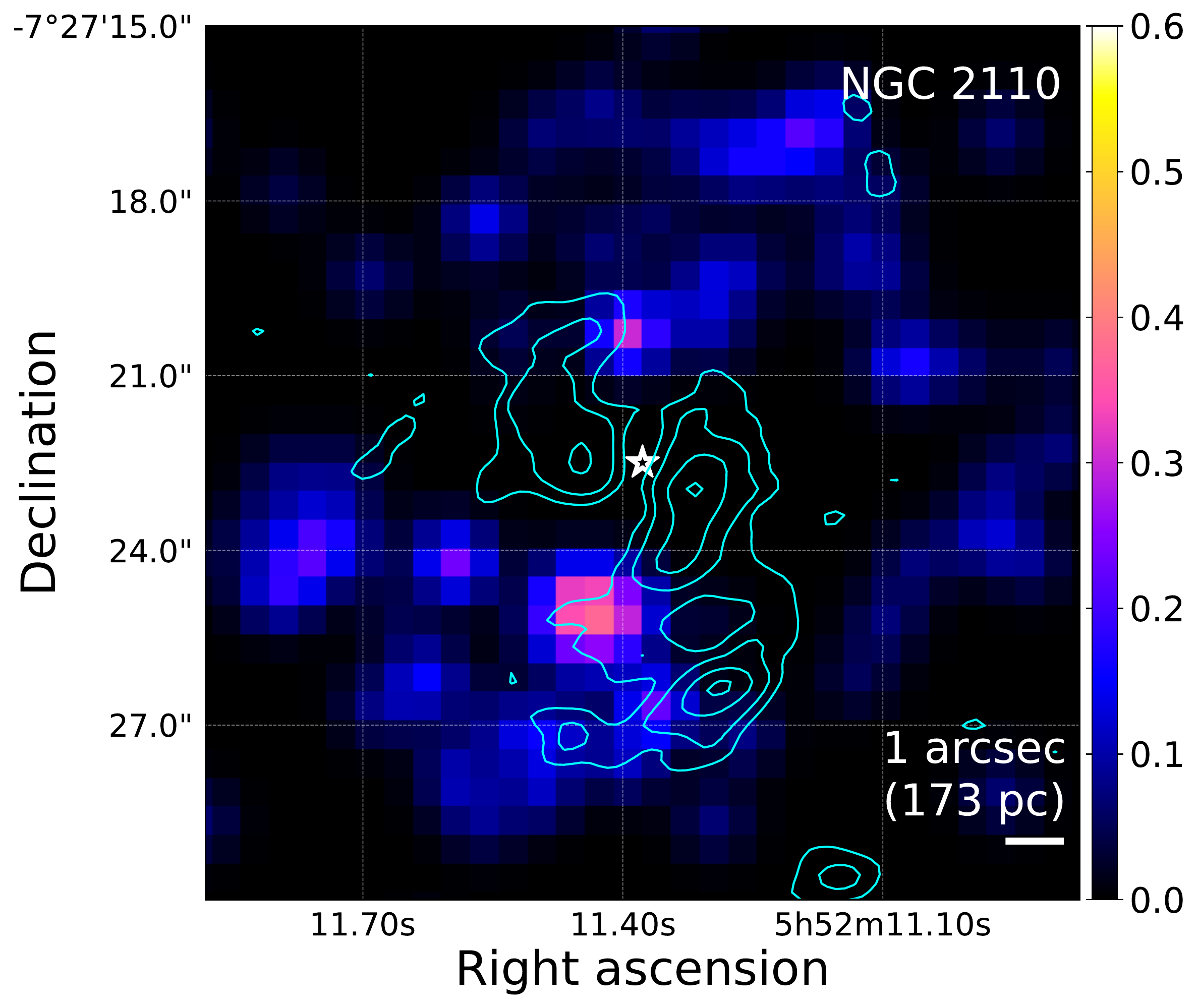}
    \includegraphics[width=5.7cm]{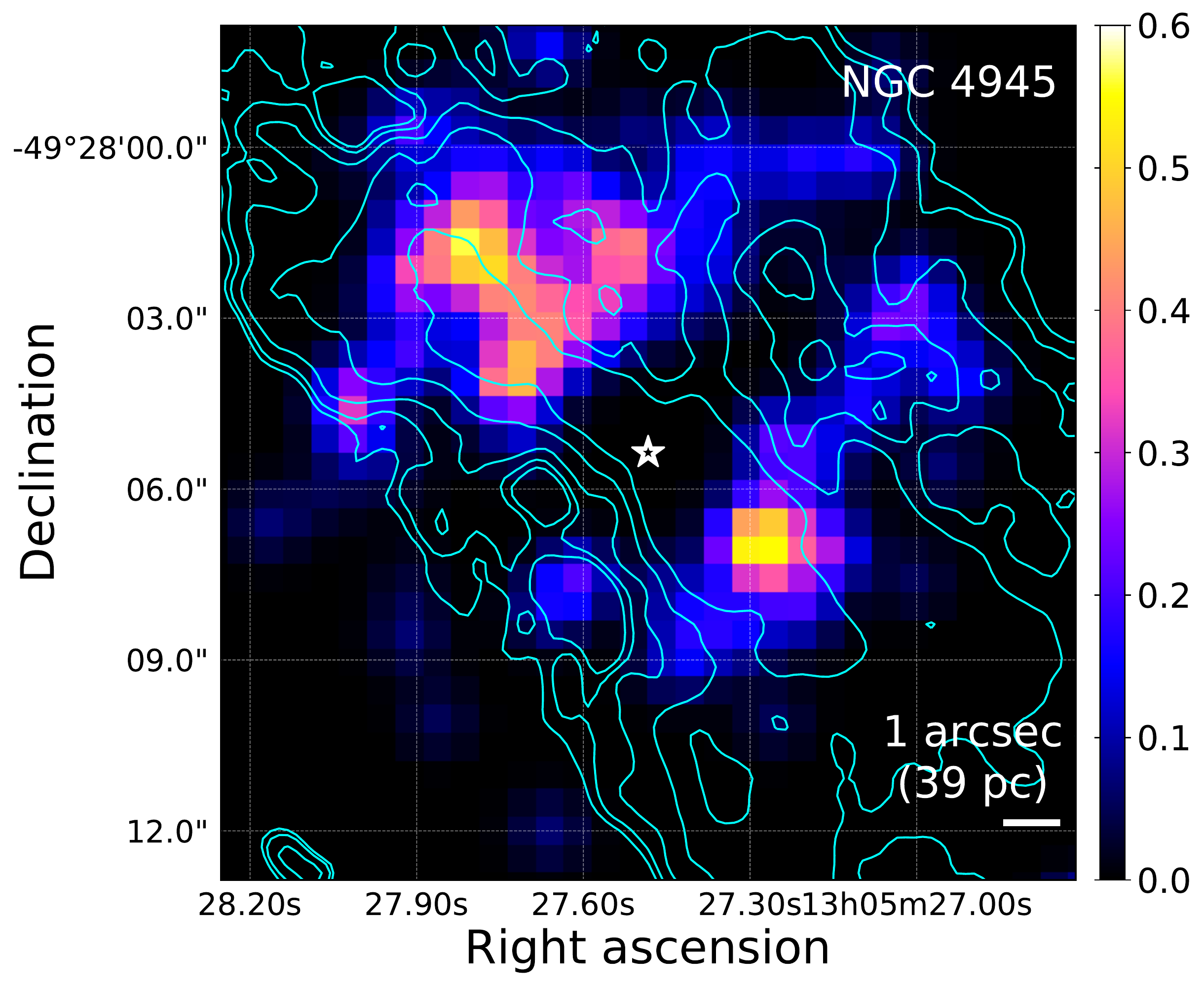}
    \caption{
    Spatial distributions of CO($J$=2--1) emission (cyan contours) and the ratio of 6--7 and 3--6\,keV images from which nuclear point sources are subtracted (color follows right bars) for NGC 1068, NGC 2110, and NGC 4945 from left to right. The panels show the $\approx$ 15\arcsec$\times$15\arcsec\ regions centered at the AGN positions (white stars). 
    North is up and east is left.
    Cyan contours represent  
    20$\sigma$, 40$\sigma$, 80$\sigma$, 160$\sigma$, and 230$\sigma$, 
    where 1$\sigma$ = 0.046 Jy km/s per beam for NGC 1068, and  1$\sigma$ = 0.071 Jy km/s per beam for NGC 4945.
    For NGC 2110, the cyan contours represent 5$\sigma$, 10$\sigma$, 20$\sigma$, and 30$\sigma$, where 1$\sigma$ = 0.060 Jy km/s per beam. 
    }
    \label{fig:fe_co21_images}
\end{figure*}

\begin{figure*}
    \centering
    \includegraphics[width=8.5cm]{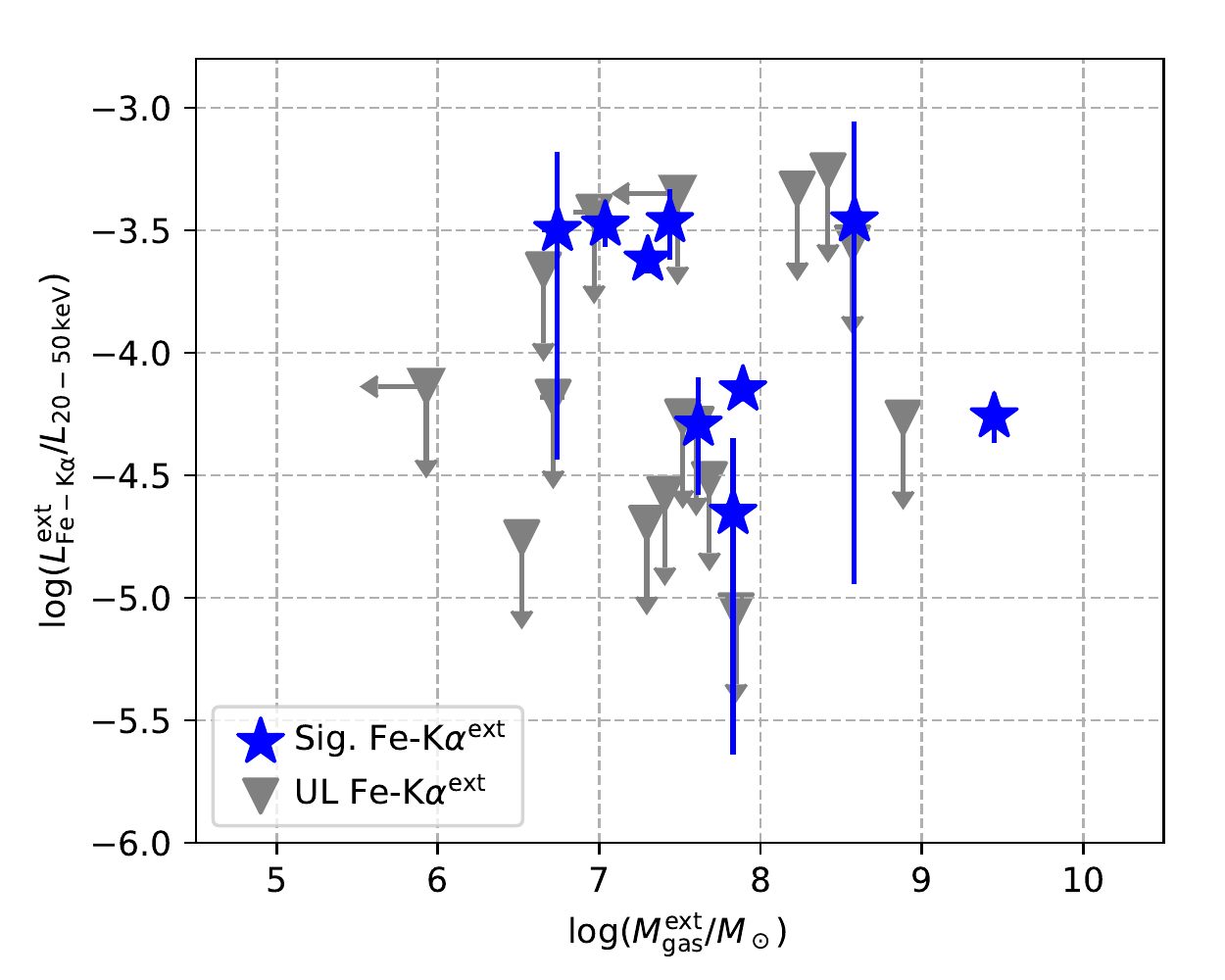}    
    \includegraphics[width=8.5cm]{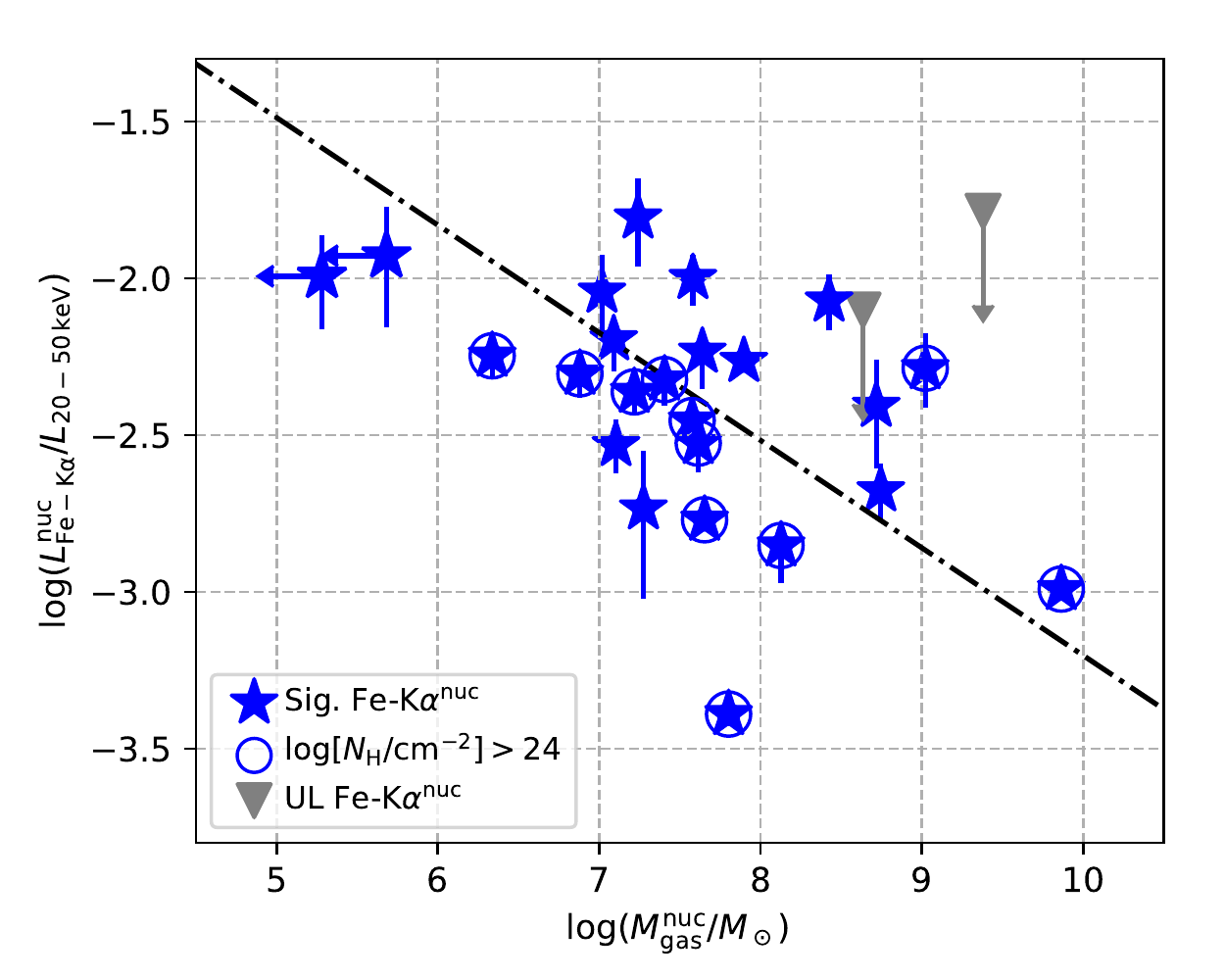}\vspace{-0.2cm}
    \caption{
    Scatter plots of the molecular gas mass and the  Fe-K$\alpha$-to-continuum ratio for the external (left) and  nuclear (right) scales. 
    AGNs with significant Fe-K$\alpha$ emission 
    ($L^{\rm ext}_{\rm Fe-K\alpha}/\Delta L^{\rm ext}_{\rm Fe-K\alpha} > 1$, 
    or $L^{\rm nuc}_{\rm Fe-K\alpha}/\Delta L^{\rm nuc}_{\rm Fe-K\alpha} > 1$)
    are indicated by blue stars. 
    In contrast, those with upper limits (UL) are shown as  gray triangles. 
    In the left figure, NGC 1052 with $\log(M^{\rm ext}_{\rm gas}/M_\odot) < 4$ is located outside.    
    In the right panel, open circles are over-plotted for AGNs whose line-of-sight X-ray absorbing column densities are above $10^{24}$ cm$^{-2}$. 
    A trend was found on the nuclear scale but not on the external scale. The dot-dashed line in the right panel is derived using a bootstrap method that incorporates the upper limits. 
    }
    \label{fig:m_gas_vs_lfe_lx}
\end{figure*}

\begin{figure*}
    \centering
    \includegraphics[width=8.5cm]{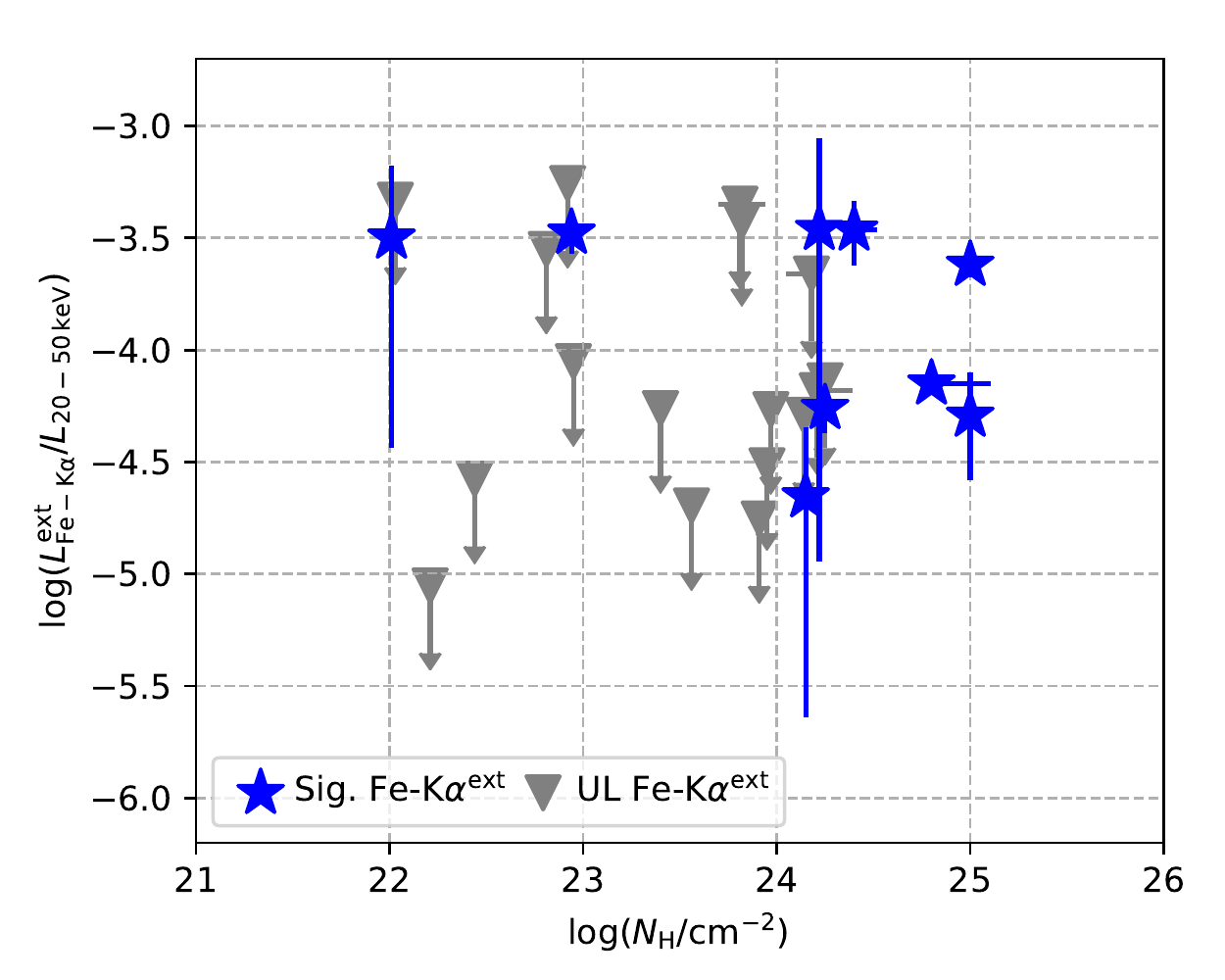}
    \includegraphics[width=8.5cm]{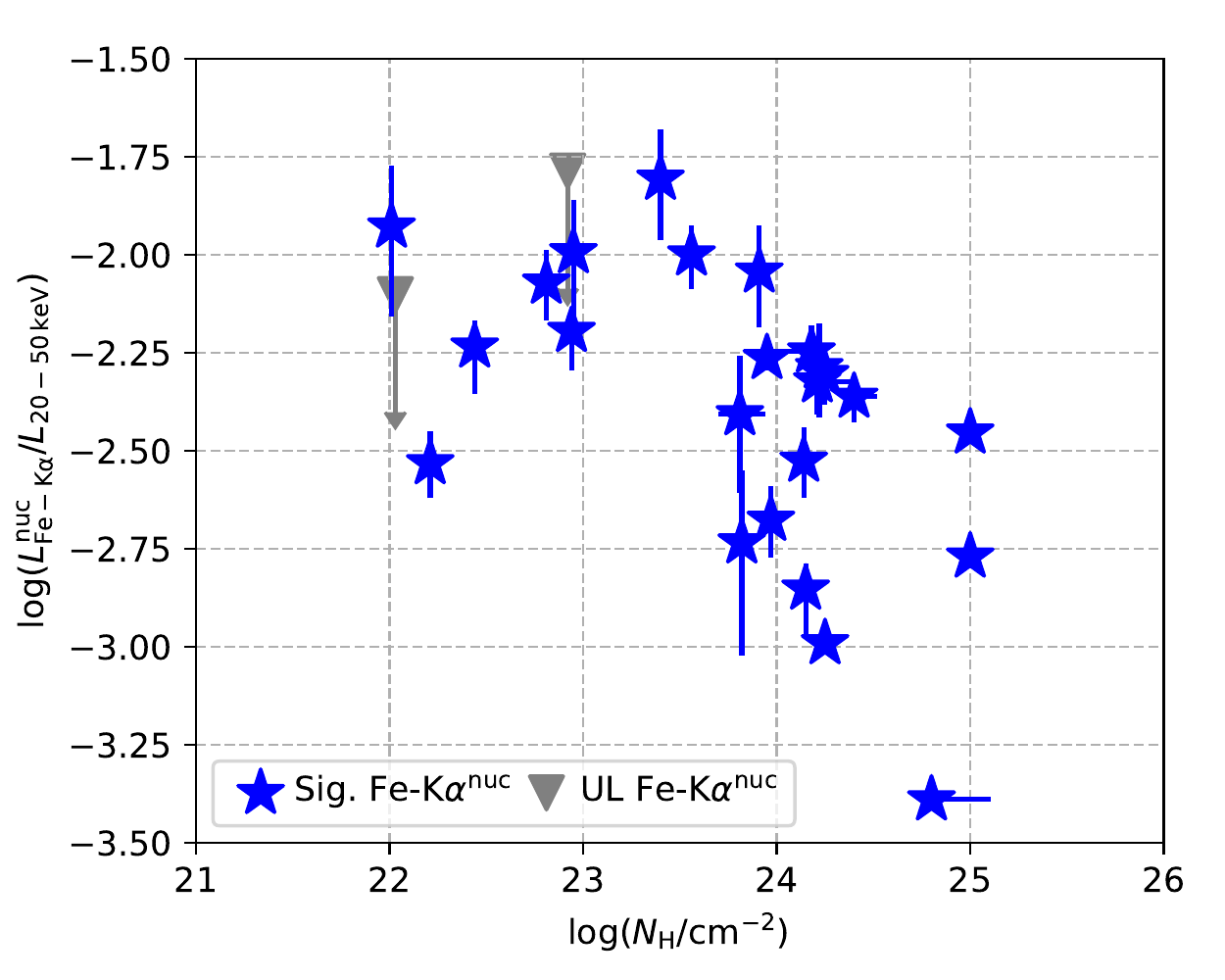}\vspace{-0.2cm} 
    \caption{(Left) Scatter plot of the X-ray column density and the ratio between the external Fe-K$\alpha$ and 20--50\,keV luminosities. 
    (Right) Same, but for nuclear Fe-K$\alpha$ emission.
    AGNs with significant Fe-K$\alpha$ emission 
    ($L^{\rm ext}_{\rm Fe-K\alpha}/\Delta L^{\rm ext}_{\rm Fe-K\alpha} > 1$ 
    or $L^{\rm nuc}_{\rm Fe-K\alpha}/\Delta L^{\rm nuc}_{\rm Fe-K\alpha} > 1$)
    are indicated by blue stars. 
    In contrast, those with upper limits (UL) are shown as gray triangles.}
    \label{fig:nh_vs_lfe_lx}
\end{figure*}

\subsection{Relation of Fe-K$\alpha$ Emitting Sites and Molecular Gas Distributions}\label{sec:ext_fe_co}

We present here that the ALMA CO($J$=2--1) data give us
interesting insights into the impact of the AGN X-ray irradiation 
on the ISM.
We take two approaches.
First, we compare the distributions of the CO($J$=2--1) and 
Fe-K$\alpha$ emission, and discuss their spatial relation. 
As the external Fe-K$\alpha$ emission is not detected for most of the objects, we particularly focus on NGC 1068, NGC 2110, and NGC 4945, for which the external Fe-K$\alpha$ emission is detected at the highest significance levels of $\approx 8\sigma$, $\approx 5\sigma$, and $\approx 10\sigma$, respectively (Table~\ref{tab:fe_data}). 
Note that although the external Fe emission of NGC 6240 is significantly detected at $\approx$ 4$\sigma$, we do not discuss this object because 
we do not correct for the presence of two nuclei in the 6--7\,keV-to-3--6\,keV image.
As detailed later in Section~\ref{sec:fe_vs_nh}, these highly significant detections are likely due to their higher signal-to-noise ratios.
Second, we statistically assess a relation of the CO($J$=2--1) 
and Fe-K$\alpha$ emission while considering all 26 targets.

\subsubsection{Spatial Comparison of the Fe and CO Emission}
\label{sec:spatial_comp}

For NGC 1068, NGC 2110, and NGC 4945, Figure~\ref{fig:fe_co21_images} compares the CO($J$=2--1) images and the X-ray ratio (6--7\,keV/3--6\,keV) images.
As discussed in Section~\ref{sec:ionization}, 
whereas the 6--7 keV spectrum of NGC 4945 includes the Fe-K$\alpha$ and Fe~{\sc xxv} lines, regardless of which is prominent, the ratio can trace photoionized iron atoms. 
This is also true for NGC 1068. 
In addition to the Fe-K$\alpha$ line, the external spectrum showed significant emission lines at 6.63$\pm0.02$ and 6.96$\pm0.03$\,keV in the rest frame (Figure~\ref{fig:xspec_ngc_1068}). Such high-energy lines have been interpreted as originating from photoionized gas by nuclear X-ray emission \citep[e.g.,][]{Bia01,Ogl03}. 
Regarding NGC 2110, only the Fe-K$\alpha$ emission with high EWs of $\sim$ 1.5--1.6\,keV, favoring photoionization of iron atoms, was detected. 
Thus, all X-ray-ratio images likely trace the gas irradiated by the AGN X-ray emission. 
We discuss the objects separately as follows.

NGC 1068: Fe emission is elongated in the northeast--southwest direction. A similar elongation was observed for the 
CO($J$=2--1) emission. 
However, while bright Fe and CO($J$=2--1) emission is observed
in the northeast region located $\sim$ 3\arcsec\ ($\sim$ 180 pc) from the nucleus, they are spatially separated.
The extended Fe emission is closer to the nucleus than the CO($J$=2--1) emission. This suggests that the ISM closer to the AGN is   preferentially irradiated by the AGN X-ray emission, and the CO($J$=2--1) emission may be suppressed.
Whereas there are a few conceivable suppressing mechanisms (e.g., CO and/or H$_2$ gas destruction or super-thermal), 
CO molecules may be dissociated into C and O, as recently proposed for a nearby luminous AGN of NGC 7469 by \cite{Izu20b}. 
However, we do not draw a conclusion regarding the dominant process in this study. 
For a robust conclusion, additional information is needed. 
Lastly, in the southwest region, CO($J$=2--1) and Fe distributions appear to be separated, similar to those in the northeast region. 
These facts suggest that the irradiated gas is seen to be 
close to the molecular gas, but they are spatially separated, likely owing to the change in the properties of the ISM therein. 

NGC 2110: This AGN host galaxy also provides a similar picture to that suggested by NGC 1068. 
Although the south-bright Fe-K$\alpha$ emission seems to be associated with faint CO($J$=2--1) emission, one can see a clear spatial separation between the brightest CO($J$=2--1) and Fe-K$\alpha$ emission.
This is consistent with the fact that AGN X-ray radiation contributes to the change in the physical properties of the ISM. 
According to \cite{Ros19}, who studied near- and mid-infrared spectroscopy data of NGC 2110, the north--south region with 
CO($J$=2--1) deficit and bright Fe-K$\alpha$ emission may be filled by ionized and warm molecular gas. One can also refer to \cite{Fab19b} and \cite{Kaw20} for more details.

NGC 4945: Because this galaxy is an almost edge-on system, 
it is quite difficult to discuss whether the Fe and CO($J$=2--1) emission is spatially separated or not, such as for NGC 1068 and NGC 2110. 
However, we remark two important points, suggesting the AGN X-ray irradiation of the ISM. 
First, the Fe emission is preferentially bright along the galaxy disk in the northeast-to-southwest direction, where CO emission is also bright. 
Second, fainter Fe emission with 6--7\,keV-to-3--6\,keV ratios of $\sim 0.15$ is seen 
over $\sim$ 4\arcsec\ ($\sim$ 160 pc) in the northwest direction, where 
the CO($J$=2--1) emission is moderately bright. The distribution appears as a double horn, and this could be due to limb brightening of a hollow, cone shaped outflow oriented close to the plane of sky that was seen and modeled with inclination angle of 75$^\circ$ in \cite{Ven17} for optical [N\,{\small II}] emission.
Seemingly, the Fe-K$\alpha$ outflow lies closer to the nucleus than [N\,{\small II}], and this may be because the Fe-K$\alpha$ emission is produced by the fluorescent process, which is preferably expected in dense nuclear regions. 
Thus, given that the outflow is driven by the AGN, it is natural that the AGN X-ray emission irradiates the gas therein.

\subsubsection{Relation between Fe-K$\alpha$ and CO($J$=2--1) Emission}\label{sec:fe_vs_co21}

To further discuss the possible spatial separation between the cold molecular gas and the Fe-K$\alpha$ emission, we examine the relation between the gas mass ($M^{\rm ext}_{\rm gas}$) and the ratio of the Fe-K$\alpha$ and continuum luminosities ($L^{\rm ext}_{\rm Fe-K\alpha}/L_{\rm 20-50~keV}$).
As explained in the second paragraph from the last
of Section~\ref{sec:alma_data}, 
the gas masses were derived from the external CO($J$=2--1) luminosities.
The left panel of Figure~\ref{fig:m_gas_vs_lfe_lx} shows a scatter plot of the two parameters. Here, because the two physical values are derived from the same regions for each object, this discussion is unaffected by differences in object distance.
No significant correlation is found with $P \approx 0.7$.  
While we cannot infer much from the plot, the absence of a positive correlation is consistent with the spatial separation. 
Thus, by considering the results in Section~\ref{sec:spatial_comp} (particularly for NGC 1068 and NGC 2110), we suggest that X-ray irradiation may 
alter the ISM in such a way that it cannot be traced well by CO($J$=2--1) emission.

\subsubsection{Relation between Fe-K$\alpha$ and Column Density}\label{sec:fe_vs_nh}

Recently, \cite{Ma20} performed a systematic study of nearby Compton-thick AGNs ($z \sim$ 0.01) to search for extended Fe-K$\alpha$ emission and found it at a relatively high rate (i.e., five out of seven). 
Following their work, we examine whether there is a relation between the sight-line X-ray absorbing column density and the Fe-line-to-continuum luminosity ratio ($L^{\rm ext}_{\rm Fe-K\alpha}/L_{\rm 20-50~keV}$), as shown in the left panel of Figure~\ref{fig:nh_vs_lfe_lx}). 
Although we can find that the detection rate of the extended Fe-K$\alpha$ emission is relatively high for Compton-thick AGNs ($\sim 50$\%), 
it is not so straightforward to conclude the correlation given that 
some Compton-thick AGNs with 
$N_{\rm H} \sim 10^{24}$ cm$^{-2}$ have low $L^{\rm ext}_{\rm Fe-K\alpha}/L_{\rm 20-50~keV}$ ratios, whereas a Compton-thin AGN of NGC 2110 with $\log(N_{\rm H}/{\rm cm}^{-2}) \approx 23$ shows a high $L^{\rm ext}_{\rm Fe-K\alpha}/L_{\rm 20-50~keV}$ ratio of $\approx 10^{-3.5}$.

\begin{figure}
    \centering
    \includegraphics[width=8.5cm]{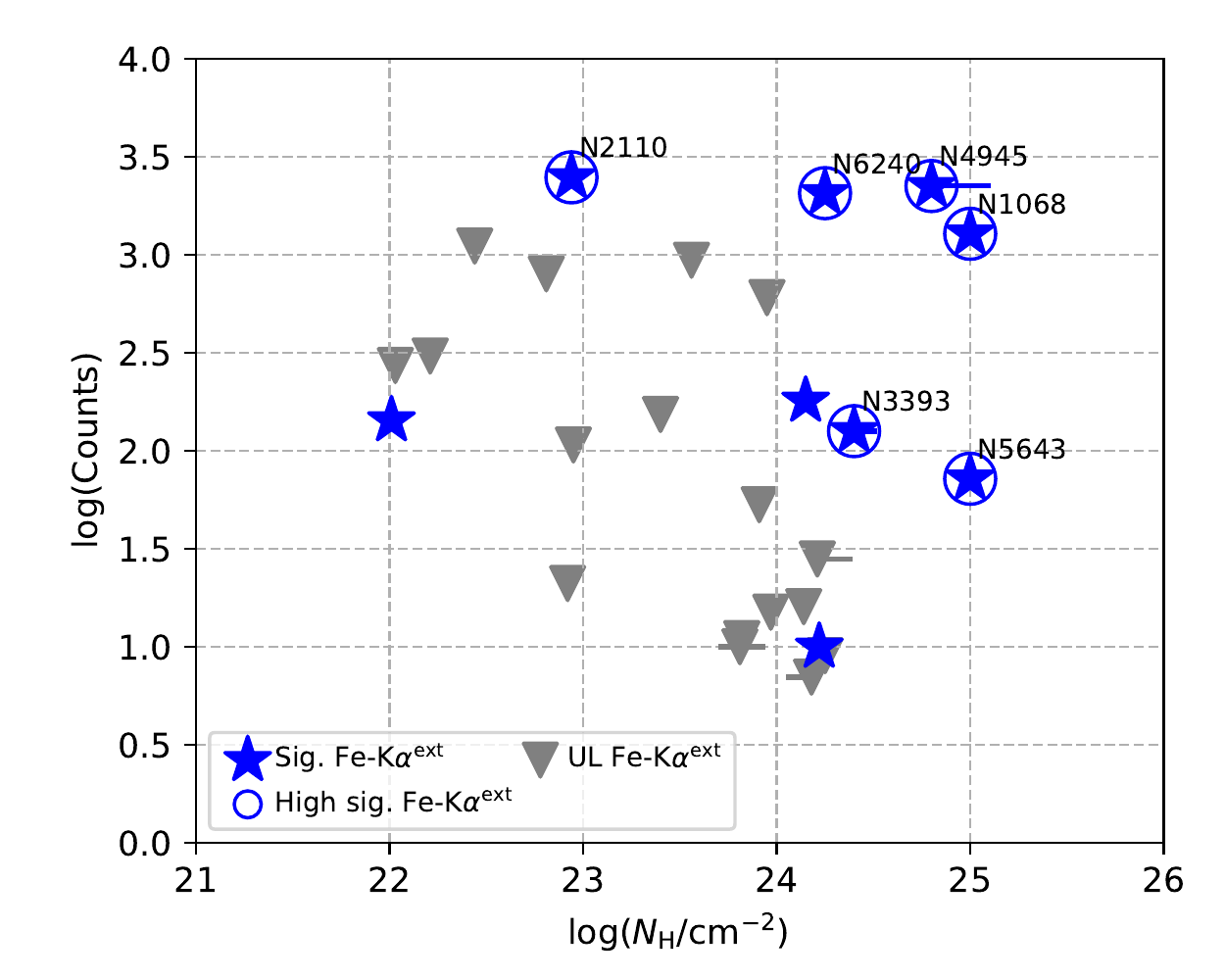}
    \vspace{-0.7cm}
    \caption{
    Photon count of the external spectrum in the 3--7\,keV band as a function of column density. AGNs with significant Fe-K$\alpha
    ^{\rm ext}$ detections ($L^{\rm ext}_{\rm Fe-K\alpha}/\Delta L^{\rm ext}_{\rm Fe-K\alpha} > 1$) are indicated by blue stars. Moreover, open circles and names are overplotted for those with  significance levels above 2$\sigma$. 
    In contrast, those with upper limits (UL) are shown as
    gray triangles.
    Note that the target names are partly omitted for clarity  (e.g., NGC 2110 is presented as N2110). 
    }
    \label{fig:f_x_vs_lfe_lx}
\end{figure}

We suspect that their spectra with high signal-to-noise ratios and high column densities help the detection, explaining the high detection rate in the Compton-thick AGNs. 
Simply, a spectrum with a higher signal-to-noise ratio allows the detection of fainter emission. In addition, a higher column density suppresses nuclear emission more, making it easier to detect
extended emission. 
Our suggestion is supported by Figure~\ref{fig:f_x_vs_lfe_lx}, where the spectral count in the 3--7\,keV band is plotted as a function of the column density. In the figure, AGNs with significant Fe-K$\alpha$ detections above 2$\sigma$ have spectra with higher spectral counts and higher column densities. The detection for the Compton-thin AGN NGC 2110 can be explained in the same way. Its spectrum has a high spectral count of $\sim 2500$.

\begin{figure*}
    \centering
    \includegraphics[width=8.5cm]{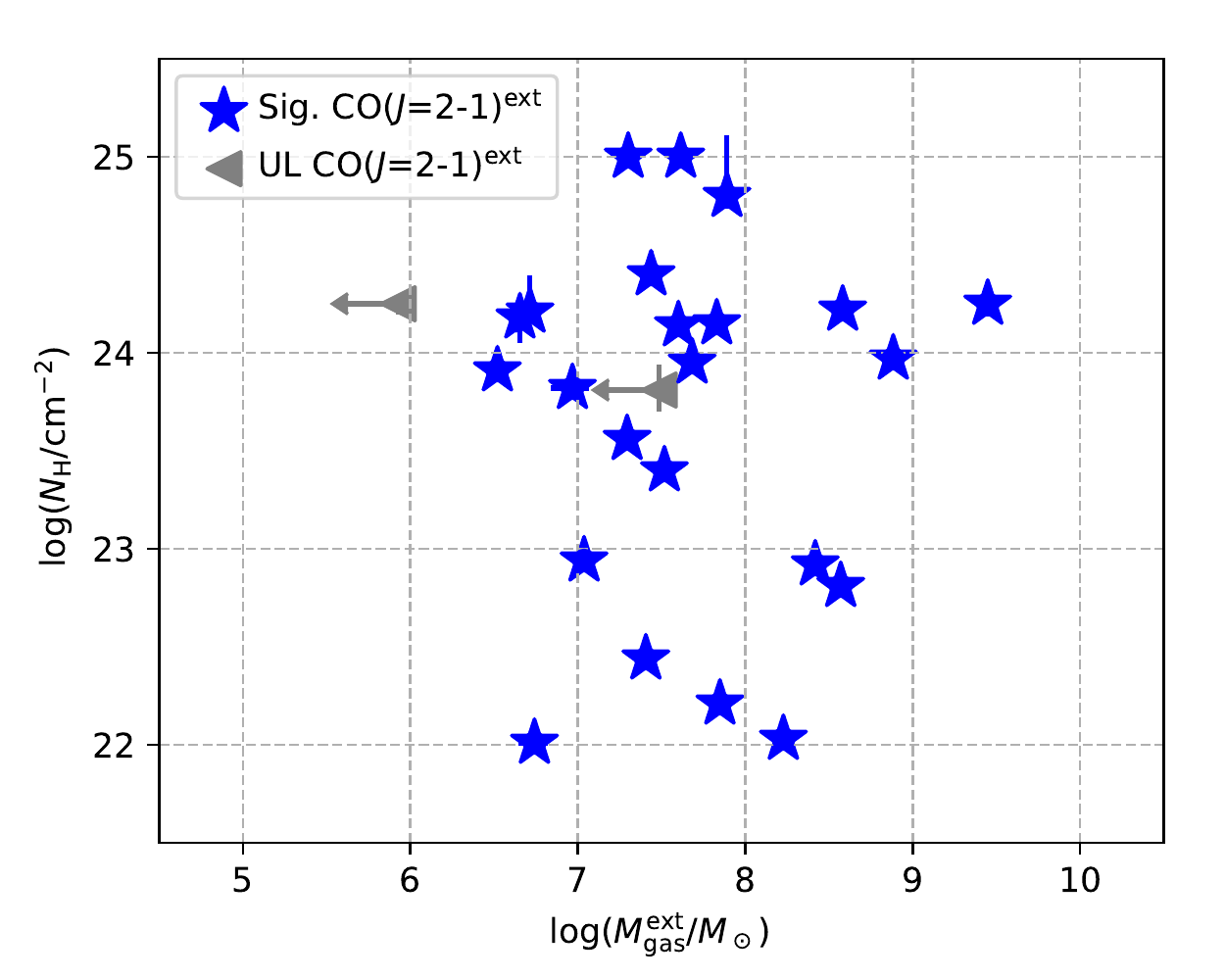}
    \includegraphics[width=8.5cm]{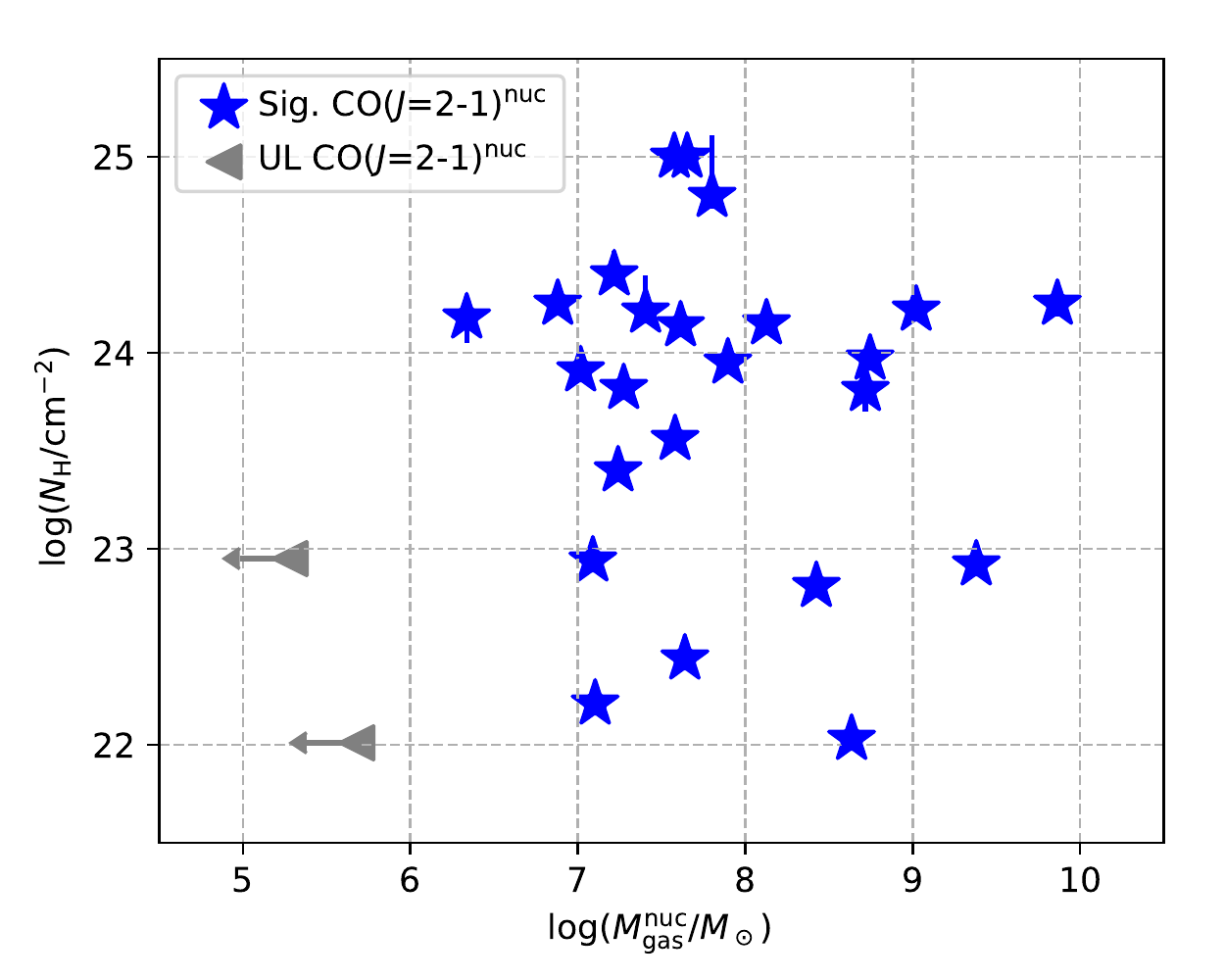}\vspace{-0.2cm}
    \caption{ (Left) Scatter plot of the external-scale molecular gas mass and the absorbing hydrogen column density derived in the X-ray band.
    NGC 1052 with $\log(M^{\rm ext}_{\rm gas}/M_\odot) < 4$ is located outside. 
    (Right) Same, but for nuclear-scale molecular gas.
    AGNs with significant CO($J$=2--1) emission 
    ($L^{\rm ext}_{\rm CO}/\Delta L^{\rm ext}_{\rm CO} > 1$ 
    or $L^{\rm nuc}_{\rm CO}/\Delta L^{\rm nuc}_{\rm CO} > 1$)
    are indicated by blue stars. 
    In contrast, those with upper limits (UL) are shown as 
    gray triangles. 
    }
    \label{fig:gas_mass_vs_nh}
\end{figure*}

In addition, we check the hypothesis that AGNs with higher column densities are in gas-rich galaxies (the right panel of Figure~\ref{fig:gas_mass_vs_nh}). With the bootstrap method, no correlation is found ($P = 0.55^{+0.14}_{-0.13}$), 
suggesting that there is no clear connection between the obscuring material and the molecular gas on scales of $\gtrsim 100$ pc. 
No correlation is confirmed also on the external scales
$(P = 0.88^{+0.10}_{-0.13}$; the left panel of Figure~\ref{fig:gas_mass_vs_nh}). 
A connection on smaller scales of $\sim 10$ pc was however suggested by \cite{Gar21}, who found a positive correlation between the line-of-sight column densities derived in the X-ray band and those derived from CO($J$=3--2).

Based on the discussion in Sections~\ref{sec:fe_origin} and \ref{sec:ext_fe_co}, we argue 
two points. First, the AGN can form an extended Fe-K$\alpha$ emission by irradiating the surrounding ISM (Section~\ref{sec:fe_origin}). 
Second, the physical and/or chemical properties of the ISM are altered, and such ISM may be missed by CO($J$=2--1) emission or cold molecular gas tracers (Section~\ref{sec:ext_fe_co}). 
Given that cold gas is the source of star formation, X-ray radiation could potentially function as a negative AGN feedback.

\begin{figure*}
    \centering
    \includegraphics[width=8.5cm]{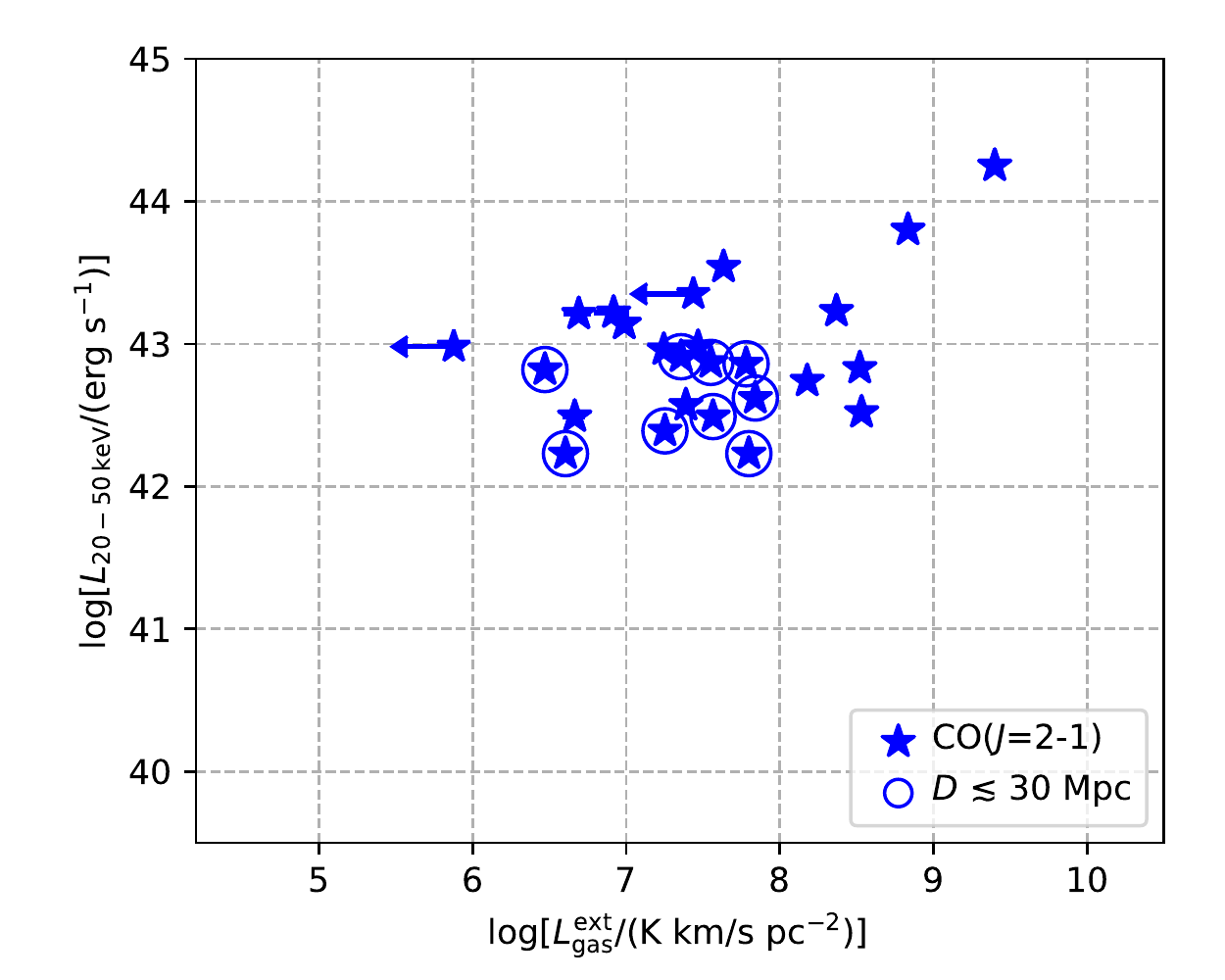}
    \includegraphics[width=8.5cm]{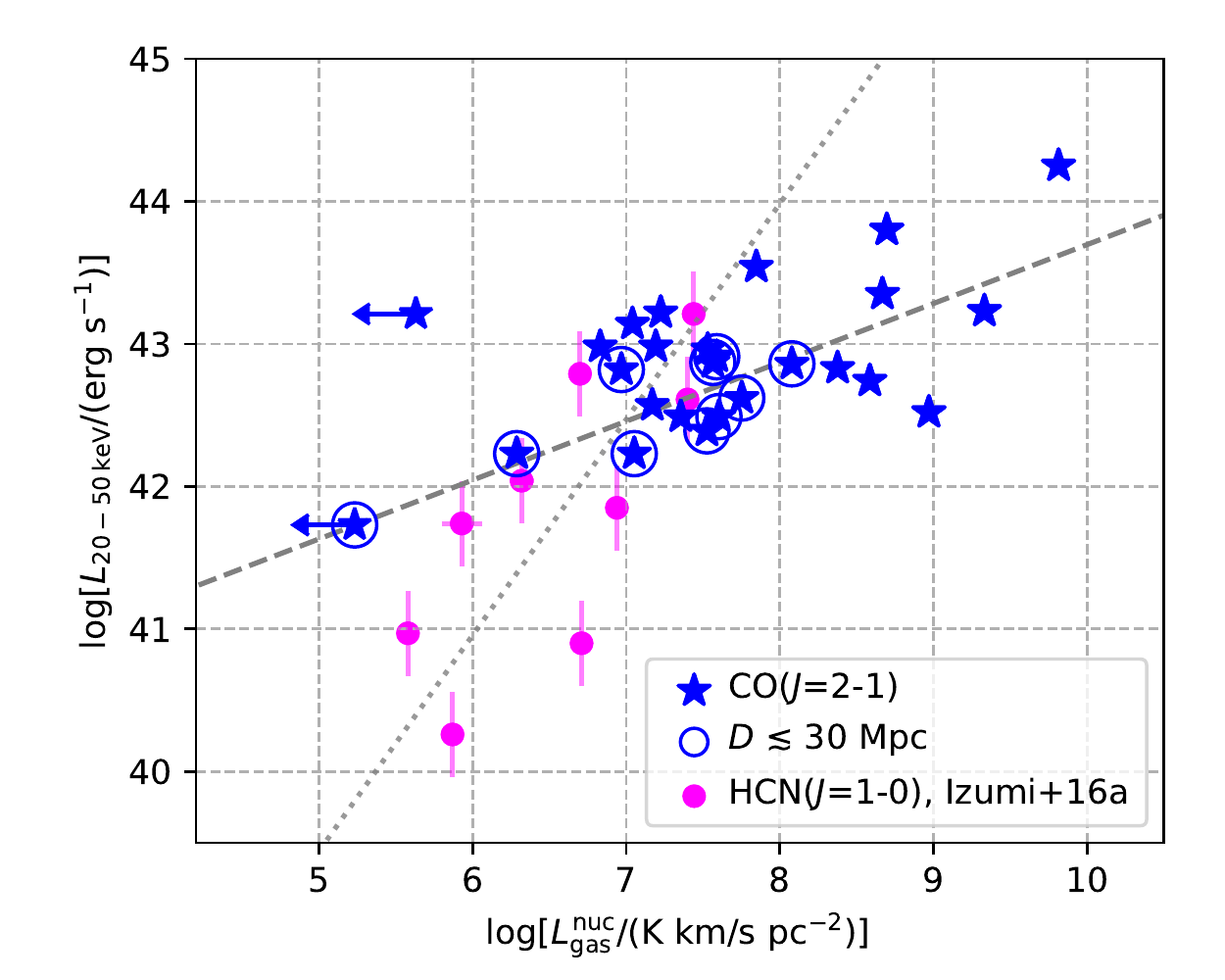}\vspace{-0.2cm}
    \caption{
    (Left) Scatter plot of the 
    CO($J$=2--1) luminosity in the external region 
    and the X-ray (20--50\,keV) luminosity. 
    For the CO($J$=2--1) data of the 10 nearby AGNs at distances  $\lesssim$ 30 Mpc, unfilled circles are overplotted. As inferred from the figure, no significant correlation is found. 
    (Right) Dependence of the X-ray (20--50\,keV) luminosity on the CO($J$=2--1) luminosity (blue) and  HCN($J$=1--0) luminosity (magenta). 
    The CO luminosities are from the nuclear regions, and 
    the HCN ones are retrieved from \cite{Izu16b}. 
    The unfilled circles are overplotted in the same way as in the left panel. 
    The regression lines derived from the CO($J$=2--1) and HCN($J$=1--0) data (Equations~\ref{eqn:co} and \ref{eqn:hcn}) are indicated by dashed and dotted lines, respectively. 
    }
    \label{fig:l_nuc_gas_vs_lx}
\end{figure*}

\subsection{Nuclear-scale Gas and Mass Accretion onto SMBHs}\label{sec:dis_nuc_scale}

In this section, we focus on the nuclear regions, where we expect X-ray irradiation to have a stronger effect on the ISM.

\subsubsection{Molecular Gas Mass and Accretion Rate}\label{sec:acc_vs_gasmass}

To reveal the potential effect of X-ray radiation on the ISM, we first investigate a link between the mass of the surrounding cold molecular gas and the mass accretion rate. 
Similar studies have already been conducted. For example, \cite{Yam94} presented a significant correlation between 
the X-ray (0.5--4.5\,keV) and CO($J$=1--0) luminosities for a nearby sample of 13 Seyfert 1 galaxies and 5 quasars, supporting the link. The correlation was further assessed using larger samples. \cite{Mon11} found a significant correlation between the CO($J$=2--1) and X-ray (0.3--8\,keV) luminosity for 45 nearby Seyfert galaxies including type-1 and type-2 objects ($P$ = $2\times10^{-7}$) \citep[see also][]{Sha20}. However, a more recent study by \cite{Kos21}, who used CO($J$=2--1) and X-ray-based bolometric luminosities, did not find a significant correlation for an ultrahard X-ray ($>$ 10\,keV) sample consisting of $\approx$ 200 AGNs. 
In this context, our study can give new insights by exploiting 
the much higher resolutions of ALMA than those of single-dish
telescopes used in previous studies (i.e., NRAO 12-m, CSO, JCMT, and APEX). Specifically, we investigate the correlation by considering the physical scale of cold molecular gas or by separating the nuclear and external regions.

First, we examine their correlation in the external scale 
by using the 20--50\,keV luminosity ($L_{\rm 20-50~keV}$) and 
the CO($J$=2--1) luminosity ($L^{\rm ext}_{\rm CO({\it J}=2-1)}$)  
as proxies of the accretion rate and the gas mass, respectively. 
The measured values are listed in Tables~\ref{tab:sample} and \ref{tab:co_data}.
A scatter plot of $L_{\rm 20-50~keV}$ and $L^{\rm ext}_{\rm CO({\it J}=2-1)}$ is shown in the left panel of Figure~\ref{fig:l_nuc_gas_vs_lx}.
Adopting the same bootstrap analysis as used previously in Section~\ref{sec:thomson}, we obtain 
\begin{eqnarray}
&\log&(L_{\rm 20-50~keV}/{\rm erg~s}^{-1}) = 39.5^{+0.3}_{-0.2} \nonumber\\
&&+ 0.46^{+0.02}_{-0.04} \log (L^{\rm ext}_{\rm CO({\it J}=2-1)}/{\rm K~km~s^{-1}~pc}^{-2})
\end{eqnarray}
with $P = 0.25^{+0.07}_{-0.04}$ and $\rho_{\rm s} = 0.23\pm0.03$. No significant correlation is found. 
Furthermore, to reduce the effect caused by differences in object distance as much as possible, 
we apply the bootstrap analysis to only 10 nearby AGNs within $D \lesssim 30$ Mpc, where  the inner radii of the external regions (2\arcsec) span a range of $\approx$ 80--300 pc with a median of $\approx$ 200 pc. 
The derived regression line is 
\begin{eqnarray}
    &\log& (L_{\rm 20-50~keV}/{\rm erg~s}^{-1}) = 40.4^{+0.3}_{-0.2} \nonumber\\
    &&+ 0.30^{+0.02}_{-0.03} \log (L^{\rm ext}_{\rm CO({\it J}=2-1)}/{\rm K~km~s^{-1}~pc}^{-2})
\end{eqnarray}
with $P = 0.31^{+0.05}_{-0.06}$ and $\rho_{\rm s} = 0.38^{+0.10}_{-0.08}$.
We cannot find significant correlation. 
These insignificant correlations can be inferred from the left panel of Figure~\ref{fig:l_nuc_gas_vs_lx}, where a flat distribution of data points is observed in the range $L^{\rm ext}_{\rm CO({\it J}=2-1)} \approx 10^{6-8.5}$ K~km~s$^{-1}$~pc$^{-2}$.
Thus, there seems to be no strong link between the accretion rate and cool, molecular gas on a scale larger than $\sim$ 200 pc.

We then focus on the nuclear scale, as shown in the right panel of Figure~\ref{fig:l_nuc_gas_vs_lx}. In the same way, we derive a regression line for the entire sample as 
\begin{eqnarray}
&\log&(L_{\rm 20-50~keV}/{\rm erg~s}^{-1}) = 39.3^{+0.3}_{-0.2} \nonumber\\
&& + 0.47^{+0.02}_{-0.03} \log (L^{\rm nuc}_{\rm CO({\it J}=2-1)}/{\rm K~km~s^{-1}~pc}^{-2}).
\end{eqnarray}
The $p$-value and the Spearman rank coefficient are 
$P = 0.048^{+0.009}_{-0.012}$ and $\rho_{\rm s} = 0.39\pm0.02$, respectively. 
This result suggests a positive trend. For the nearby 10 AGNs, we obtain 
\begin{eqnarray}\label{eqn:co}
    &\log&(L_{\rm 20-50~keV}/{\rm erg~s}^{-1}) = 39.6^{+0.5}_{-0.2} \nonumber \\
    &&+ 0.41^{+0.03}_{-0.06} \log (L^{\rm nuc}_{\rm CO({\it J}=2-1)}/{\rm K~km~s^{-1}~pc}^{-2})
\end{eqnarray}
with $P = 0.047^{+0.034}_{-0.032}$ and $\rho_{\rm s} = 0.64^{+0.10}_{-0.06}$. 
The trend is then found to still remain, even for the limited sample for which the median of the outer radii of the nuclear regions (2\arcsec) is $\approx$ 200 pc. 
These results suggest that the molecular gas on a scale smaller than $\sim$ 200 pc is well linked to mass accretion onto the central SMBH.

To extend the discussion, we present a correlation analysis
between the HCN($J$=1--0) and the X-ray luminosities. The HCN($J$=1--0) line traces much denser gas than CO($J$=2--1), because its critical density of $\sim 10^{6}$ cm$^{-3}$ is much higher than that of CO($J$=2--1) ($\sim 10^{3}$ cm$^{-3}$ )\footnote{https://home.strw.leidenuniv.nl/\~{}moldata/}. 
\cite{Izu16b} compiled HCN data obtained with 
interferometers (i.e., Plateau de Bure Interferometer, Nobeyama Millimeter Array, and ALMA) for 10 nearby AGNs ($D \sim$ 20 Mpc, except for one at 70~Mpc).
The median of their resolutions or the beam sizes was $\approx$ 220 pc. They found a positive trend with the 2--10\,keV luminosity ($P = 0.058$). 
To compare the HCN($J$=1--0) result with our CO($J$=2--1) result as consistently as possible, we searched the literature for the 20--50\,keV luminosities of the 10 AGNs with the HCN luminosities. We found X-ray luminosities for all but NGC 6951.
By performing the bootstrap analysis on the updated HCN sample, 
a regression line is obtained as 
\begin{eqnarray}\label{eqn:hcn}
    &\log&(L_{\rm 20-50~keV}/{\rm erg~s}^{-1}) = 31.9\pm1.1 \nonumber\\
    &&+ 1.5\pm0.2 \log (L_{\rm HCN({\it J}=1-0)}/{\rm K~km~s^{-1}~pc}^{-2})
\end{eqnarray}
with $P = 0.036^{+0.063}_{-0.026}$ and $\rho_{\rm s} = 0.7^{+0.1}_{-0.2}$, suggesting a positive trend. 
Here, the uncertainties in the HCN($J$=1--0) luminosities are taken from \cite{Izu16b}. 
Those in the 20--50\,keV luminosities are set to 0.3 dex uniformly by following \cite{Izu16b}.
While some luminosities were derived based on Swift/BAT average spectra \citep{Ric17c}, others were estimated by a single epoch hard X-ray observation \citep[i.e., NGC~1097, NGC 4579, and NGC 2273;][]{Nem06,Mas16,You19}. Therefore, to account for the short-time flux variability, we adopt the uncertainty of 0.3 dex. For simplicity, we consider the same uncertainty even if a luminosity was derived from a long-term averaged BAT spectrum. 

An intriguing fact can be suggested from the possibility that the ratio of CO-to-HCN luminosities increases with $L_{\rm 20-50\,keV}$.
However, we note that our argument relies on the extrapolation of the regression lines due to a small overlap of the CO and HCN samples in the X-ray luminosity range, which suggests that the ratio of the CO($J$=2--1) and  HCN($J$=1--0) luminosities increases with X-ray luminosity. 
Quantitatively, for an increase in X-ray luminosity 
by an order of magnitude, the ratio changes by a factor of $\approx$ 30--60.

Whether the above result is inconsistent with the HCN-enhancement in AGN host galaxies discussed in previous studies \citep[e.g., ][]{Koh05,Kri08,Ima09,Izu16a} may be debated. 
We argue that there is no clear discrepancy with our result. 
For example, \cite{Koh05} compiled the data of CO($J$=1--0) and HCN($J$=1--0) on scales of a few hundred parsecs using the Nobeyama Millimeter Array. He then proposed that pure AGNs showed  
higher HCN($J$=1--0)-to-CO($J$=1--0) ratios ($\gtrsim 0.2$) than  composite objects (i.e., an AGN co-exists with nuclear starburst). 
Here, by paying attention to the X-ray luminosities of his targeted AGNs\footnote{The absorption-corrected 2--10 keV luminosities of the four pure-AGNs in \cite{Koh05} are as follows on a logarithmic scale: 42.9 for NGC 1068, 41.0 for NGC 1097, 40.89 for NGC 5033, and 40.6 for NGC 5194 
\citep[][]{Iyo96,Xu16,Ric17c}. 
Also, those of the composite objects are as follows: 
41.3 for NGC 3079, 42.1 for NGC 3227, 41.3 for NGC 4051, 
39.9 for NGC 6764, 42.1 for NGC 7479, 43.2 for NGC 7469 \citep[][]{Ric17c,Lah18}.}, 
we notice an interesting point. Three of the four pure AGNs are X-ray faint as $L_{\rm 2-10\,keV} = 10^{40-41}$ \ergs, with the remaining object of NGC 1068 having $L_{\rm 2-10\,keV} \approx 10^{43}$ \ergs. 
On the contrary, all composite galaxies, except NGC 6764 with $L_{\rm 2-10\,keV} = 10^{39.9}$ \ergs, are X-ray luminous with $L_{\rm 2-10\,keV} = 10^{41-43}$ \ergs.
Thus, a decreasing trend of the HCN($J$=1--0)/CO($J$=1--0) ratio with X-ray luminosity is observed, which is consistent with our results. Although we focus on the X-ray luminosity to discuss the HCN enhancement, there are other energetic phenomena to be considered, such as the AGN jet and the outflow. Thus, a more detailed discussion of individual objects is encouraged in future papers.

\subsubsection{Explanation for Trends of X-ray and Molecular Line Emission Luminosities}\label{sec:exp4mols}

To explain the increasing CO($J$=2--1)-to-HCN($J$=1--0) ratio with X-ray luminosity, we suggest that the decrease in gas density must be taken into consideration in addition to other possible causes, such as 
a change in the abundance ratio between HCN and CO molecules and the effective heating of CO molecules. 
For the discussion, we refer to an extensive numerical study of \cite{Mei07}. 
While considering the thermal and chemical balances of gas, they calculated the intensities of emission lines from molecular and atomic gases exposed to FUV and X-ray radiation over a large range of gas densities ($\sim 10^{2-6}$ cm$^{-3}$). Here, we focus particularly on the results in the range $10^{5-6}$ cm$^{-3}$, which covers a critical density of $\sim 10^6$ cm$^{-3}$ for HCN($J$=1--0) and also those of observationally suggested HCN($J$=1--0) emitting gas
($\sim 10^{5}$ cm$^{-3}$; e.g., \citealt{Kru07}).
As shown in their Figure 12, they found that for the  density range, the ratio of CO($J$=1--0)/HCN($J$=1--0) changes insignificantly over a wide range of incident X-ray flux of 1--100 erg cm$^{-2}$ s$^{-1}$.
The result can be applied to our AGNs, as their 1--100\,keV fluxes are $\approx$ 2--20 erg cm$^{-2}$ s$^{-1}$ at 1\arcsec\ from the center, or $\sim$ 200 pc as the median value. 
Here, we assume that the X-ray photons from an AGN are not heavily absorbed until they reach a gas cloud. 
Under such circumstances, \cite{Mei07} also suggested that the CO($J$=2--1)/CO($J$=1--0) ratio is almost unchanged with an increase in X-ray luminosity by one order of magnitude (their Figure 4). 
Thus, combined with almost no change in the CO($J$=1--0)/HCN($J$=1--0) ratio with X-ray luminosity, X-ray heating is insufficient to explain the observed rapid increase in CO($J$=2--1) luminosity. 
Hence, a decrease in the gas density remains the explanation for the rapid increase in the CO($J$=2--1)/HCN($J$=1--0) ratio.
Indeed, \cite{Mei07} predicted that the CO($J$=1--0)/HCN($J$=1--0) 
ratio could be rapidly increased (their Figure~12) 
with decreasing gas density because with decreasing 
density, HCN($J$=1--0) with a high critical density becomes weaker. Quantitatively, to explain the factor of $\approx$ 30–-60 decrease in HCN($J$=1--0)/CO($J$=2--1), the gas density must decrease by a factor of $\sim$ 10.
In summary, whereas the X-ray heating may have an impact on the line ratio, a decrease in the molecular gas density must be considered. 

\subsubsection{Physical Mechanism to Reduce Gas Density around the AGN}\label{sec:phys4gas}

Why can the gas density decrease around an X-ray luminous AGN? This may be due to the evaporation of gas clouds by harsh X-ray radiation.

As a supportive study for the evaporation picture, we refer to \cite{Hoc11}, who performed numerical simulations to investigate dynamical evolution of 
gas clouds located 10 pc from central mass-accreting SMBHs. 
The gas clouds were exposed to AGN X-ray radiation, and 
UV emission and cosmic rays originating from star-formation activity were also considered as the other possible mechanisms affecting the gas clouds \cite[see also][]{Hoc10}. 
The incident X-ray fluxes into gas clouds 
range over 0--160 erg cm$^{-2}$ s$^{-1}$, and encompass the expected fluxes for our AGNs, as mentioned previously (i.e., 2--20 erg cm$^{-2}$ s$^{-1}$). In contrast, some of our SMBH masses 
are larger by $\sim$ one order of magnitude than the range considered in their work (10$^{6-8} M_\odot$). Specifically, among 15 AGNs whose SMBH masses were estimated using stellar dispersions in \cite{Kos17}, nine have masses larger than $10^8~M_\odot$ and the others have masses within the range of $10^{7-8}~M_\odot$. 
However, we consider molecular clouds located further away than 10~pc (a fixed distance used by the authors in their modeling), therefore we may simply refer to models where gravity from the BH is weak (i.e., models with BH mass of $10^6 M_\odot$).

An intriguing result of \cite{Hoc11} is that 
stronger X-ray emission evaporates a larger portion of the gas cloud. In harsh conditions, even a dense gas 
($> 10^{5}$ cm$^{-3}$) can be affected despite its strong self-shielding. Eventually, the mass of gas with high densities ($> 10^5$ cm$^{-3}$) decreases 
(e.g., see the simulation results with M04, M10, M16, and M22, in their Figure~7). 
Given that the resultant gas whose density is globally reduced would not be favored to excite HCN($J$=1--0) but would still manage to excite CO($J$=2--1), the prediction is qualitatively consistent with our finding (i.e., the  CO($J$=2--1)-to-HCN($J$=1--0) ratio increases with X-ray luminosity).

The above evaporation, or outflow, picture is supported by a negative trend between the ratio of $\log(L^{\rm nuc}_{\rm Fe-K\alpha}/L_{\rm 20-50~keV}$) and the molecular gas mass, as shown in the right panel of Figure~\ref{fig:m_gas_vs_lfe_lx}. 
The bootstrap analysis suggests their negative trend with $P = 0.05^{+0.11}_{-0.04}$ and $\rho_{\rm s} = -0.39\pm0.11$. 
A regression line is derived as 
\begin{eqnarray}\label{eqn:fefx2mgas}
      \log&&(L^{\rm nuc}_{\rm Fe-K\alpha}/L_{\rm 20-50~keV}) \nonumber \\
     &&  =  0.23^{+0.28}_{-0.22} 
      -0.34^{+0.03}_{-0.04} \log (M^{\rm nuc}_{\rm gas}/M_\odot). 
\end{eqnarray}
The $L^{\rm nuc}_{\rm Fe-K\alpha}/L_{\rm 20-50~keV}$ ratio can be used as a proxy for the solid angle of gas irradiated by the AGN X-ray emission. Also, given the positive trend of $L_{\rm 20-50~keV}$ and $M^{\rm nuc}_{\rm gas}$, as inferred from the right panel of Figure~\ref{fig:l_nuc_gas_vs_lx}, the equation can suggest that more luminous AGNs are less covered by dense gas that can be traced by Fe-K$\alpha$ emission. This is qualitatively consistent with a radiation-driven outflowing model \cite[e.g., ][]{Fab09,Ric17nat,Ogw21}, where more surrounding material is blown away by stronger AGN emission.

Furthermore, the radiation-driven model can be inferred from the right panel of Figure~\ref{fig:nh_vs_lfe_lx}, showing a scatter plot of
$L^{\rm nuc}_{\rm Fe-K\alpha}/L_{\rm 20-50~keV}$, or the covering factor, versus $N_{\rm H}$. Their rank coefficient 
is negative as $\rho_{\rm S} = -0.36^{+0.09}_{-0.10}$, although  the significance is low with $P = 0.07^{+0.11}_{-0.05}$. 
This negative $\rho_{\rm S}$ value is indeed consistent with a model prediction of \cite{Fab09} (e.g., see their Figure~1) that AGNs with higher Eddington ratios tend to have higher absorbing column densities and lower covering factors \citep{Ric17c}, thus inferring a negative trend of covering factor and column density.

One might think that the decrease of $L^{\rm nuc}_{\rm Fe-K\alpha}/L_{\rm 20-50~keV}$ with $N_{\rm H}$ suggests the absorption of Fe-K$\alpha$ emission, especially in Compton-thick AGNs with $N_{\rm H} > 10^{24}$ cm$^{-2}$ \citep[e.g.,][]{Ric14b}, and wonder whether the absorption affects the discussion on the negative trend between $L^{\rm nuc}_{\rm Fe-K\alpha}/L_{\rm 20-50~keV}$ and $M^{\rm nuc}_{\rm gas}$, or the outflow picture. 
To see the possibility of the absorption effect, regression lines between $L^{\rm nuc}_{\rm Fe-K\alpha}/L_{\rm 20-50~keV}$ and $M^{\rm nuc}_{\rm gas}$ are constrained for Compton-thin and Compton-thick AGN samples, and are found to have normalizations of $0.30^{+0.05}_{-0.06}$ and $0.045^{+0.022}_{-0.021}$, respectively. 
Here, the slopes are fixed at that of Equation  (\ref{eqn:fefx2mgas}) to focus on the normalization.
The lower normalization for the Compton-thick AGNs is in fact consistent with their Fe-K$\alpha$ emission being absorbed. 
However, even if there is the effect of the absorption, that does not affect our argument. To reduce the influence of the absorption, 
we derive regression lines for the Compton-thin and Compton-thick samples by leaving slope and normalization as free parameters. For these samples, we obtain  slopes of $-0.26^{+0.05}_{-0.07}$ and $-0.38\pm0.02$, respectively, and both negative values
are consistent with the outflow picture.

\begin{figure}[!h]
    \centering
    \includegraphics[width=9cm]{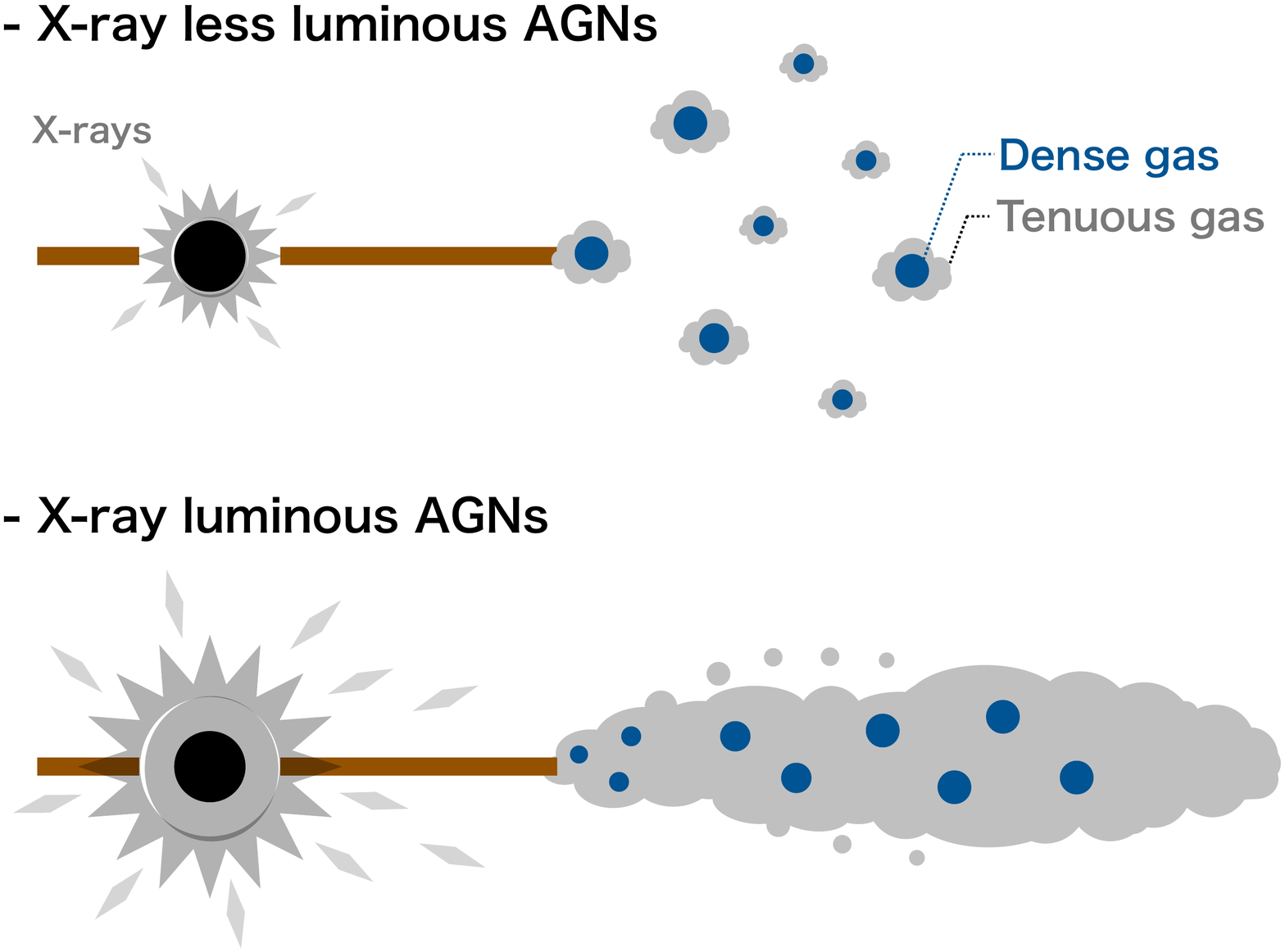}
    \caption{Schematic of the nuclear region. 
    This presents three key points. 
    (i) X-ray emission is enhanced with increasing the molecular gas mass.
    (ii) A larger fraction of surrounding gas is
    evaporated by stronger X-ray radiation, which reduces the global (averaged) gas density. 
    (iii) Stronger X-ray radiation also induces outflow, leading to a decreased gas covering factor.}
    \label{fig:schematic_pic} 
\end{figure}

In Figure~\ref{fig:schematic_pic}, we schematically illustrate the association between the central X-ray radiation and the surrounding molecular gas based on the above discussion. The figure shows that the mass accretion rate increases with the surrounding gas mass, and the resultant stronger AGN X-ray radiation evaporates a larger fraction of dense gas, yielding tenuous gas, and blows it out, reducing the gas covering fraction.

A natural prediction from this picture is that the AGN X-ray irradiation affects the surrounding star formation because dense gas is the site for this process. 
As discussed in the right panel of Figure~\ref{fig:l_nuc_gas_vs_lx}, the dense gas fraction would decrease with the X-ray luminosity. Given that the dense gas fraction is a key parameter determining the star-formation efficiency \cite[e.g.,][]{Gao04,Big15}, the AGN may have the potential to reduce the star-formation efficiency. 
This was also suggested from the hydrodynamical simulations of \cite{Hoc11}. 
Consequently, this theoretical work further suggests that even the initial mass function of stars may be altered.

Motivated by Figure 4 of \cite{Gao04}, showing that 
a dramatic increase of HCN($J$=1--0)-to-CO($J$=1--0) luminosity ratio for luminous and ultra-luminous infrared galaxies (i.e., very actively star-forming galaxies), we assess the correlation 
between $L^{\rm nuc}_{\rm Fe-K\alpha}/L_{\rm 20-50~keV}$ and 
SFR (the right panel of Figure~\ref{fig:sfr_vs_lfelx}).
A negative trend is favored from $\rho_{\rm S} = -0.32^{+0.07}_{-0.11}$, while no significant correlation is found ($P$ = $0.12^{+0.12}_{-0.09}$).
We note that for a sample of the nearest AGNs with $D \lesssim$ 30 Mpc, $\rho_{\rm S} = -0.47^{+0.14}_{-0.07}$ and $P$ = $0.21^{+0.08}_{-0.07}$. 
Thus, with our data, we cannot explicitly suggest that the star formation is being suppressed by X-ray irradiation. However, given that the SFRs  were measured at larger apertures than 2\arcsec\, of the nuclear regions, it would be too early to conclude the absence of the effect.

In summary, as our second response to the main question, we suggest that the AGN hard X-ray radiation has the potential to suppress the surrounding star formation by evaporating as well as blowing out the surrounding gas. Thus, this type of negative AGN feedback may be triggered by AGN hard X-ray emission.

\section{Summary}\label{sec:sum}

Our main goal in this study is to understand \textit{
what impact the hard X-ray irradiation has on the ISM and star formation}.
For this purpose, we have performed 
a systematic analysis of ALMA and Chandra data taken for 26 nearby ($z < 0.05$ or $D \lesssim$ 200 Mpc) AGNs, selected from the 
70-month Swift/BAT ultrahard X-ray catalog \citep{Bau13}.
By exploiting their high-resolution data, we defined the nuclear region with a central radius of 2\arcsec\ and the external annular region of 2\arcsec--4\arcsec.
The 2\arcsec\ scale corresponds to $\sim$ 100--600 pc for the majority of our AGNs (19 out of 26 AGNs). 

First, by focusing on the external regions, we suggest that the AGN can form extended hard X-ray emission by irradiating the surrounding ISM, 
and it is difficult to trace the X-ray-irradiated gas by the CO($J$=2--1) line (Section~\ref{sec:fe_origin}; Section~\ref{sec:ext_fe_co}). Thus, AGN X-ray irradiation has the potential to alter the physical and chemical properties of ISM. This suggestion is based on the following findings. 

\begin{itemize}
    \item We find that the $L^{\rm ext}_{\rm Fe-K\alpha}/L_{\rm 20-50~keV}$ ratio, a proxy for the solid angle of gas irradiated by AGN X-ray emission, does not correlate with the scattering fraction ($f_{\rm scat}$), disfavoring Thomson scattering as the general mechanism that forms extended Fe-K$\alpha$ emission 
    (Figure~\ref{fig:fscat_vs_lfelx} and Section~\ref{sec:thomson}). 
    \item We find that, particularly for some AGNs with bright Fe lines, the properties of the Fe lines (e.g., high EW and highly ionized Fe lines) suggest 
    photoionized gas as their origin (Section~\ref{sec:ionization}). 
    \item No correlation is found between the external Fe-K$\alpha$ luminosity ($L^{\rm ext}_{\rm Fe-K\alpha}$) and the galaxy-scale SFR. 
    This supports the idea that HMXBs, one of the most plausible classes of stellar objects that emit Fe-K$\alpha$ photons, do not strongly contribute to the extended Fe-K$\alpha$ emission
    (Section~\ref{sec:hxrb4fe} and Figure~\ref{fig:sfr_vs_lfe}).     
    \item By comparing the CO($J$=2--1) and X-ray ratio (6--7\,keV/3--6\,keV) images in the two AGNs with the well-detected Fe-K$\alpha$ emission (NGC 1068 and NGC 2110), we find that 
    regions with bright Fe emission appear to be separated from those with CO($J$=2--1) emission. For the other galaxy of NGC 4945 with significant Fe-K$\alpha$ emission ($\approx 10\sigma$), it is difficult to discuss the spatial relation between the Fe and CO($J$=2--1) emission because of the edge-on view. 
    However, the brighter Fe emission in its galaxy disk and fainter one in an outflow region would suggest that X-ray irradiation occurs  (Section~\ref{sec:spatial_comp} and Figure~\ref{fig:fe_co21_images}).
    \item We find no significant correlation between the 
    the extended Fe-K$\alpha$ emission ($L^{\rm ext}_{\rm Fe-K\alpha}/L_{\rm 20-50~keV}$) and the mass of the cold molecular gas ($M^{\rm ext}_{\rm gas}$). This is consistent with the spatial separation
    (Section~\ref{sec:fe_vs_co21} and Figure~\ref{fig:m_gas_vs_lfe_lx}).  
\end{itemize}

Furthermore, we have suggested that the AGN hard X-ray radiation has the potential to suppress the surrounding star formation by evaporating and blowing out the surrounding gas. Thus, negative AGN feedback may be triggered by AGN hard X-ray emission. The results on which our suggestion is based are as follows:

\begin{itemize}
    \item We find a positive trend between the 20--50\,keV luminosity ($L_{\rm 20-50~keV}$) and the nuclear CO($J$=2--1) luminosity ($L^{\rm nuc}_{\rm CO}$). Furthermore, we confirm a positive trend between the 20--50\,keV luminosity ($L_{\rm 20-50~keV}$) and the nuclear HCN($J$=1--0) luminosity ($L^{\rm nuc}_{\rm HCN}$) as well 
    (Section~\ref{sec:acc_vs_gasmass} and Figure~\ref{fig:l_nuc_gas_vs_lx}).
    \item Based on the above obtained trends, we further find that the ratio of CO($J$=2--1)-to-HCN($J$=1--0)     luminosities increases with the 20--50\,keV luminosity, 
    suggesting a decrease in the dense gas fraction with X-ray luminosity (Section~\ref{sec:exp4mols}). 
    \item A negative trend is found between $L^{\rm nuc}_{\rm Fe-K\alpha}/L_{\rm 20-50~keV}$ and $M^{\rm nuc}_{\rm gas}$. Given the positive trend between 
    $L_{\rm 20-50~keV}$ and $M^{\rm nuc}_{\rm gas}$ (Figure~\ref{fig:l_nuc_gas_vs_lx}),
    this is consistent with a radiation-driven AGN outflow model \citep[e.g., ][]{Fab09,Ric17nat}. This 
    is supportive evidence for 
    the evaporation of gas in the nuclear region (Section~\ref{sec:phys4gas}).
\end{itemize}

Future studies using other molecular and atomic emission lines will help to constrain the physical parameters, such as density, and kinetic temperature, and the chemical abundance. These are valuable for a better understanding of the relation between the AGN and its circumnuclear material. 
In addition, our sample is limited to nearby AGNs ($\lesssim$ 200 Mpc).
However, future high-resolution X-ray telescopes \citep[e.g., AXIS and Lynx;][]{Lyn18,Mus19} will improve this situation and provide  opportunities to study distant and much more luminous AGNs. 
They are likely to be interesting targets because they are expected to affect the surrounding ISM and star formation on larger scales ($\gg$ 100 pc). 
Finally, much sharper images anticipated from a super-resolution of MIXIM down to $\lesssim 0\arcsec.01$--0.\arcsec1 will provide valuable opportunities to resolve nuclear structures 
of nearby AGNs at $< $ 10 pc \citep{Hay18,Asa20}.

\newpage

\begin{acknowledgments} 

We thank the anonymous referee for careful reading and valuable suggestions that improved the manuscript. 
Also, we thank Kohei Ichikawa for providing the SFR data in a machine readable form. We acknowledge support from Fondecyt postdoctral fellowship 3200470 (T.K.), and Fondecyt Iniciacion grant 11190831 (C.R.). 
T.K., T.I. and M.I. are supported by JSPS KAKENHI grant numbers JP20K14529, JP20K14531, and JP21K03632, respectively. 
Part of this work was financially supported by the Grant-in-Aid for JSPS Fellows for young researchers JP19J00892 (S.B.). The scientific results reported in this article are based on data obtained from the Chandra Data Archive. This research has made use of software provided by the Chandra X-ray Center (CXC) in the application packages CIAO. 
This paper makes use of the following ALMA data: 
ADS/JAO.ALMA\#2012.1.00474.S,\\
ADS/JAO.ALMA\#2013.1.00525.S,\\
ADS/JAO.ALMA\#2013.1.00623.S,\\
ADS/JAO.ALMA\#2013.1.01161.S,\\
ADS/JAO.ALMA\#2013.1.01225.S,\\
ADS/JAO.ALMA\#2015.1.00086.S,\\
ADS/JAO.ALMA\#2015.1.00370.S,\\
ADS/JAO.ALMA\#2015.1.01572.S,\\
ADS/JAO.ALMA\#2016.1.00232.S,\\
ADS/JAO.ALMA\#2016.1.00254.S,\\
ADS/JAO.ALMA\#2016.1.01279.S,\\
ADS/JAO.ALMA\#2016.1.01553.S,\\
ADS/JAO.ALMA\#2017.1.00236.S,\\
ADS/JAO.ALMA\#2017.1.00255.S,\\
ADS/JAO.ALMA\#2017.1.00904.S,\\
ADS/JAO.ALMA\#2017.1.01439.S,\\
ADS/JAO.ALMA\#2018.1.00538.S,\\
ADS/JAO.ALMA\#2018.1.00699.S.
ALMA is a partnership of ESO (representing its member states), NSF (USA) and NINS (Japan), together with NRC (Canada), MOST and ASIAA (Taiwan), and KASI (Republic of Korea), in cooperation with the Republic of Chile. The Joint ALMA Observatory is operated by ESO, AUI/NRAO and NAOJ. We appreciate the JVO portal (http://jvo.nao.ac.jp/portal/) operated by ADC/NAOJ for the quick look at the ALMA archive data. 
Data analysis was in part carried out on the Multi-wavelength Data Analysis System operated by the Astronomy Data Center (ADC), National Astronomical Observatory of Japan.
This work is supported by a publication committee in NAOJ. 

\end{acknowledgments}

\clearpage

\appendix 
\restartappendixnumbering

\section{Negligible Impact of Absorption on Nuclear CO($J$=2--1) Luminosity}\label{app:co_analysis} 

\begin{figure}[!h]
    \centering
    \includegraphics[width=13.5cm]{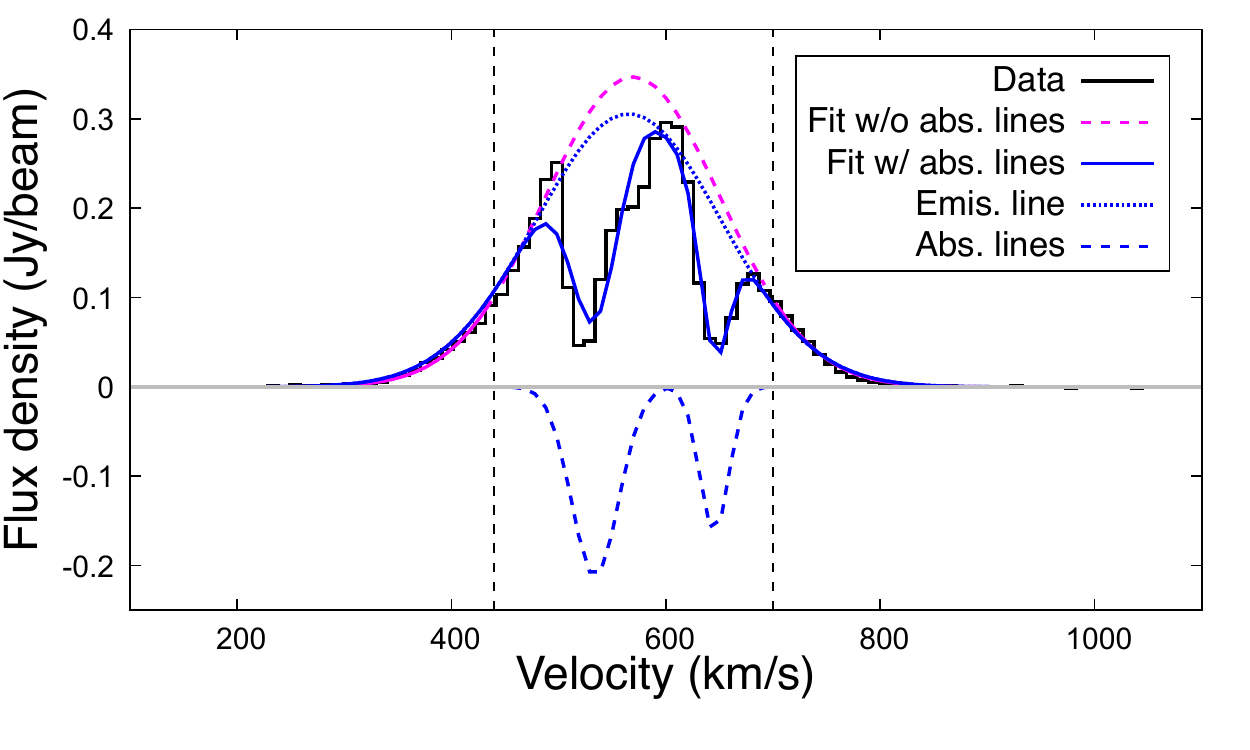}
    \caption{CO($J$=2--1) spectrum (black solid line) extracted from the AGN position of NGC 4945 with a beam of 0\arcsec.34$\times$0\arcsec.31. 
    The blue solid line indicates the best-fit model that considers one emission line (blue dotted line) and two absorption lines (blue dashed lines). The magenta line indicates the emission model fitted to the velocity ranges of 100--440\,km/s and 700--1100\,km/s, where absorption is negligible. The two black vertical dashed lines indicate 440\,km/s and 700\,km/s. 
    }
    \label{app:fig:co_spec_NGC4945}
\end{figure}

Here, we show that CO($J$=2--1) absorption inferred in some objects (e.g., see Figure~\ref{app:co_image} of NGC 3081, NGC 4945, NGC 5506, and NGC 5728) would not have a strong impact on the emission flux estimate. This suggestion is based on a detailed analysis of the CO spectrum of NGC 4945. This object is selected owing to its proximity ($\sim$ 8 Mpc); the nuclear spectrum is expected to be easily affected by nuclear absorption lines. 
The spectrum is extracted from the AGN position with a beam of 0\arcsec.34$\times$0\arcsec.31 (Figure~\ref{app:fig:co_spec_NGC4945}). The central spectrum is  subject to absorption most severely and is adopted to see the strongest impact. 
In the spectrum of Figure~\ref{app:fig:co_spec_NGC4945}, two absorption lines appear, and accordingly, we fit three Gaussian functions. One is used to reproduce the emission, and the other two functions are used to consider the absorption lines. Such an analysis was performed for nearby AGNs with absorption lines around the nuclei \citep[e.g.,][]{Rose19}. Figure~\ref{app:fig:co_spec_NGC4945} shows that the adopted model captures the overall shape of the observed spectrum. With this model, we find that the flux density of the absorption-corrected emission line is $\approx$ 66 Jy/beam km/s, and given the observed flux density of $\approx$ 49 Jy/beam km/s, we may underestimate the nuclear CO luminosity of NGC 4945 by $\approx$ 25\%. 
Furthermore, as a less model-independent method, we fit a single Gaussian function to the side velocity bands of 100--440 km/s and 700--1100 km/s, where the absorption lines are not significant. Thus, the fitted function is expected to trace the intrinsic shape of the emission, and we derive its flux to be $\sim$ 70 Jy/beam, which is almost the same as that derived by using the three functions. 
In both estimates, the degree of underestimate is small on a logarithmic scale, as it corresponds to $\sim$ 0.1 dex. 
A smaller effect is expected in the outer regions. 
Thus, there would not be a strong impact on the flux estimates and,  consequently, the overall discussion.

\clearpage

\section{Nuclear and external X-ray spectra}\label{app:xspec}

\begin{figure*}[!h]
    \centering
    \includegraphics[width=4.8cm]{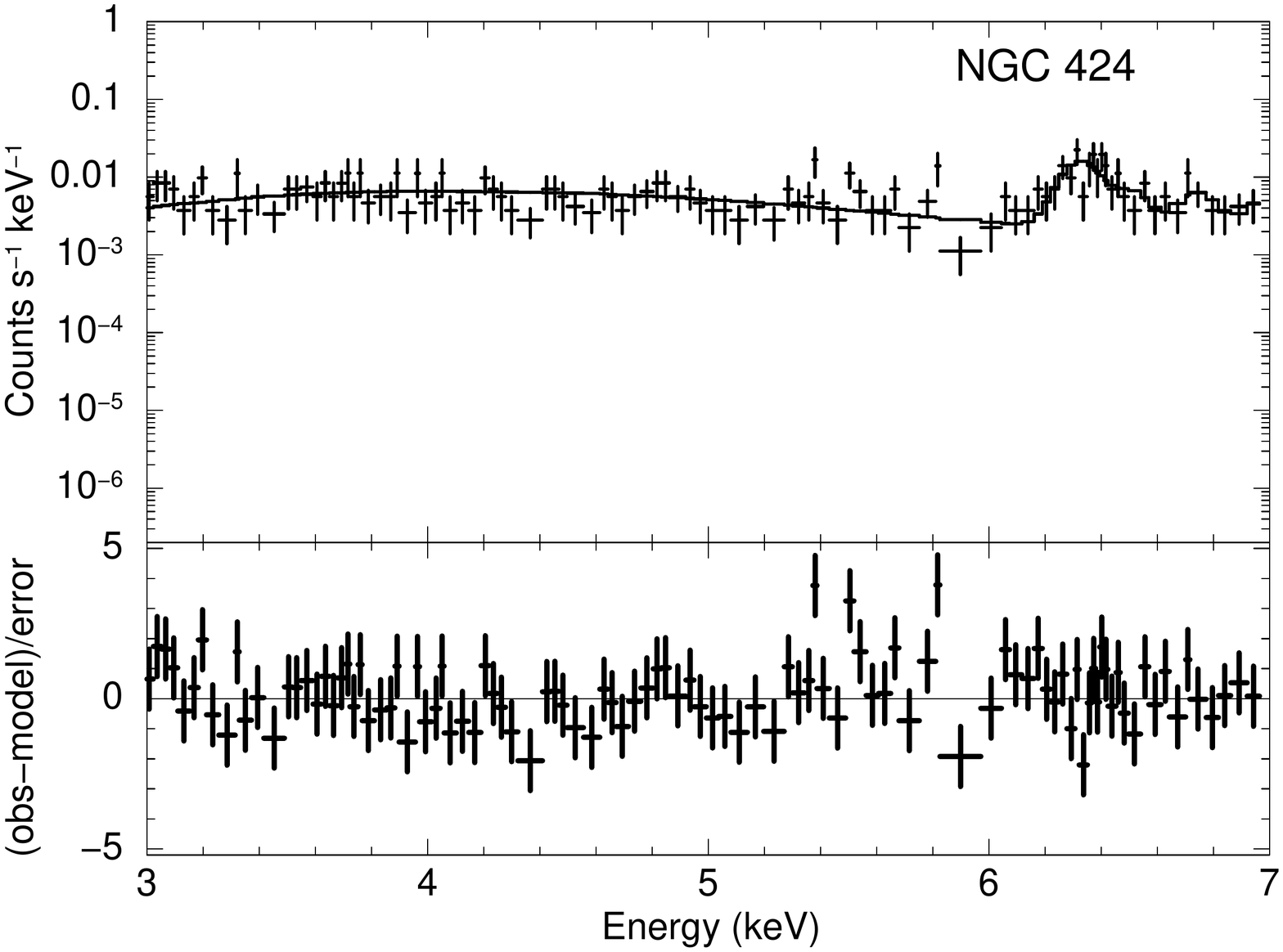}\hspace{-.5cm}
    \includegraphics[width=4.8cm]{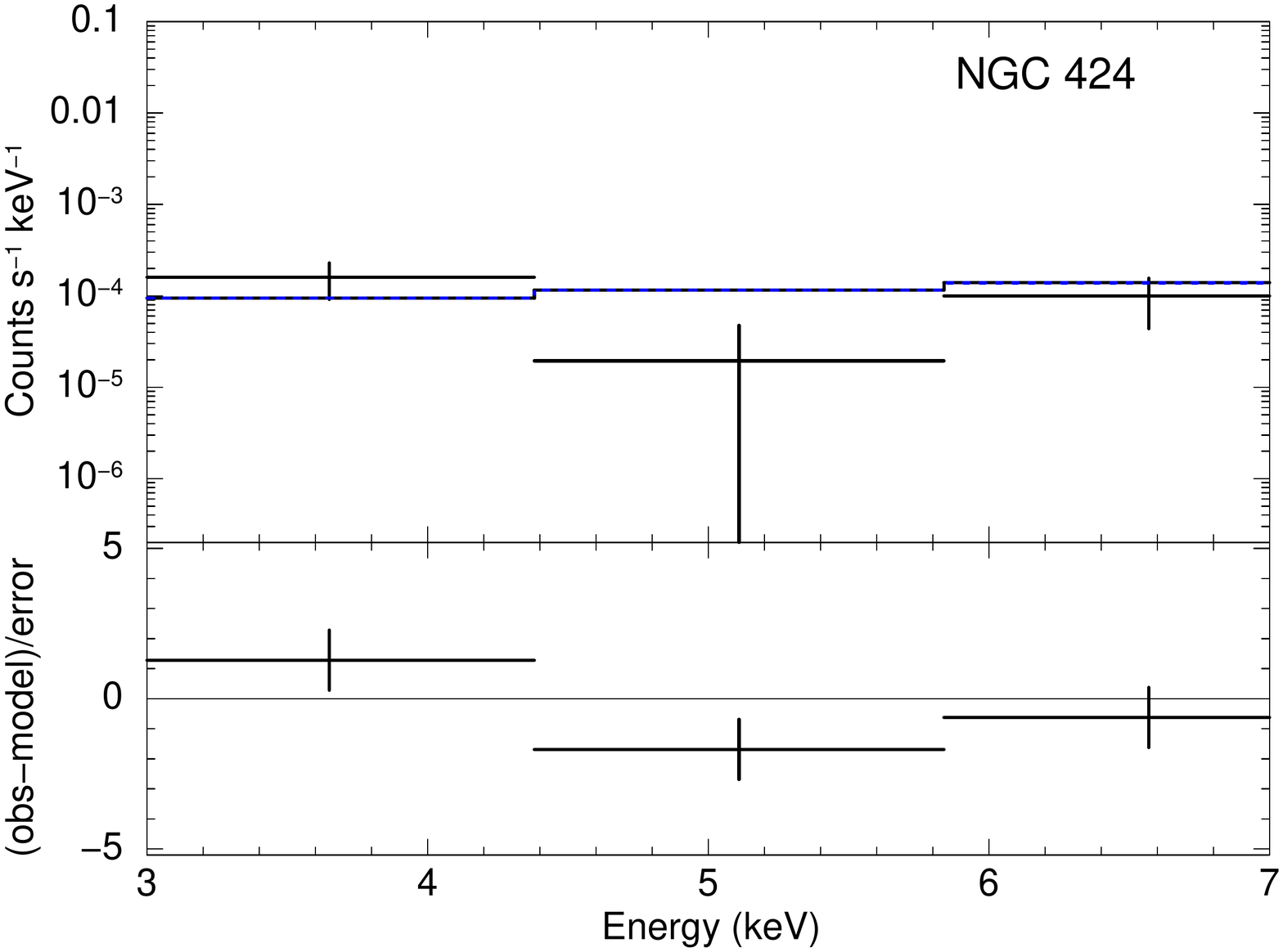}\hspace{-.5cm}
    \includegraphics[width=4.8cm]{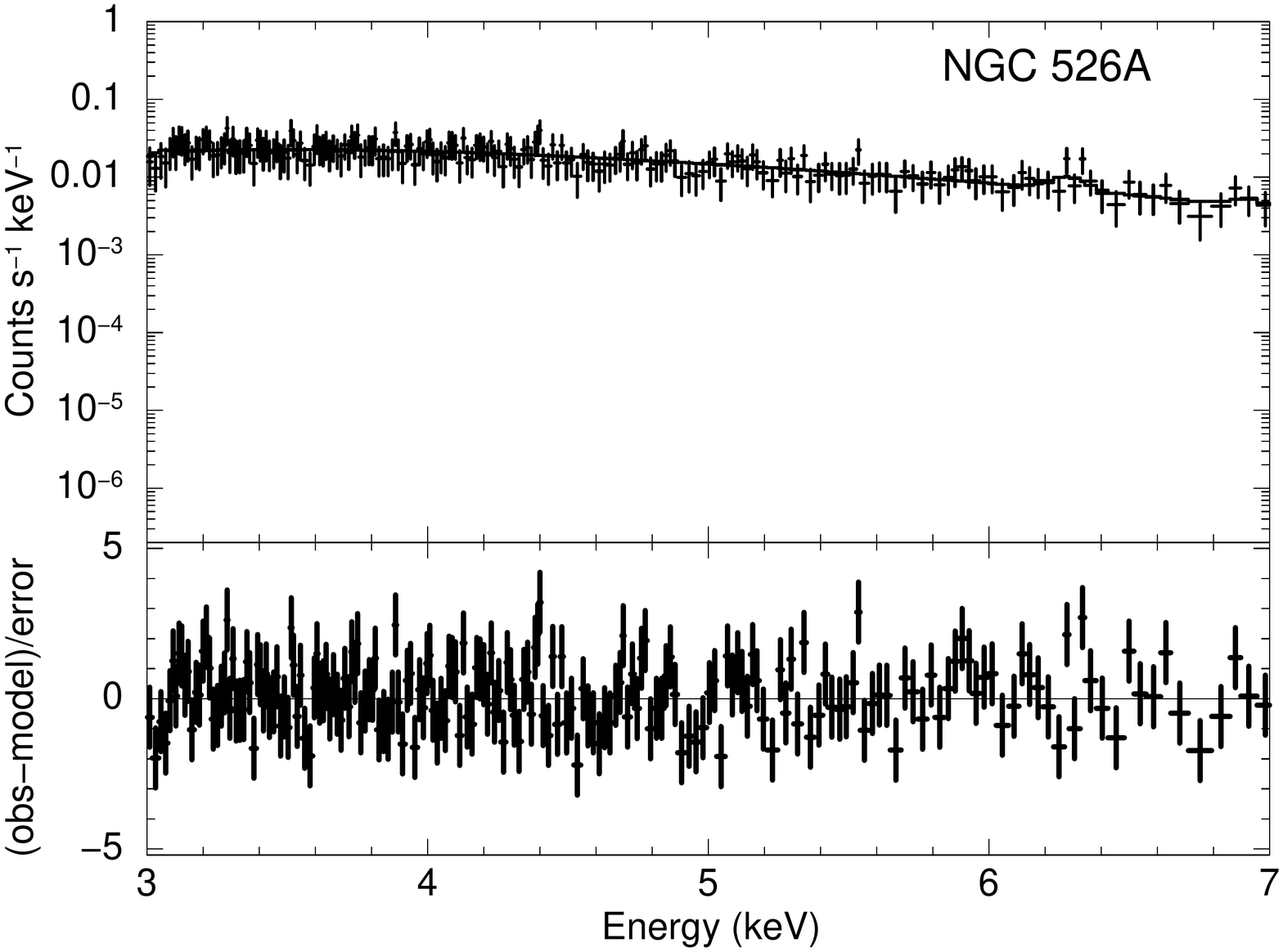}\hspace{-.5cm}
    \includegraphics[width=4.8cm]{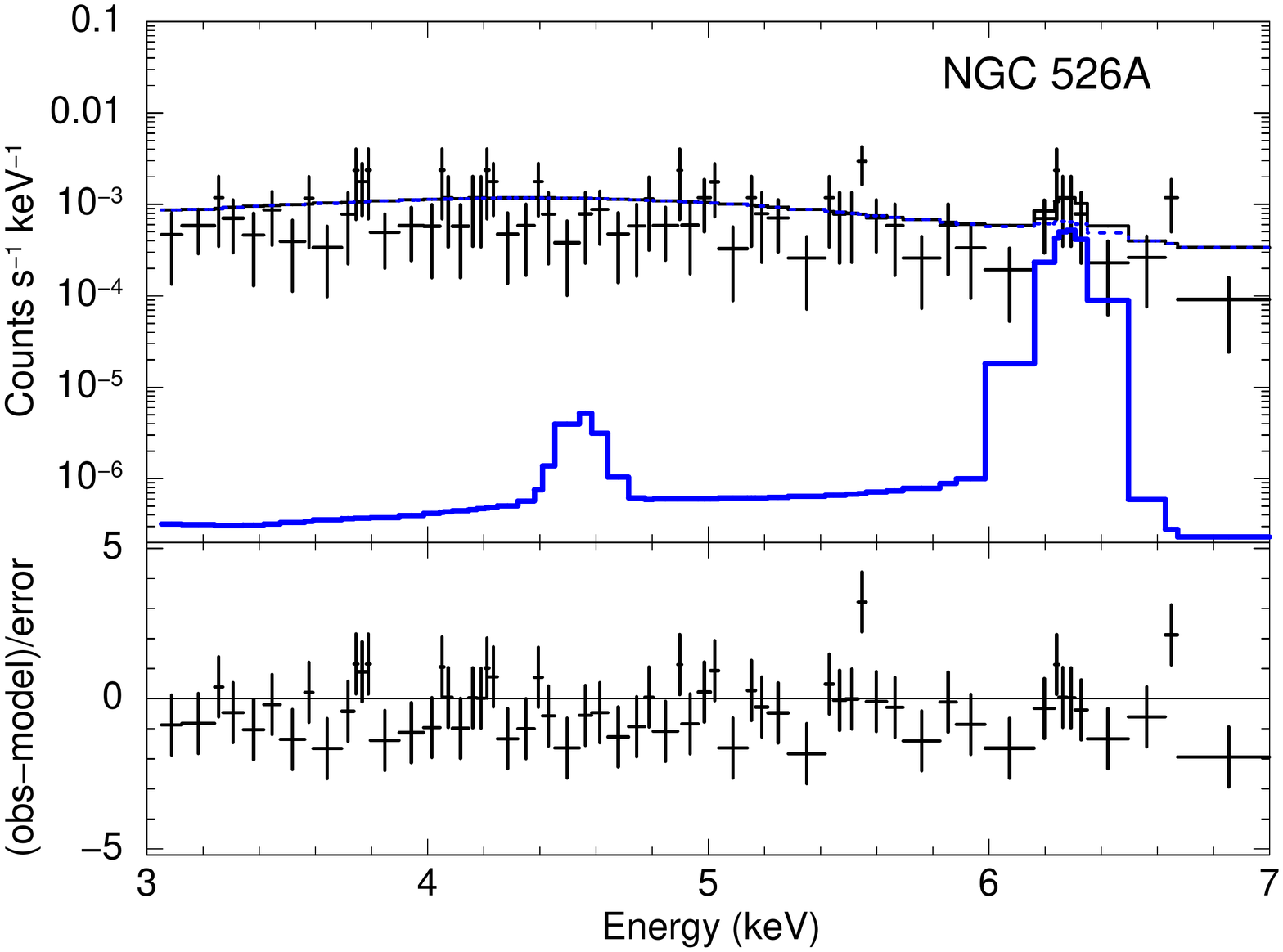}\\ \vspace{-0.5cm}
    \includegraphics[width=4.8cm]{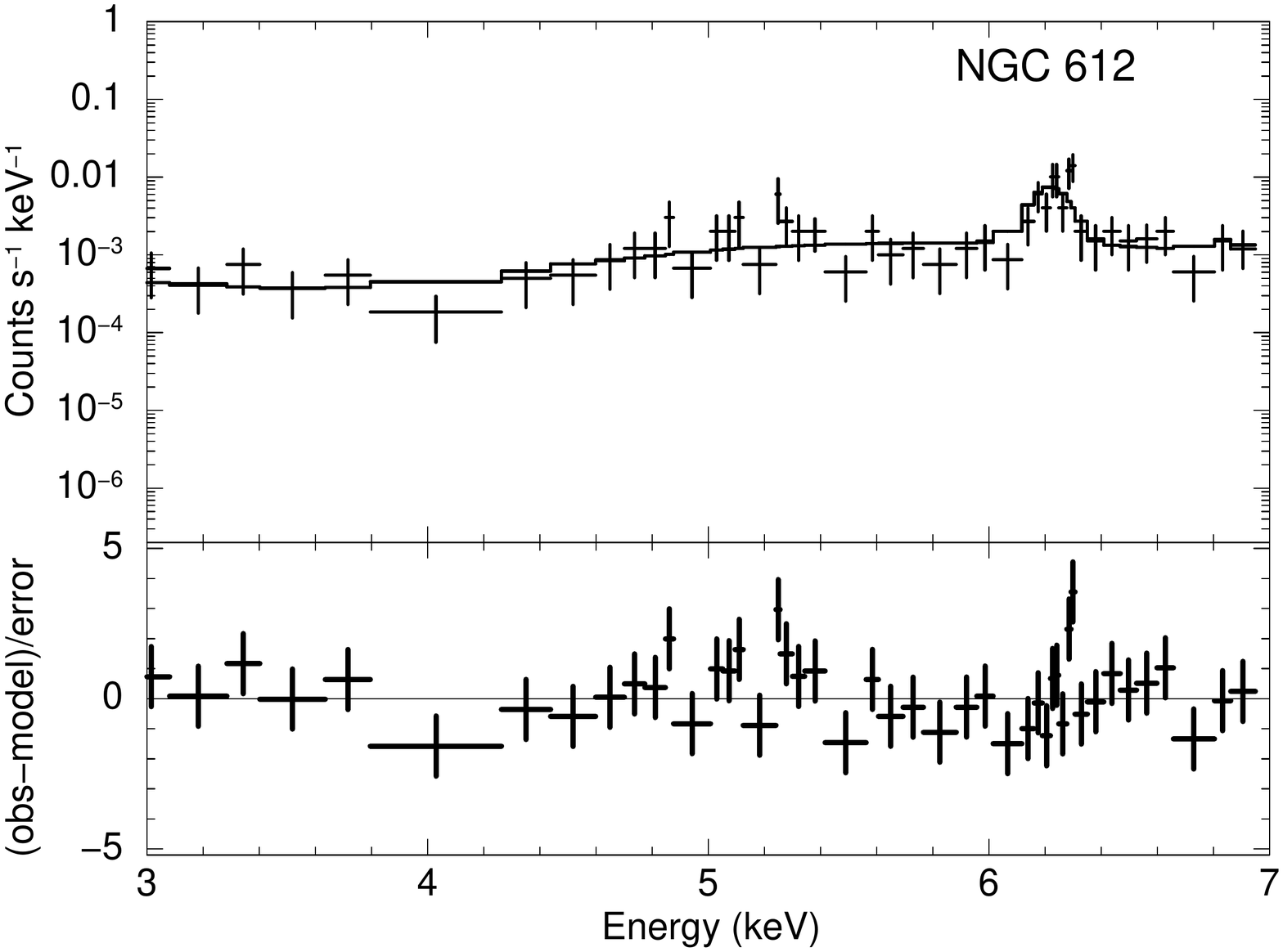}\hspace{-.5cm}
    \includegraphics[width=4.8cm]{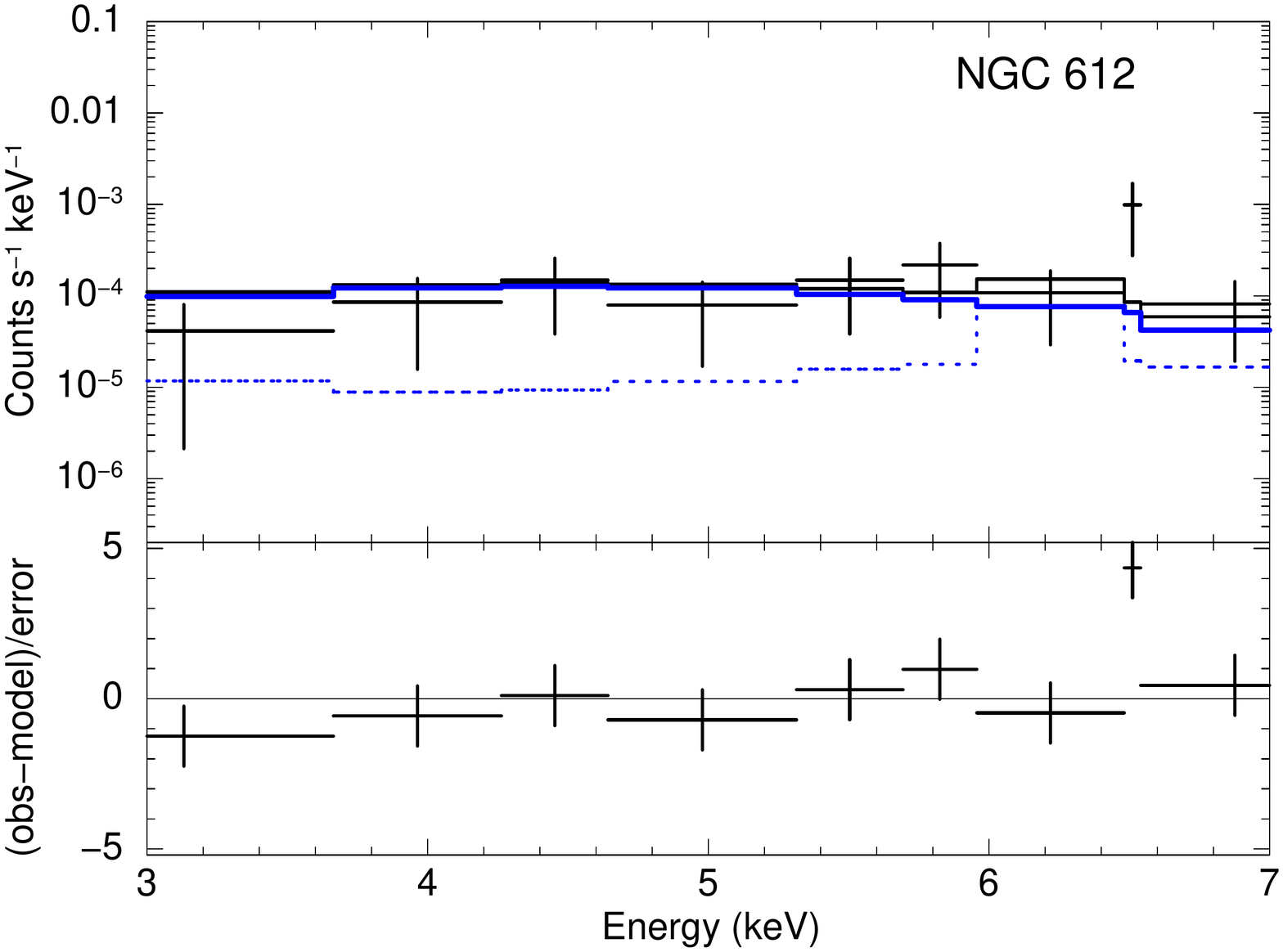}\hspace{-.5cm}
    \includegraphics[width=4.8cm]{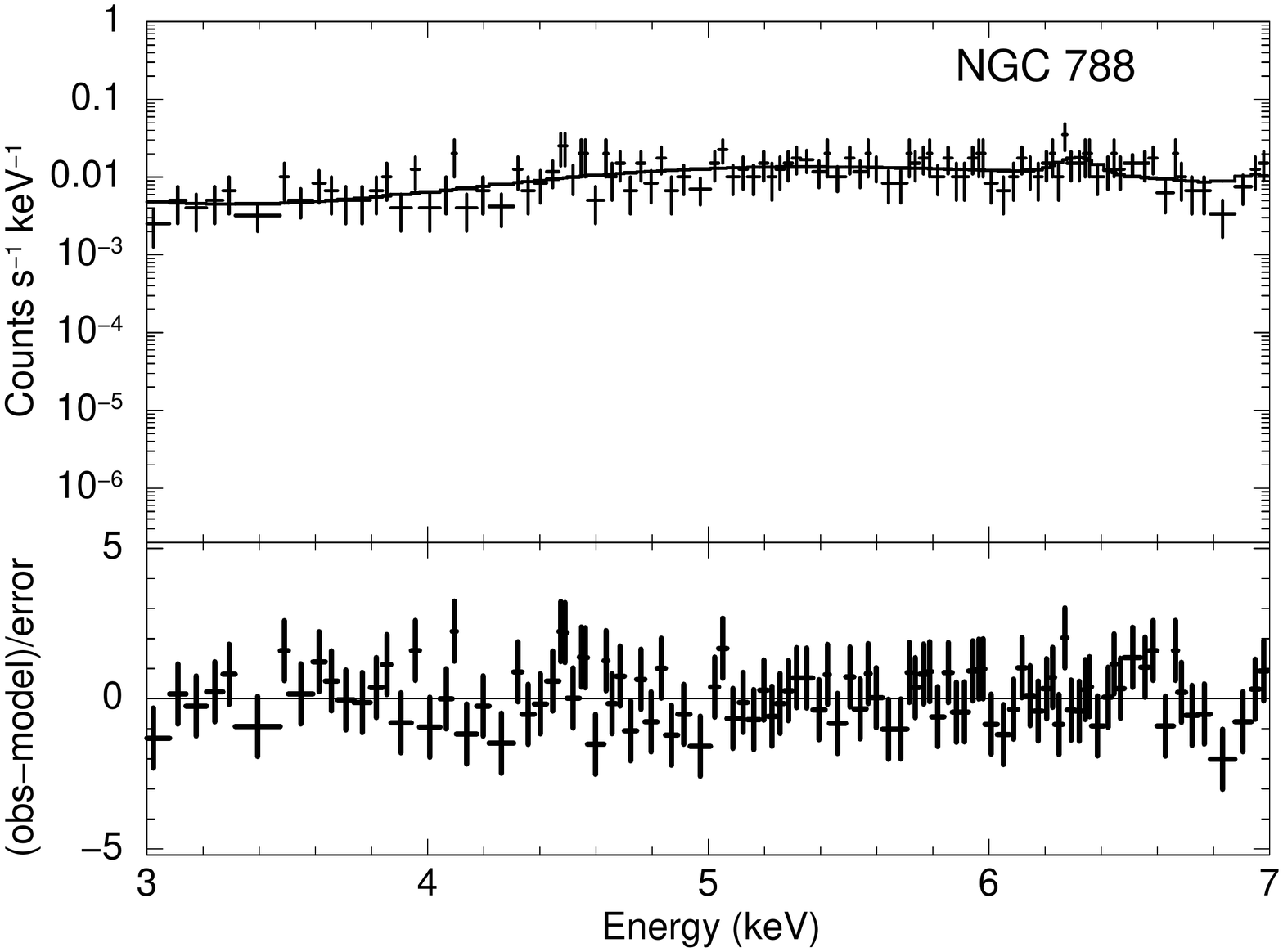}\hspace{-.5cm}
    \includegraphics[width=4.8cm]{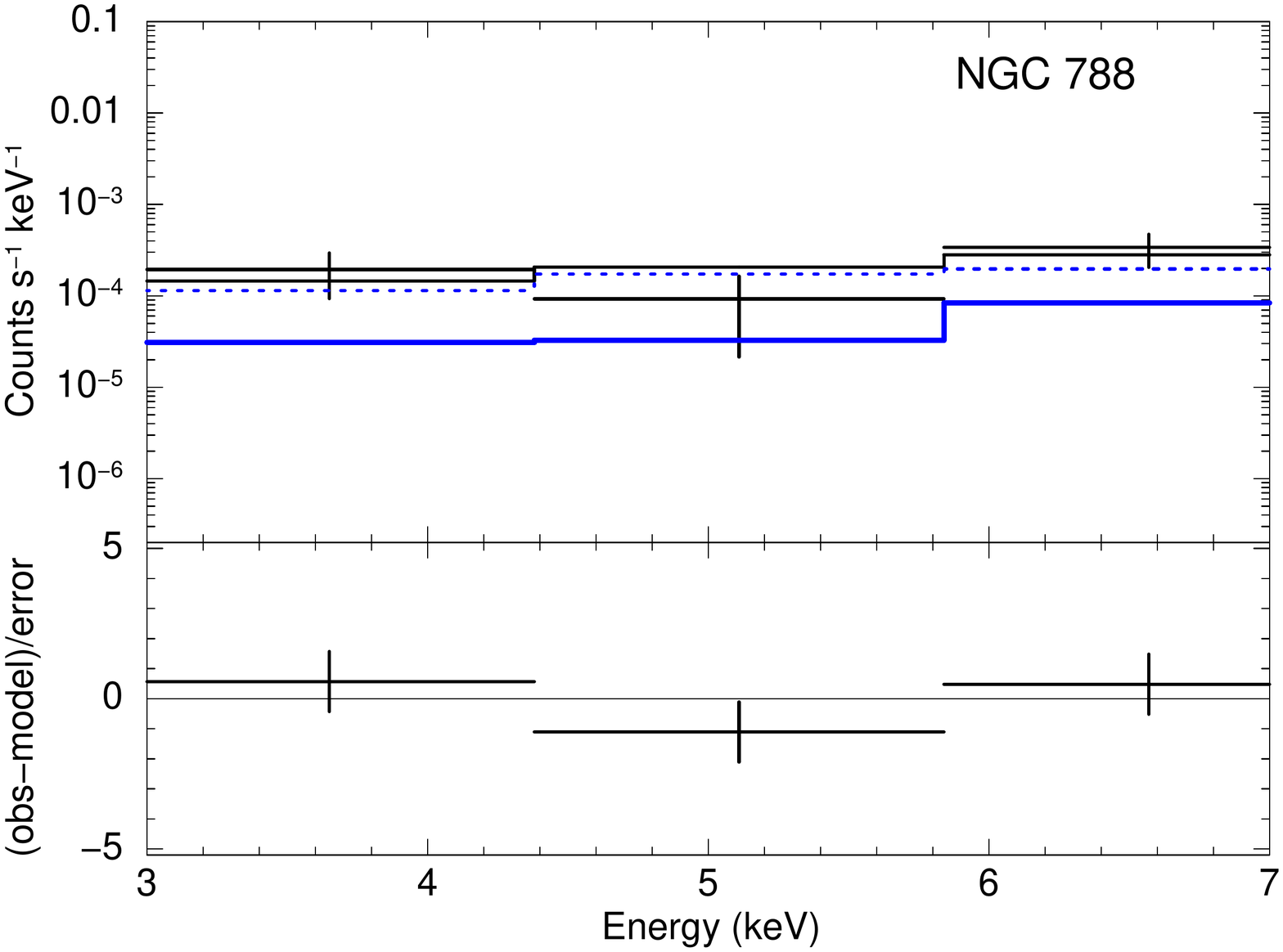}\\ \vspace{-0.5cm}
    \includegraphics[width=4.8cm]{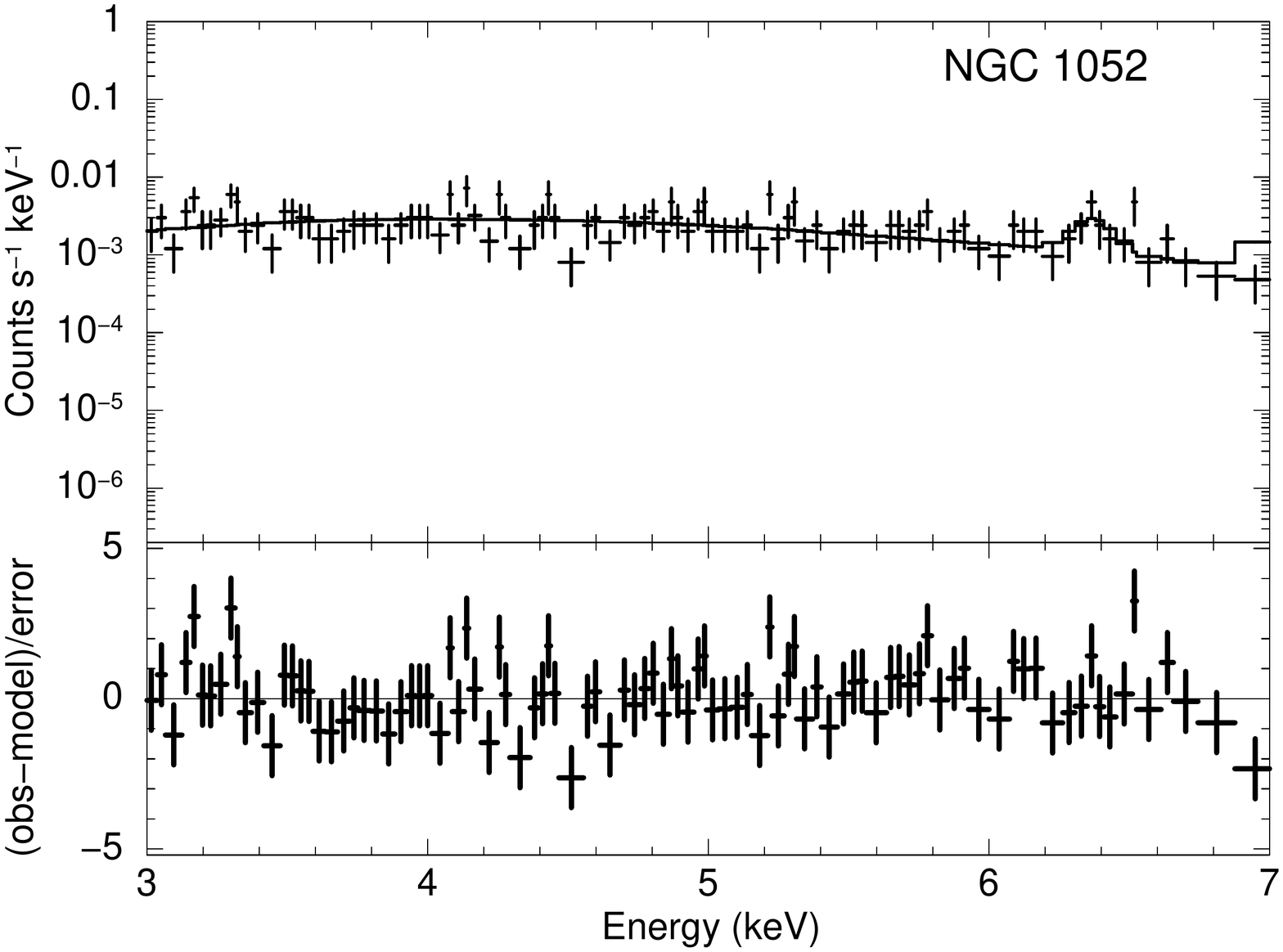}\hspace{-.5cm}
    \includegraphics[width=4.8cm]{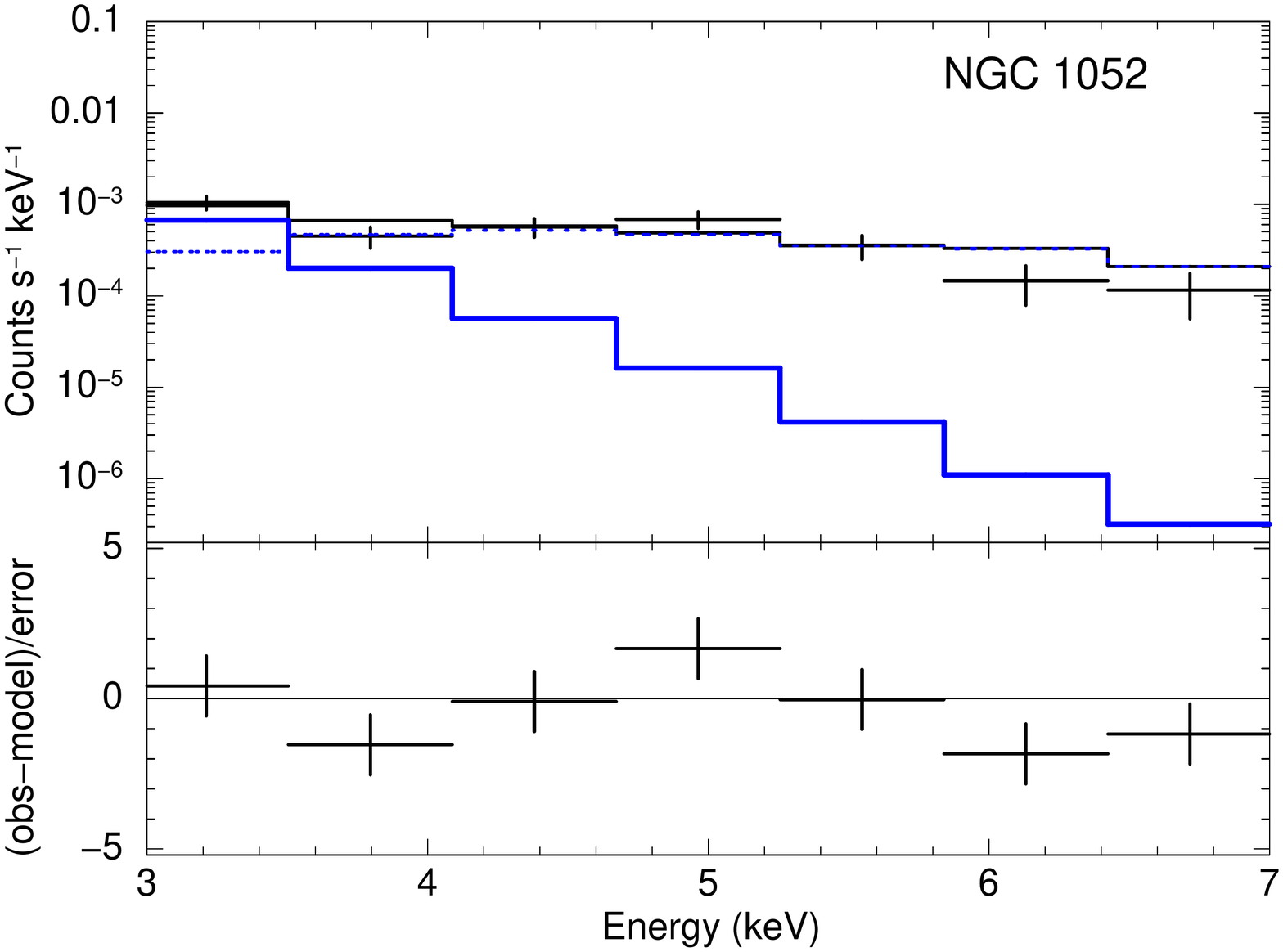}\hspace{-.5cm}
    \includegraphics[width=4.8cm]{06_NGC_1068_nuc_spec.pdf}\hspace{-.5cm}
    \includegraphics[width=4.8cm]{06_NGC_1068_ext_spec.pdf}\\ \vspace{-0.5cm}
    \includegraphics[width=4.8cm]{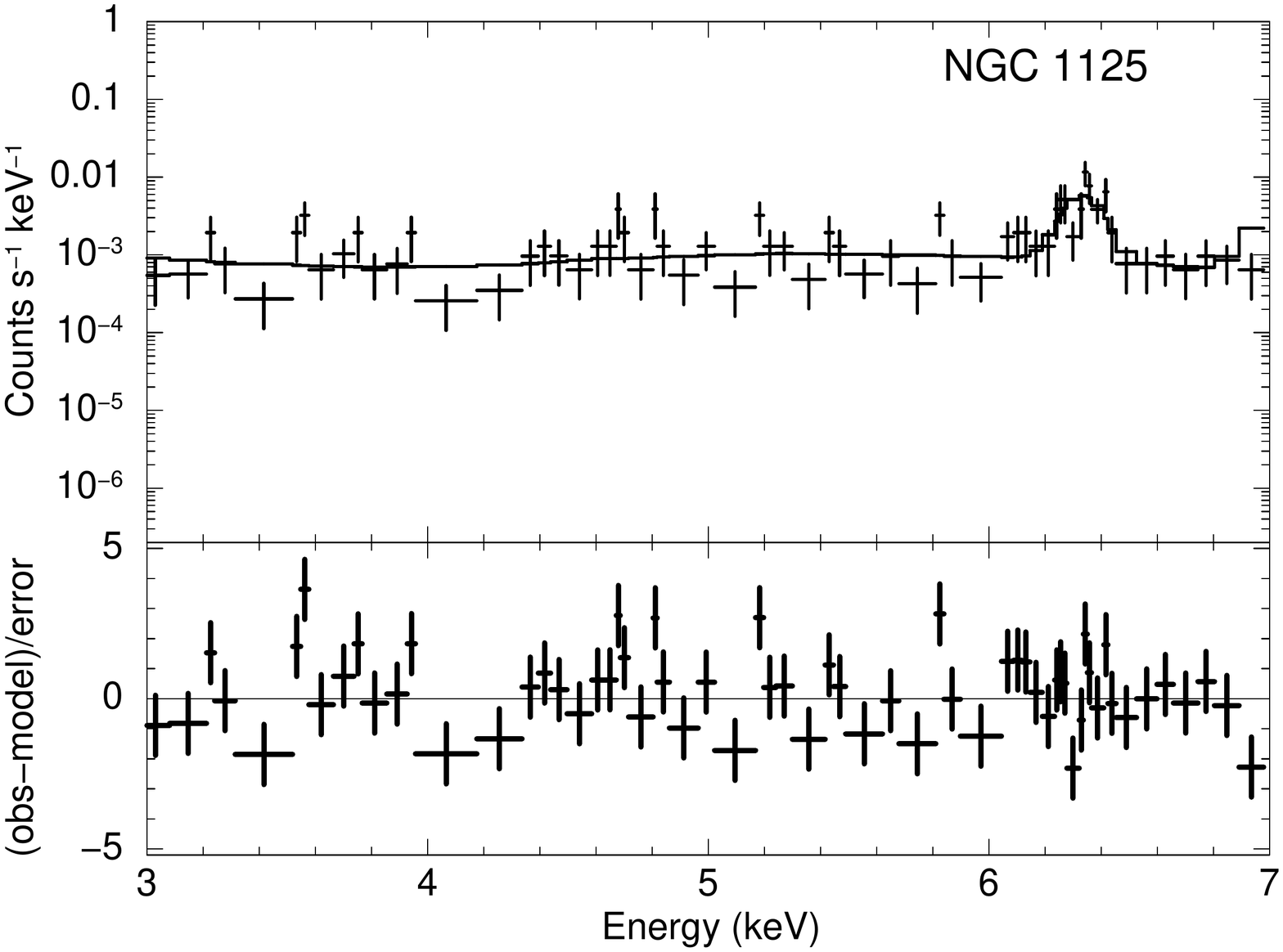}\hspace{-.5cm}
    \includegraphics[width=4.8cm]{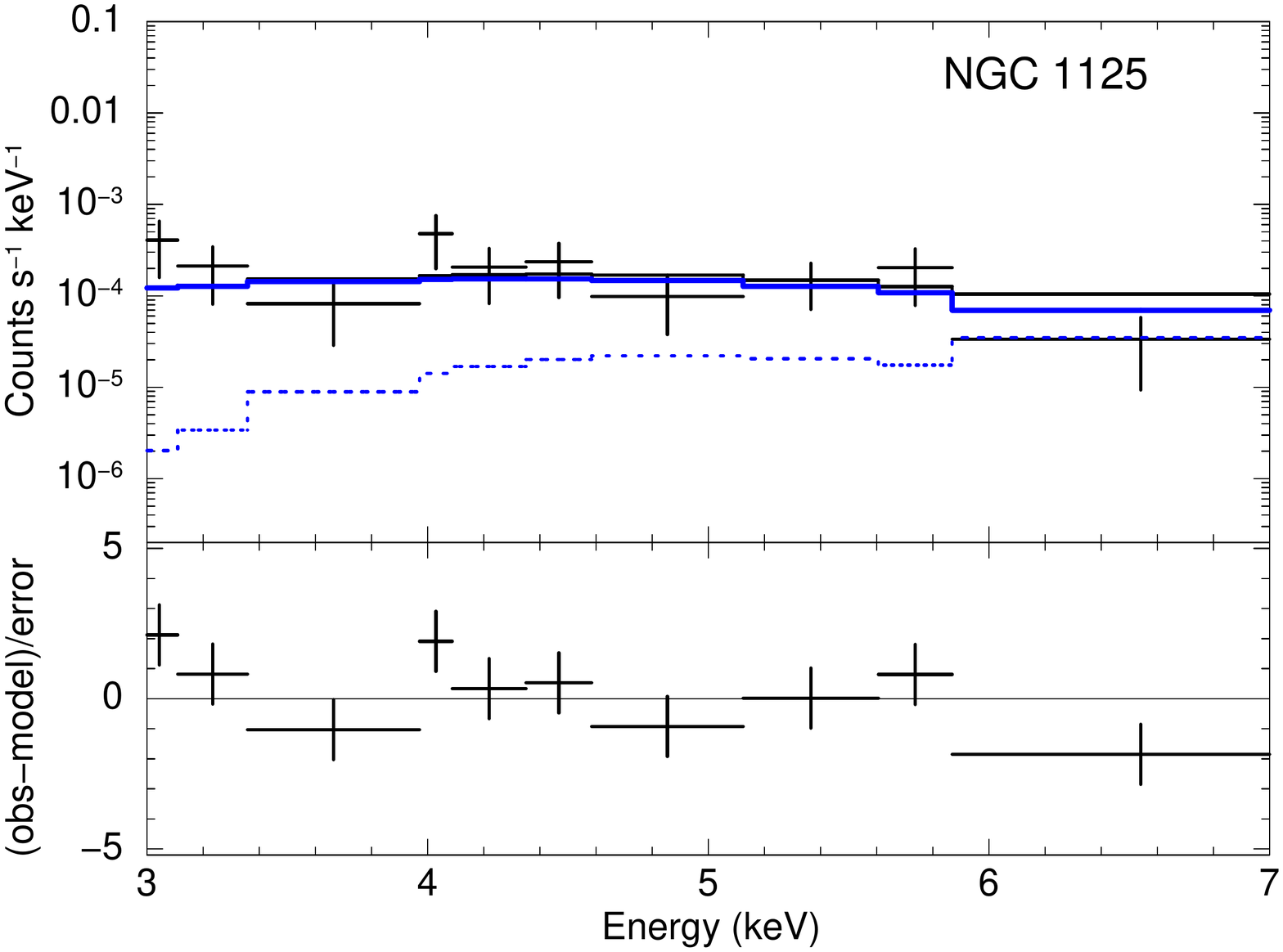}\hspace{-.5cm}
    \includegraphics[width=4.8cm]{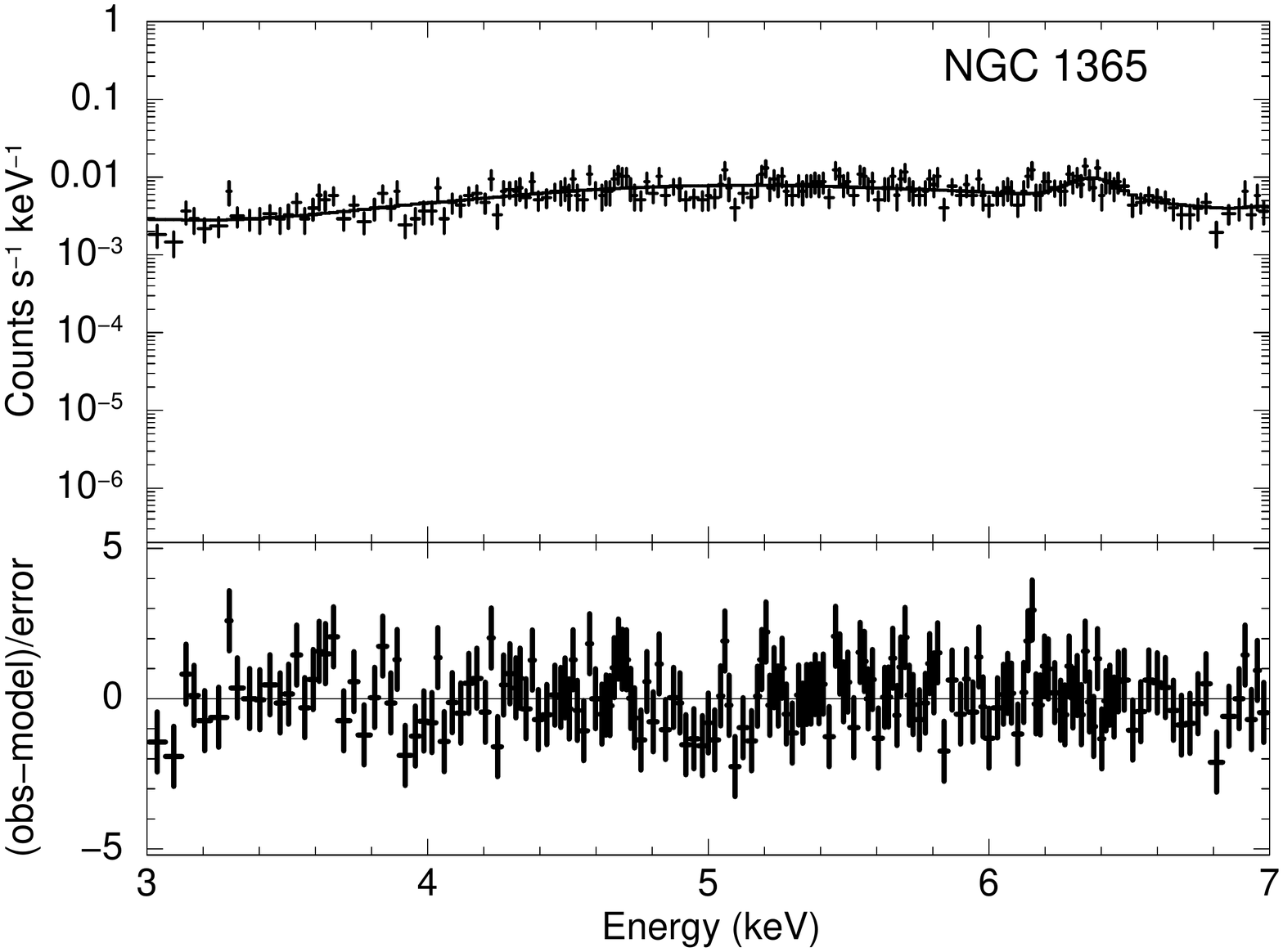}\hspace{-.5cm}
    \includegraphics[width=4.8cm]{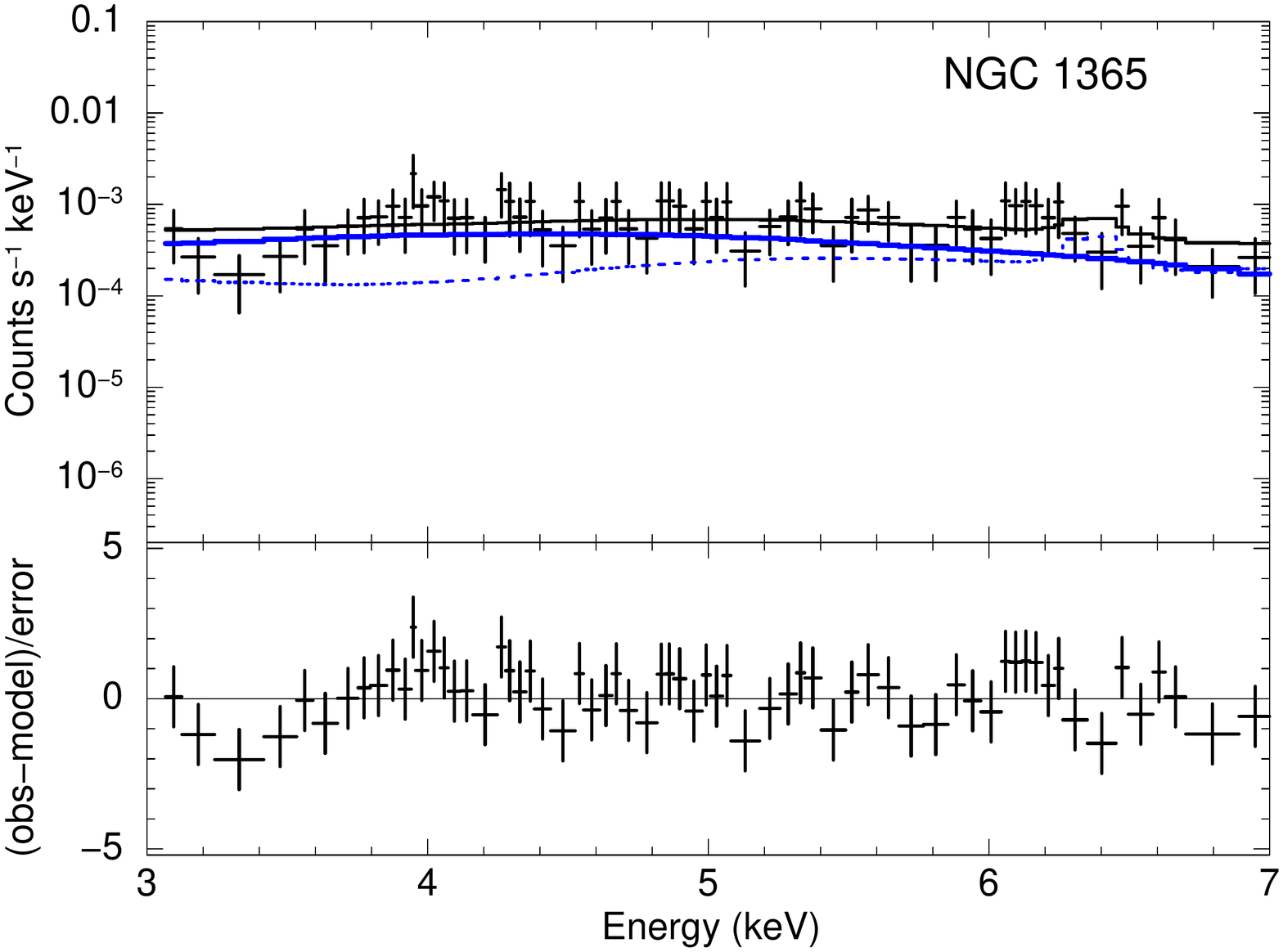}\\ \vspace{-0.5cm}
    \includegraphics[width=4.8cm]{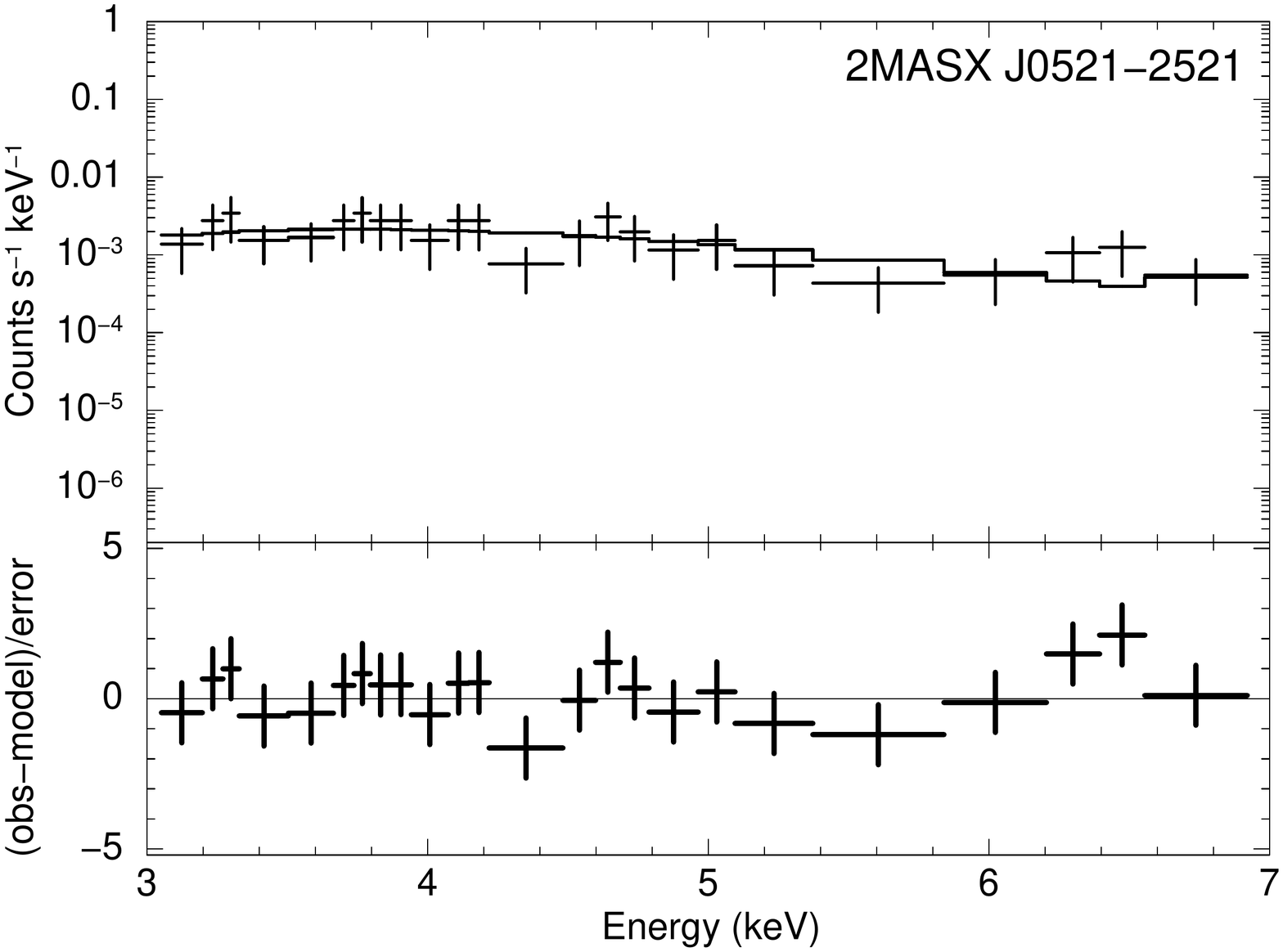}\hspace{-.5cm}
    \includegraphics[width=4.8cm]{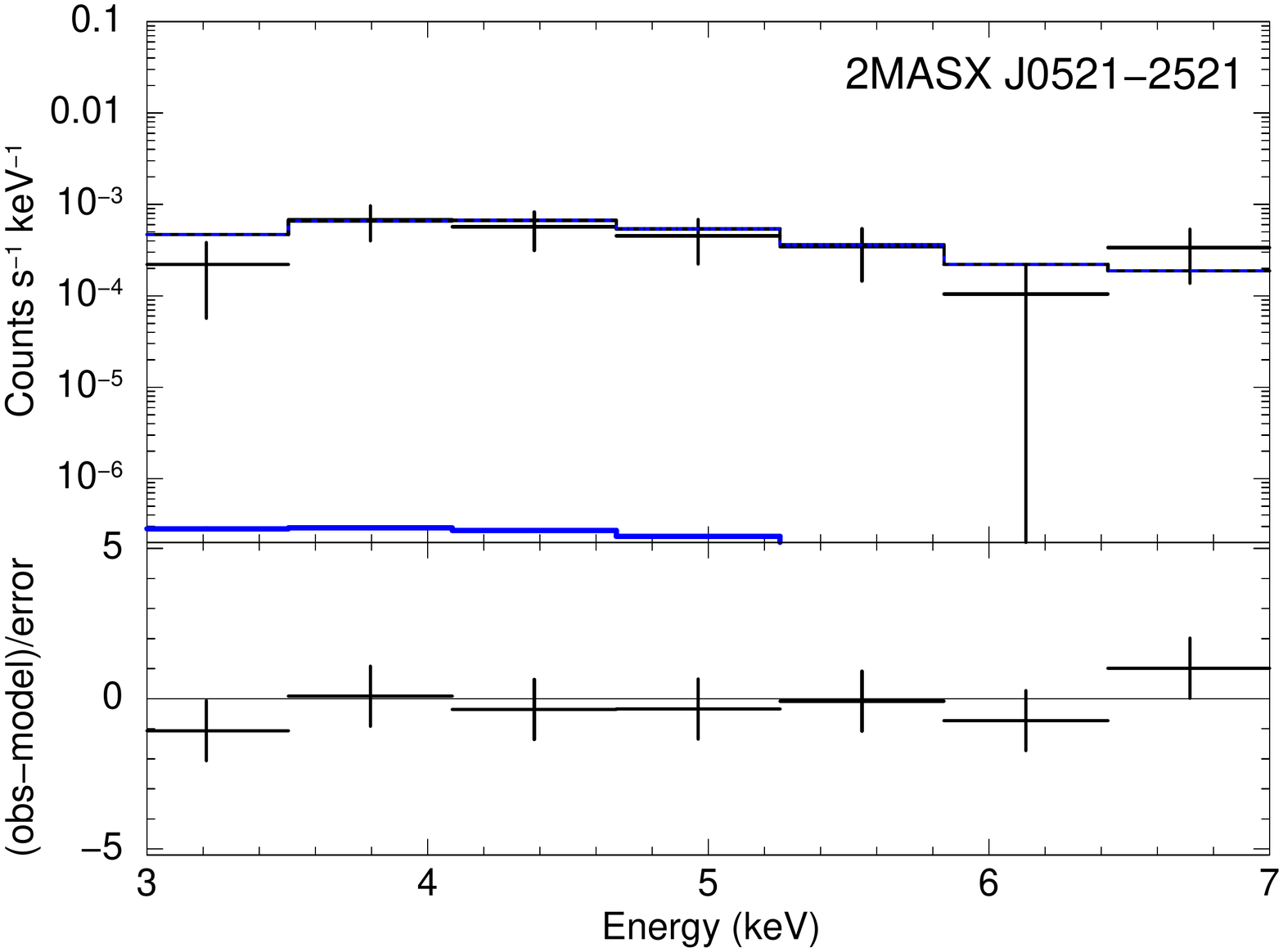}\hspace{-.5cm}
    \includegraphics[width=4.8cm]{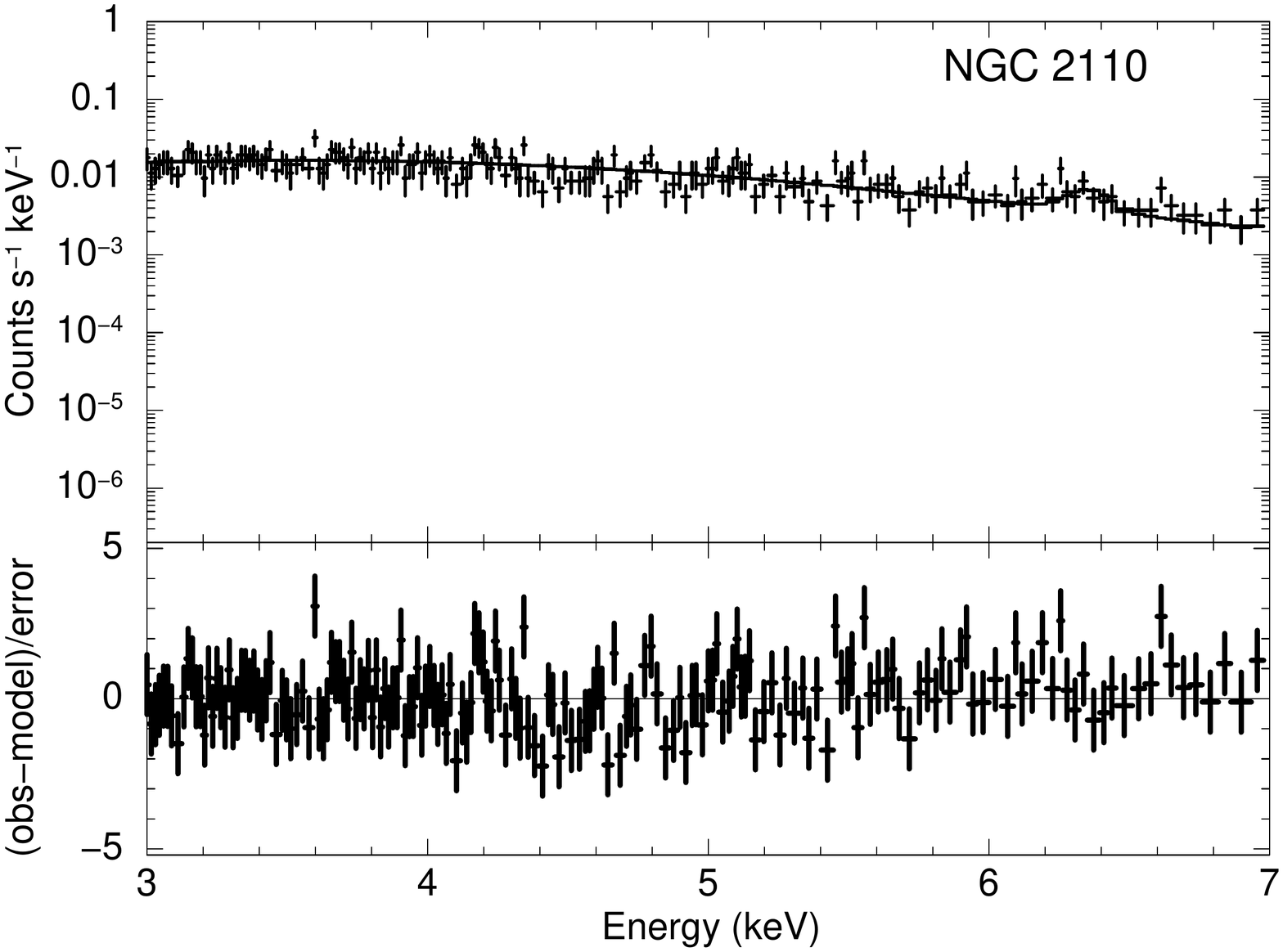}\hspace{-.5cm}
    \includegraphics[width=4.8cm]{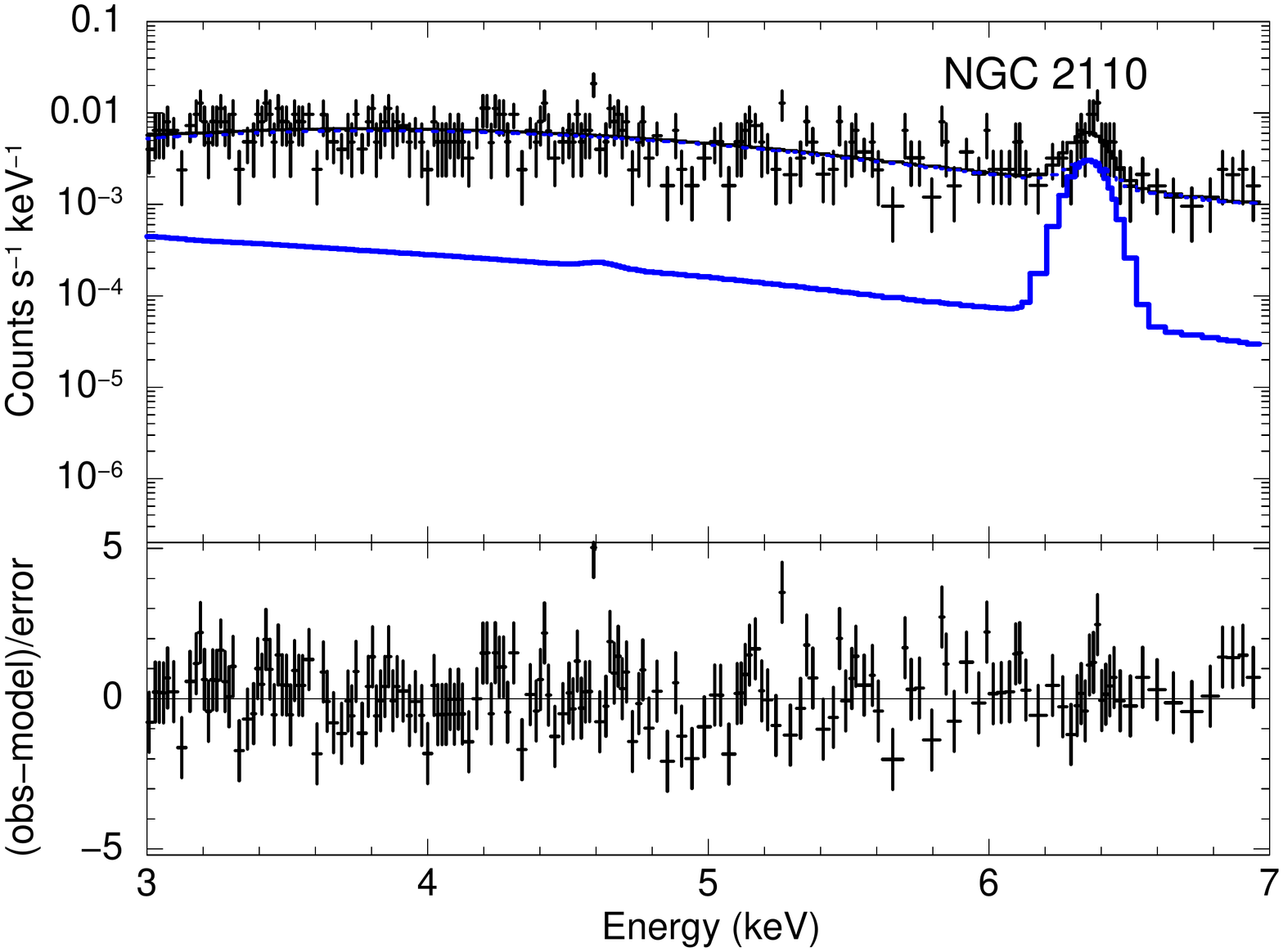}\\  \vspace{-0.5cm}   
    \includegraphics[width=4.8cm]{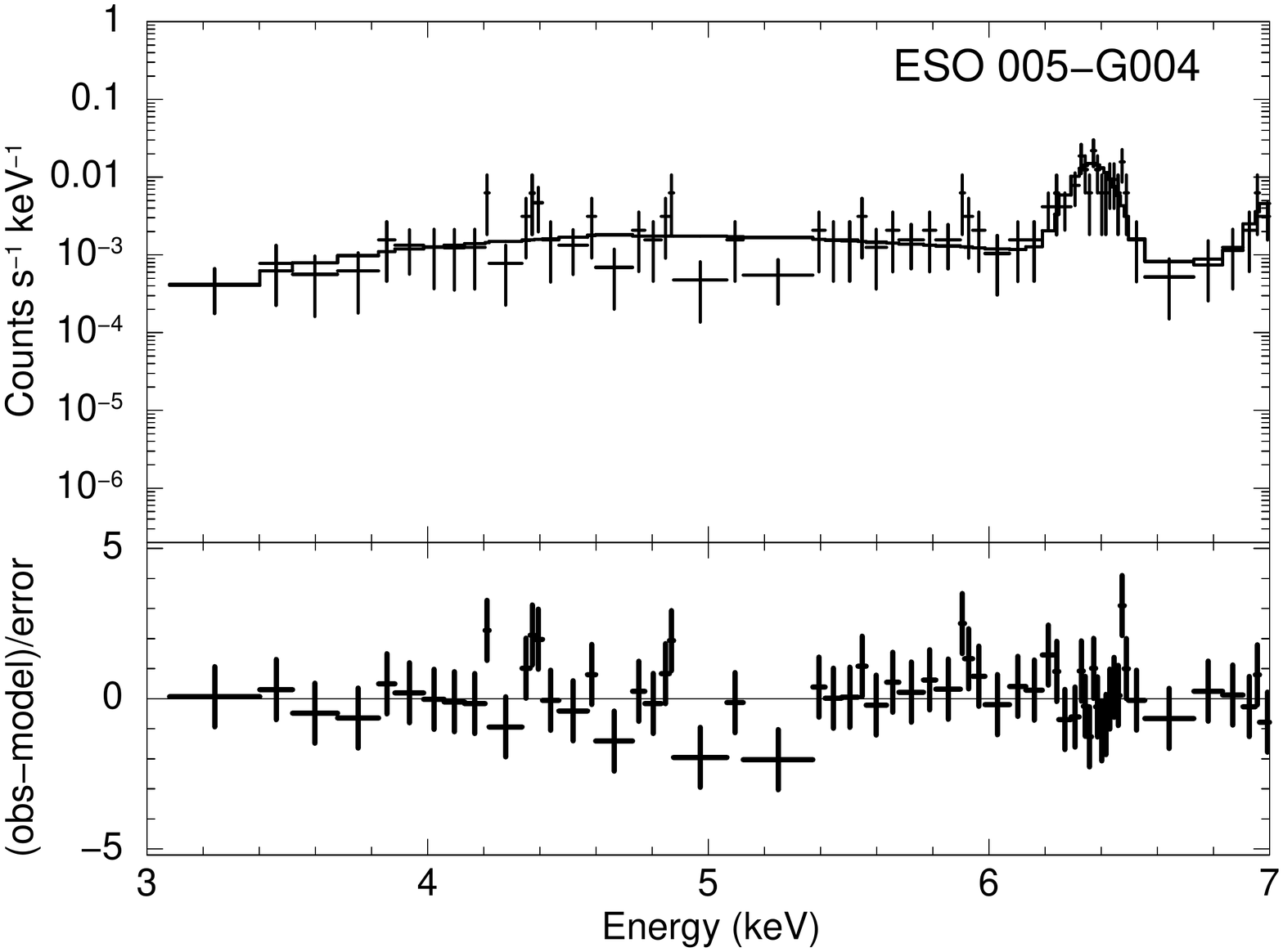}\hspace{-.5cm}
    \includegraphics[width=4.8cm]{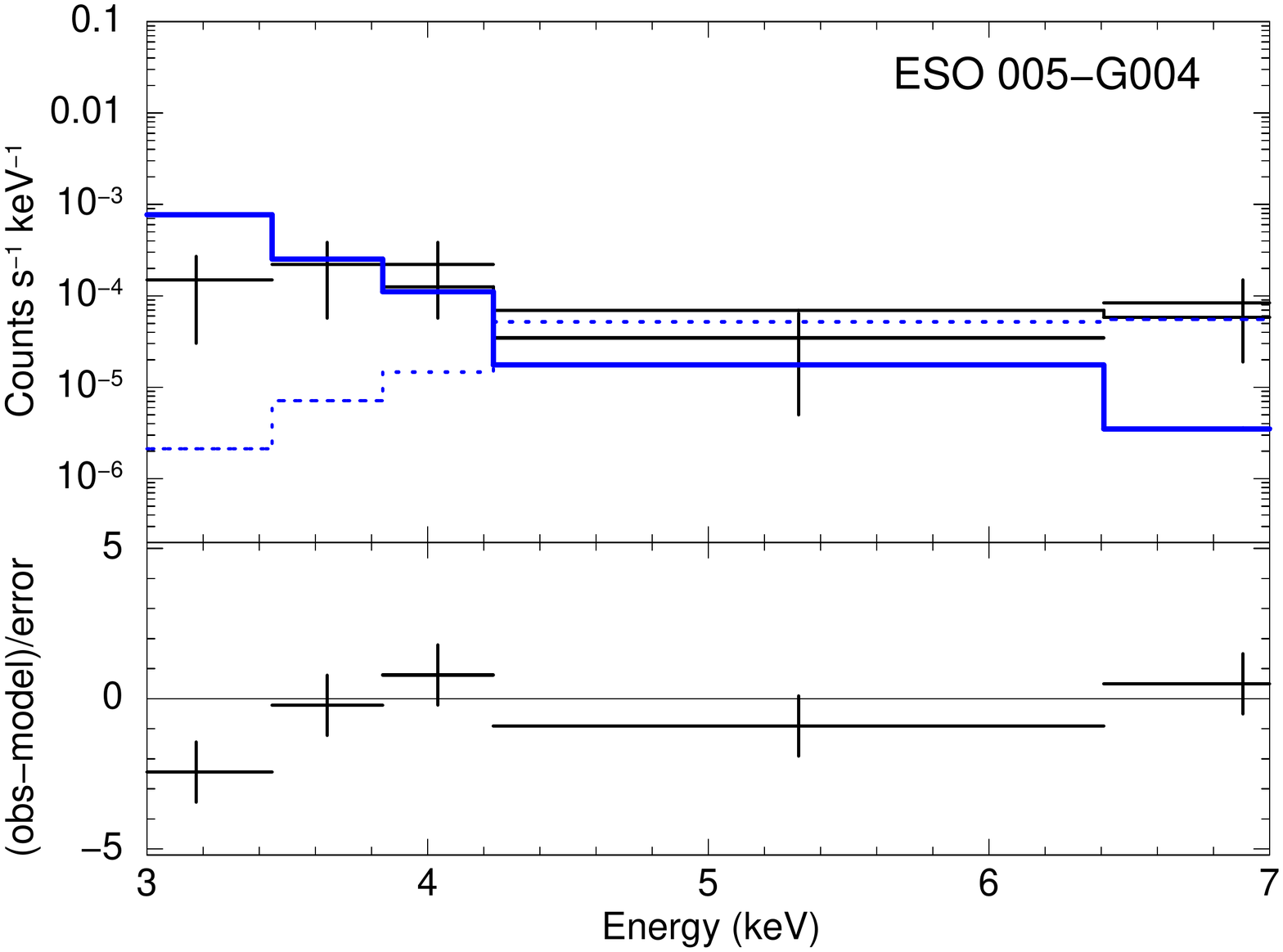}\hspace{-.5cm}
    \includegraphics[width=4.8cm]{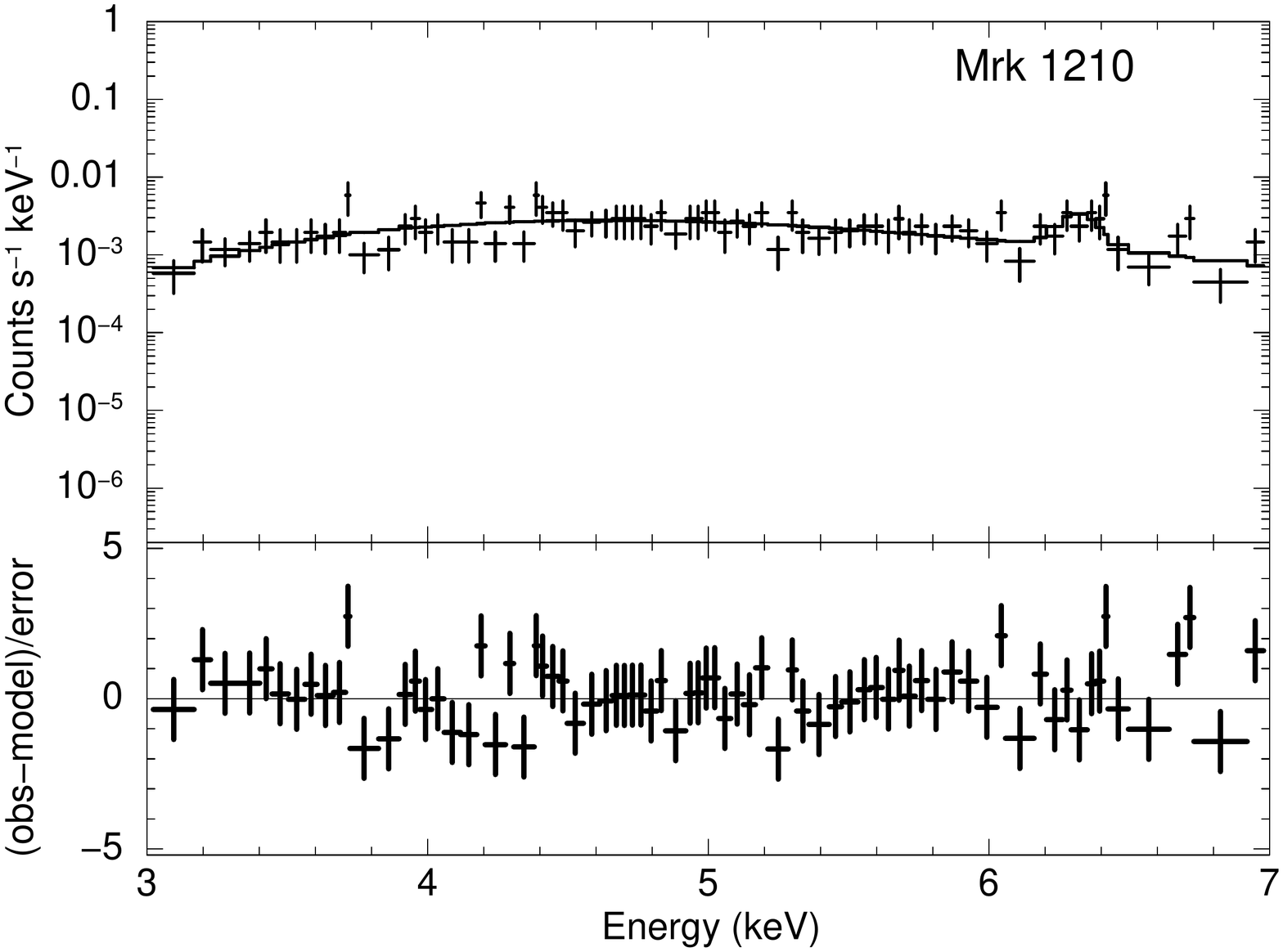}\hspace{-.5cm}
    \includegraphics[width=4.8cm]{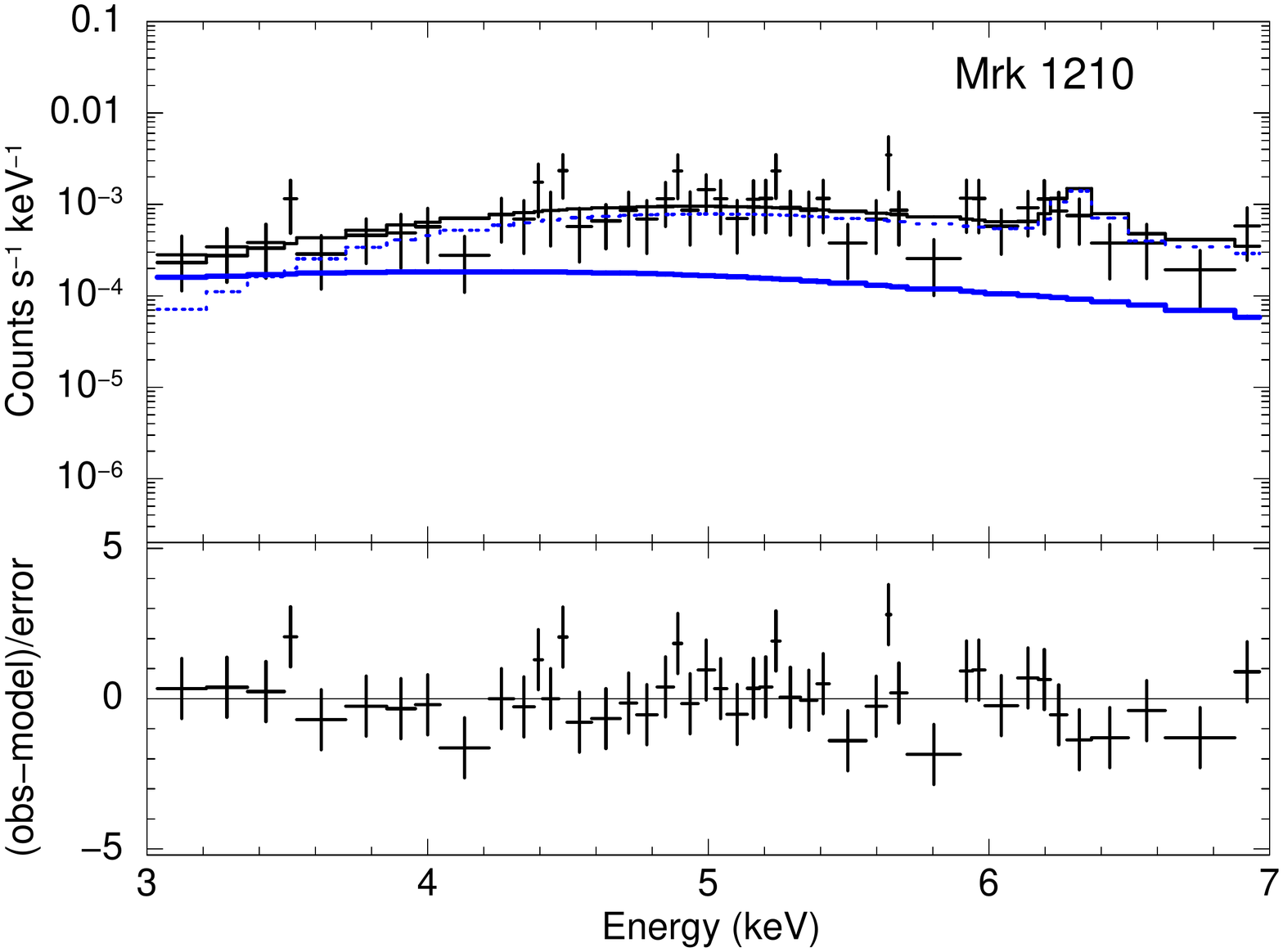}
    \caption{
    X-ray spectra from the nuclear 2\arcsec\ regions (first and third columns),
    and those taken from the external annulus regions with radii of 2\arcsec--4\arcsec\ (2nd and 4th columns).
    In each nuclear spectral figure, 
    the upper window shows observed data (crosses) and a best-fit model (solid line), and the lower one shows the residuals. For each external spectrum, 
    an in-situ emission and a contaminating emission from a nuclear source
    (Sections~\ref{sec:fe_vs_co21} and \ref{sec:acc_vs_gasmass})
    are additionally shown by the blue solid and dotted lines, respectively.
    }
    \label{app:fig:xspec}
\end{figure*}

\begin{figure*}\addtocounter{figure}{-1}
    \centering
    \includegraphics[width=4.8cm]{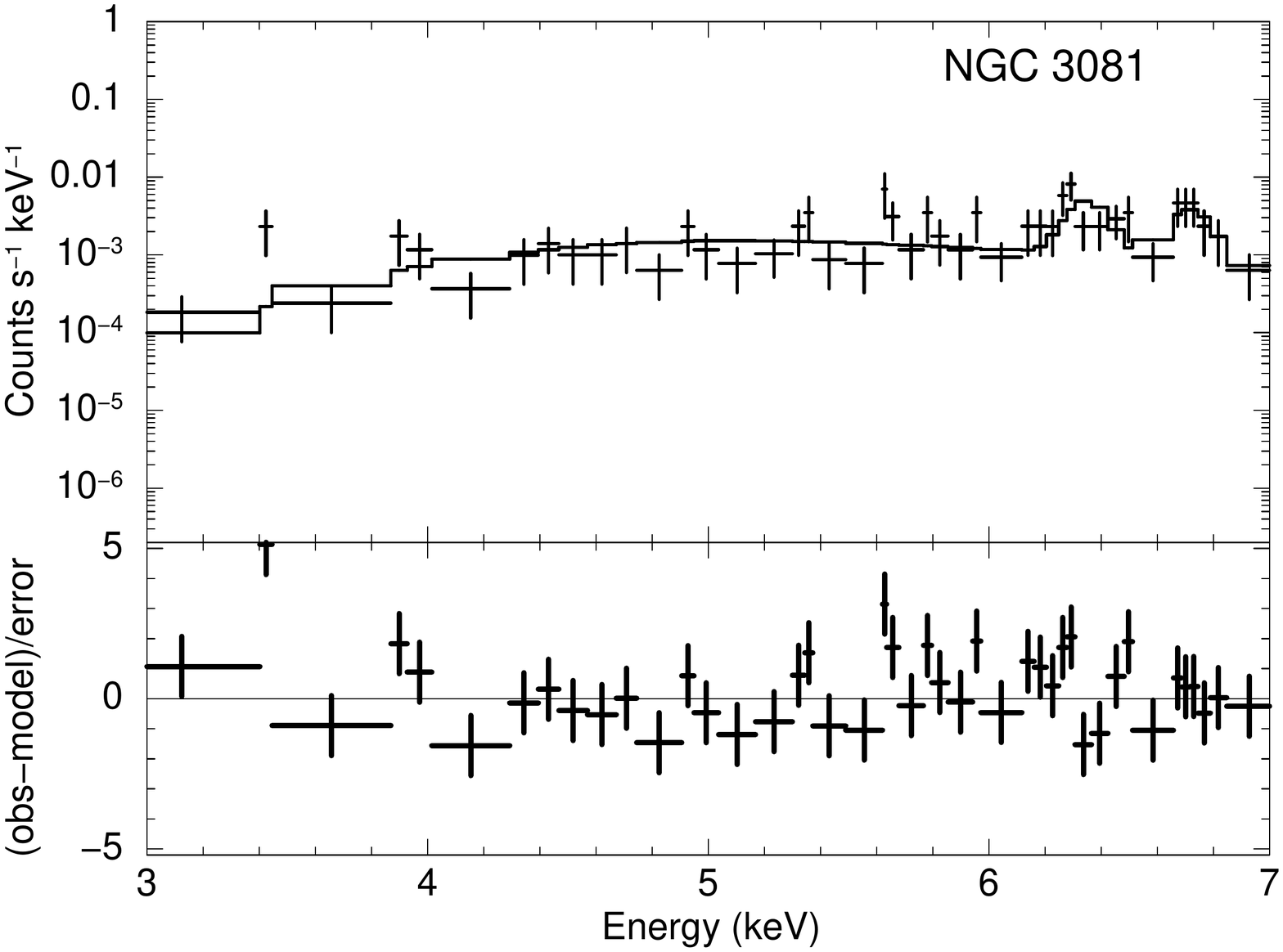}\hspace{-.5cm}
    \includegraphics[width=4.8cm]{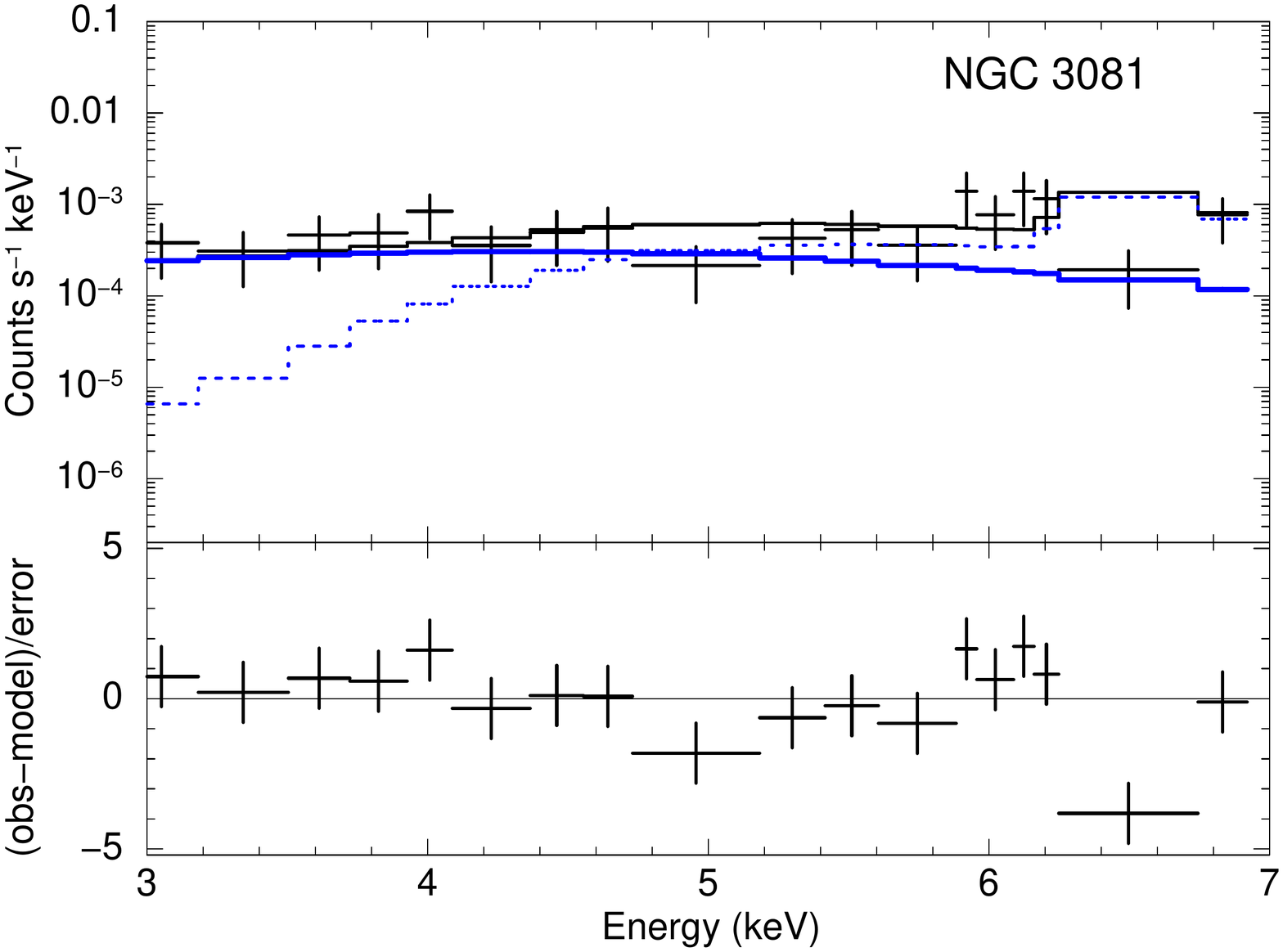}\hspace{-.5cm}
    \includegraphics[width=4.8cm]{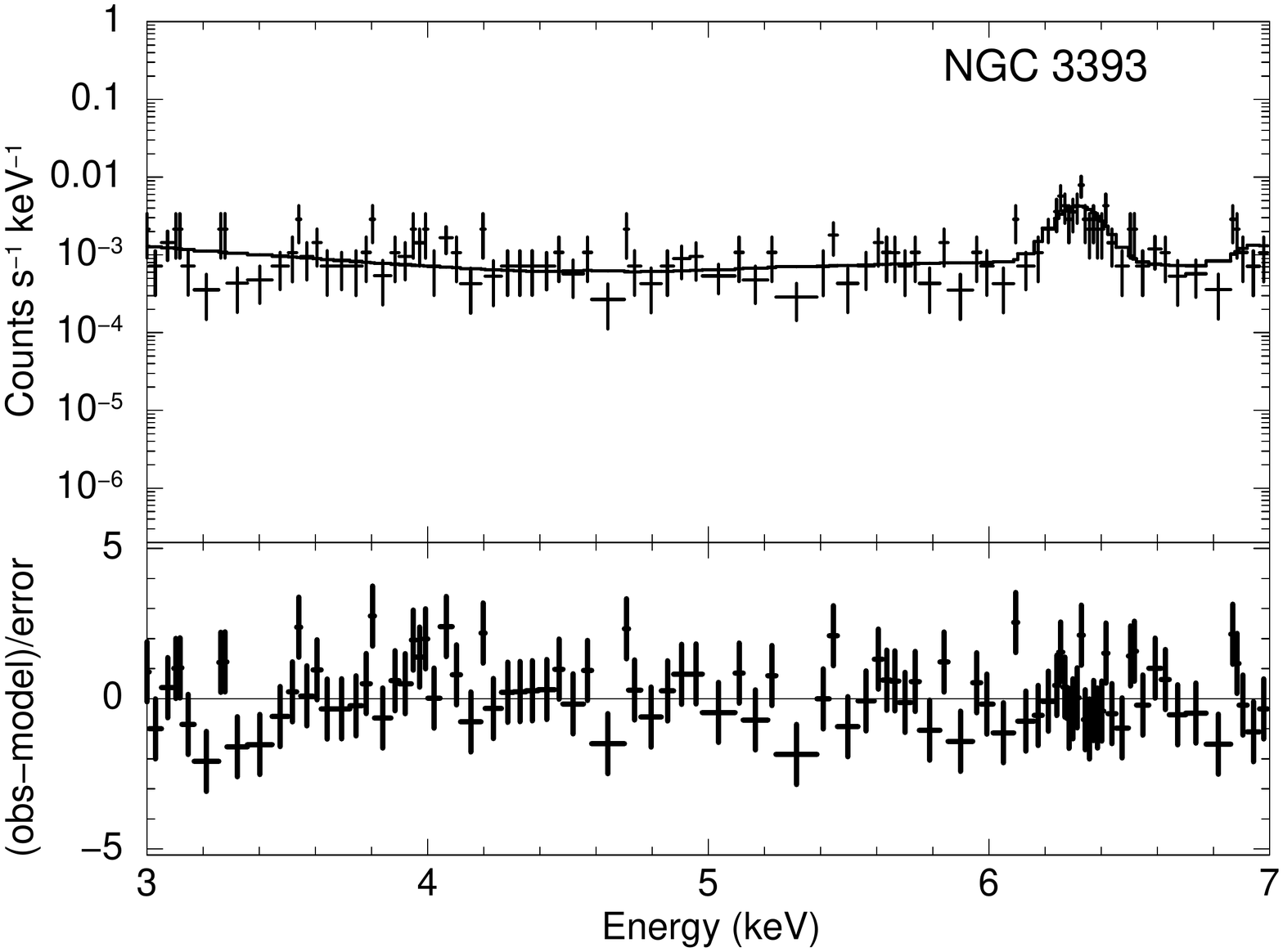}\hspace{-.5cm}
    \includegraphics[width=4.8cm]{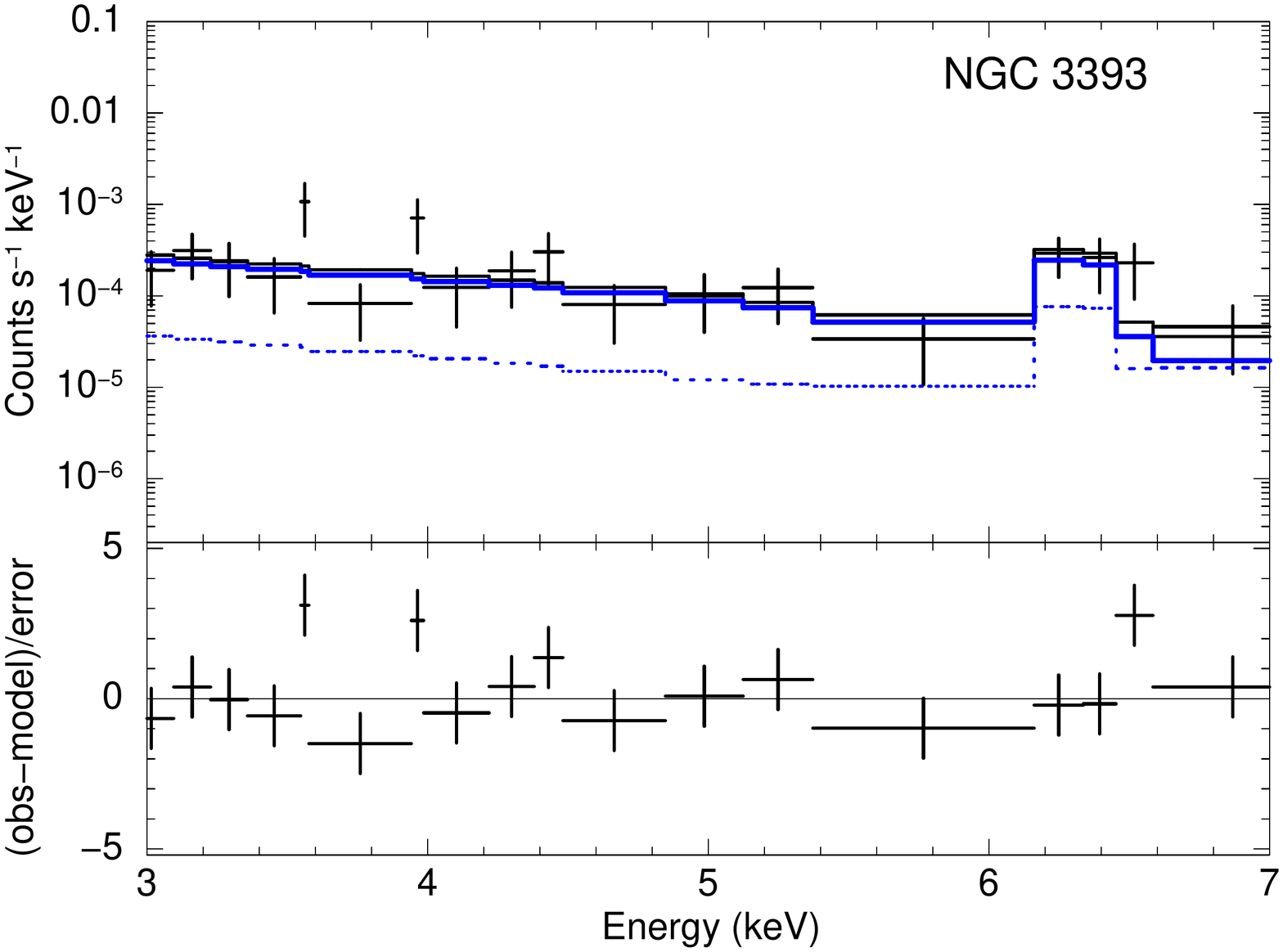}\\ \vspace{-0.5cm}
    \includegraphics[width=4.8cm]{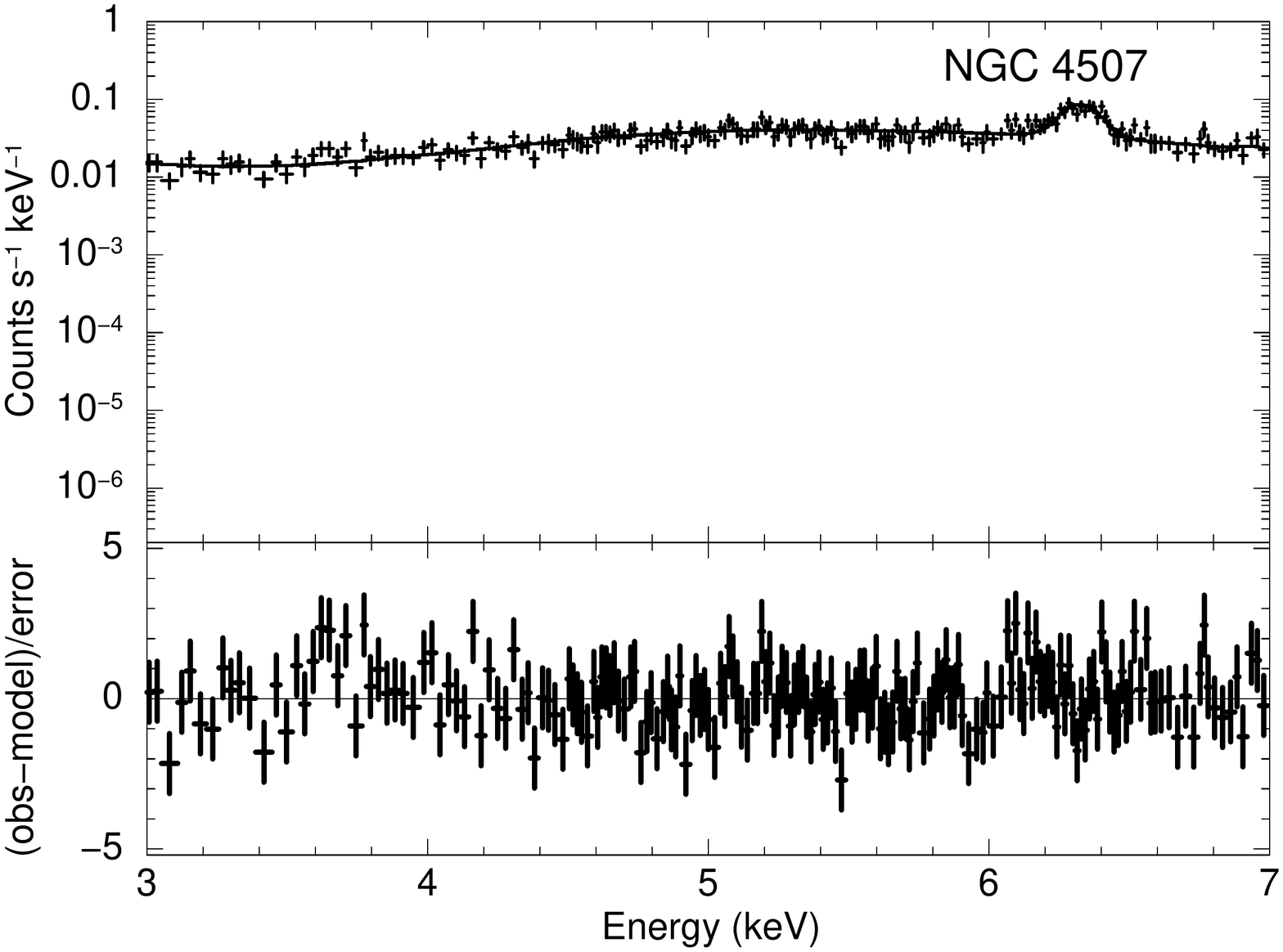}\hspace{-.5cm}
    \includegraphics[width=4.8cm]{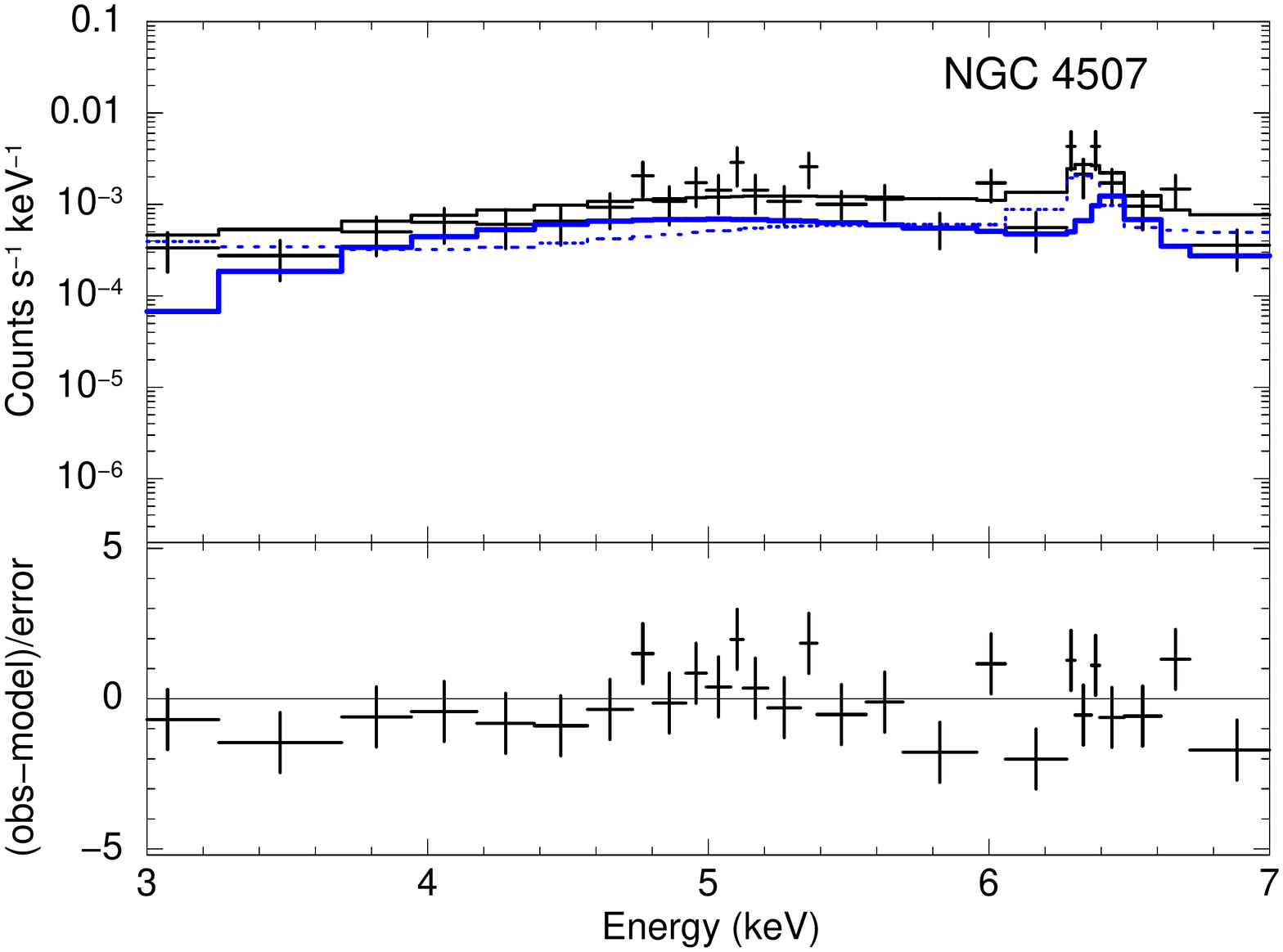}\hspace{-.5cm}
    \includegraphics[width=4.8cm]{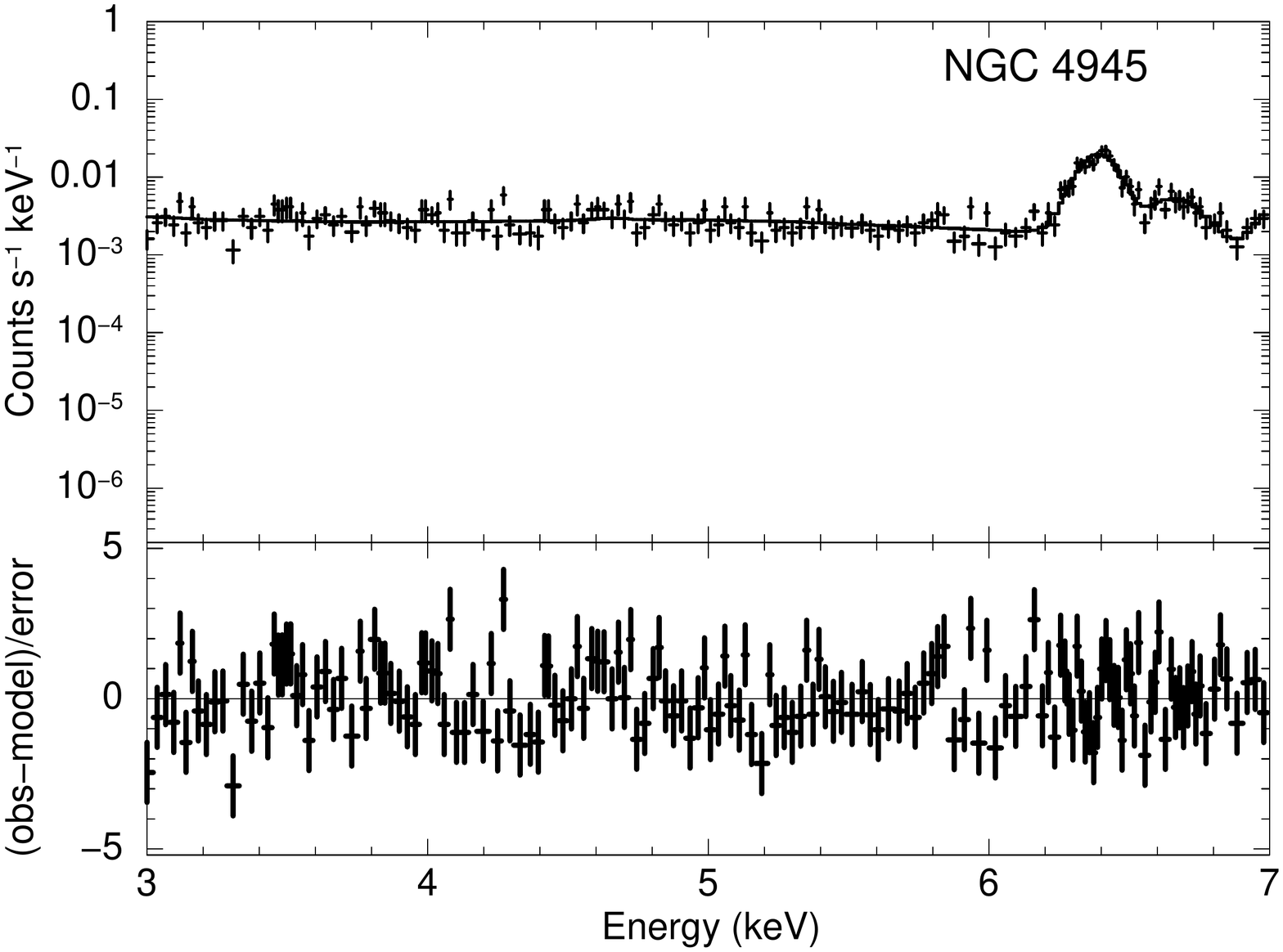}\hspace{-.5cm}
    \includegraphics[width=4.8cm]{15_NGC_4945_ext_spec.pdf}\\ \vspace{-0.5cm}
    \includegraphics[width=4.8cm]{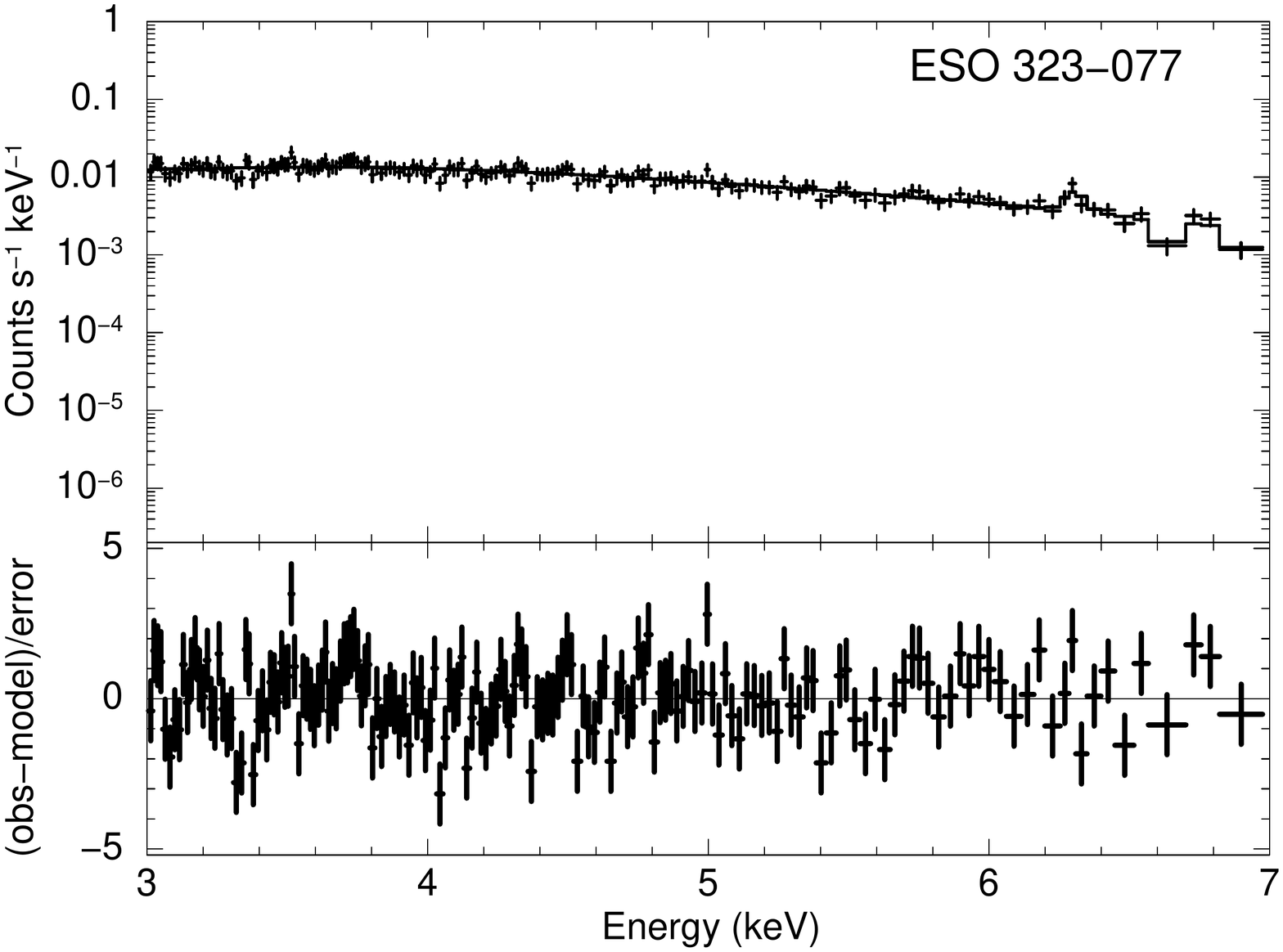}\hspace{-.5cm}
    \includegraphics[width=4.8cm]{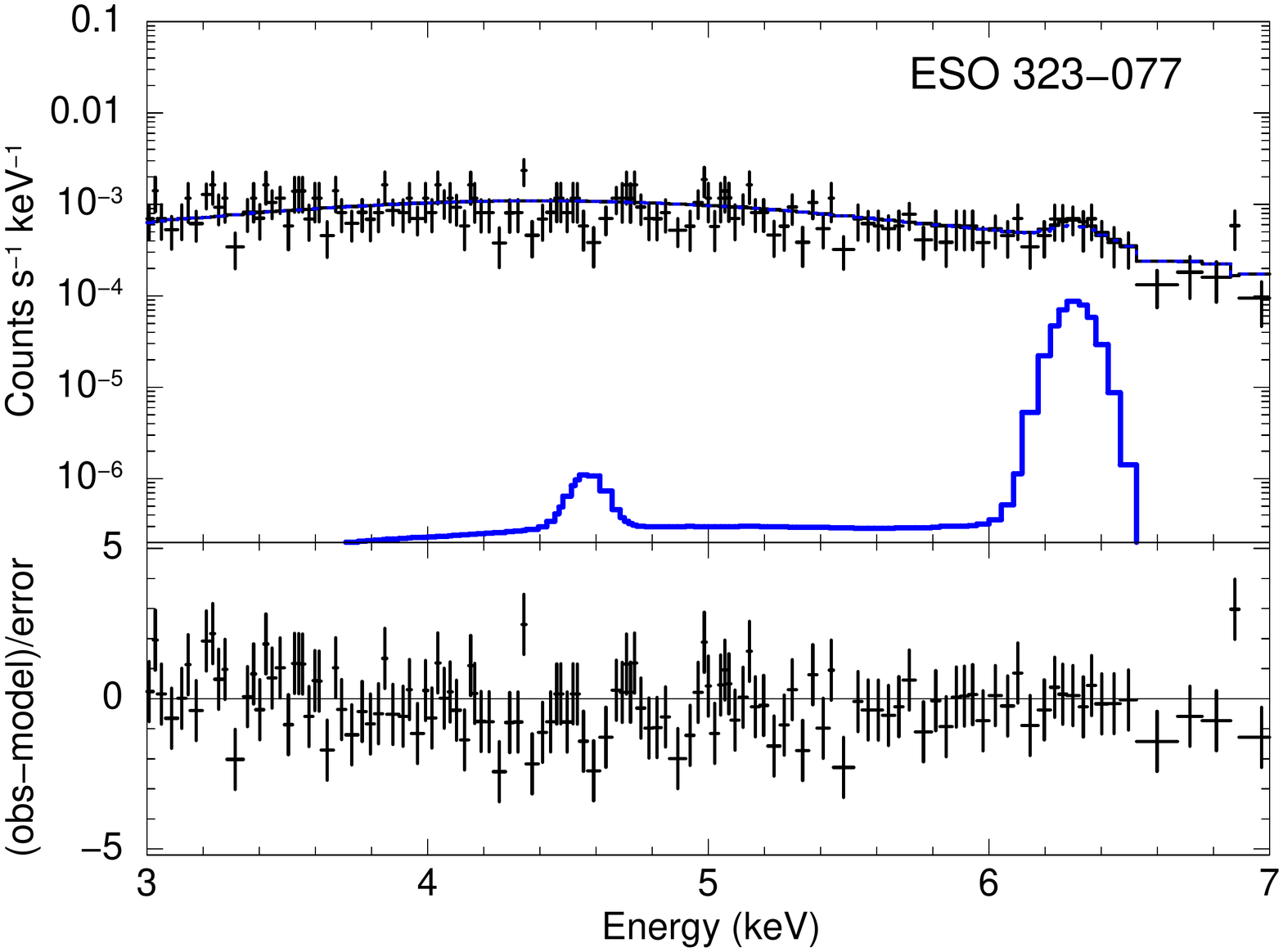}\hspace{-.5cm}
    \includegraphics[width=4.8cm]{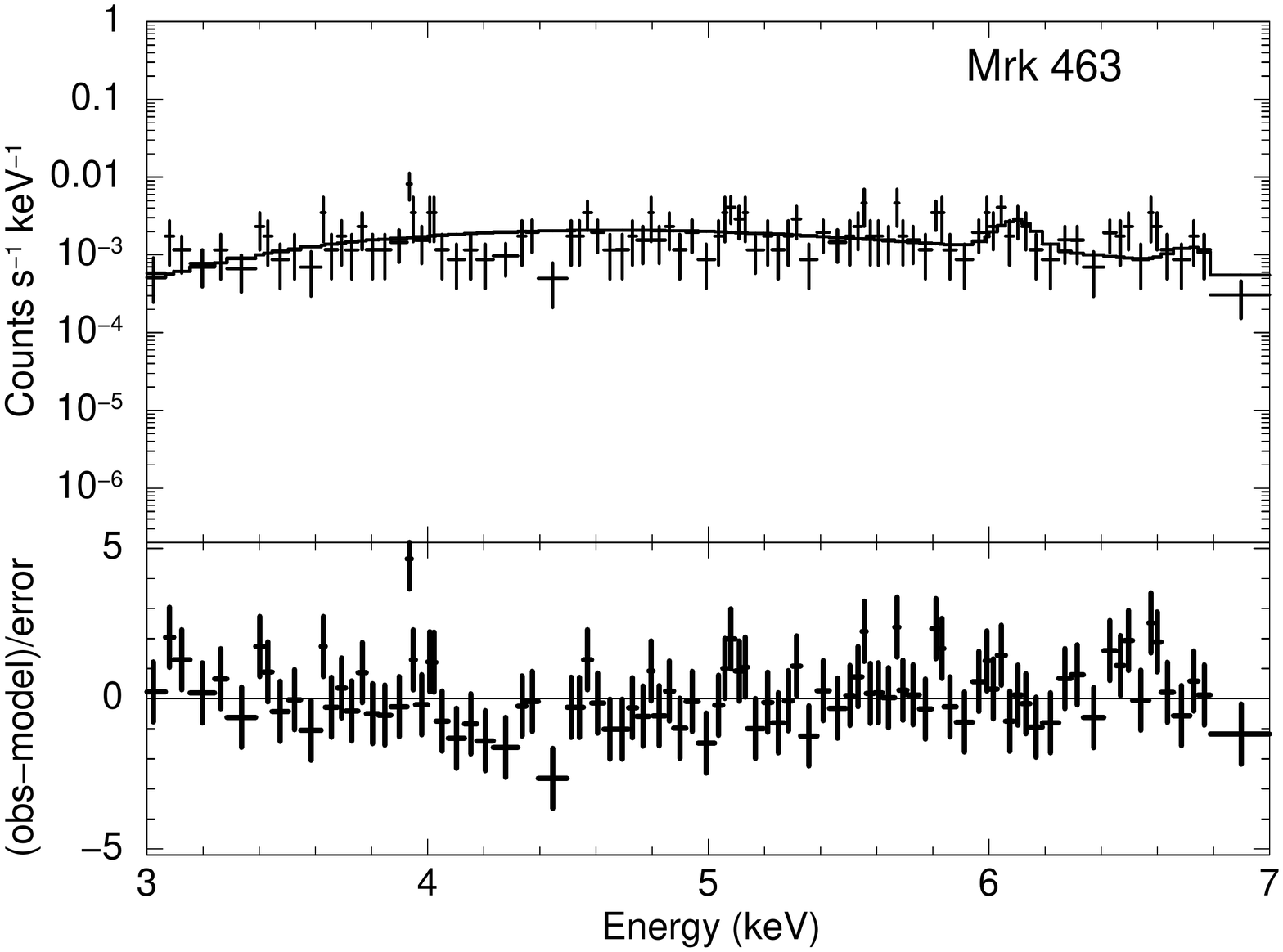}\hspace{-.5cm}
    \includegraphics[width=4.8cm]{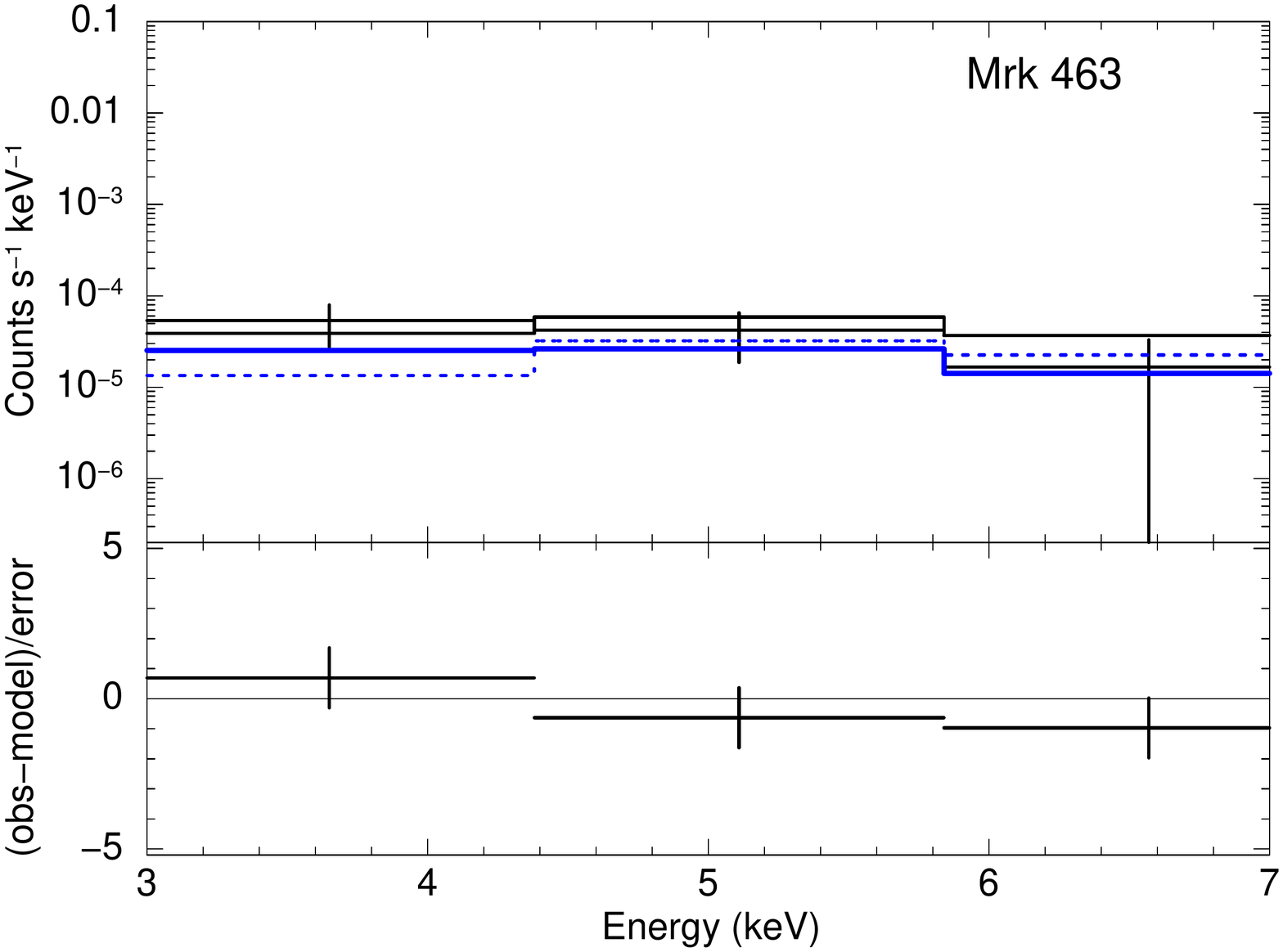}\\ \vspace{-0.5cm}
    \includegraphics[width=4.8cm]{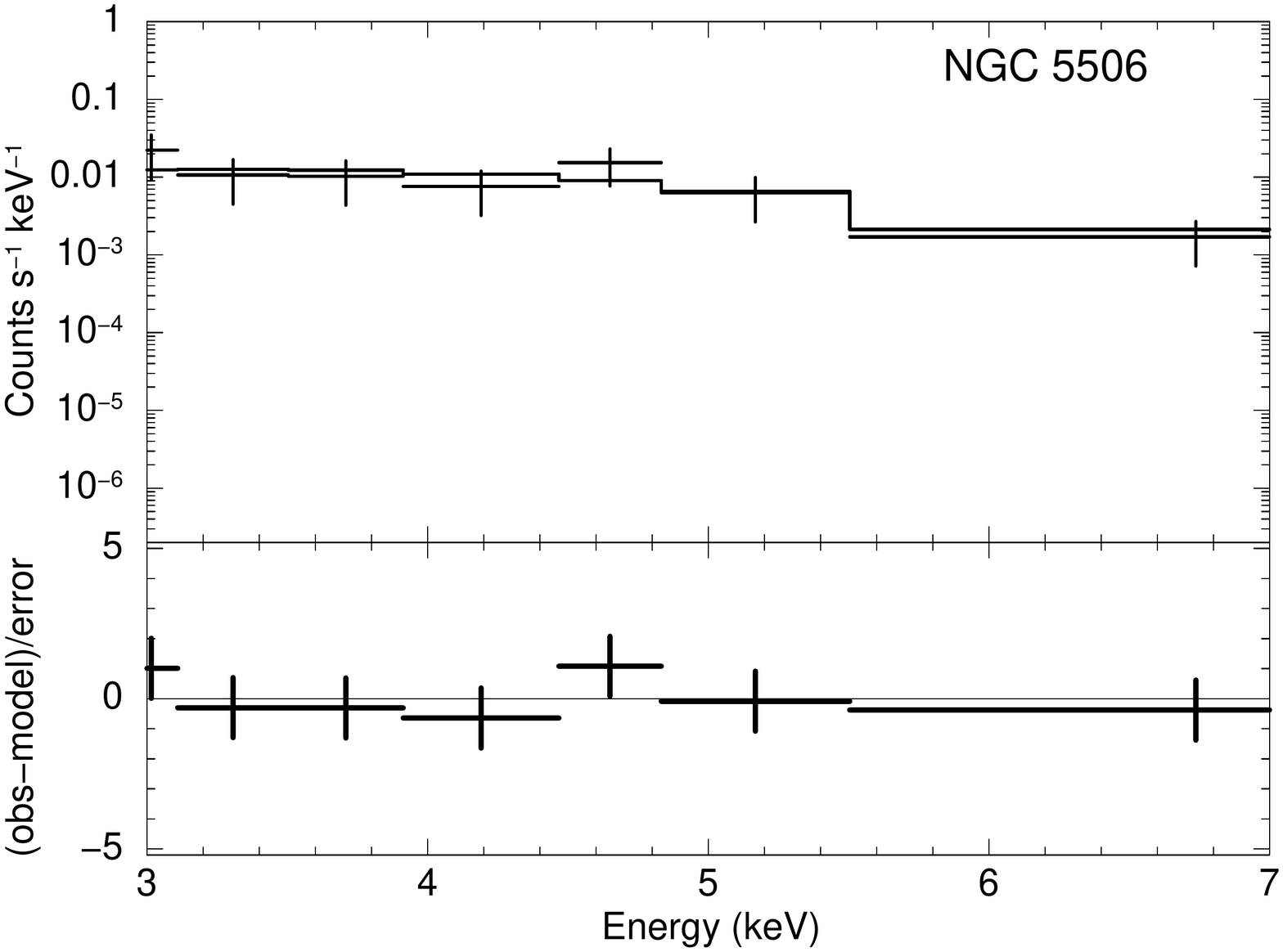}\hspace{-.5cm}
    \includegraphics[width=4.8cm]{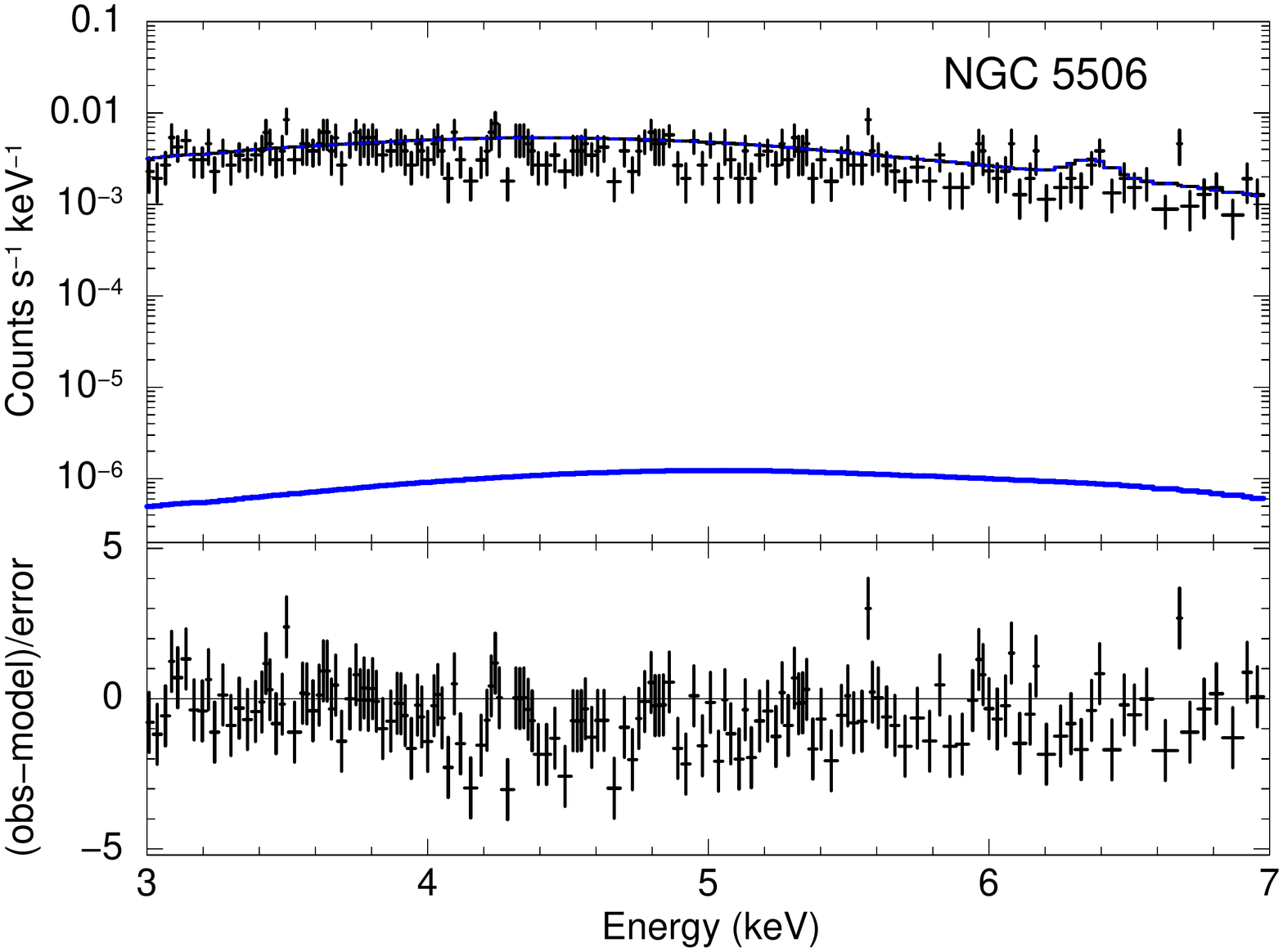}\hspace{-.5cm}
    \includegraphics[width=4.8cm]{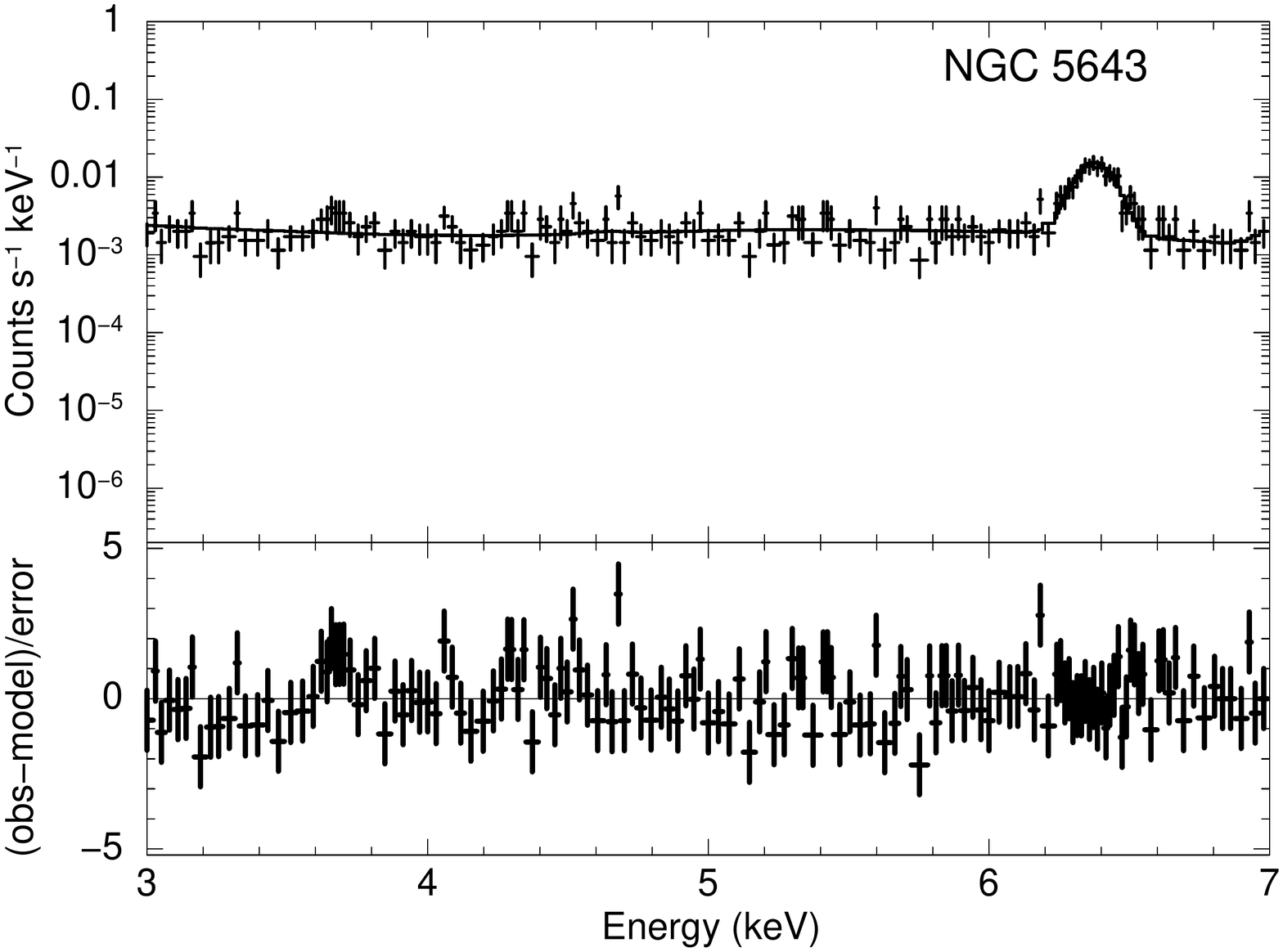}\hspace{-.5cm}
    \includegraphics[width=4.8cm]{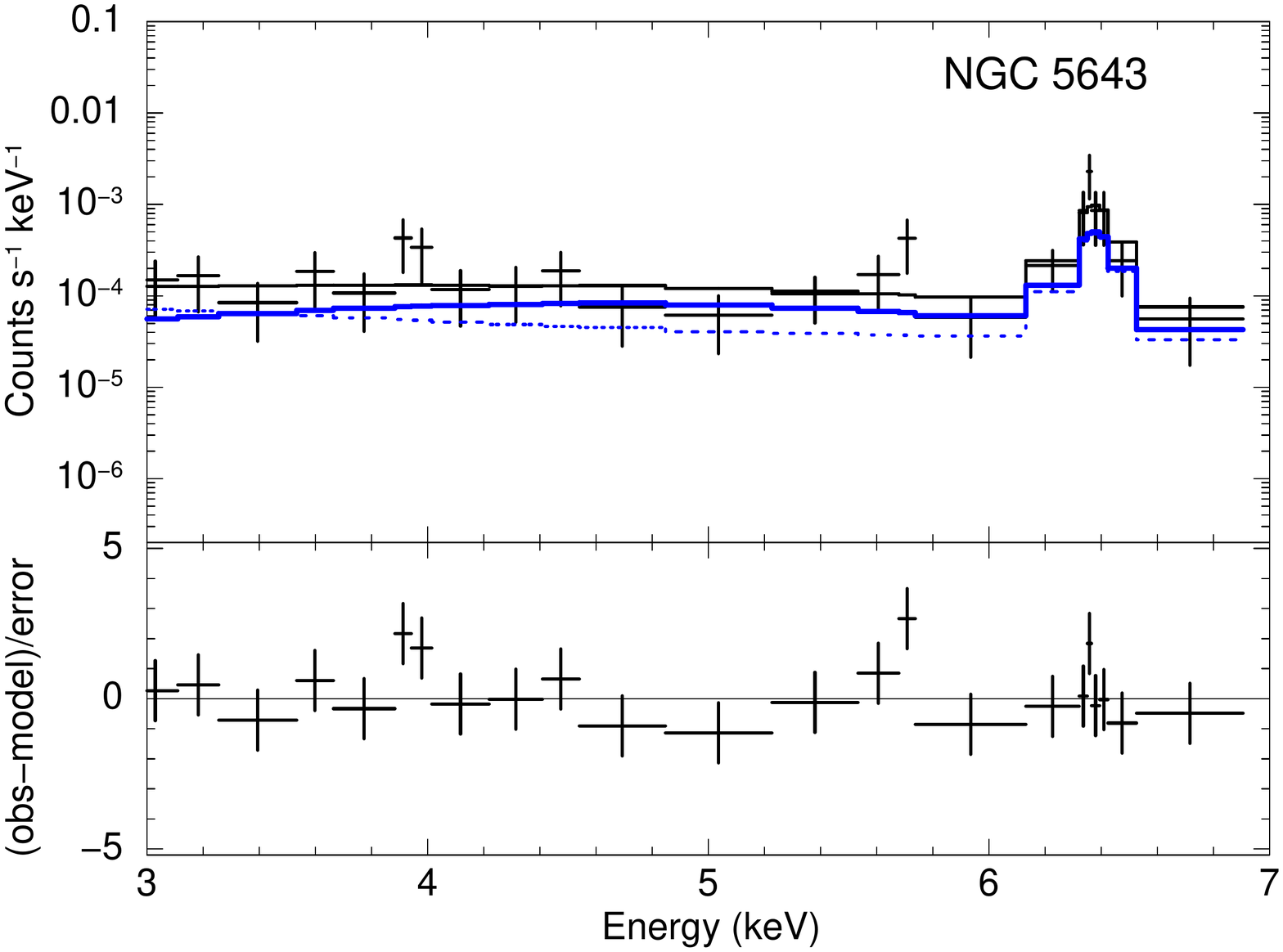}\\ \vspace{-0.5cm}
    \includegraphics[width=4.8cm]{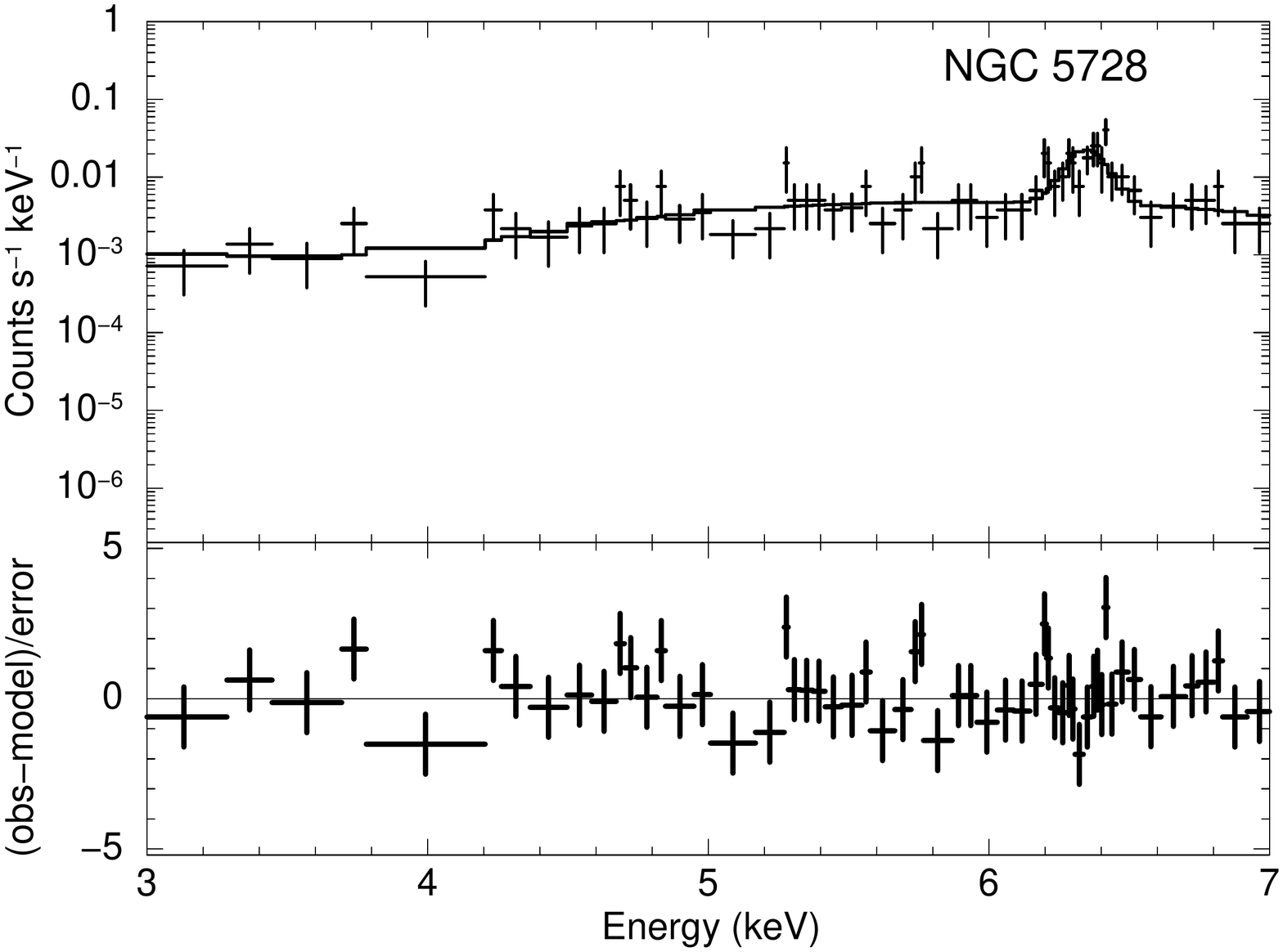}\hspace{-.5cm}
    \includegraphics[width=4.8cm]{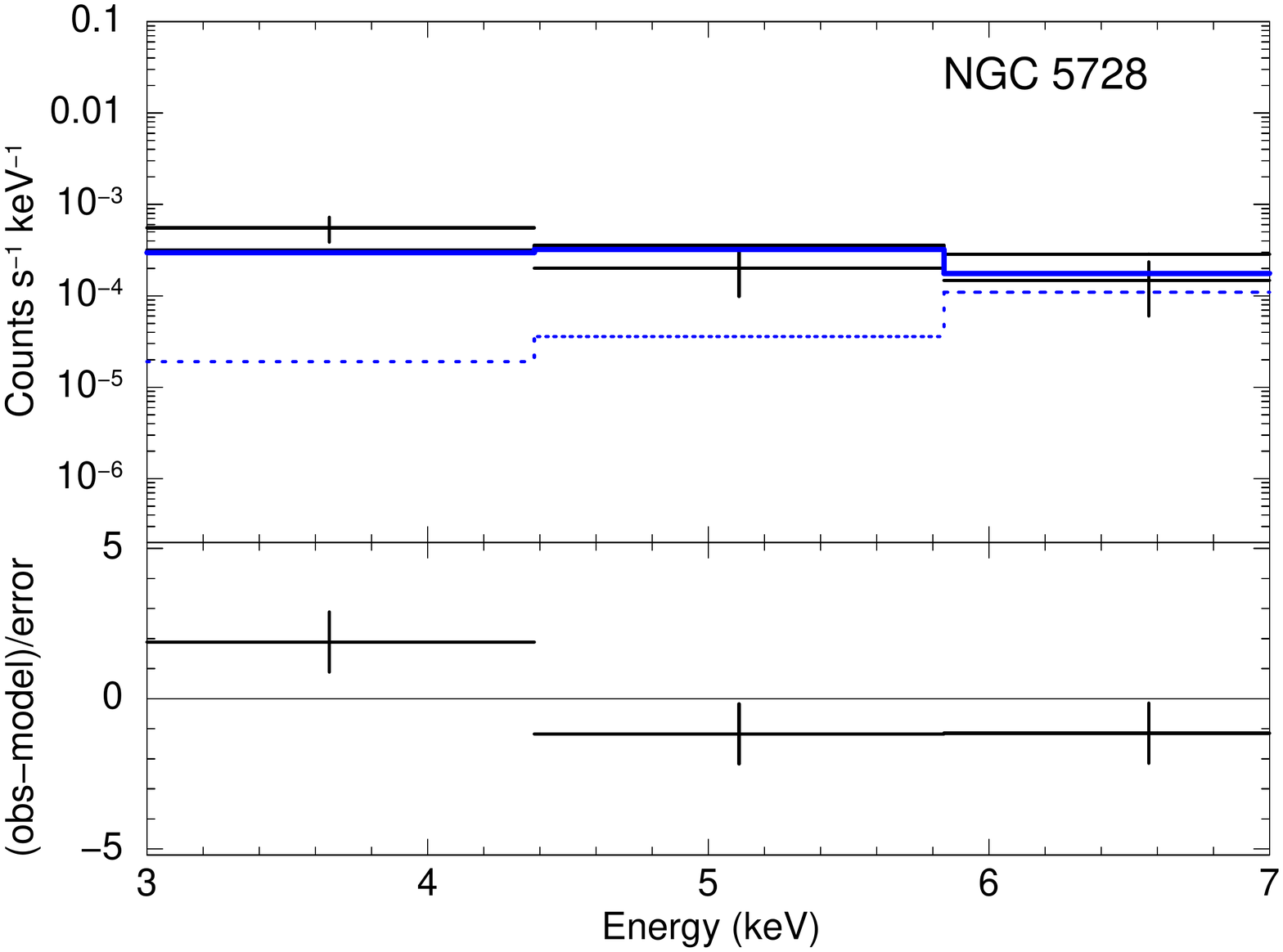}\hspace{-.5cm}
    \includegraphics[width=4.8cm]{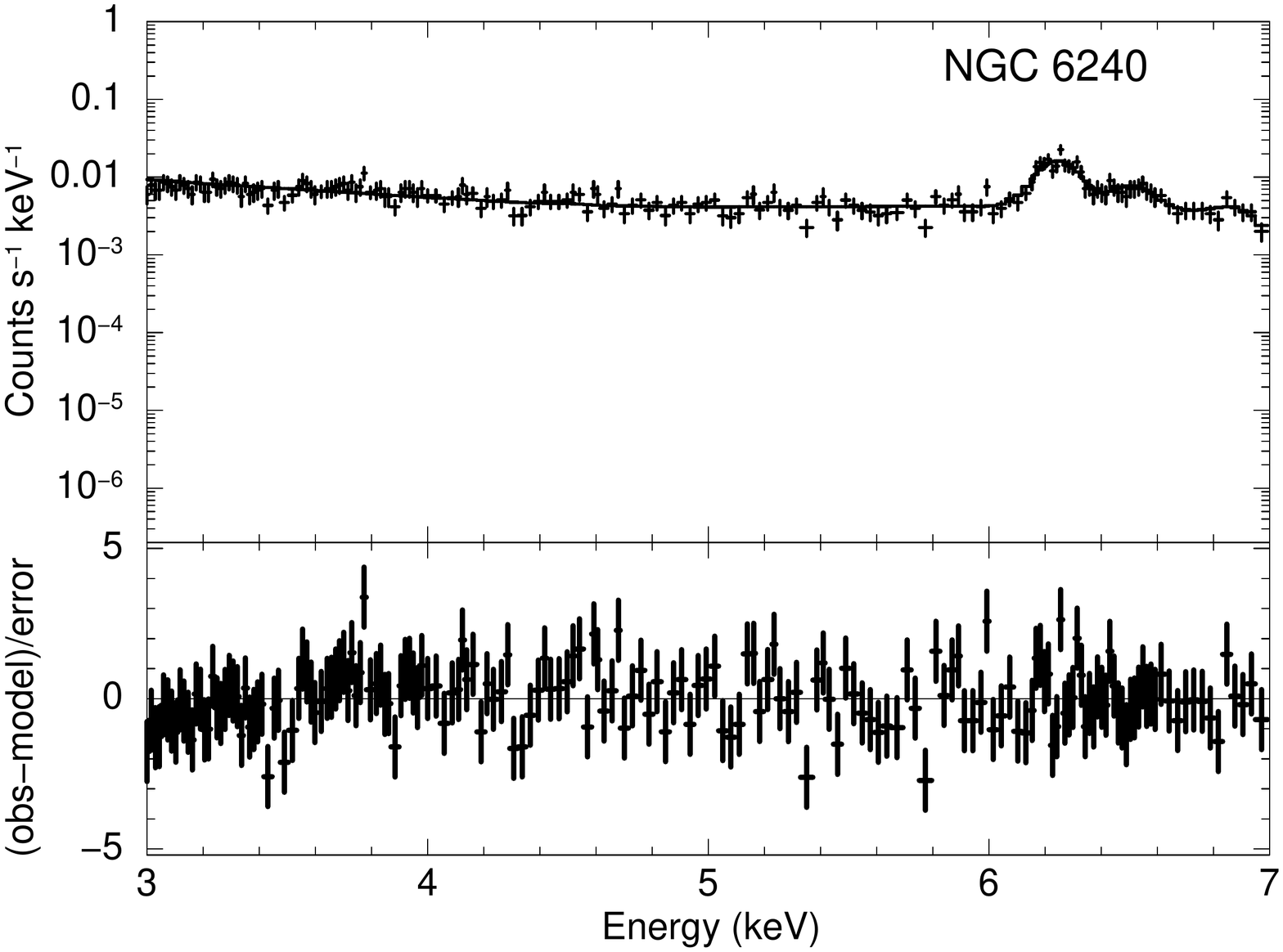}\hspace{-.5cm}
    \includegraphics[width=4.8cm]{22_NGC_6240_ext_spec.pdf}\\ \vspace{-0.5cm}
    \includegraphics[width=4.8cm]{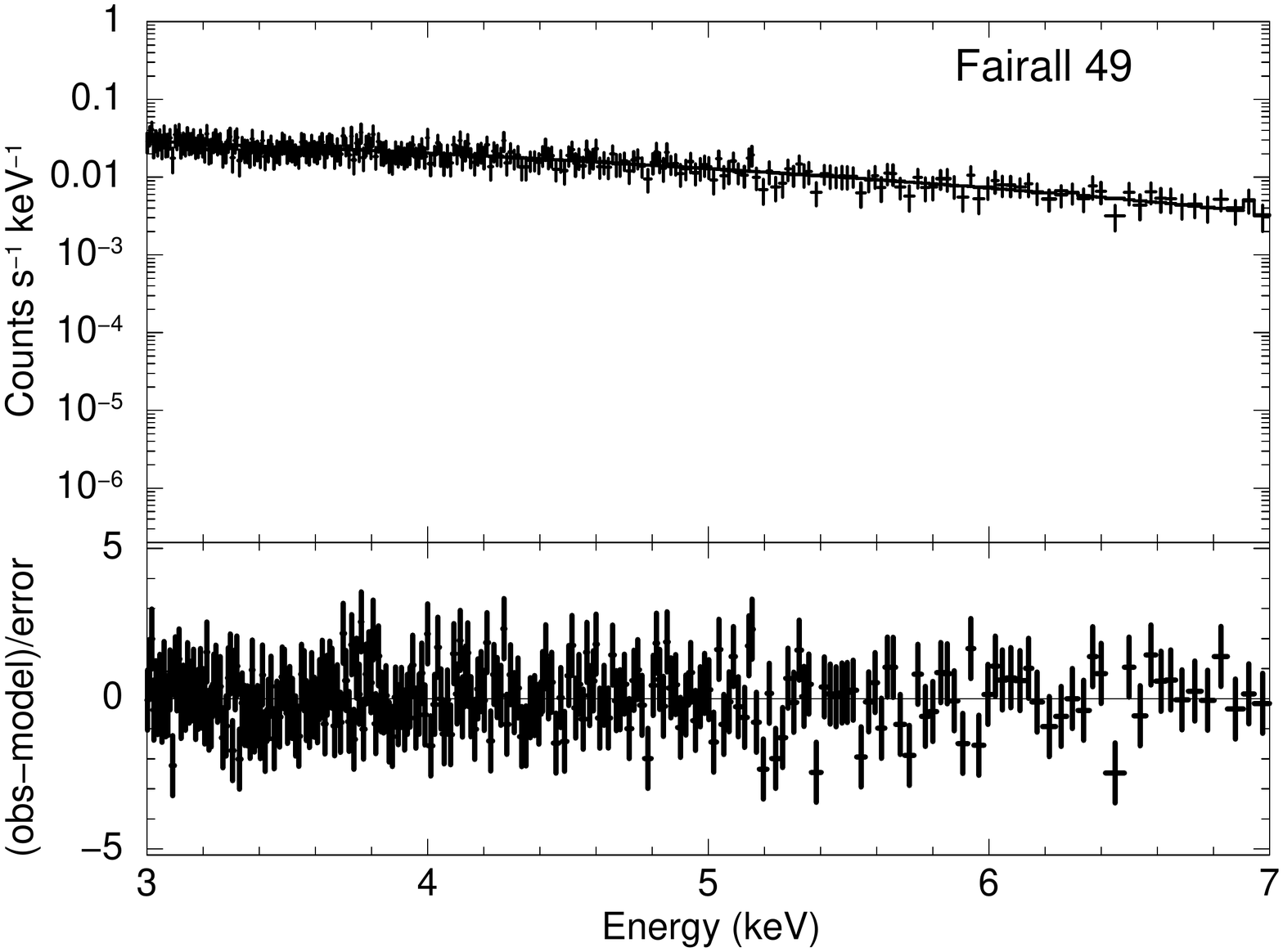}\hspace{-.5cm}
    \includegraphics[width=4.8cm]{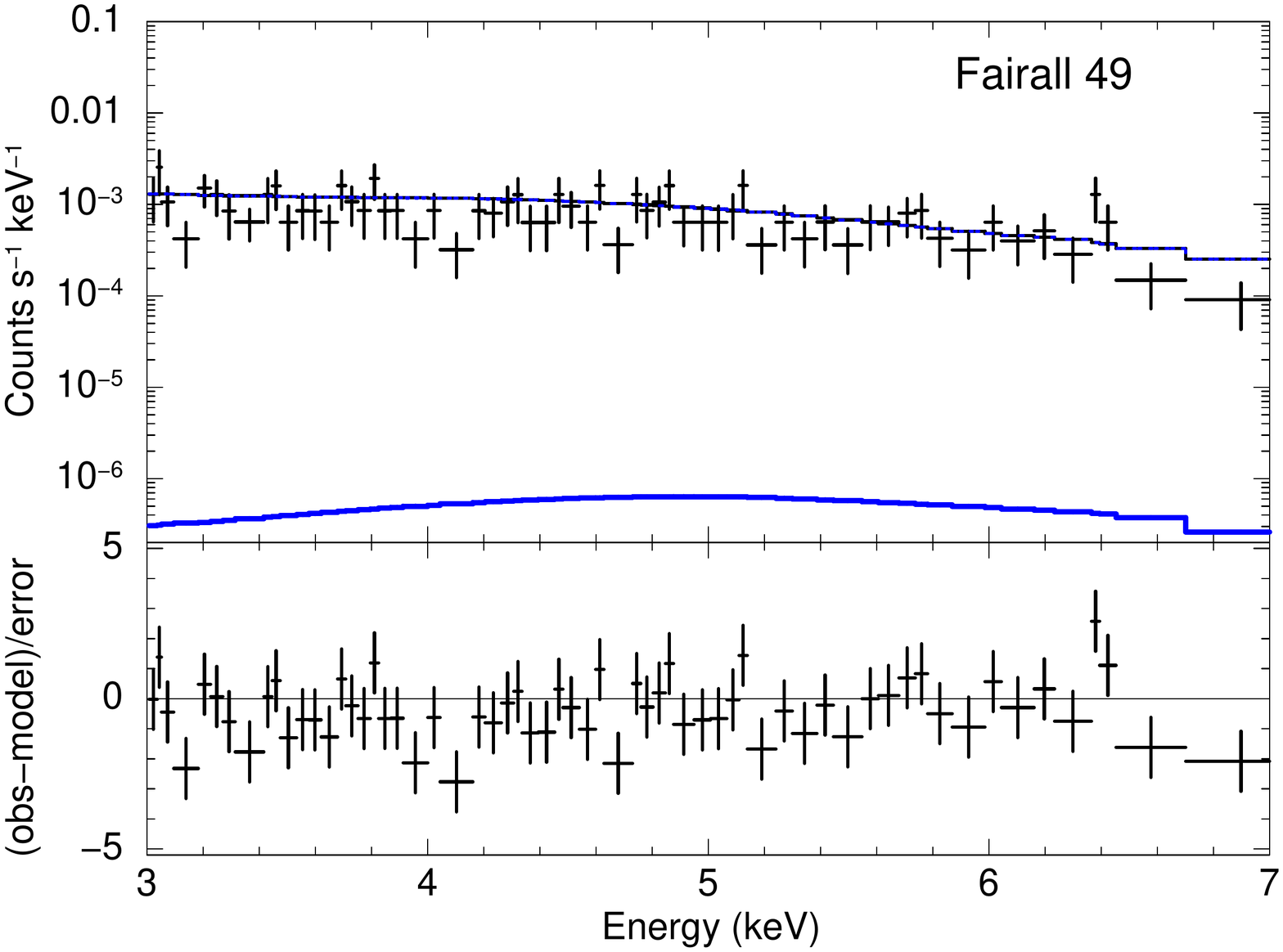}\hspace{-.5cm}
    \includegraphics[width=4.8cm]{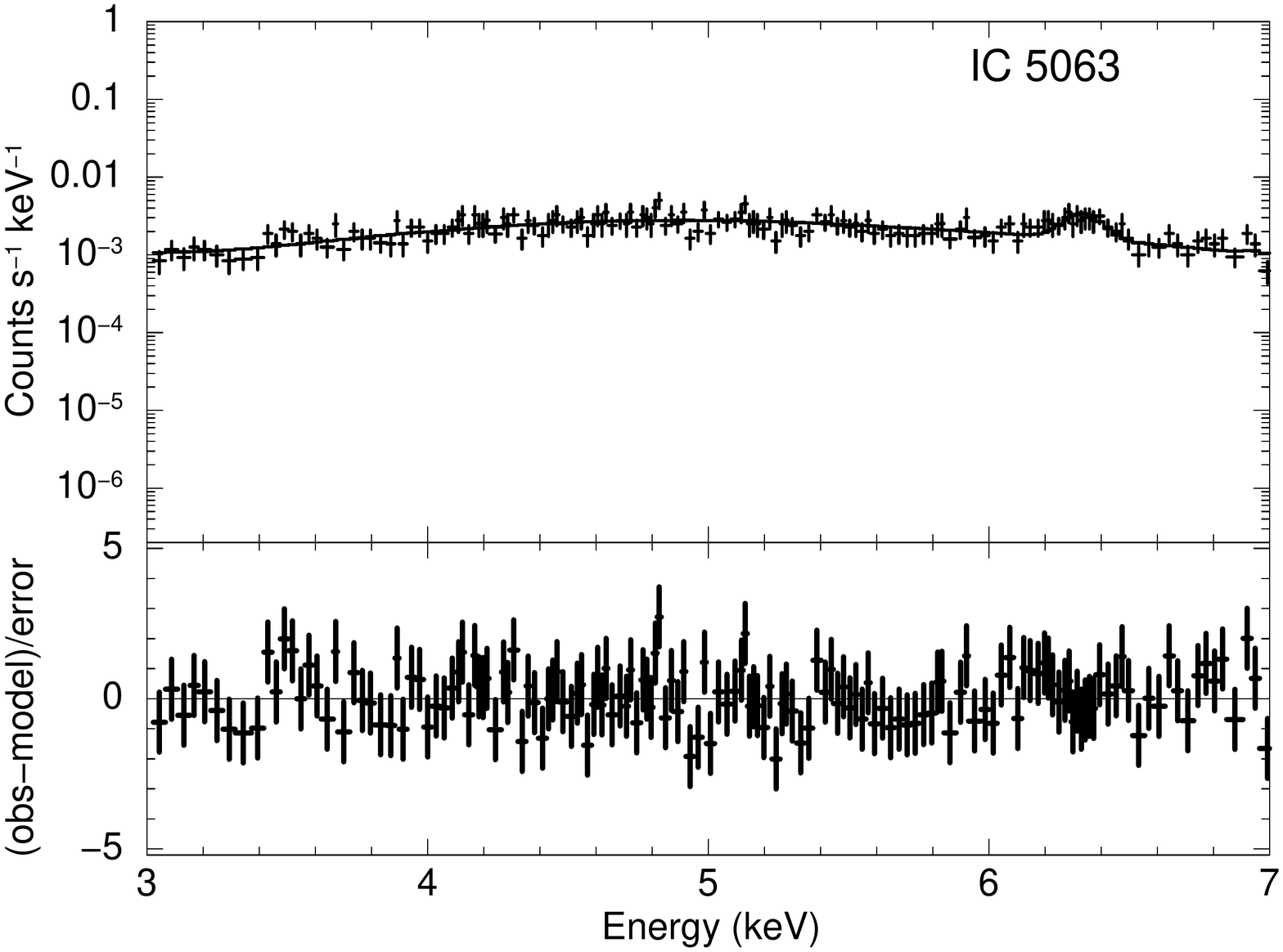}\hspace{-.5cm}
    \includegraphics[width=4.8cm]{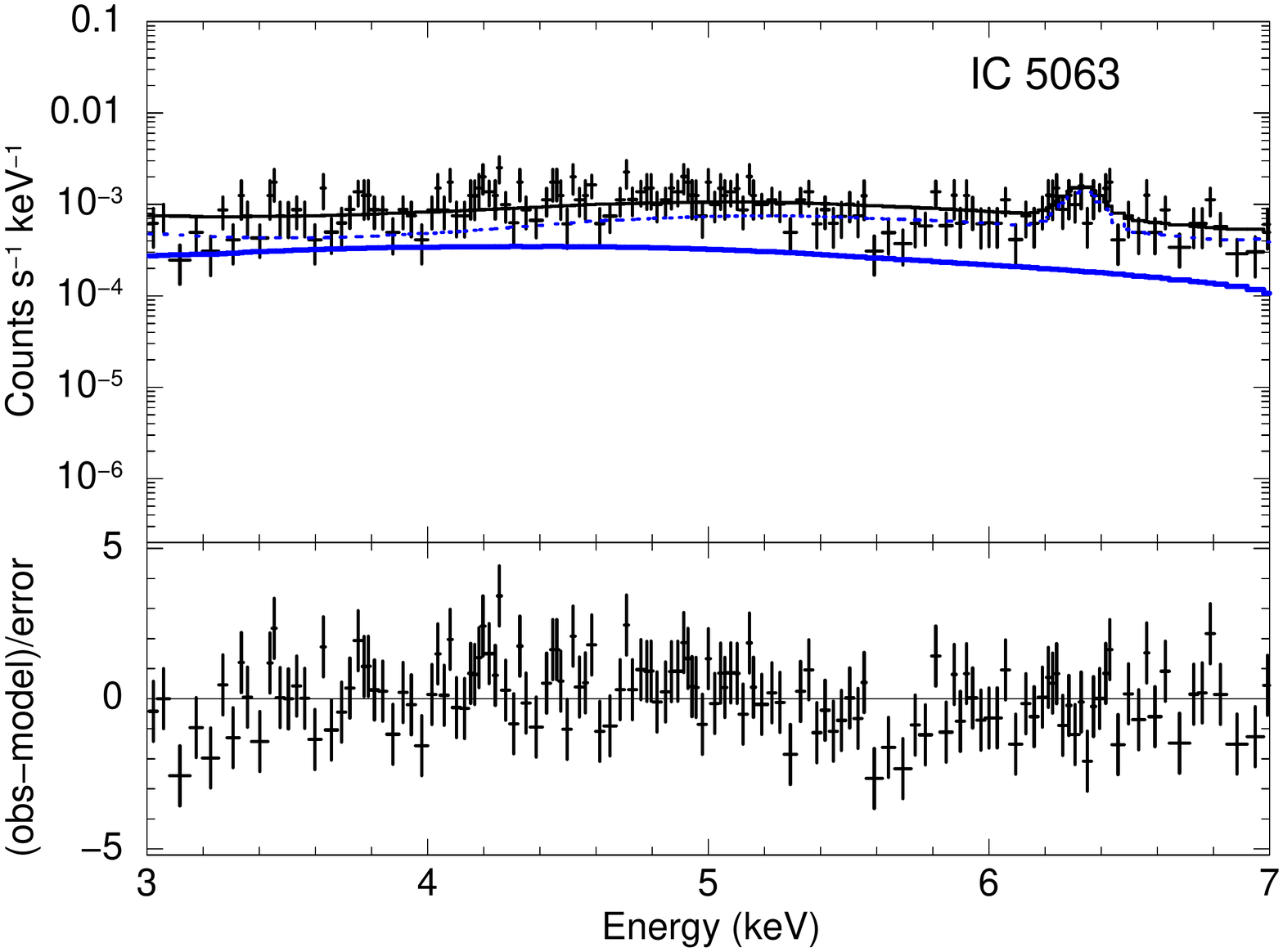}\\ \vspace{-0.5cm}
    \includegraphics[width=4.8cm]{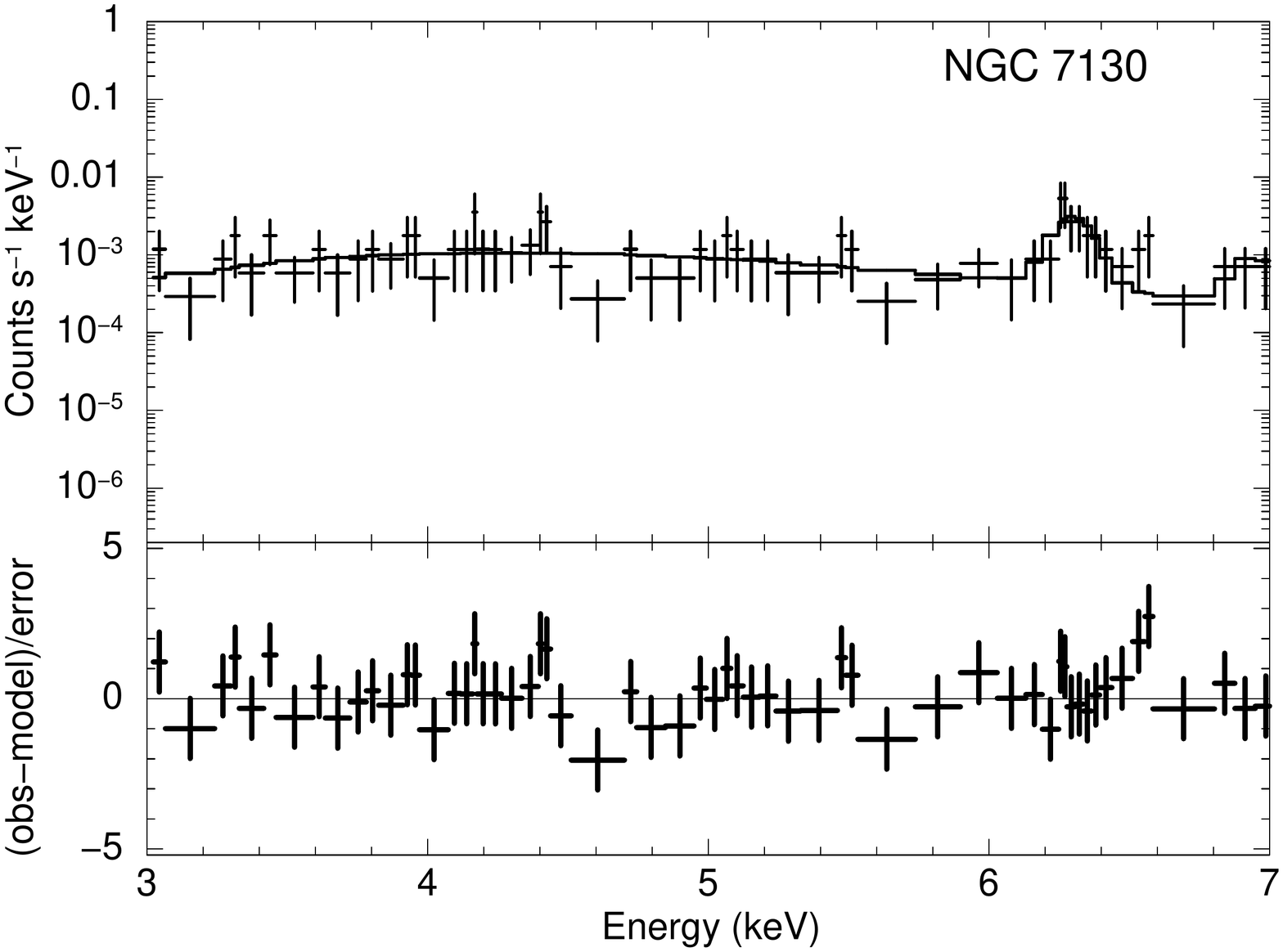}\hspace{-.5cm}
    \includegraphics[width=4.8cm]{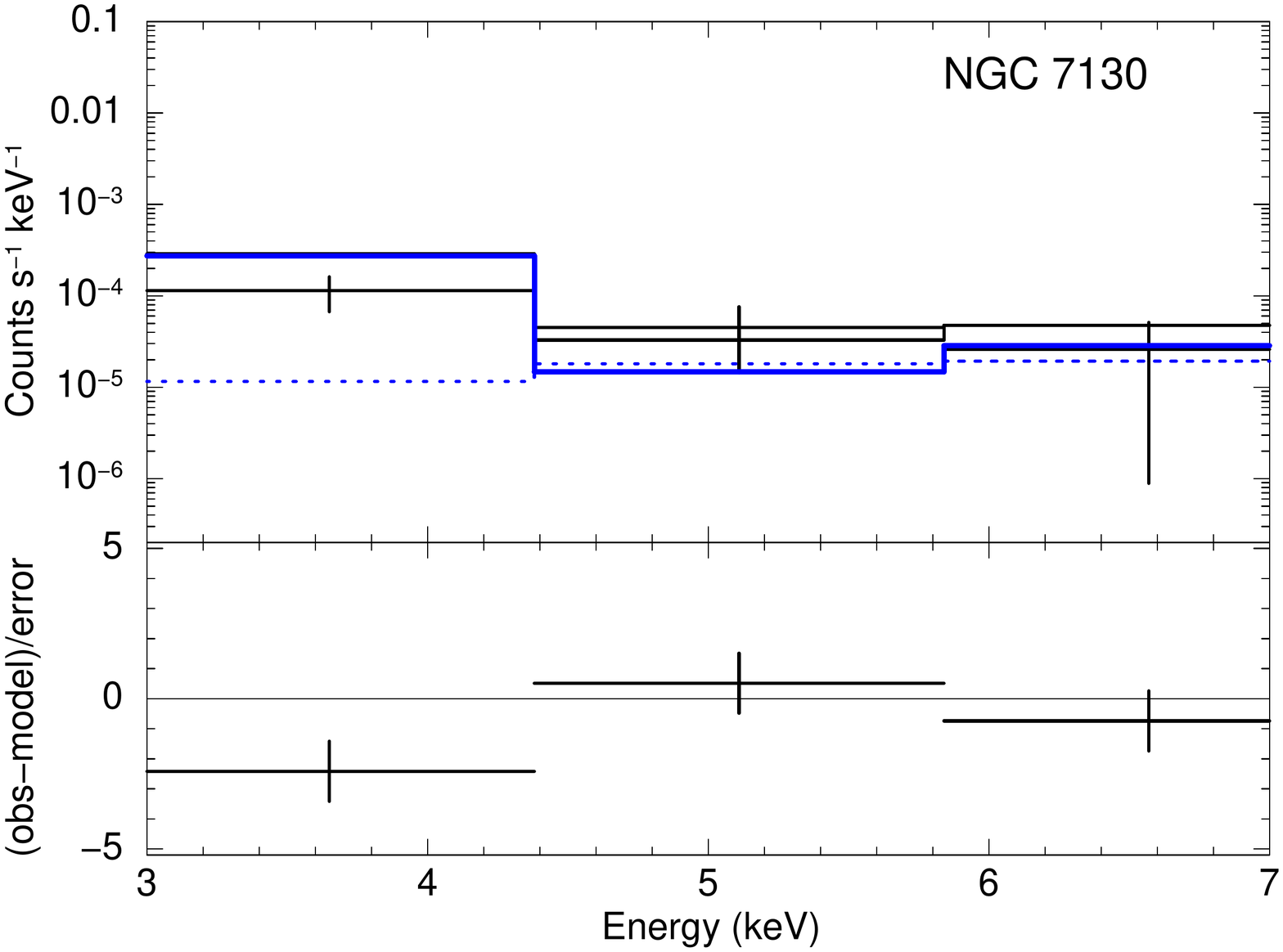}\hspace{-.5cm}
    \includegraphics[width=4.8cm]{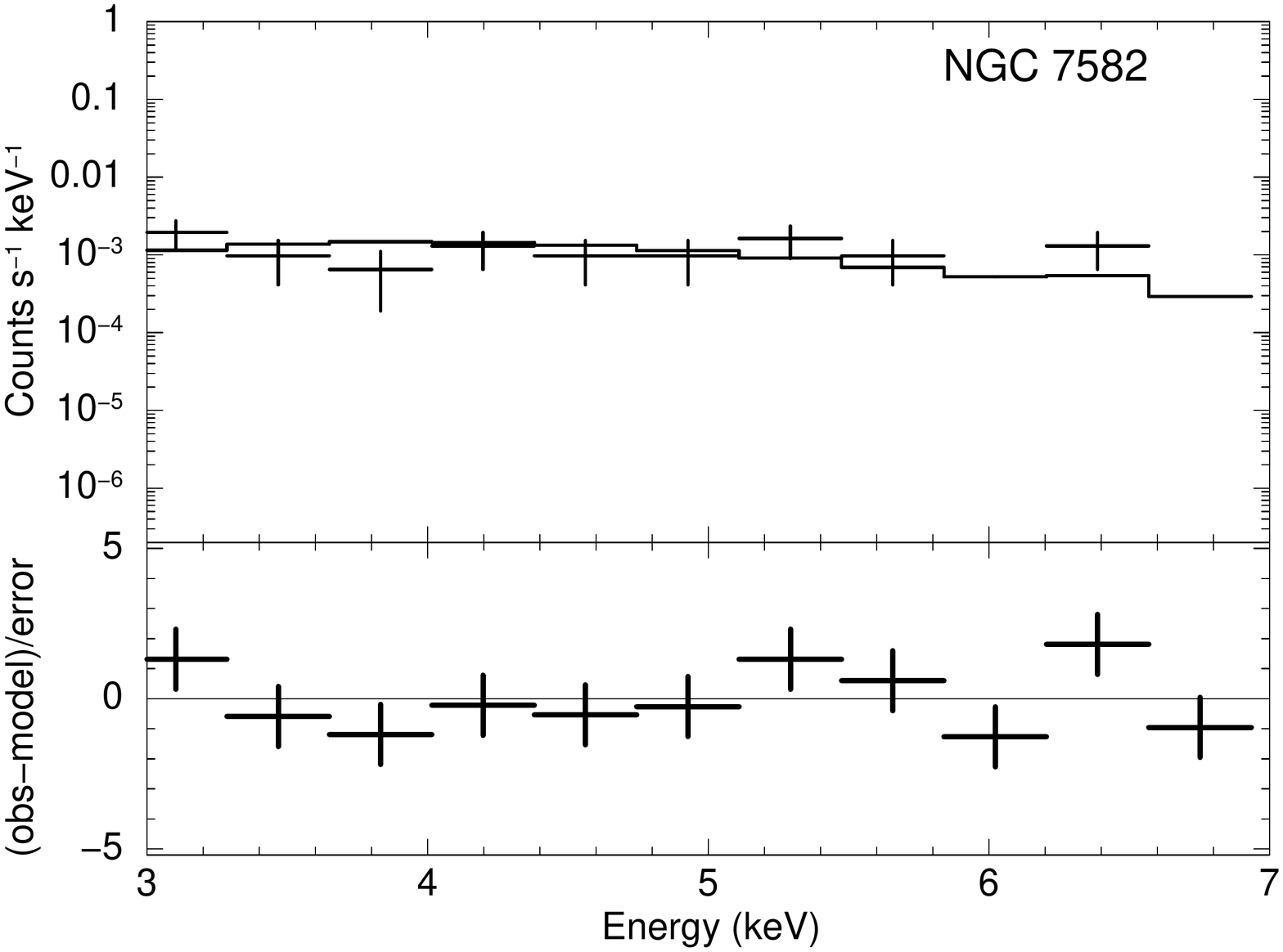}\hspace{-.5cm}
    \includegraphics[width=4.8cm]{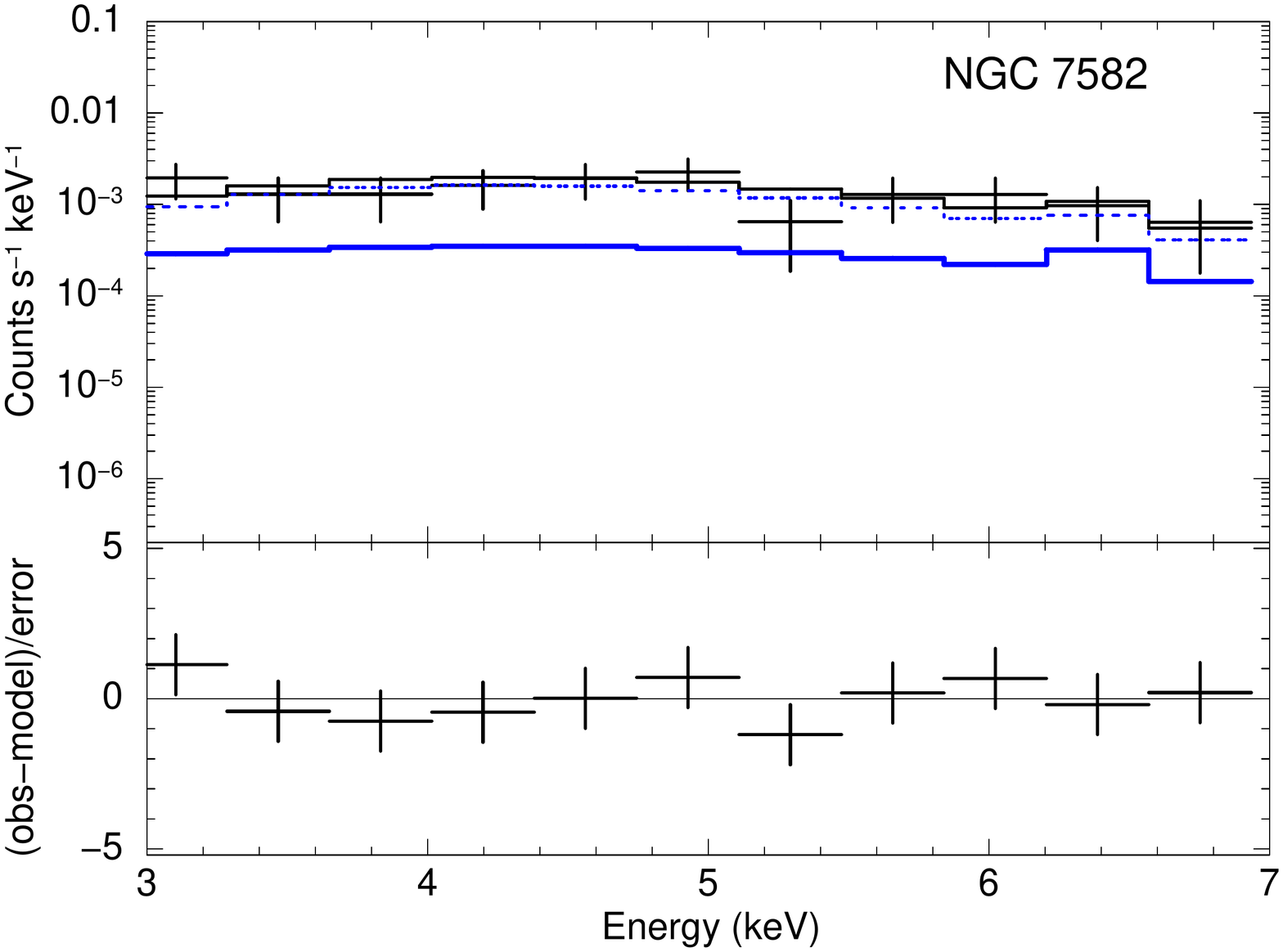}    
    \caption{Continued.}
\end{figure*}

\clearpage

\section{Radial distributions of X-ray emission}\label{app:xrad}

\begin{figure*}[!h]
    \centering
    \includegraphics[width=6cm]{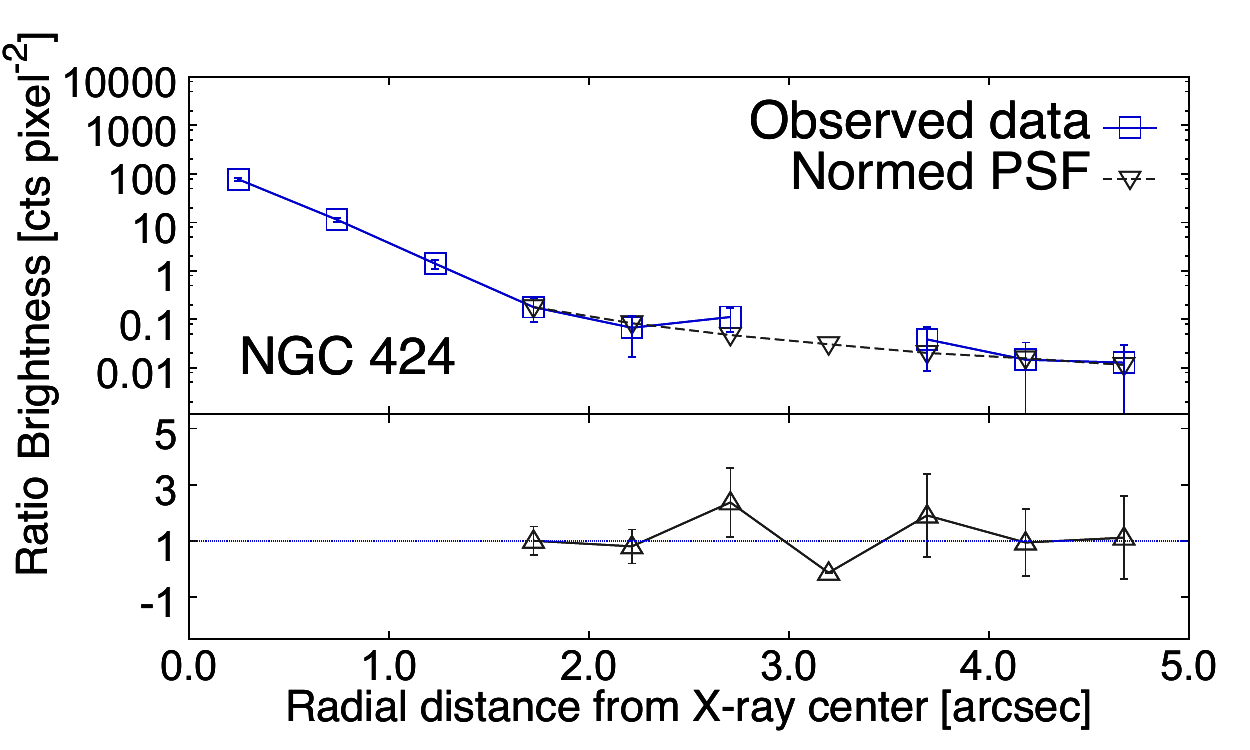}\hspace{-.3cm}
    \includegraphics[width=6cm]{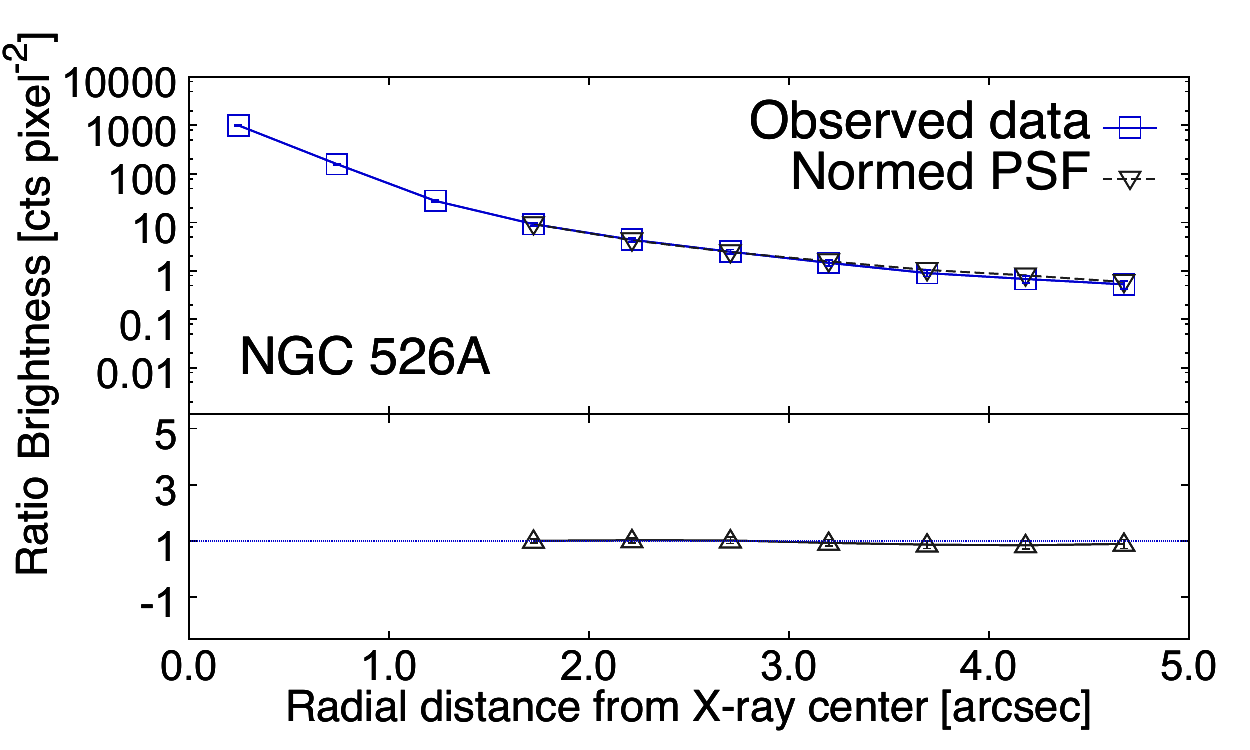}\hspace{-.3cm}
    \includegraphics[width=6cm]{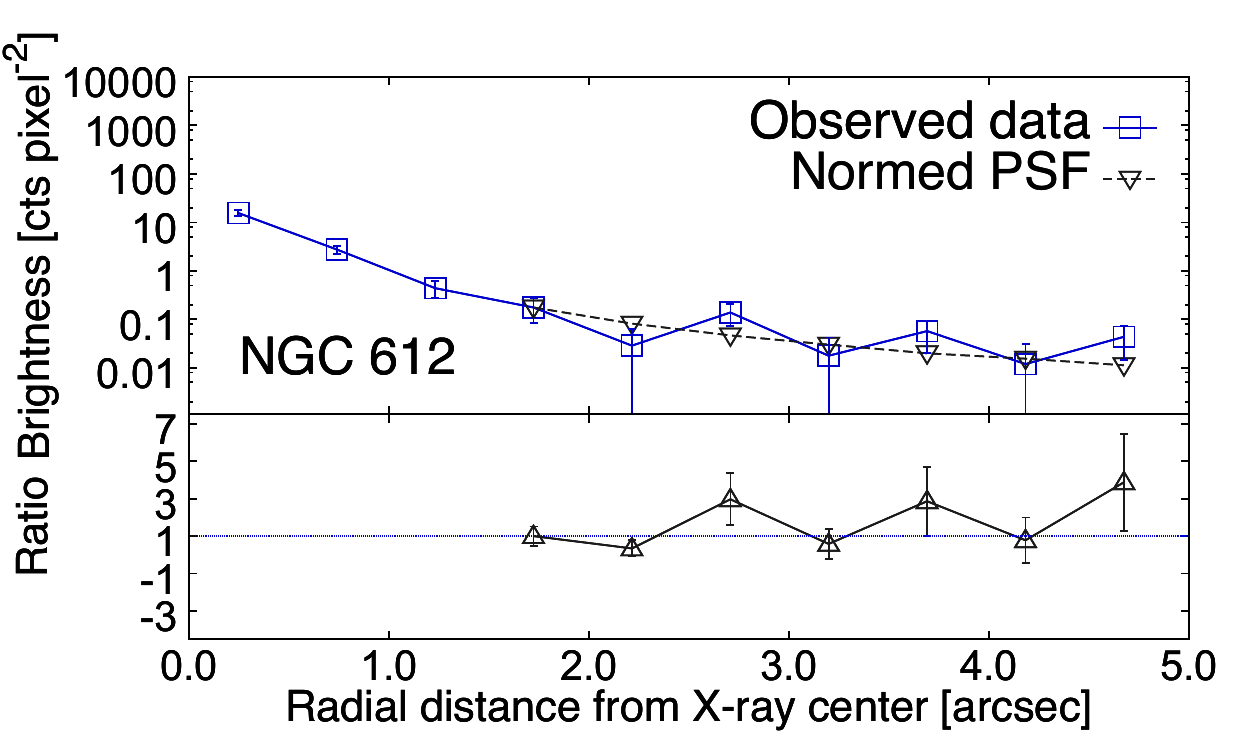}\\
    \includegraphics[width=6.0cm]{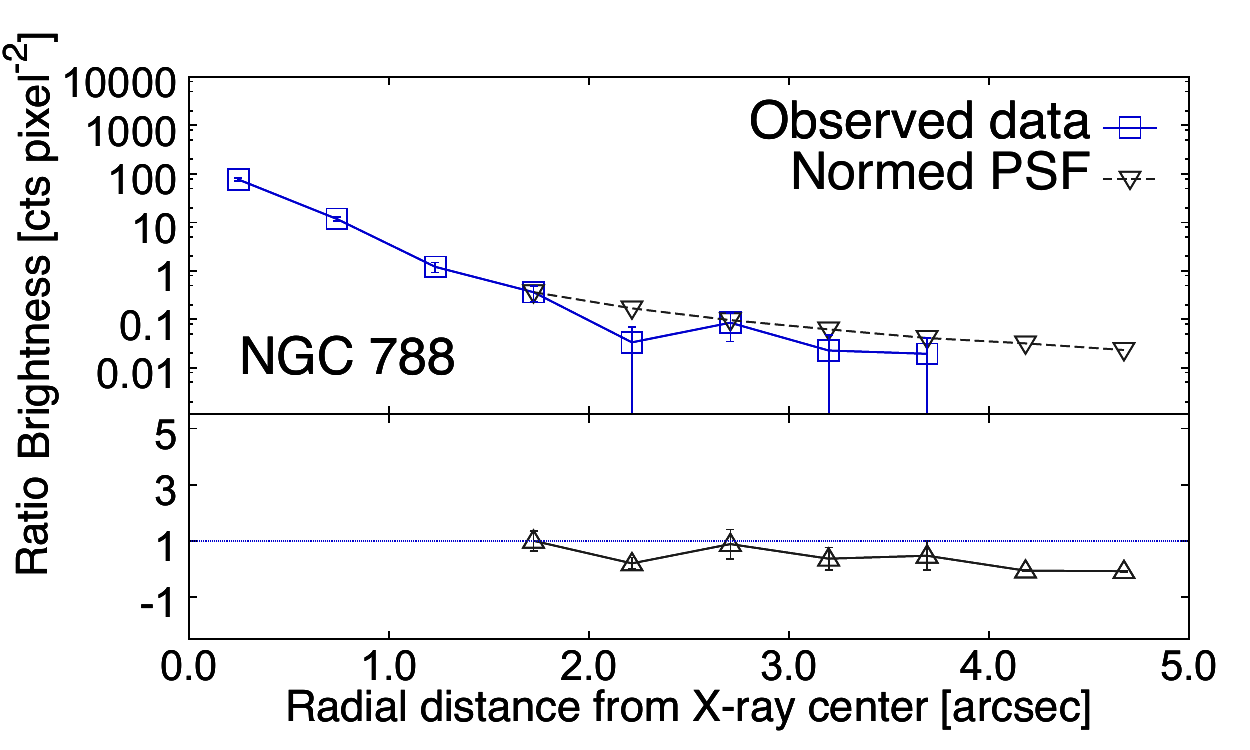}\hspace{-.3cm}
    \includegraphics[width=6.0cm]{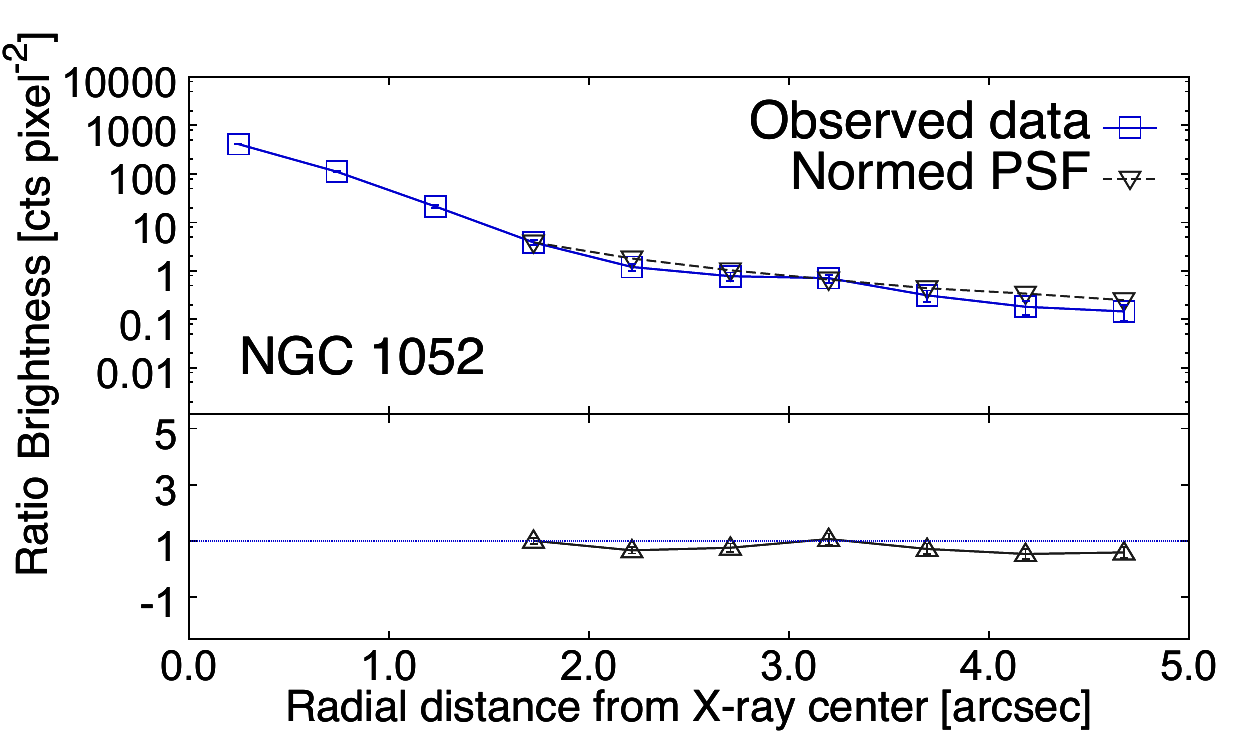}\hspace{-.3cm}
    \includegraphics[width=6.0cm]{3-6keV_06_NGC_1068_rad.pdf}\\
    \includegraphics[width=6.0cm]{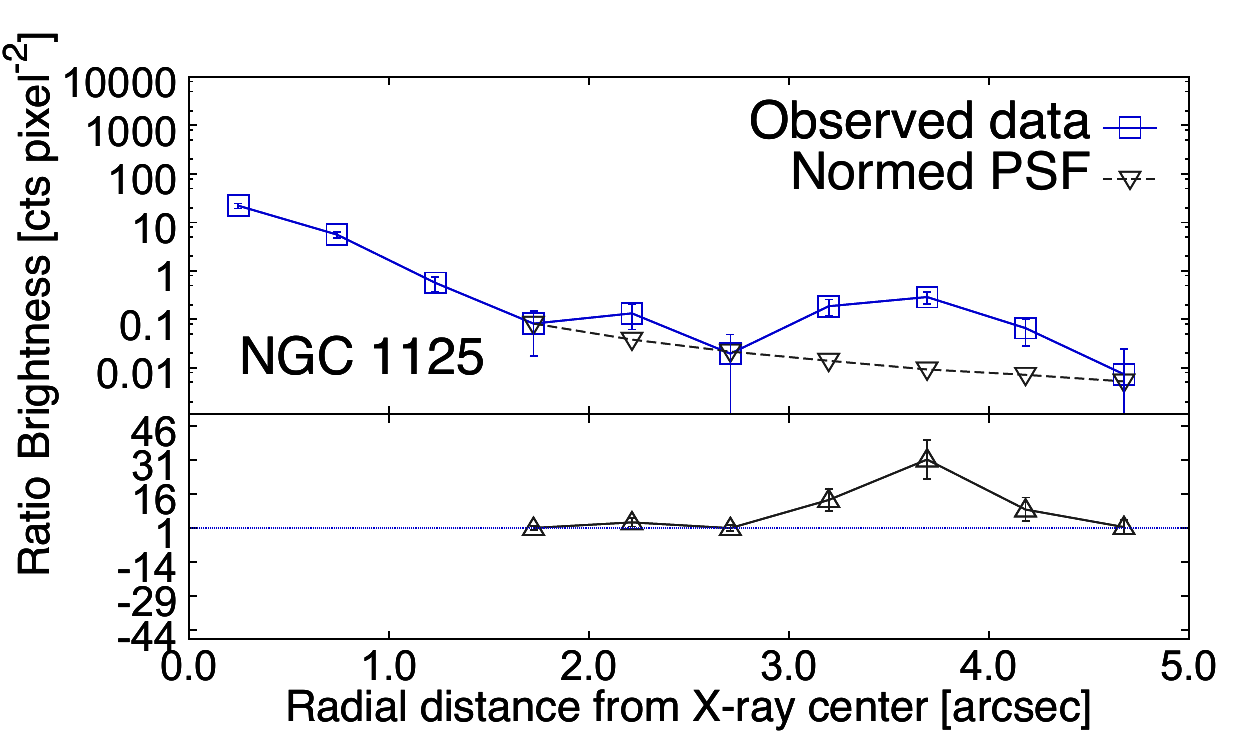}\hspace{-.3cm}
    \includegraphics[width=6.0cm]{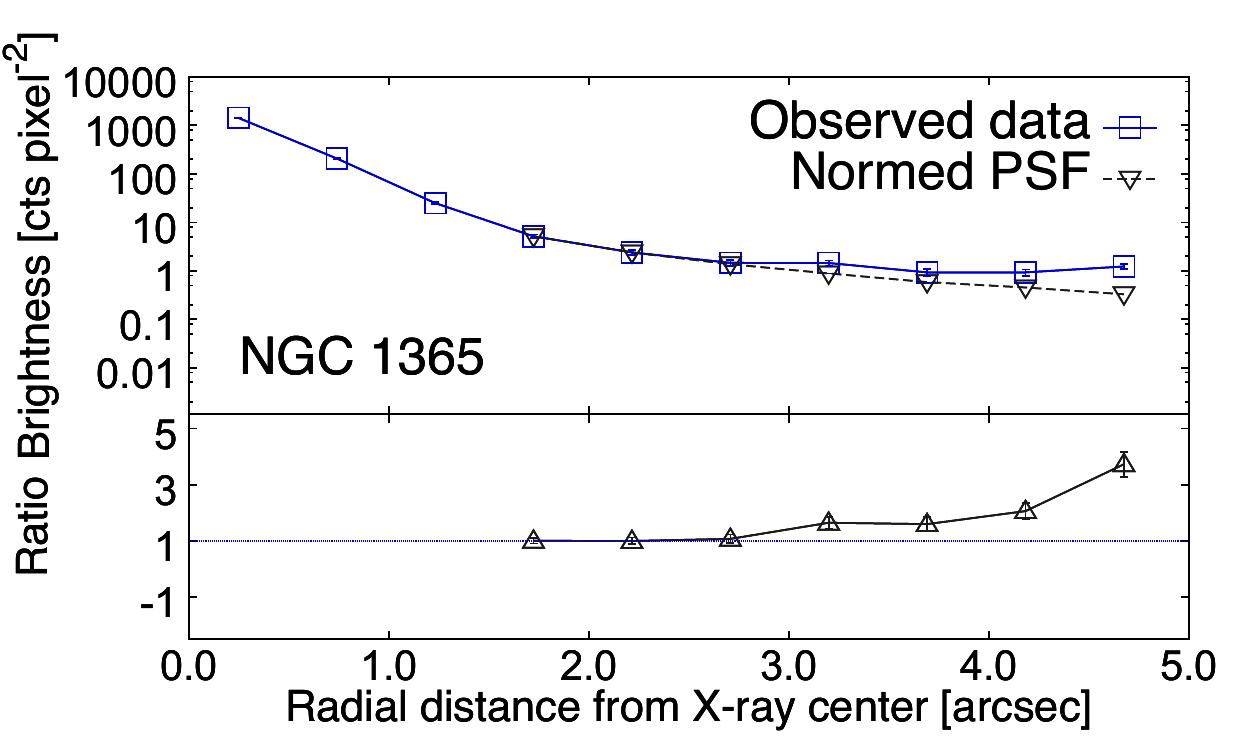}\hspace{-.3cm}
    \includegraphics[width=6.0cm]{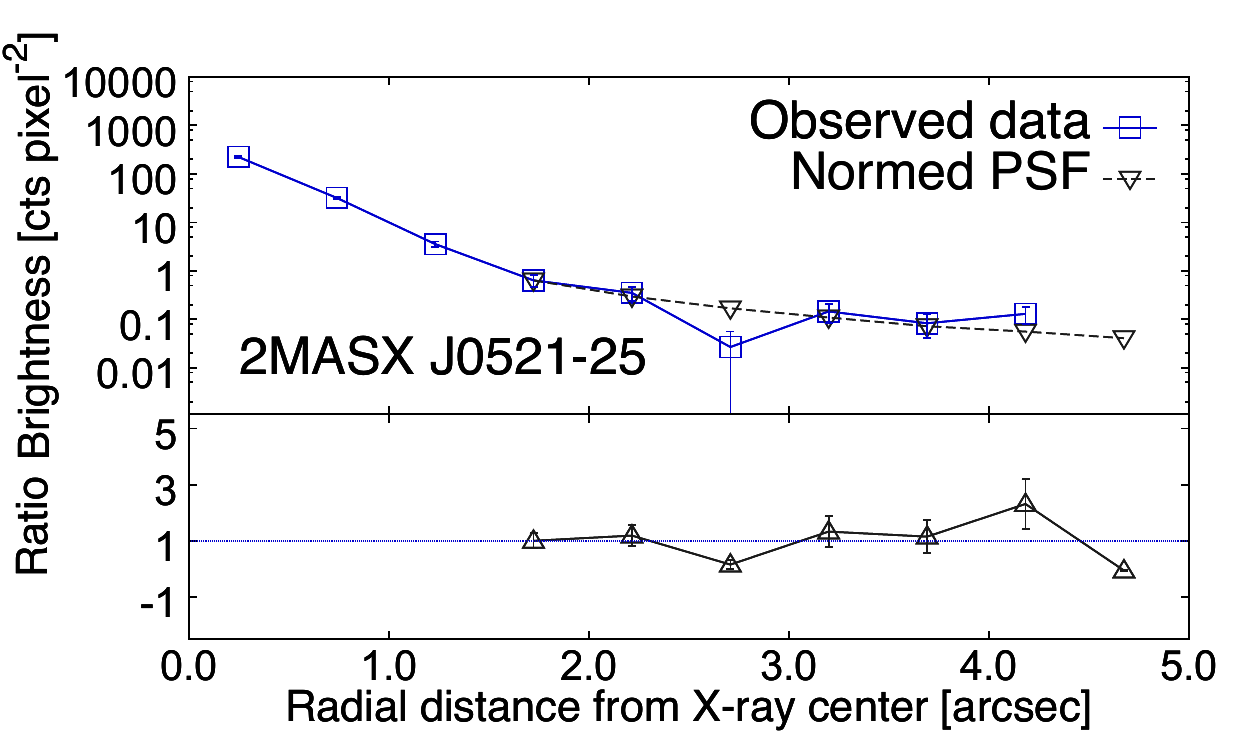}\\
    \includegraphics[width=6.0cm]{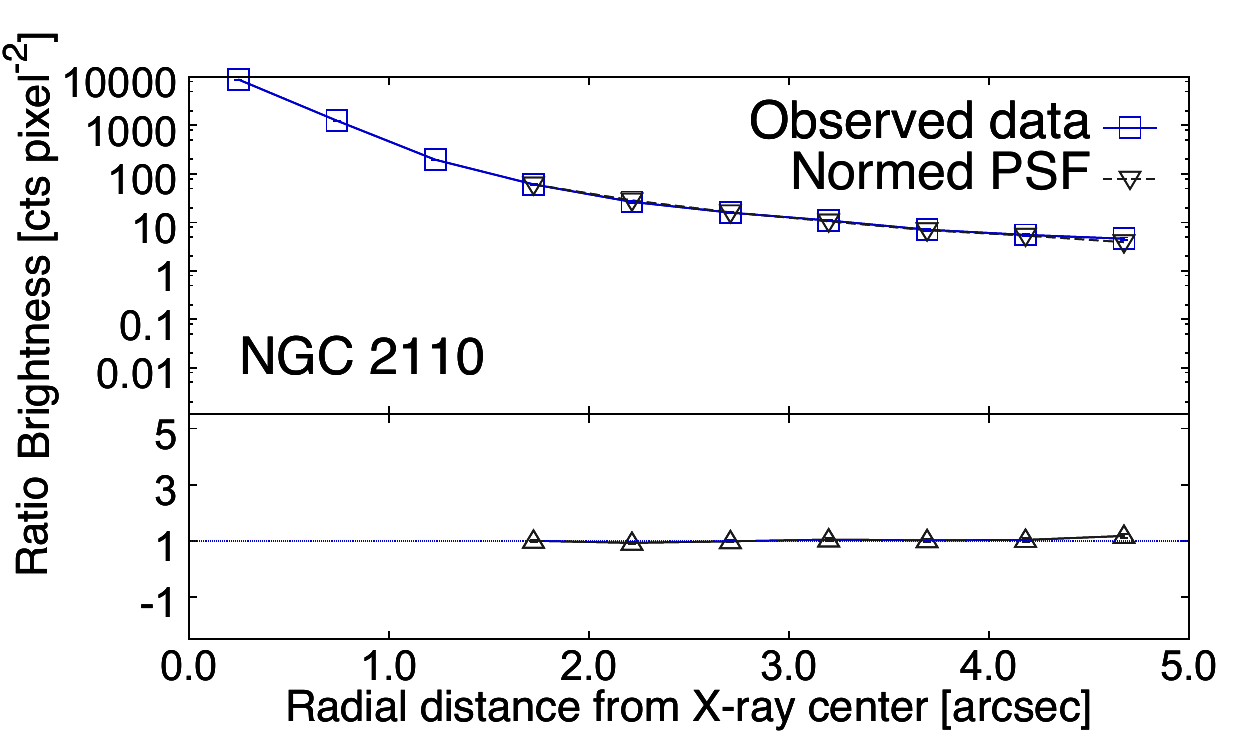}\hspace{-.3cm}
    \includegraphics[width=6.0cm]{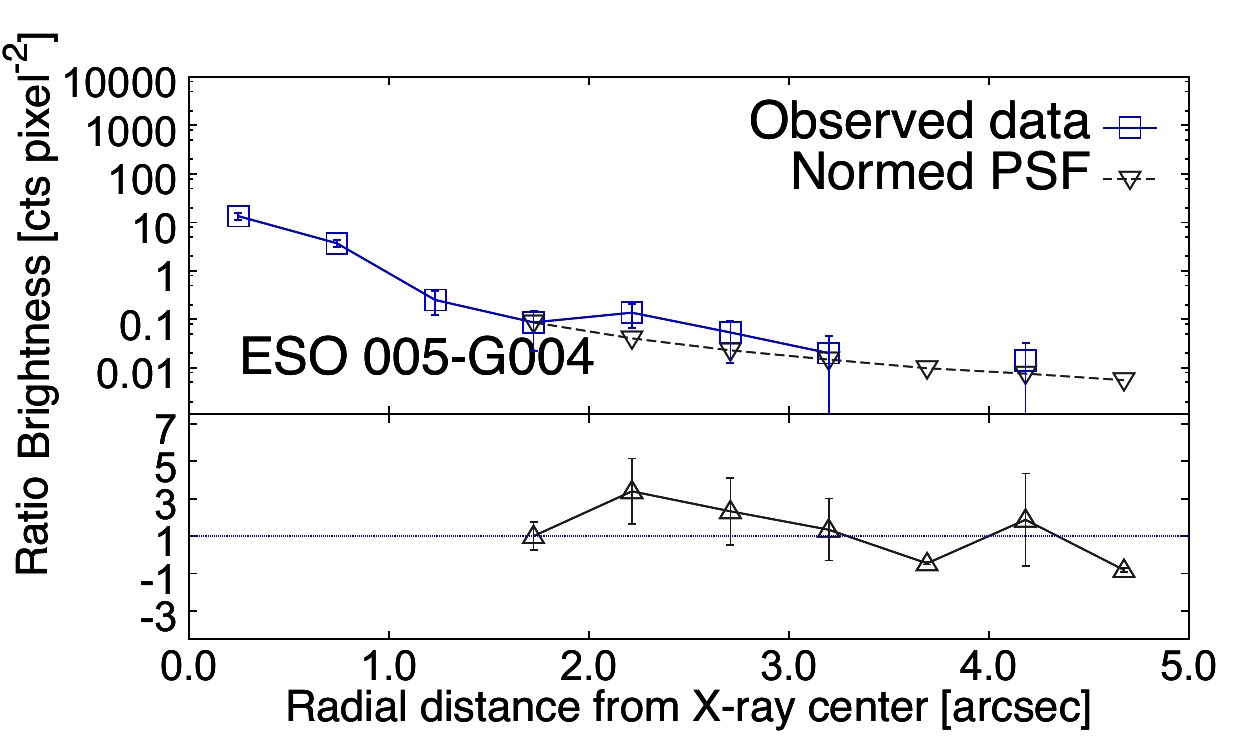}\hspace{-.3cm}
    \includegraphics[width=6.0cm]{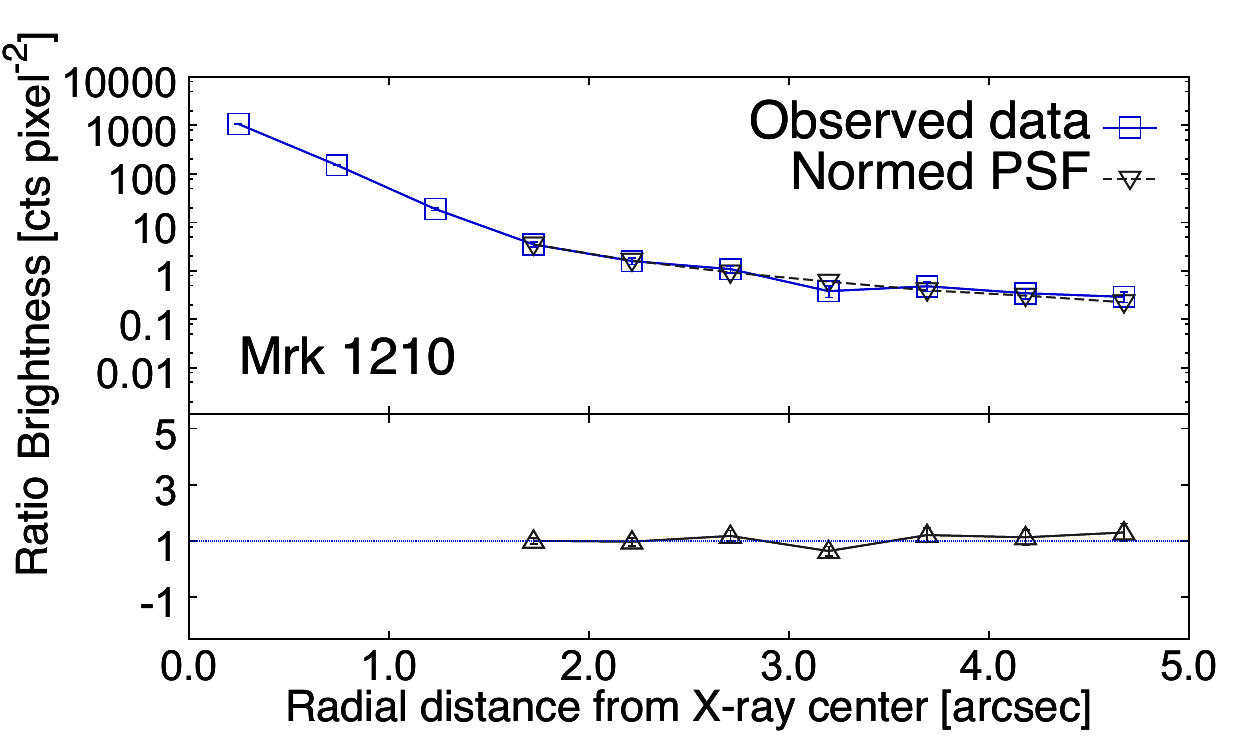}\\
    \includegraphics[width=6.0cm]{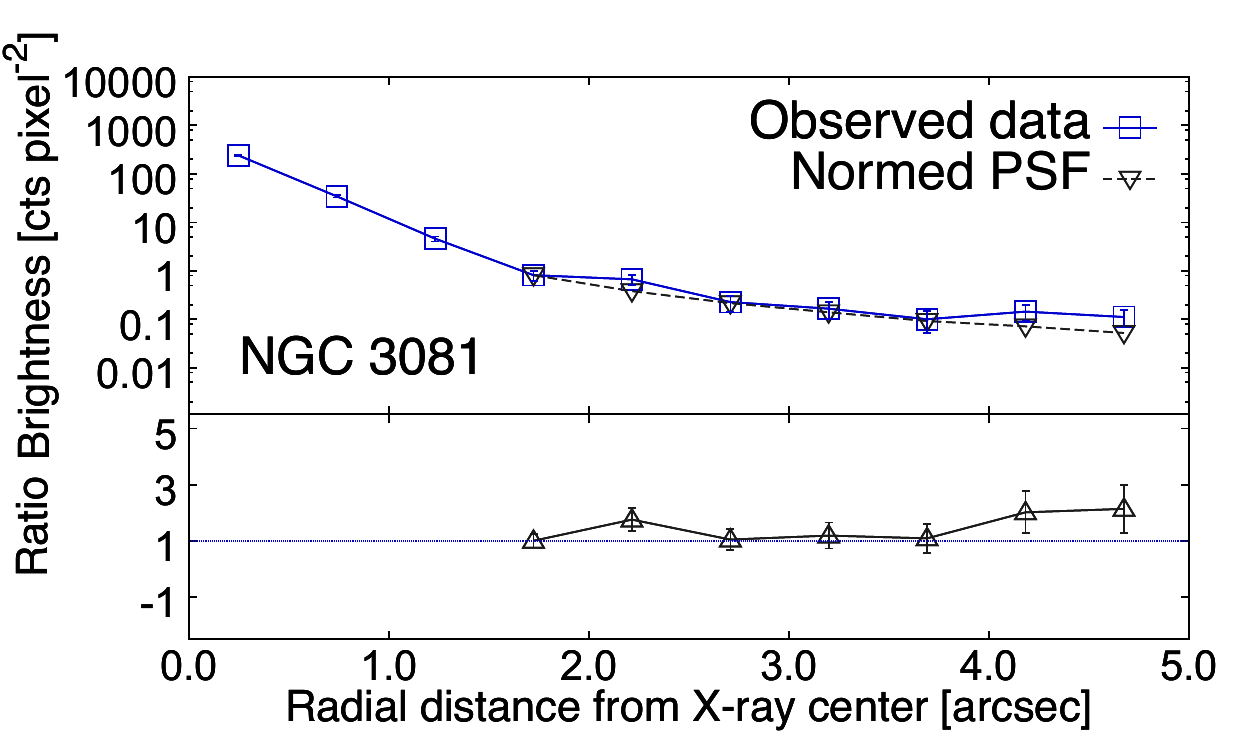}\hspace{-.3cm}
    \includegraphics[width=6.0cm]{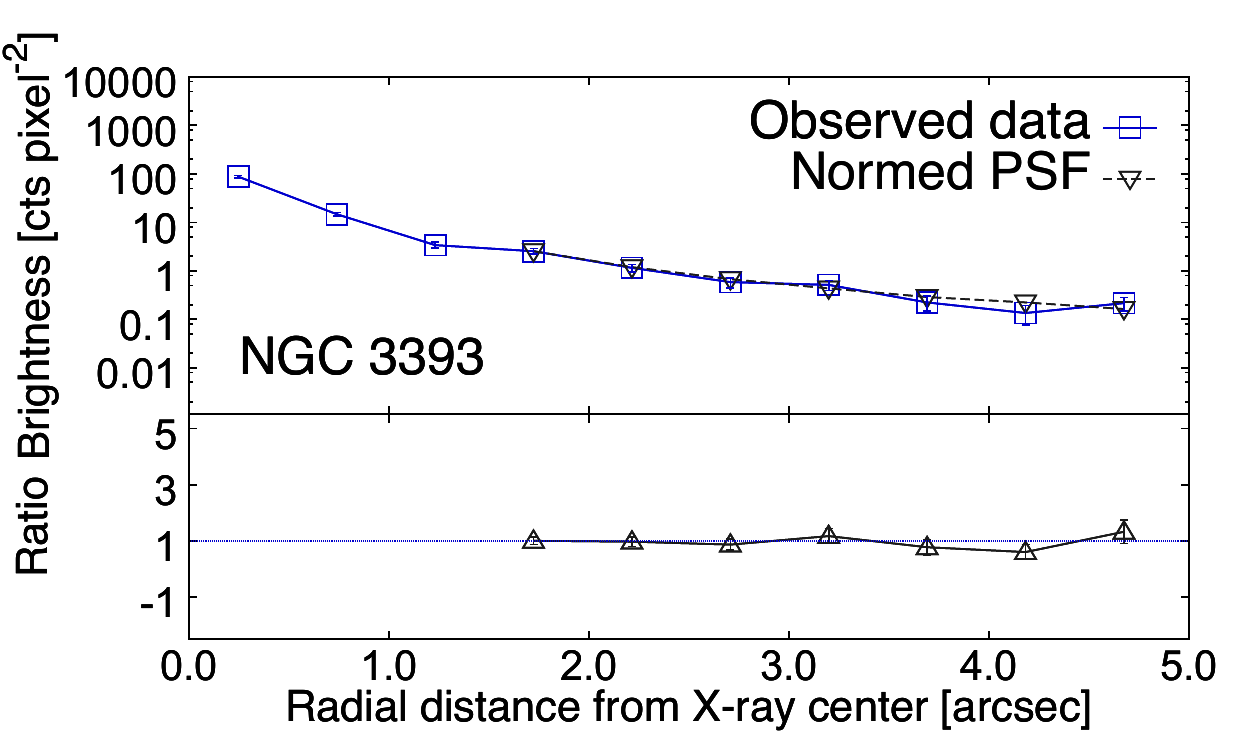}\hspace{-.3cm}
    \includegraphics[width=6.0cm]{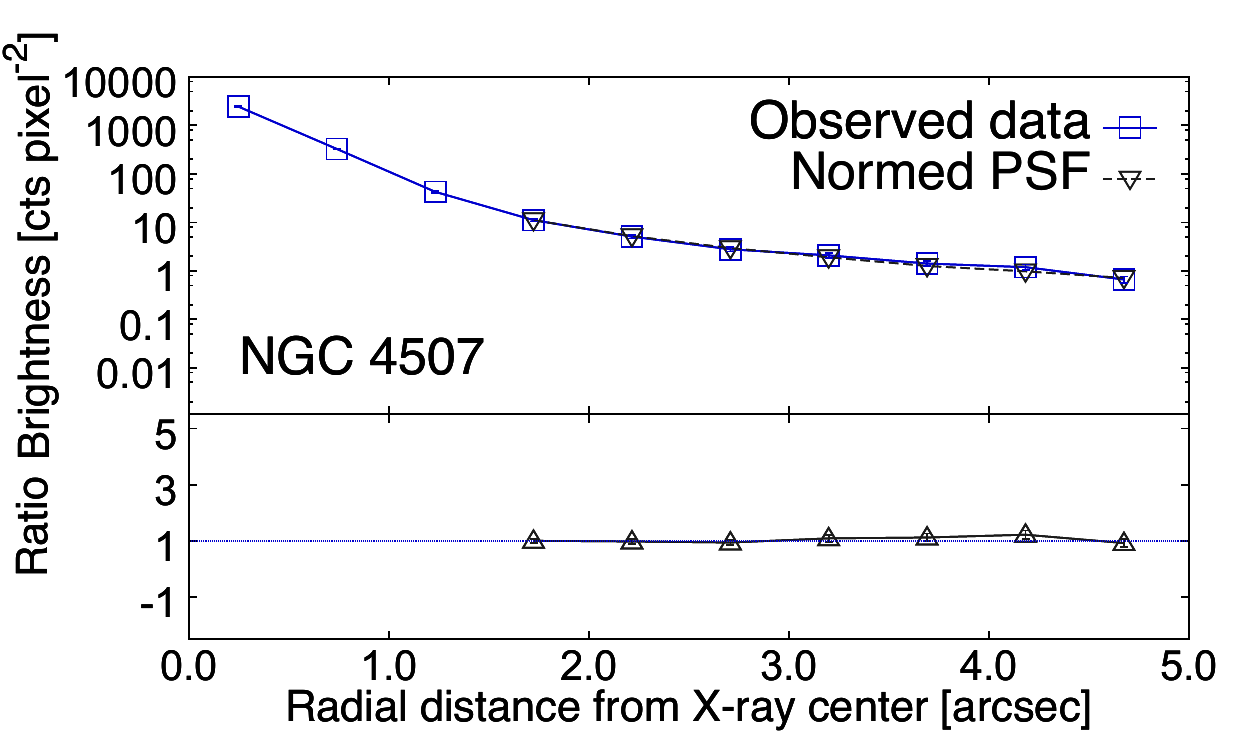}\\
    \includegraphics[width=6.0cm]{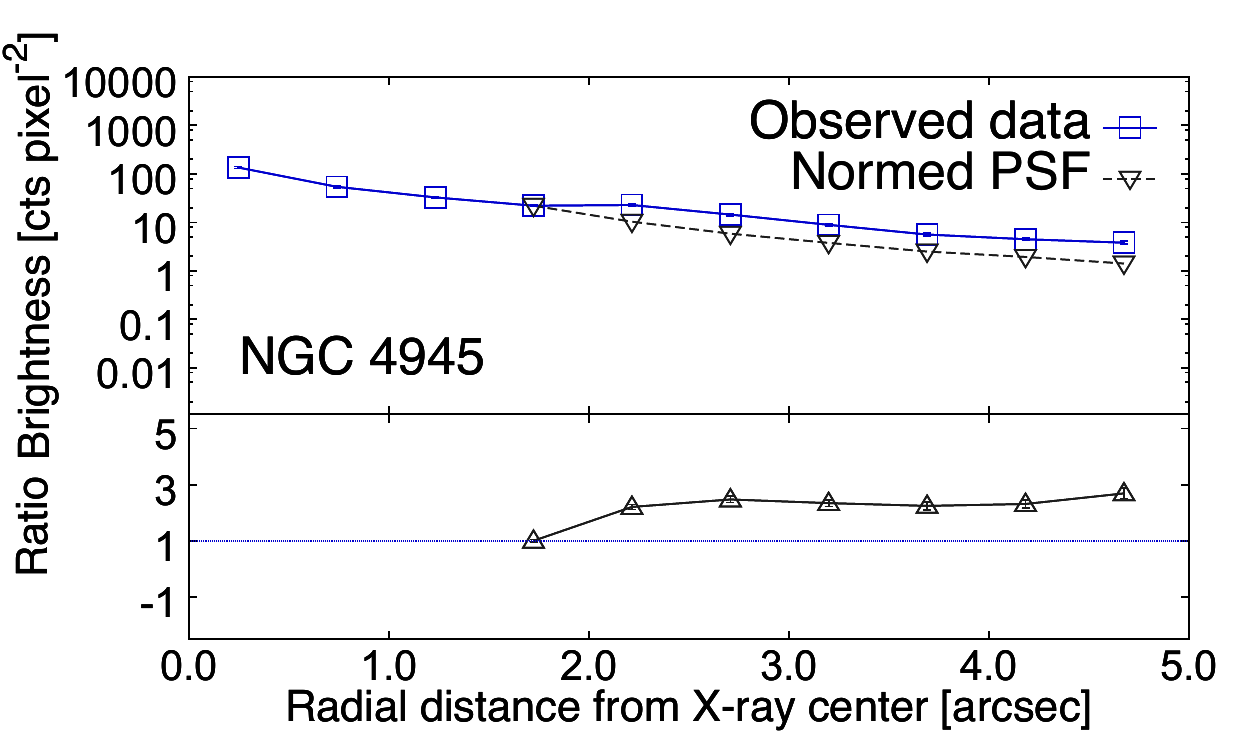}\hspace{-.3cm}
    \includegraphics[width=6.0cm]{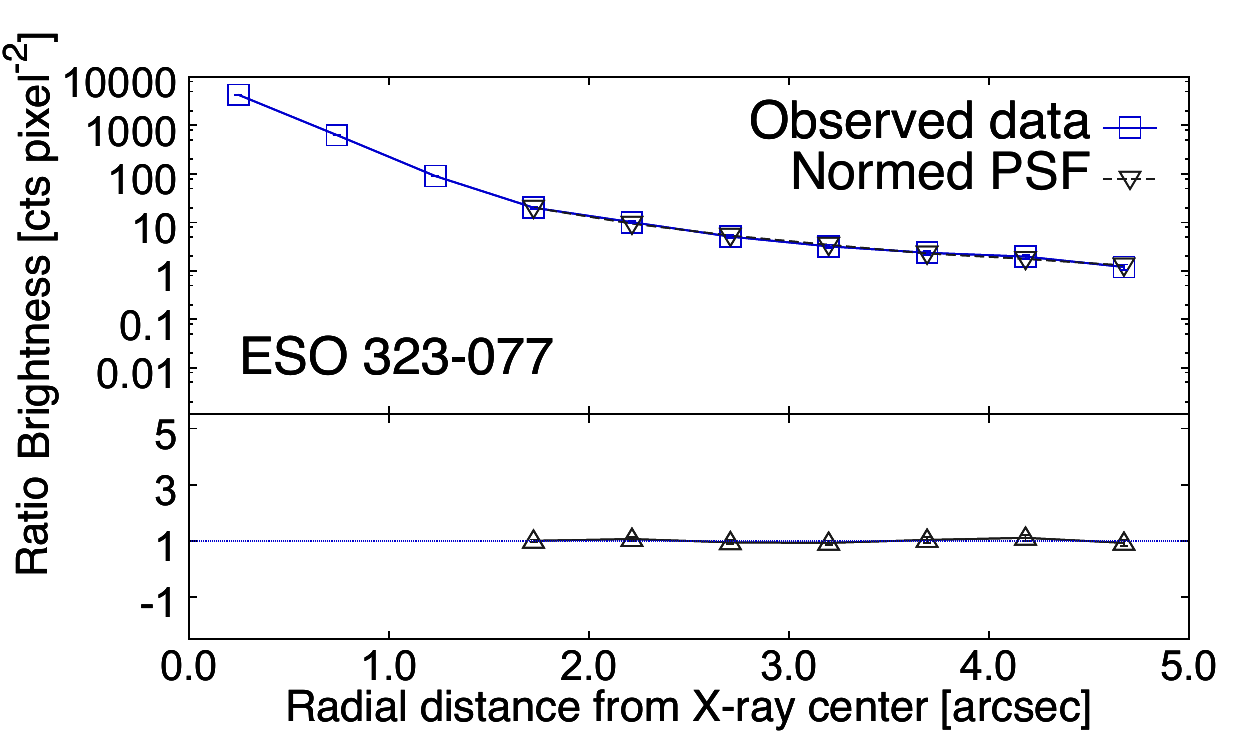}\hspace{-.3cm}
    \includegraphics[width=6.0cm]{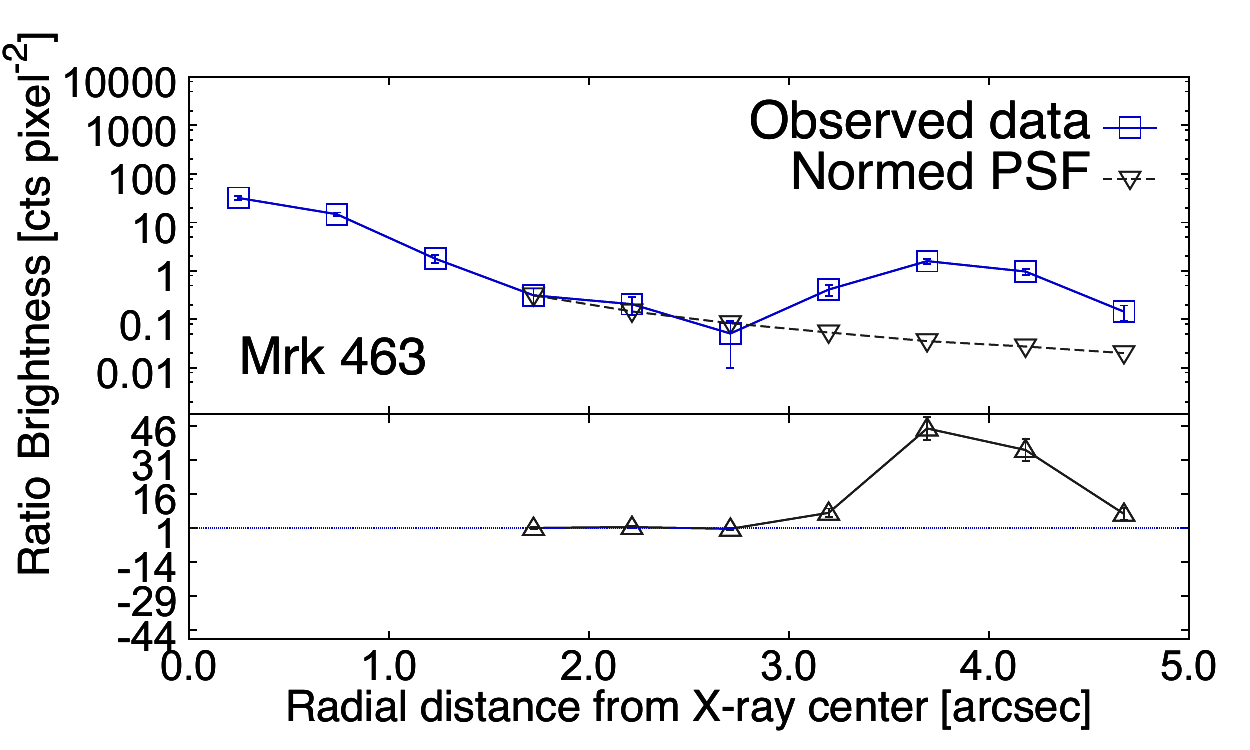}\\    
\caption{Comparisons of observed (blue) and modeled (black) radial profiles in the 3--6\,keV band. 
The lower windows show the ratios between the observed and modeled radial profiles.
    }
    \label{app:fig:rad_profile_3-6}
\end{figure*}    

\begin{figure*}\addtocounter{figure}{-1}
    \centering    
    \includegraphics[width=6.0cm]{3-6keV_19_NGC_5506_rad.pdf}\hspace{-.3cm}
    \includegraphics[width=6.0cm]{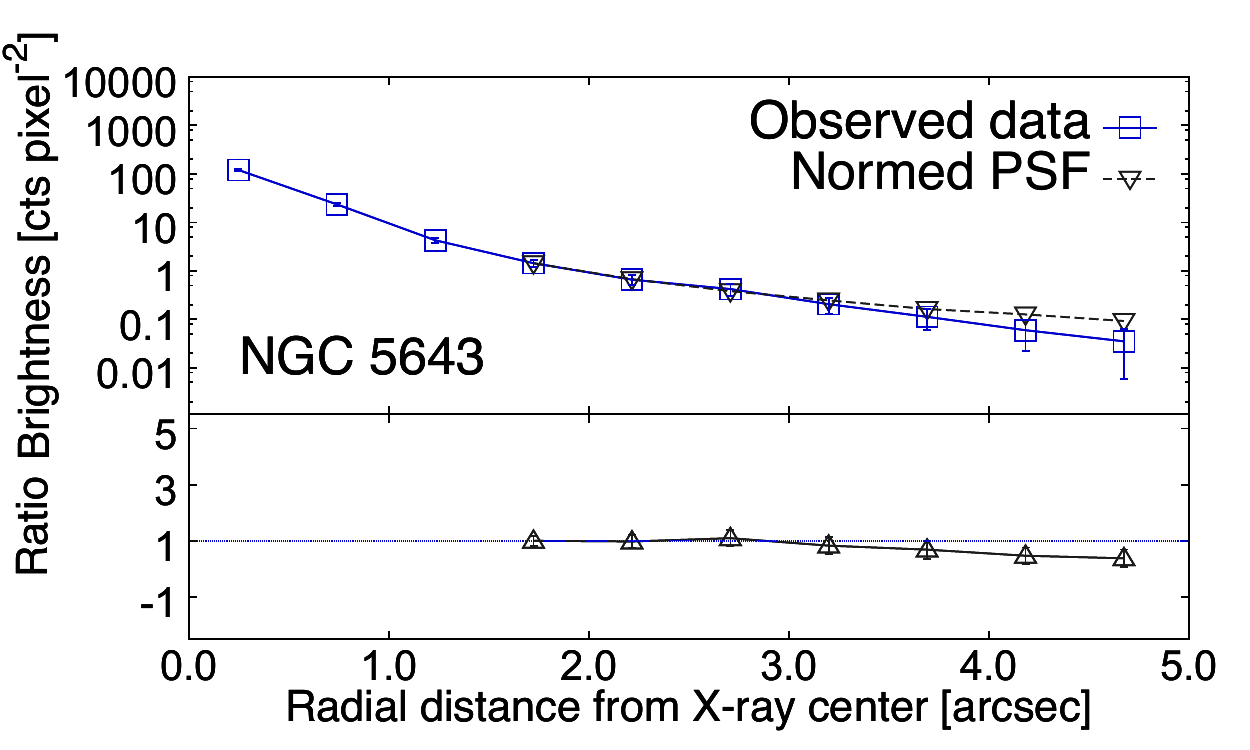}\hspace{-.3cm}
    \includegraphics[width=6.0cm]{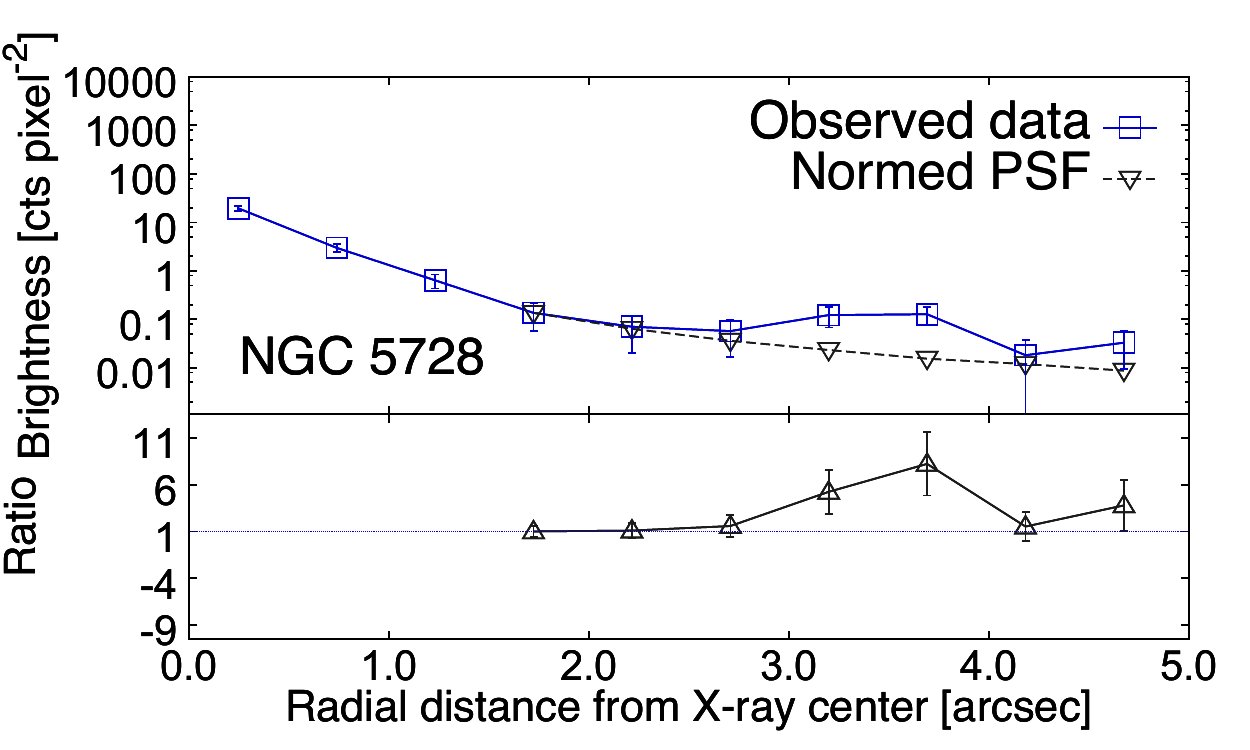}\\
    \includegraphics[width=6.0cm]{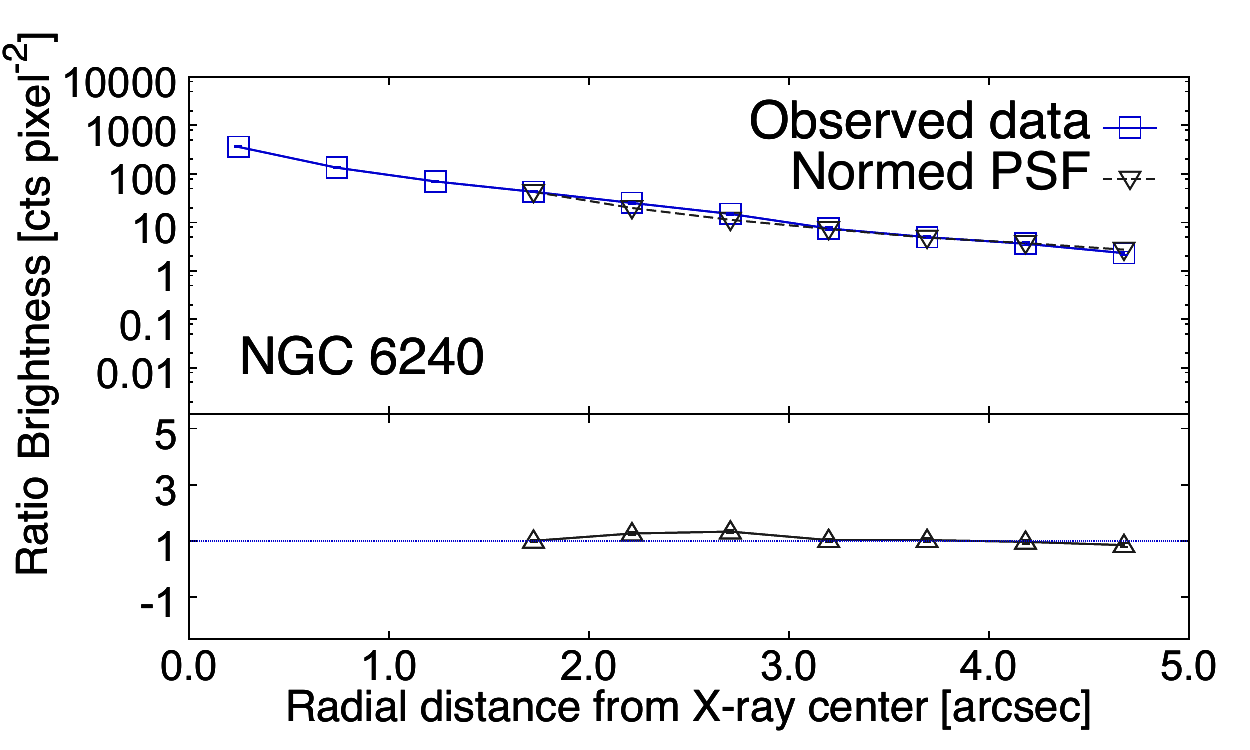}\hspace{-.3cm}
    \includegraphics[width=6.0cm]{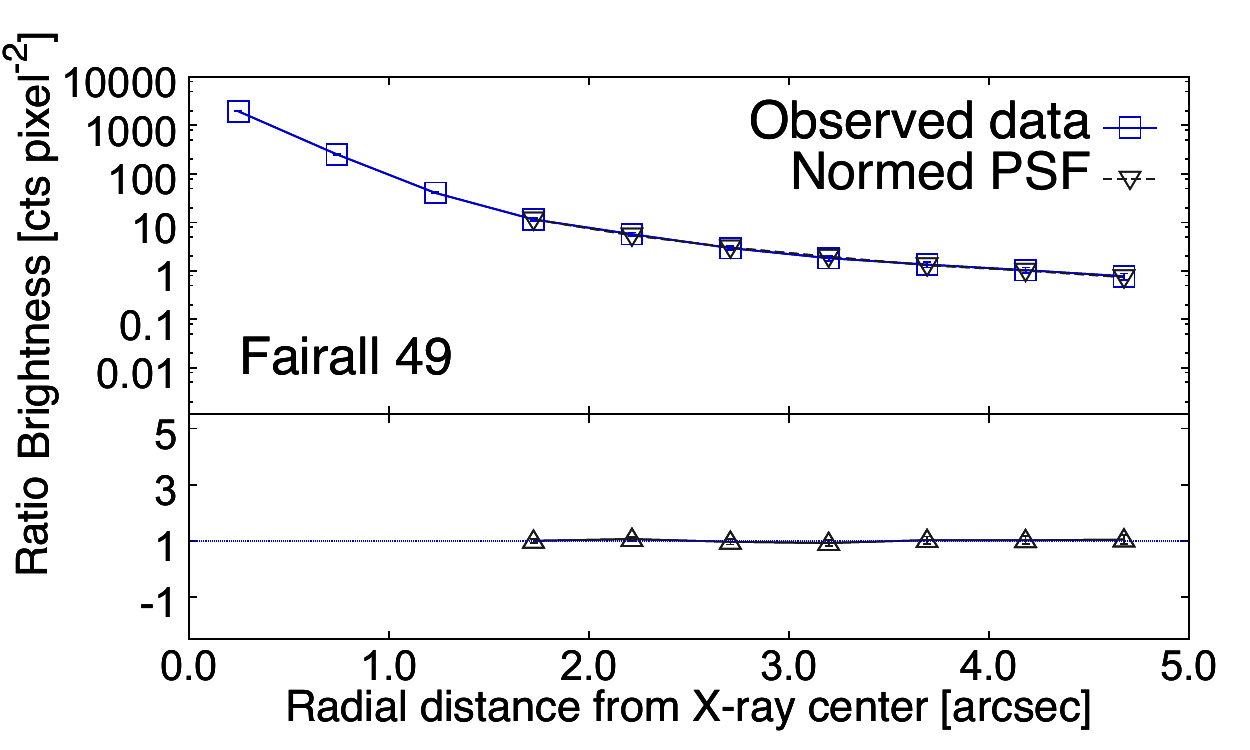}\hspace{-.3cm}
    \includegraphics[width=6.0cm]{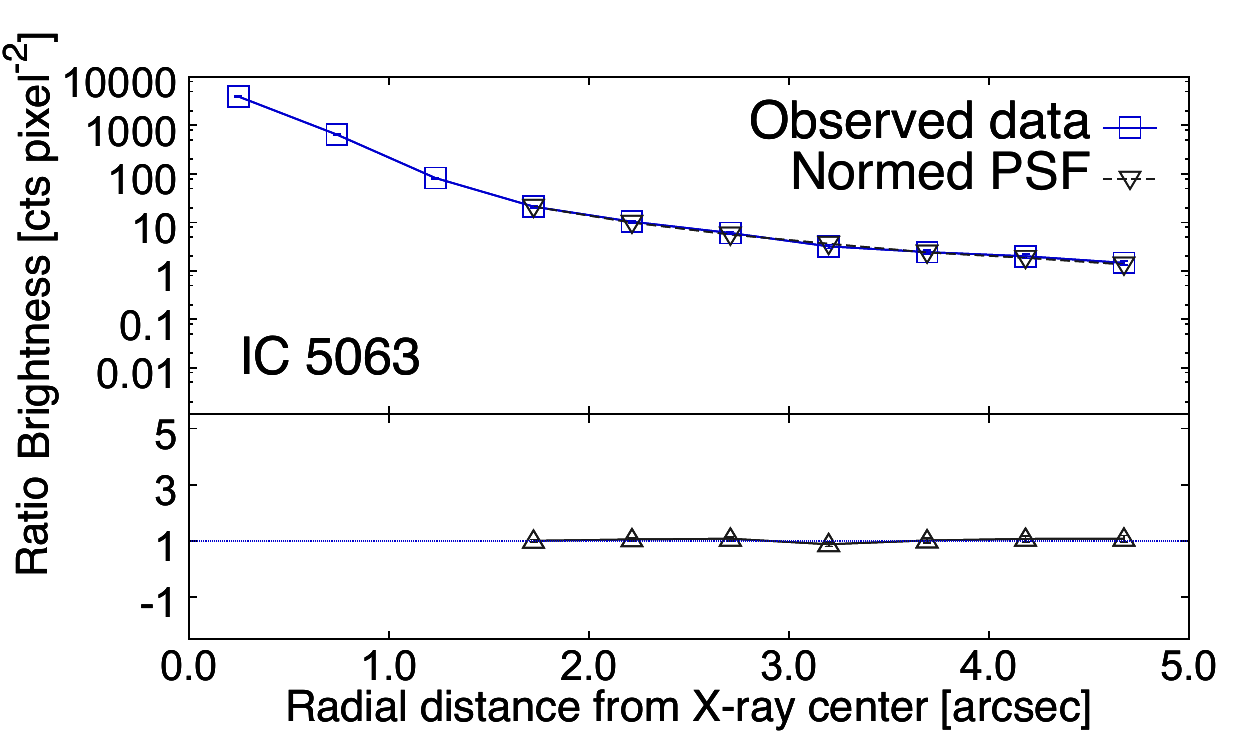}\\
    \includegraphics[width=6.0cm]{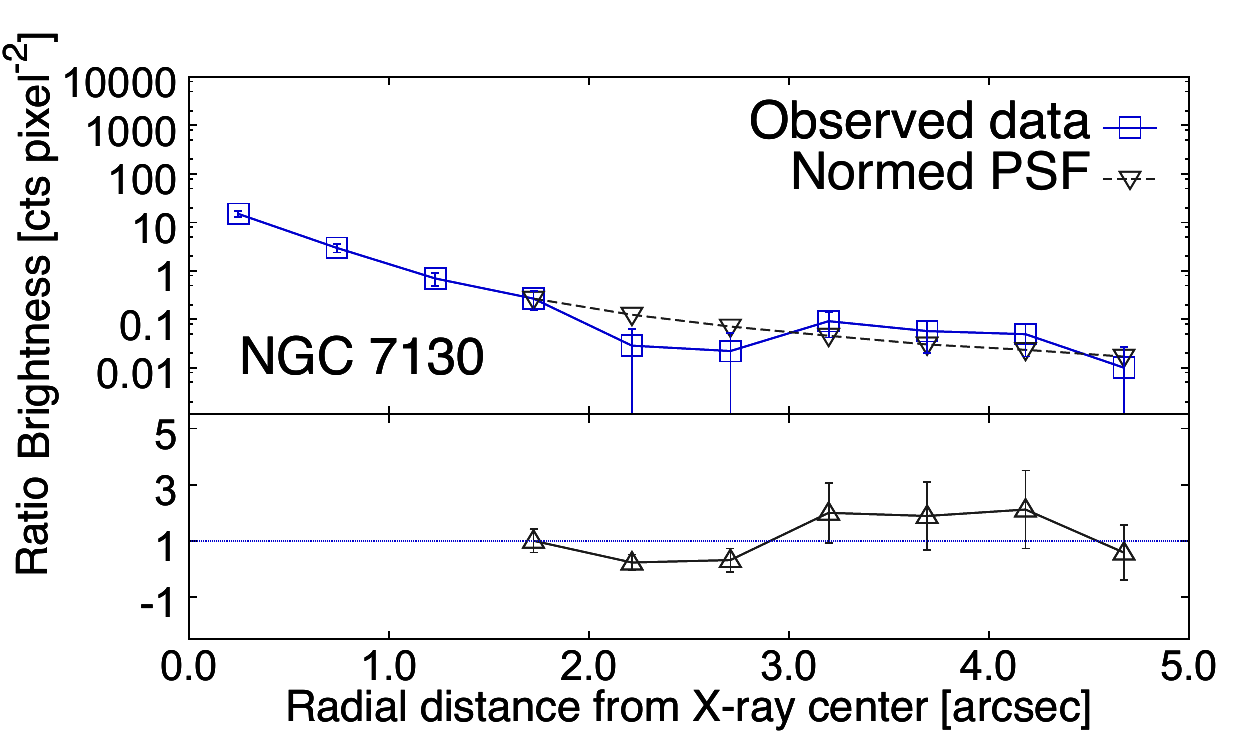}\hspace{-.3cm}    
    \includegraphics[width=6.0cm]{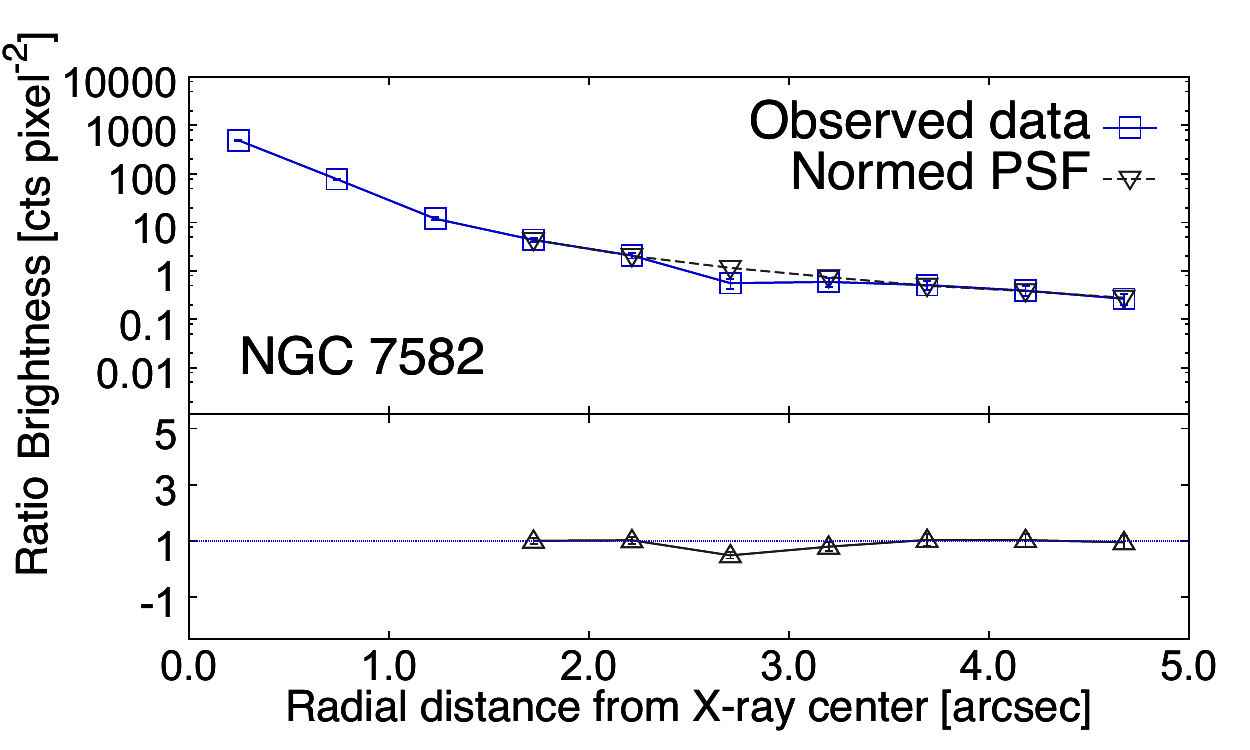}\hspace{-.3cm}   
    \caption{Continued.
    }
    \label{app:fig:rad_profile_3-6}
\end{figure*}

\begin{figure*}[!h]
    \centering
    \includegraphics[width=6.0cm]{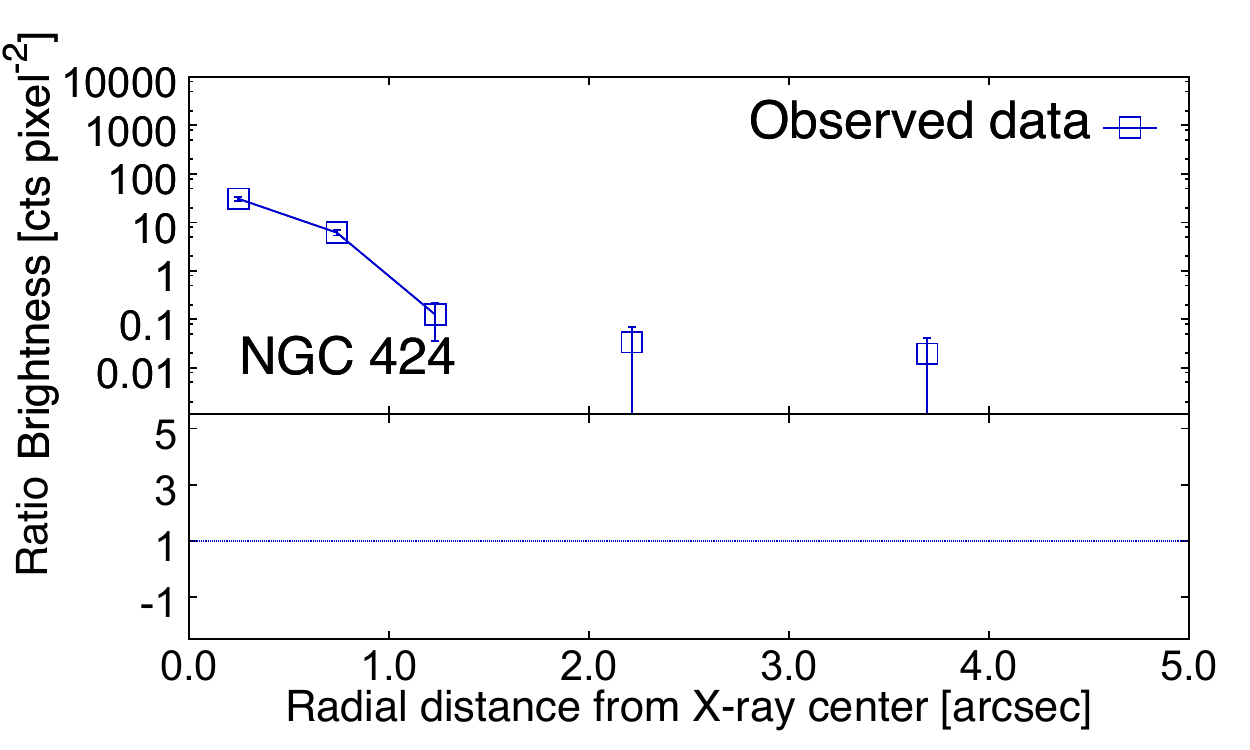}\hspace{-.3cm}
    \includegraphics[width=6.0cm]{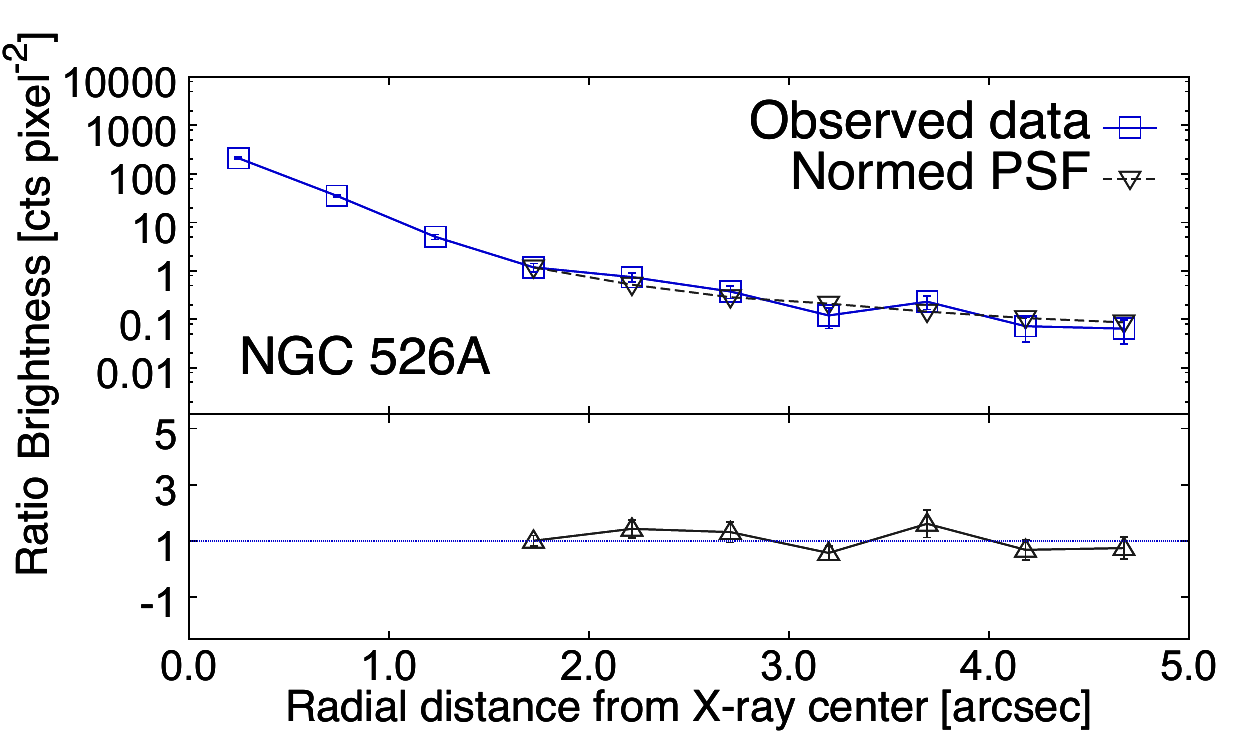}\hspace{-.3cm}
    \includegraphics[width=6.0cm]{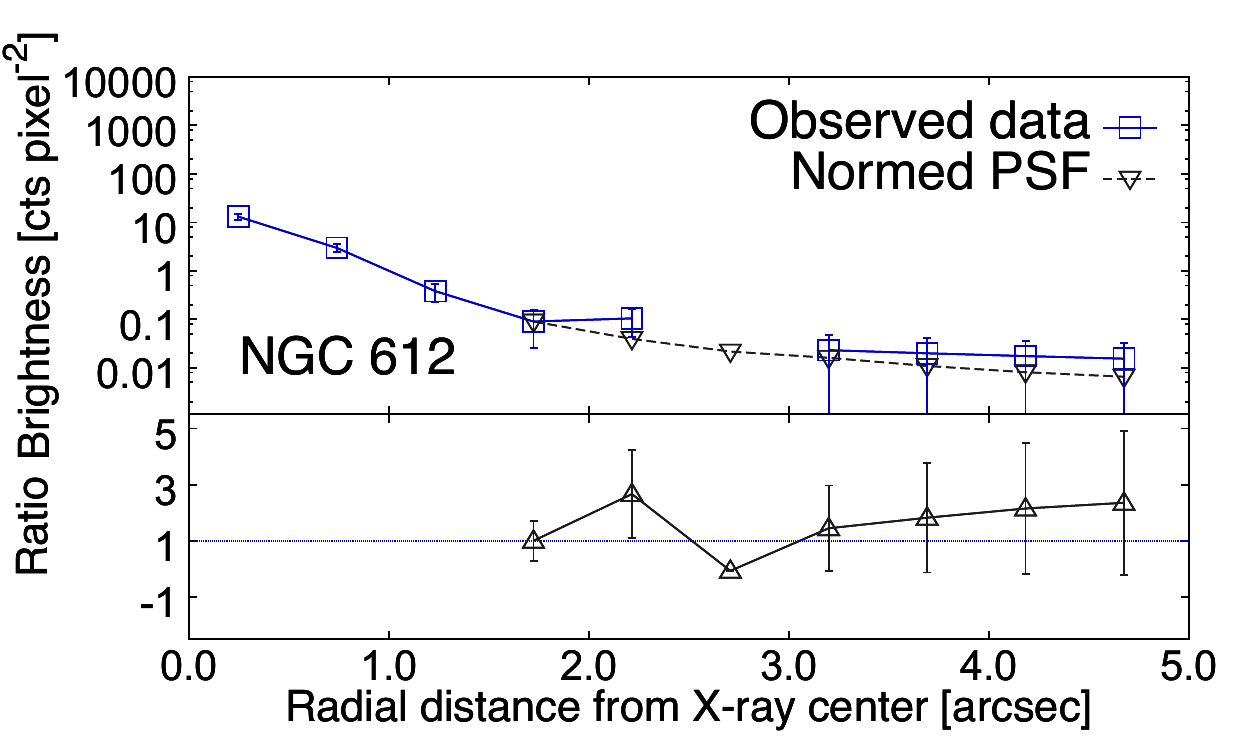}\\
    \includegraphics[width=6.0cm]{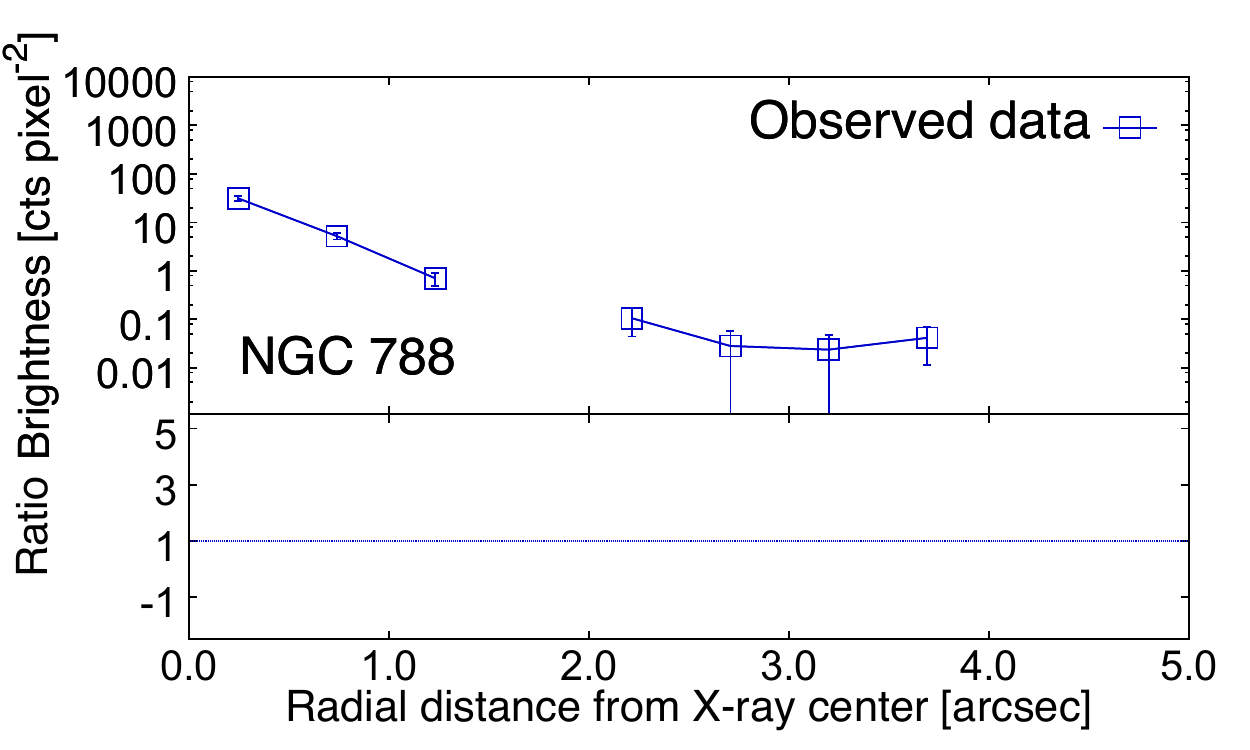}\hspace{-.3cm}
    \includegraphics[width=6.0cm]{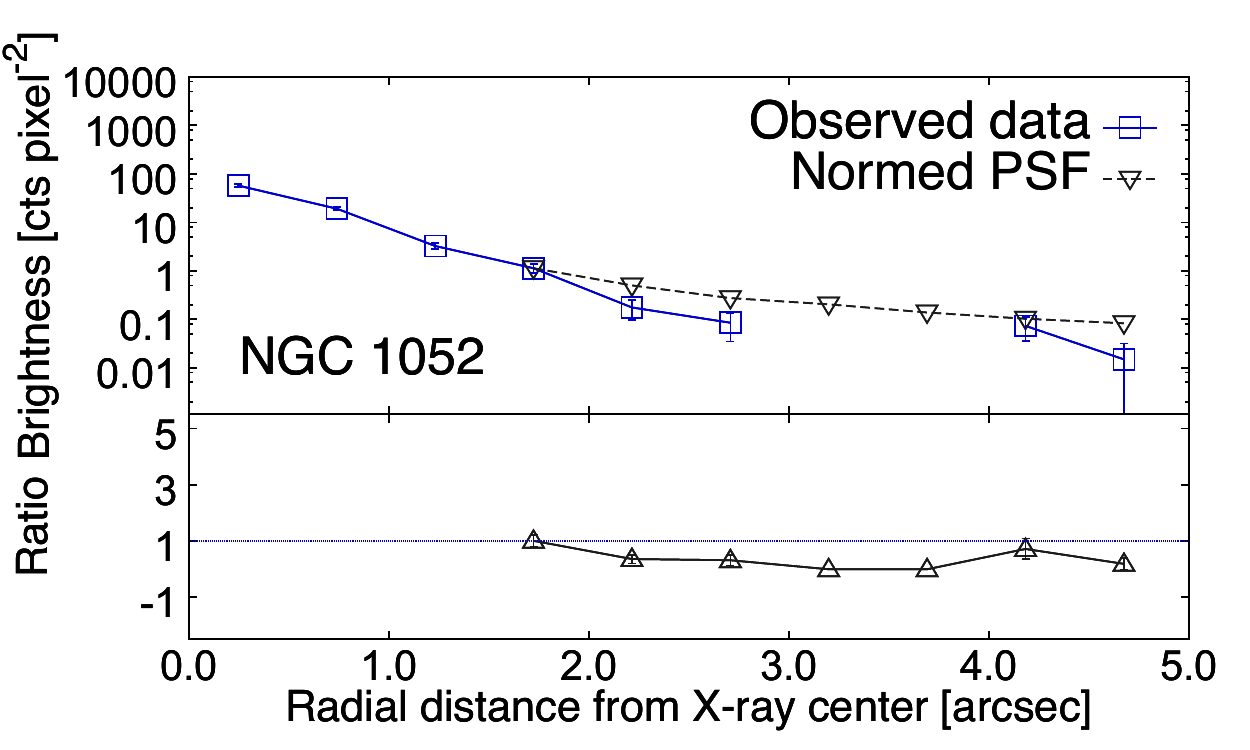}\hspace{-.3cm}
    \includegraphics[width=6.0cm]{6-7keV_06_NGC_1068_rad.pdf}\\
    \includegraphics[width=6.0cm]{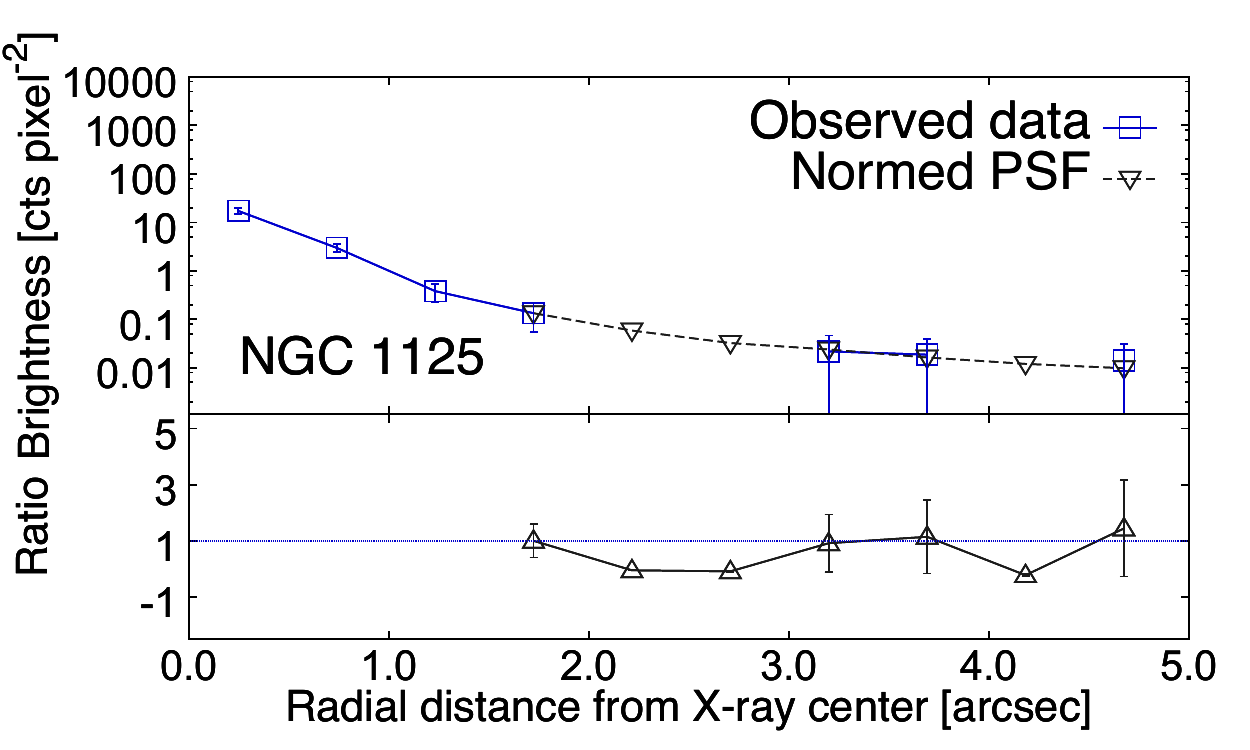}\hspace{-.3cm}
    \includegraphics[width=6.0cm]{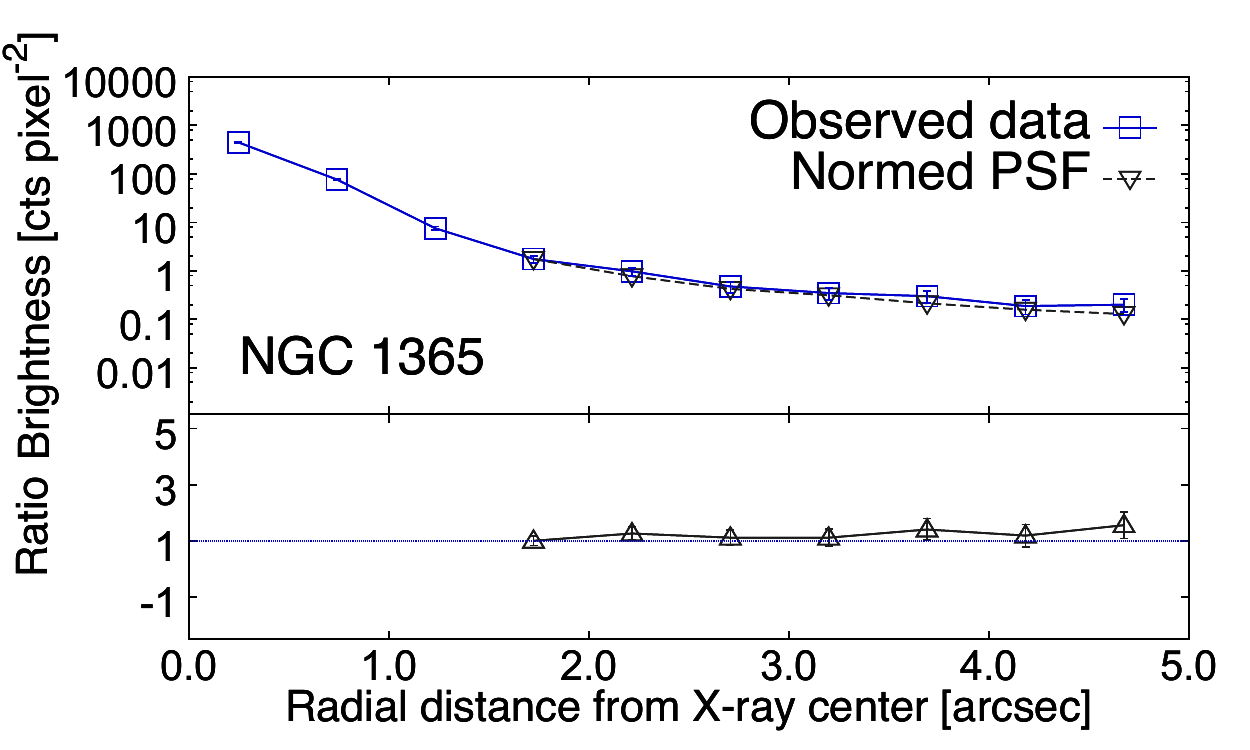}\hspace{-.3cm}
    \includegraphics[width=6.0cm]{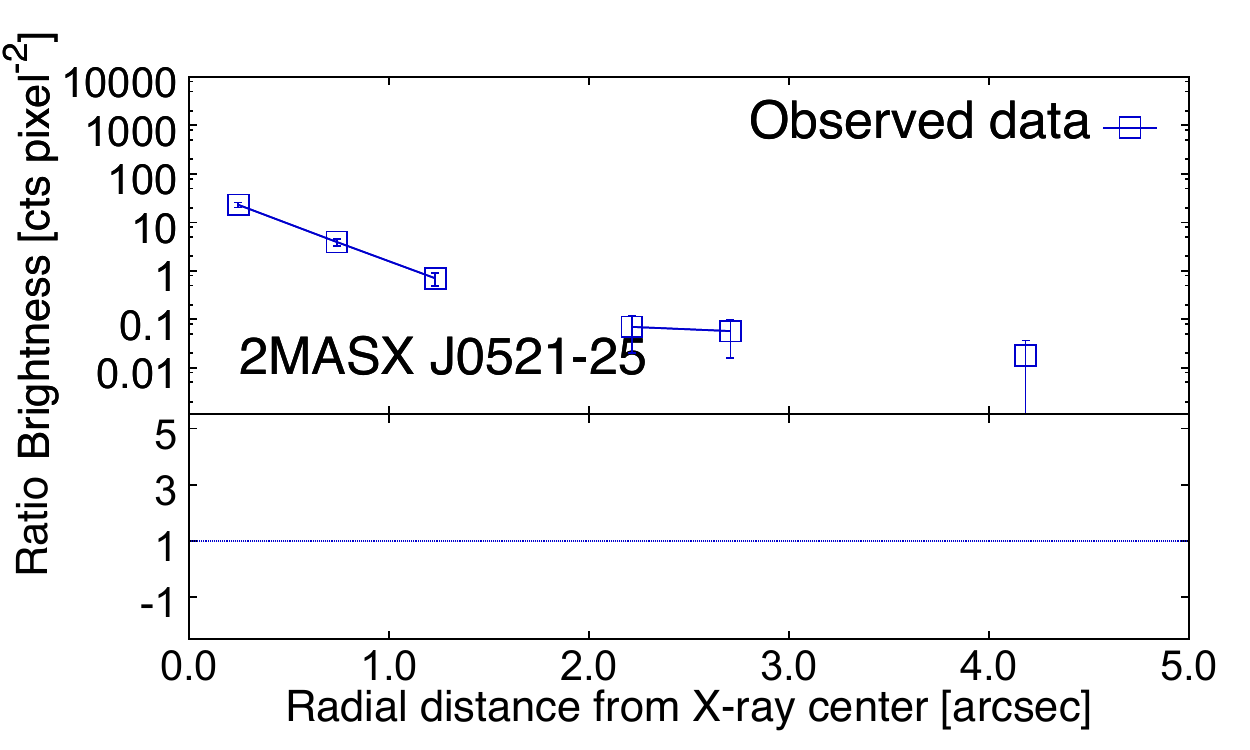}\\
    \includegraphics[width=6.0cm]{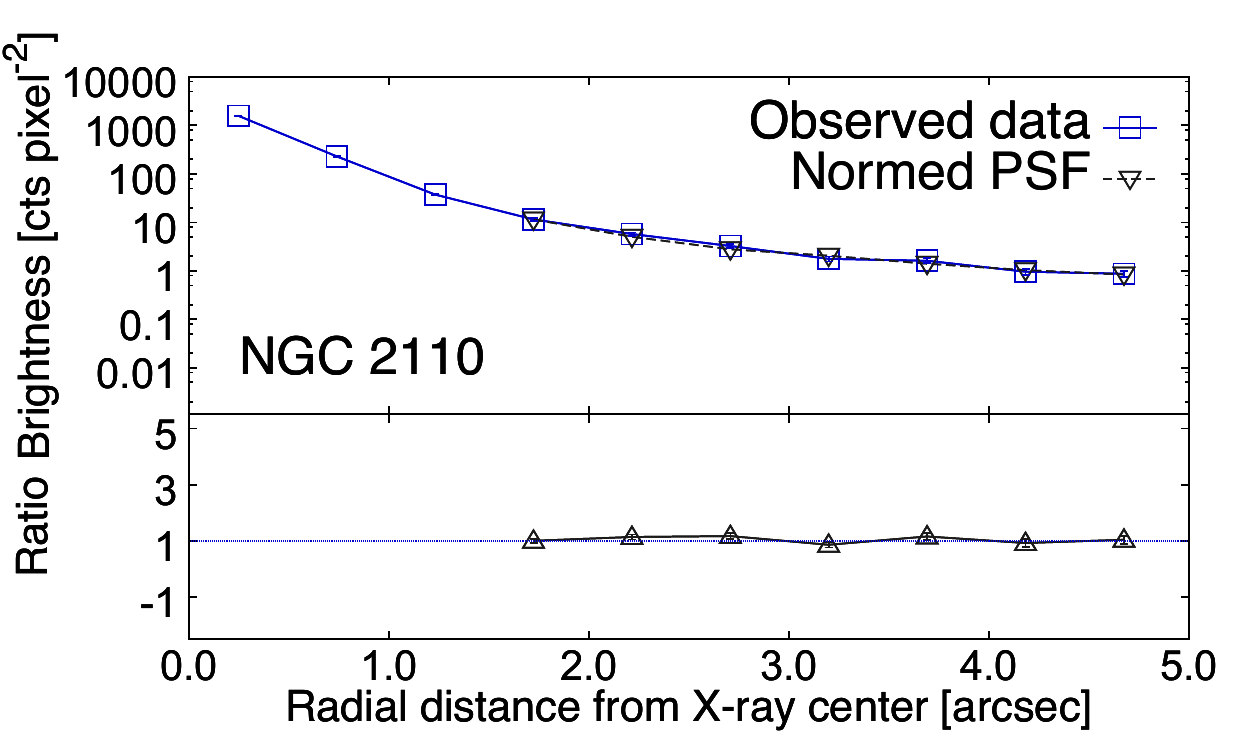}\hspace{-.3cm}
    \includegraphics[width=6.0cm]{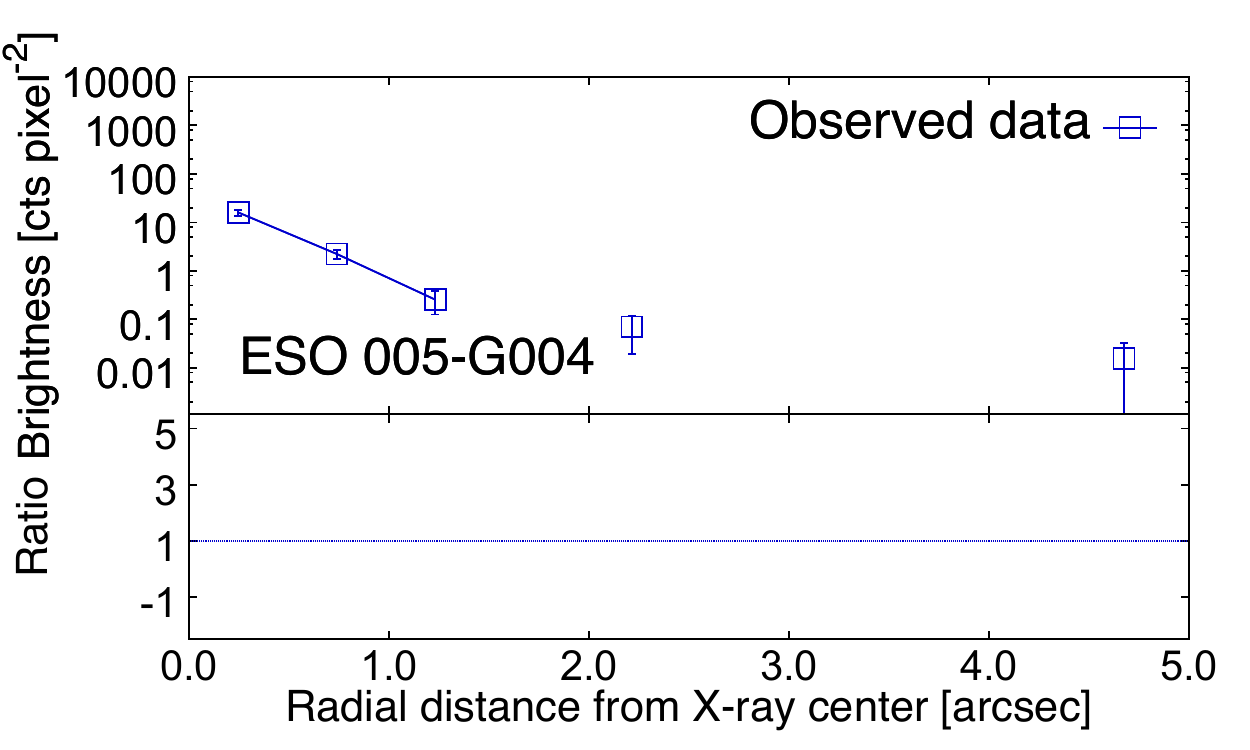}\hspace{-.3cm}
    \includegraphics[width=6.0cm]{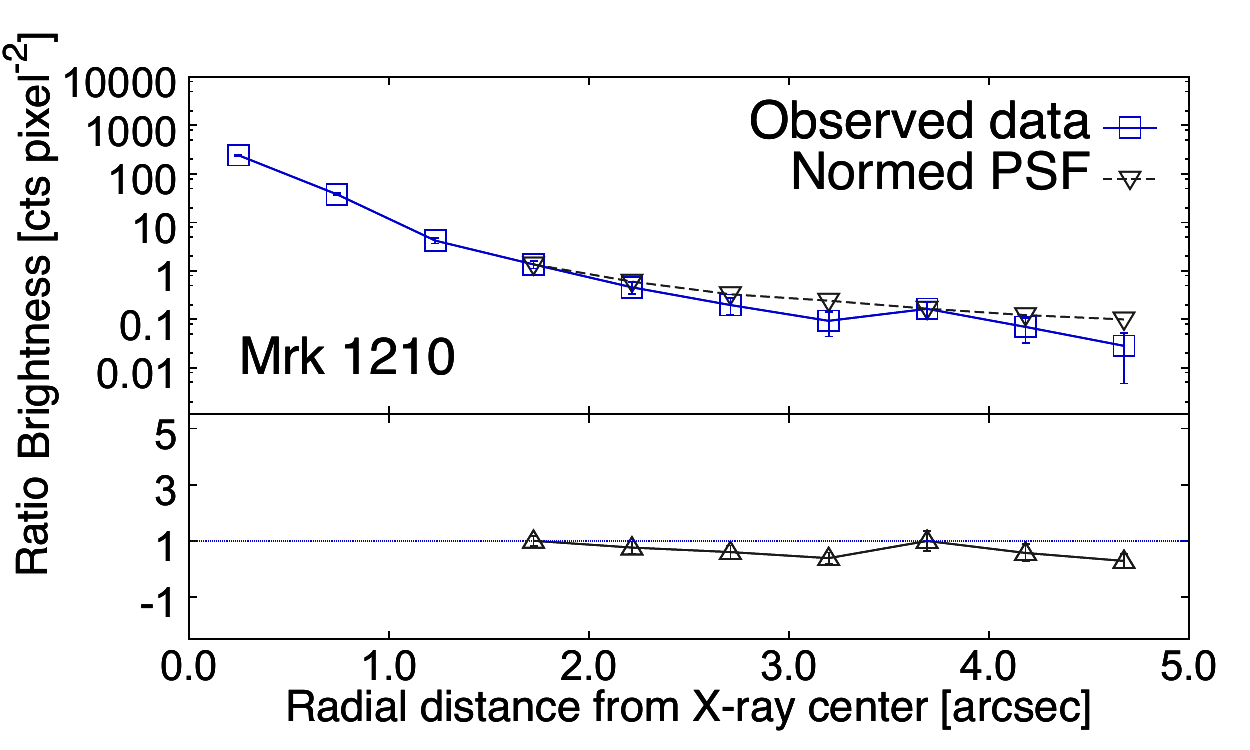}\\
    \includegraphics[width=6.0cm]{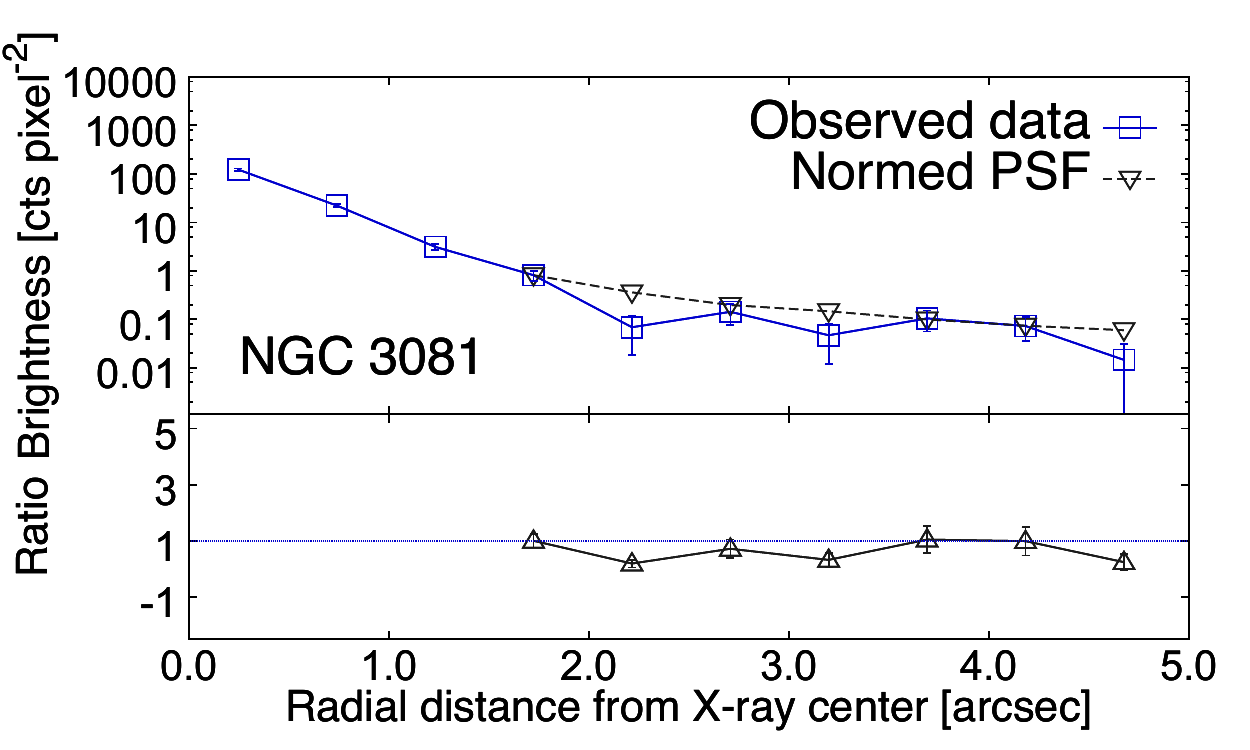}\hspace{-.3cm}
    \includegraphics[width=6.0cm]{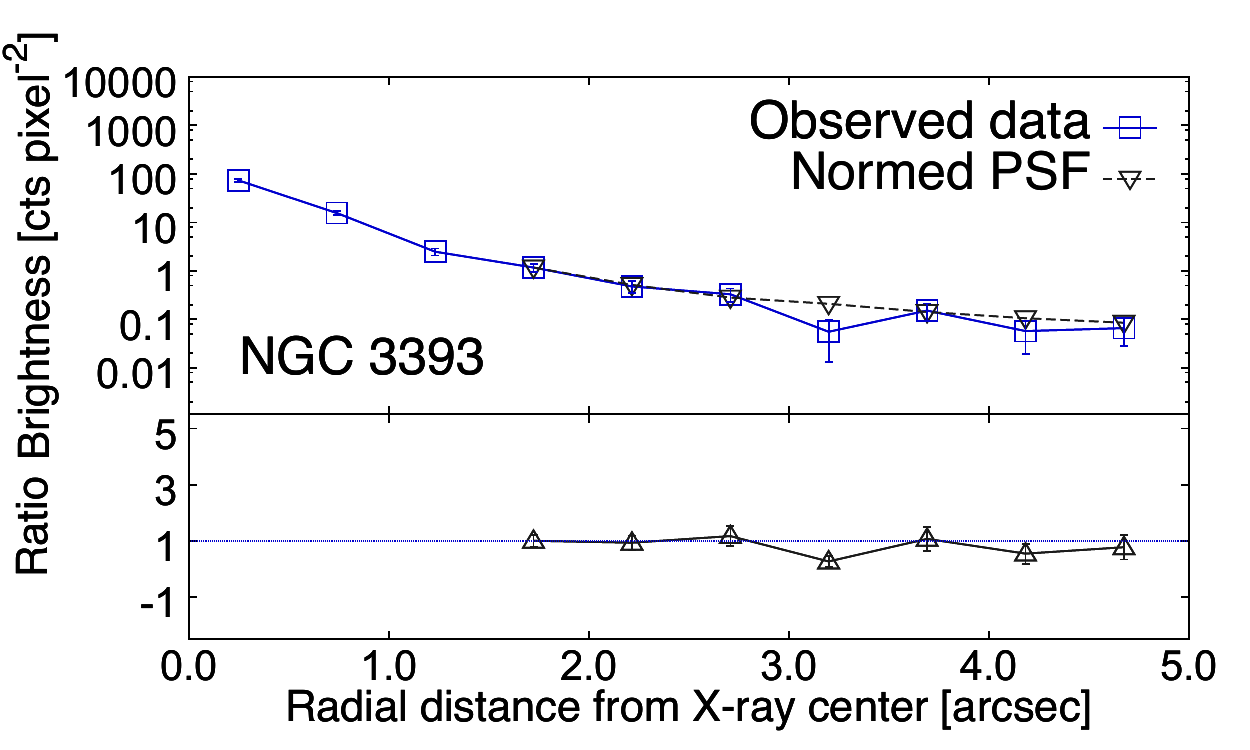}\hspace{-.3cm}
    \includegraphics[width=6.0cm]{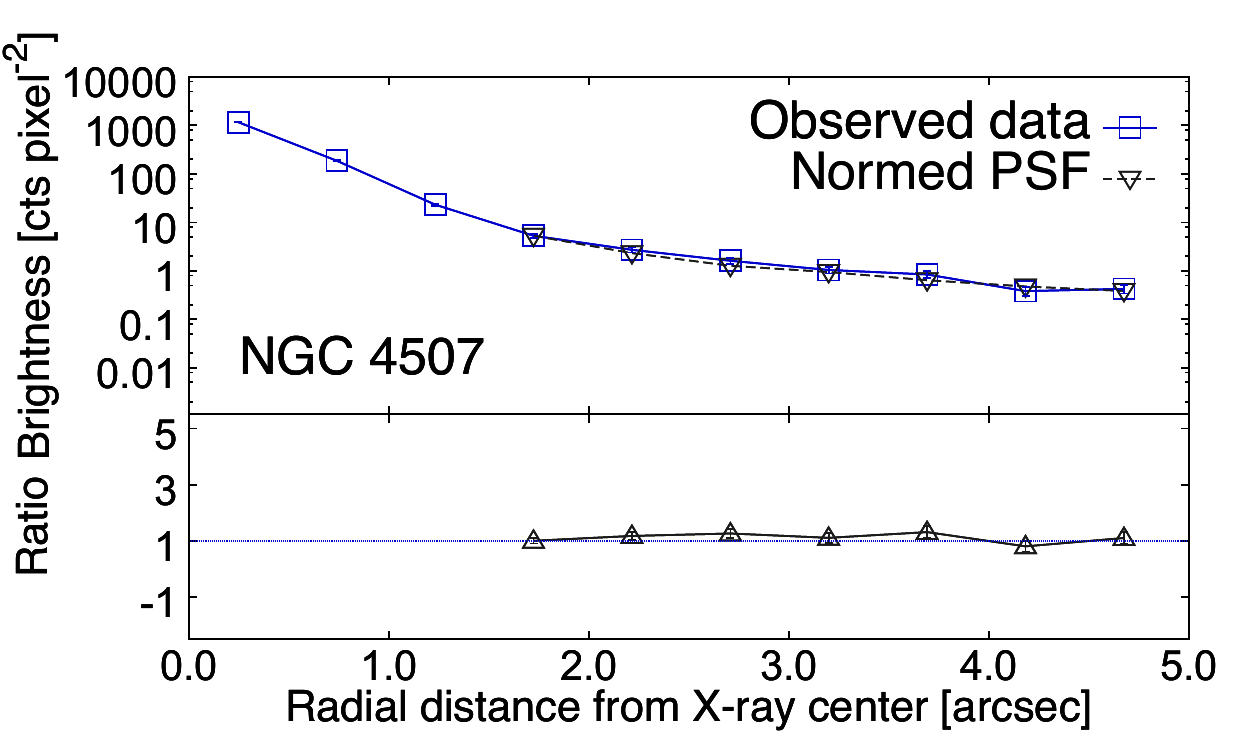}\\
    \includegraphics[width=6.0cm]{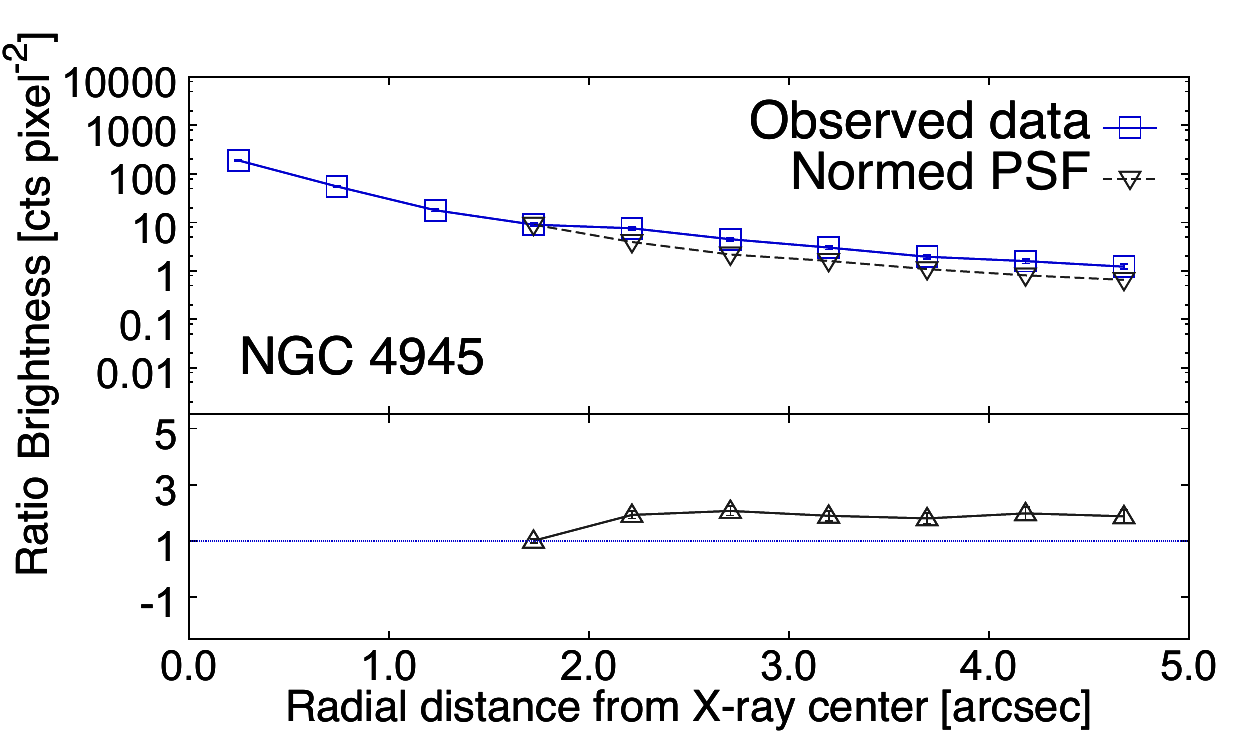}\hspace{-.3cm}
    \includegraphics[width=6.0cm]{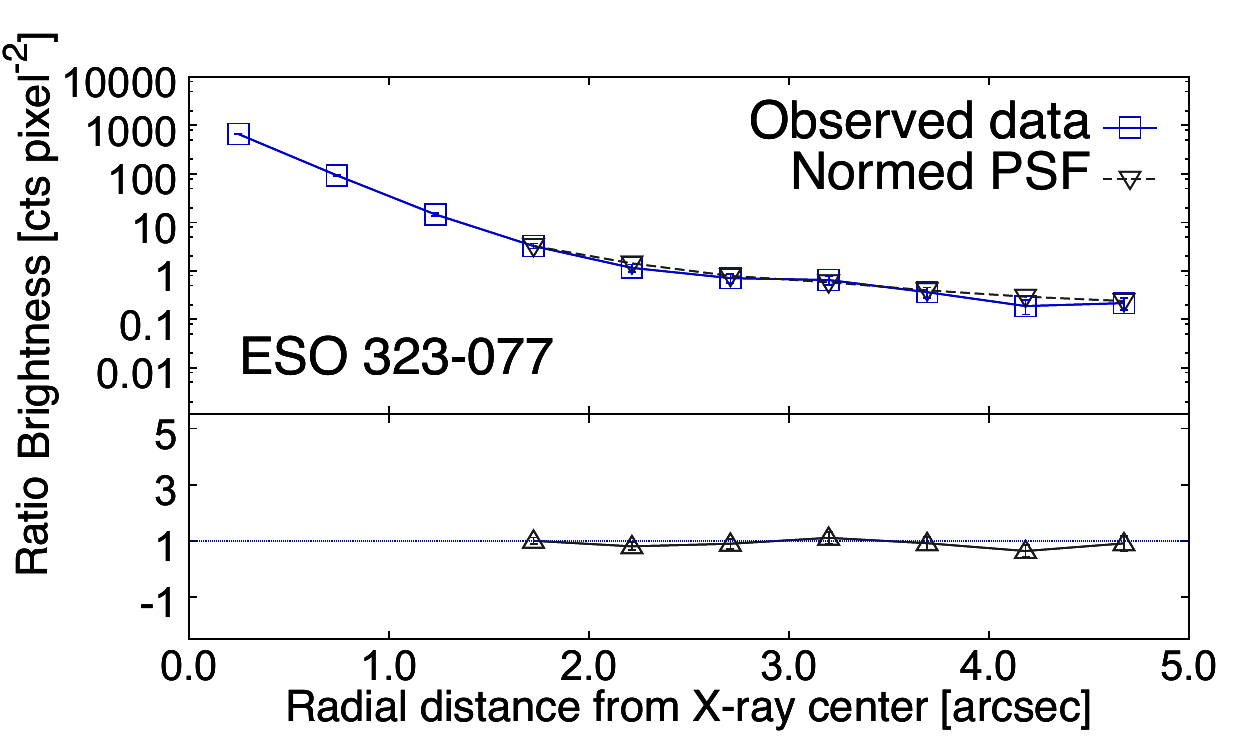}\hspace{-.3cm}
    \includegraphics[width=6.0cm]{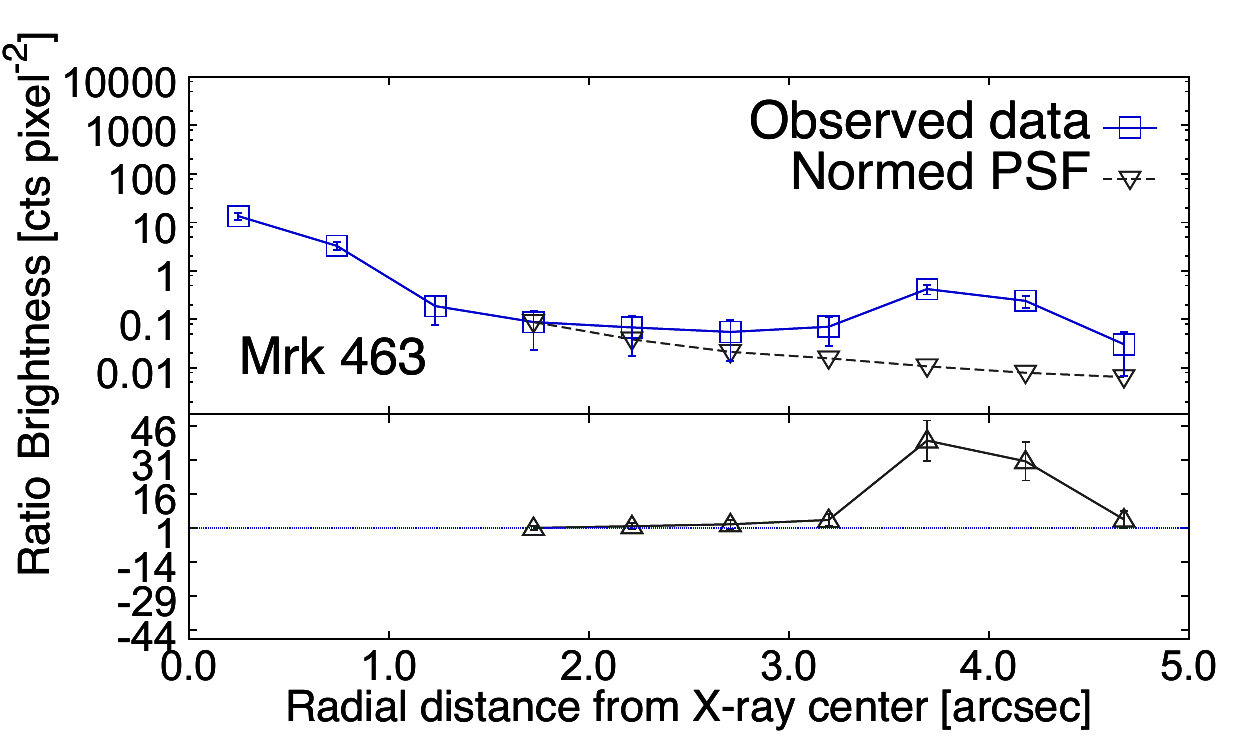}\\
\caption{Same as Figure~\ref{app:fig:rad_profile_3-6}, but for the 6--7\,keV band. 
    Only observed data are shown for the targets for which we cannot model a profile by normalizing the PSF model at 3.5 pixels, or the 4th pixel, from the center. }
    \label{app:fig:rad_profile_6-7}
\end{figure*}    

\begin{figure*}\addtocounter{figure}{-1}
    \centering        
    \includegraphics[width=6.0cm]{6-7keV_19_NGC_5506_rad.pdf}\hspace{-.3cm}
    \includegraphics[width=6.0cm]{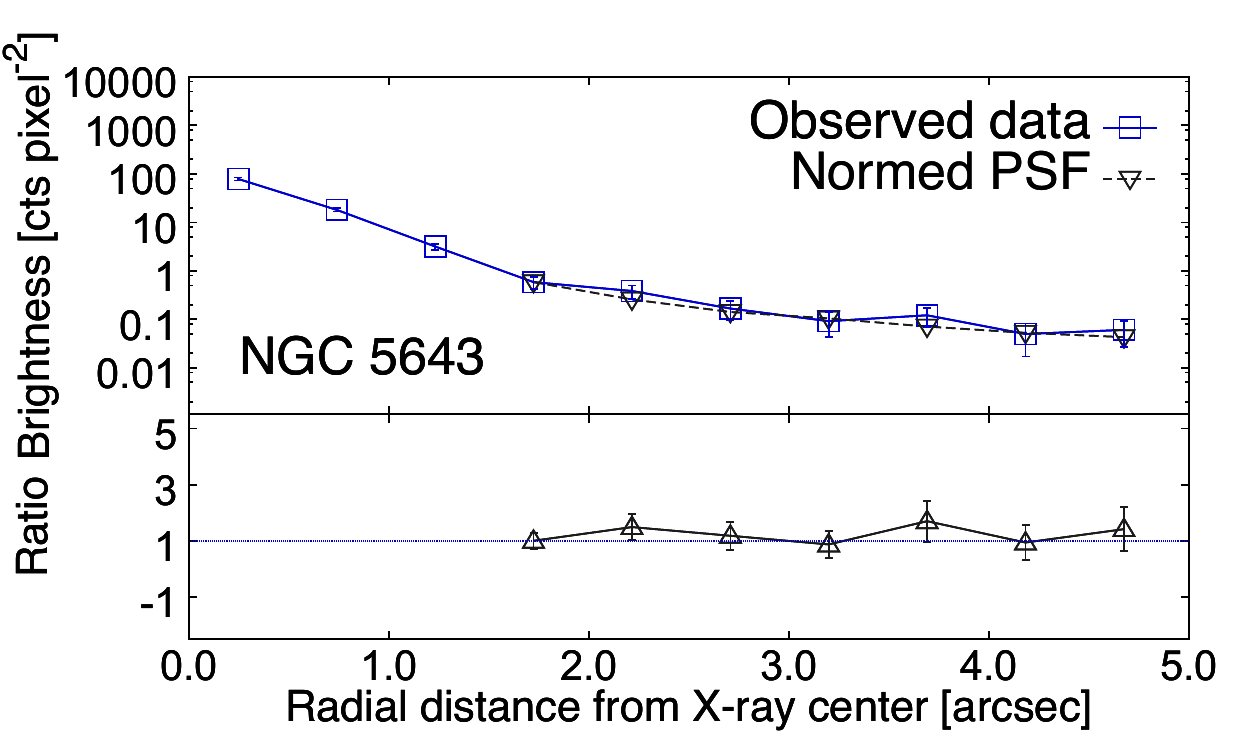}\hspace{-.3cm}
    \includegraphics[width=6.0cm]{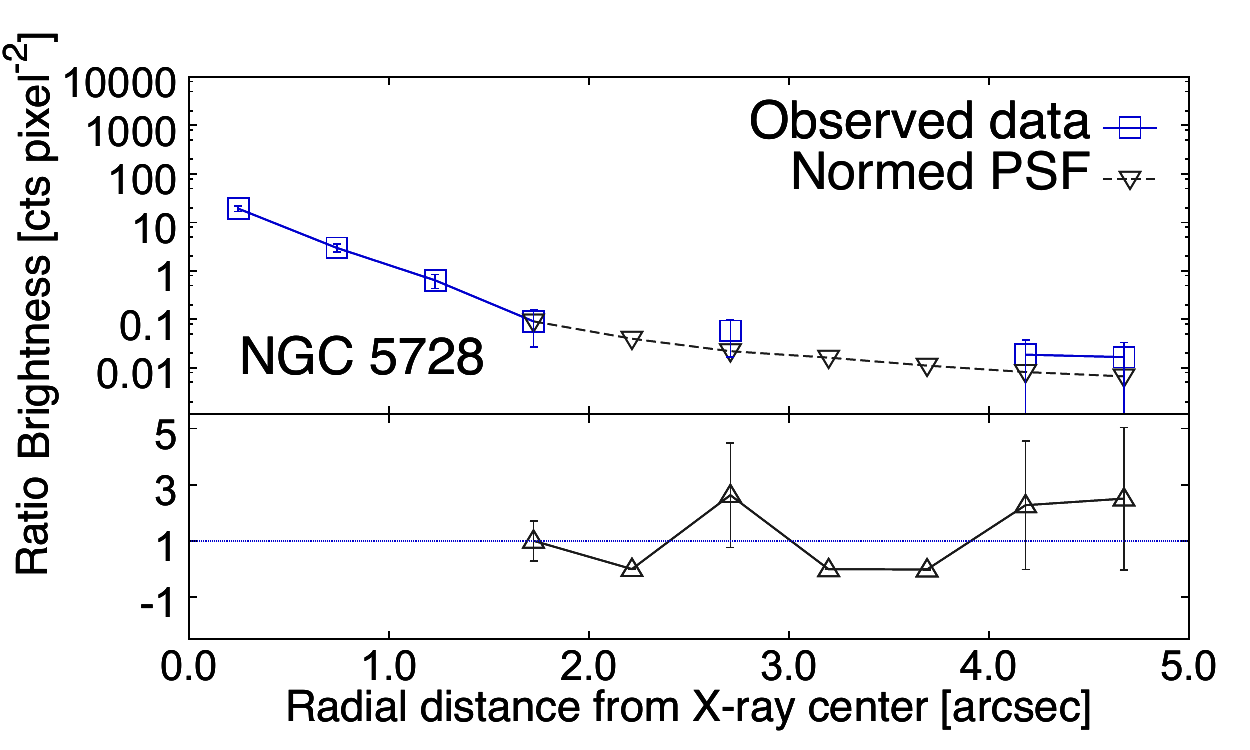}\\
    \includegraphics[width=6.0cm]{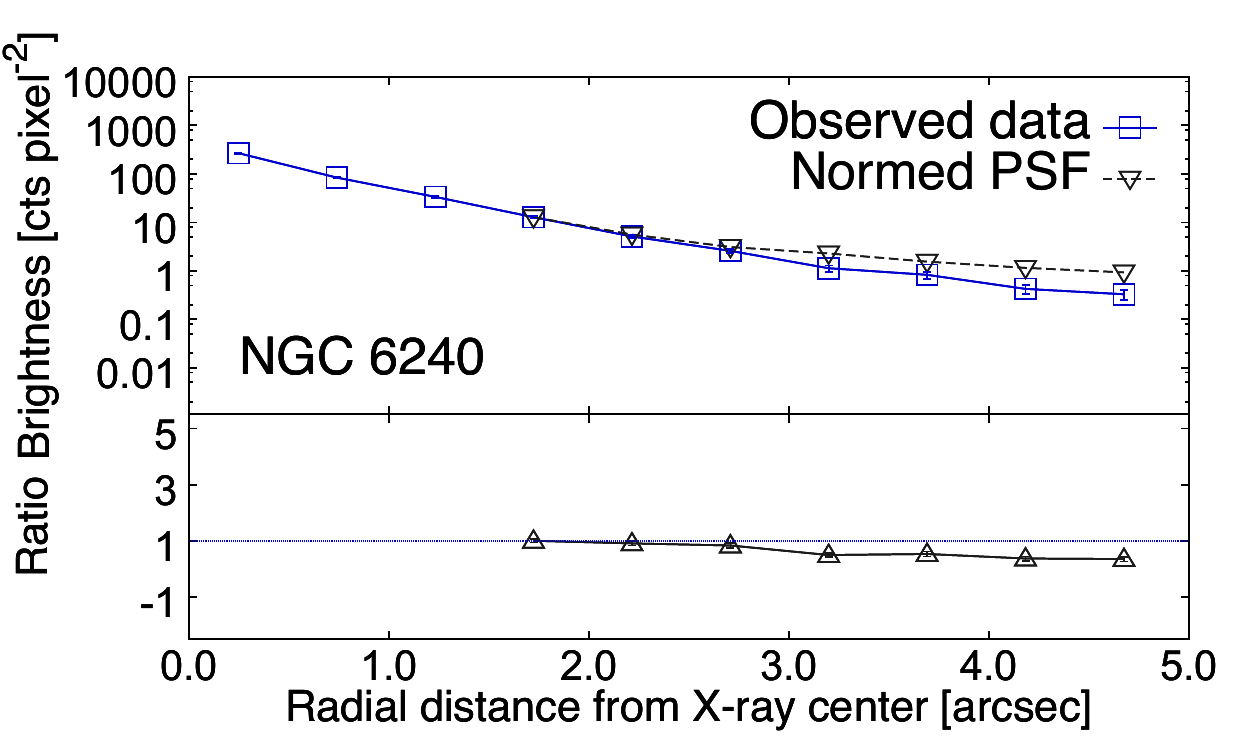}\hspace{-.3cm}
    \includegraphics[width=6.0cm]{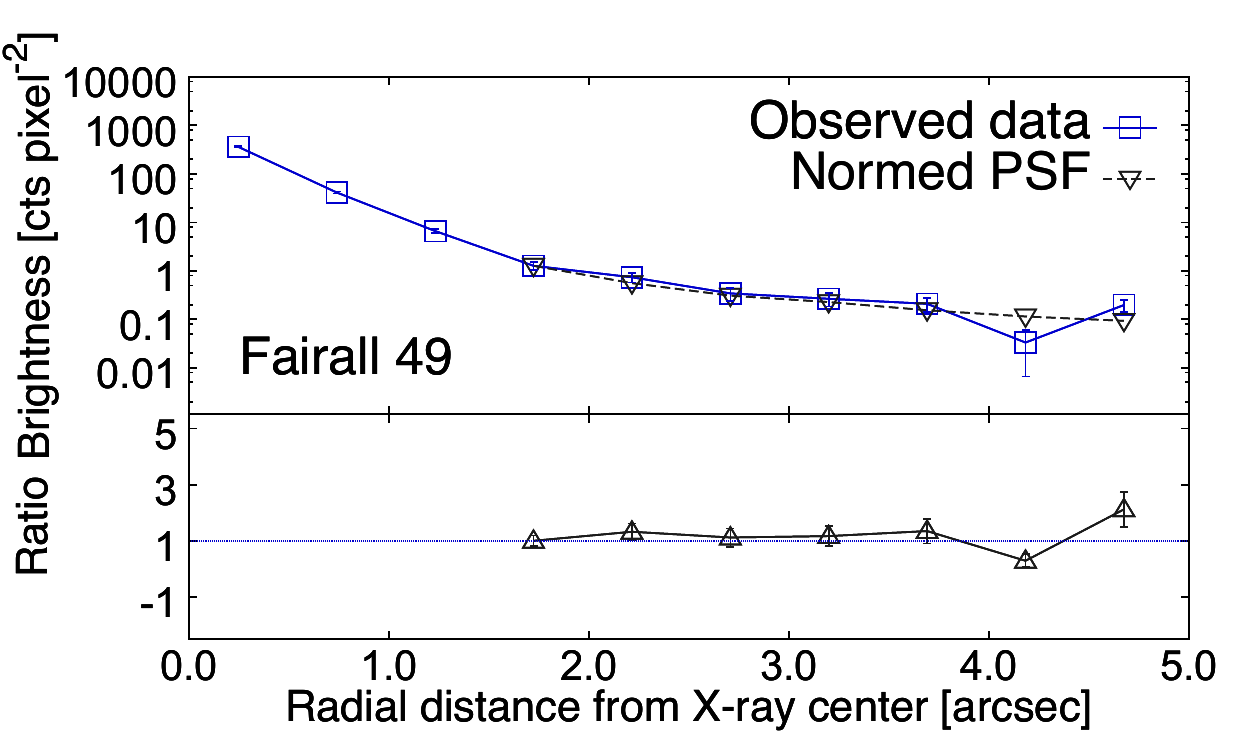}\hspace{-.3cm}
    \includegraphics[width=6.0cm]{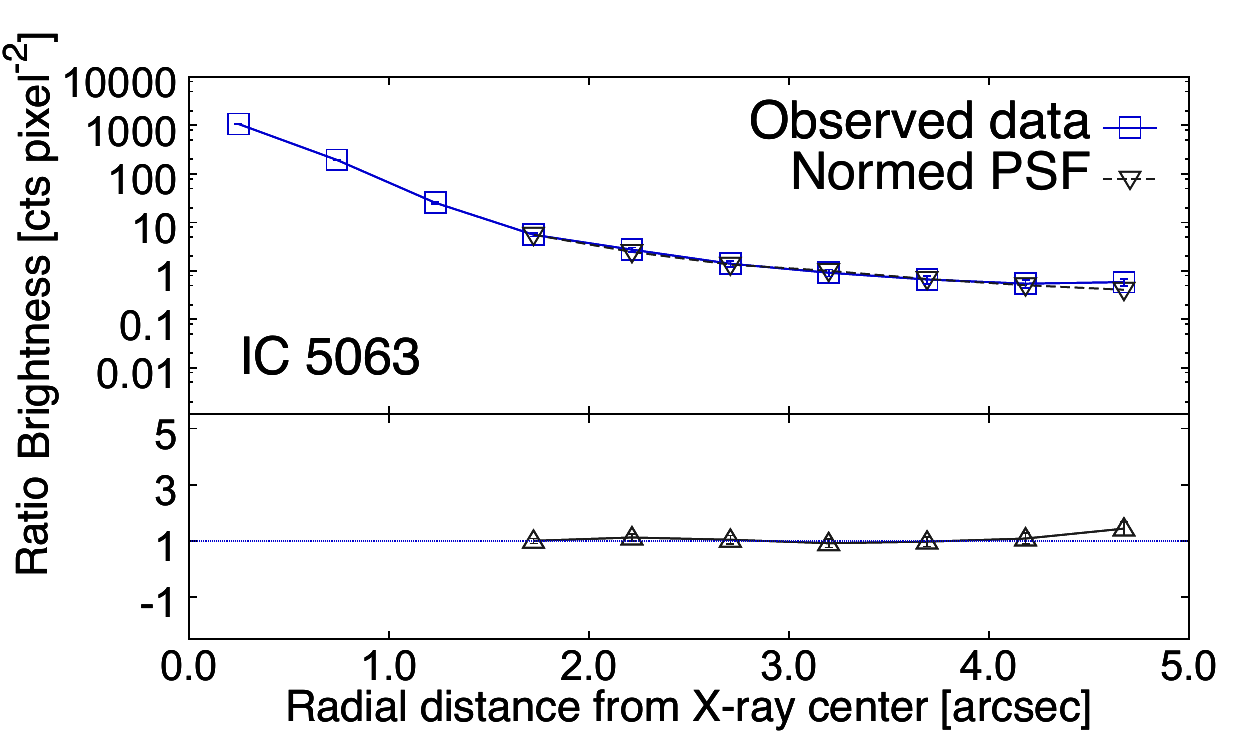}\\
    \includegraphics[width=6.0cm]{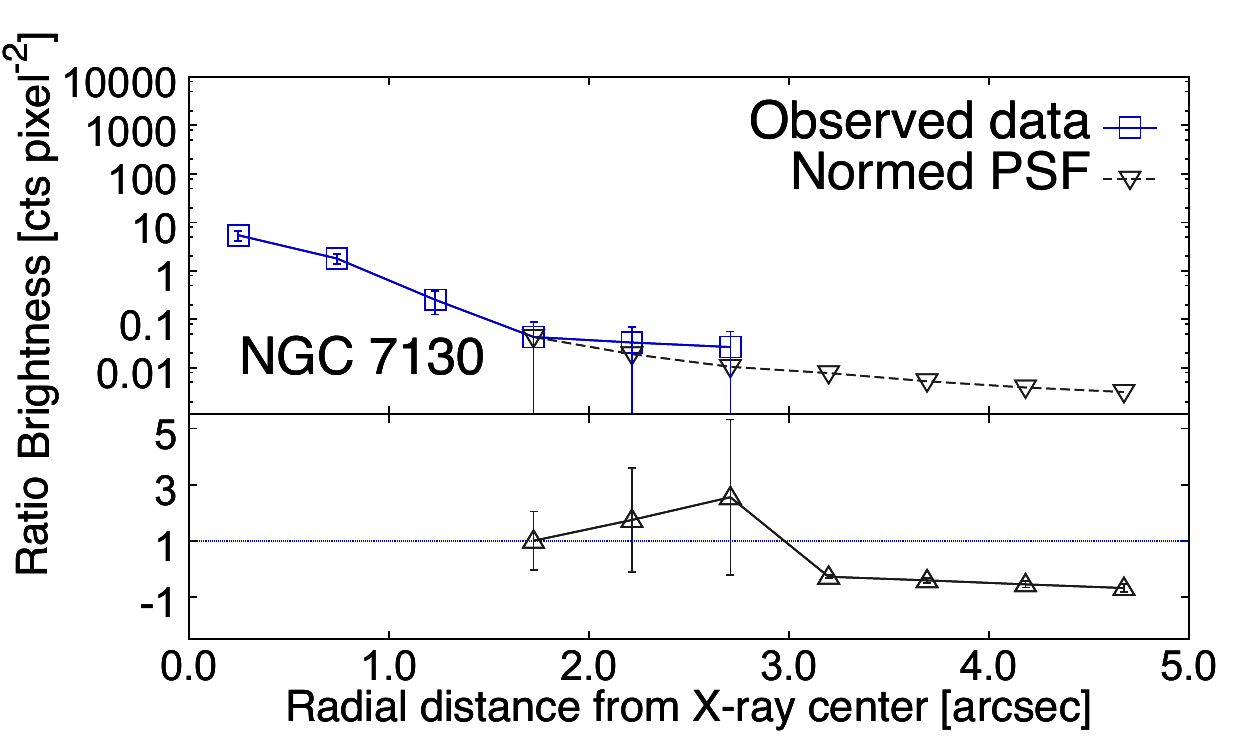}\hspace{-.3cm}    
    \includegraphics[width=6.0cm]{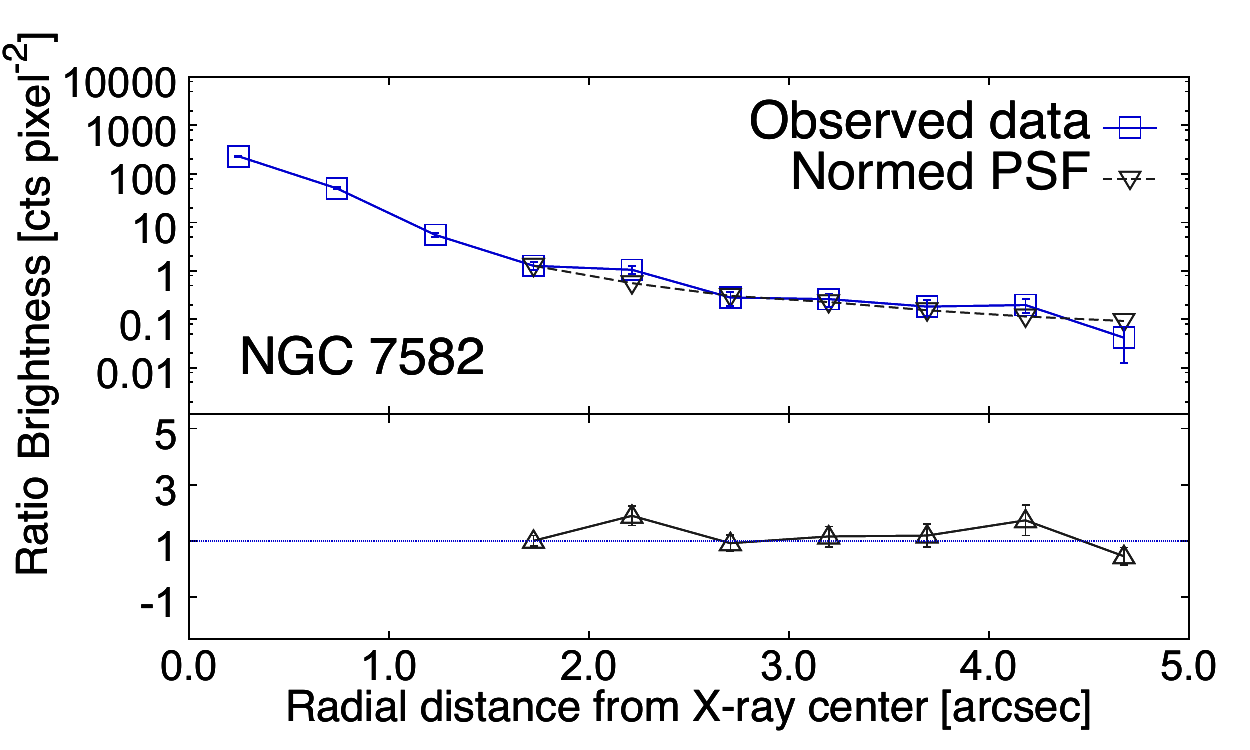}\hspace{-.3cm}   
    \caption{Continued.}
    \label{app:fig:rad_profile_6-7}
\end{figure*}

\clearpage

\section{X-ray images at 3--6\,keV and 6--7\,keV and their ratios}\label{app:ximages}

\begin{figure*}[!h]
    \centering
    \includegraphics[width=5.5cm]{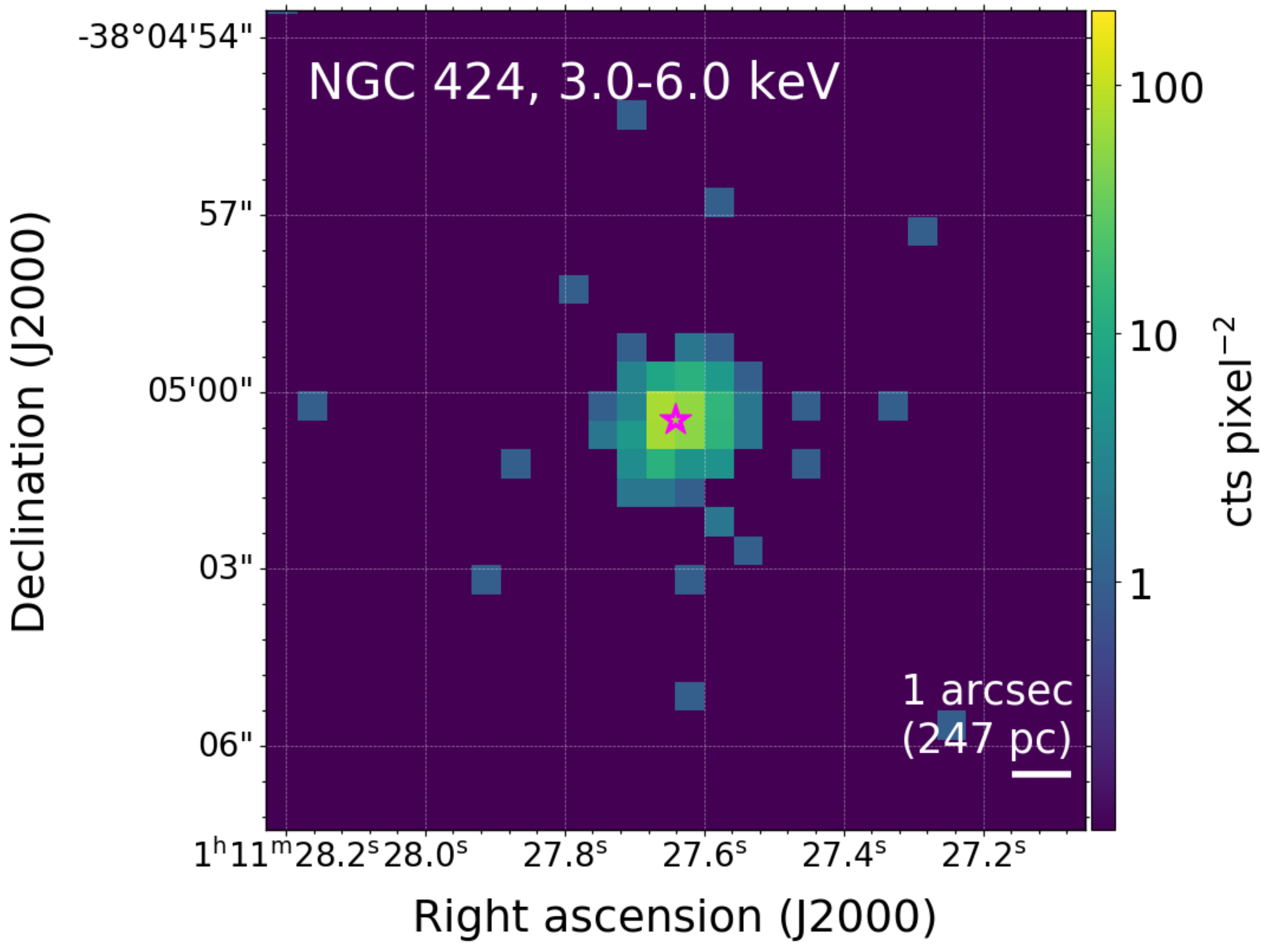}
    \includegraphics[width=5.5cm]{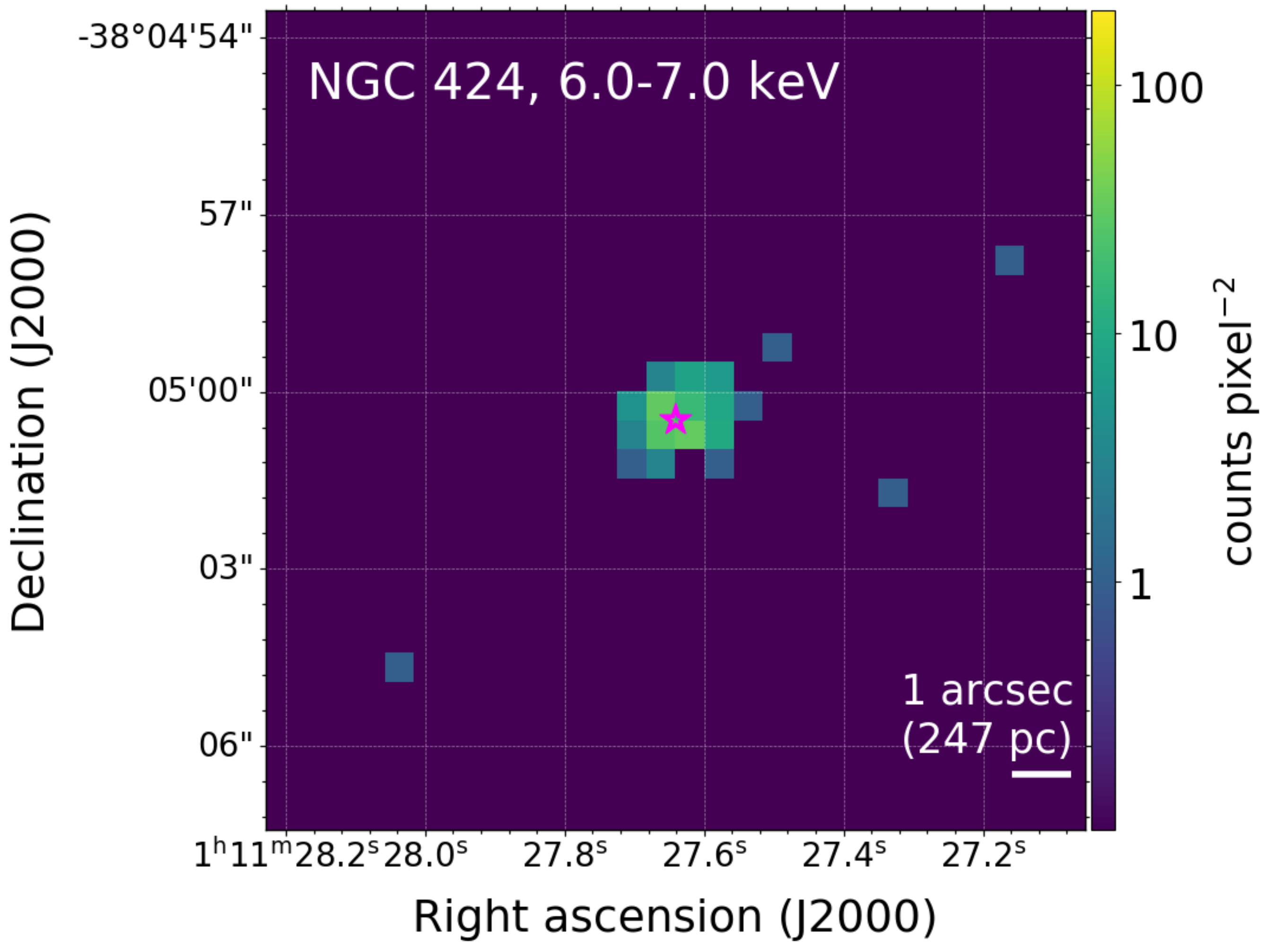} 
    \includegraphics[width=5.35cm]{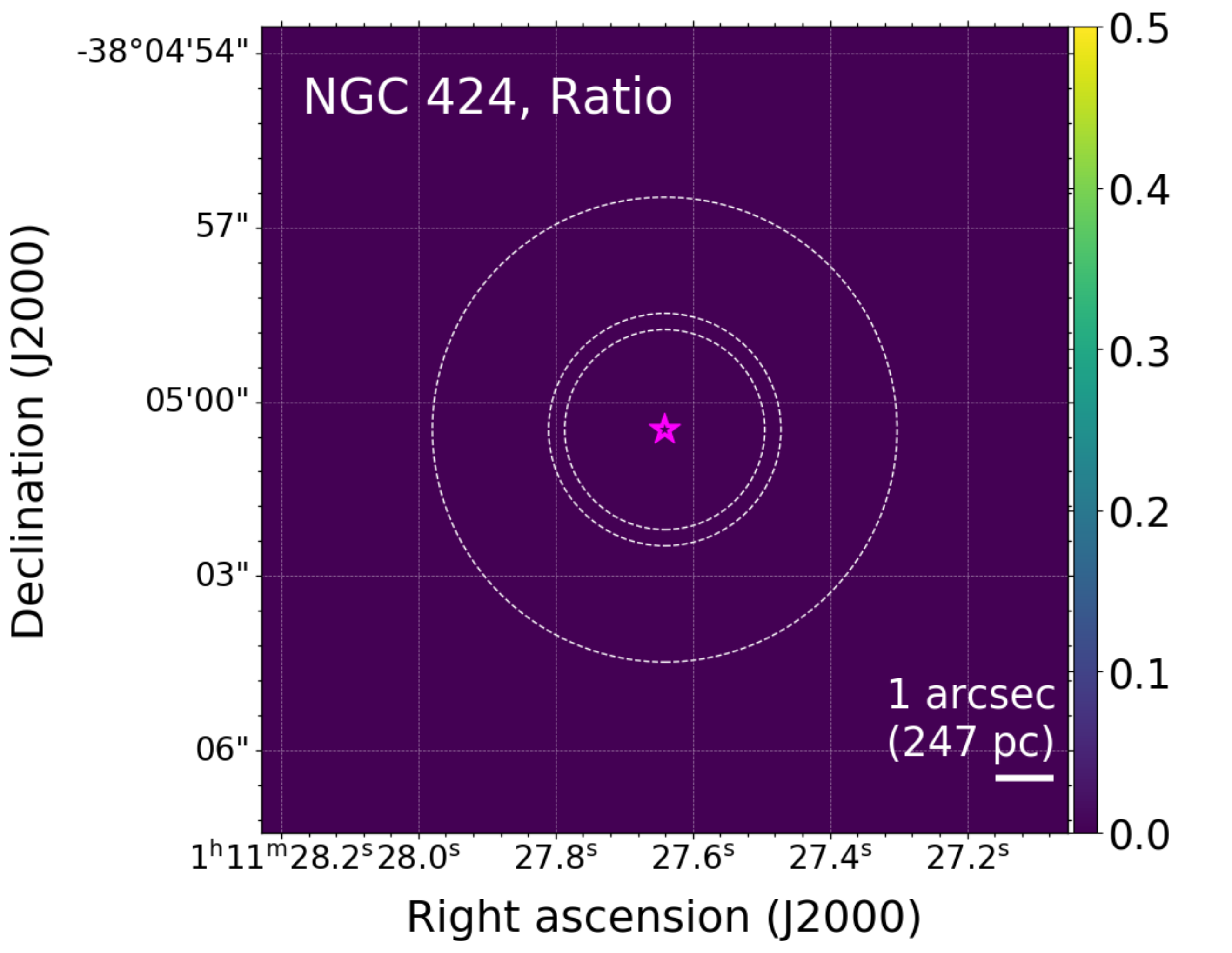}
    \\ 
    \includegraphics[width=5.5cm]{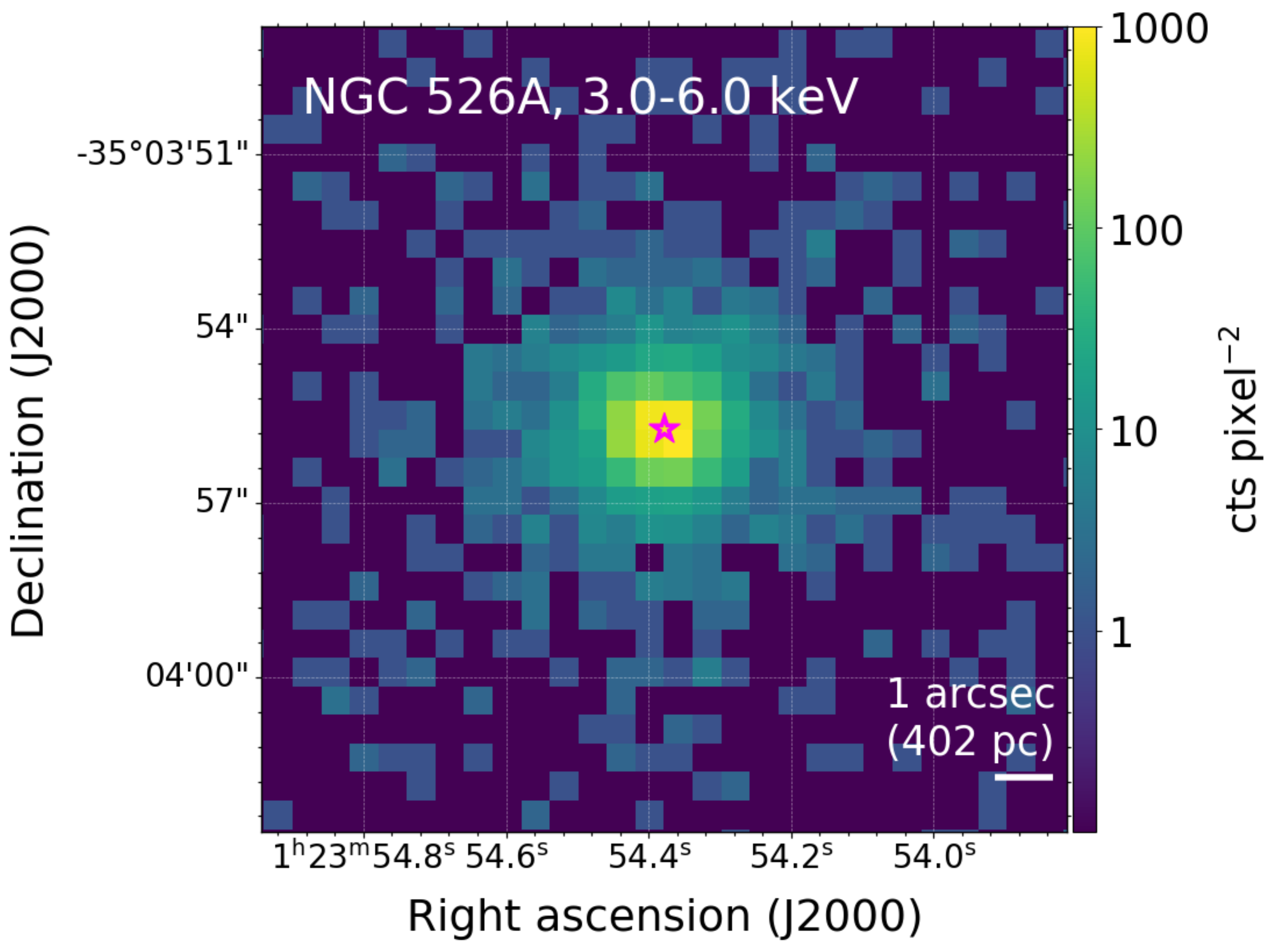}
    \includegraphics[width=5.5cm]{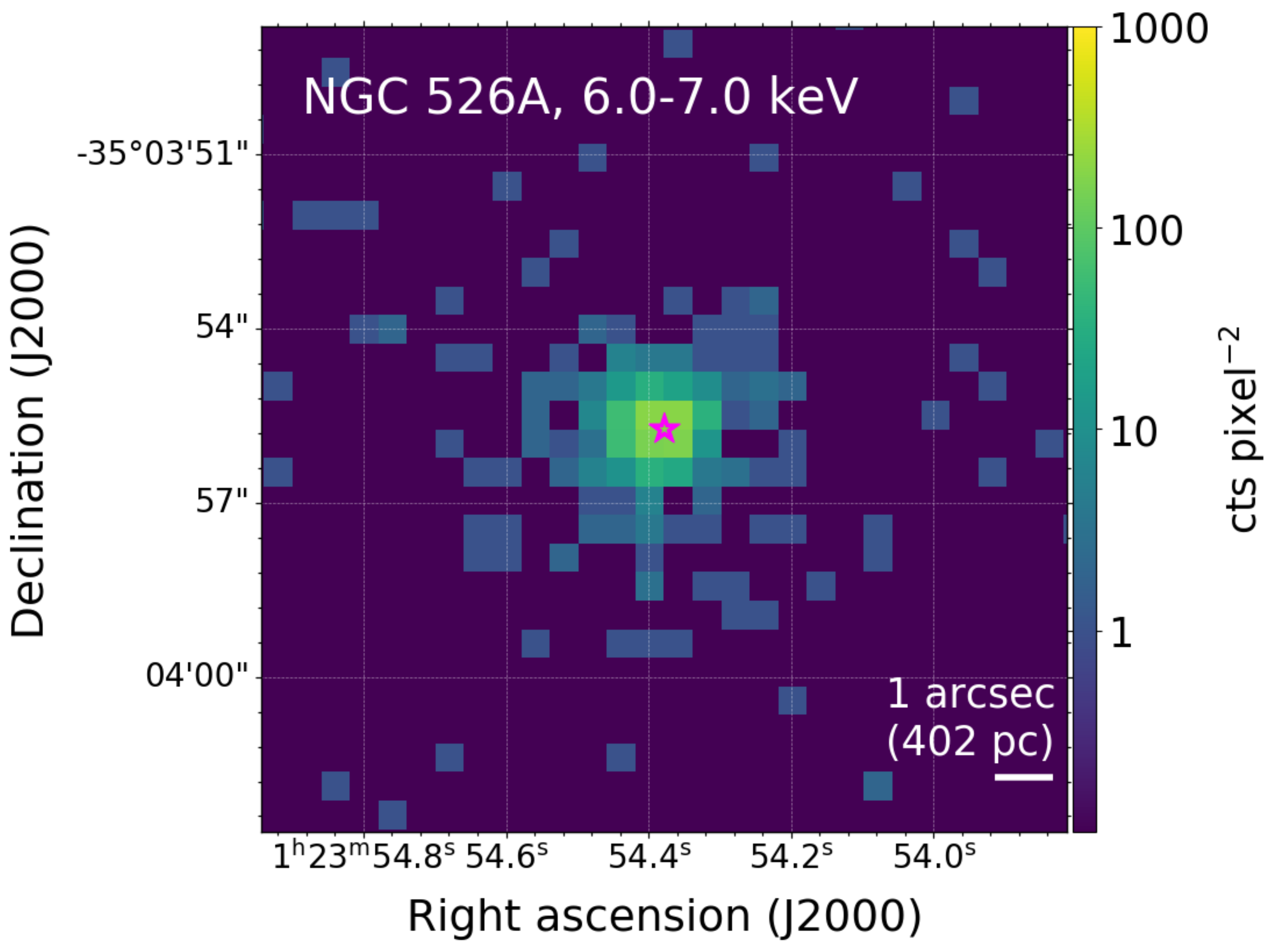}
    \includegraphics[width=5.35cm]{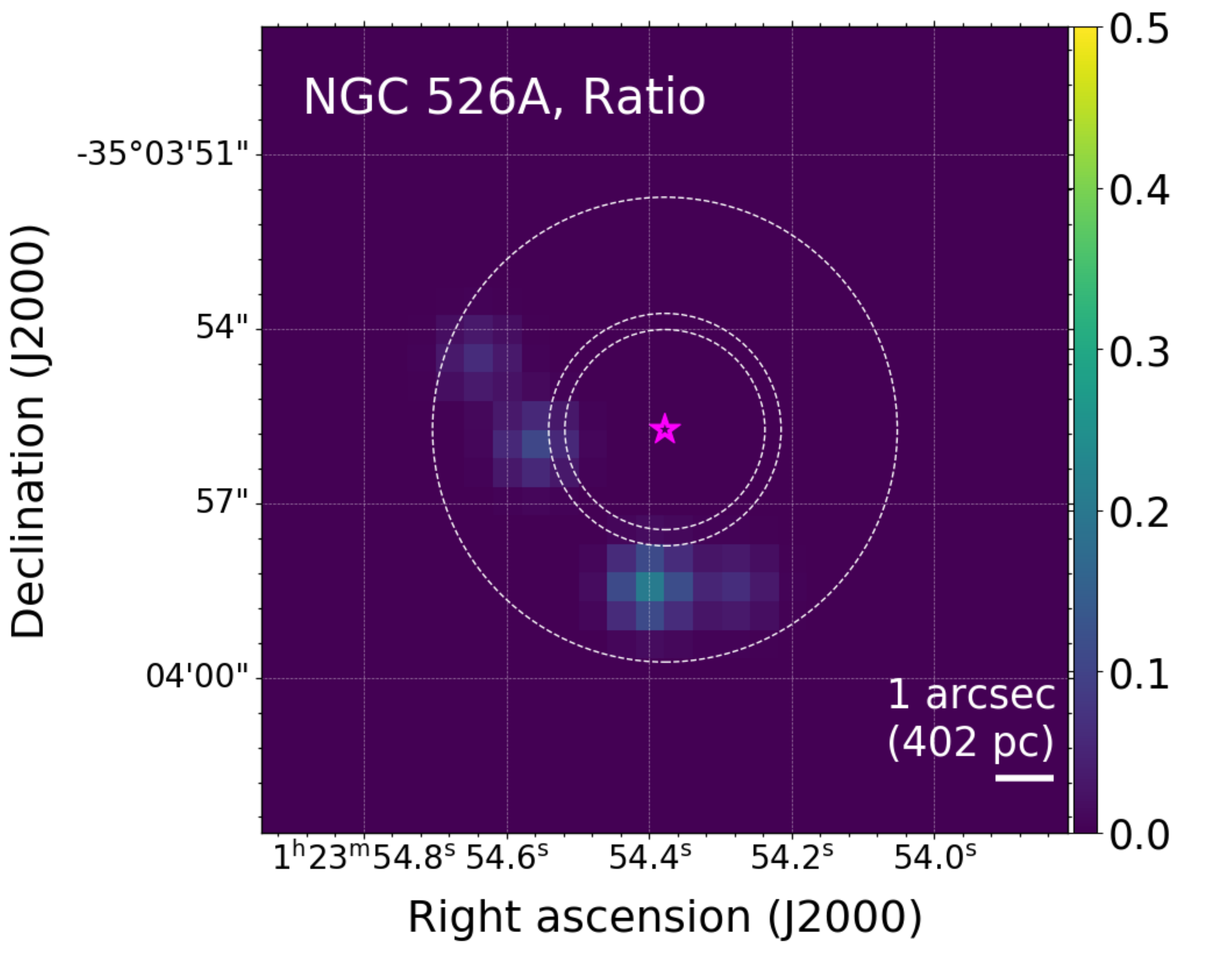}   
    \\
    \includegraphics[width=5.5cm]{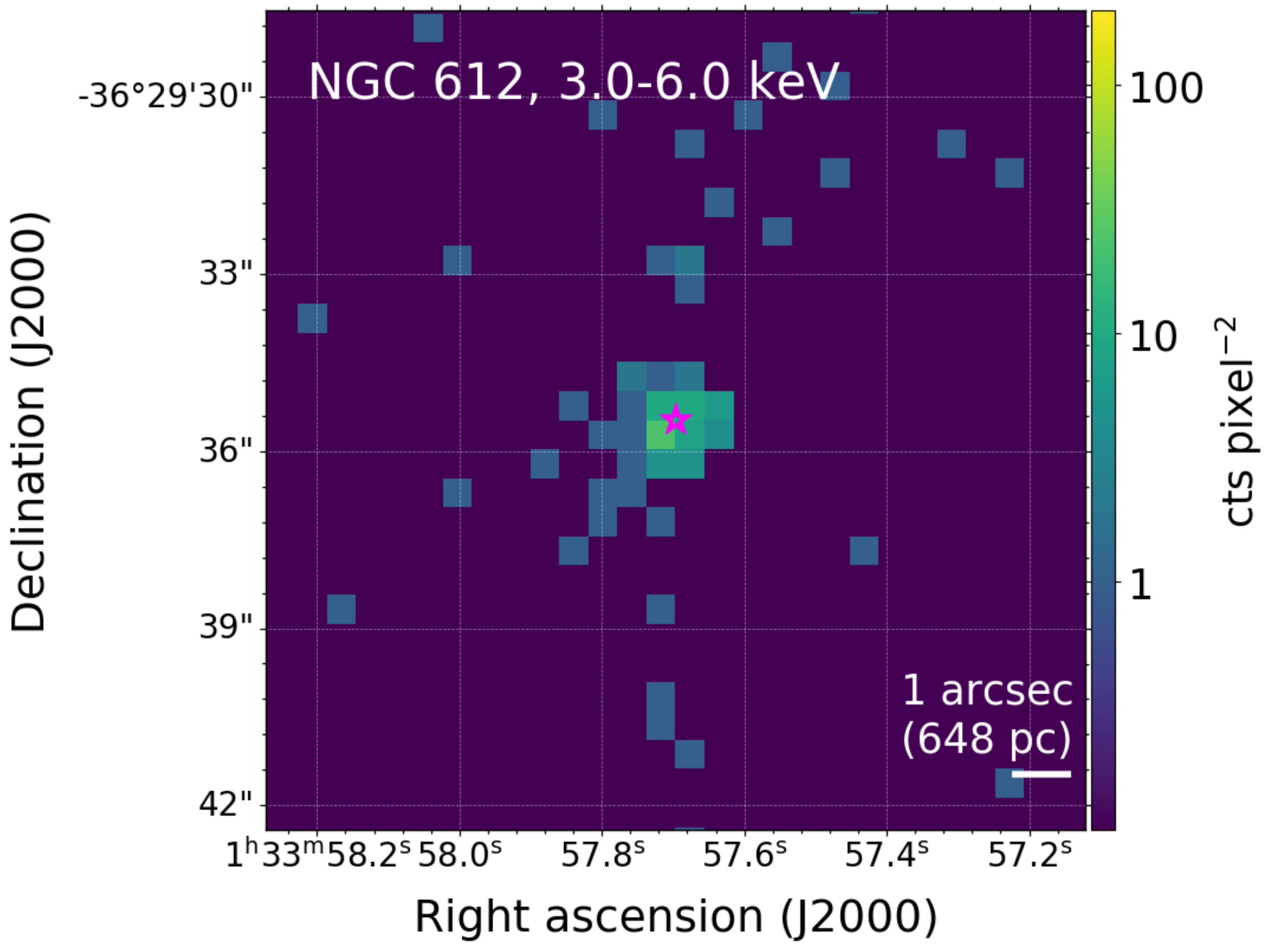}
    \includegraphics[width=5.5cm]{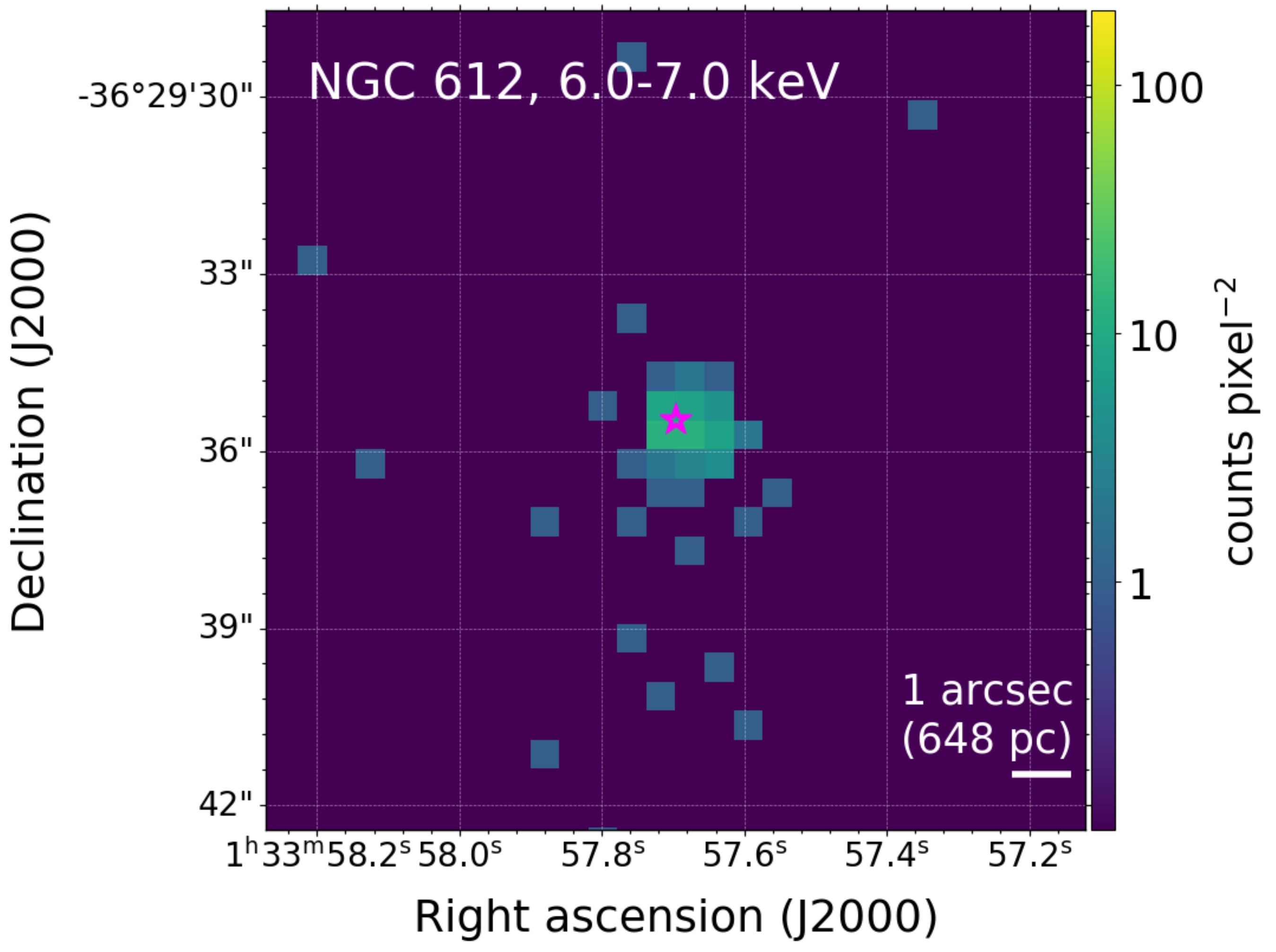}
    \includegraphics[width=5.35cm]{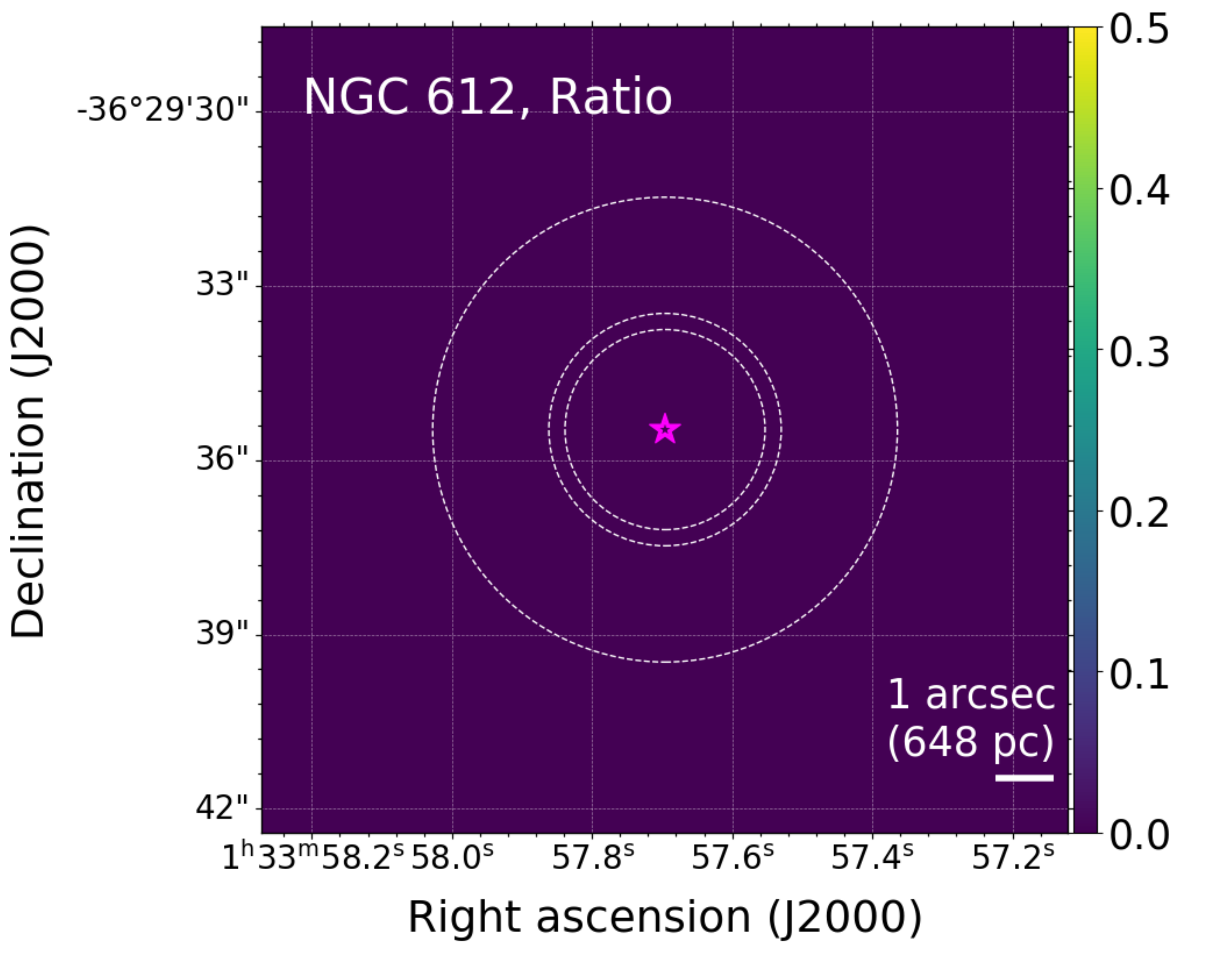}
    \\
    \includegraphics[width=5.5cm]{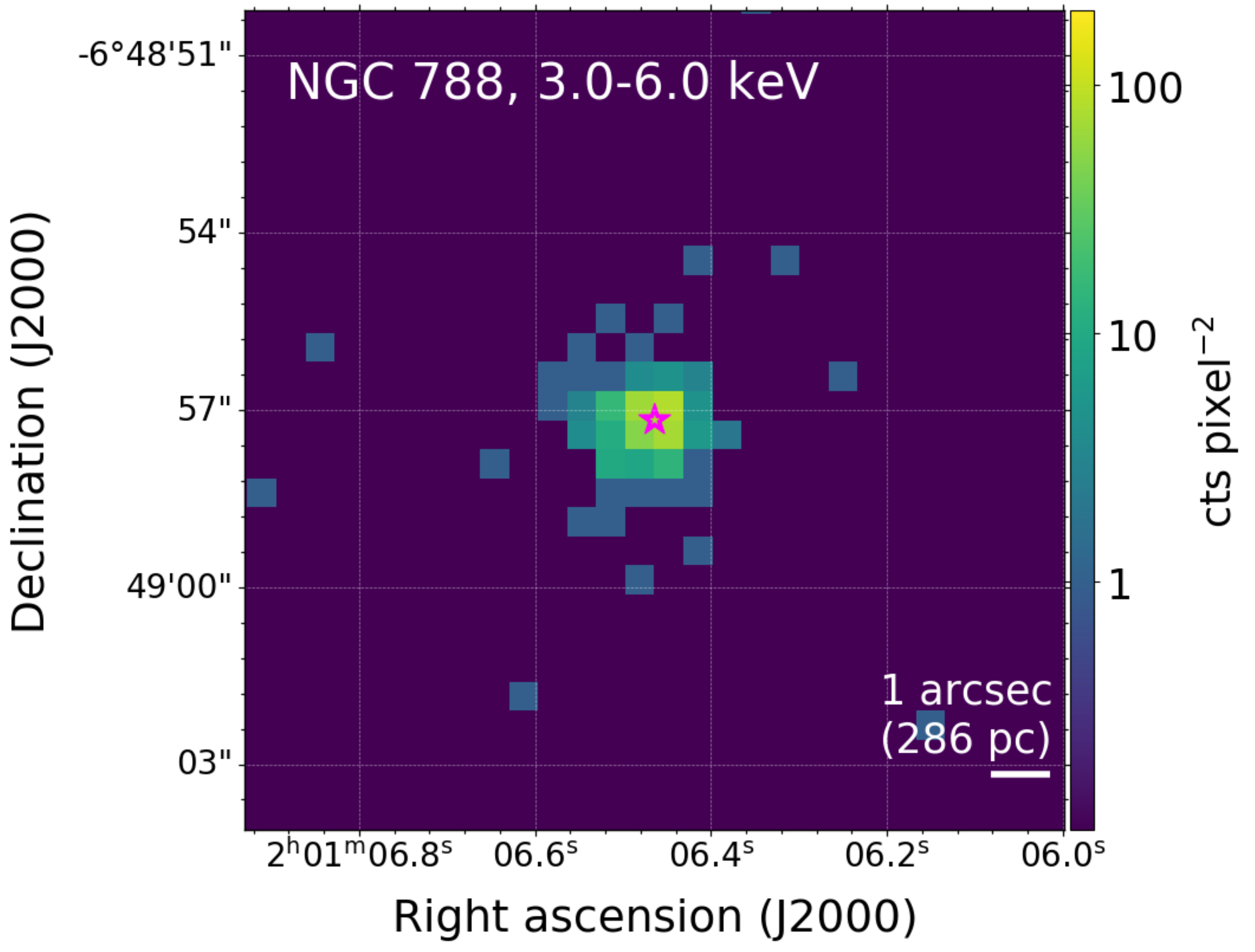}
    \includegraphics[width=5.5cm]{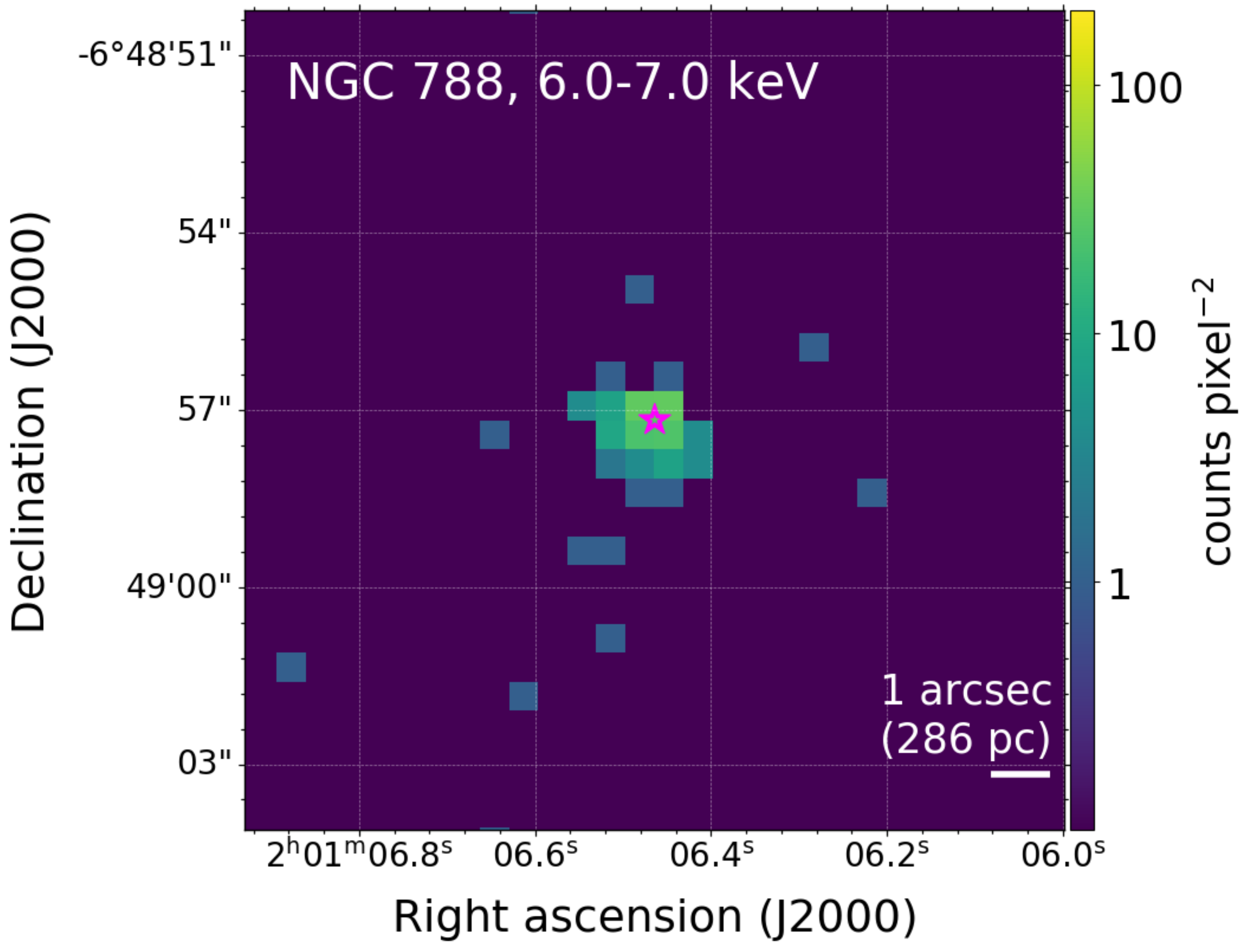} 
    \includegraphics[width=5.35cm]{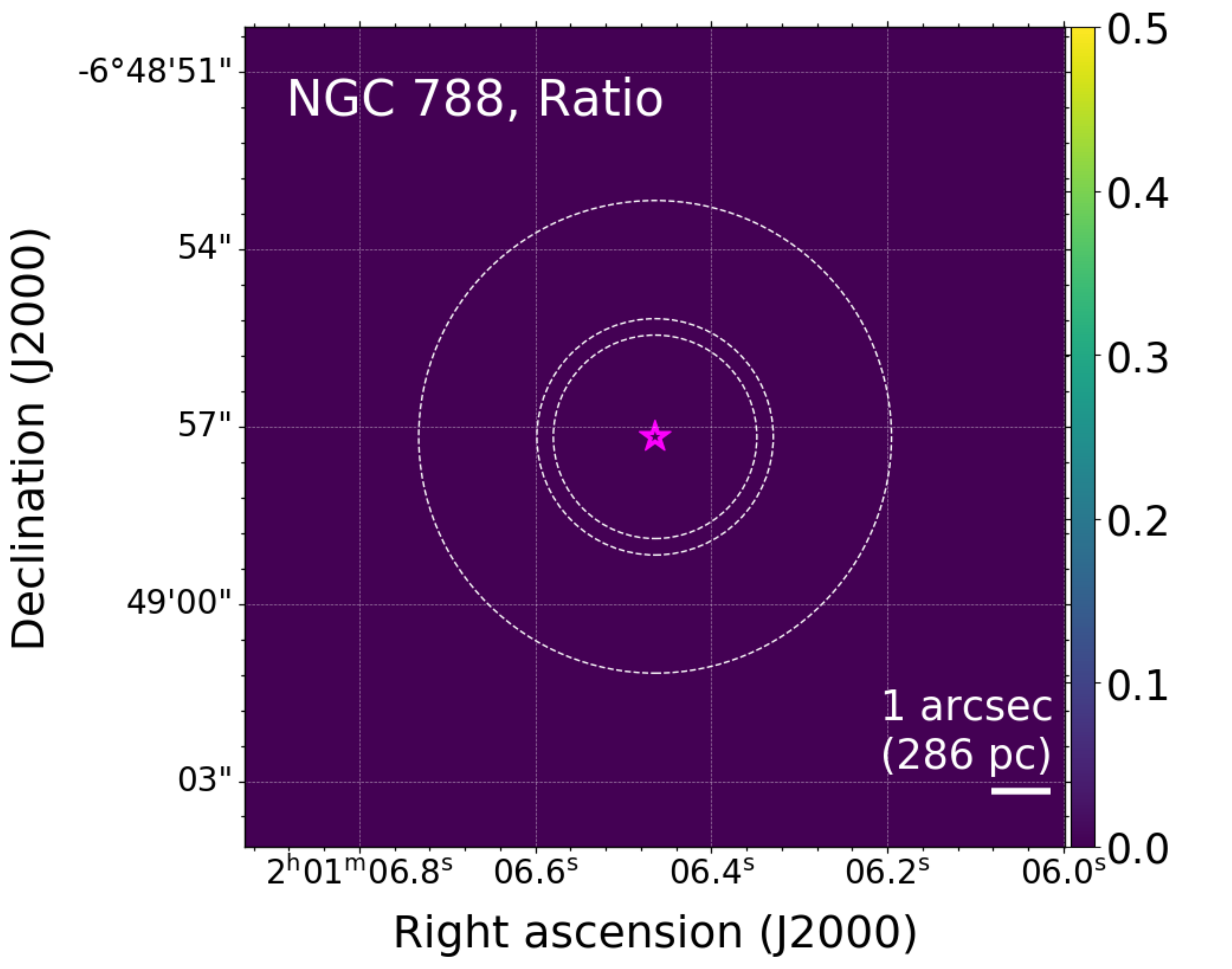}
    \\
    \caption{From left to right columns, 3.0--6.0\,keV and 6.0--7.0\,keV images in units of counts pix$^{-2}$ and 
    the ratios of the 6.0--7.0\,keV and 3.0--6.0\,keV images from which nuclear components are subtracted.
    North is up and east is left.
    }
    \label{app:fig:xray_images}
\end{figure*}

\begin{figure*}\addtocounter{figure}{-1}
    \centering
    \includegraphics[width=5.5cm]{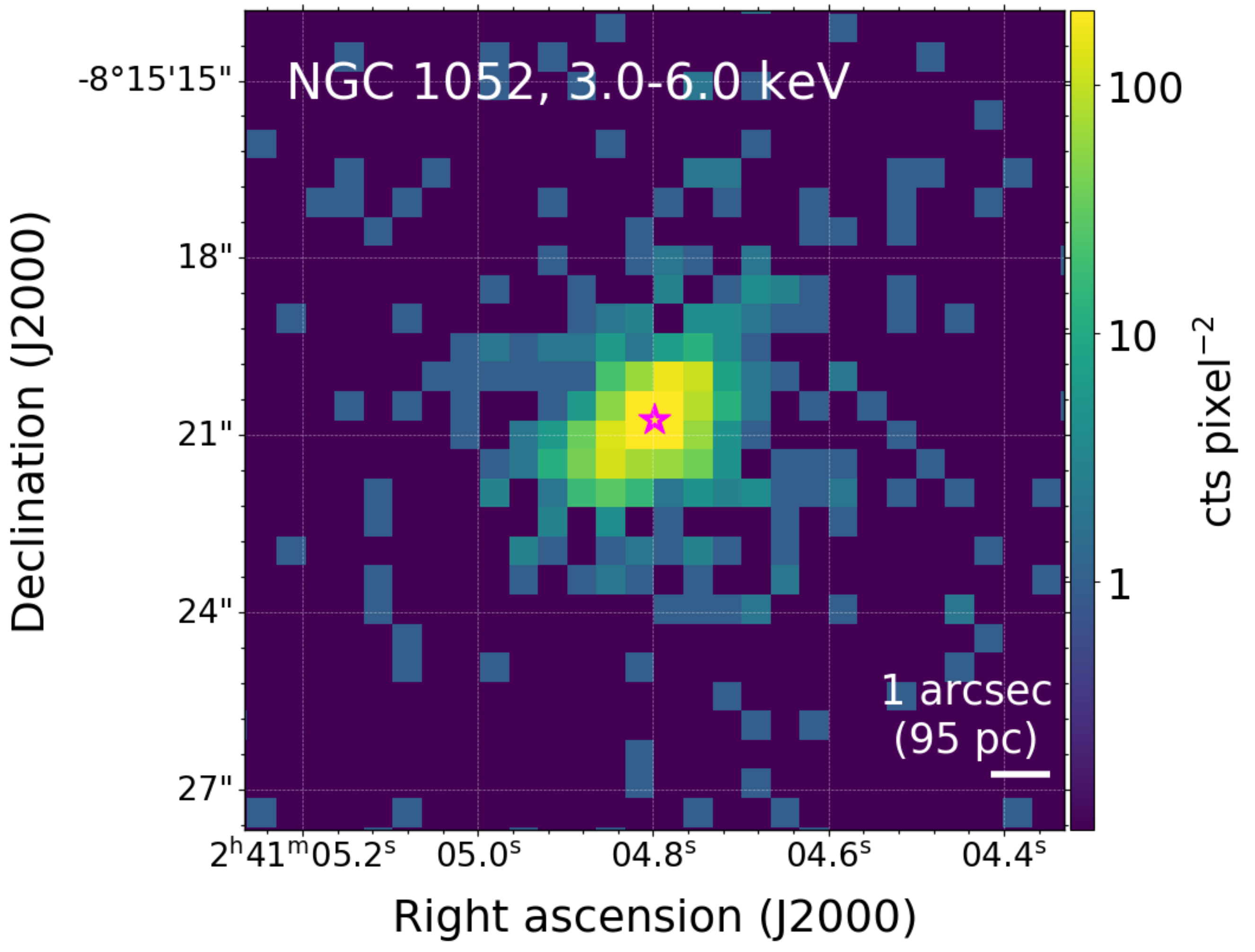}
    \includegraphics[width=5.5cm]{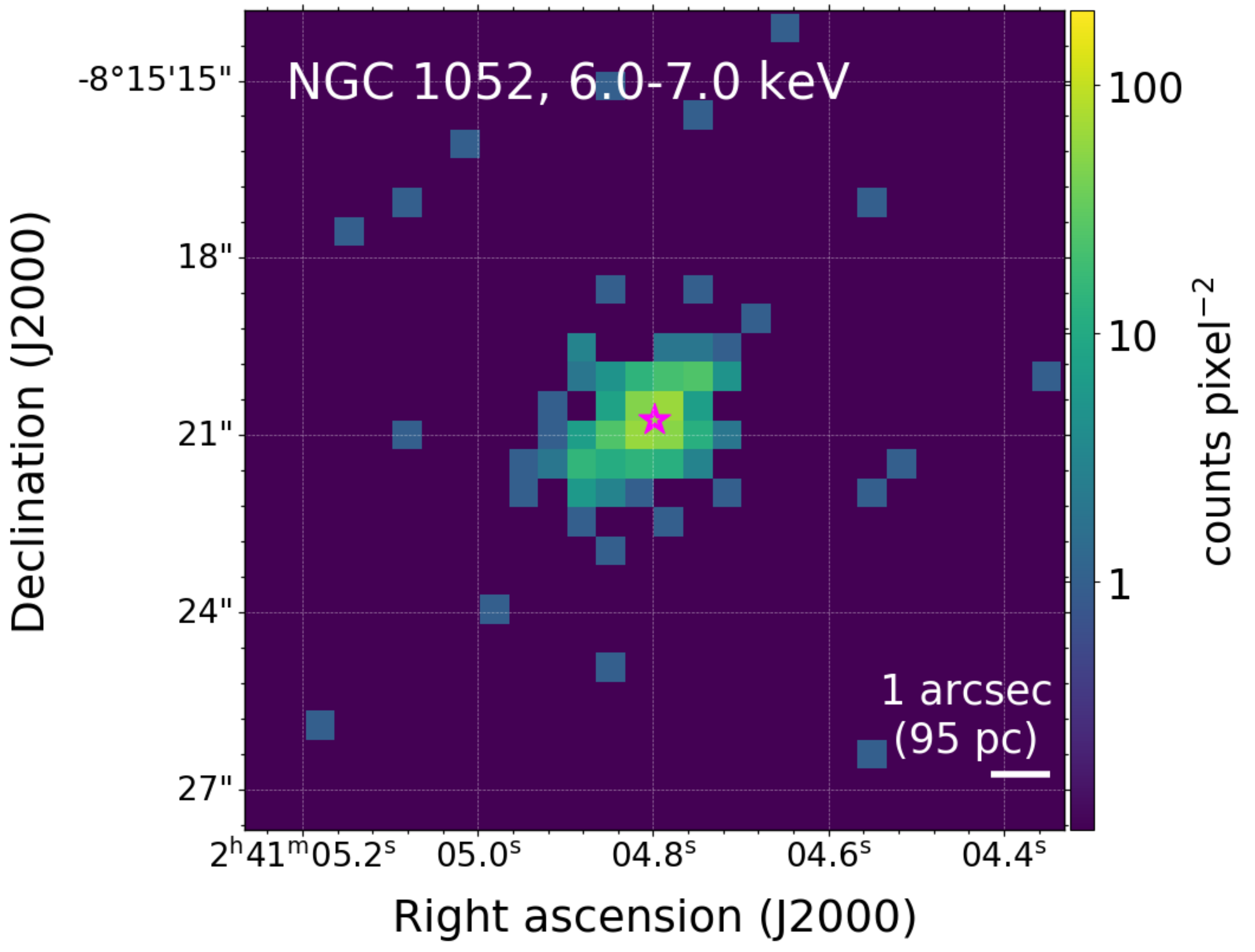}
    \includegraphics[width=5.35cm]{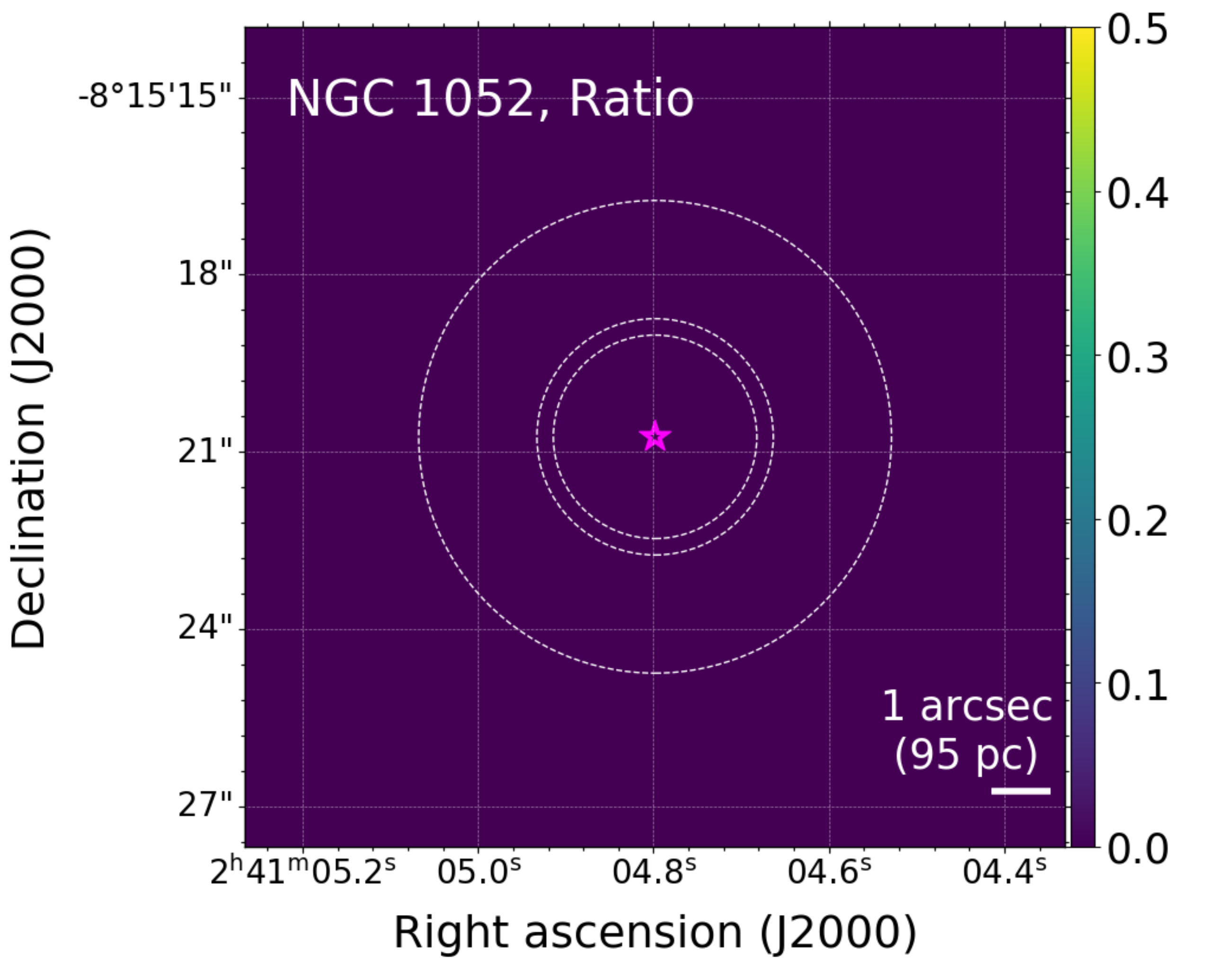}
    \\
    \includegraphics[width=5.5cm]{3-6keV_06_NGC_1068.pdf}
    \includegraphics[width=5.5cm]{6-7keV_06_NGC_1068.pdf}
    \includegraphics[width=5.35cm]{ratio_06_NGC_1068.pdf}
    \\ 
    \includegraphics[width=5.5cm]{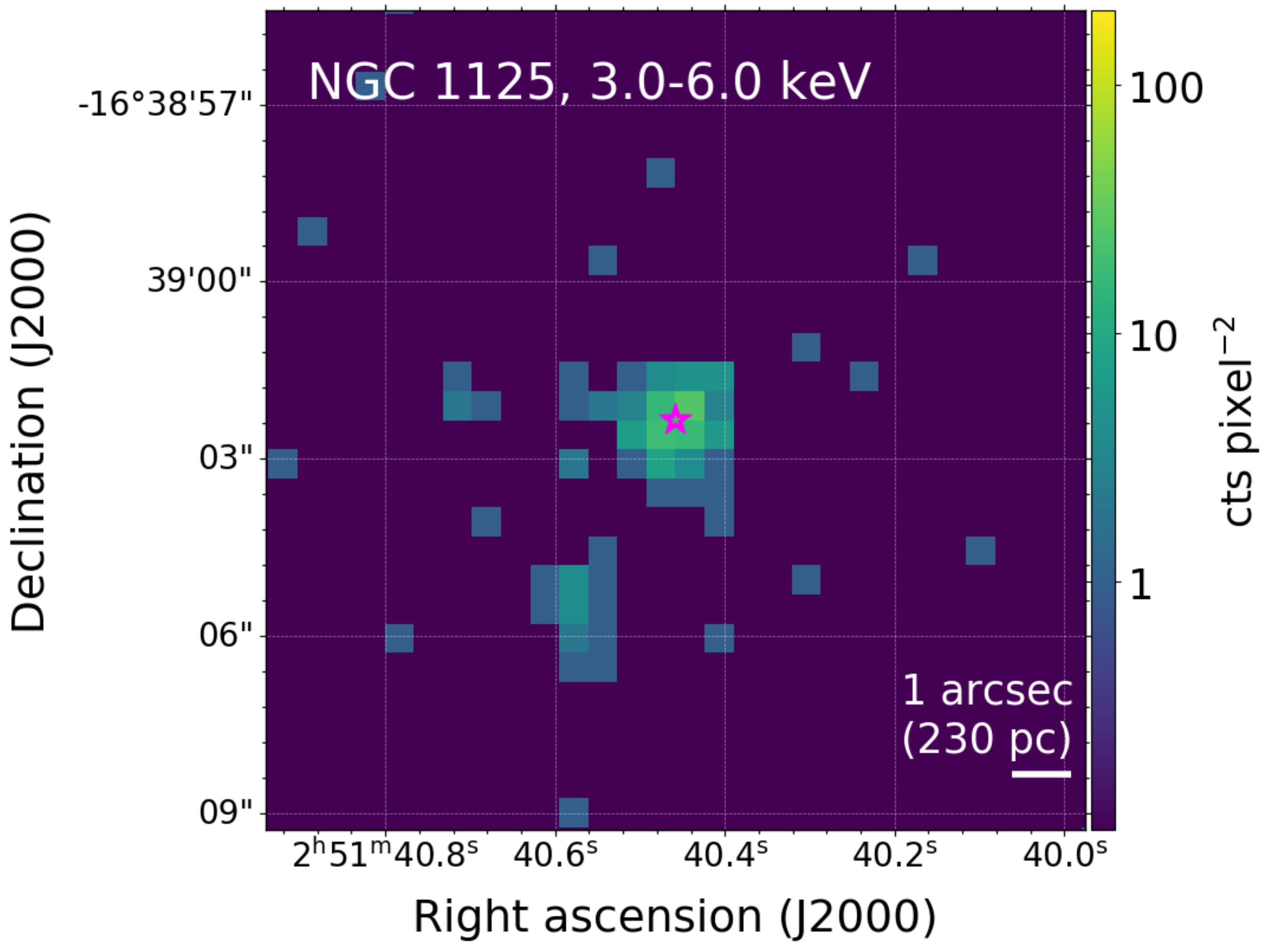}
    \includegraphics[width=5.5cm]{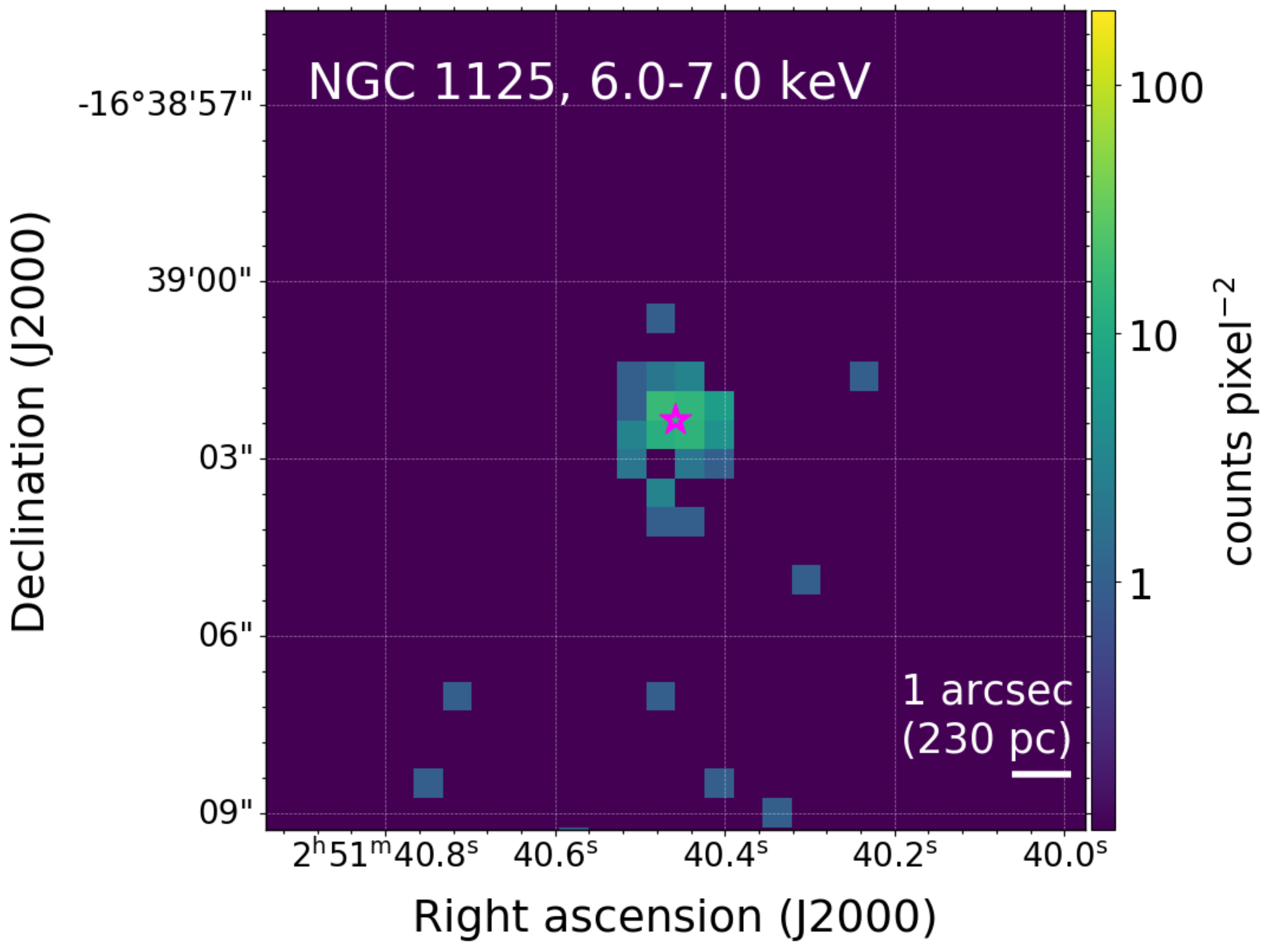}
    \includegraphics[width=5.35cm]{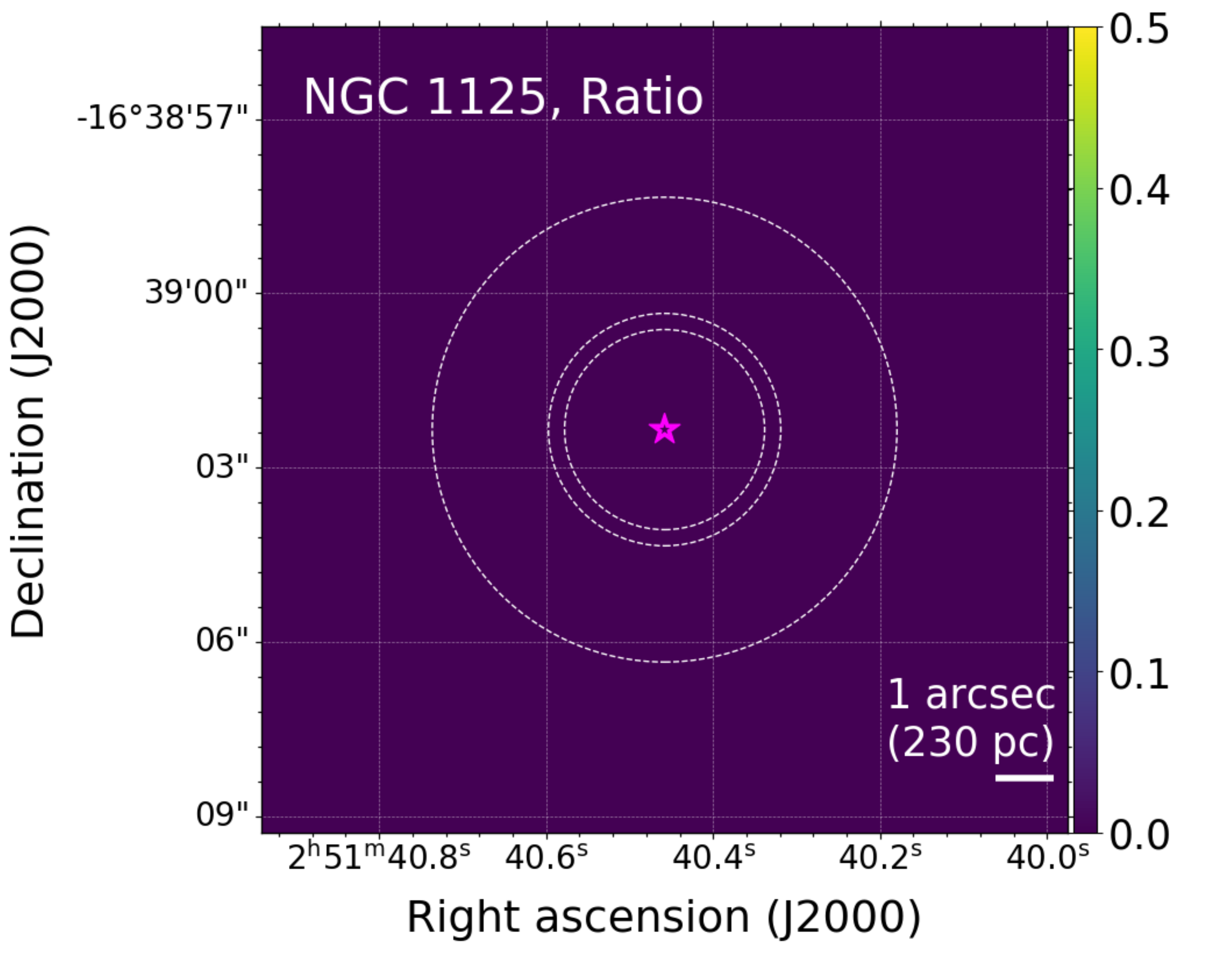}   
    \\
    \includegraphics[width=5.5cm]{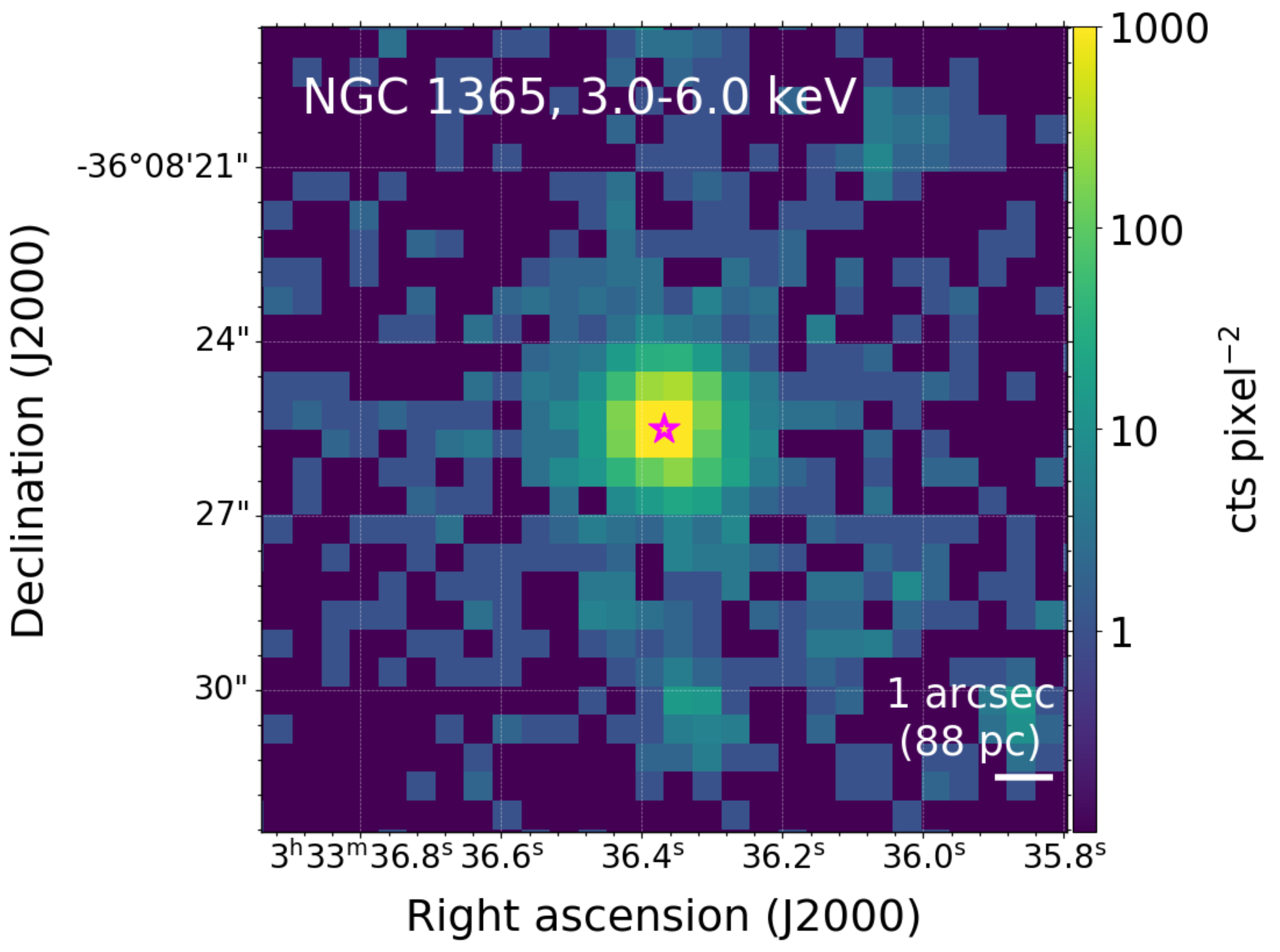}
    \includegraphics[width=5.5cm]{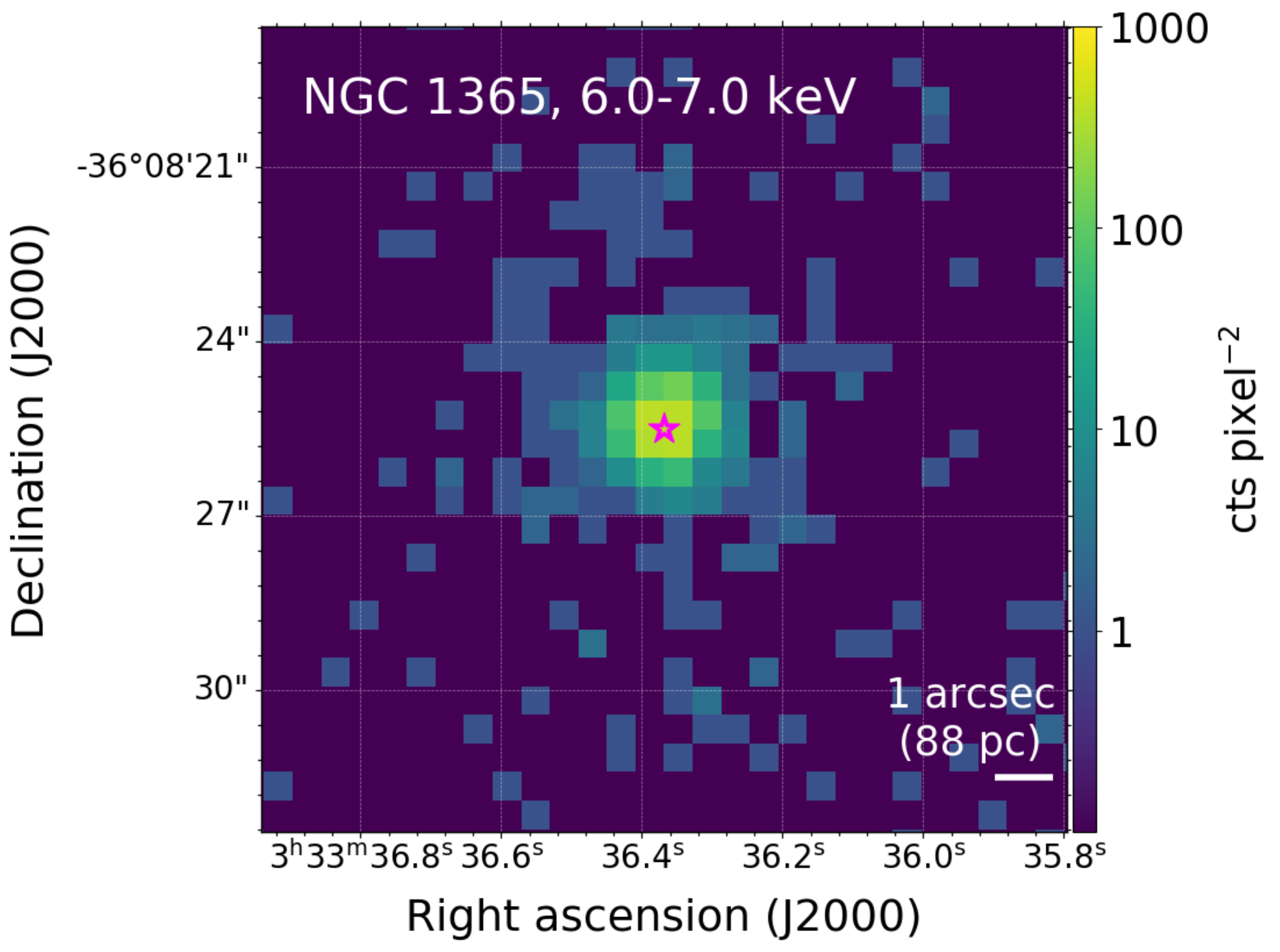}
    \includegraphics[width=5.35cm]{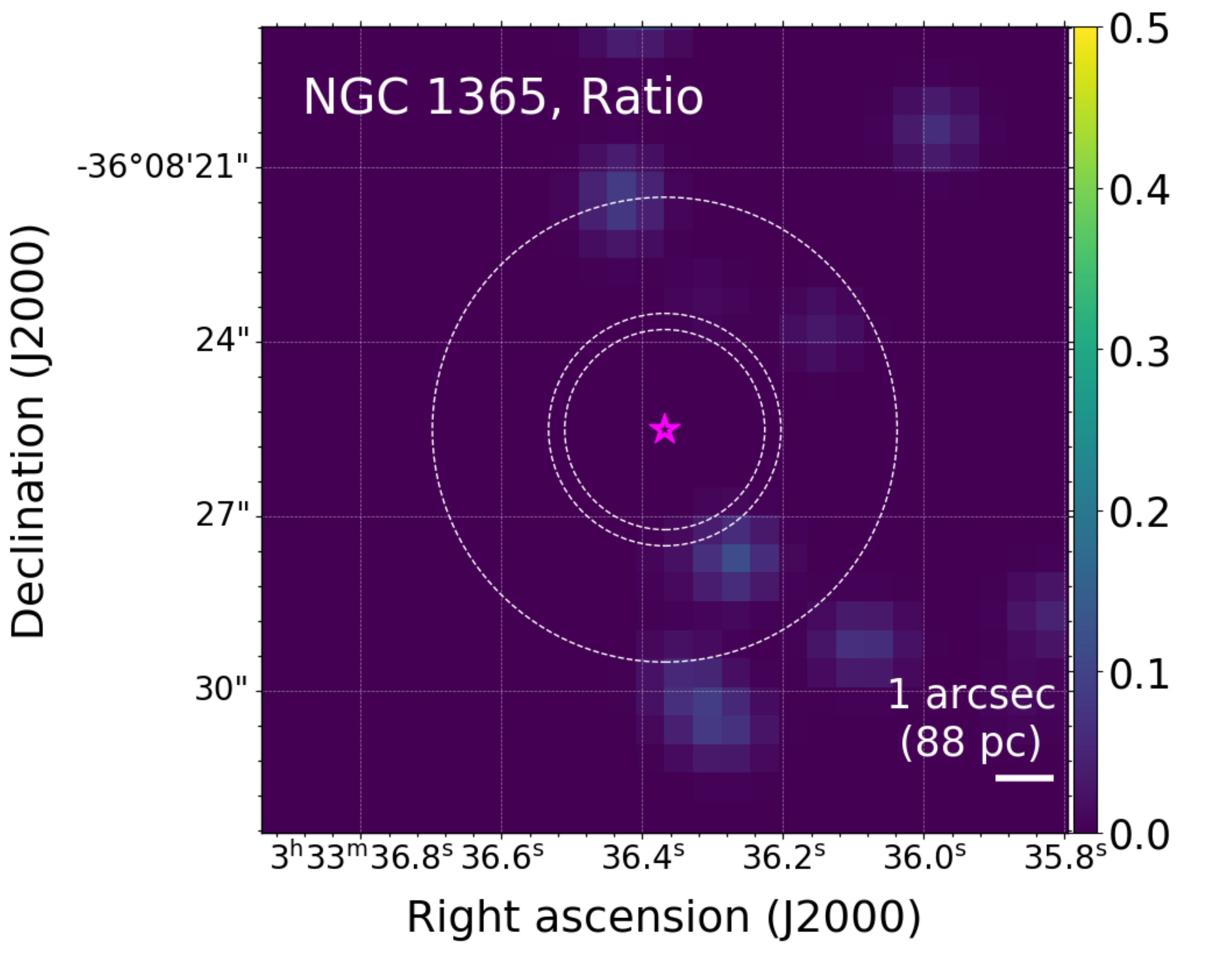}
    \\
    \includegraphics[width=5.5cm]{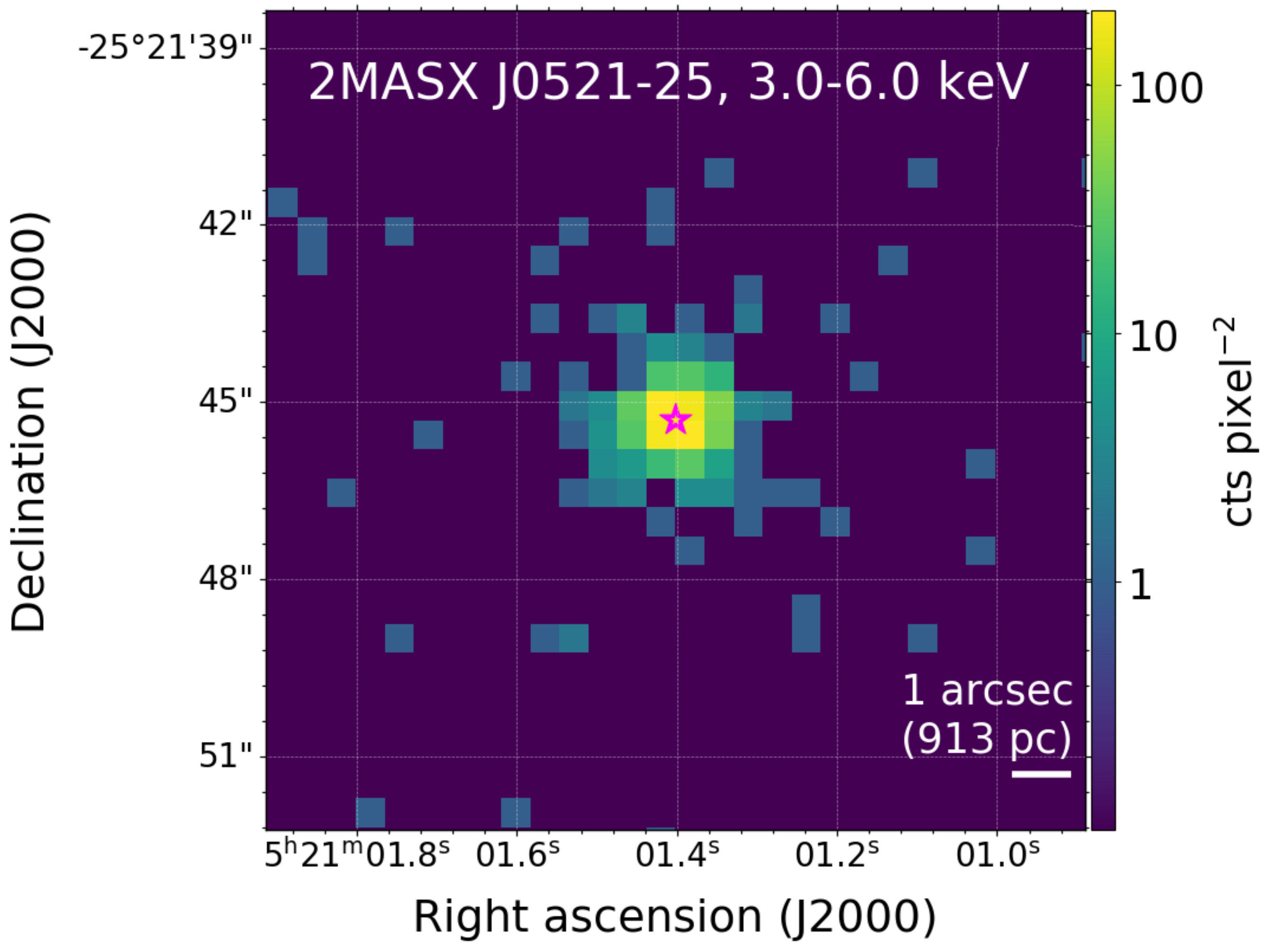}
    \includegraphics[width=5.5cm]{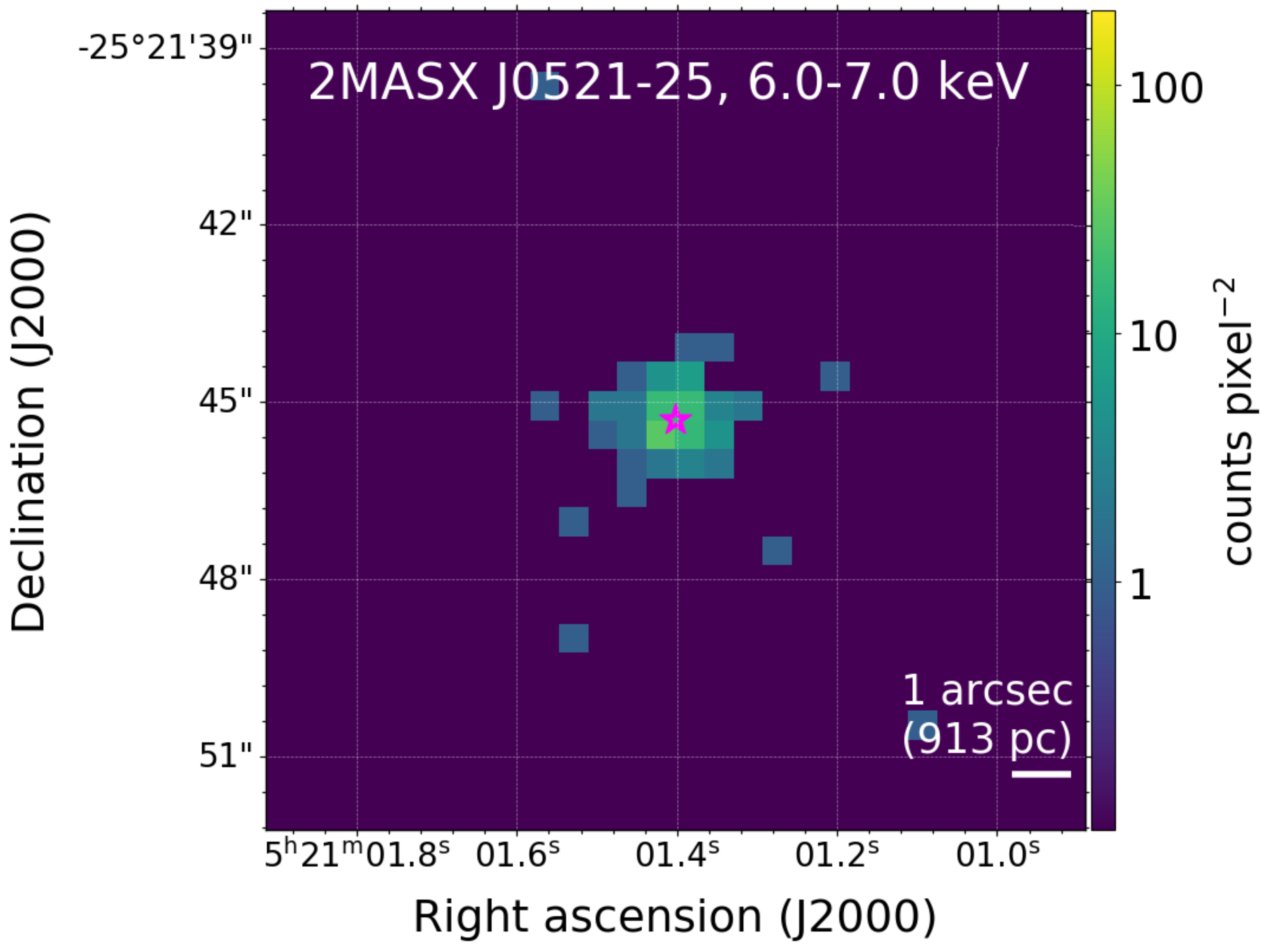}
    \includegraphics[width=5.35cm]{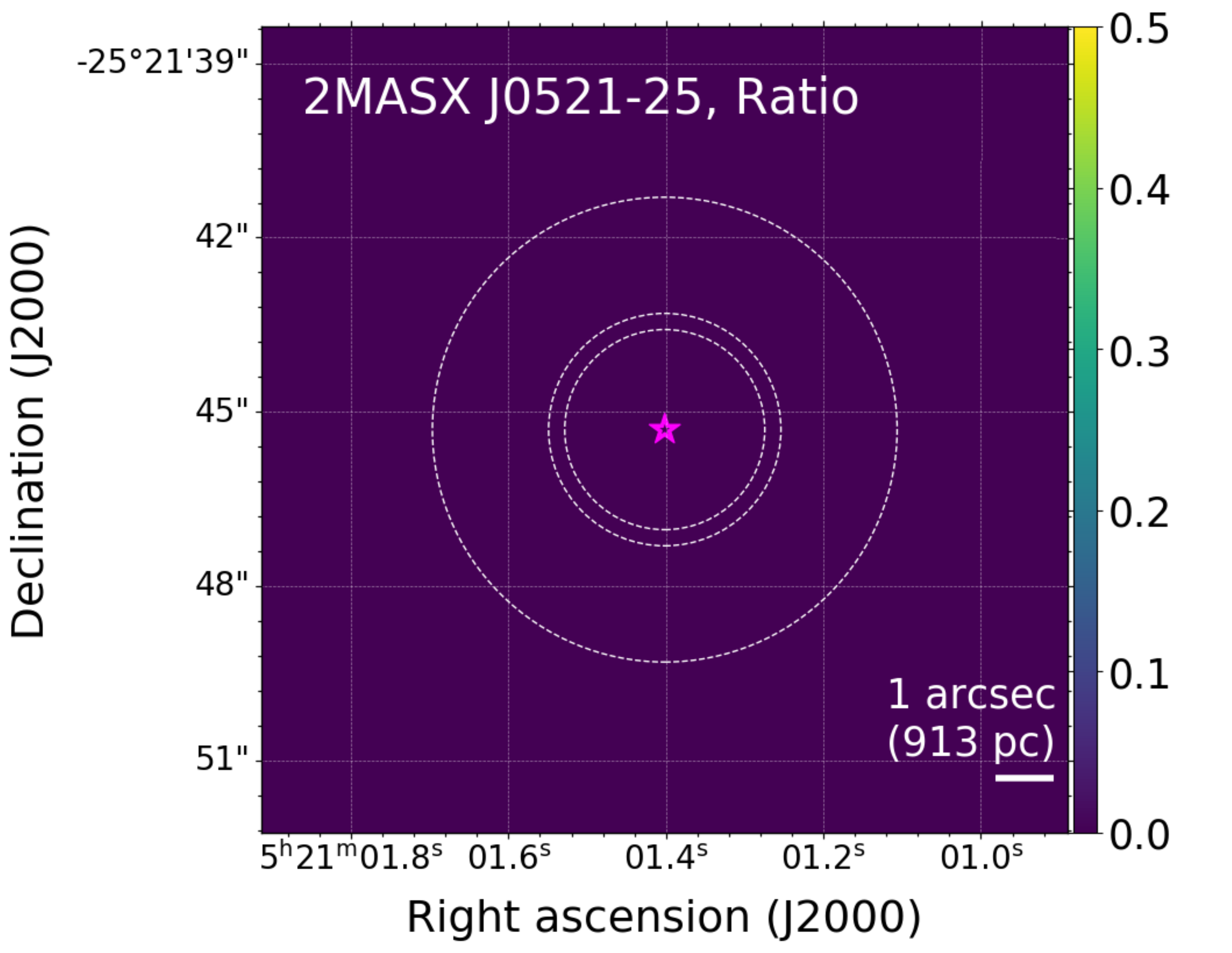}
    \caption{Continued.}
\end{figure*}

\begin{figure*}\addtocounter{figure}{-1}
    \centering
    \includegraphics[width=5.5cm]{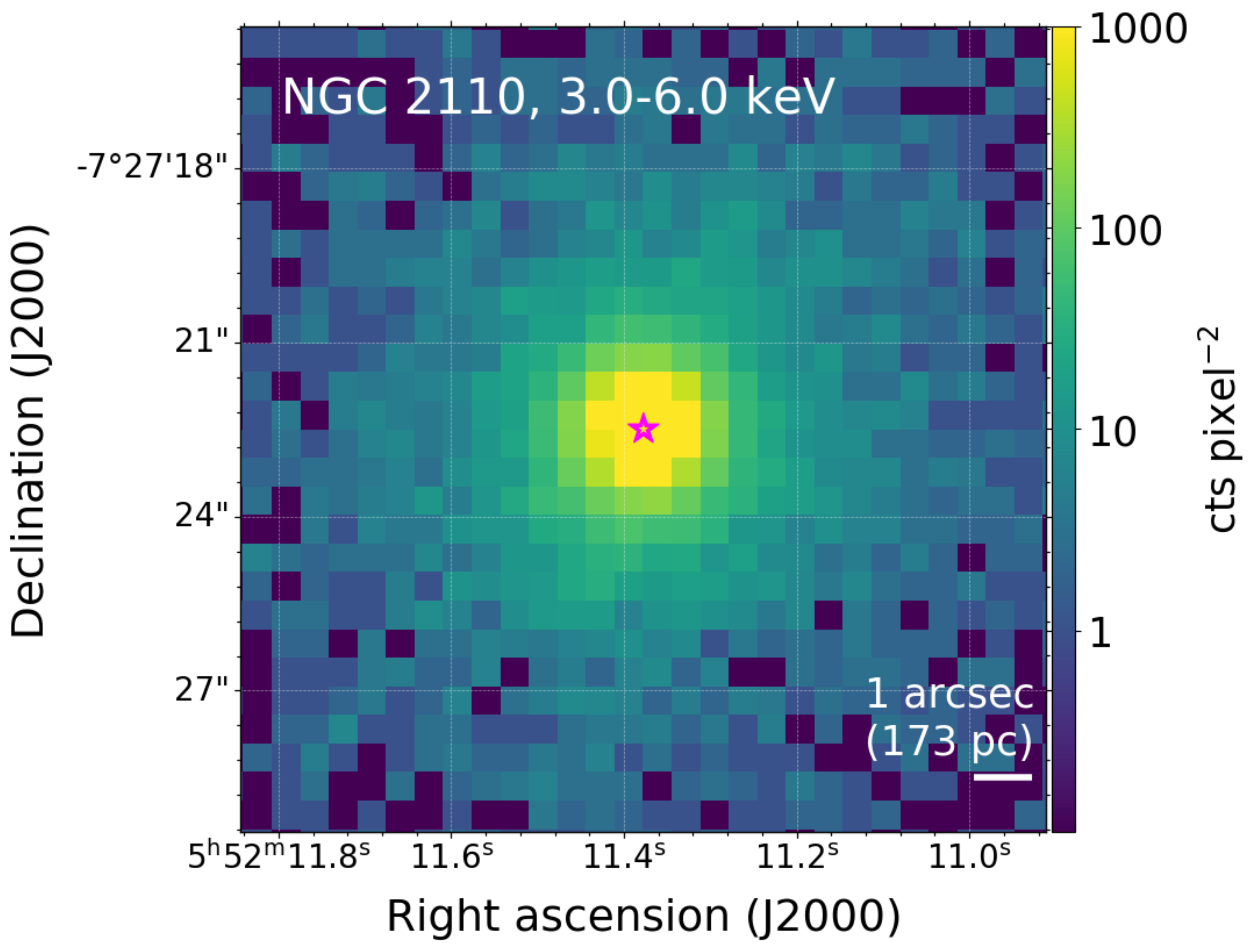}
    \includegraphics[width=5.5cm]{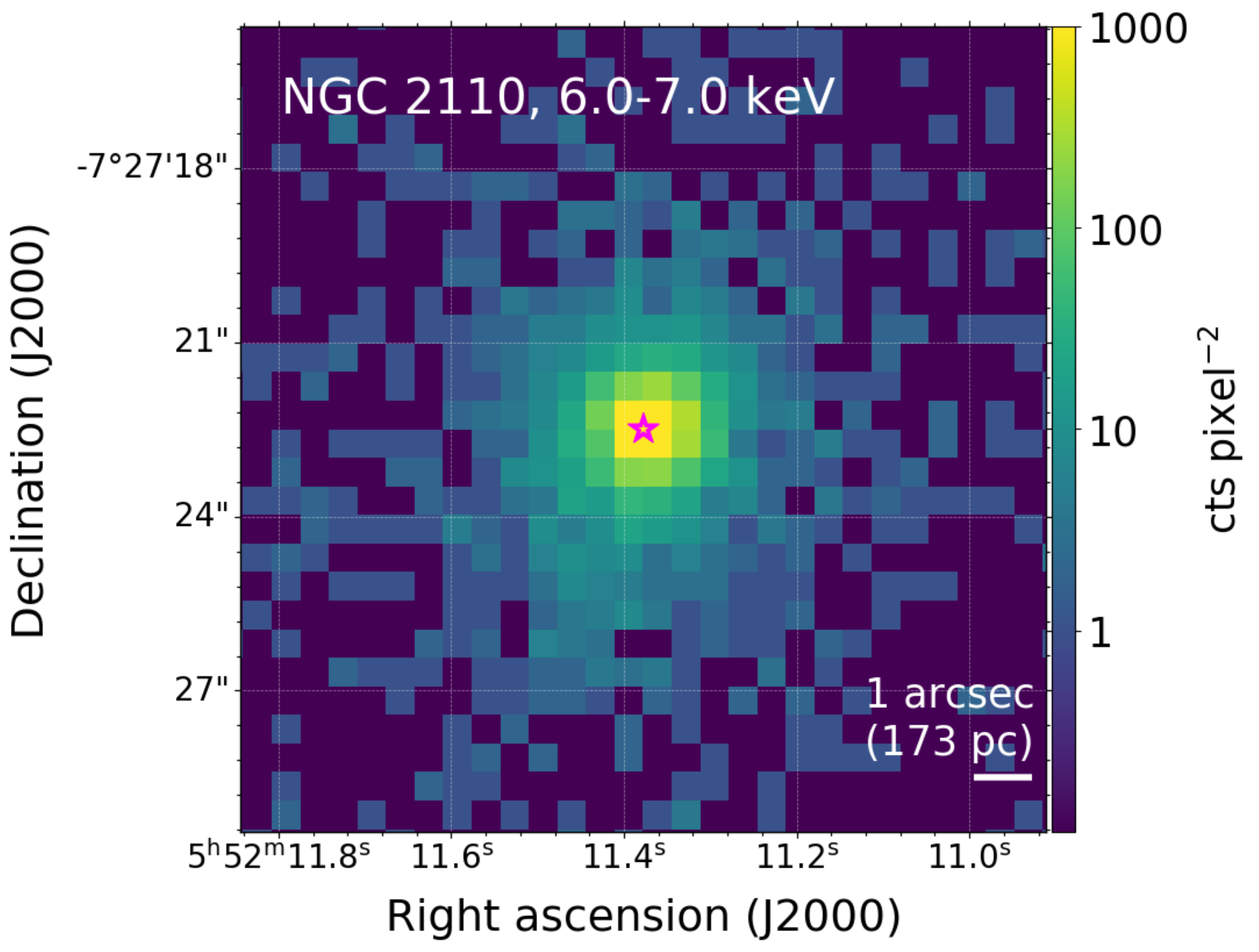}
    \includegraphics[width=5.35cm]{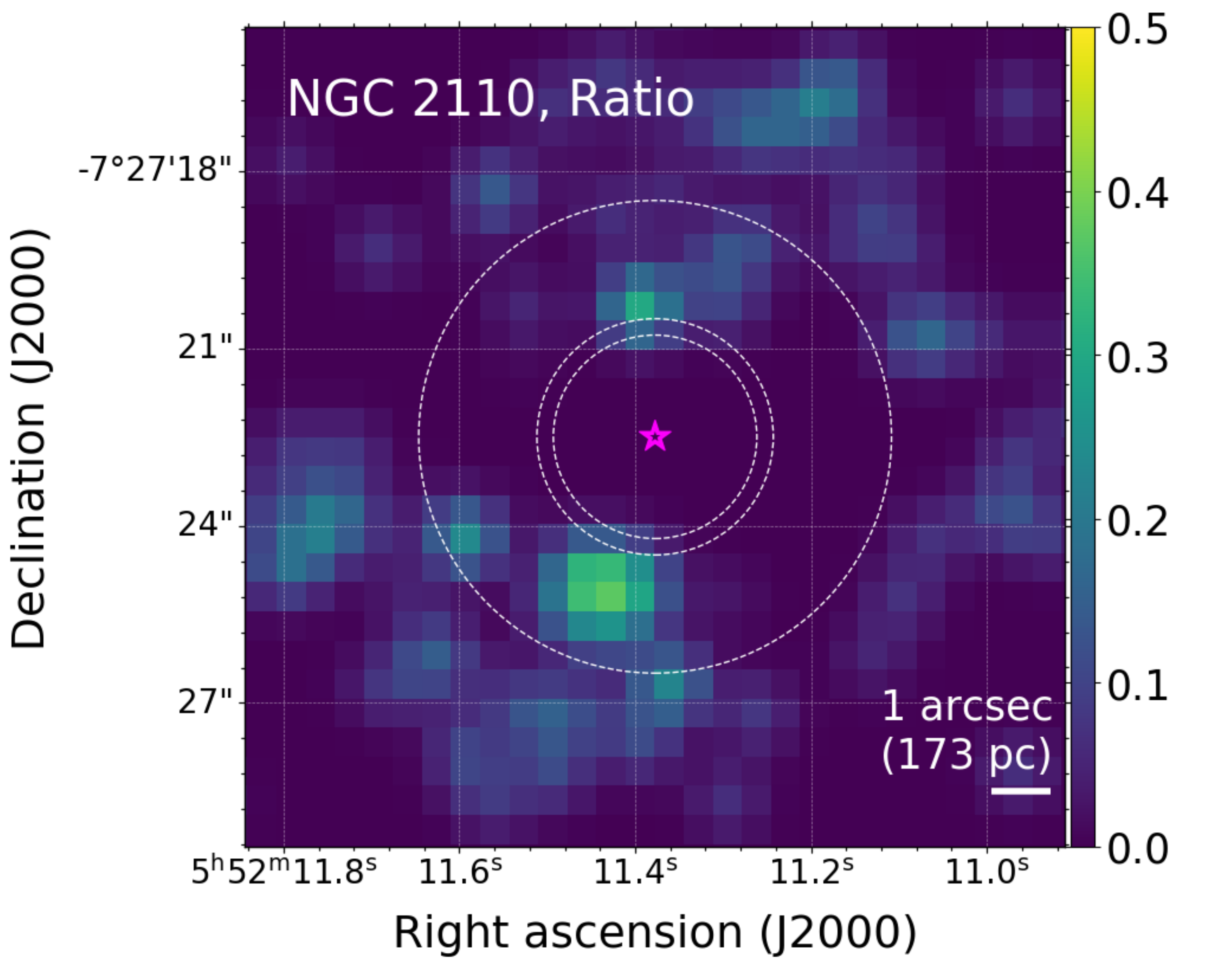}
    \\
    \includegraphics[width=5.5cm]{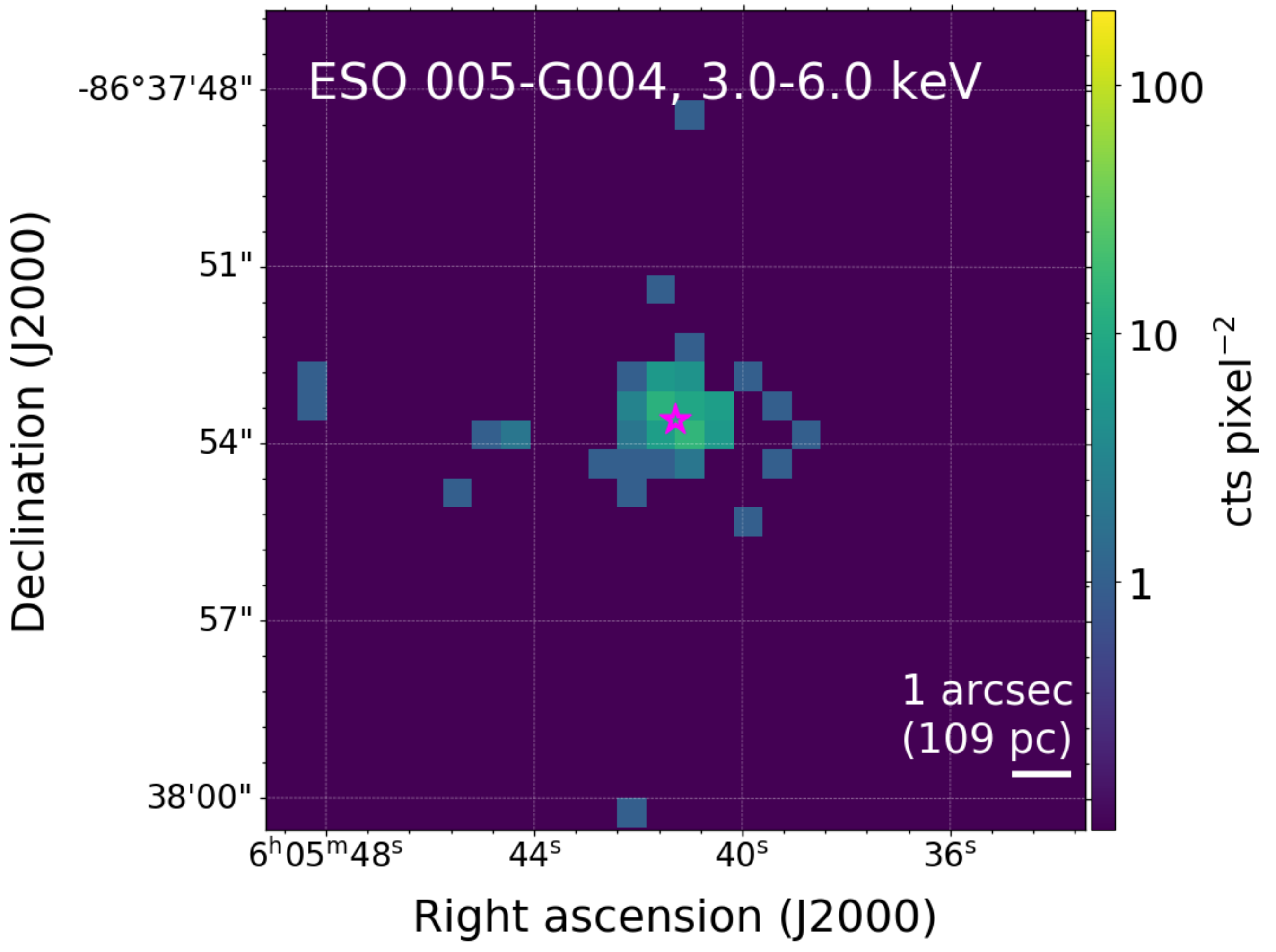}
    \includegraphics[width=5.5cm]{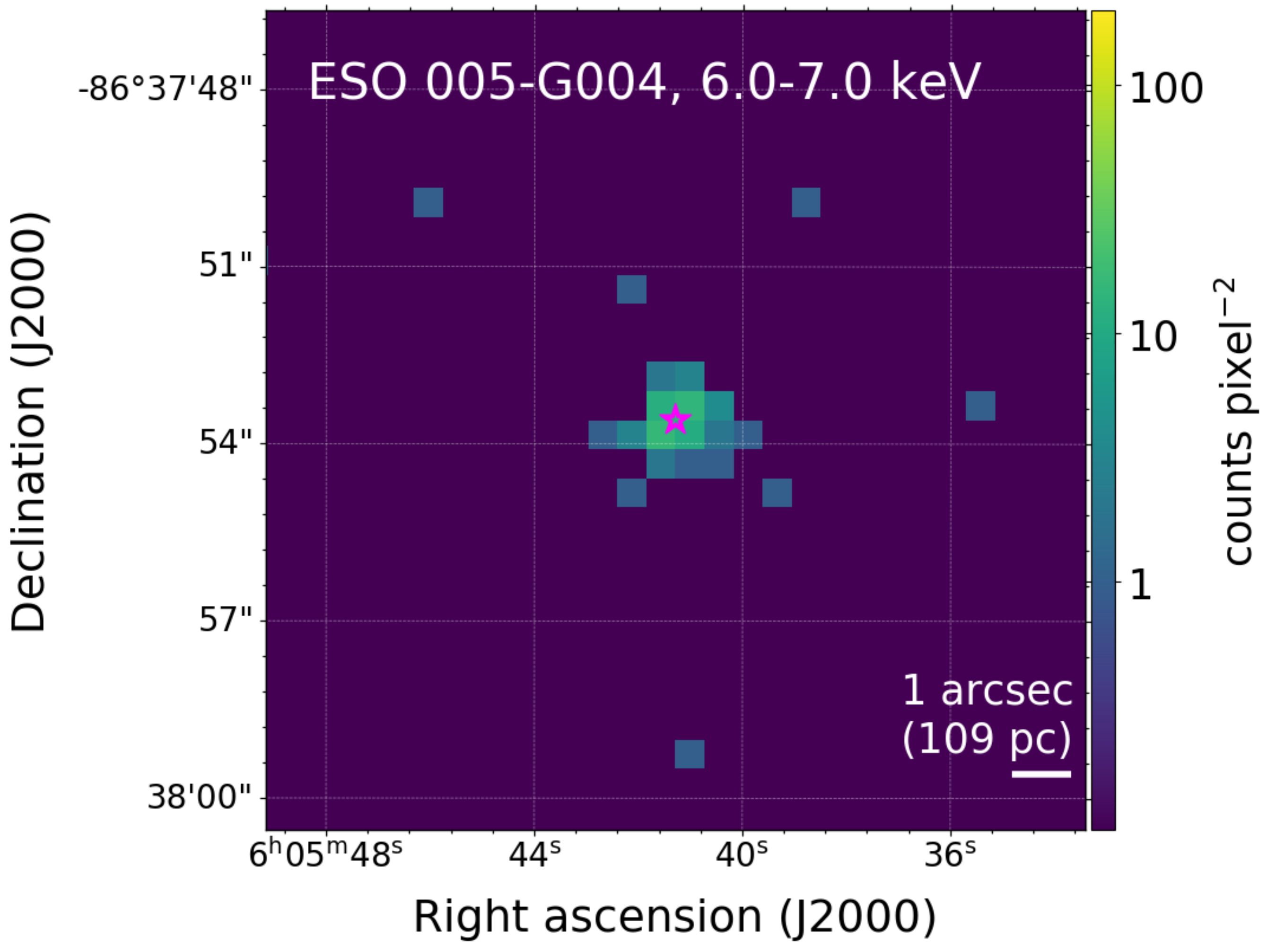}
    \includegraphics[width=5.35cm]{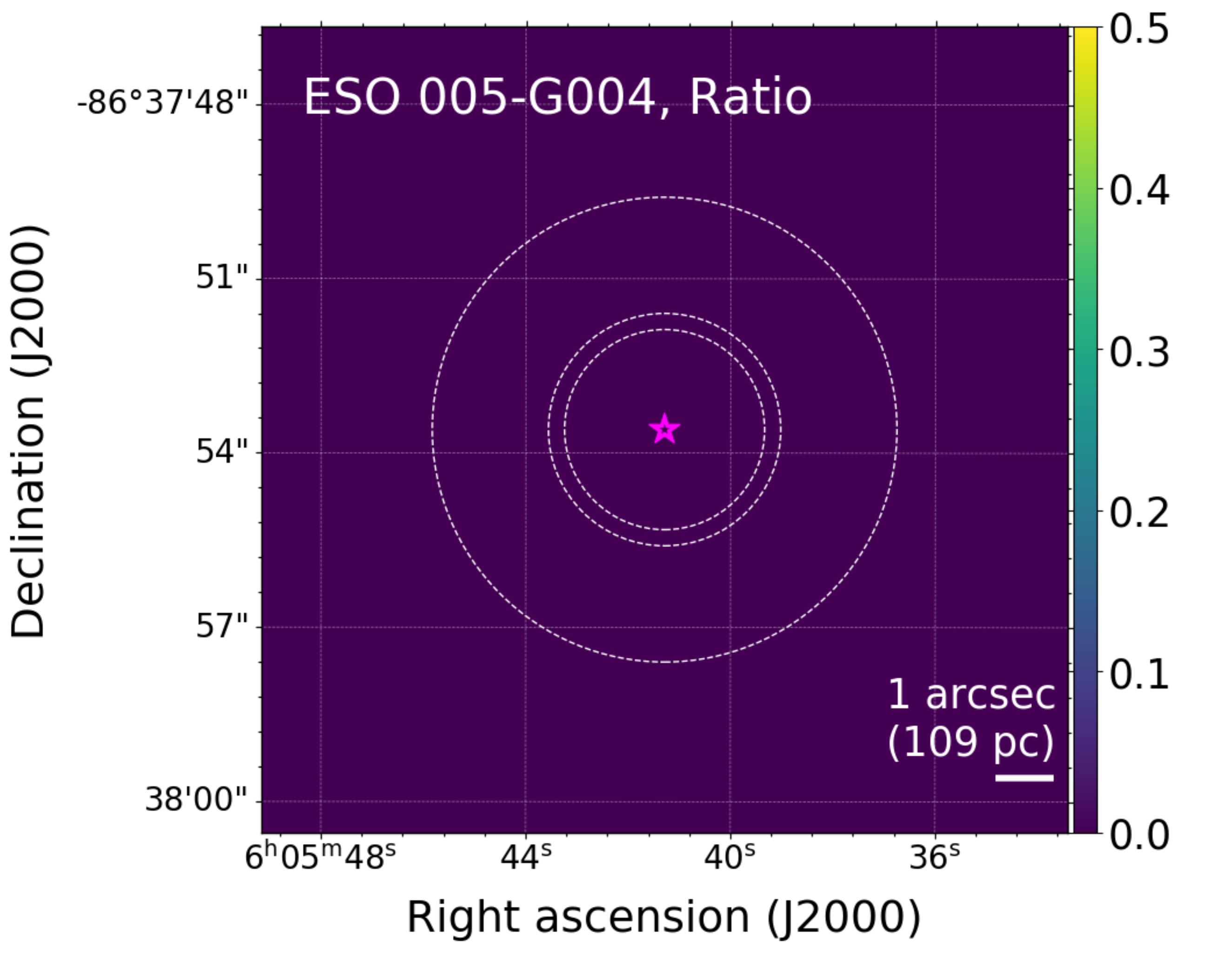}
    \\      
    \includegraphics[width=5.5cm]{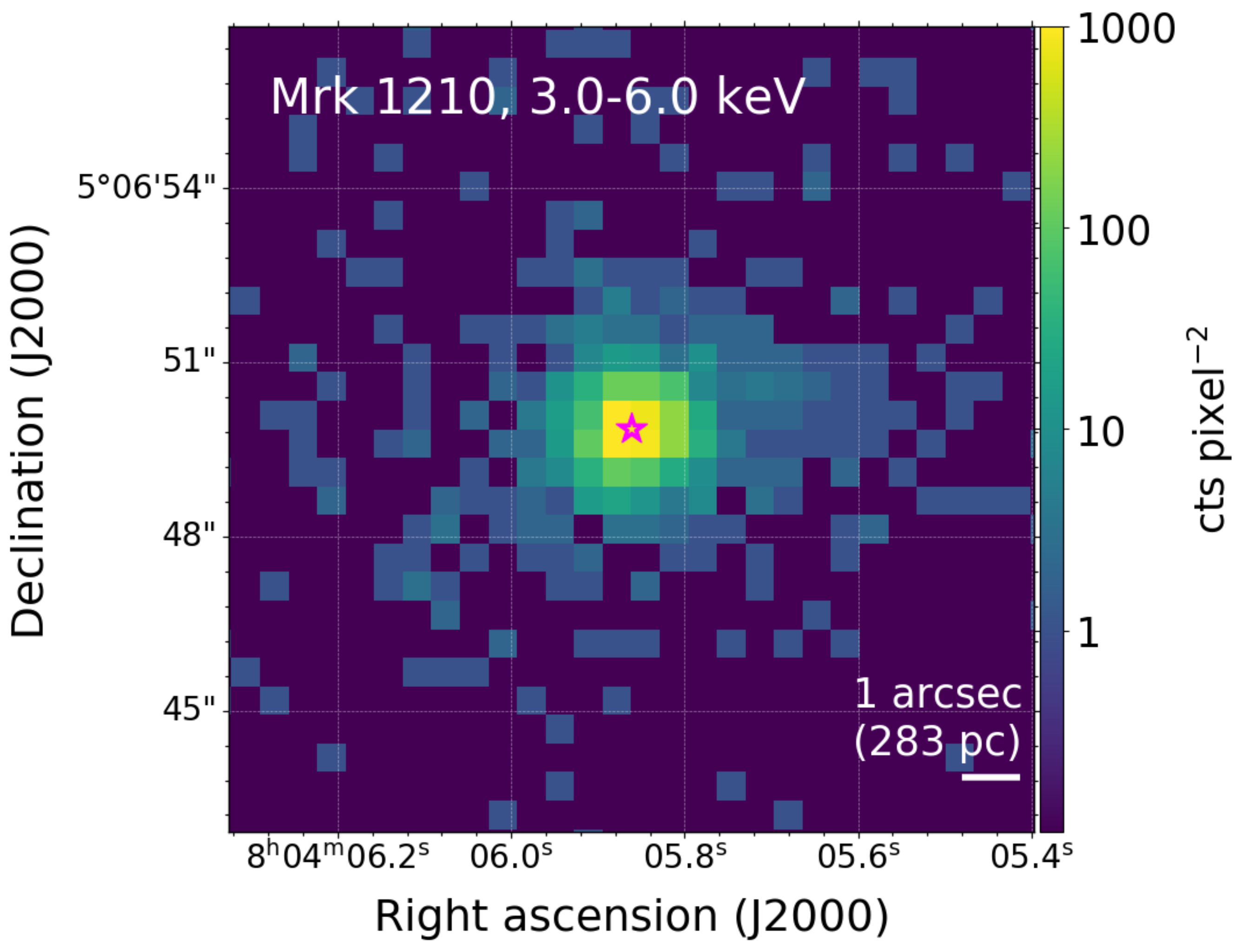}
    \includegraphics[width=5.5cm]{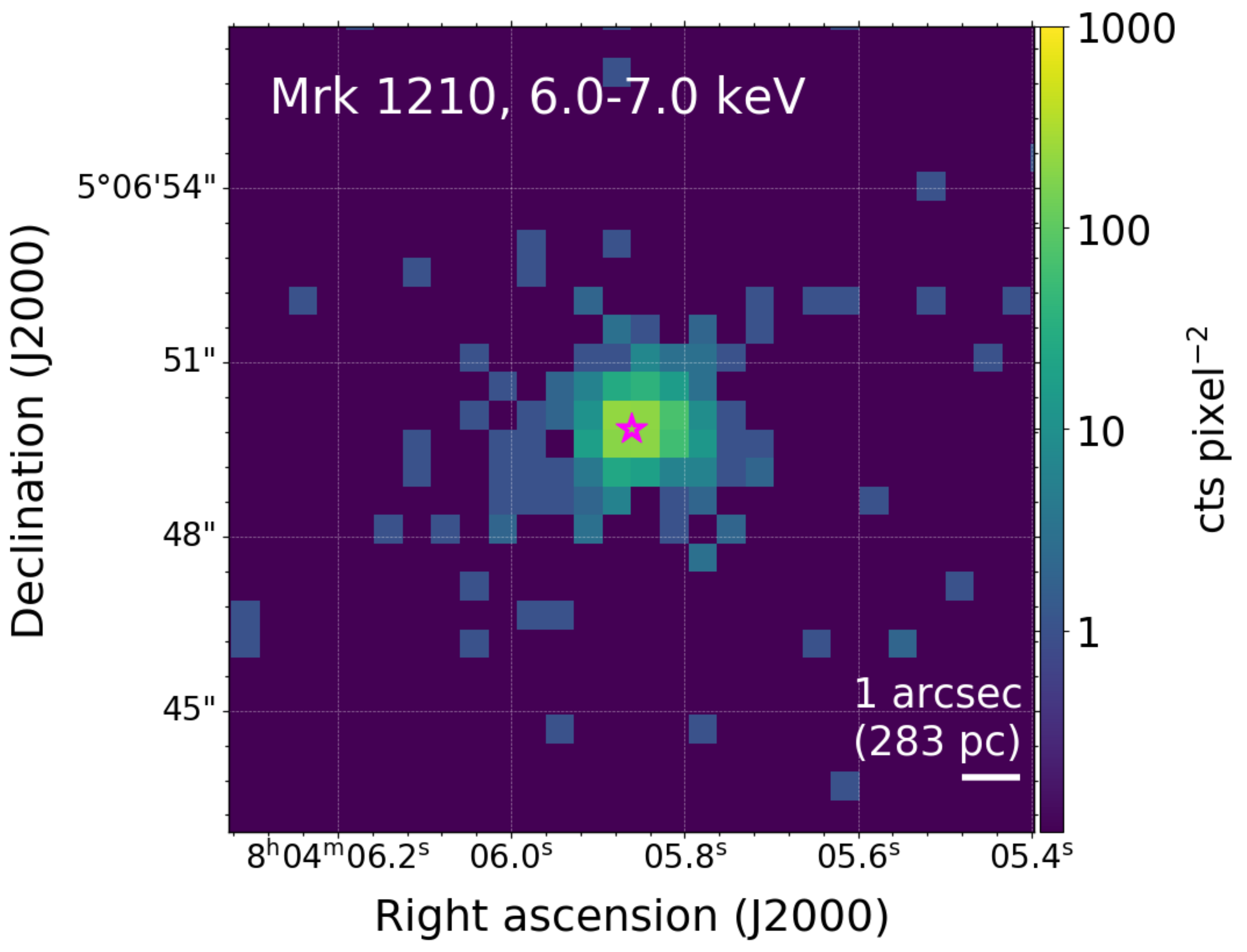}
    \includegraphics[width=5.35cm]{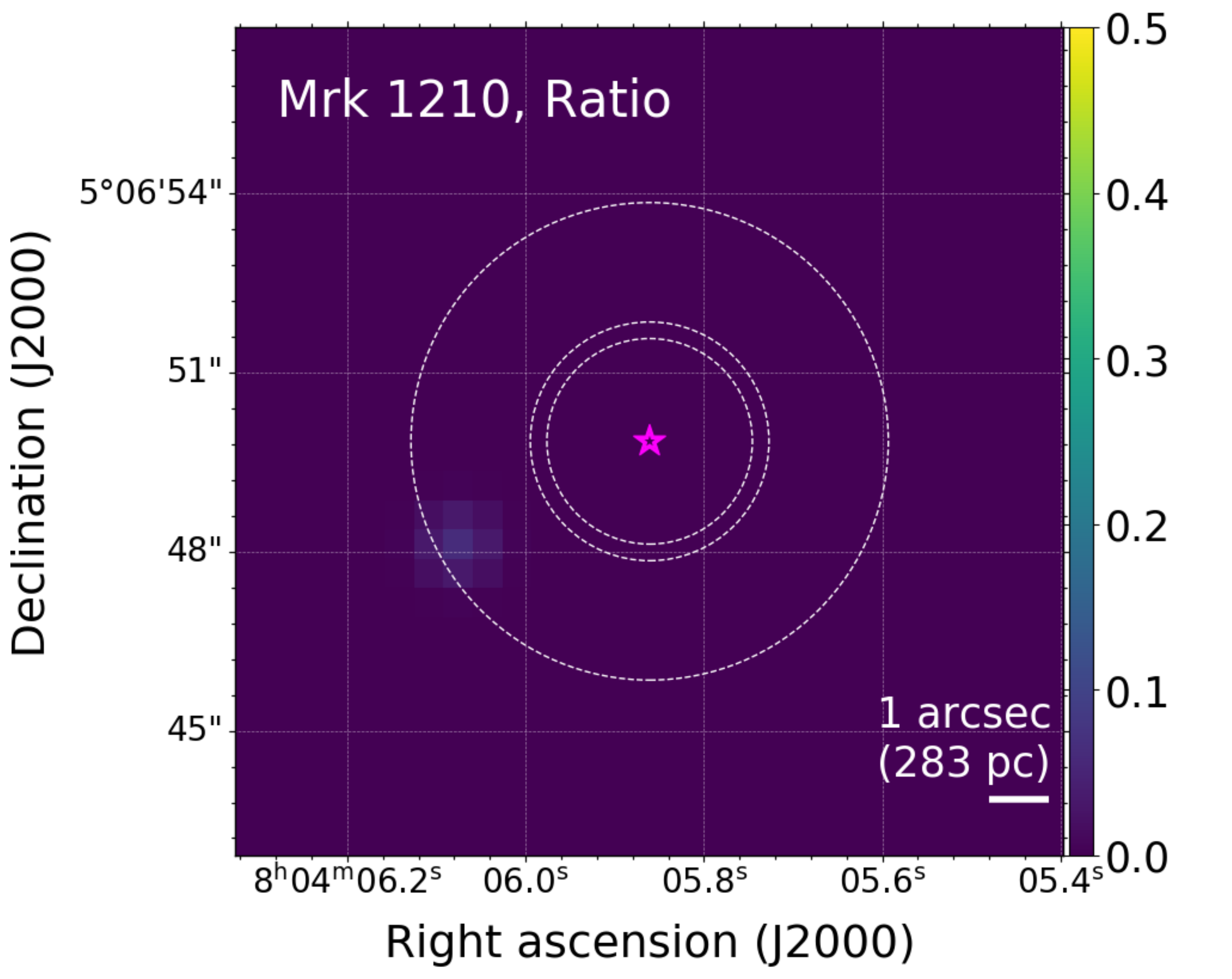}
    \\ 
    \includegraphics[width=5.5cm]{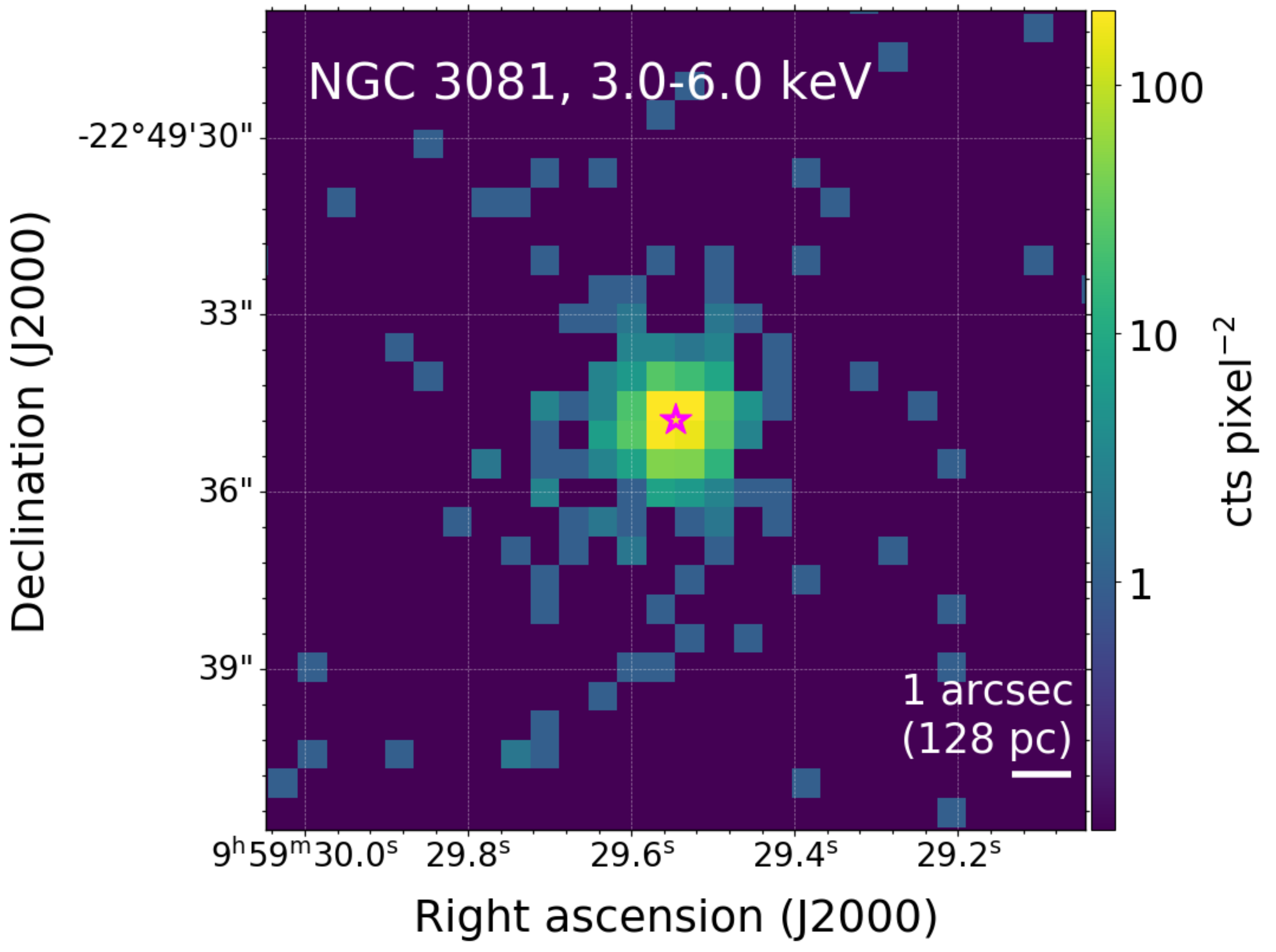}
    \includegraphics[width=5.5cm]{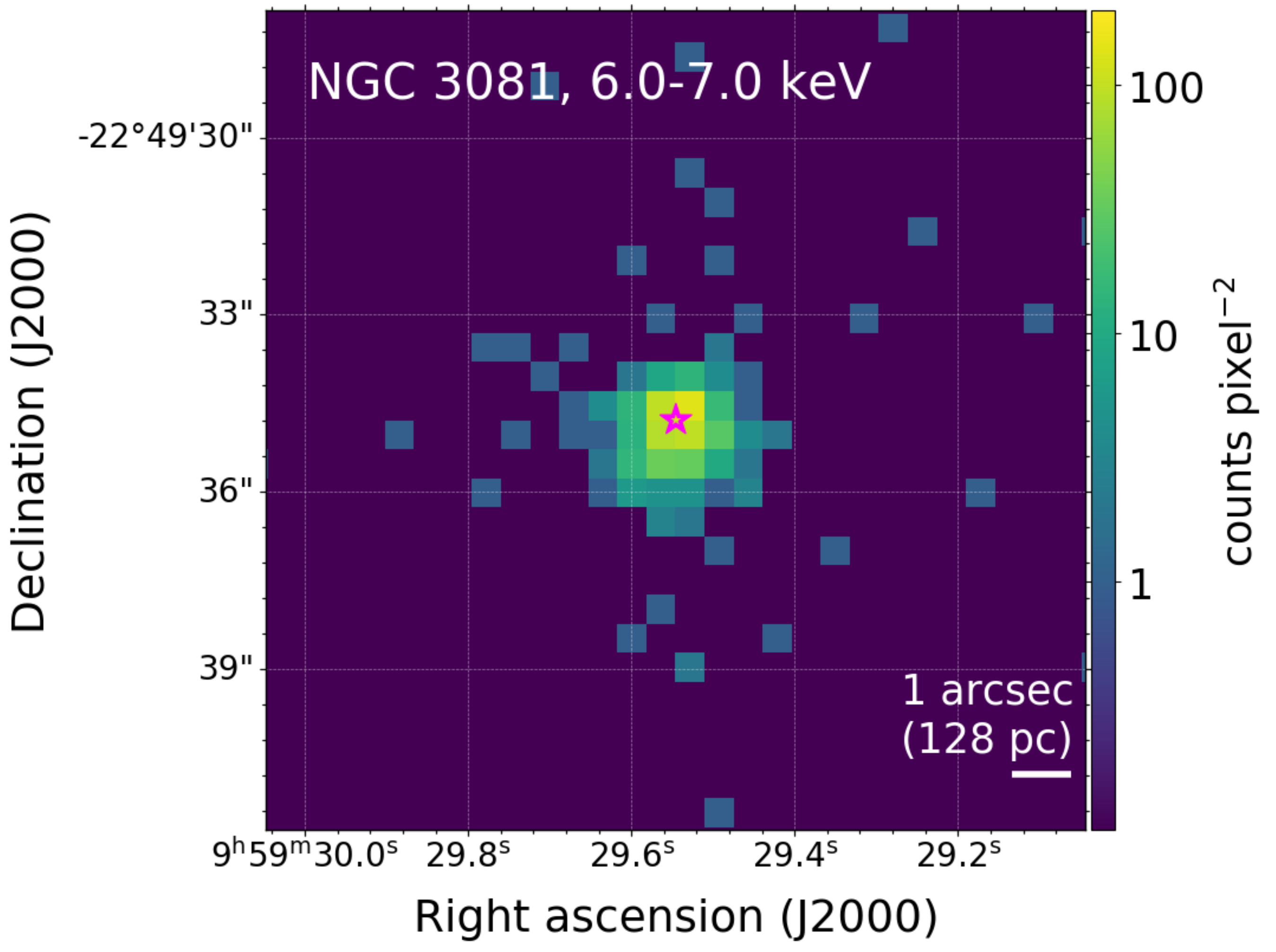}
    \includegraphics[width=5.35cm]{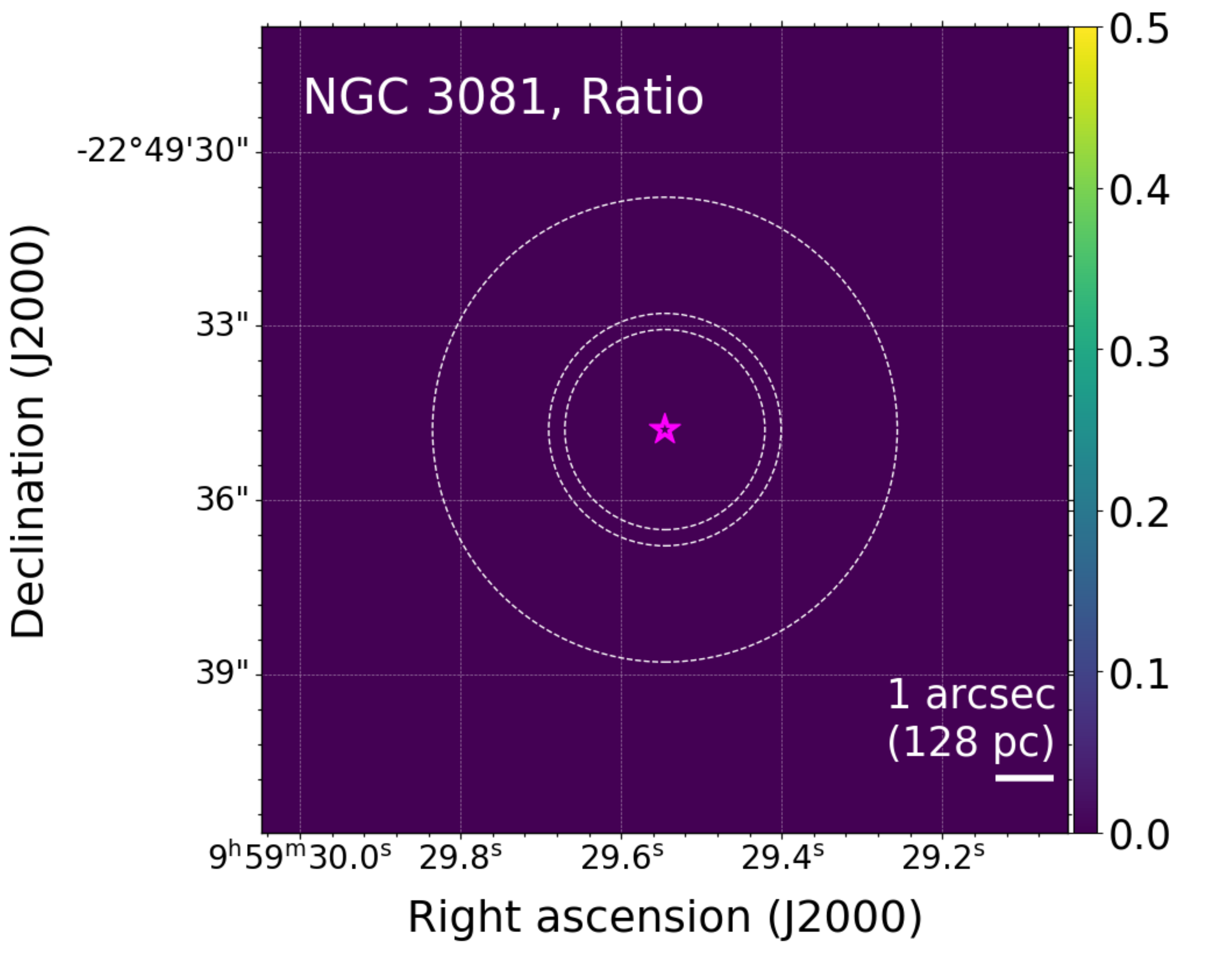} 
    \\
    \includegraphics[width=5.5cm]{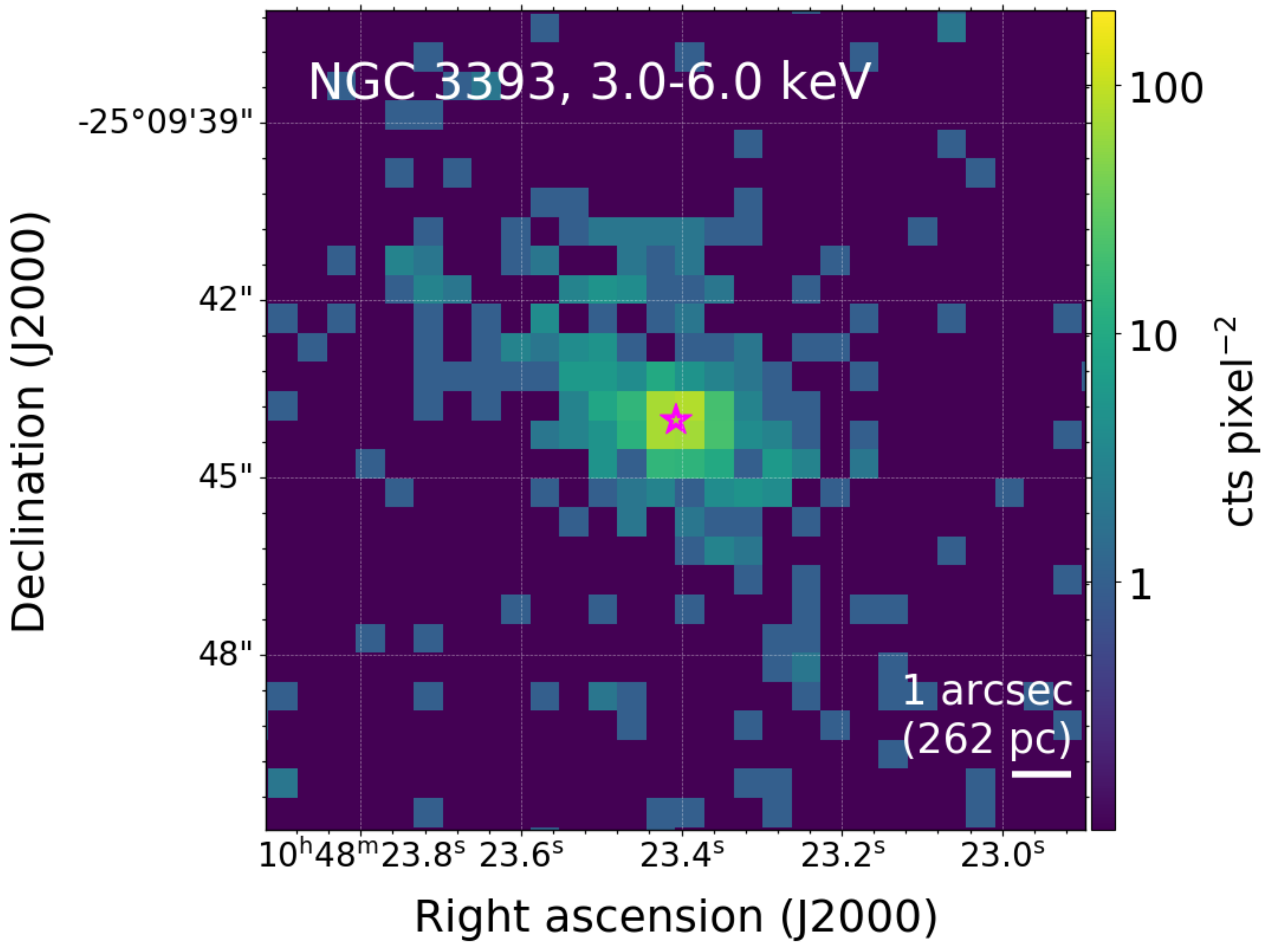}
    \includegraphics[width=5.5cm]{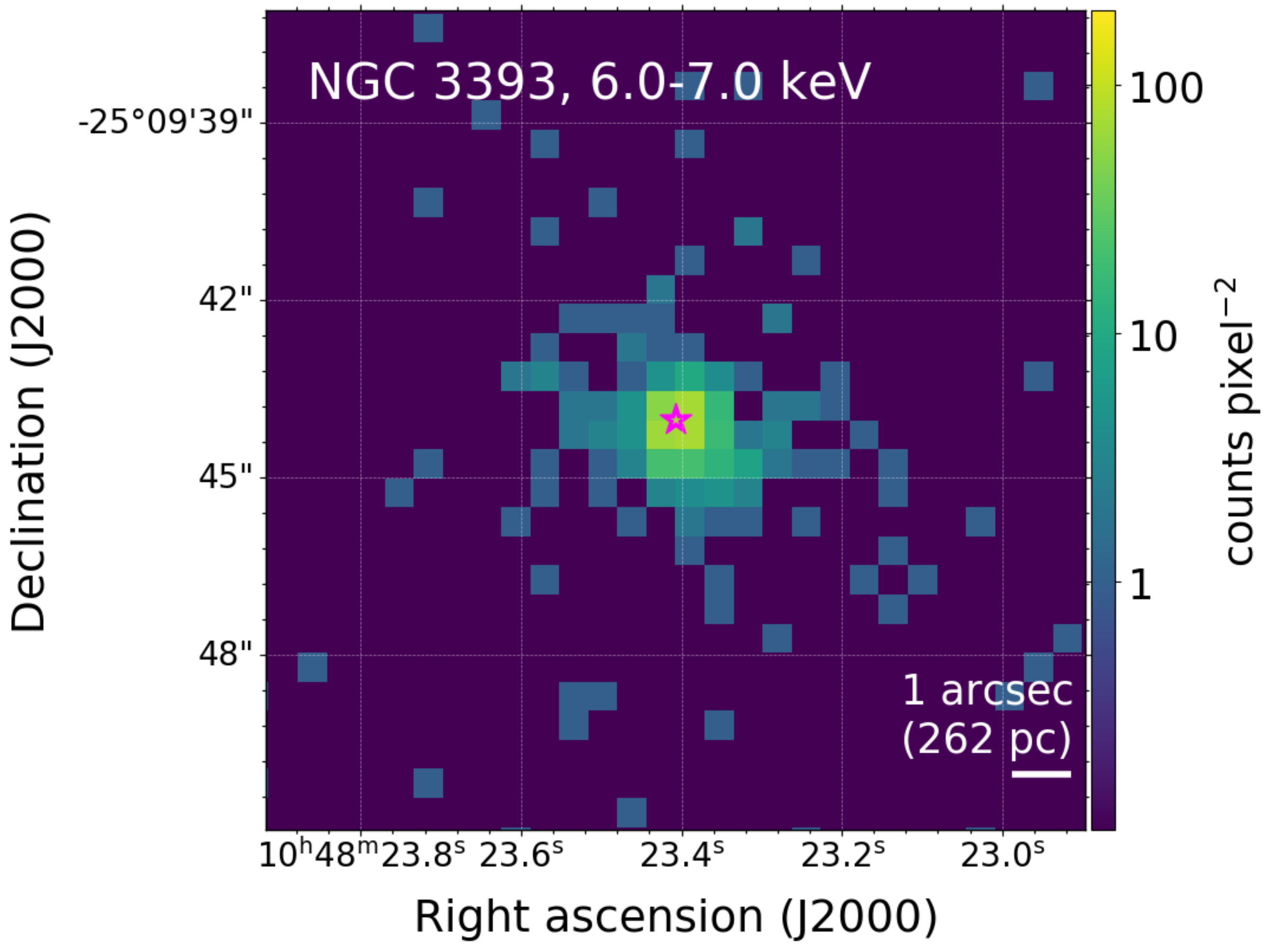}
    \includegraphics[width=5.35cm]{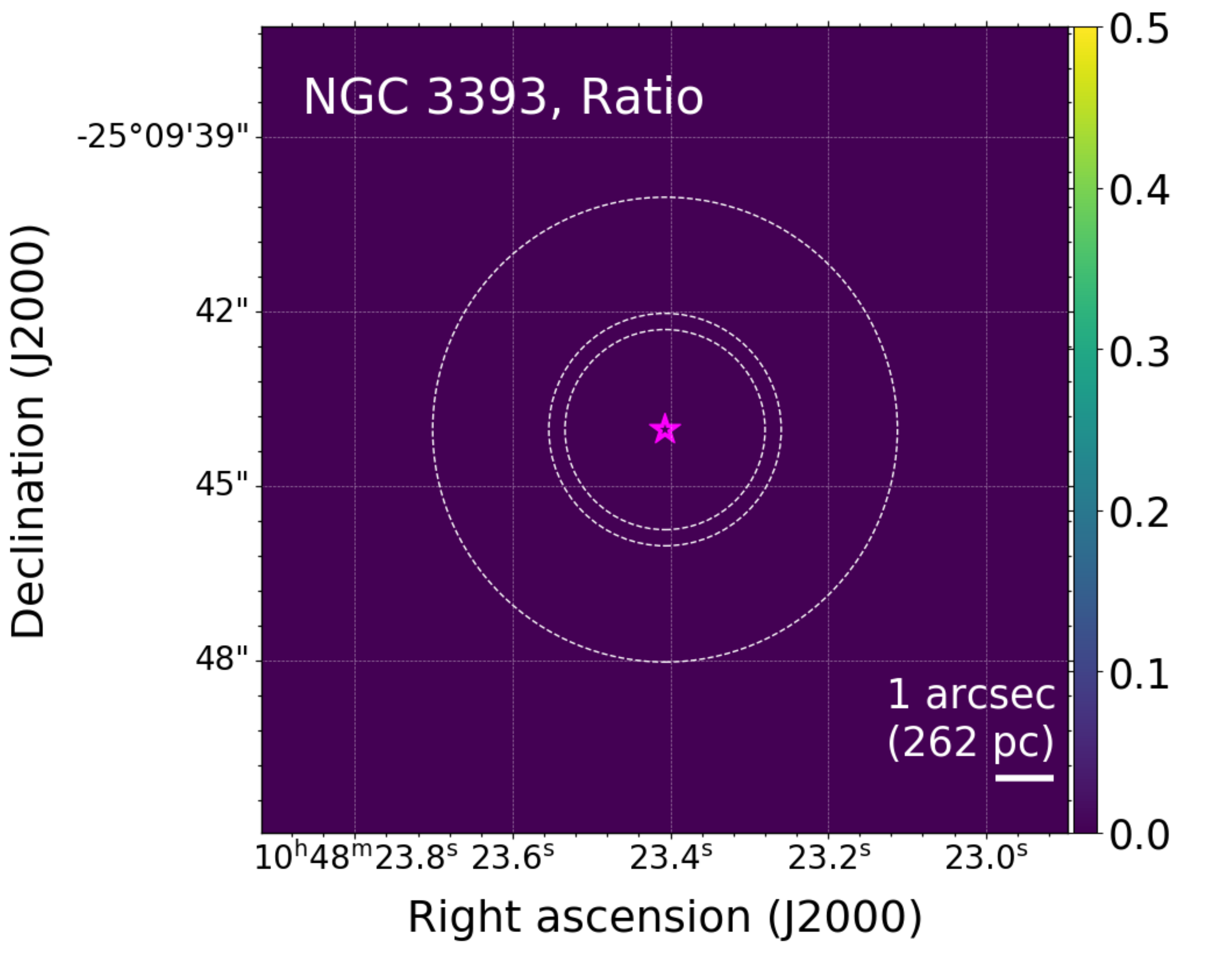}
    \caption{Continued.}
\end{figure*}

\begin{figure*}\addtocounter{figure}{-1}
    \centering
    \includegraphics[width=5.5cm]{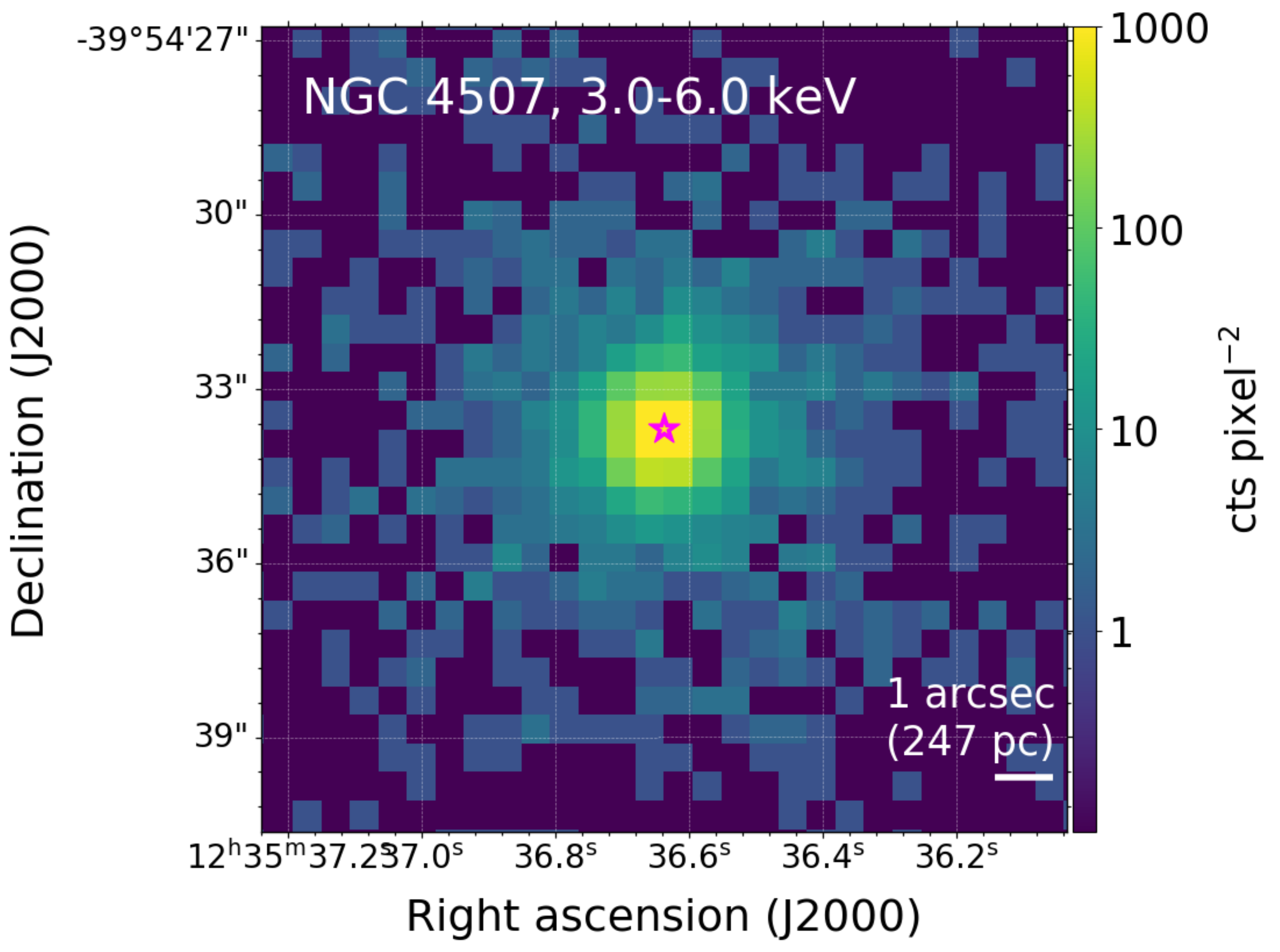}
    \includegraphics[width=5.5cm]{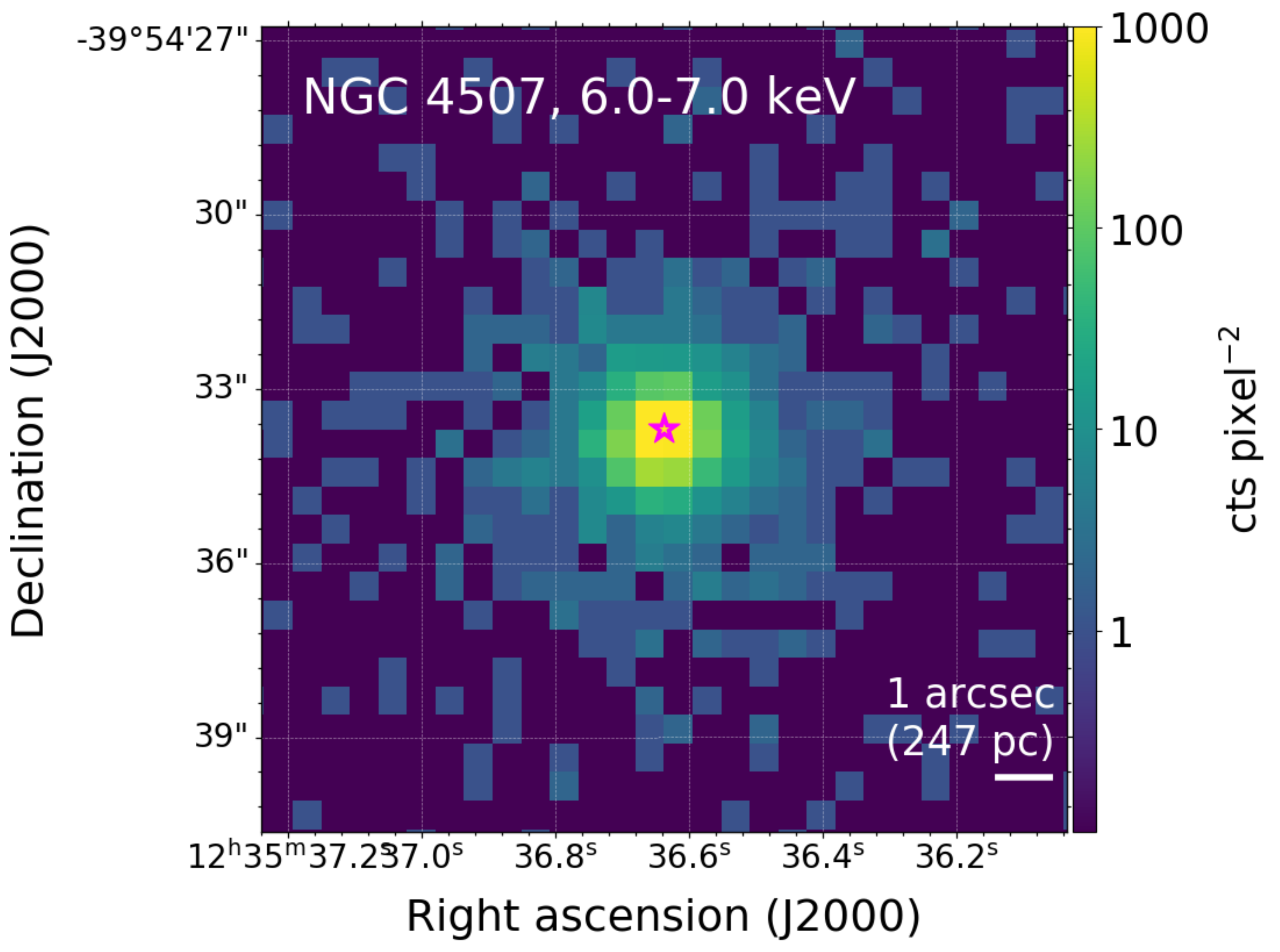}
    \includegraphics[width=5.35cm]{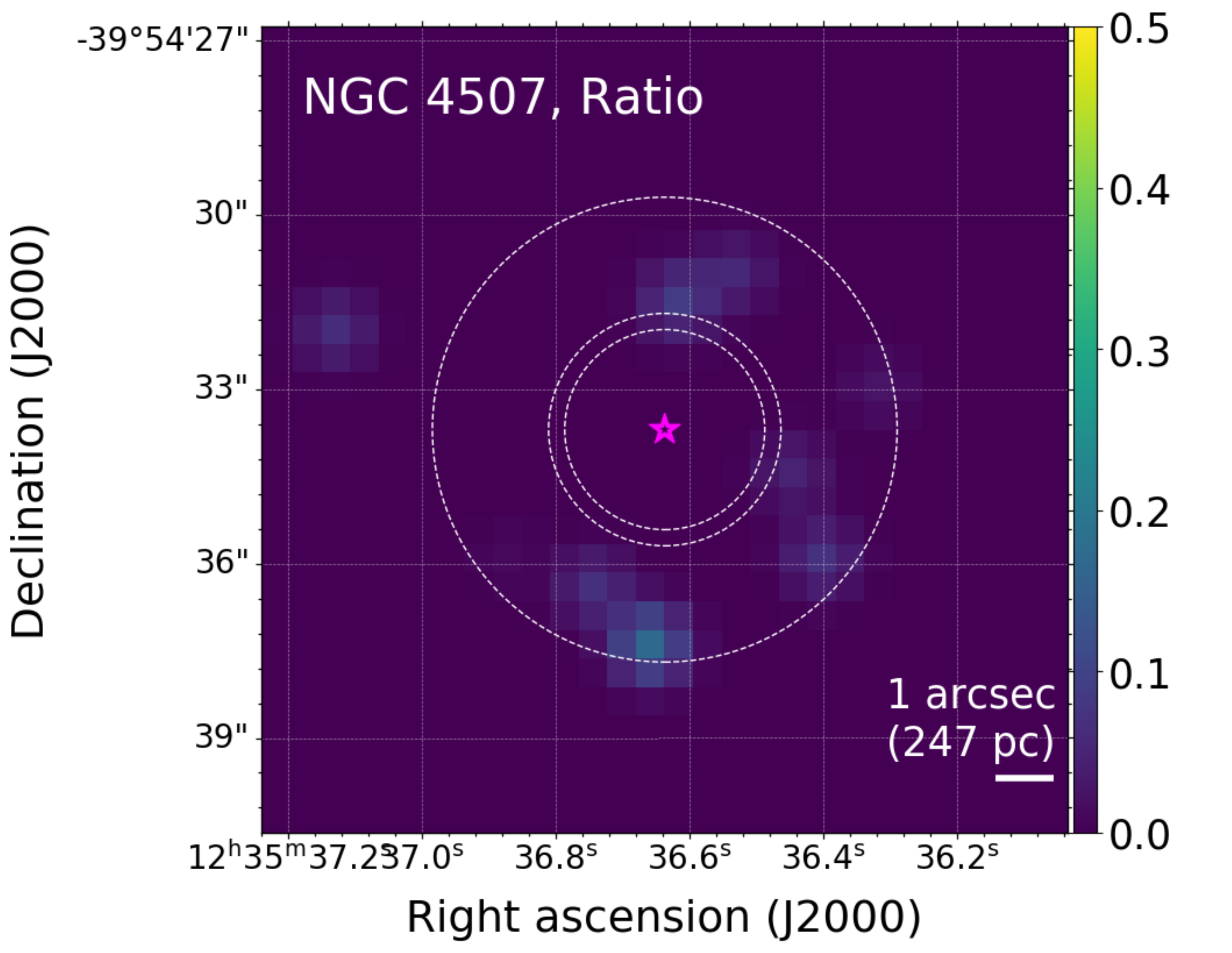}
    \\
    \includegraphics[width=5.5cm]{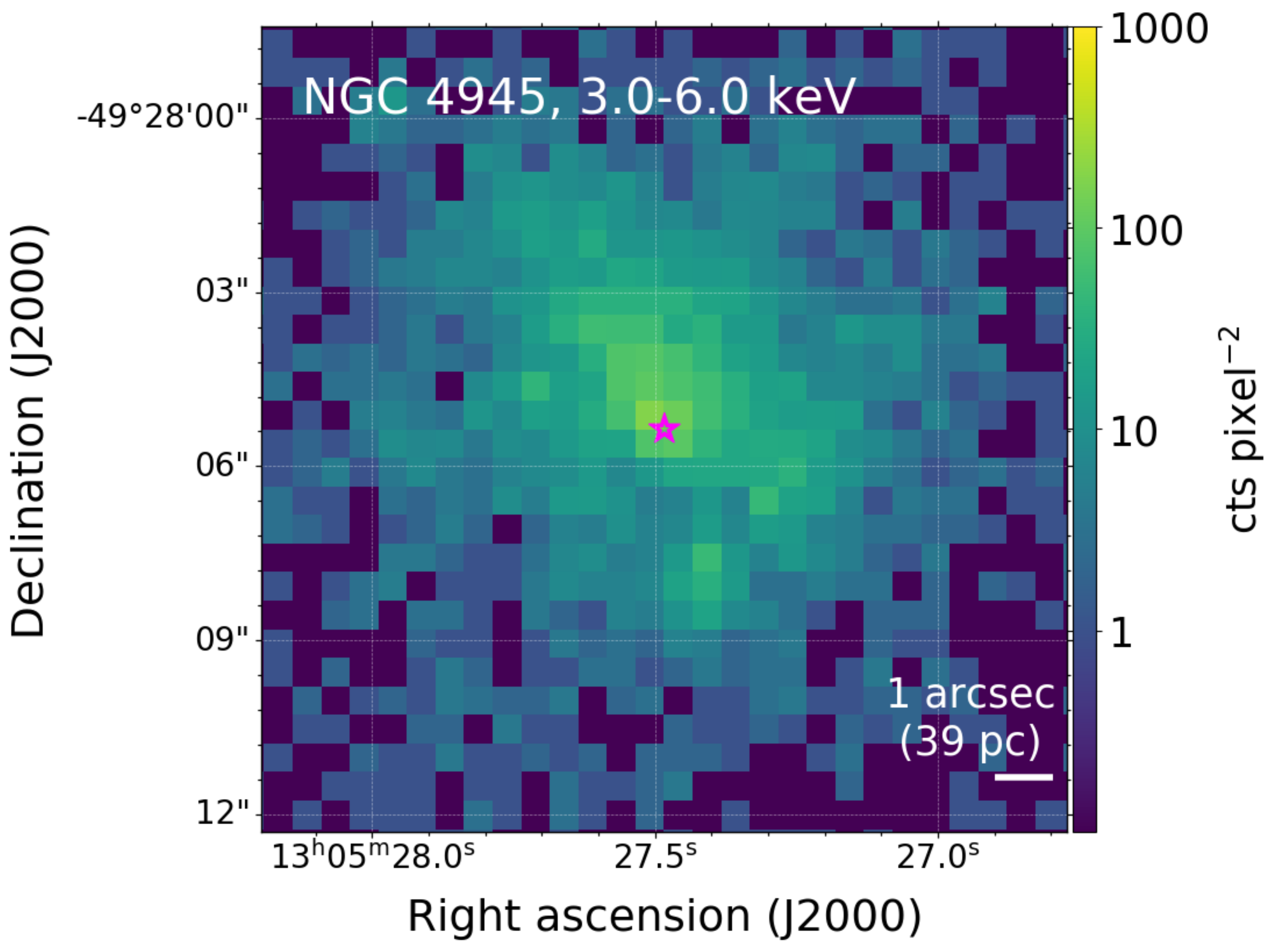}
    \includegraphics[width=5.5cm]{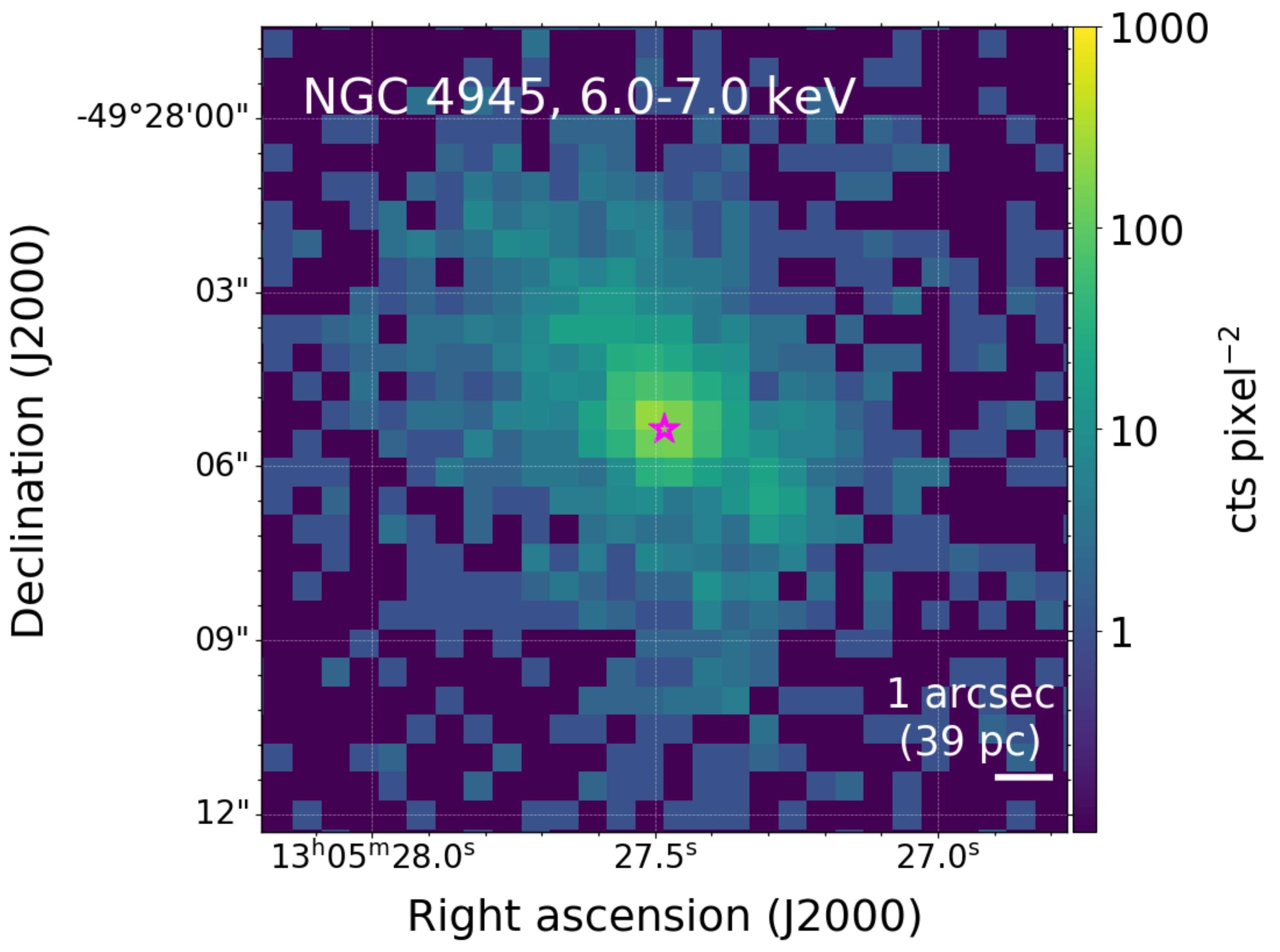}
    \includegraphics[width=5.35cm]{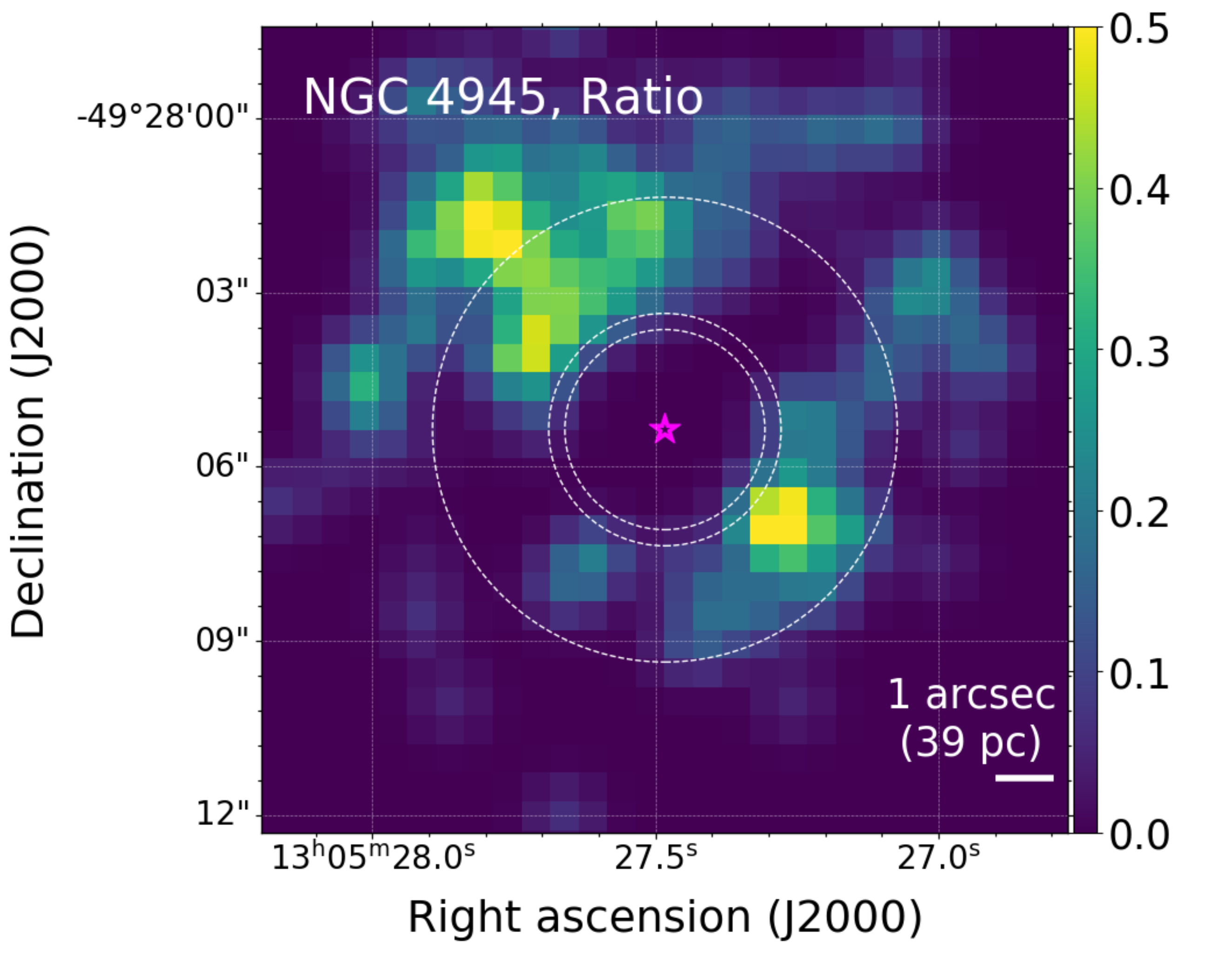}
    \\      
    \includegraphics[width=5.5cm]{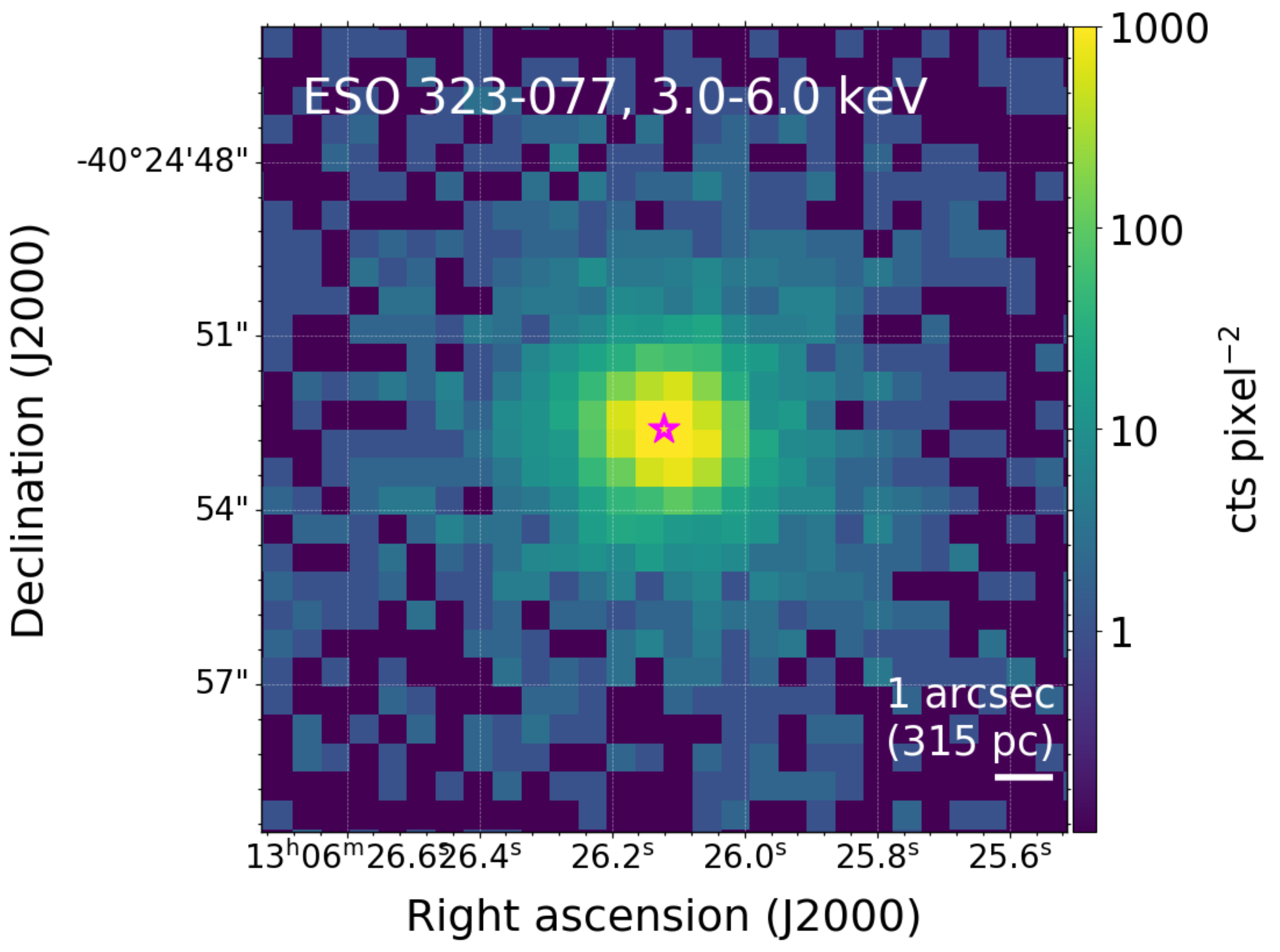}
    \includegraphics[width=5.5cm]{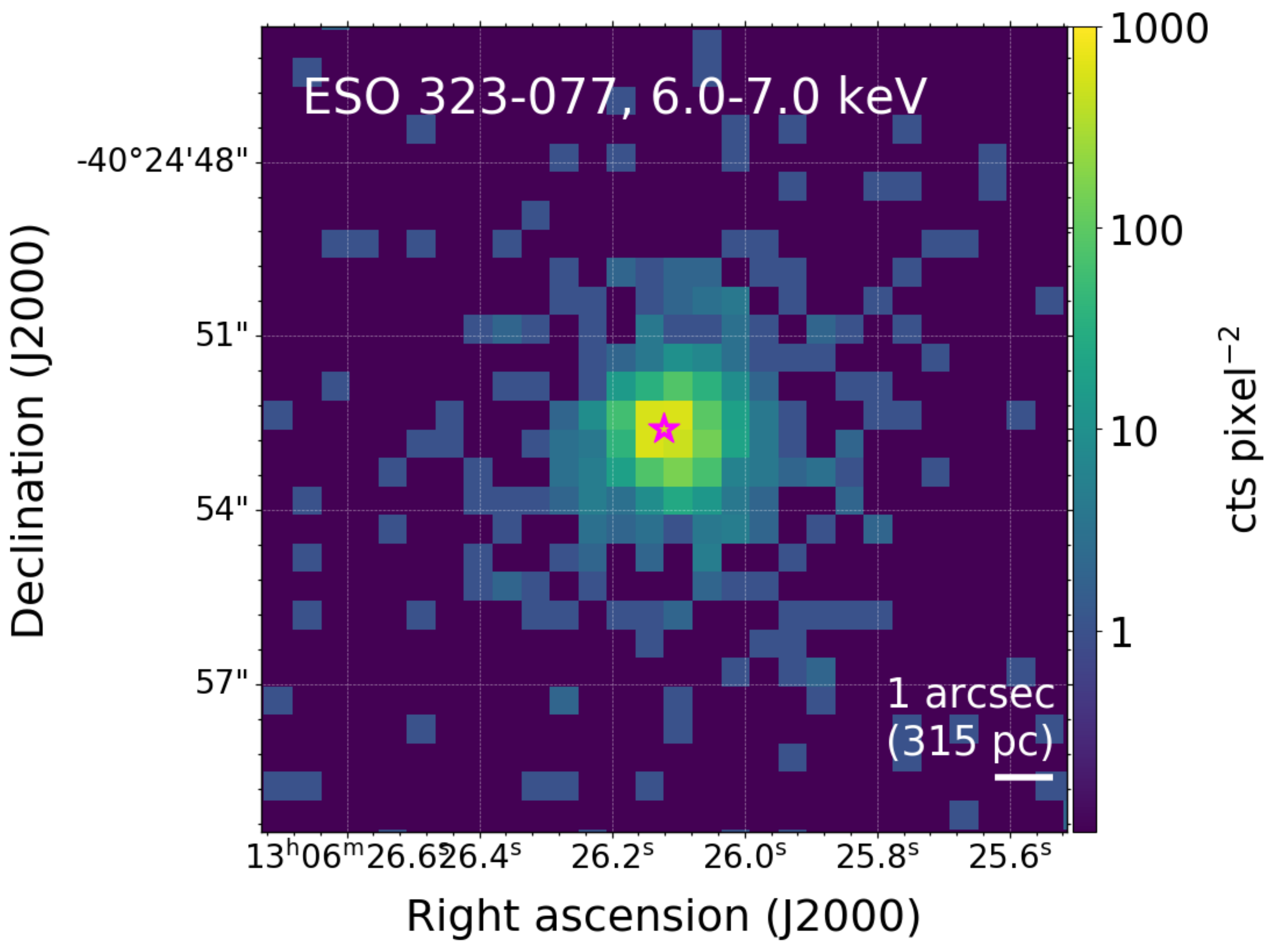}
    \includegraphics[width=5.35cm]{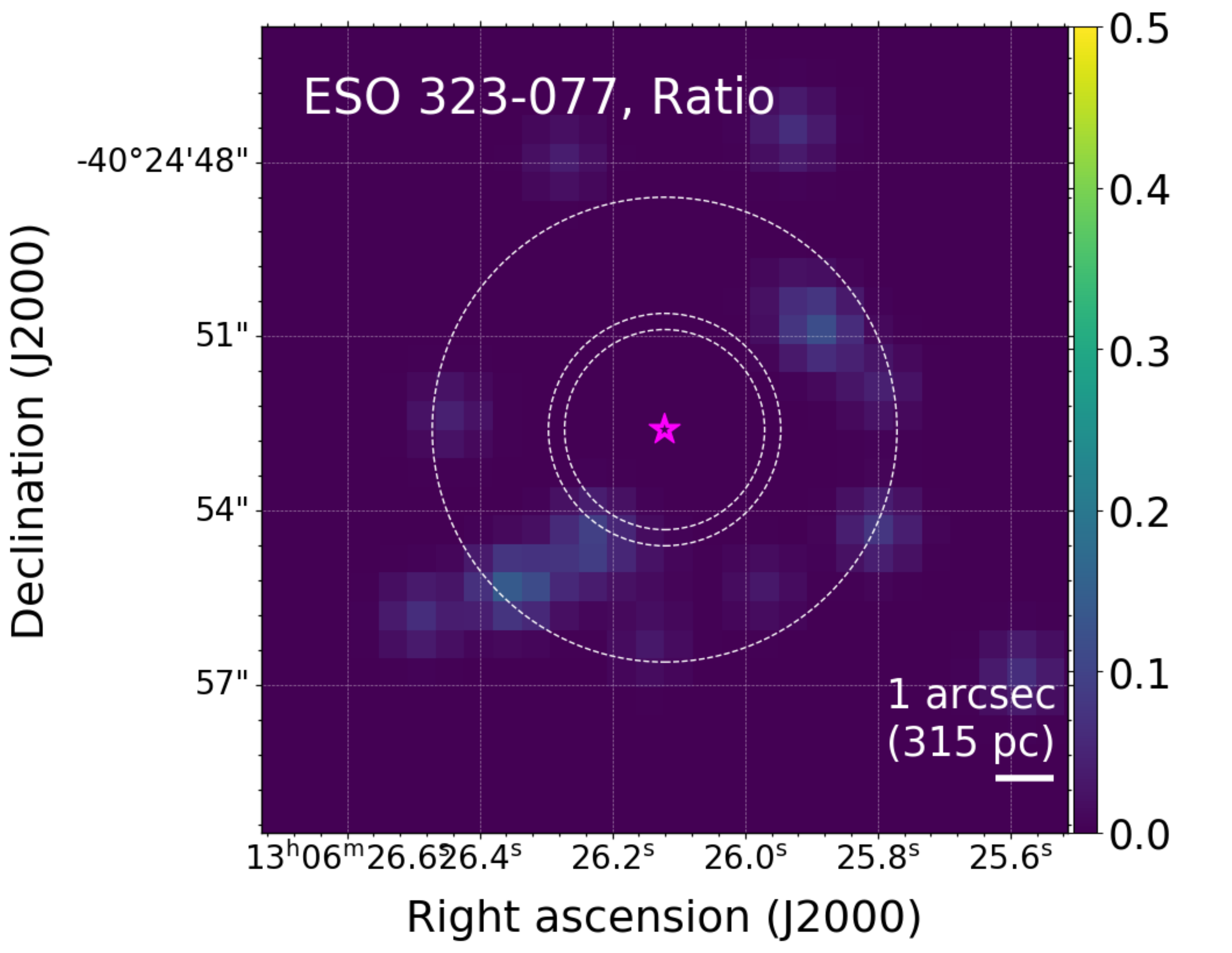}
    \\ 
    \includegraphics[width=5.5cm]{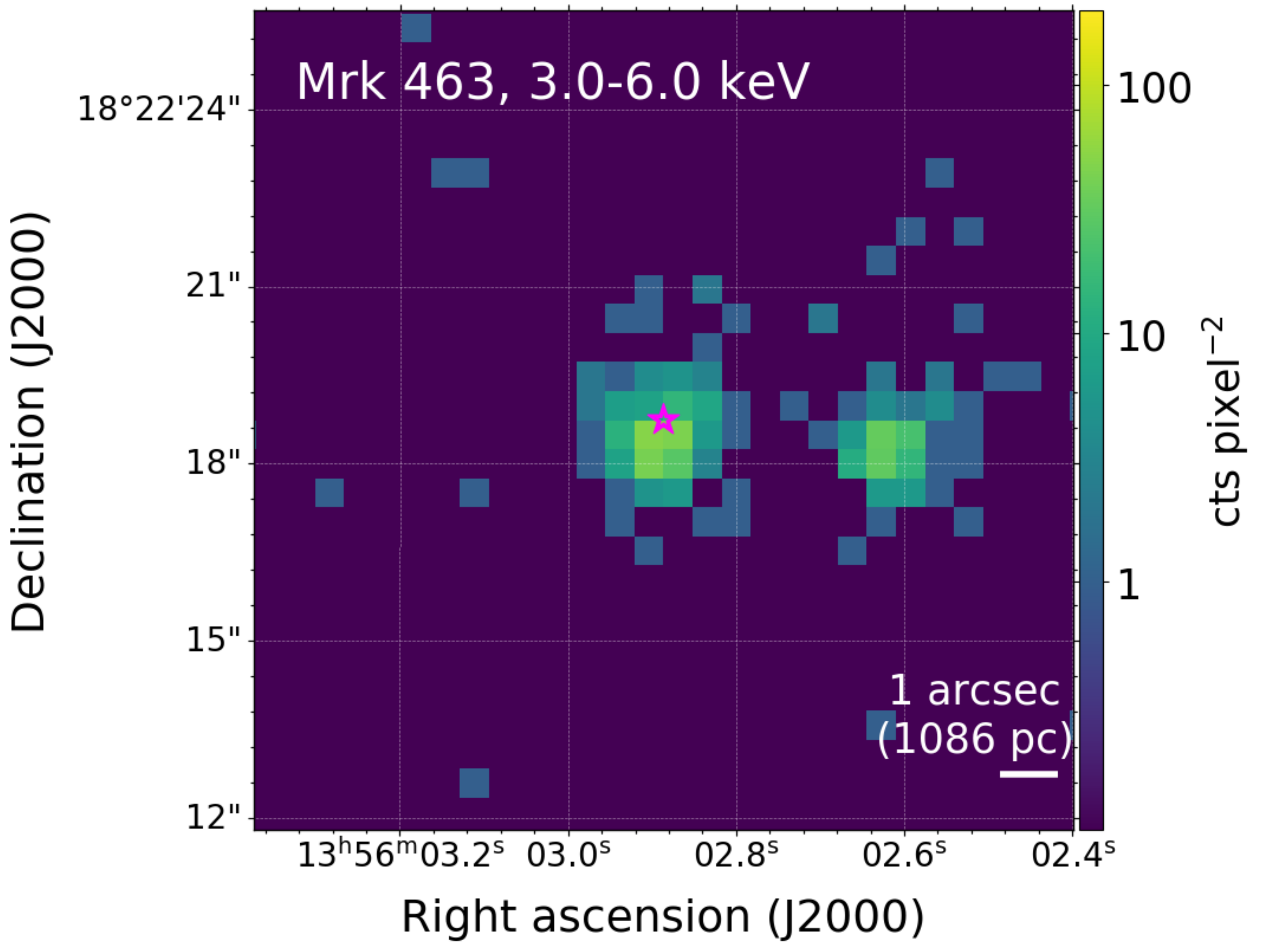}
    \includegraphics[width=5.5cm]{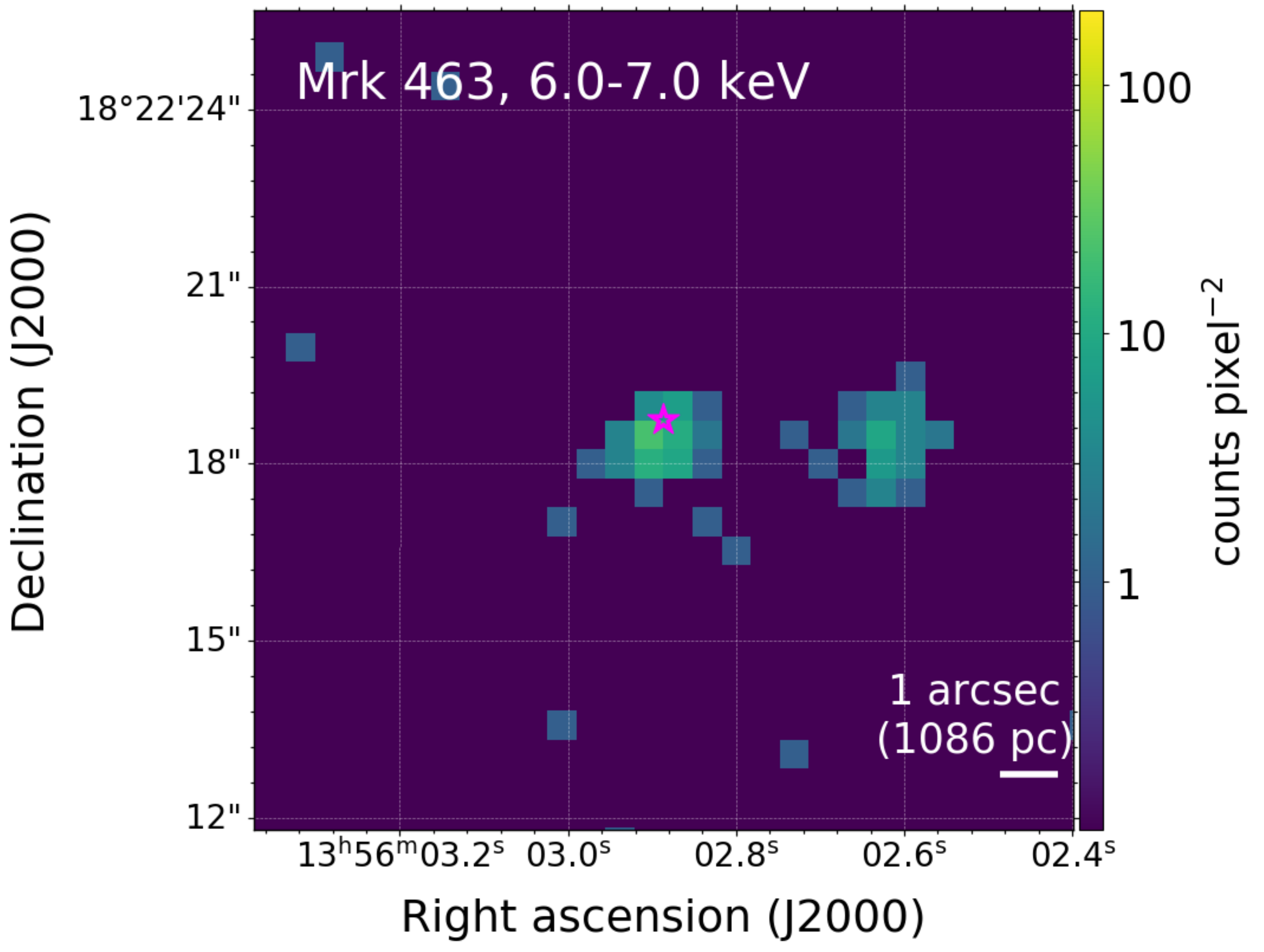}
    \includegraphics[width=5.35cm]{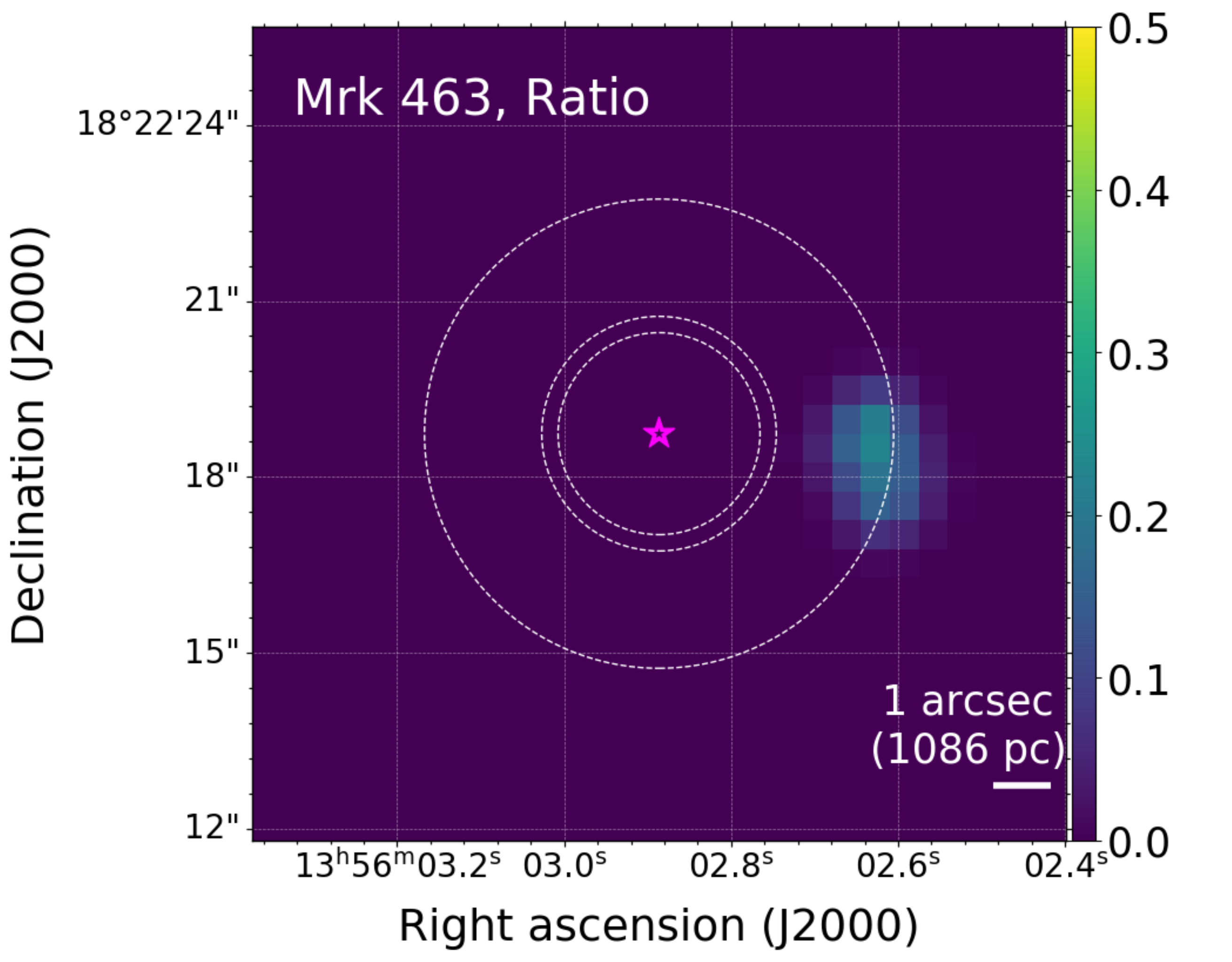}
    \\
    \includegraphics[width=5.5cm]{3-6keV_19_NGC_5506.pdf}
    \includegraphics[width=5.5cm]{6-7keV_19_NGC_5506.pdf}
    \includegraphics[width=5.35cm]{ratio_19_NGC_5506.pdf}
    \caption{Continued.}
\end{figure*}

\begin{figure*}\addtocounter{figure}{-1}
    \centering
    \includegraphics[width=5.5cm]{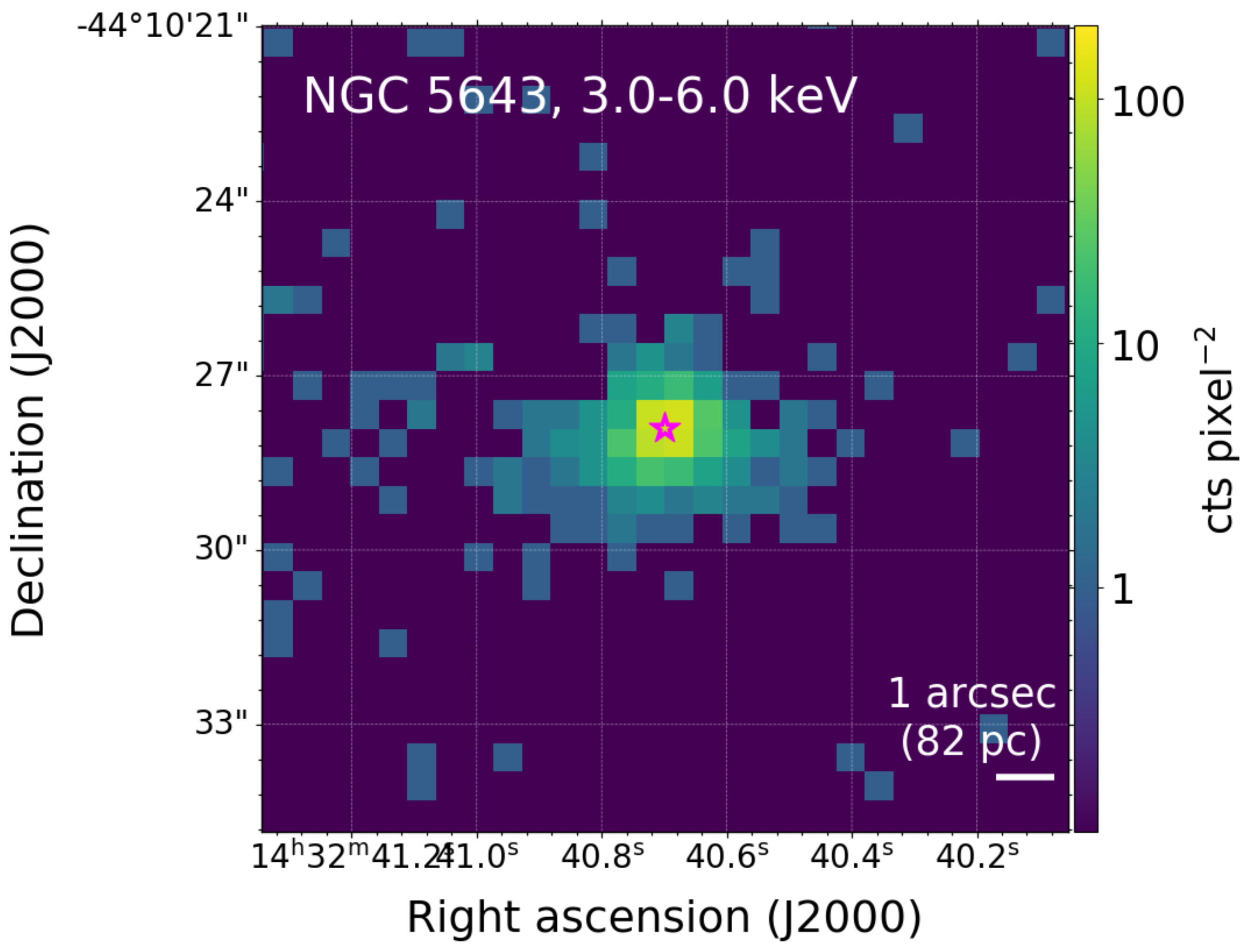}
    \includegraphics[width=5.5cm]{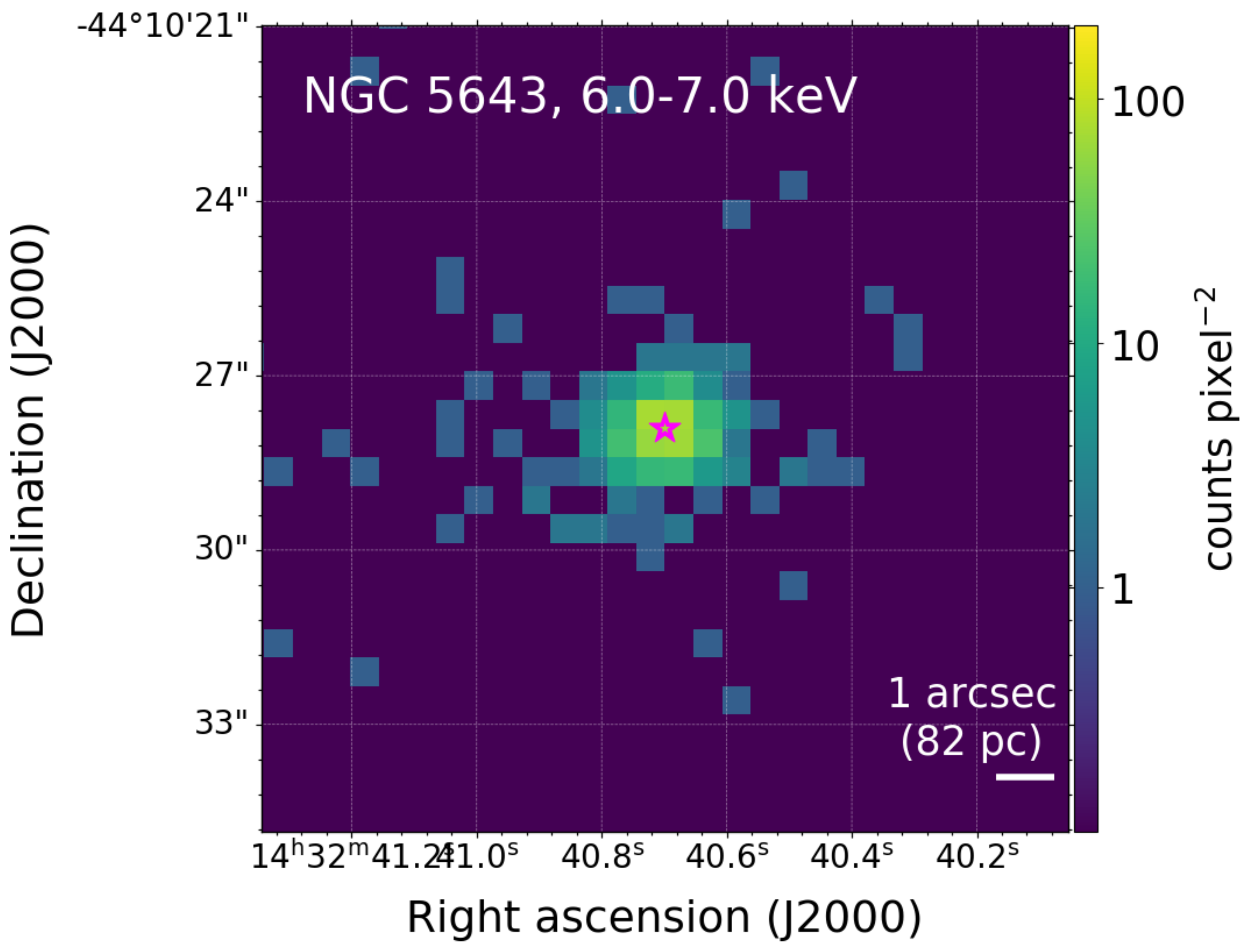}
    \includegraphics[width=5.35cm]{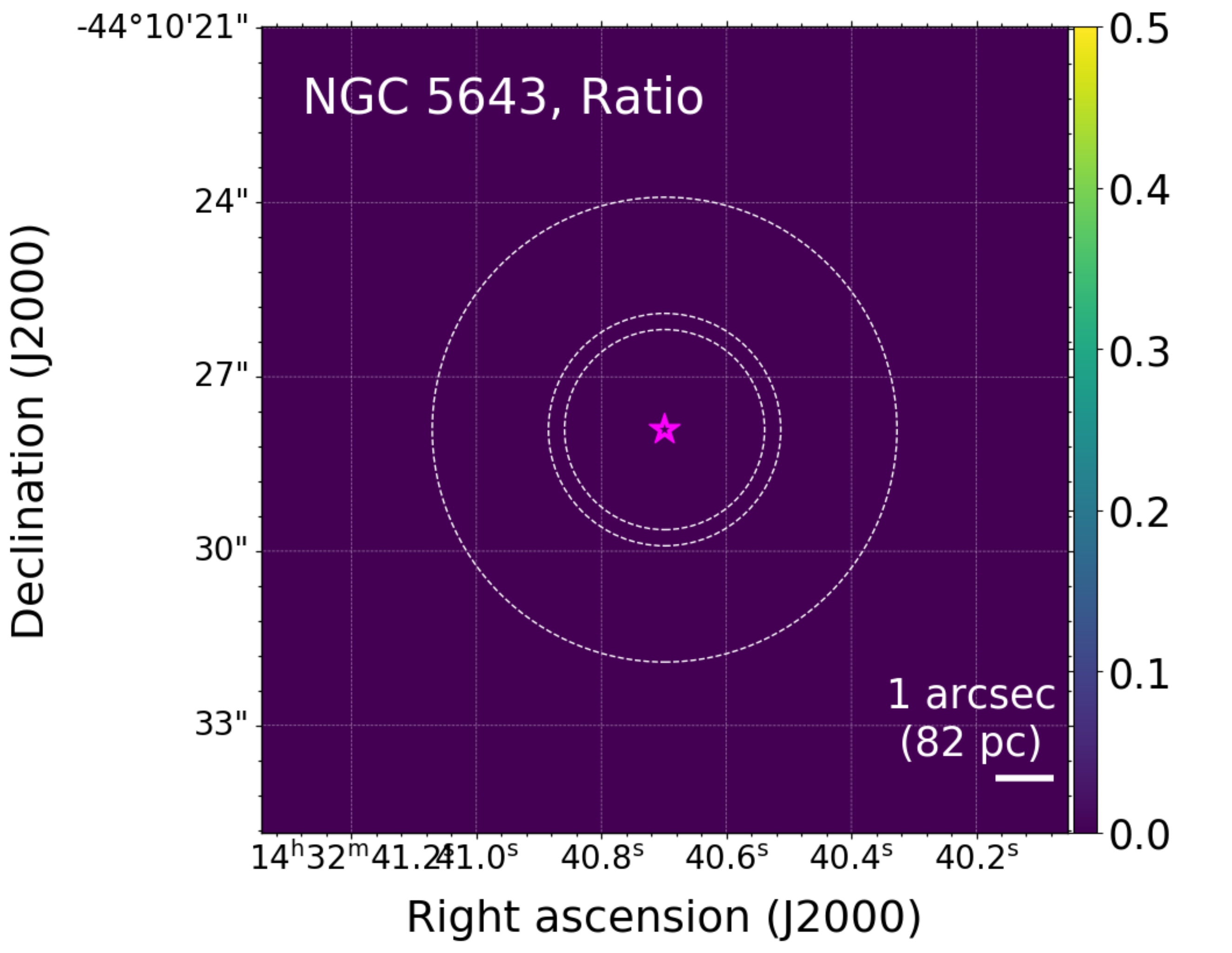}
    \\
    \includegraphics[width=5.5cm]{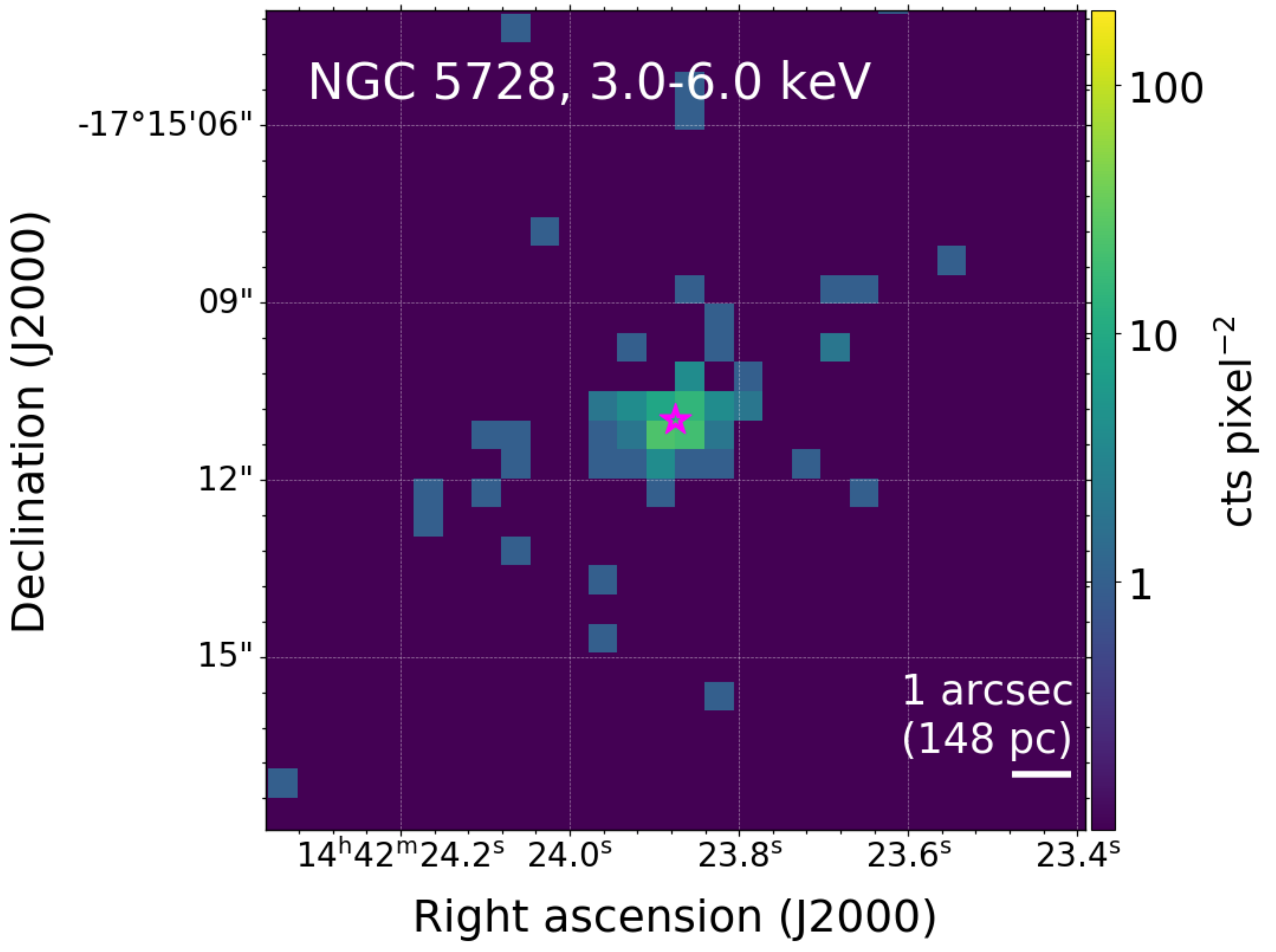}
    \includegraphics[width=5.5cm]{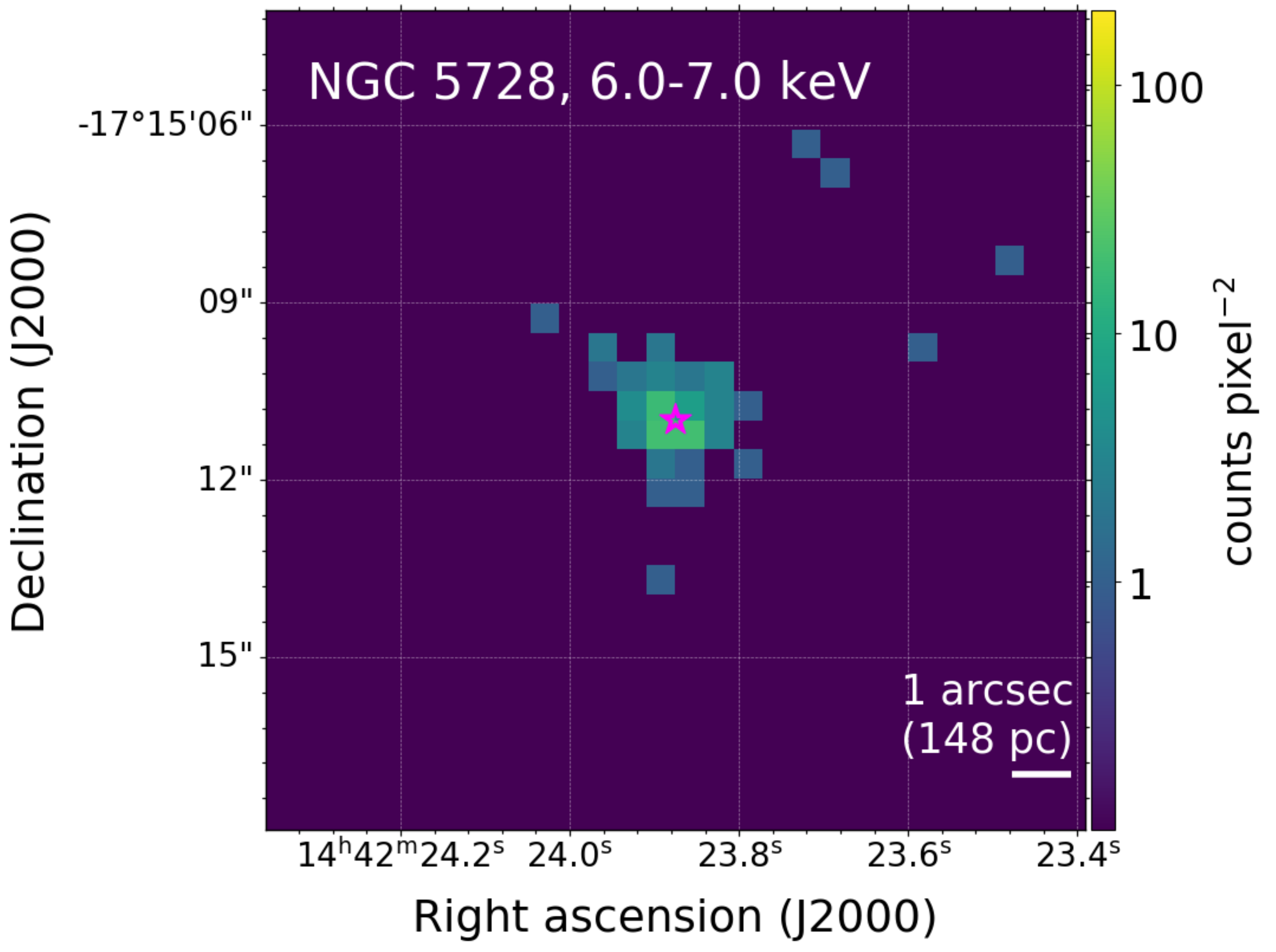}
    \includegraphics[width=5.35cm]{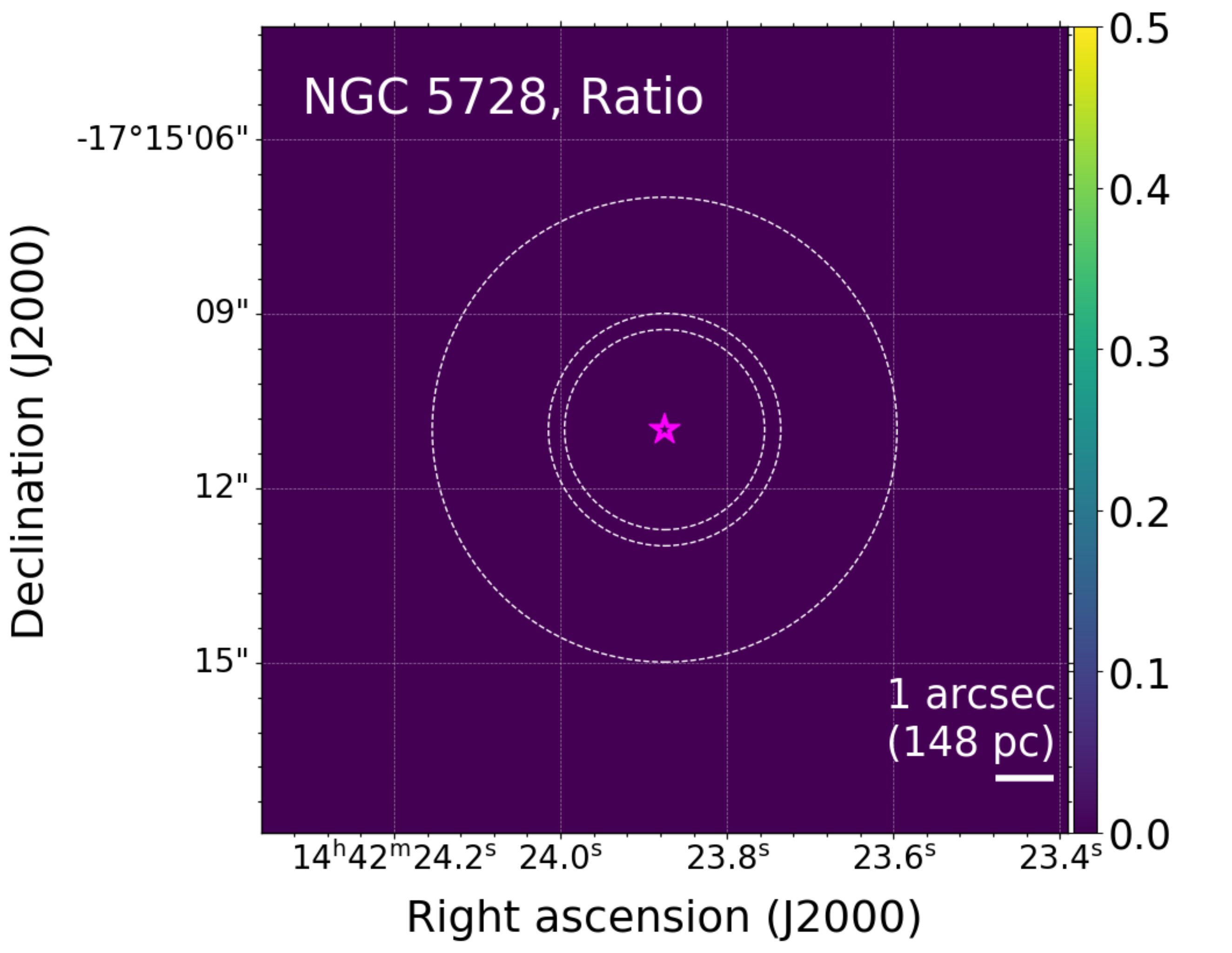}
    \\      
    \includegraphics[width=5.5cm]{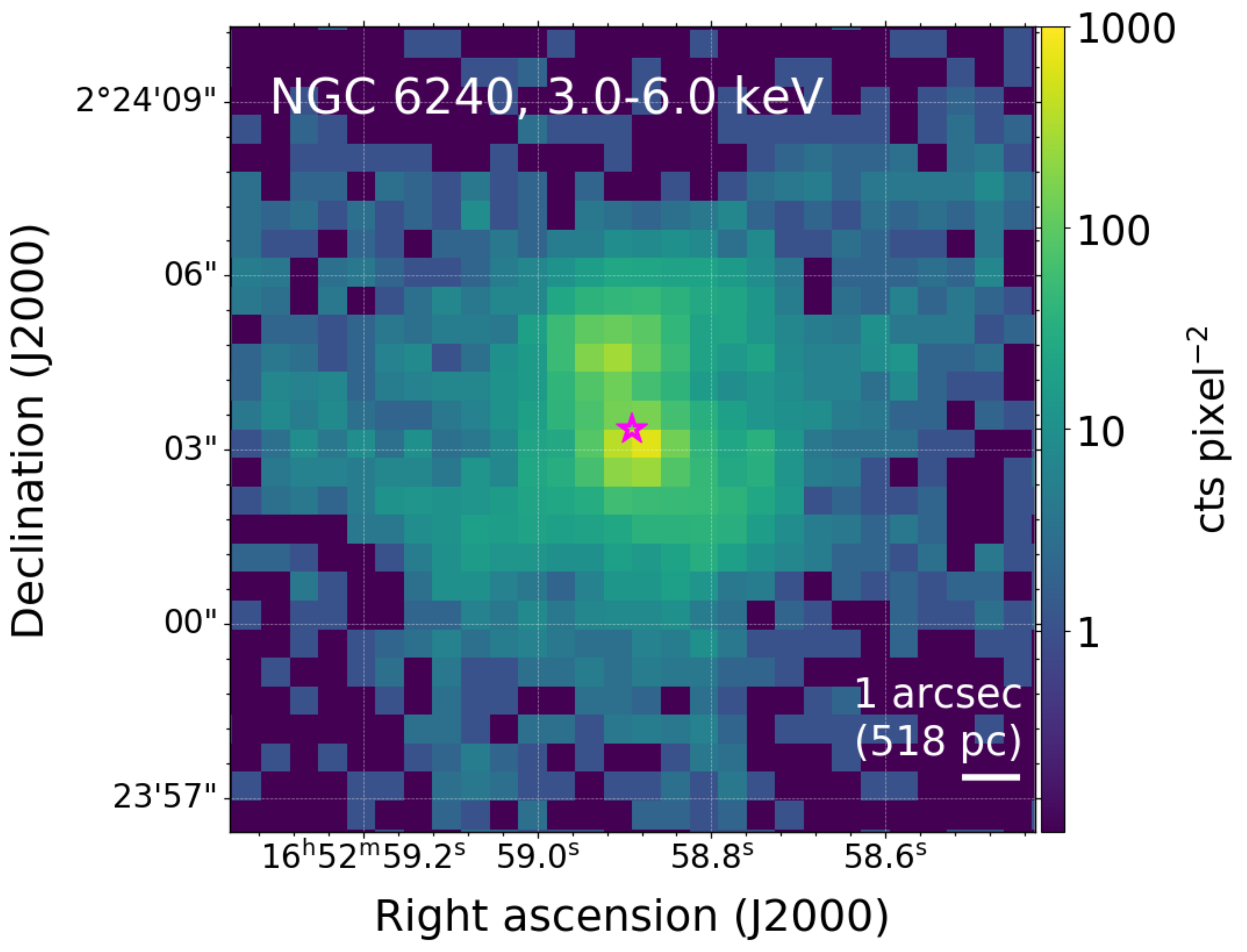}
    \includegraphics[width=5.5cm]{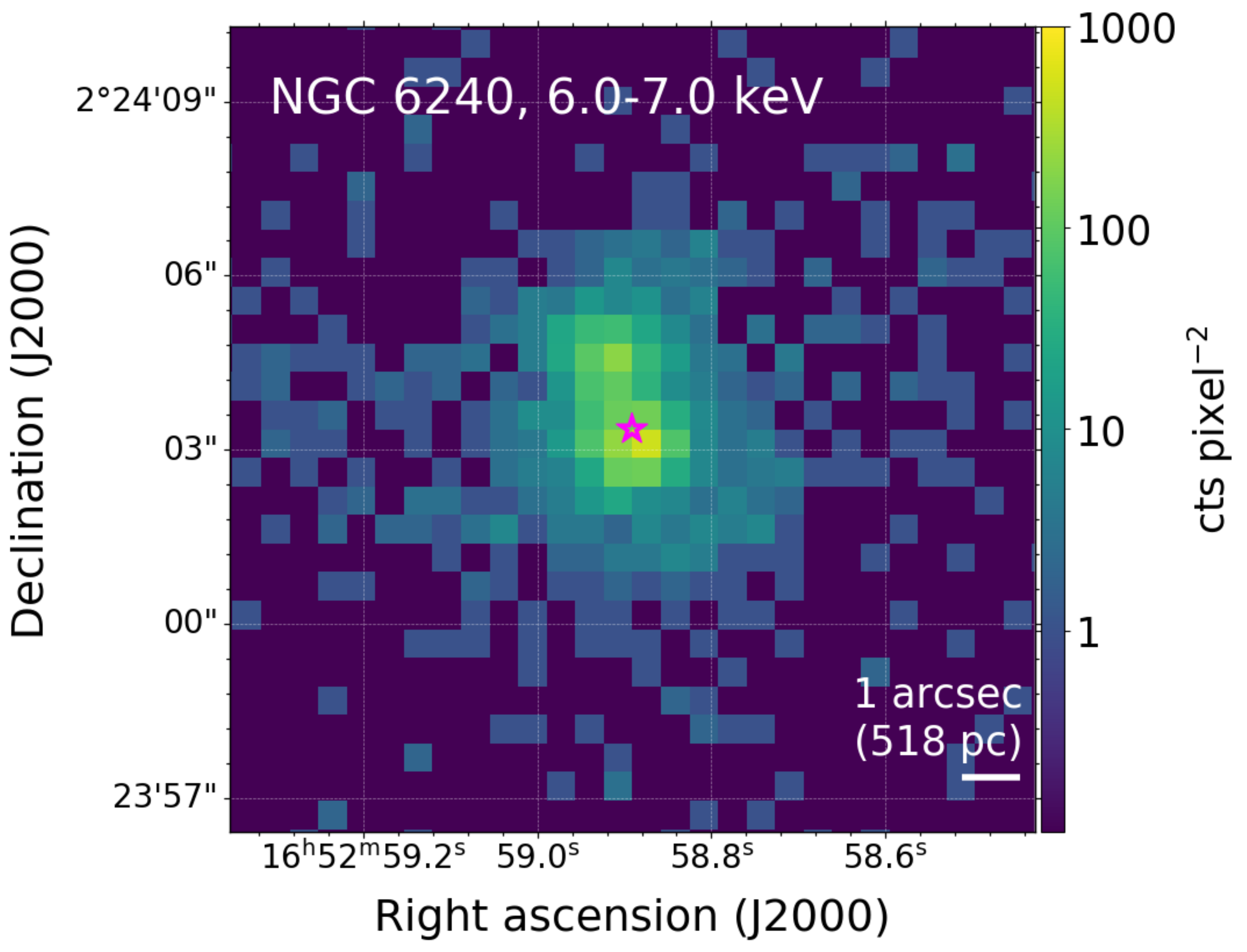}
    \includegraphics[width=5.35cm]{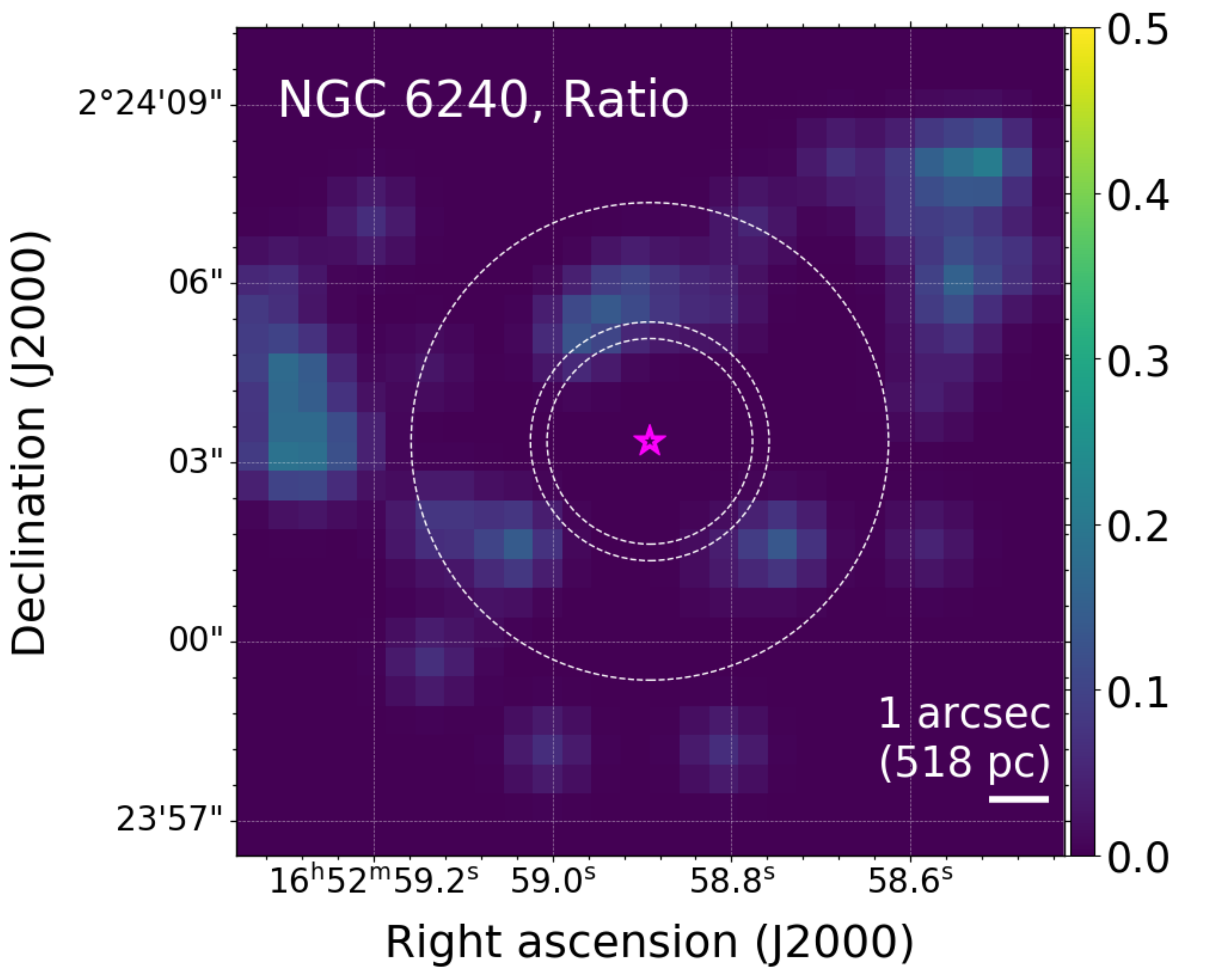}
    \\ 
    \includegraphics[width=5.5cm]{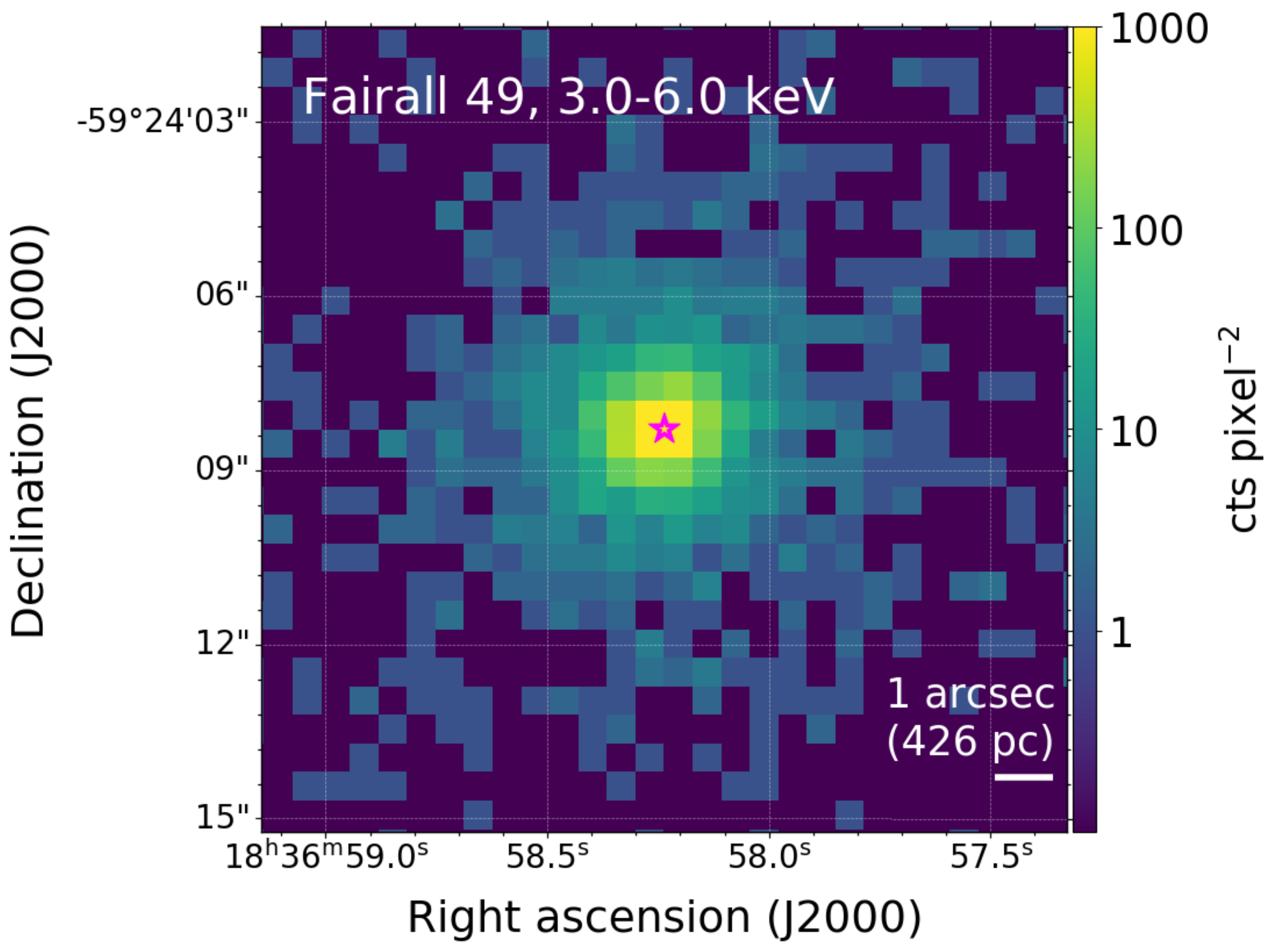}
    \includegraphics[width=5.5cm]{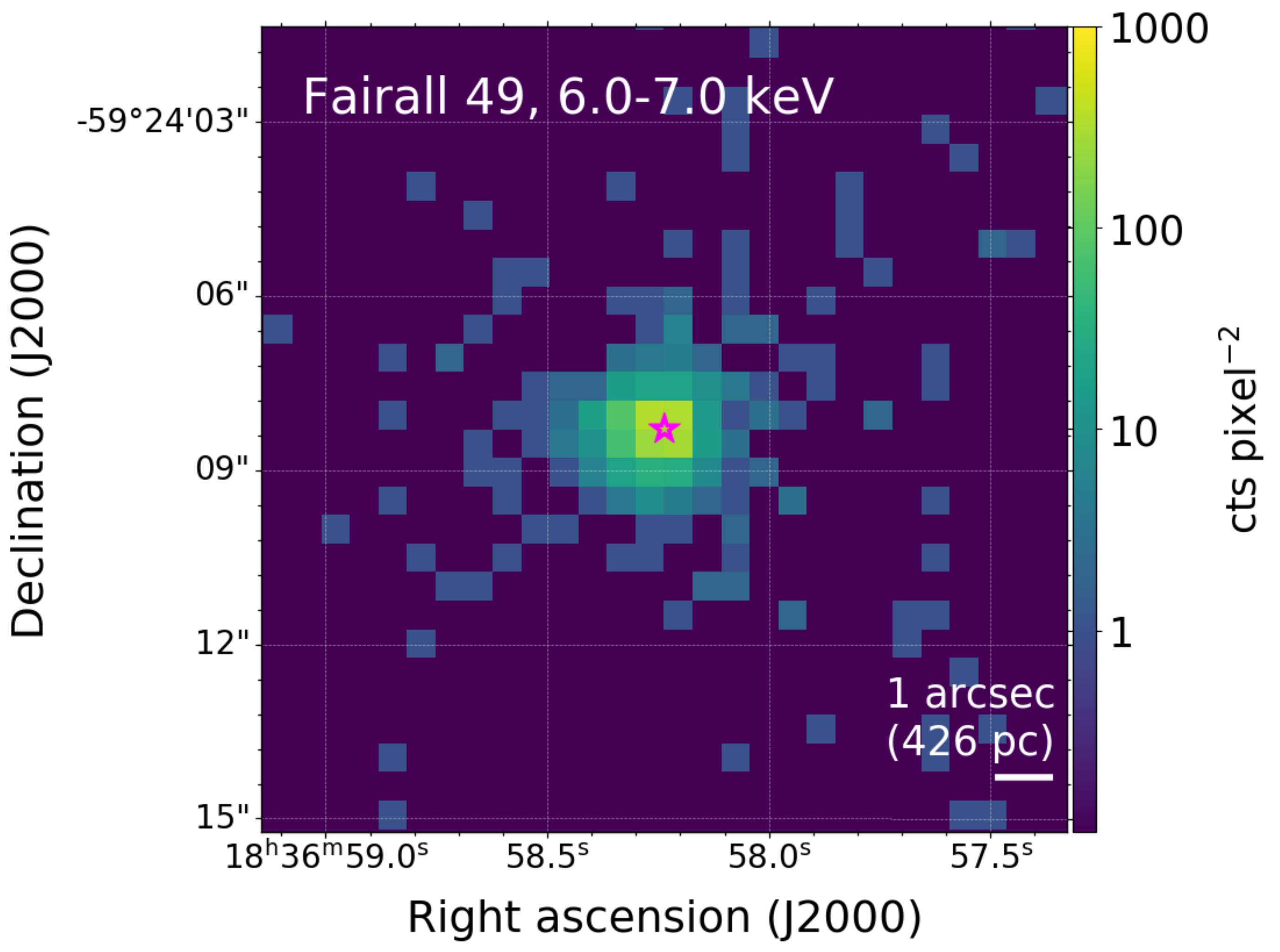}
    \includegraphics[width=5.35cm]{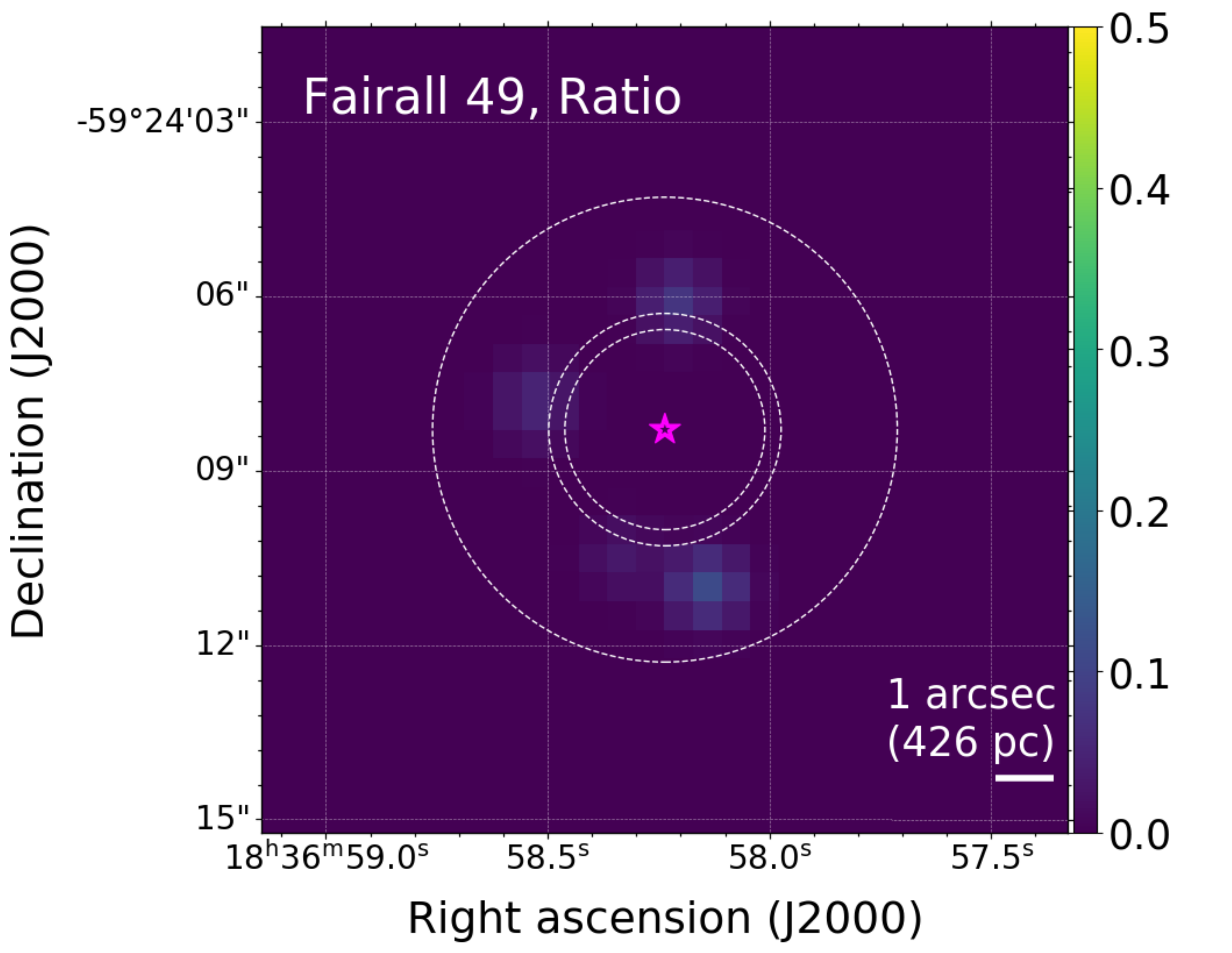}
    \\
    \includegraphics[width=5.5cm]{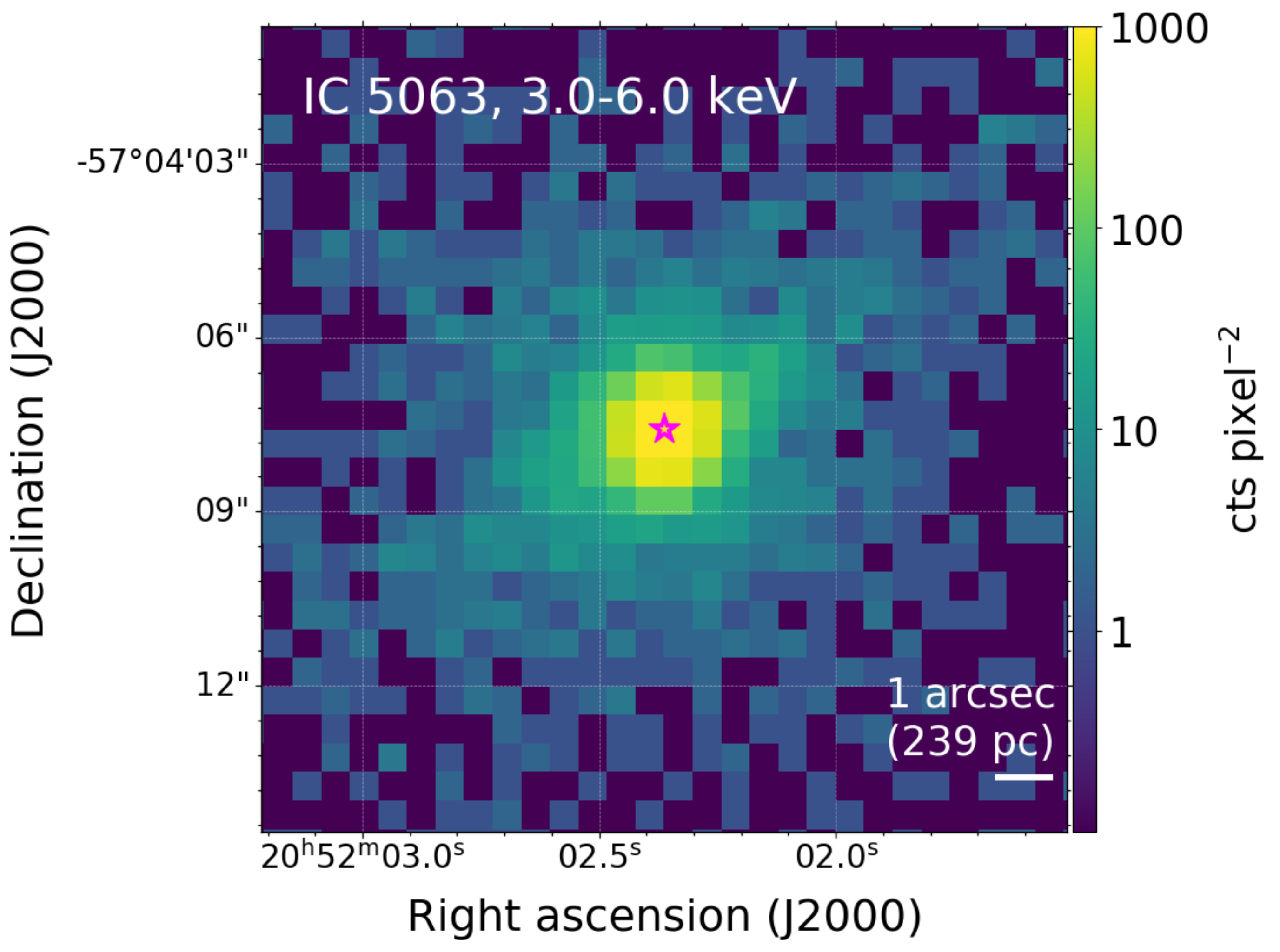}
    \includegraphics[width=5.5cm]{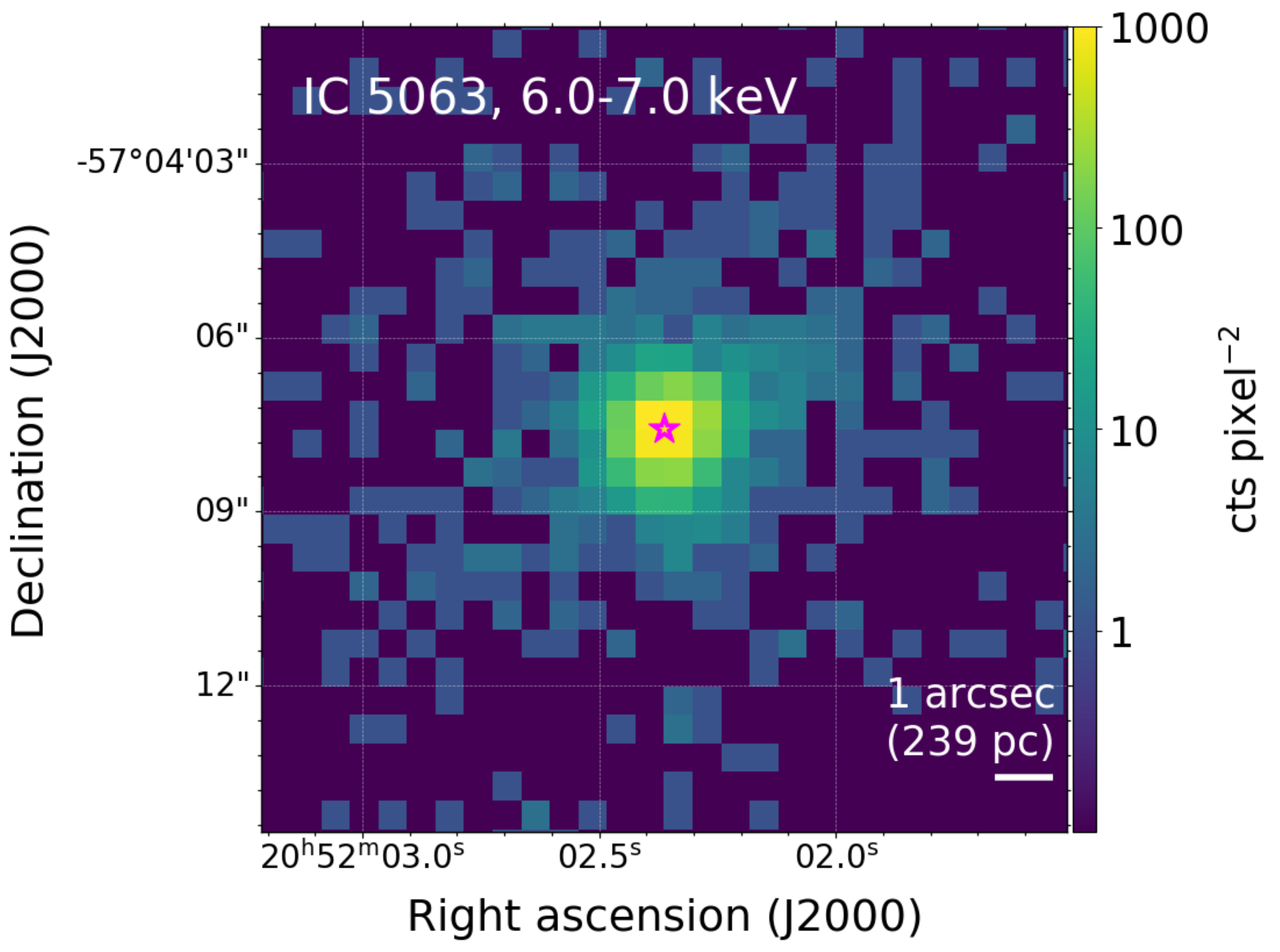}
    \includegraphics[width=5.35cm]{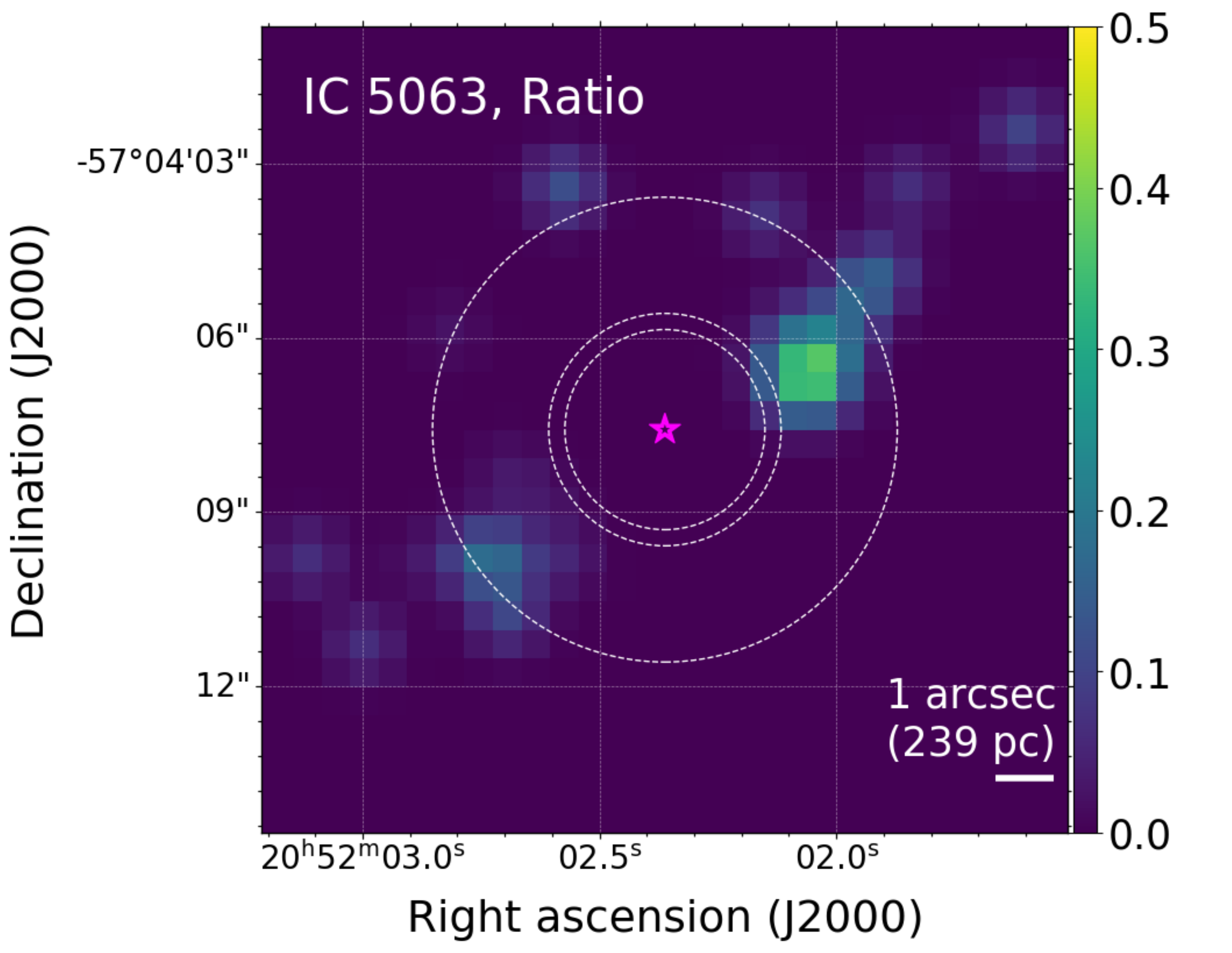}
    \caption{Continued.}
\end{figure*}

\begin{figure*}\addtocounter{figure}{-1}
    \centering
    \includegraphics[width=5.5cm]{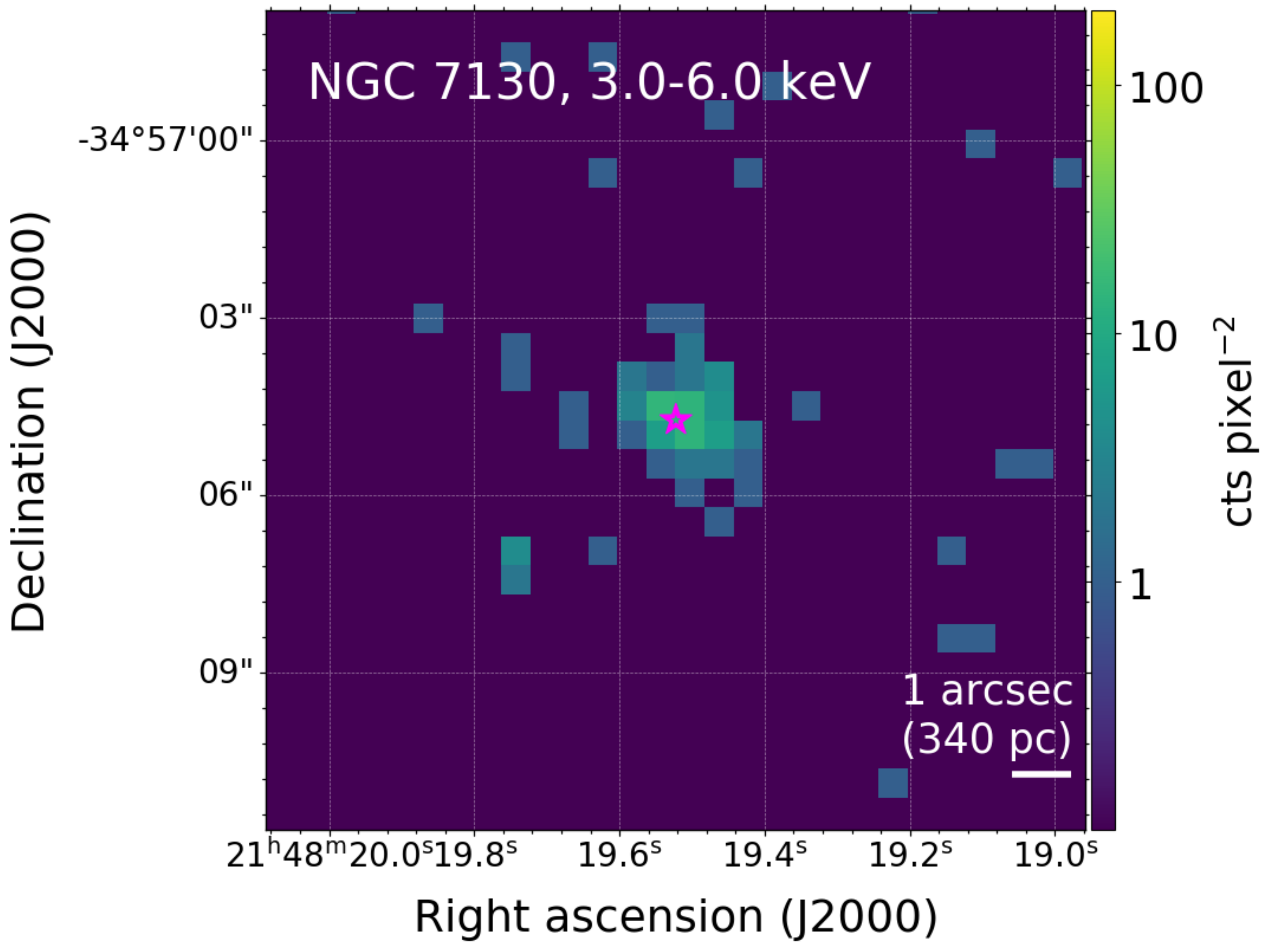}
    \includegraphics[width=5.5cm]{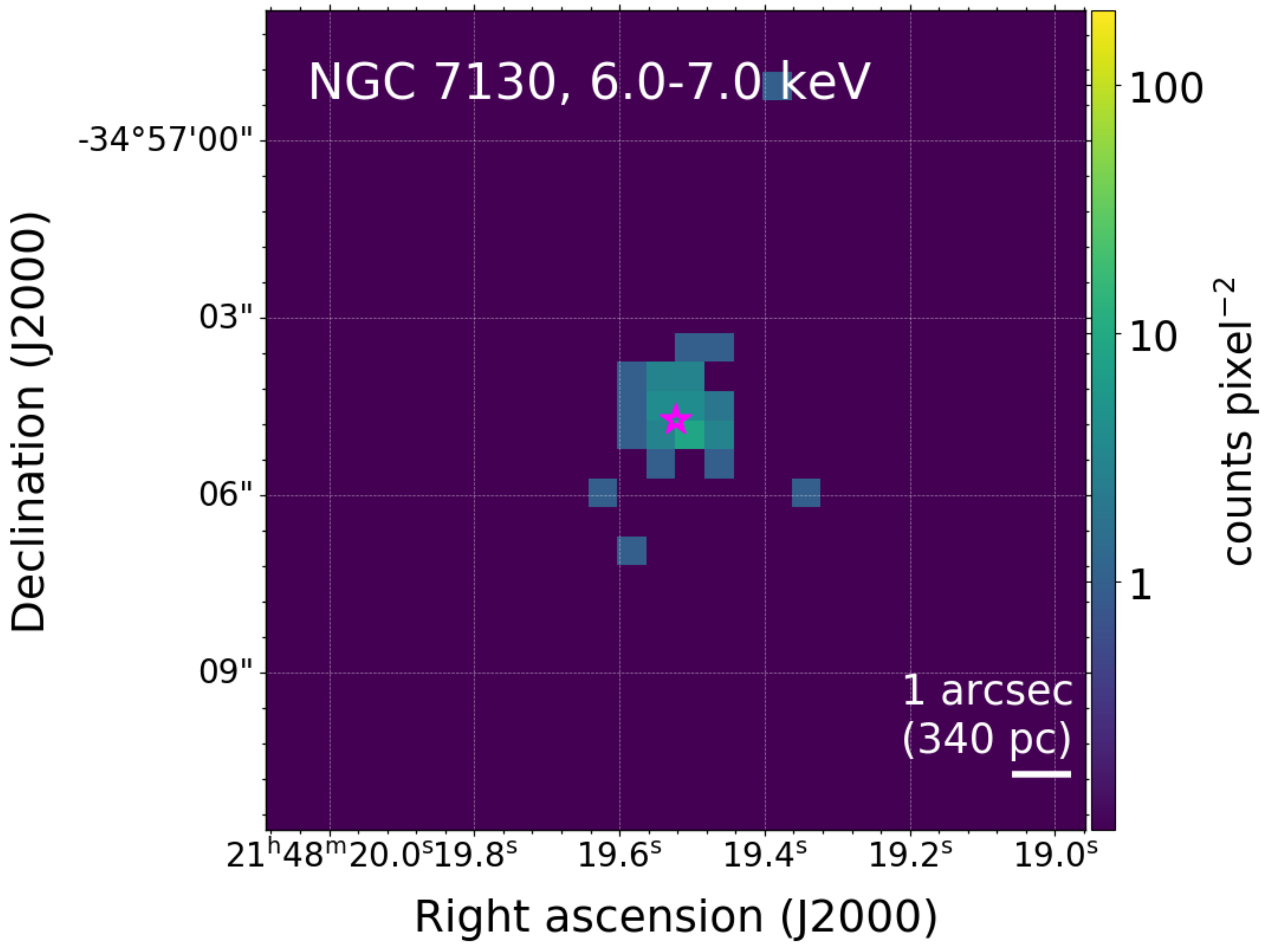}
    \includegraphics[width=5.35cm]{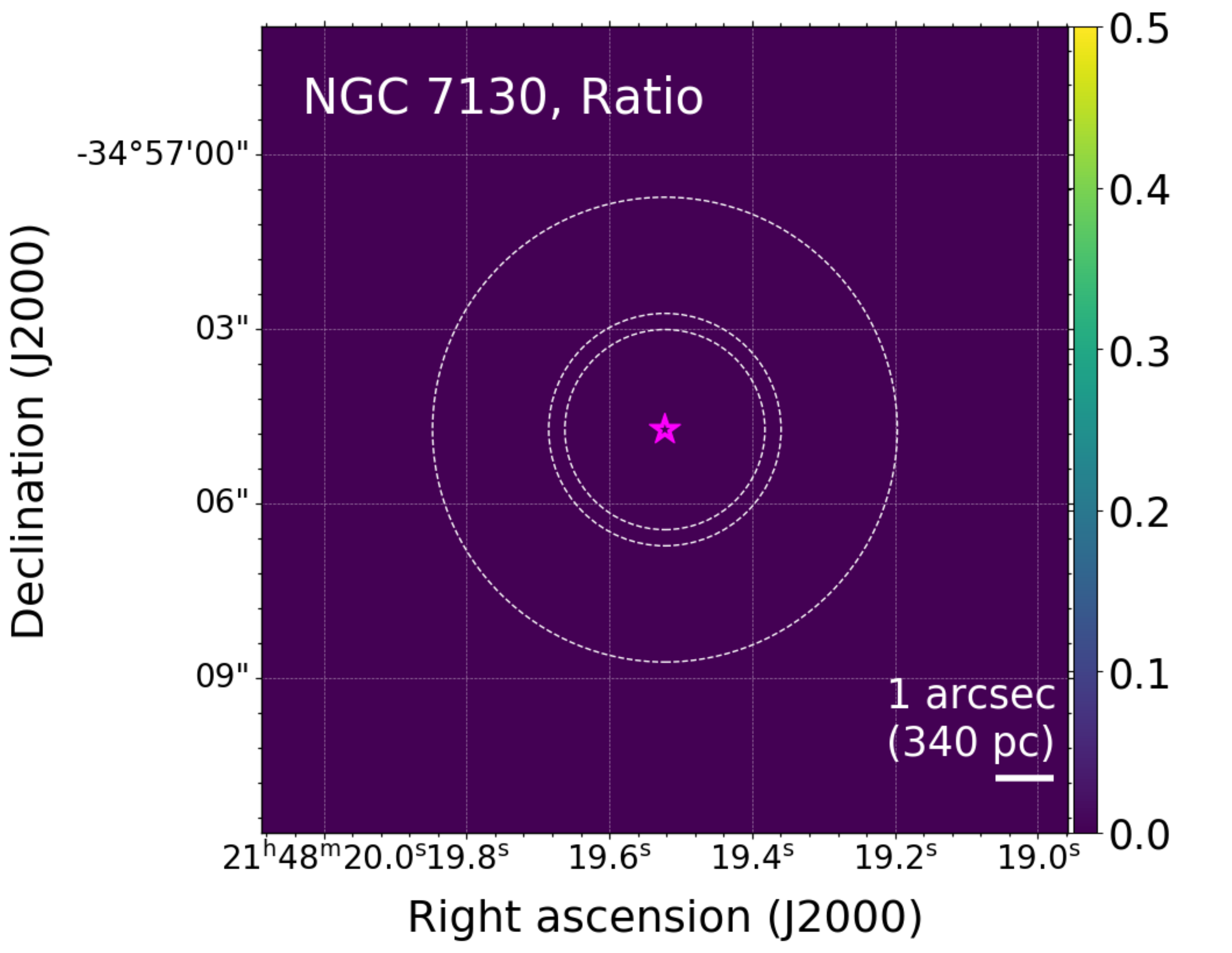}
    \\
    \includegraphics[width=5.5cm]{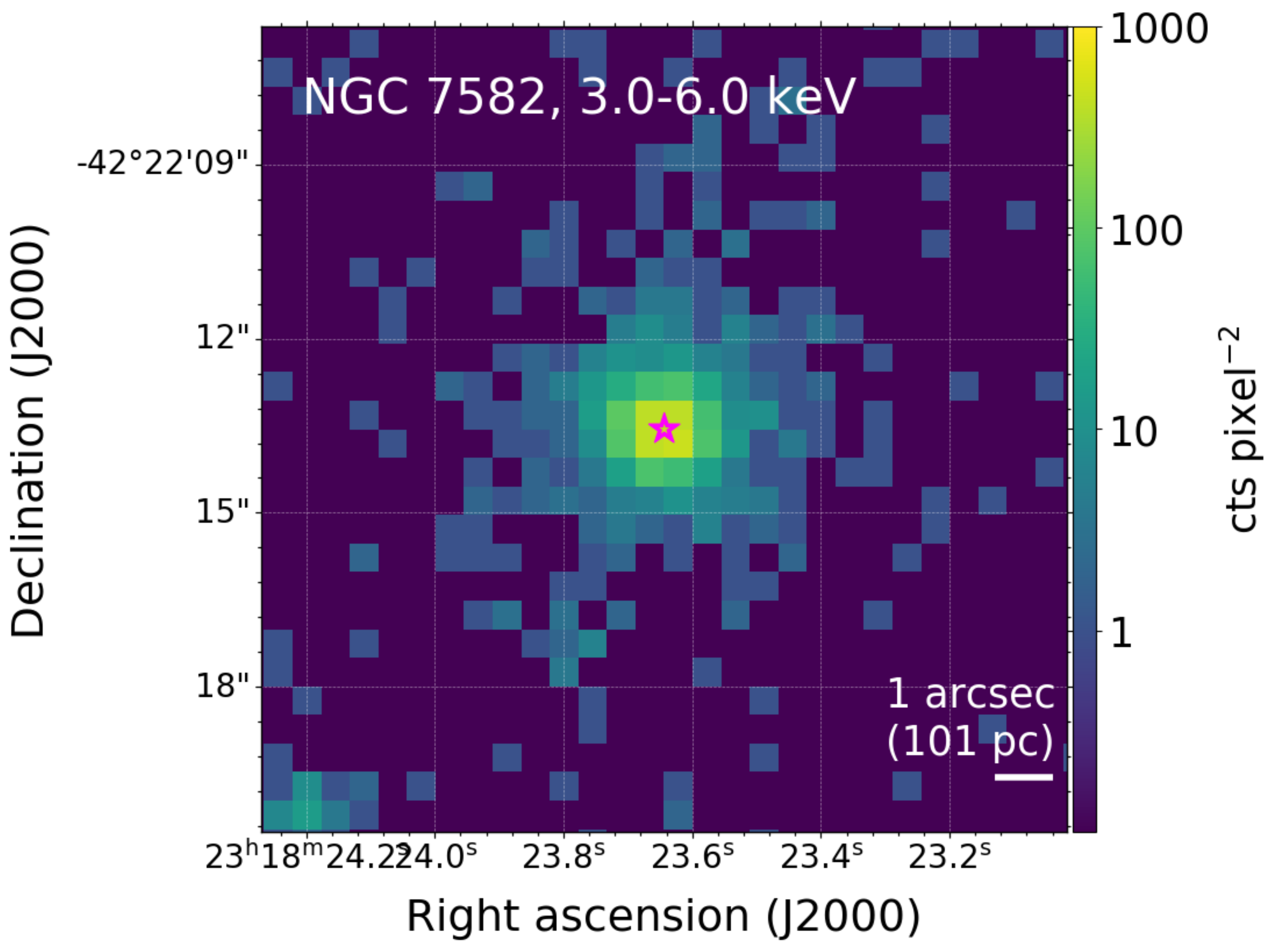}
    \includegraphics[width=5.5cm]{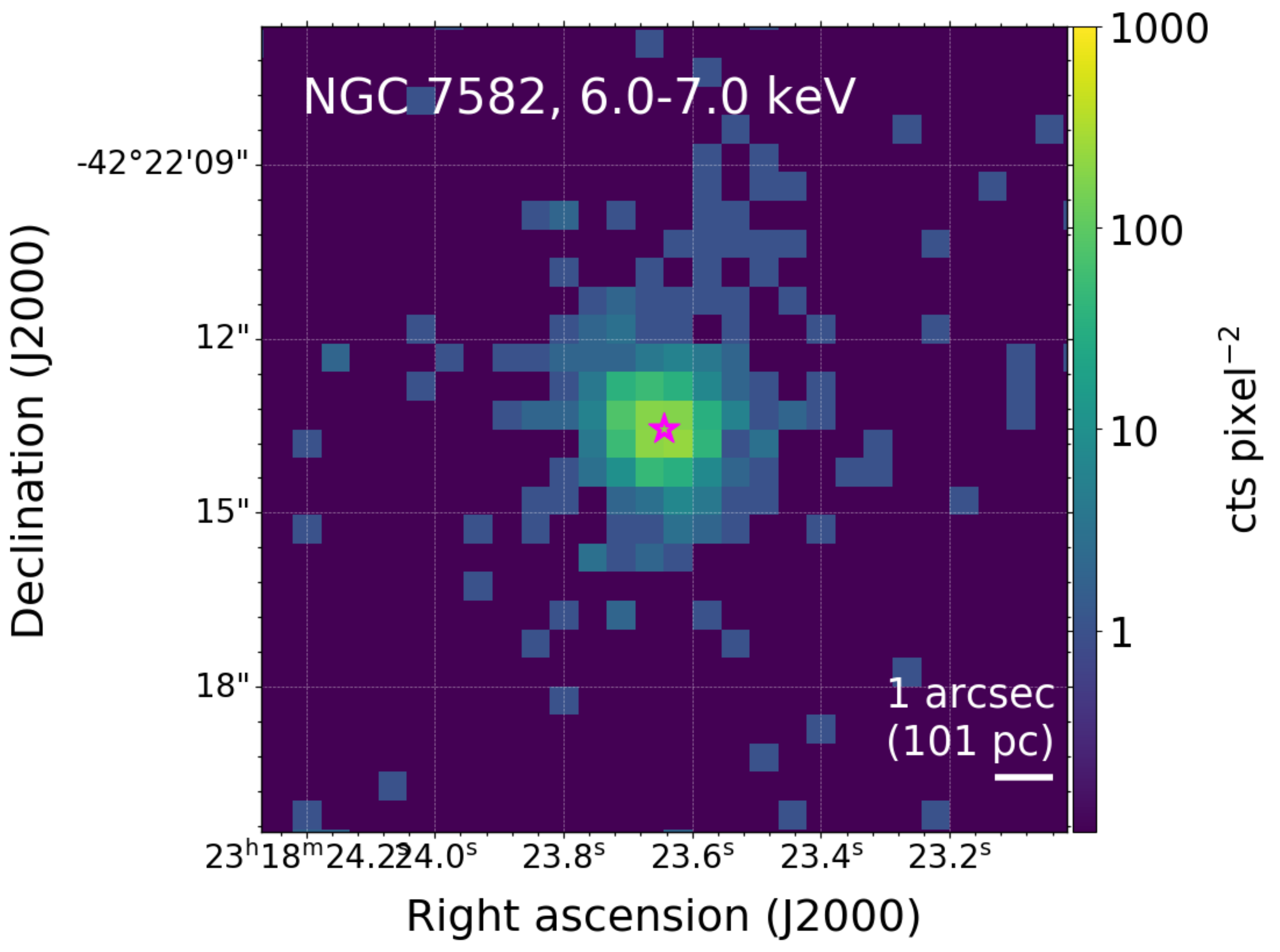}
    \includegraphics[width=5.35cm]{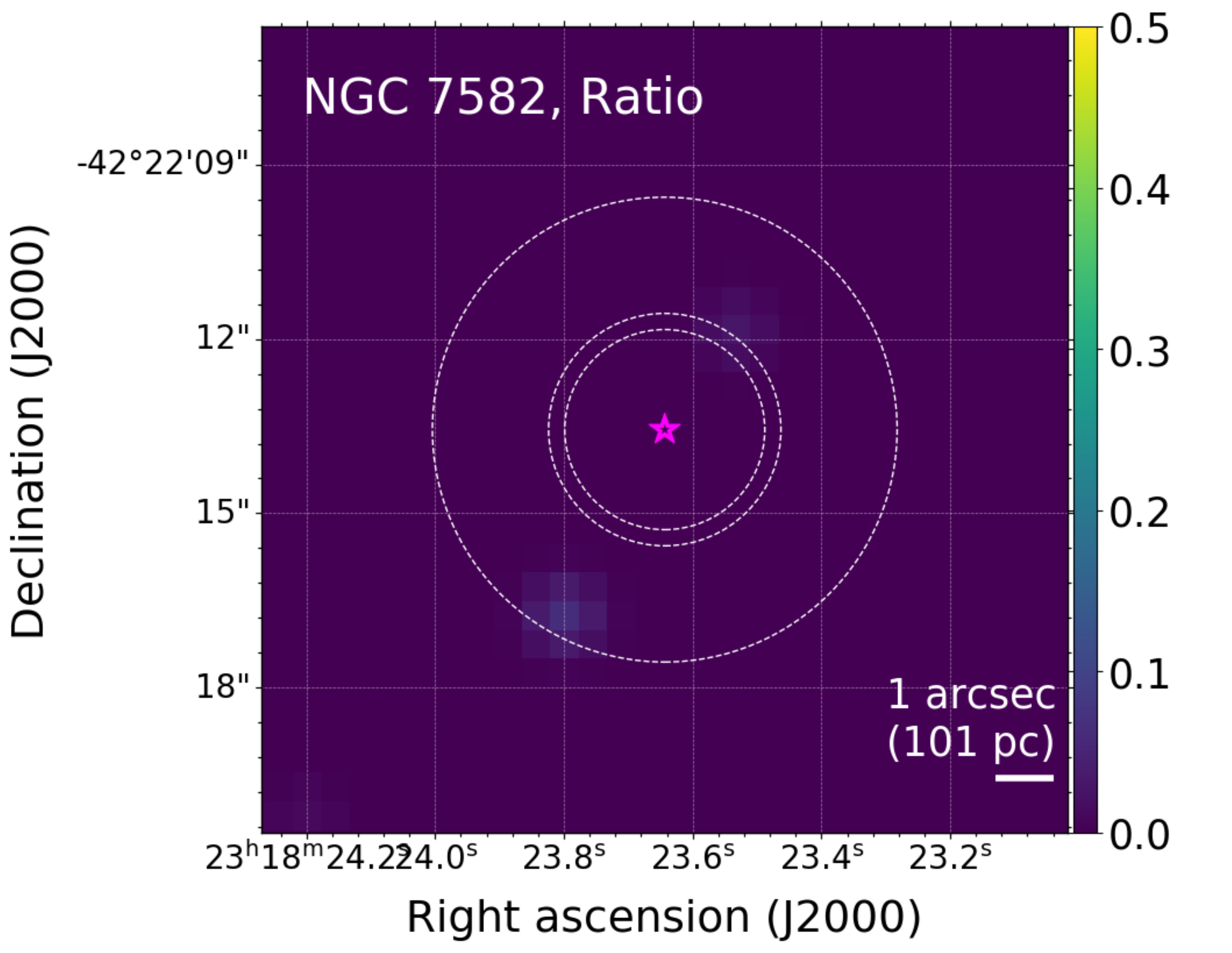}
    \caption{Continued.}
\end{figure*}

\clearpage

\section{Spectra of CO($J$=2--1) emission}\label{app:co_spec}

\begin{figure*}[!h]
    \centering
    \includegraphics[width=4.3cm]{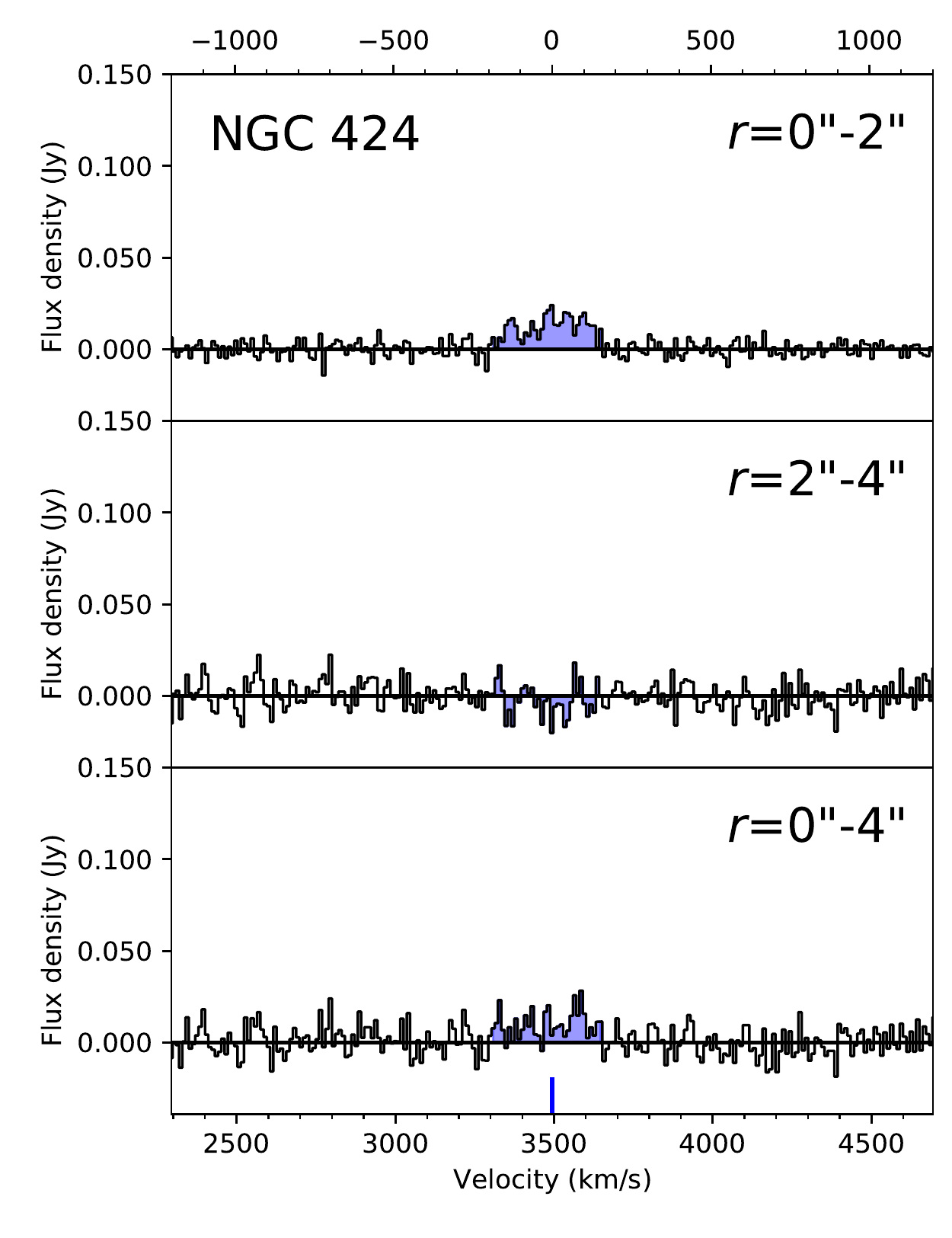}
    \includegraphics[width=4.3cm]{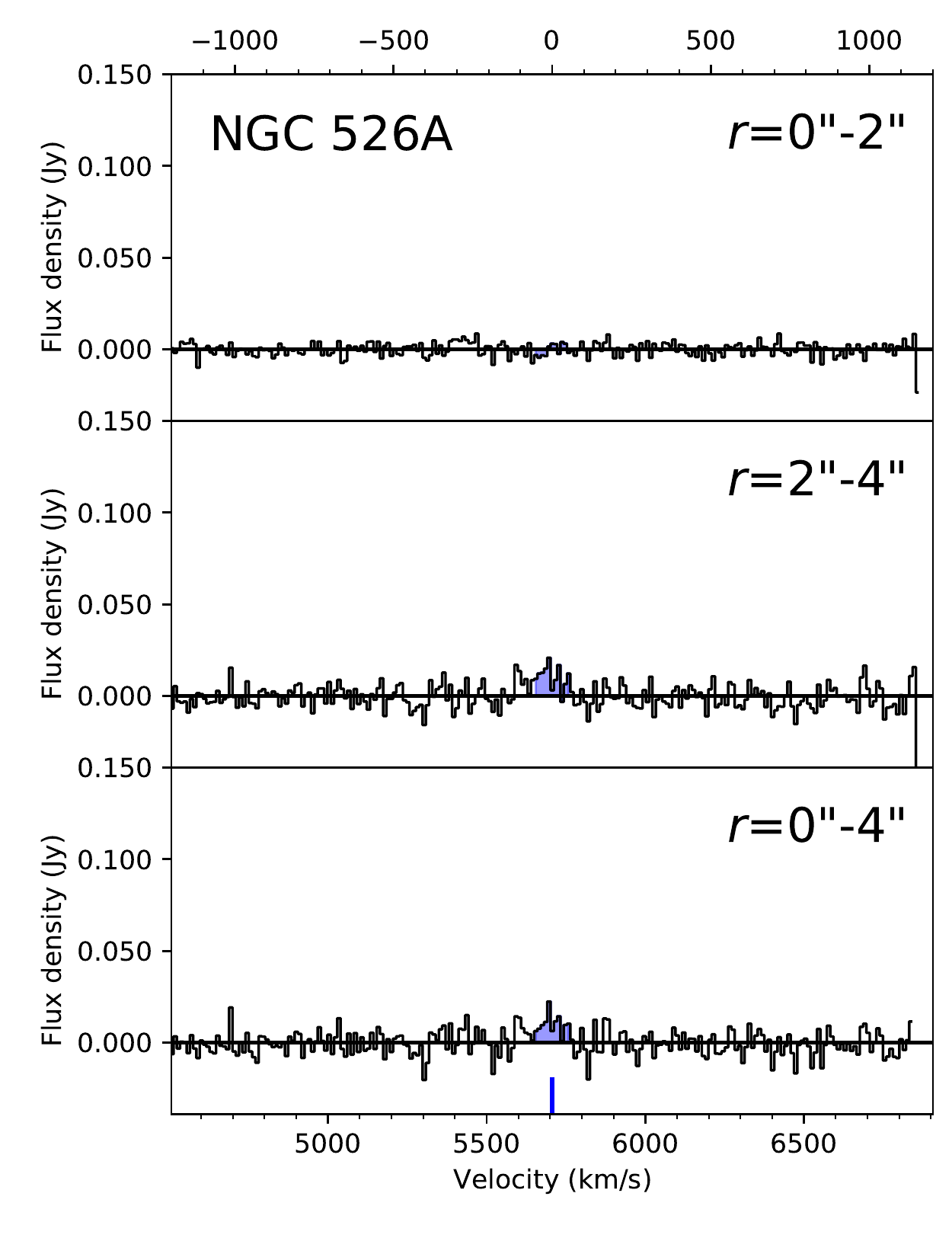}
    \includegraphics[width=4.3cm]{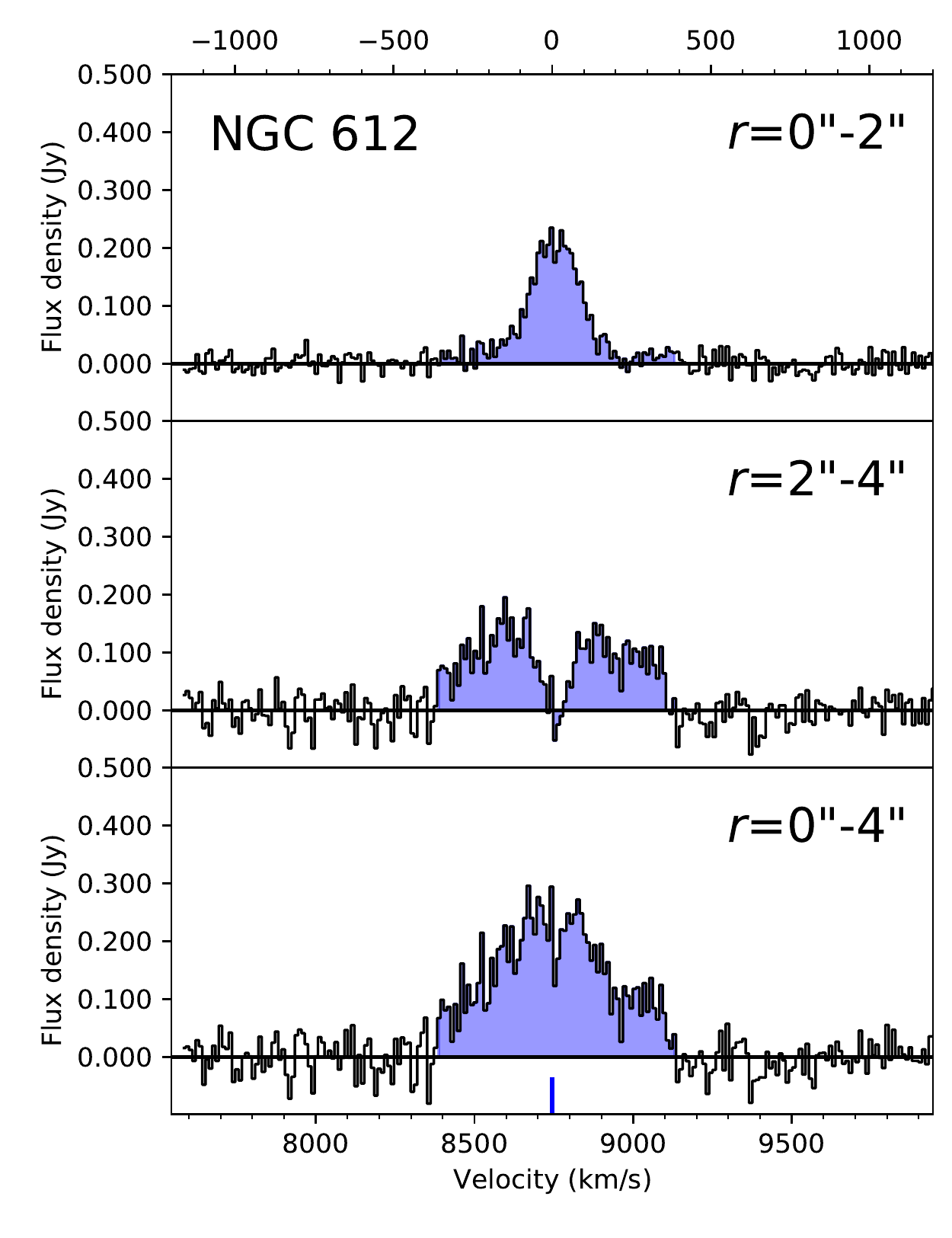}
    \includegraphics[width=4.3cm]{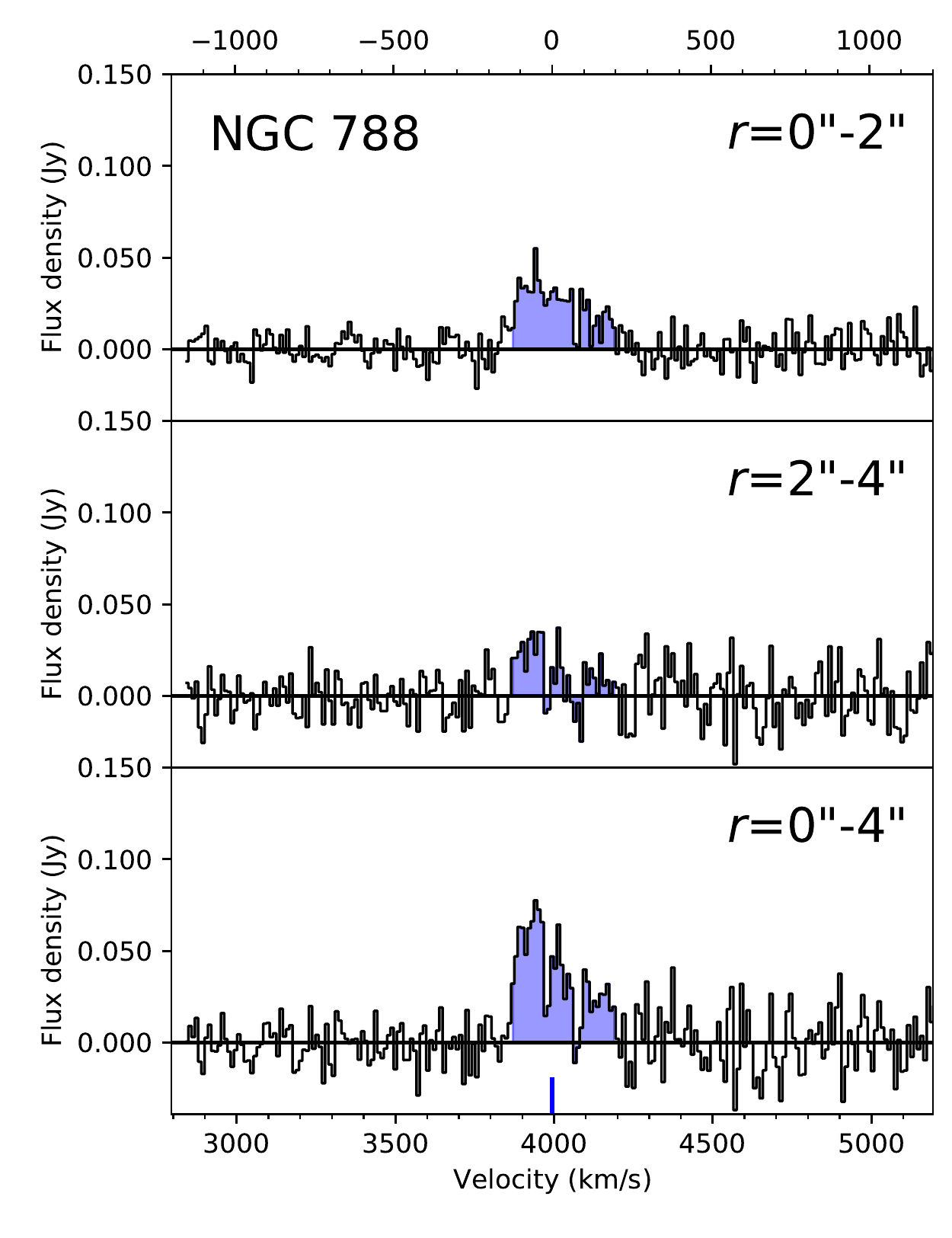}
    \includegraphics[width=4.3cm]{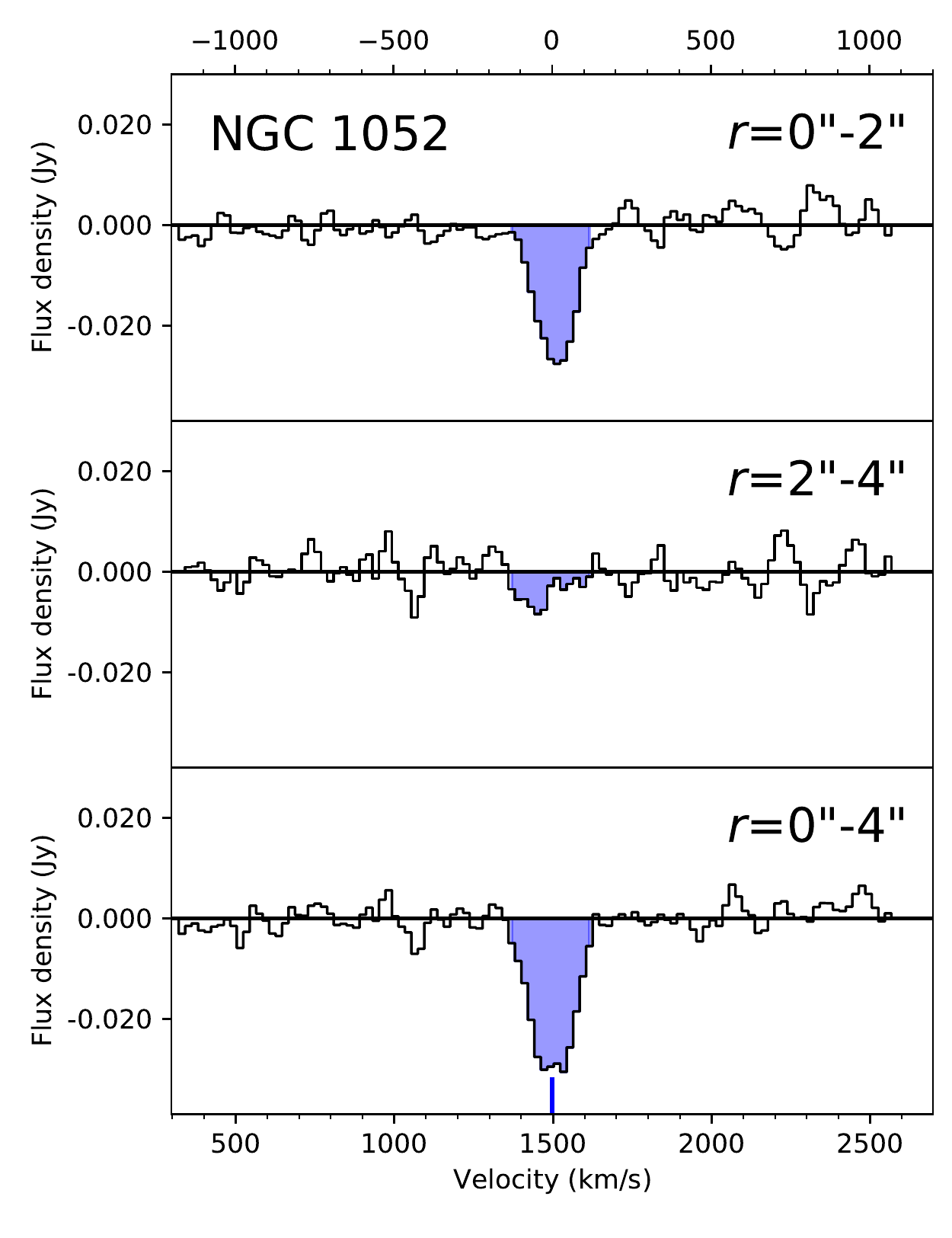}
    \includegraphics[width=4.3cm]{06_NGC_1068_co_spec.pdf}
    \includegraphics[width=4.3cm]{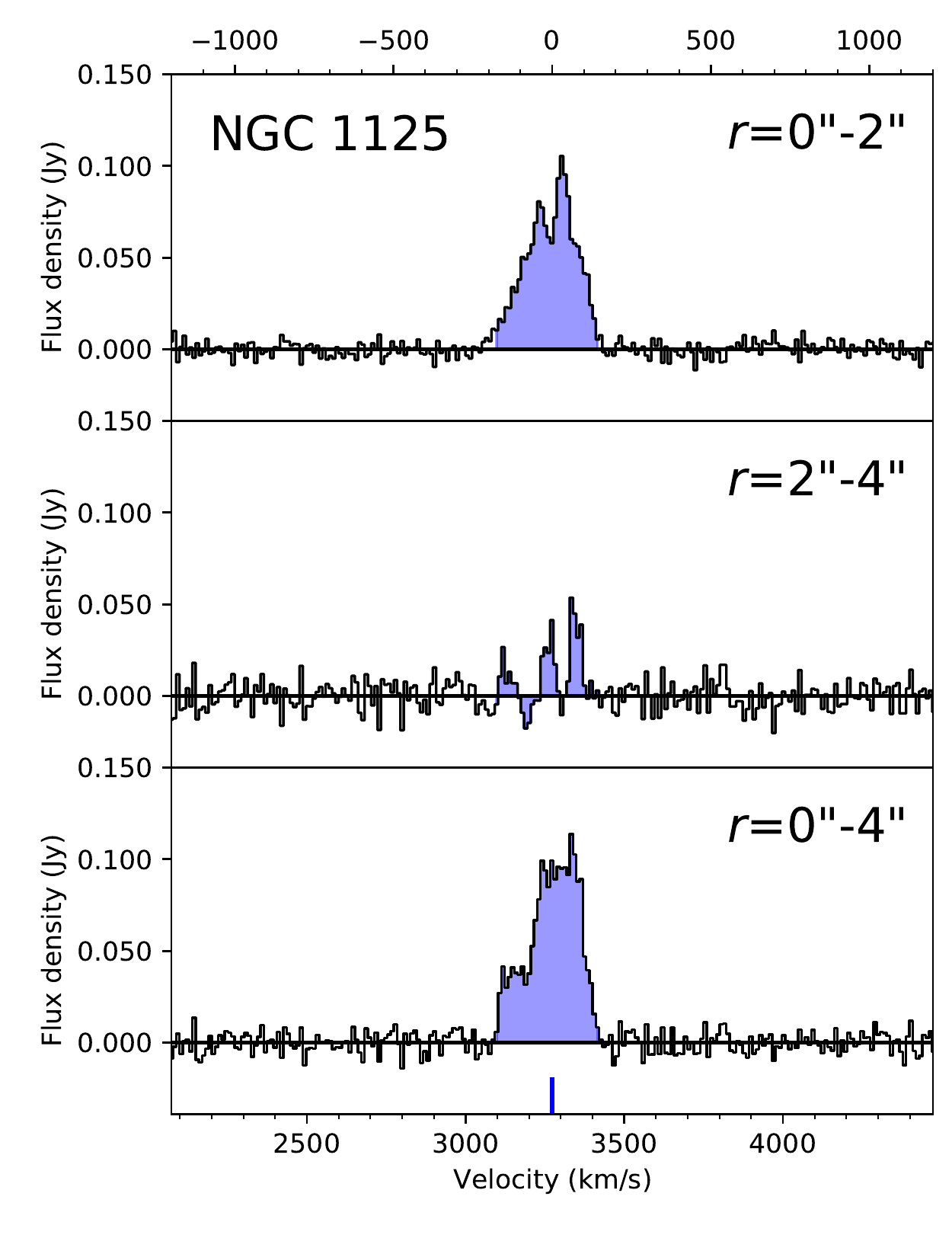}
    \includegraphics[width=4.3cm]{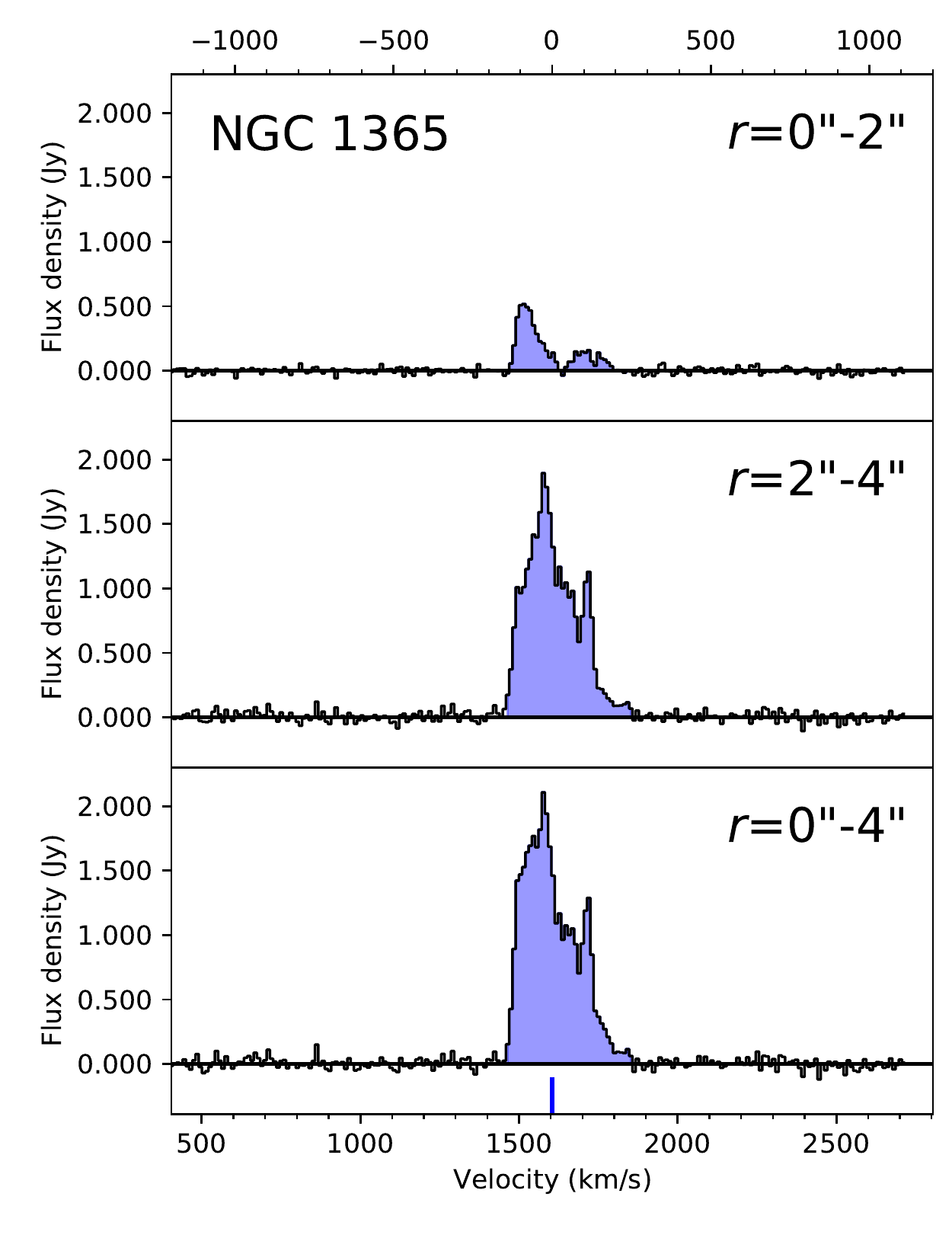}
    \includegraphics[width=4.3cm]{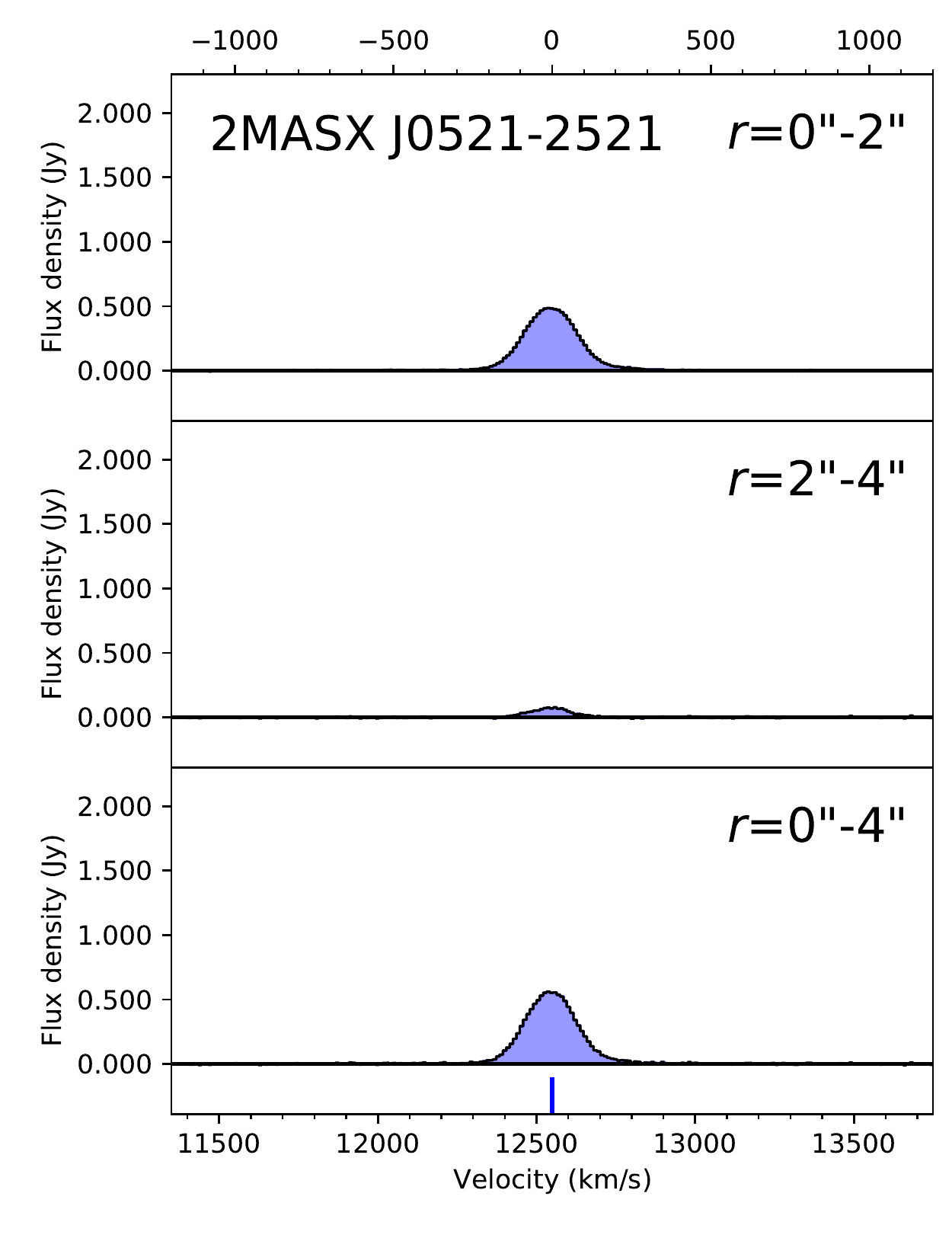}
    \includegraphics[width=4.3cm]{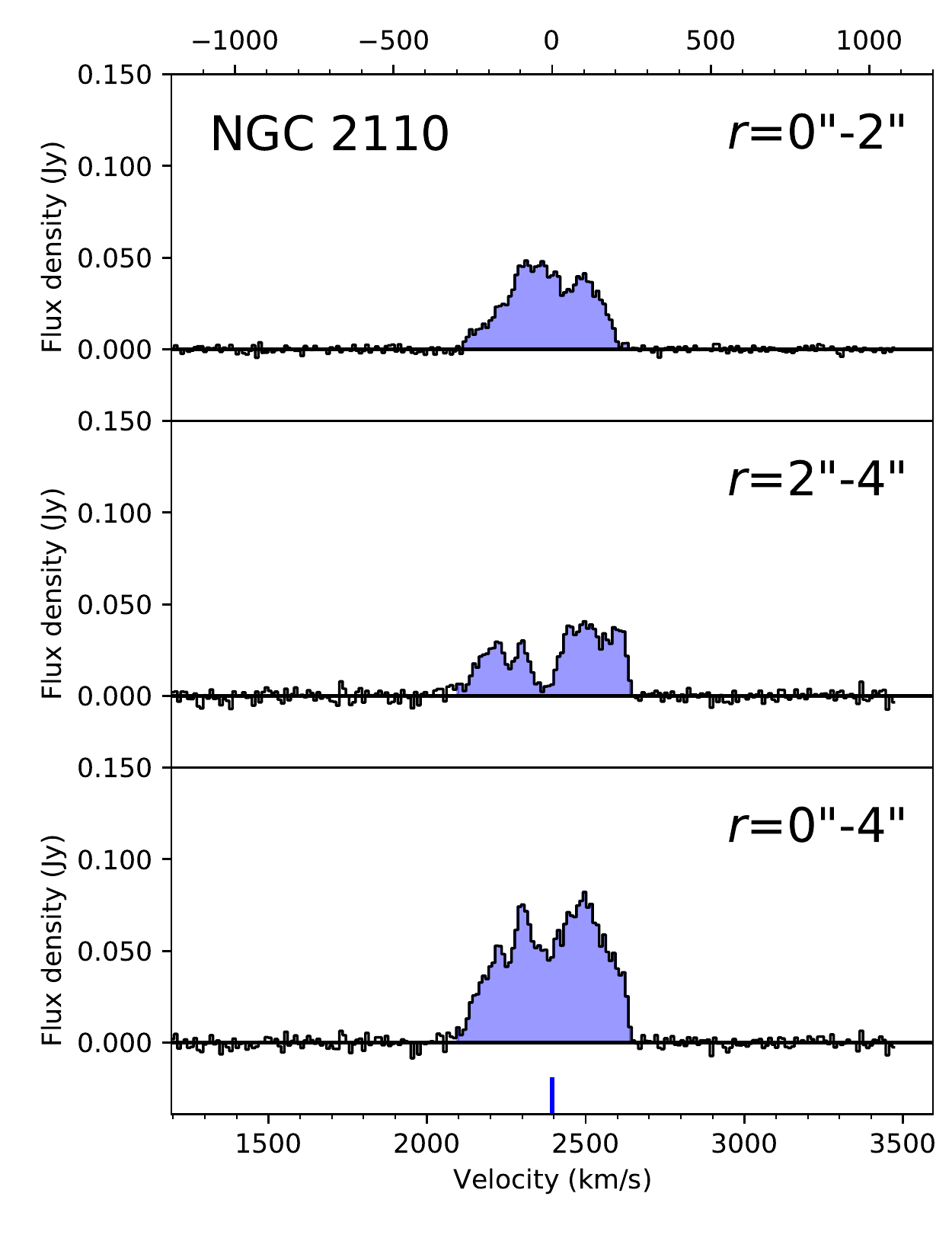}
    \includegraphics[width=4.3cm]{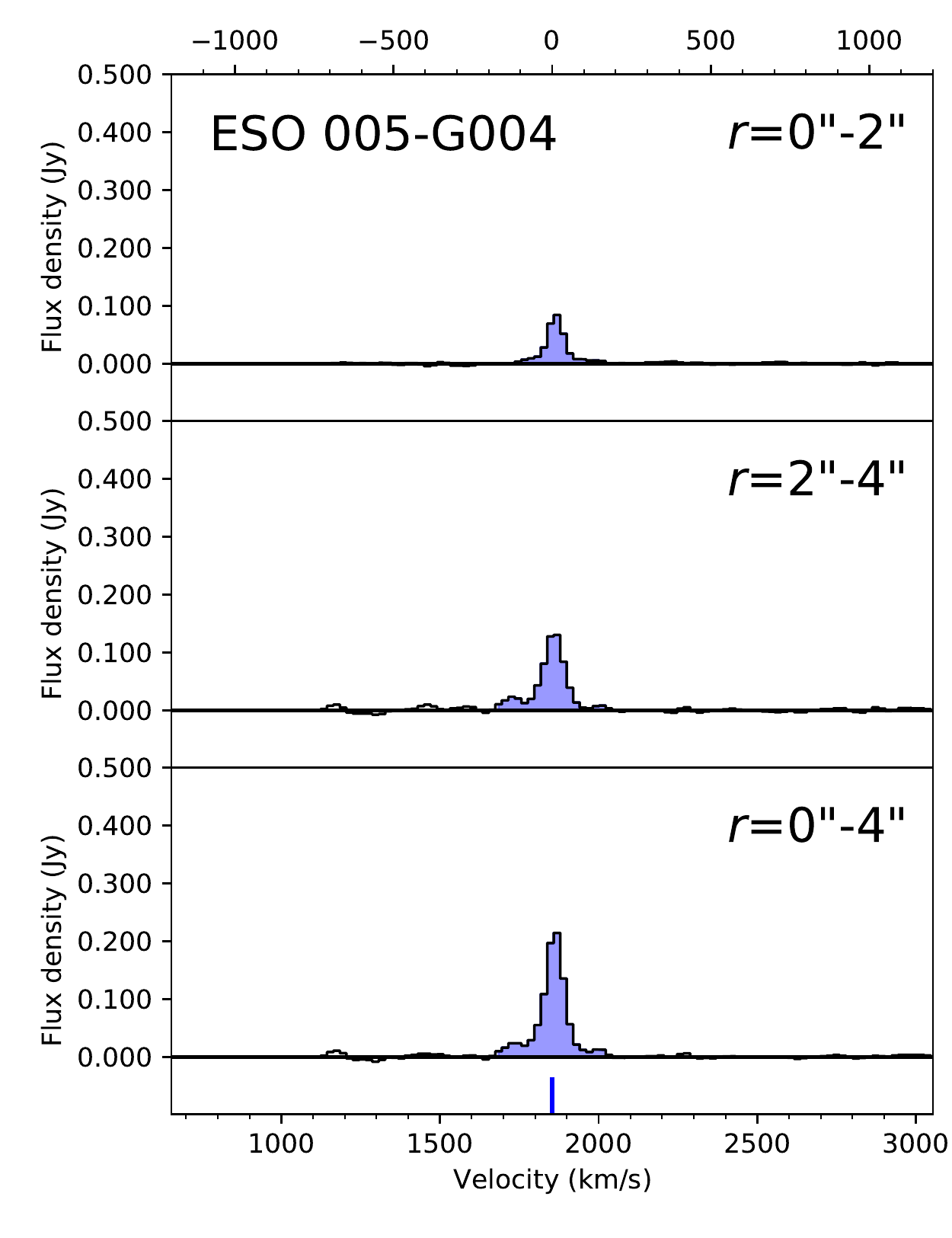}
    \includegraphics[width=4.3cm]{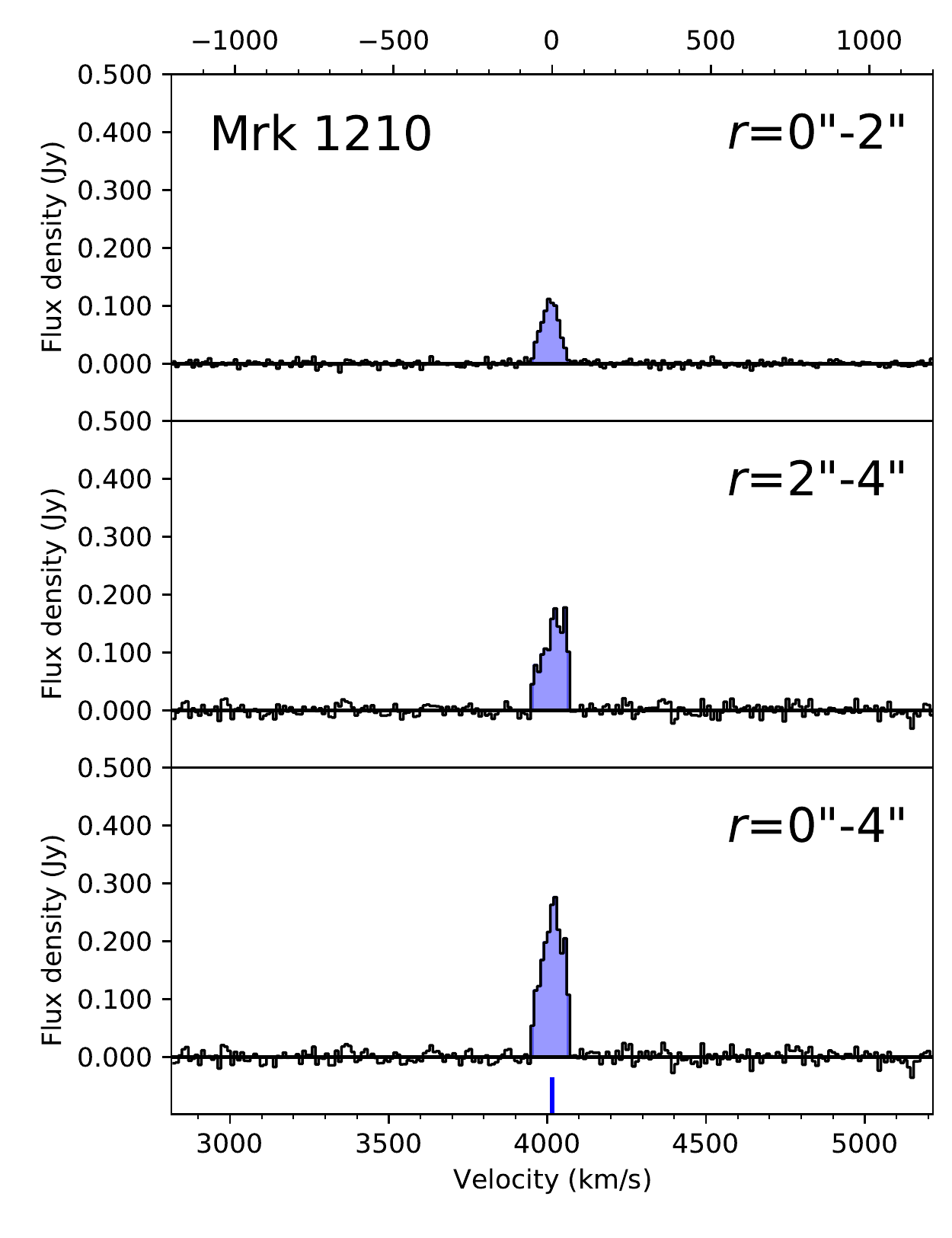} 
    \caption{Spectra around the CO($J$=2--1) emission centered at flux-weighted velocities indicated by blue vertical lines. Each figure shows three spectra taken from a nuclear ($<$ 2\arcsec) region, an external (2\arcsec--4\arcsec) region, and an entire ($<$ 4\arcsec) region, from top to bottom. 
    The CO($J$=2--1) luminosities in Table~\ref{tab:co_data} are derived from the
    blue shaded areas. 
    }
    \label{app:fig:co_spec}
\end{figure*}

\begin{figure*}\addtocounter{figure}{-1}
    \centering
    \includegraphics[width=4.3cm]{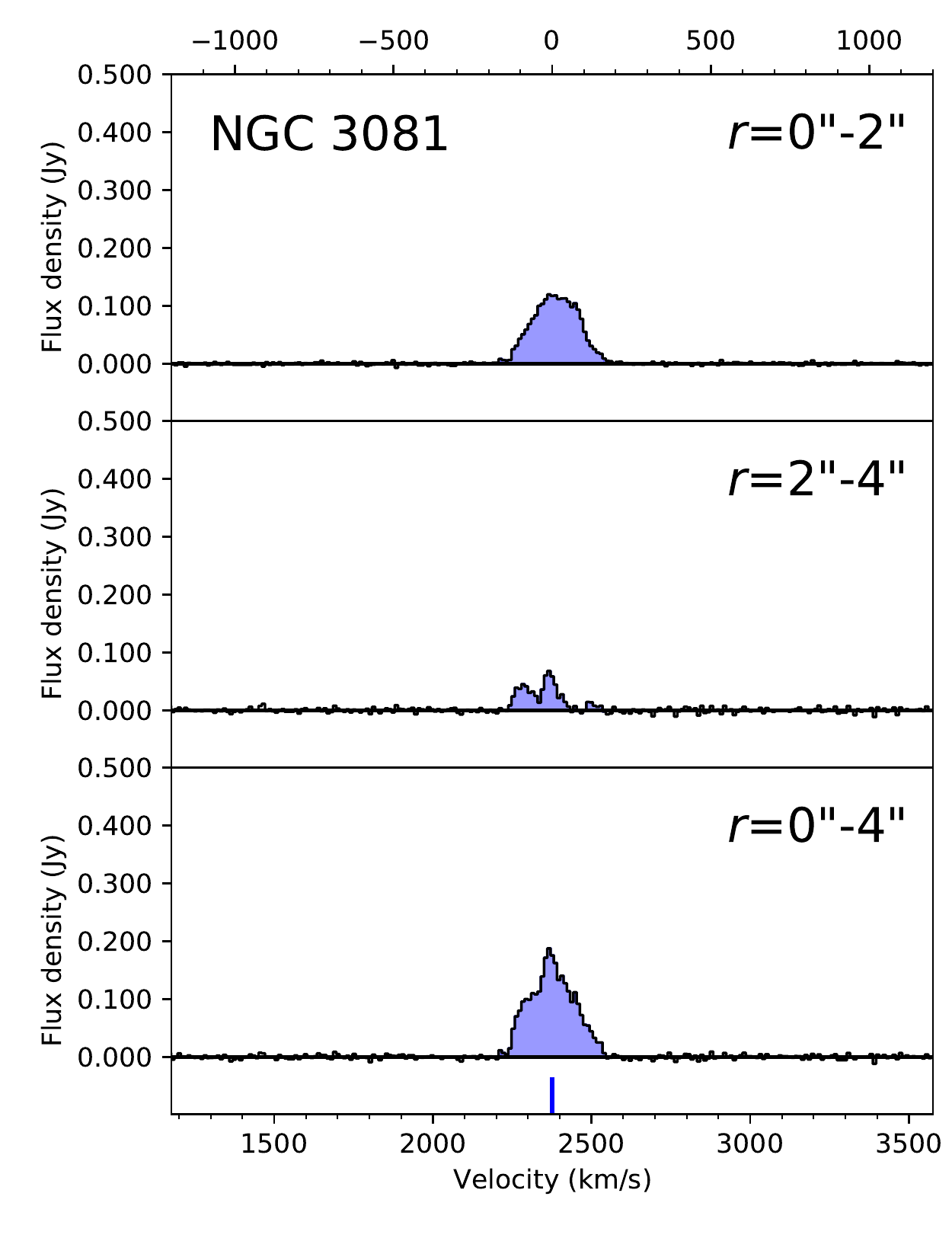}
    \includegraphics[width=4.3cm]{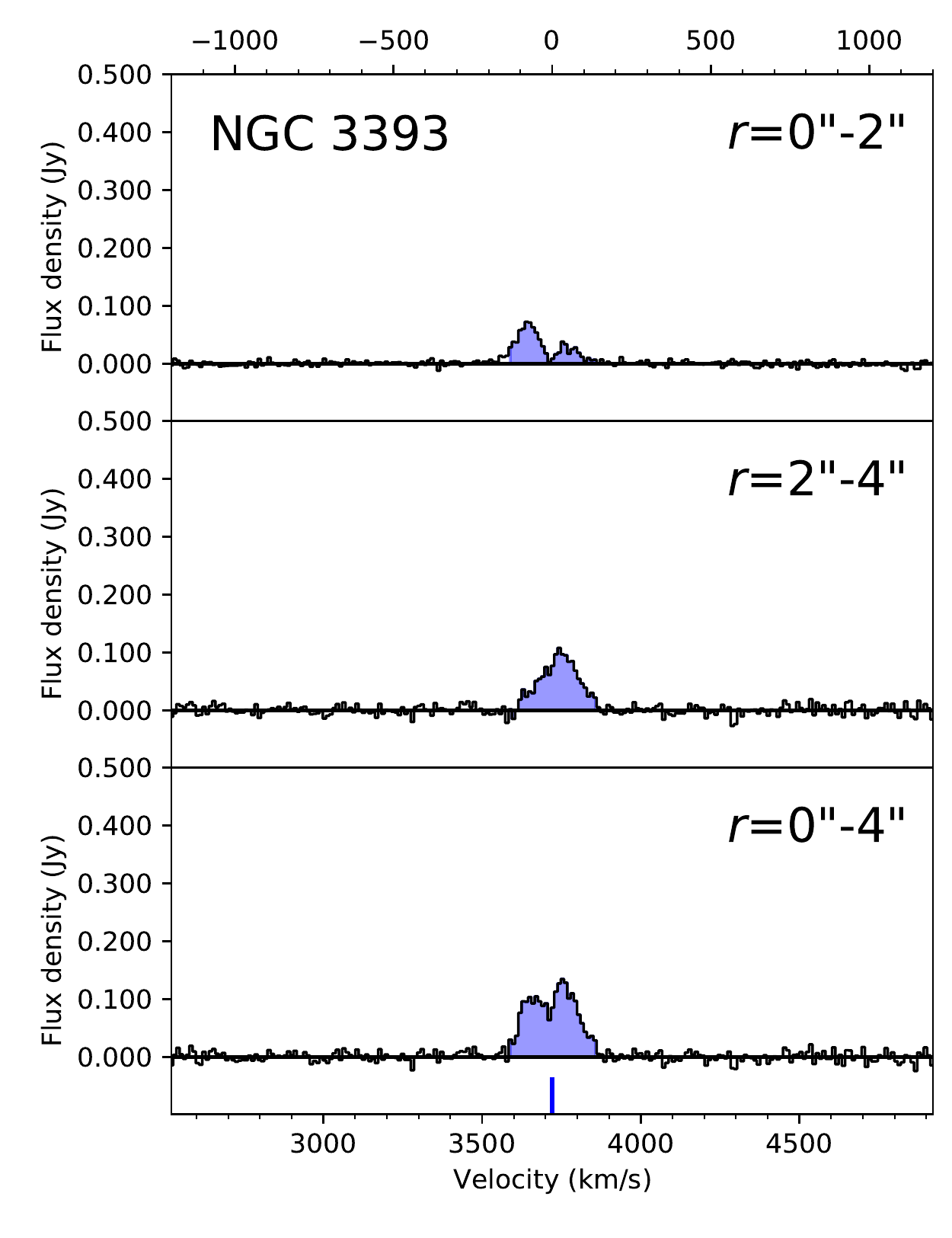}
    \includegraphics[width=4.3cm]{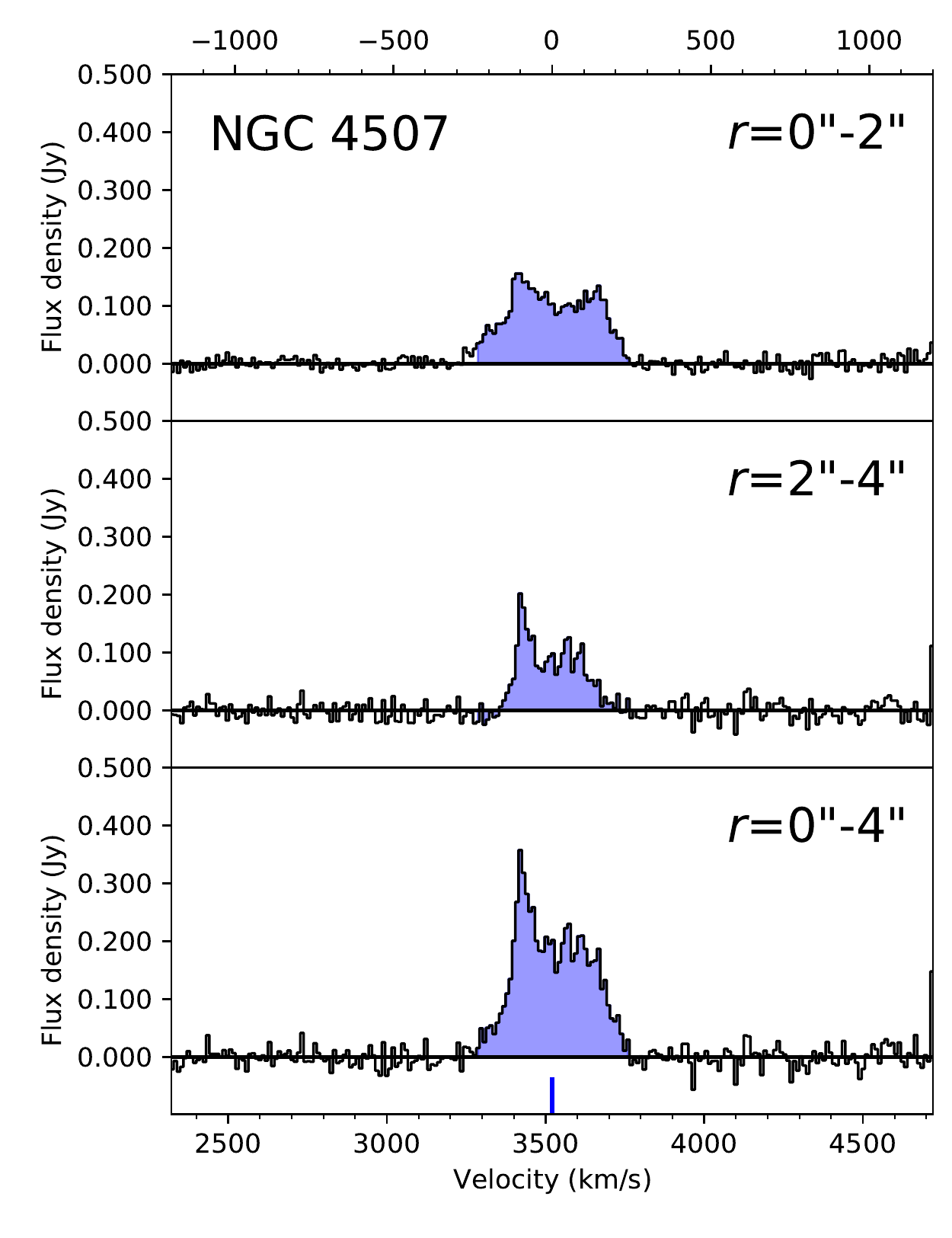}
    \includegraphics[width=4.3cm]{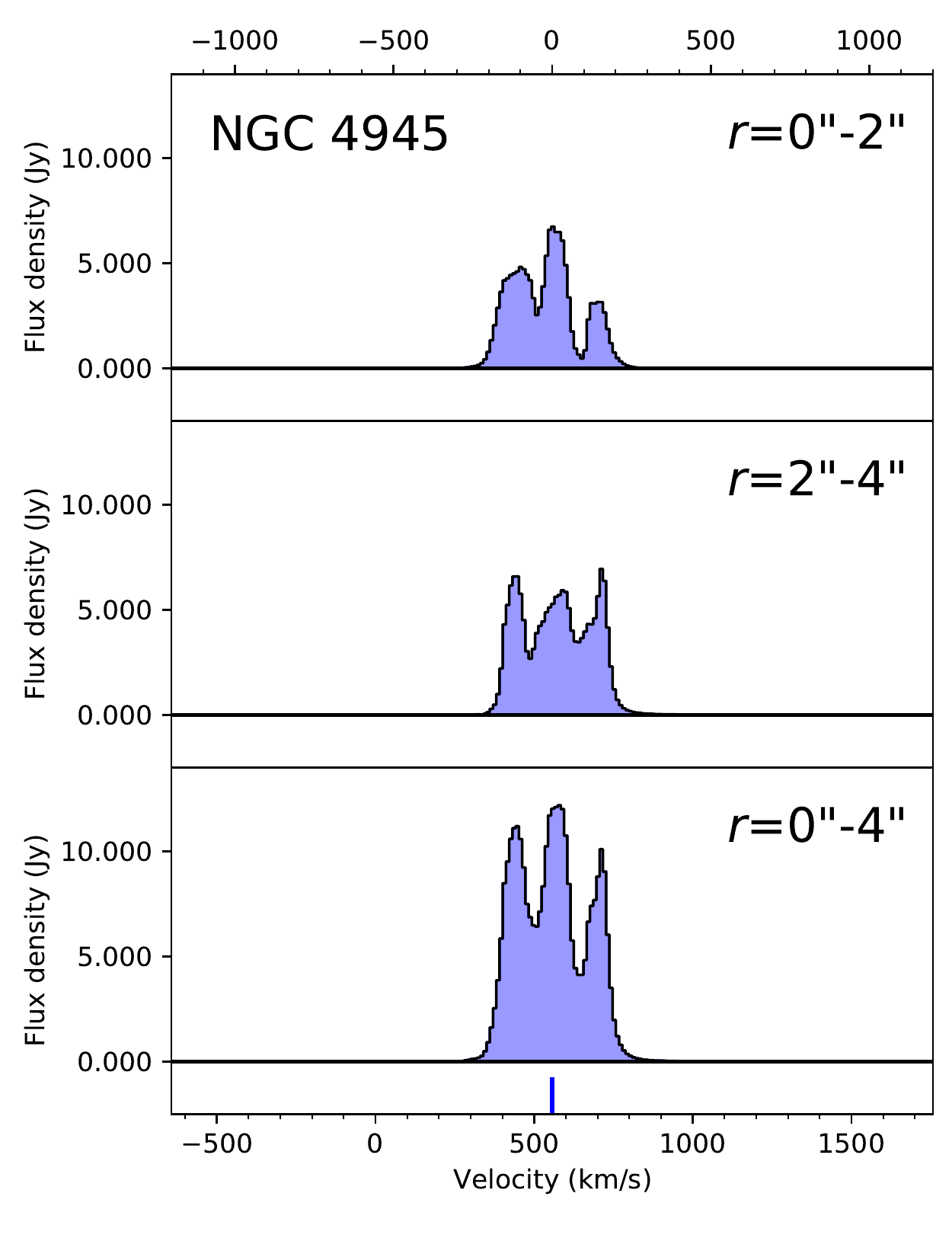}    
    \includegraphics[width=4.3cm]{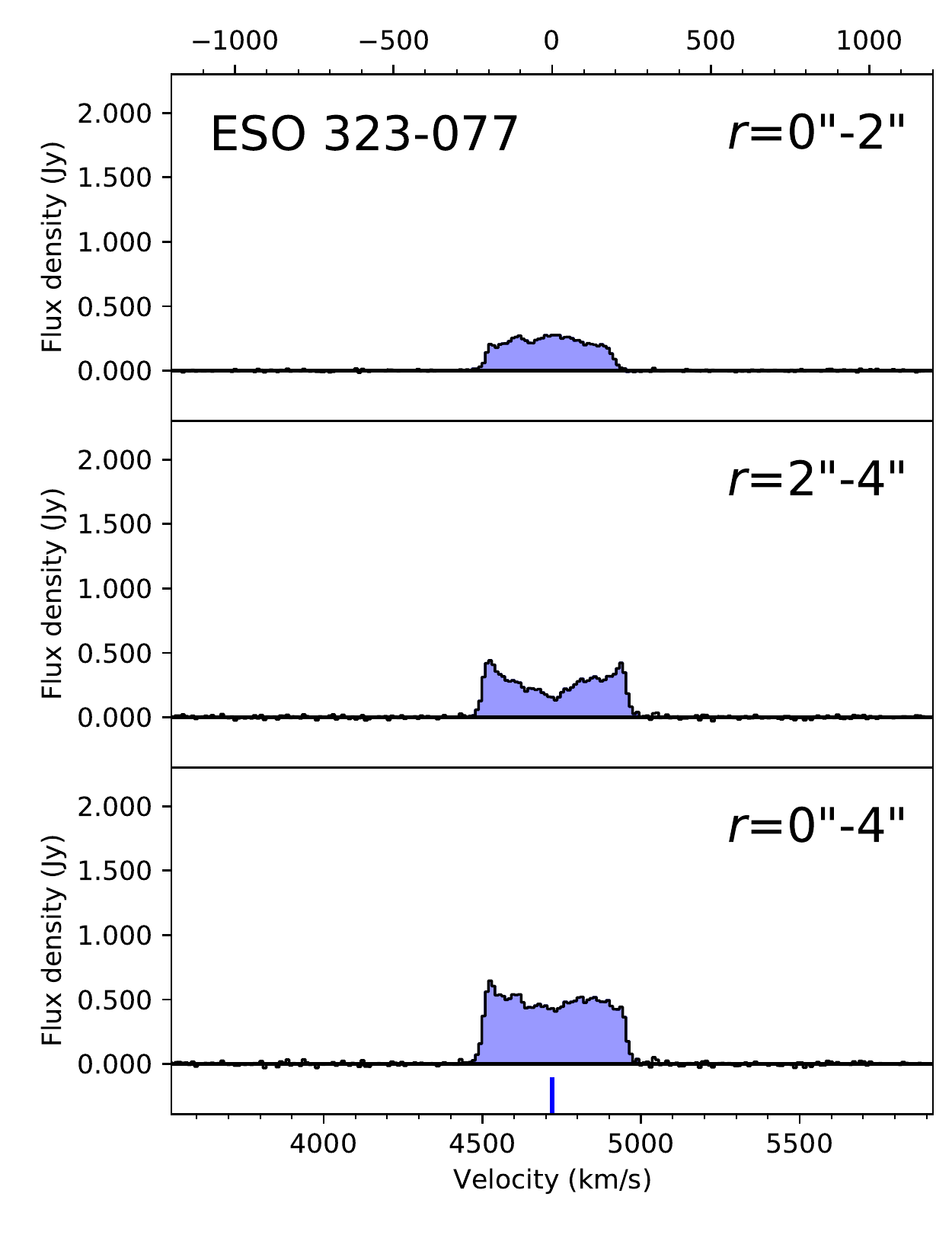}
    \includegraphics[width=4.3cm]{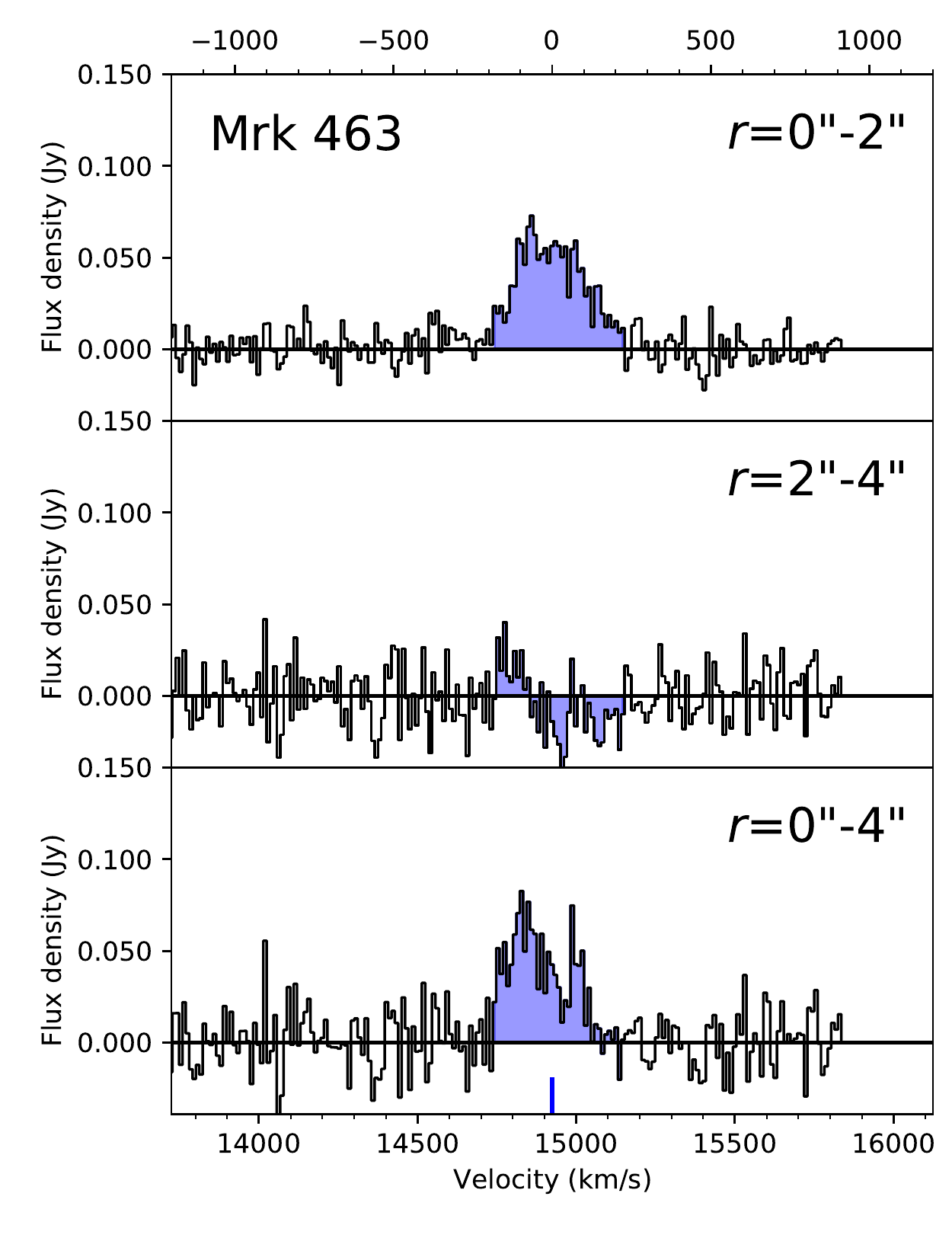}
    \includegraphics[width=4.3cm]{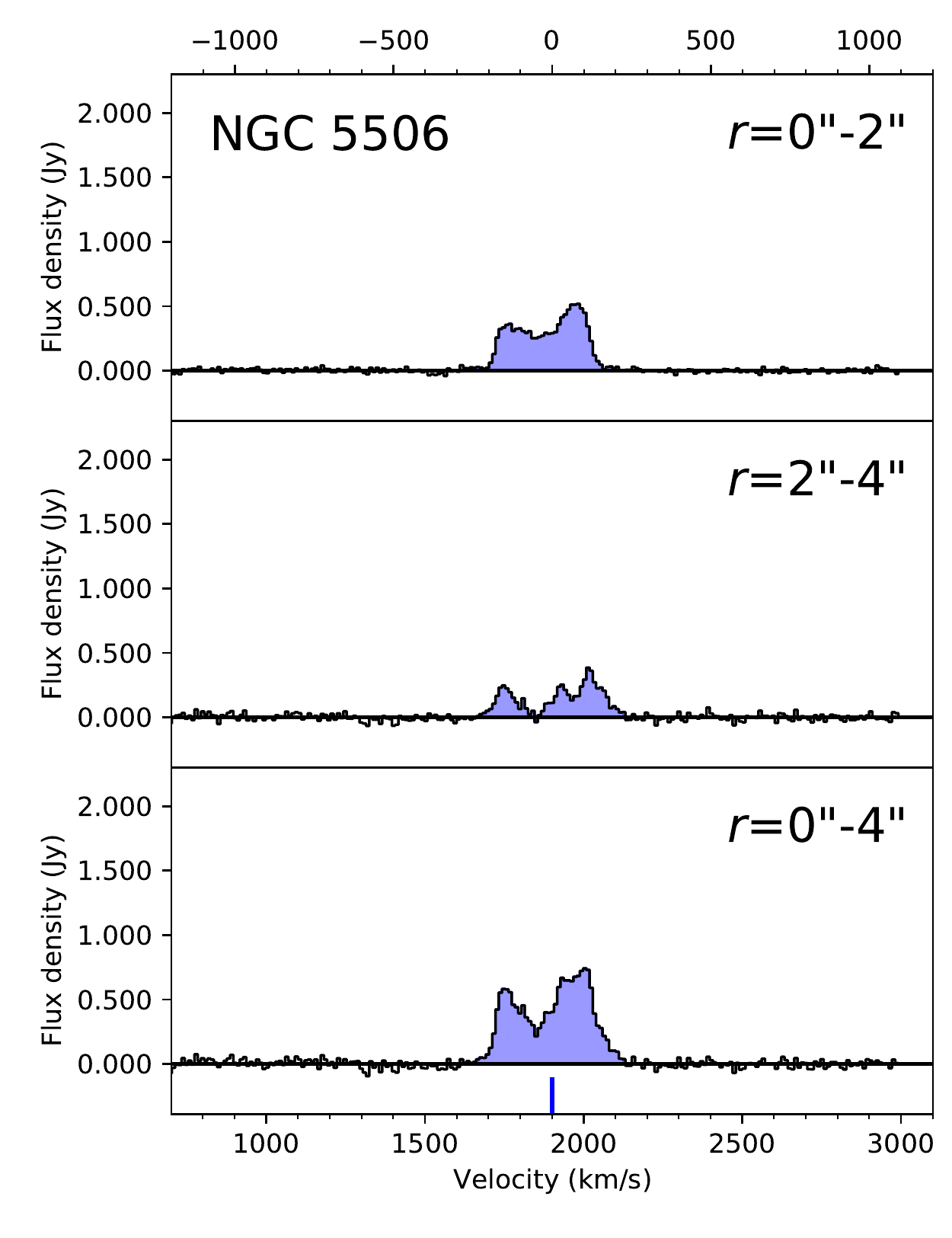}
    \includegraphics[width=4.3cm]{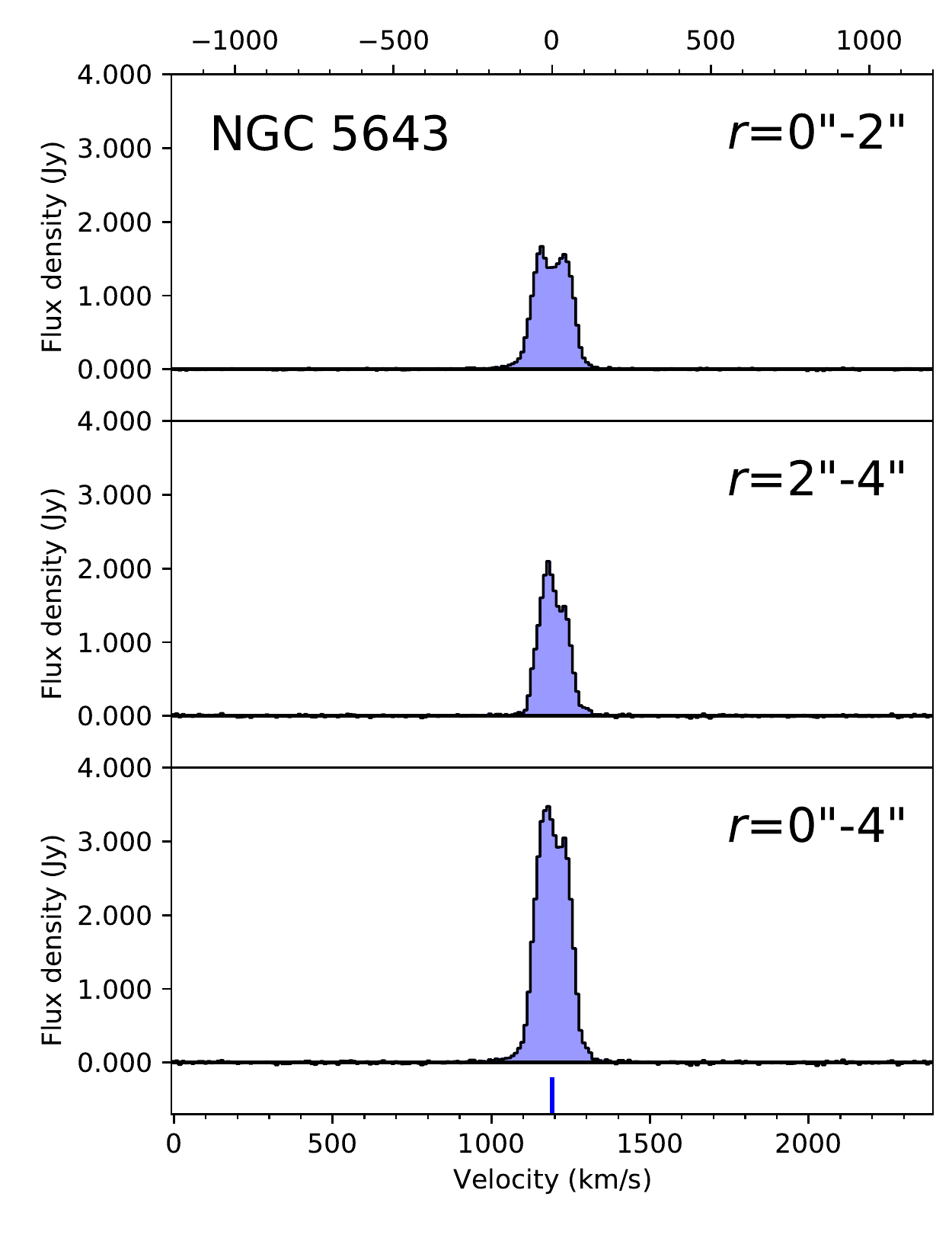}
    \includegraphics[width=4.3cm]{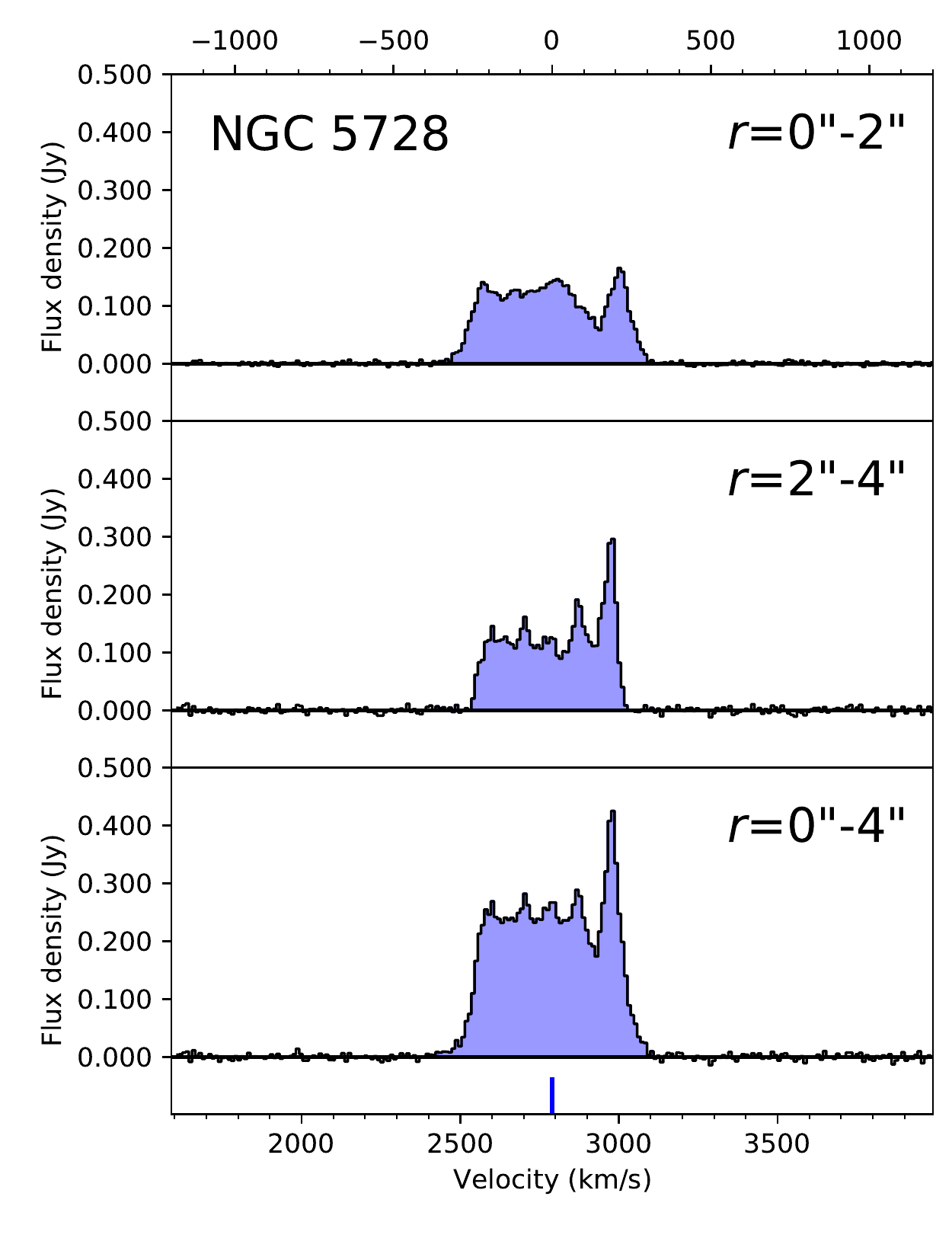}
    \includegraphics[width=4.3cm]{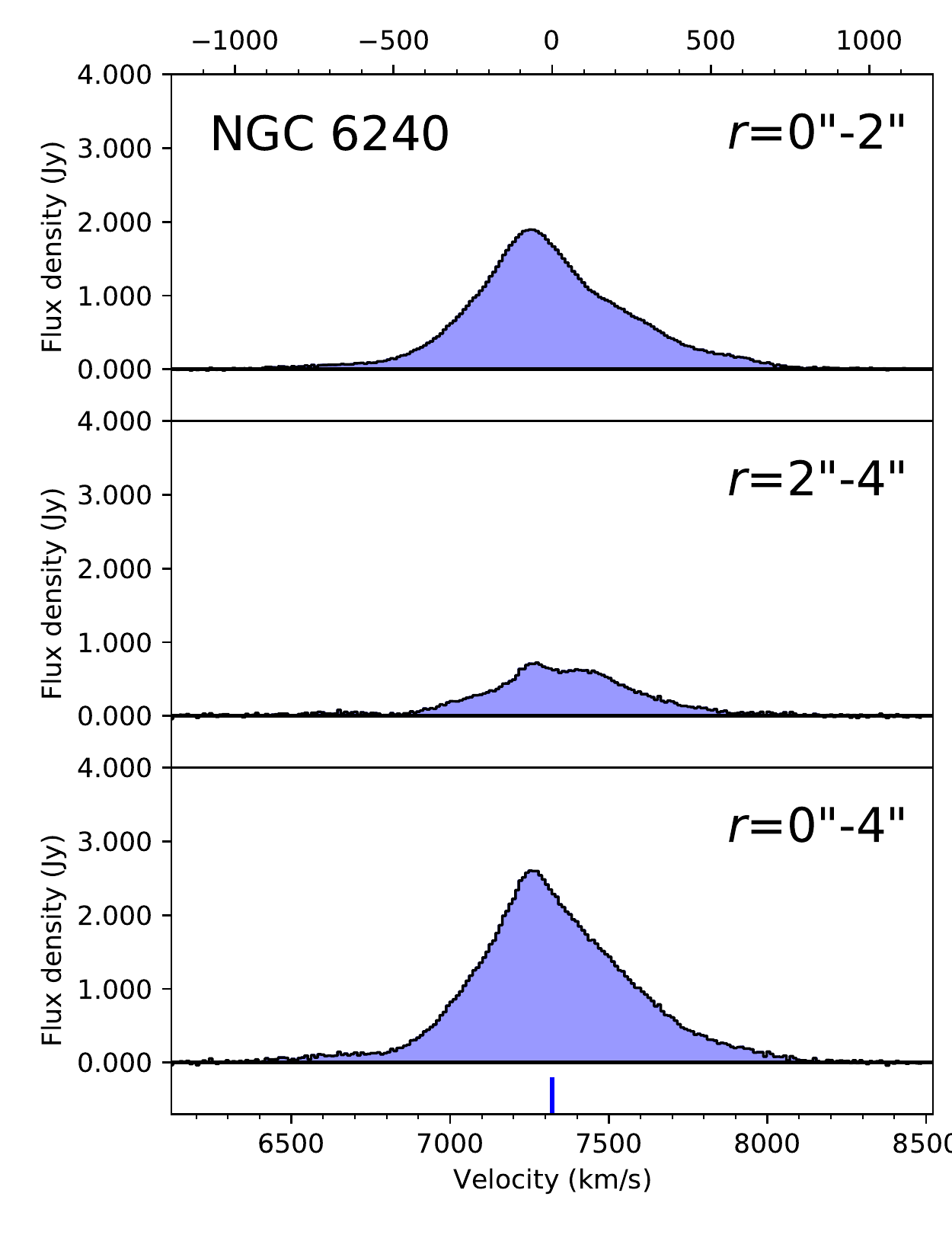}
    \includegraphics[width=4.3cm]{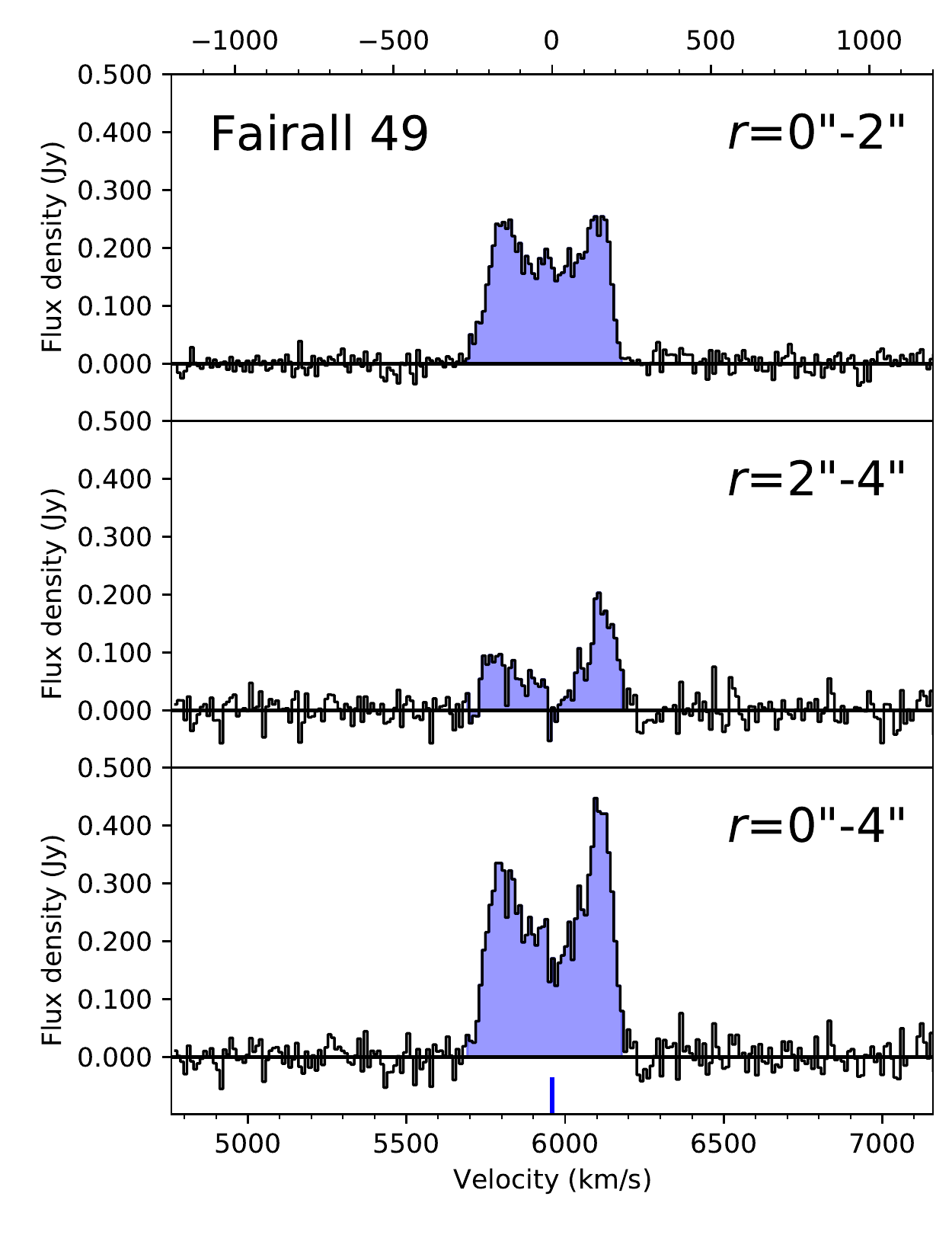}
    \includegraphics[width=4.3cm]{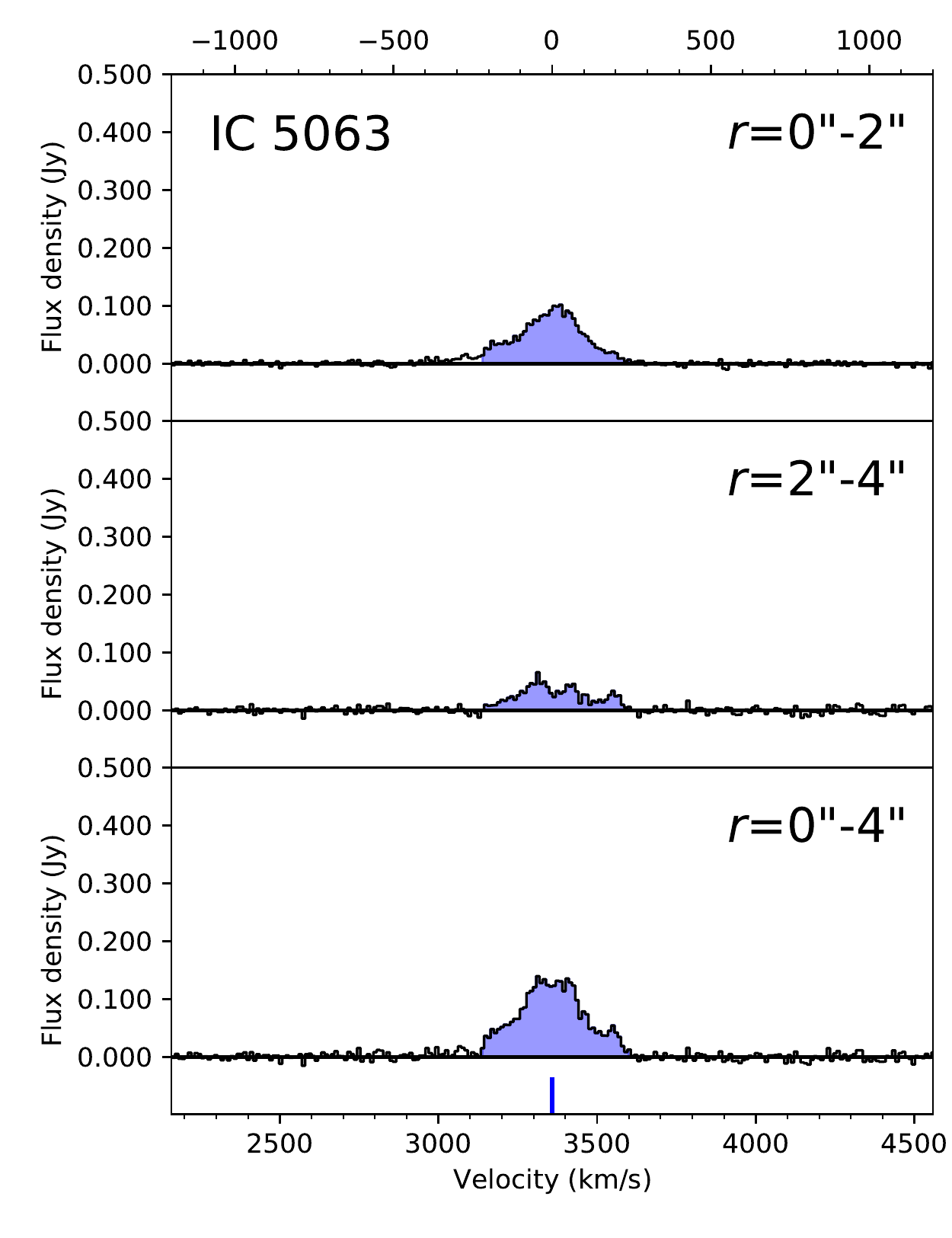}
    \includegraphics[width=4.3cm]{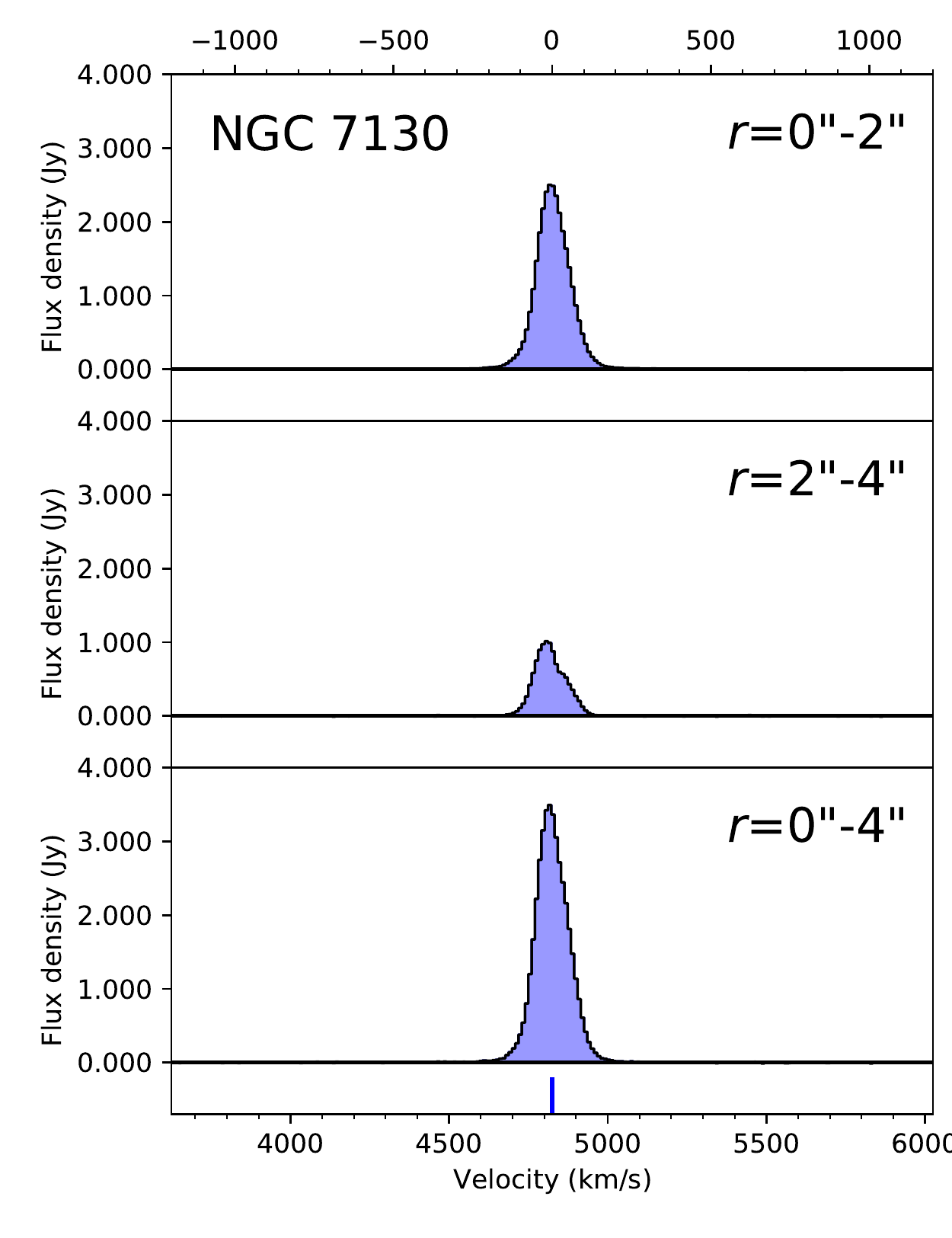}
    \includegraphics[width=4.3cm]{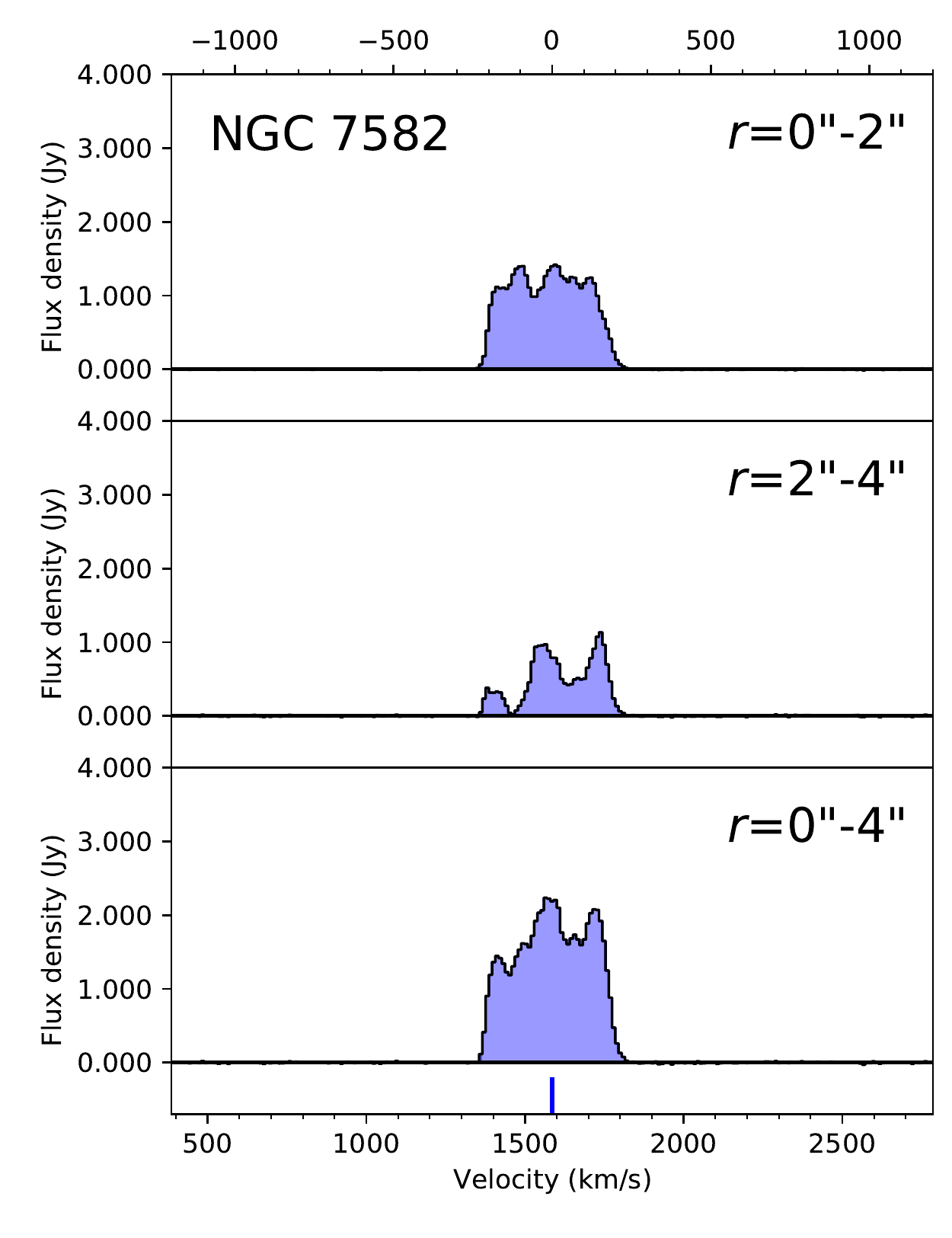}
    \caption{Continued.}
\end{figure*}

\clearpage

\section{CO($J$=2--1) images}\label{app:co_image}

\begin{figure*}[!h]
    \centering
    \includegraphics[width=5.8cm]{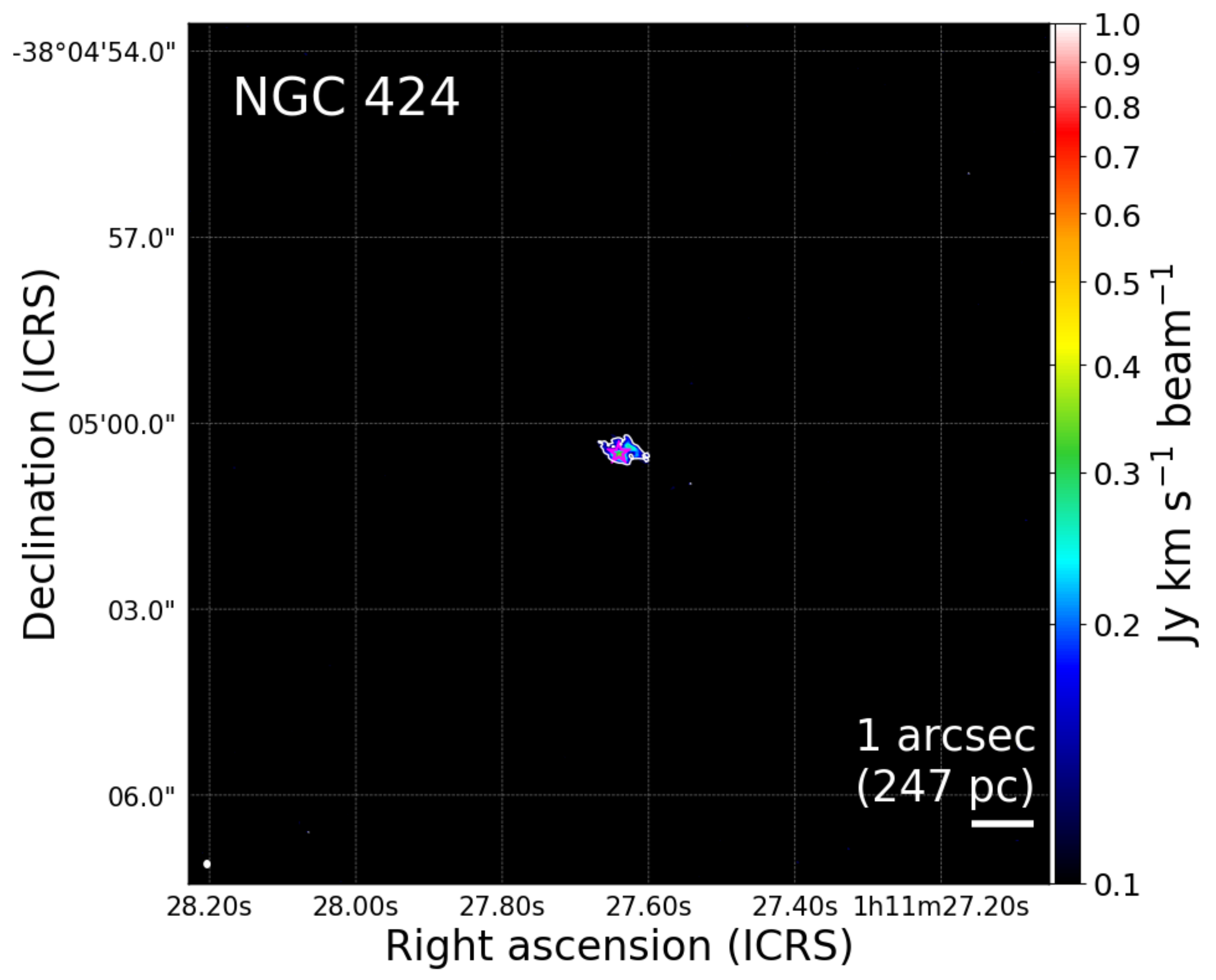}
    \includegraphics[width=5.8cm]{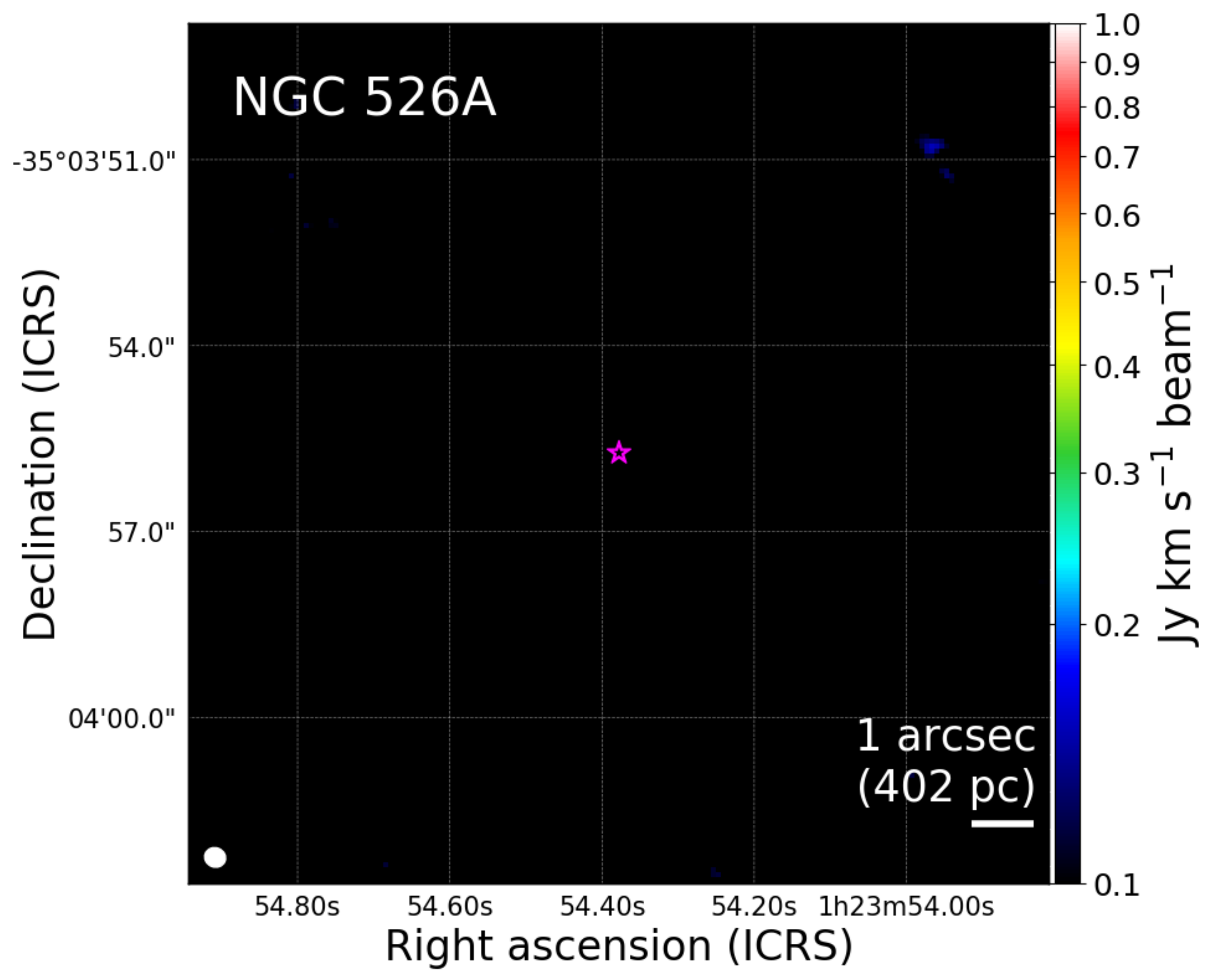}
    \includegraphics[width=5.8cm]{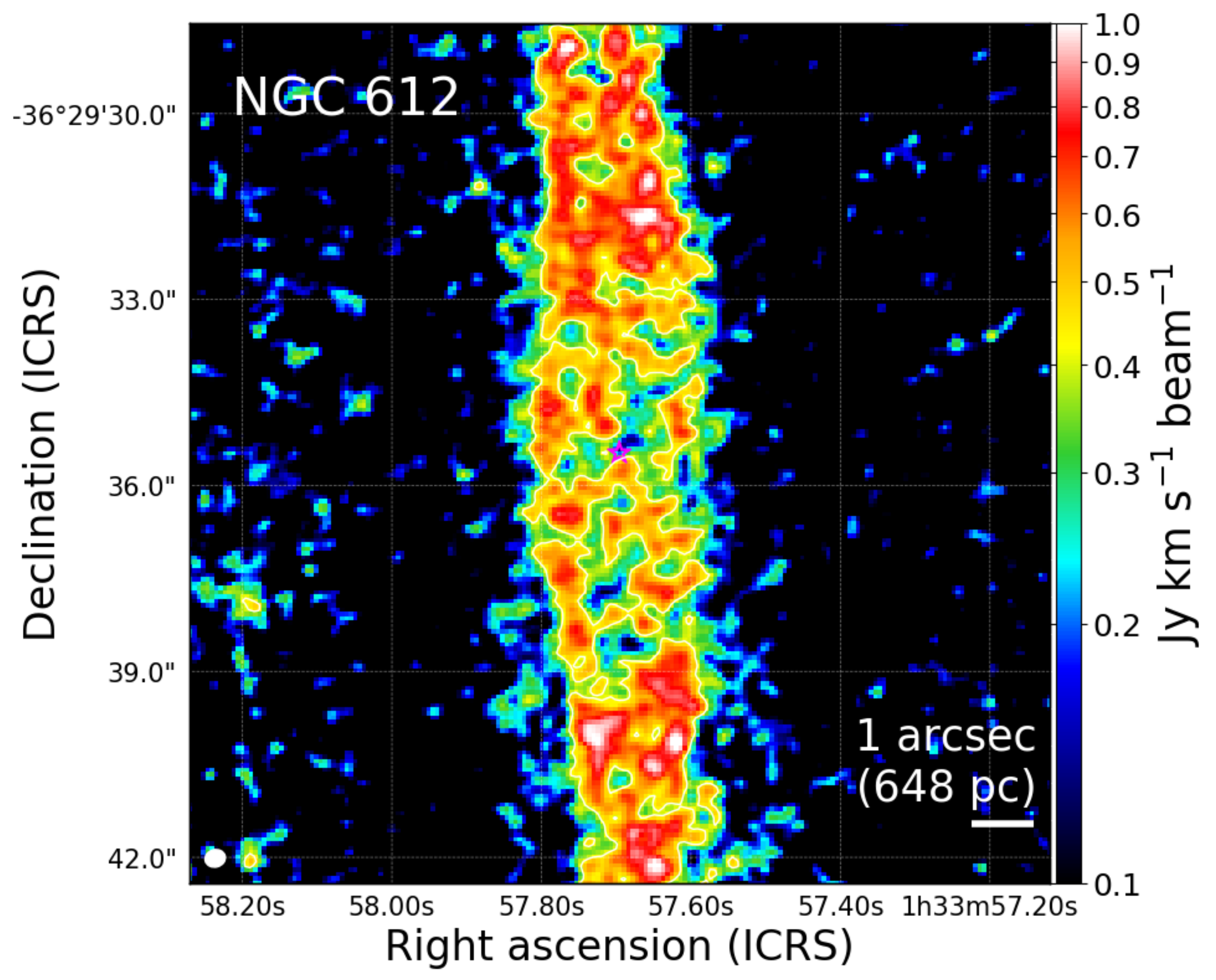}
    \includegraphics[width=5.8cm]{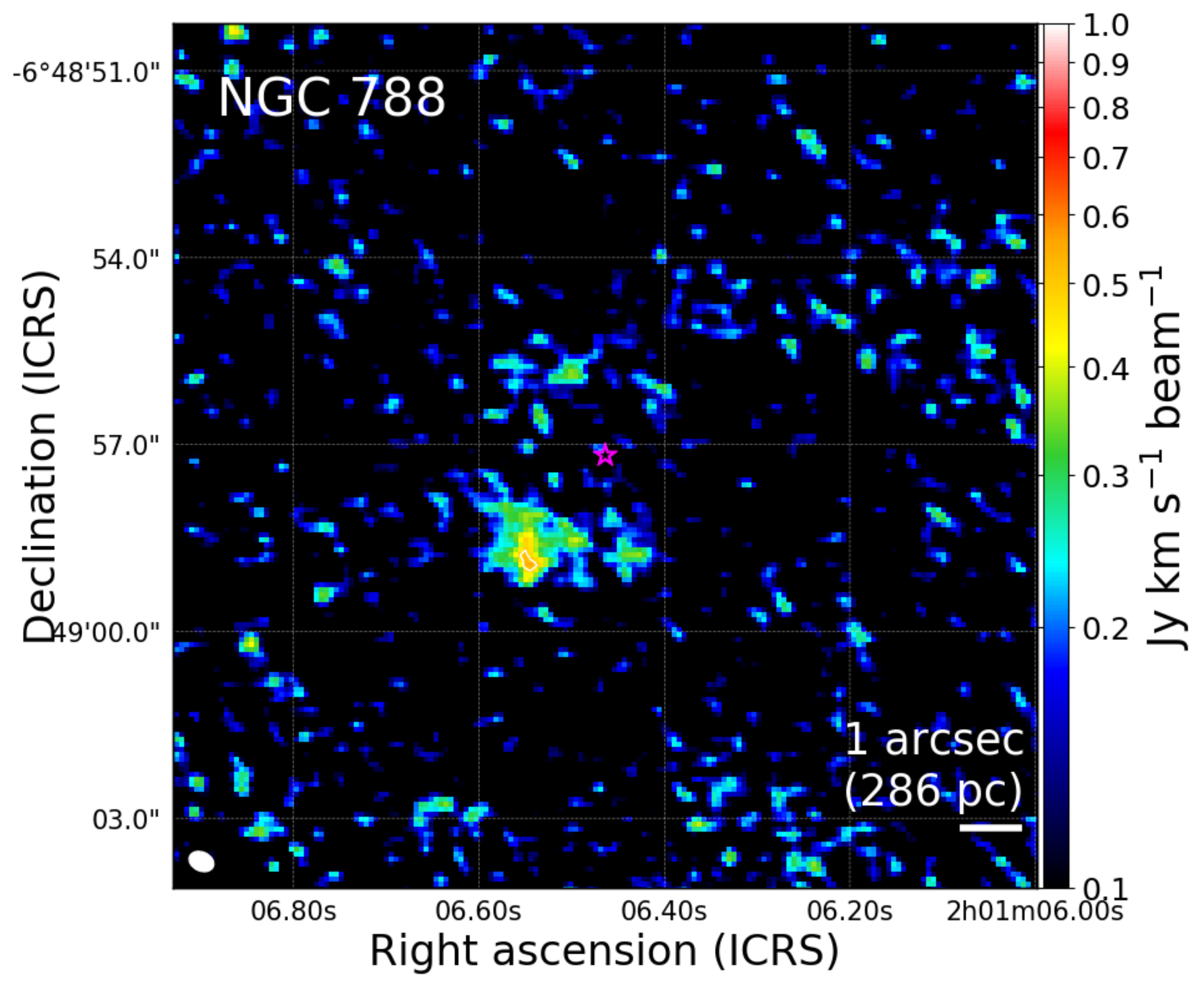}
    \includegraphics[width=5.8cm]{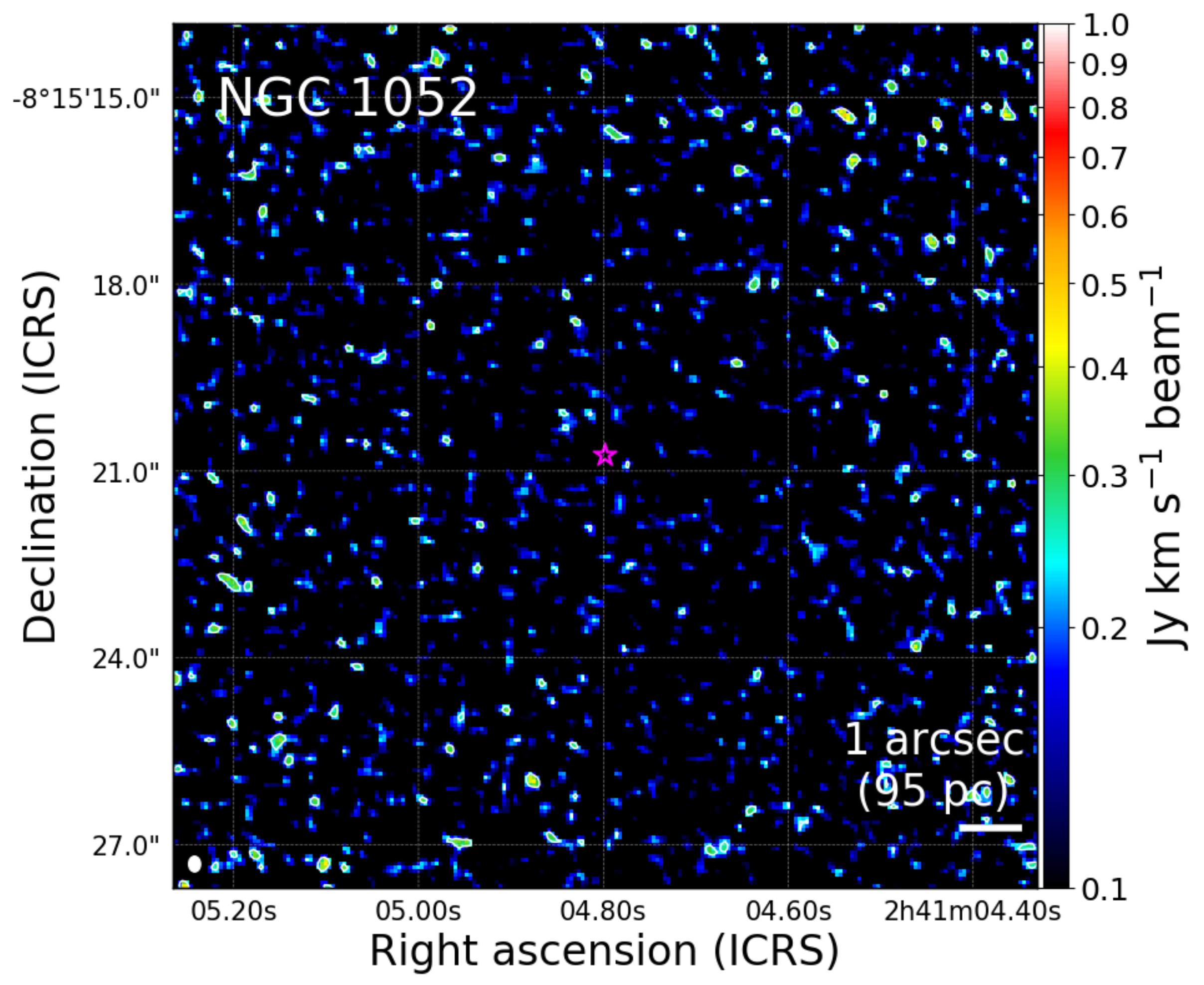}
    \includegraphics[width=5.8cm]{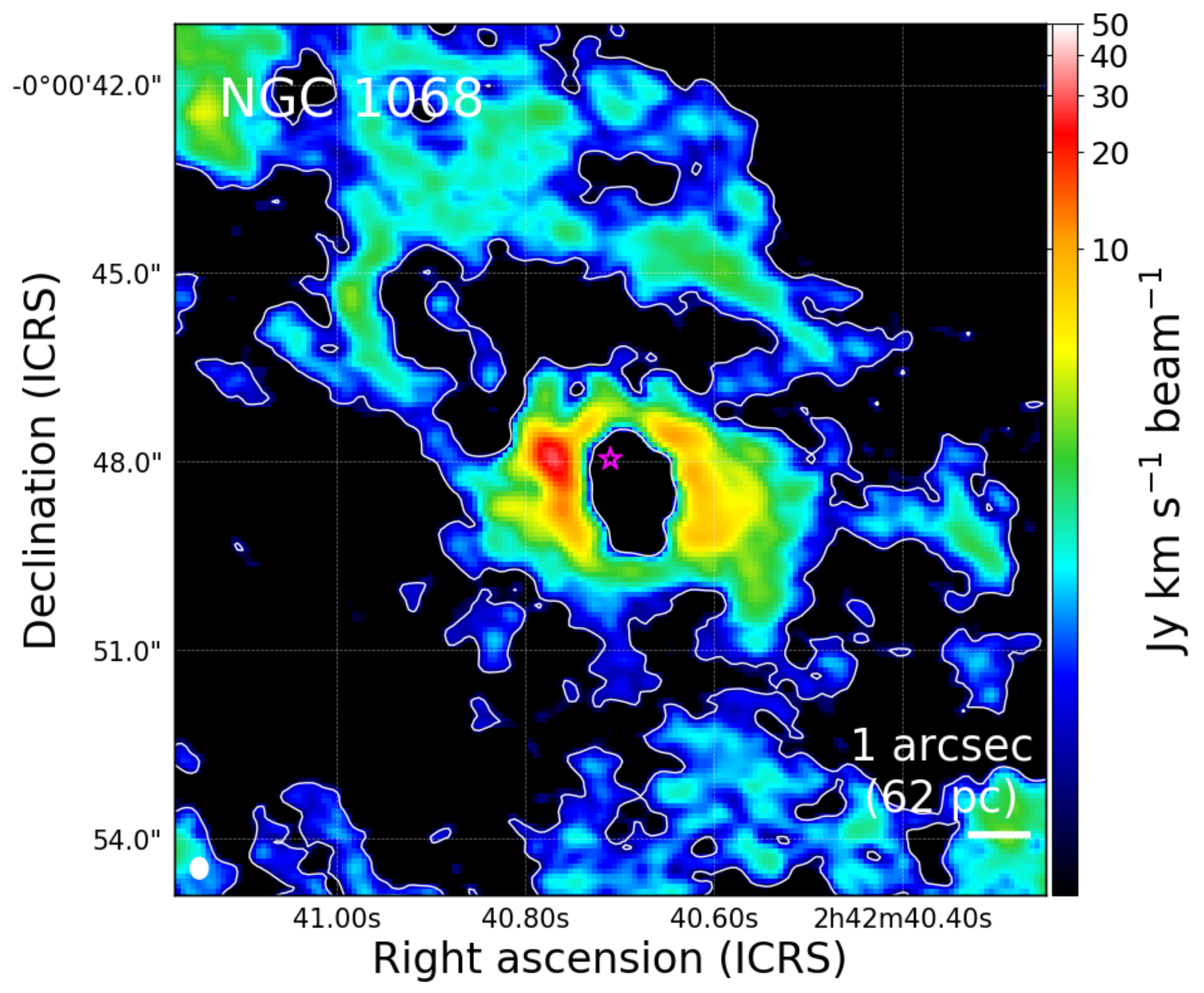}
    \includegraphics[width=5.8cm]{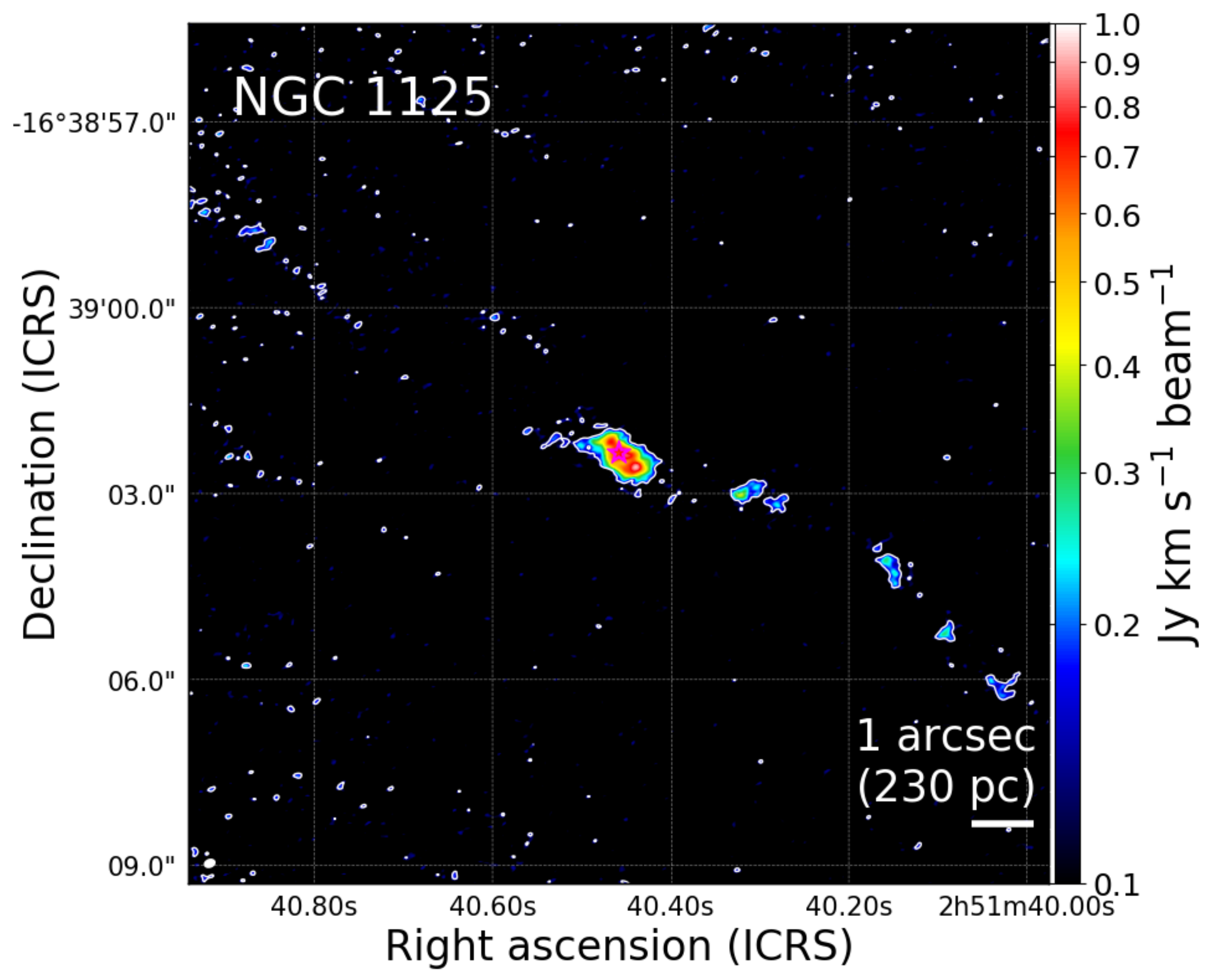}
    \includegraphics[width=5.8cm]{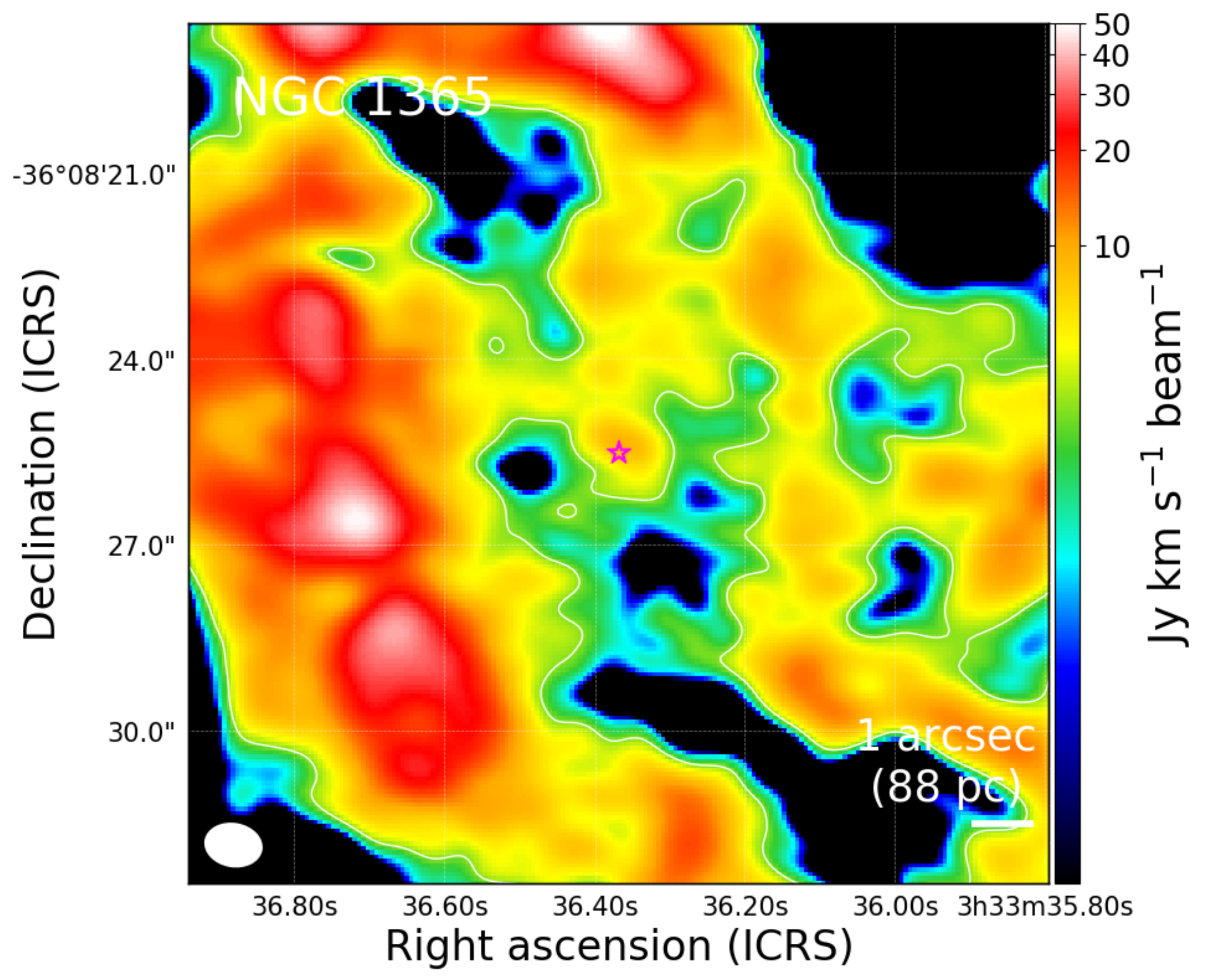}
    \includegraphics[width=5.8cm]{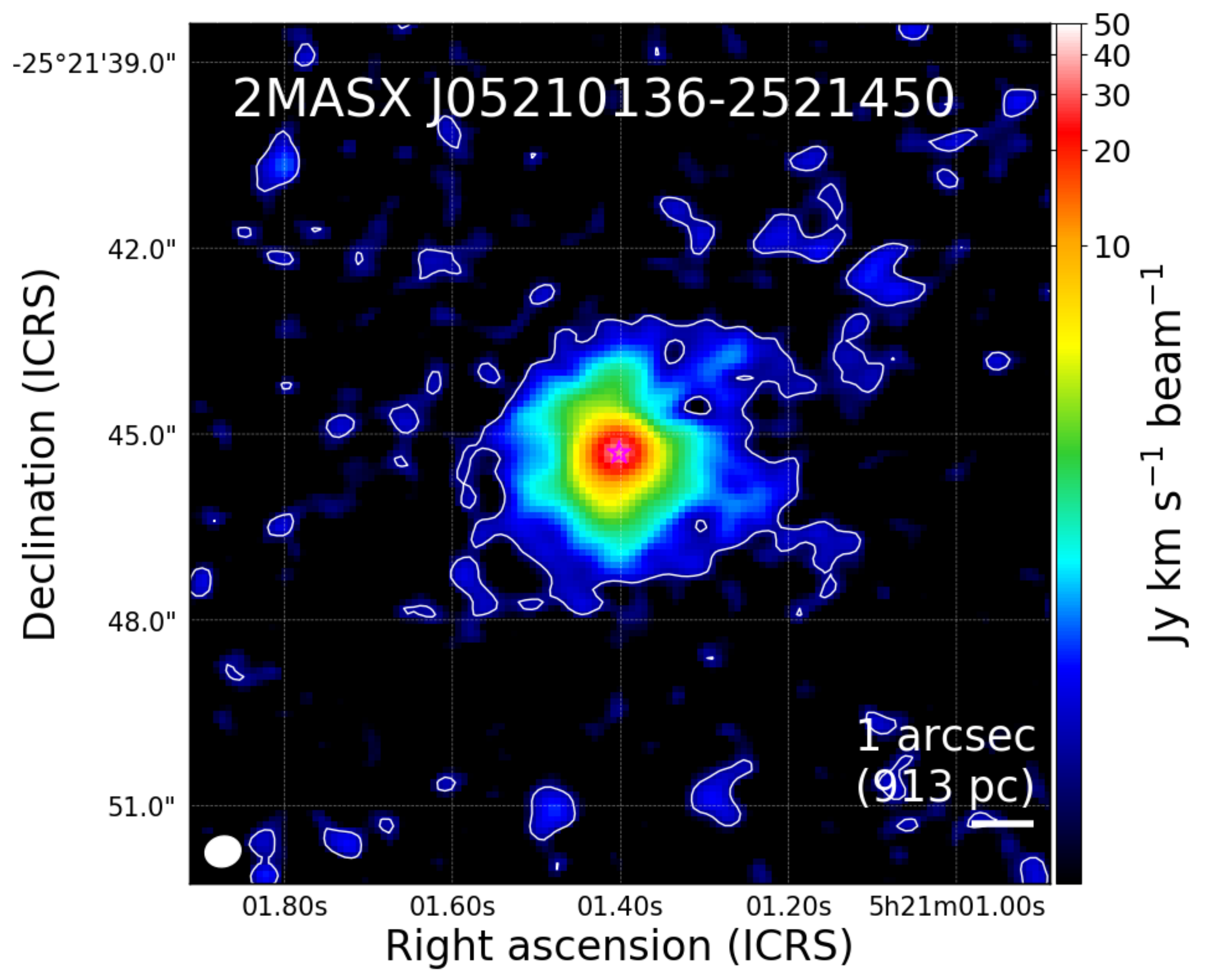}
    \includegraphics[width=5.8cm]{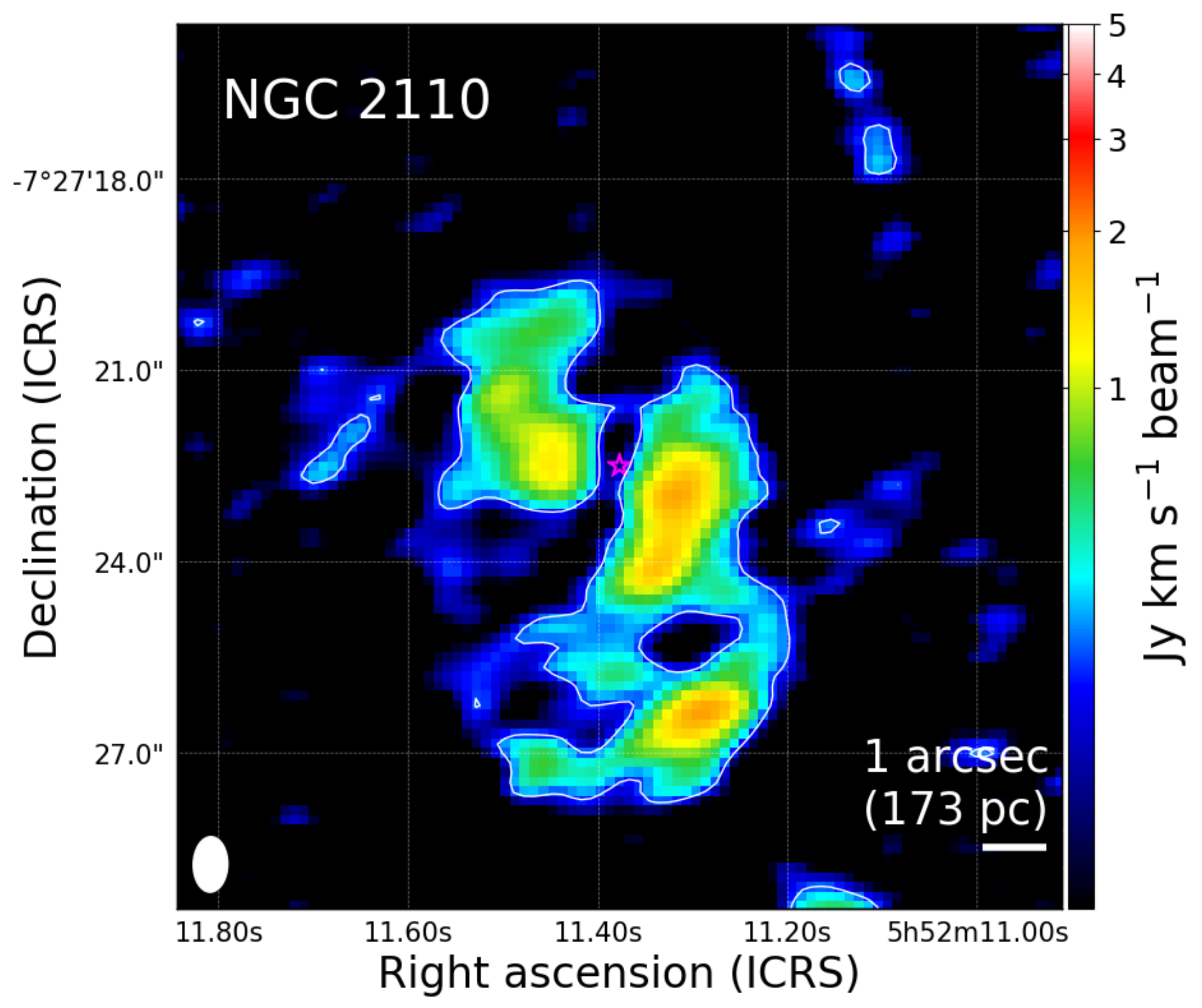}
    \includegraphics[width=5.8cm]{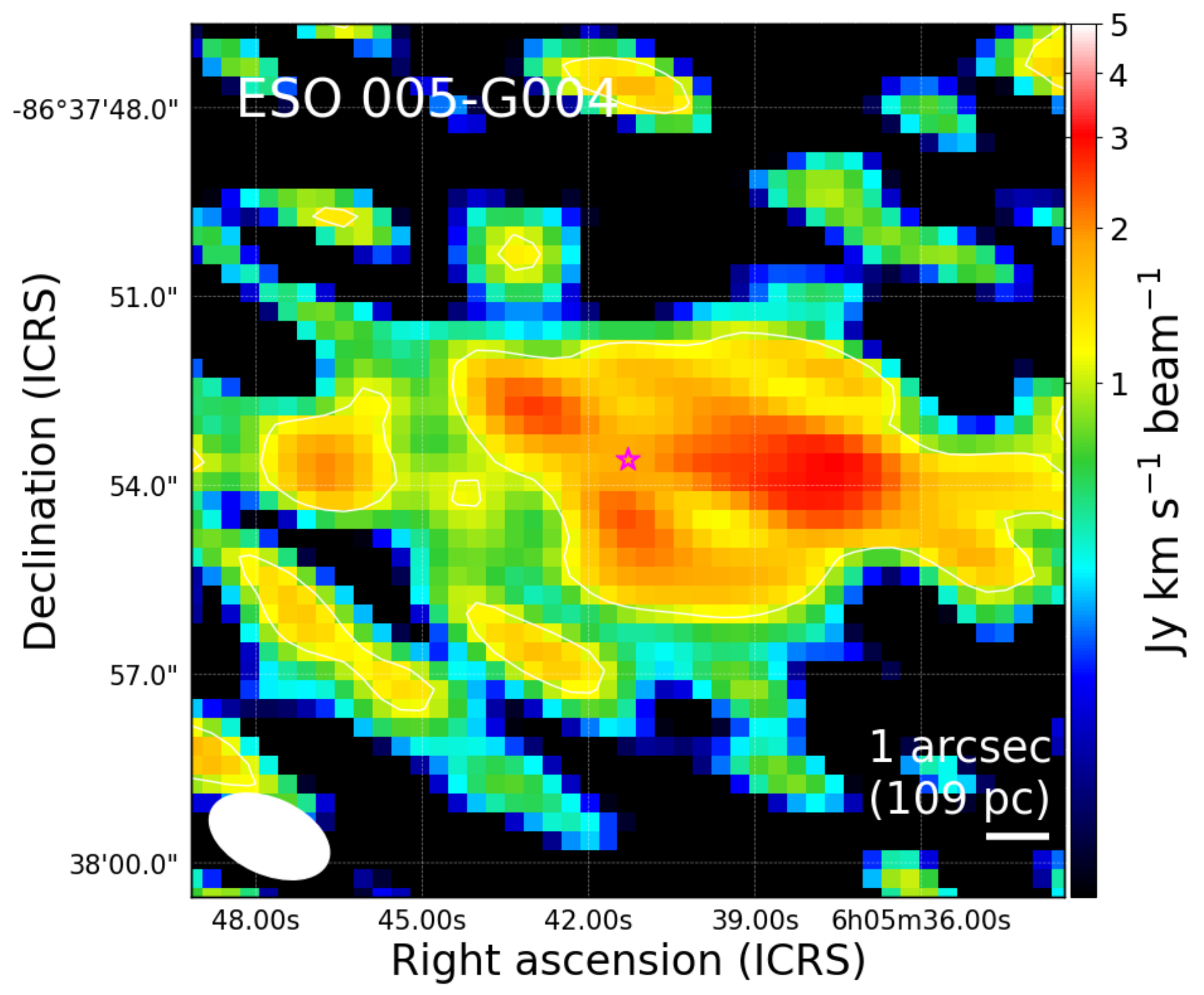}
    \includegraphics[width=5.8cm]{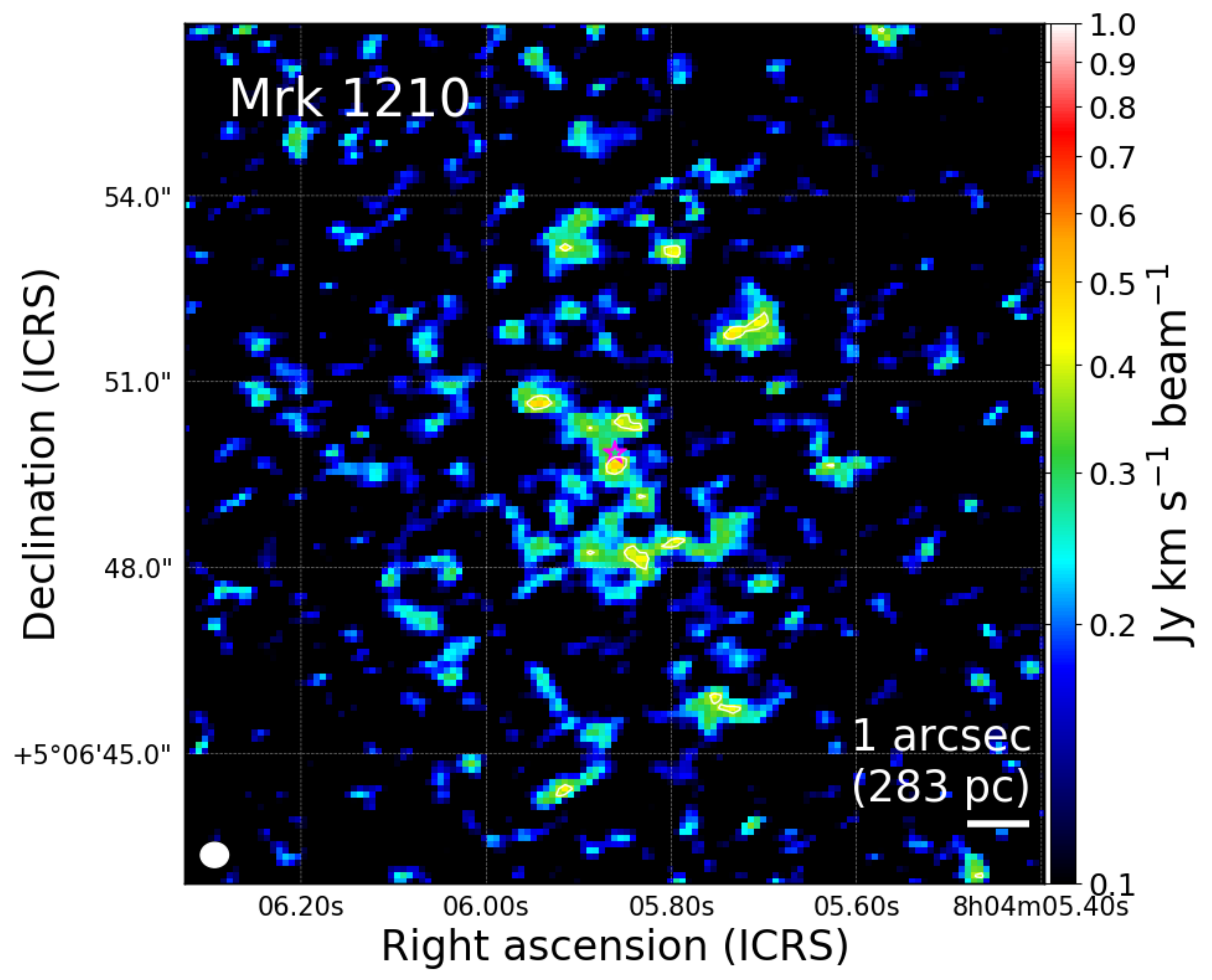}
    \caption{Moment 0 maps of the CO($J$=2--1) emission. 
    North is up and east is left.
    White contours are plotted to indicate 5$\sigma$, where 1$\sigma$ flux densities in units of Jy km s$^{-1}$ beam$^{-1}$ are listed in the column (8) of Table~\ref{tab:alma_data}. In addition, beam sizes are plotted 
    at the bottom-left corners.
    }
    \label{app:fig:co_images}
\end{figure*}

\begin{figure*}\addtocounter{figure}{-1}
    \centering
    \includegraphics[width=5.8cm]{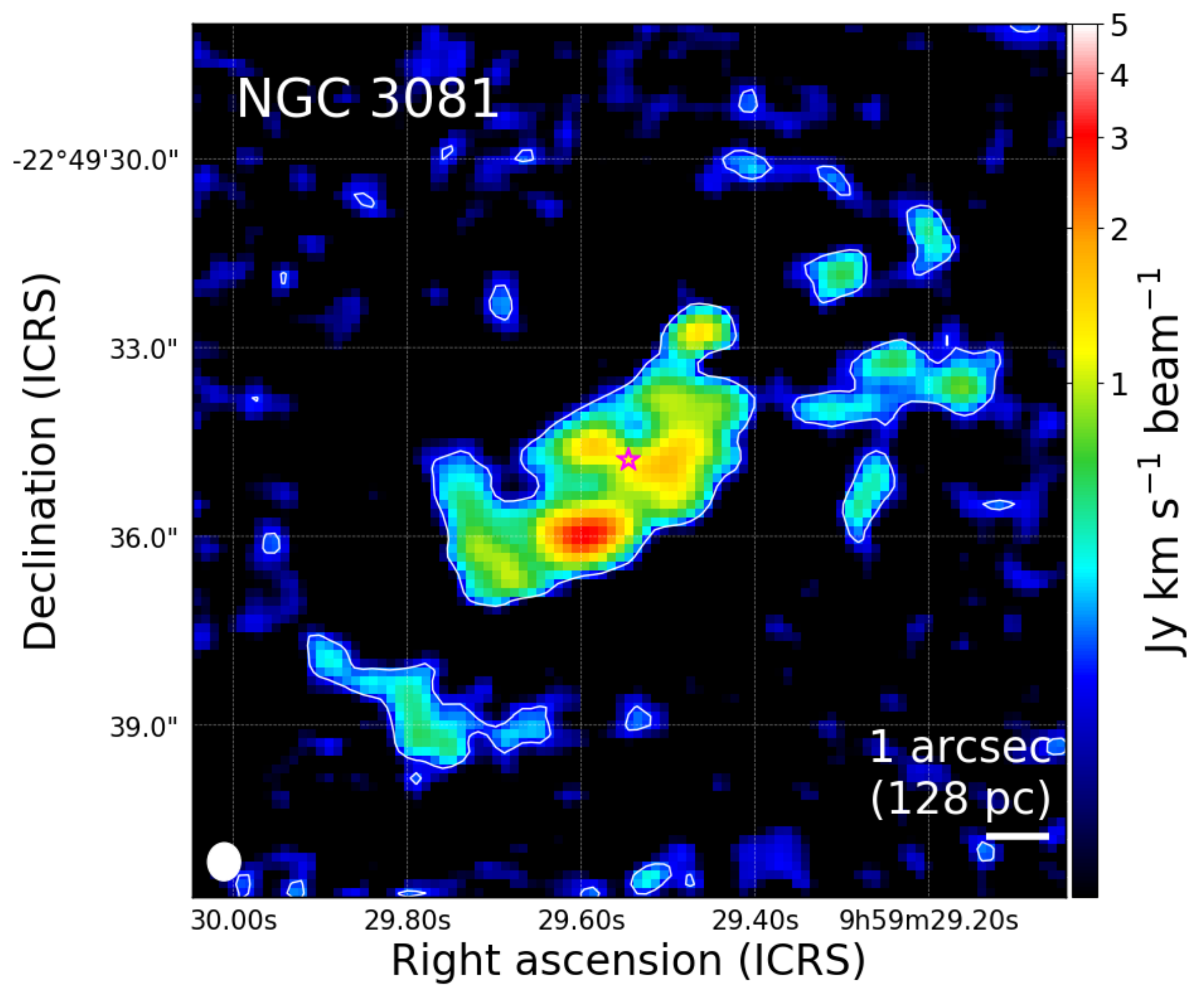}
    \includegraphics[width=5.8cm]{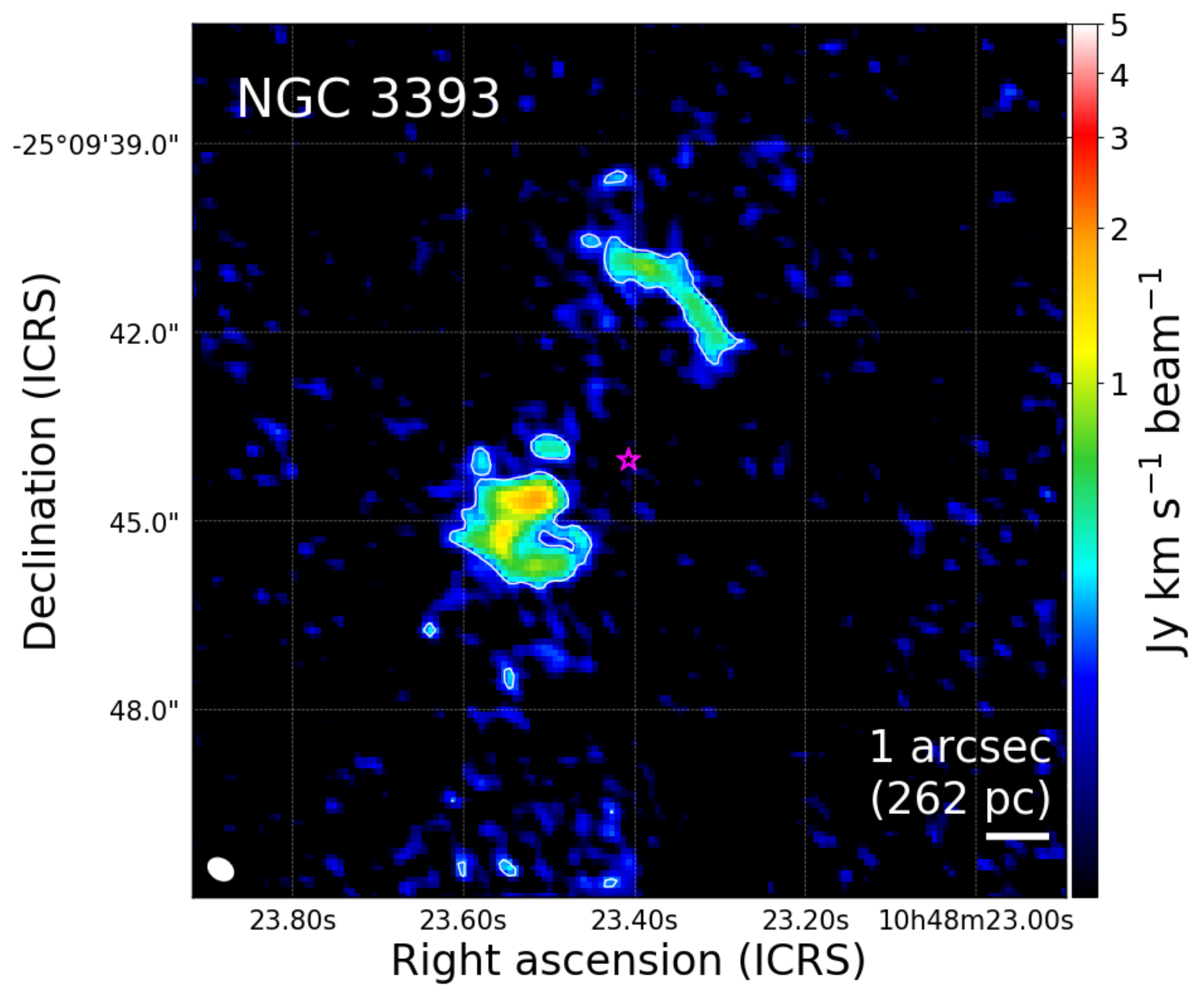}
    \includegraphics[width=5.8cm]{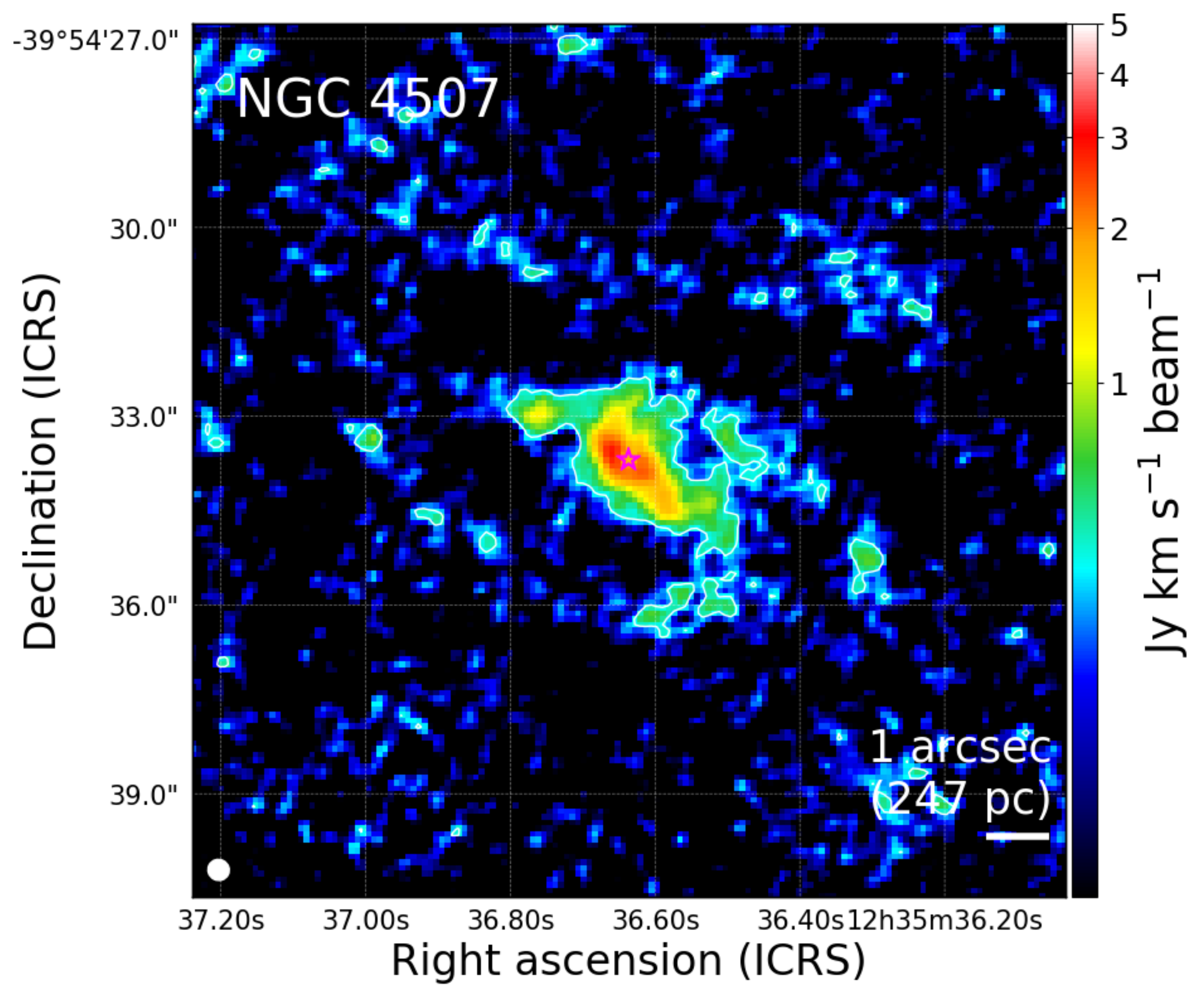}
    \includegraphics[width=5.8cm]{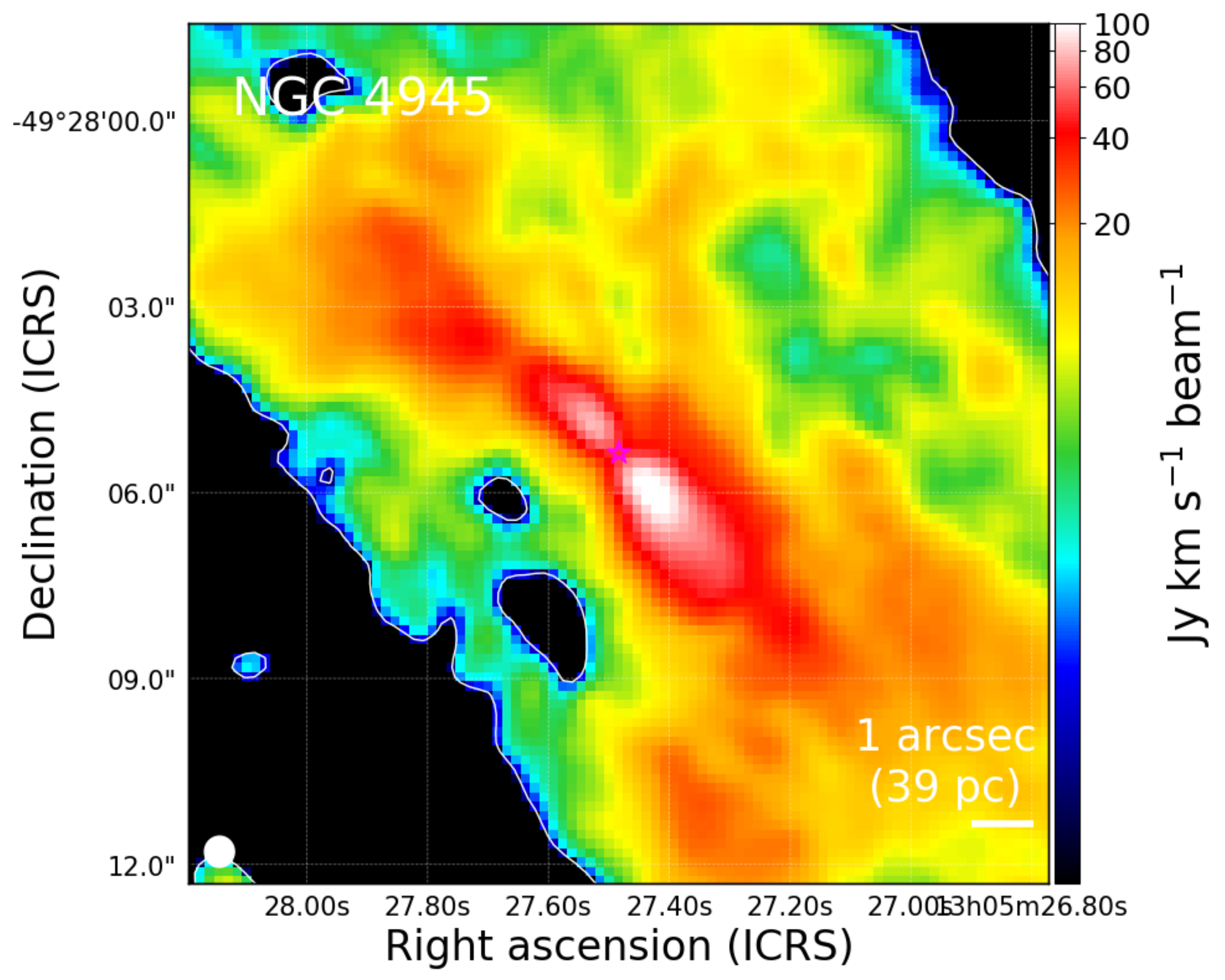}
    \includegraphics[width=5.8cm]{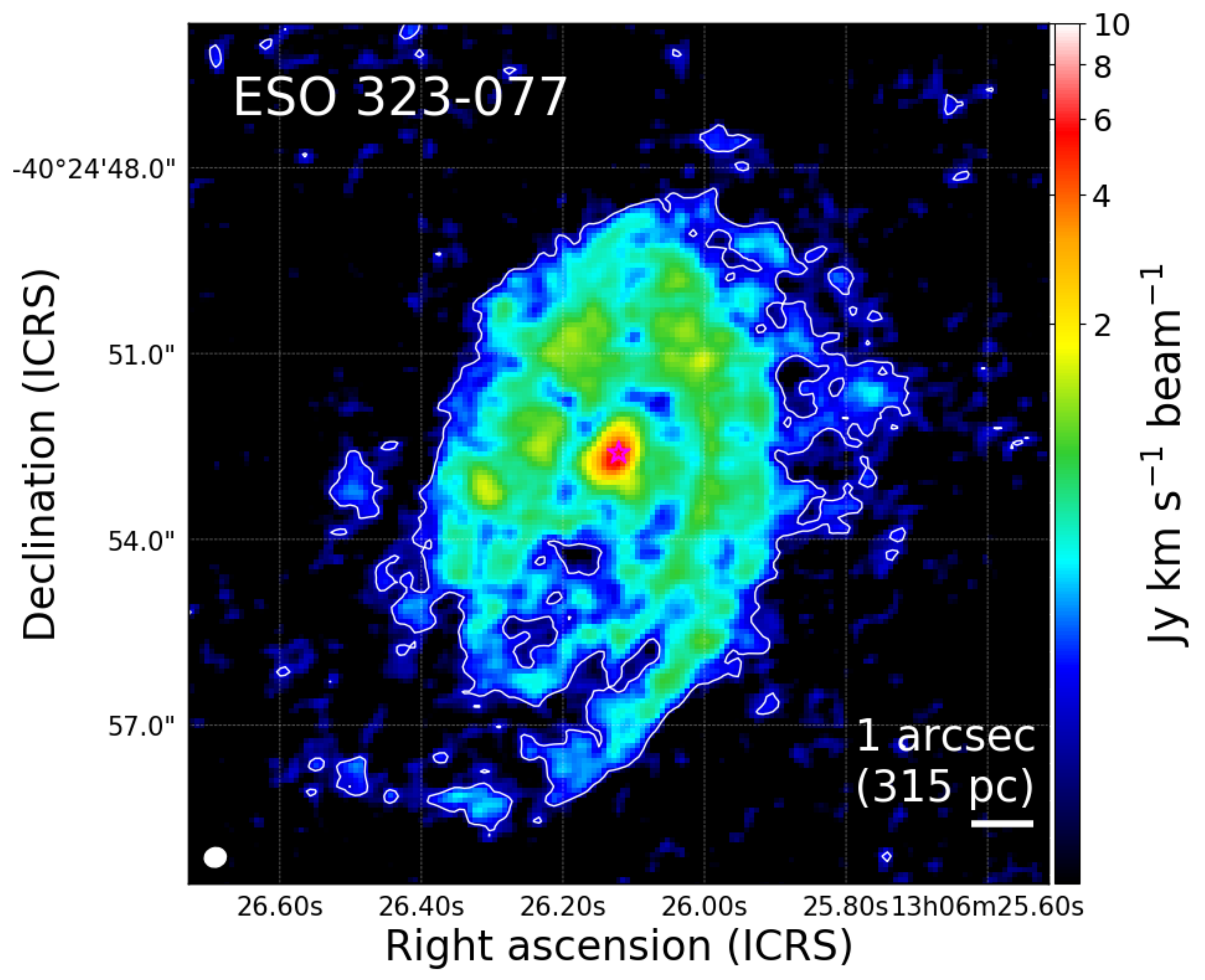}    
    \includegraphics[width=5.8cm]{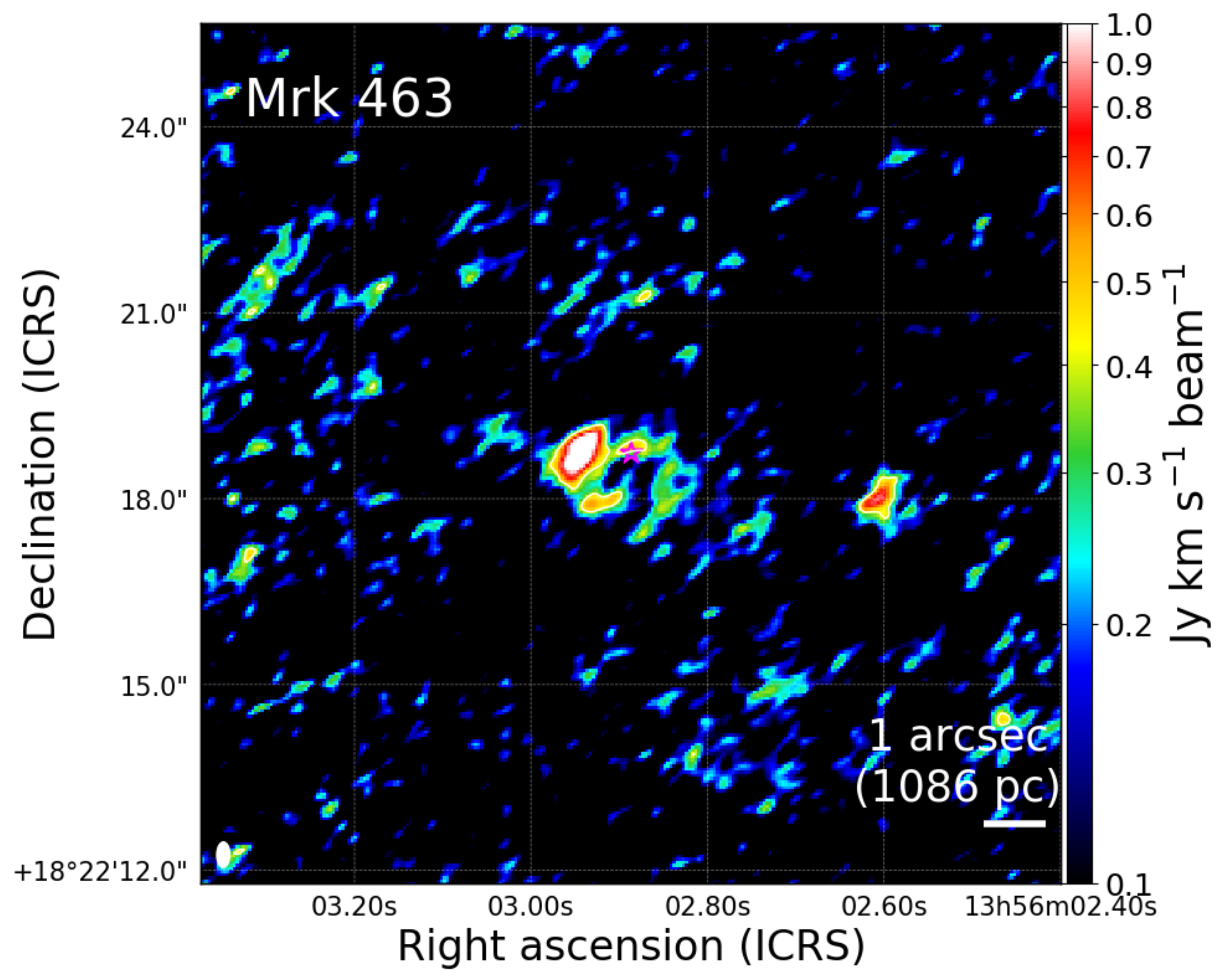}
    \includegraphics[width=5.8cm]{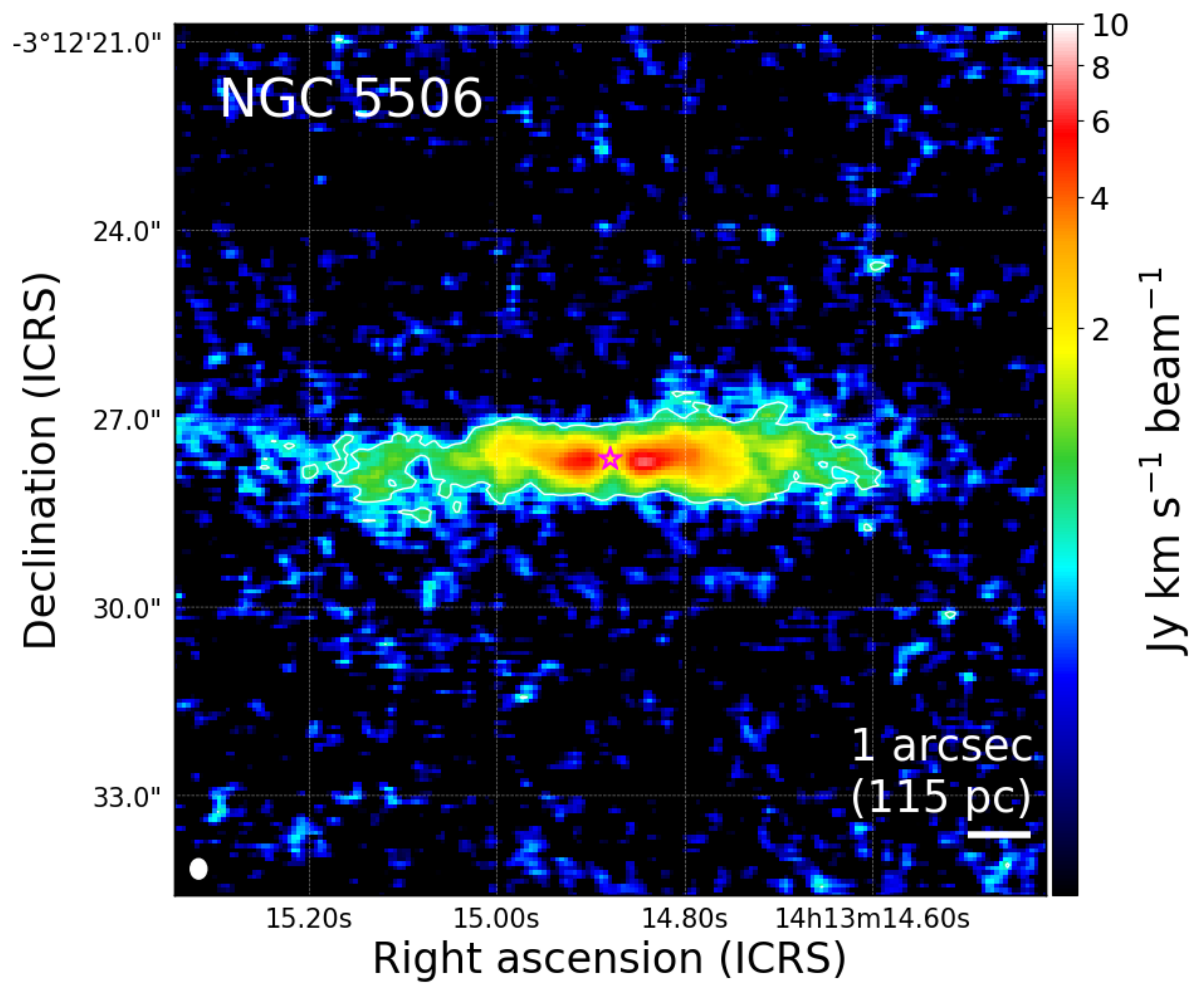}
    \includegraphics[width=5.8cm]{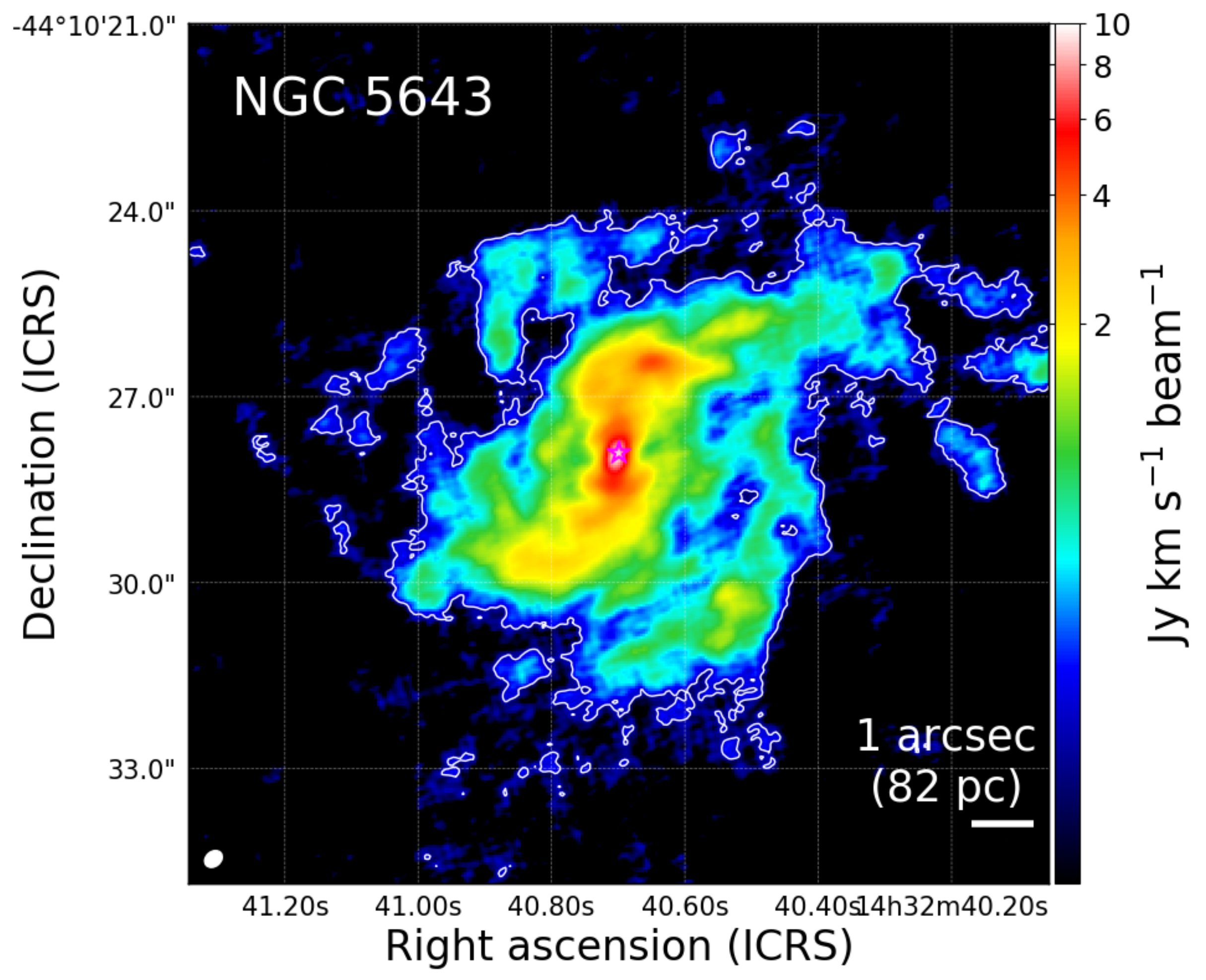}        
    \includegraphics[width=5.8cm]{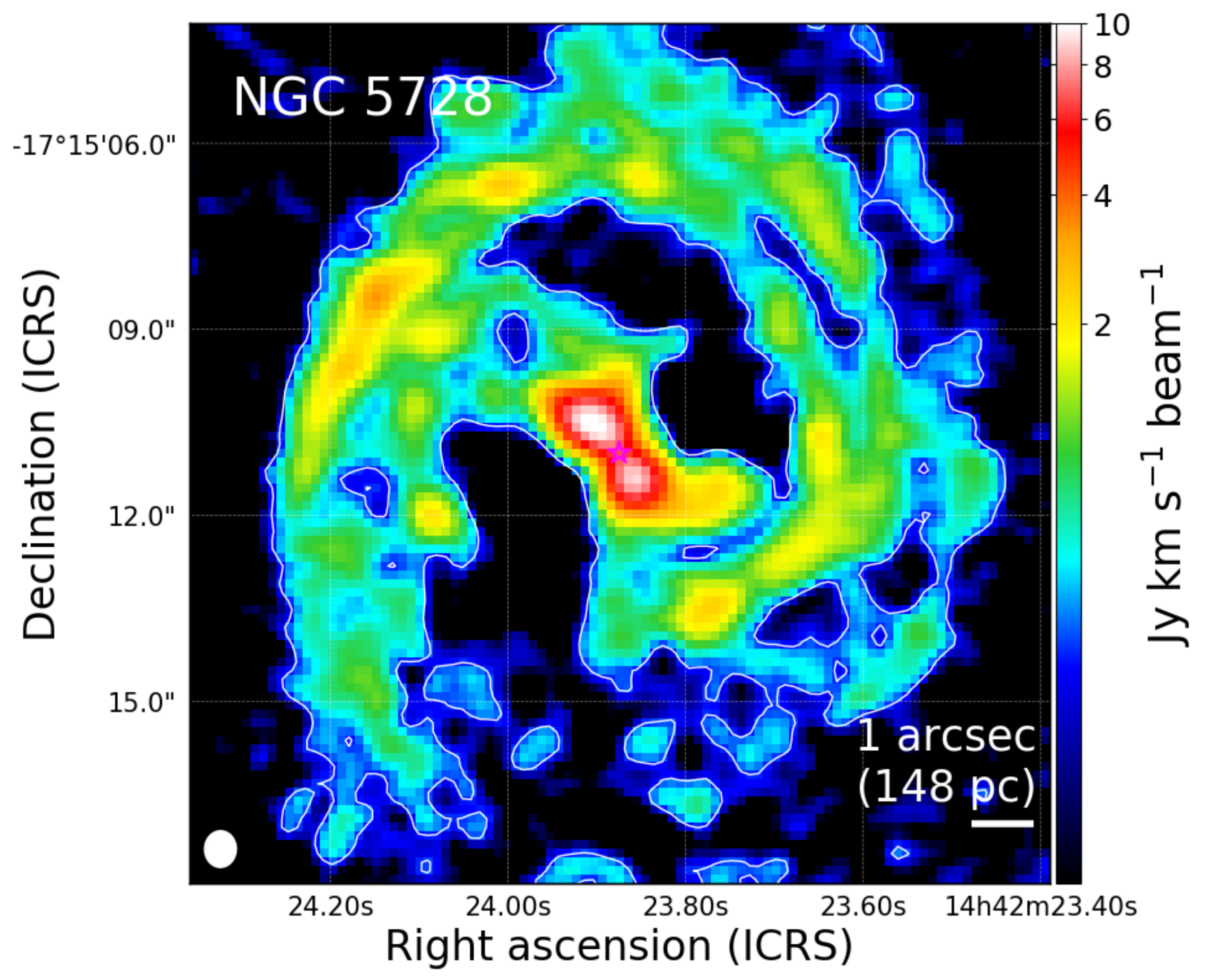}
    \includegraphics[width=5.8cm]{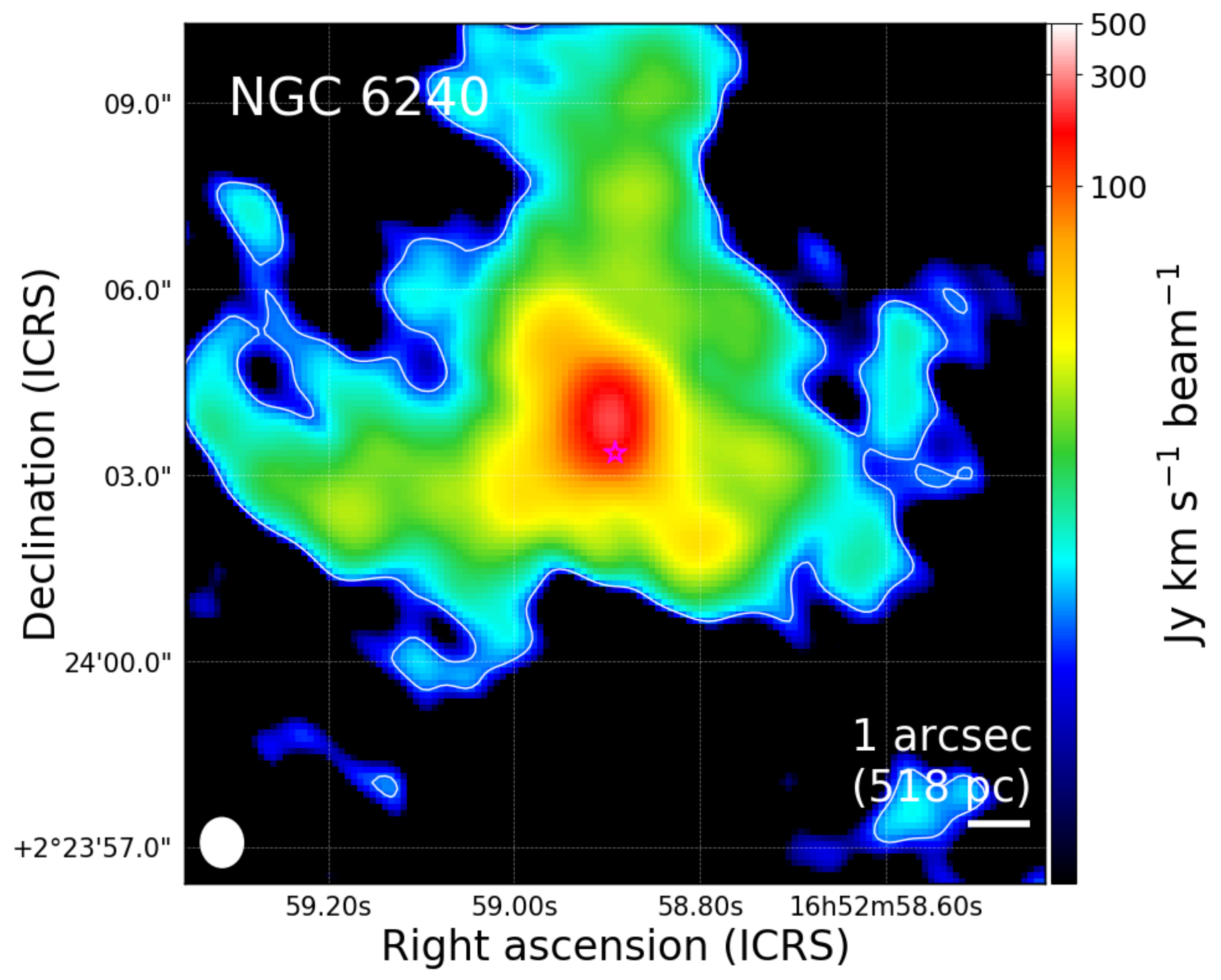}    
    \includegraphics[width=5.8cm]{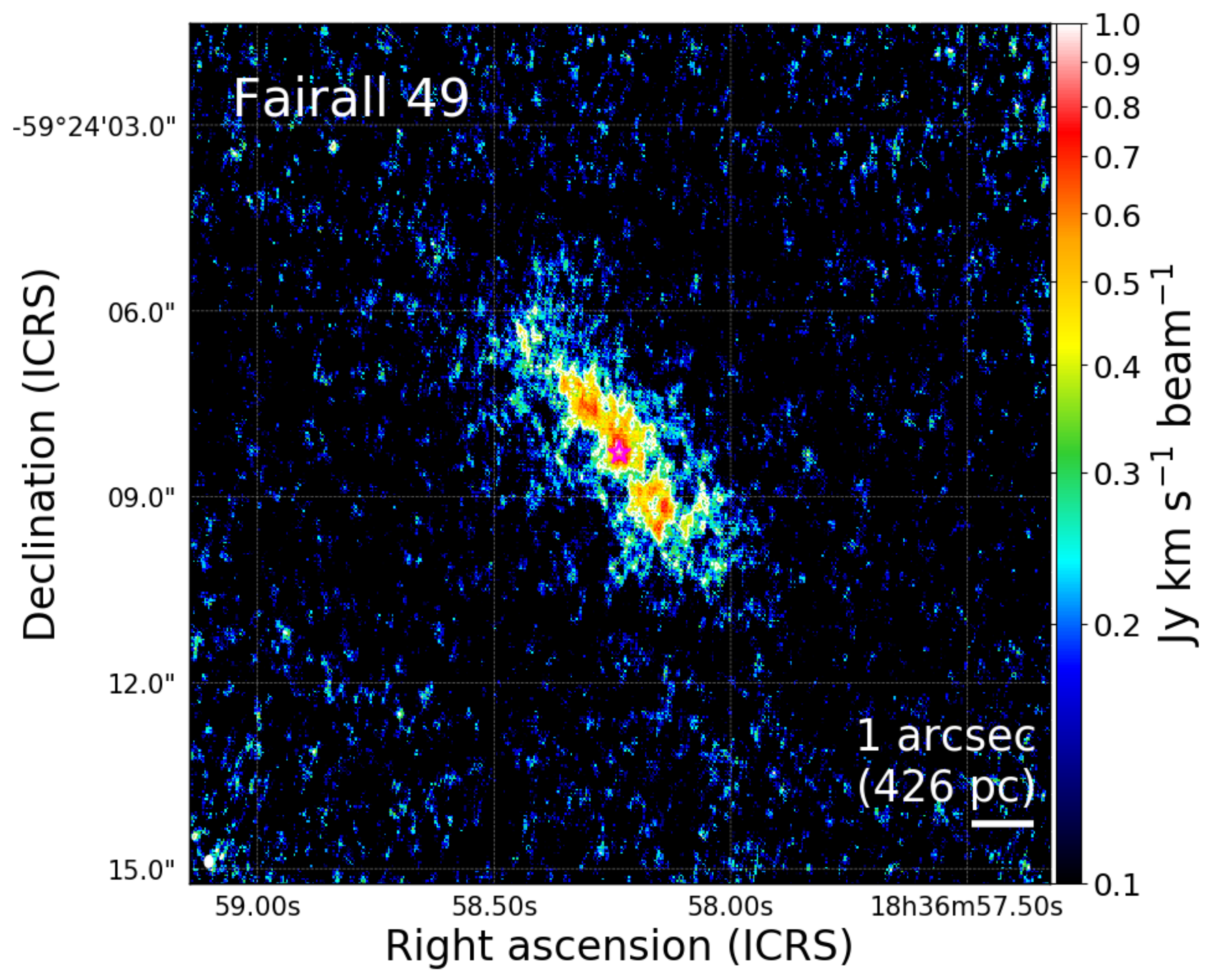}    
    \includegraphics[width=5.8cm]{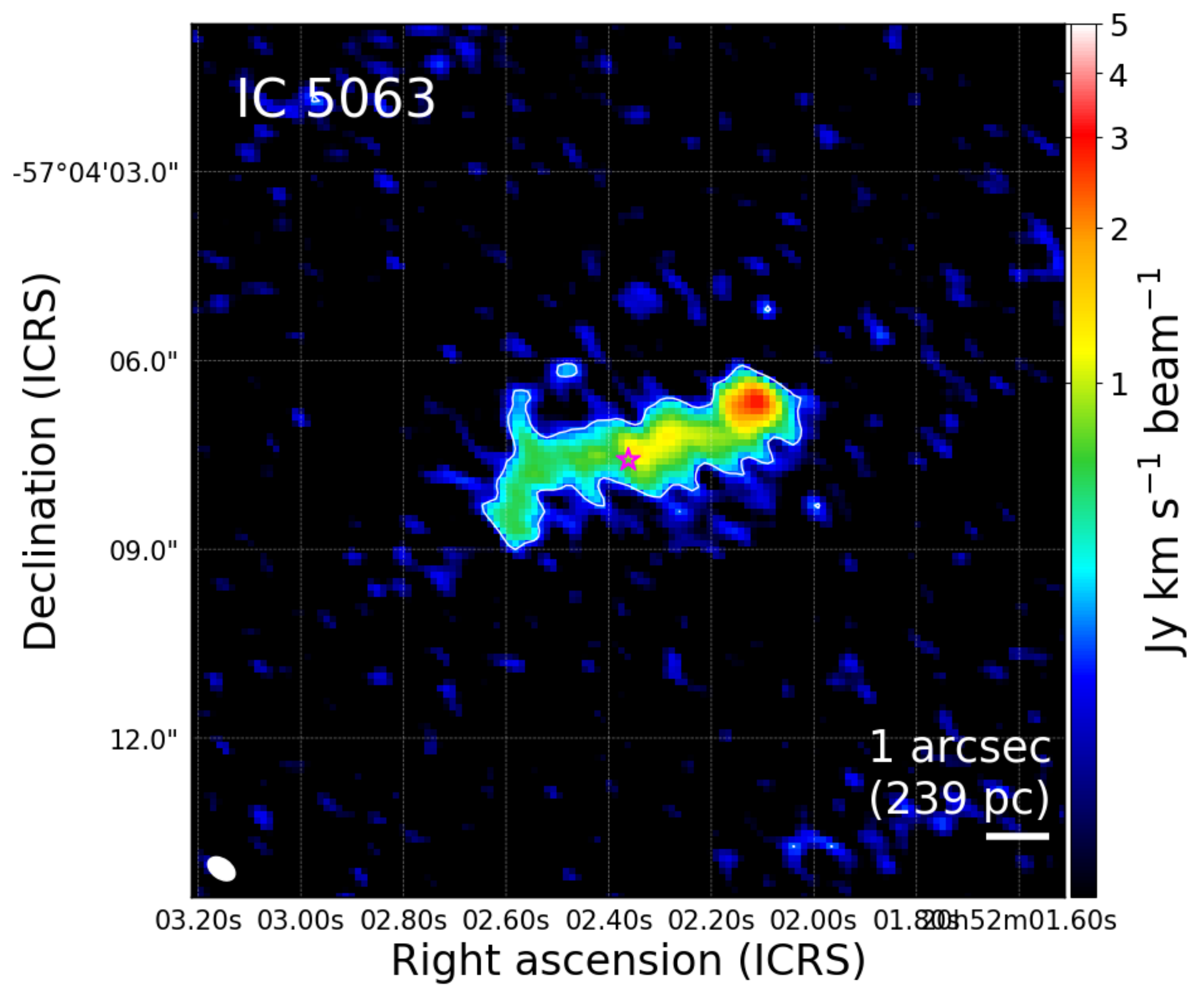}
    \caption{Continued.}
\end{figure*}

\begin{figure*}\addtocounter{figure}{-1}
    \centering
    \includegraphics[width=5.8cm]{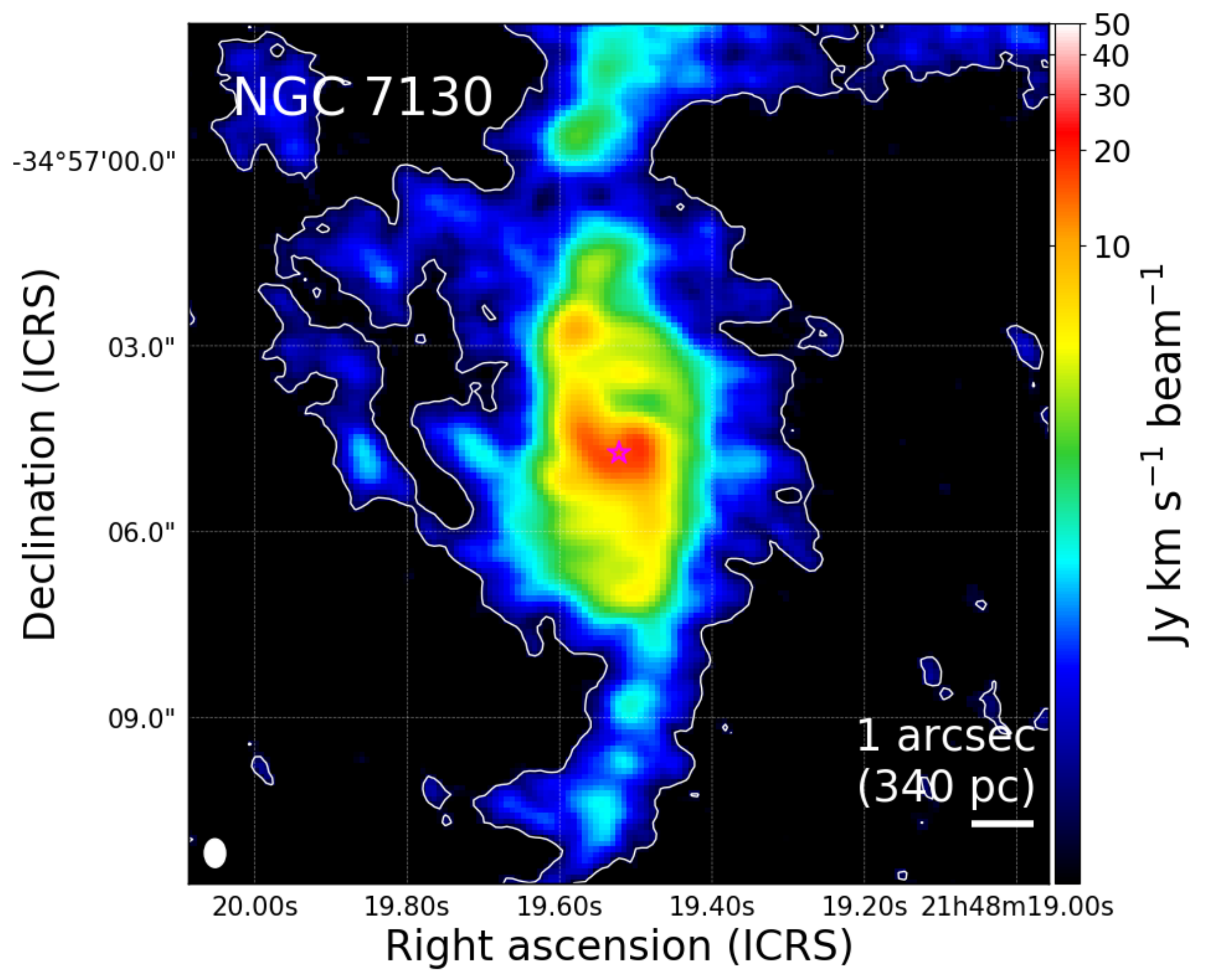}
    \includegraphics[width=5.8cm]{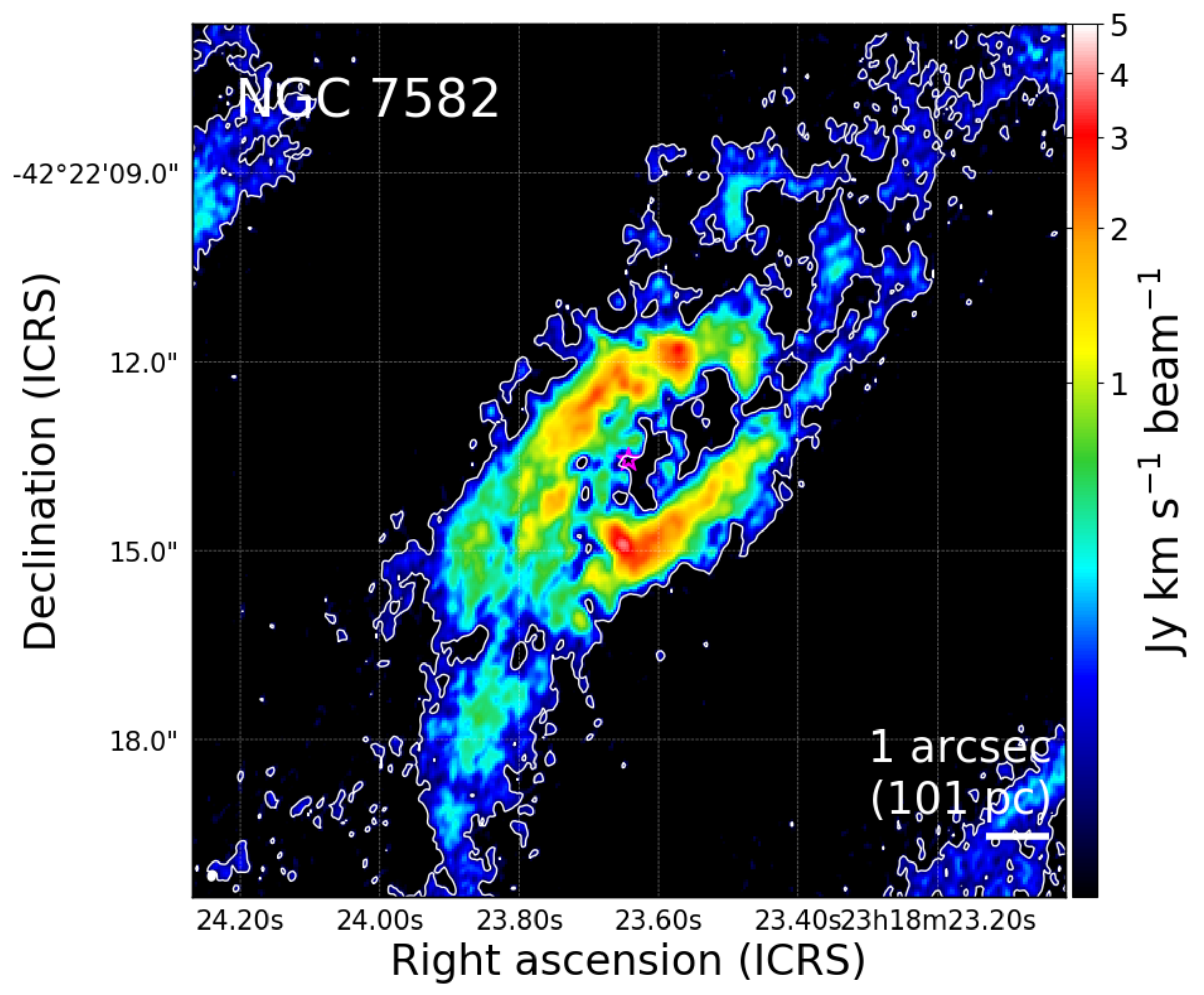} 
    \caption{Continued.}
\end{figure*}           

\bibliographystyle{aasjournal}
\bibliography{ref}

\begin{thebibliography}{}
\expandafter\ifx\csname natexlab\endcsname\relax\def\natexlab#1{#1}\fi
\providecommand{\url}[1]{\href{#1}{#1}}
\providecommand{\dodoi}[1]{doi:~\href{http://doi.org/#1}{\nolinkurl{#1}}}
\providecommand{\doeprint}[1]{\href{http://ascl.net/#1}{\nolinkurl{http://ascl.net/#1}}}
\providecommand{\doarXiv}[1]{\href{https://arxiv.org/abs/#1}{\nolinkurl{https://arxiv.org/abs/#1}}}

\bibitem[{{Asakura} {et~al.}(2020){Asakura}, {Hayashida}, {Hanasaka},
  {Kawabata}, {Yoneyama}, {Noda}, {Sakuma}, {Okazaki}, {Ishikura}, {Hanaoka},
  {Ide}, {Hattori}, {Matsumoto}, {Tsunemi}, {Awaki}, {Nakajima}, \&
  {Hiraga}}]{Asa20}
{Asakura}, K., {Hayashida}, K., {Hanasaka}, T., {et~al.} 2020, in Society of
  Photo-Optical Instrumentation Engineers (SPIE) Conference Series, Vol. 11444,
  Society of Photo-Optical Instrumentation Engineers (SPIE) Conference Series,
  114441D, \dodoi{10.1117/12.2560772}

\bibitem[{{Baek} {et~al.}(2019){Baek}, {Chung}, {Schawinski}, {Oh}, {Wong},
  {Koss}, {Ricci}, {Trakhtenbrot}, {Smith}, \& {Ueda}}]{Bae19}
{Baek}, J., {Chung}, A., {Schawinski}, K., {et~al.} 2019, \mnras, 488, 4317,
  \dodoi{10.1093/mnras/stz1995}

\bibitem[{{Baumgartner} {et~al.}(2013){Baumgartner}, {Tueller}, {Markwardt},
  {Skinner}, {Barthelmy}, {Mushotzky}, {Evans}, \& {Gehrels}}]{Bau13}
{Baumgartner}, W.~H., {Tueller}, J., {Markwardt}, C.~B., {et~al.} 2013, \apjs,
  207, 19, \dodoi{10.1088/0067-0049/207/2/19}

\bibitem[{{Behar} {et~al.}(2015){Behar}, {Baldi}, {Laor}, {Horesh}, {Stevens},
  \& {Tzioumis}}]{Beh15}
{Behar}, E., {Baldi}, R.~D., {Laor}, A., {et~al.} 2015, \mnras, 451, 517,
  \dodoi{10.1093/mnras/stv988}

\bibitem[{{Behar} {et~al.}(2018){Behar}, {Vogel}, {Baldi}, {Smith}, \&
  {Mushotzky}}]{Beh18}
{Behar}, E., {Vogel}, S., {Baldi}, R.~D., {Smith}, K.~L., \& {Mushotzky}, R.~F.
  2018, \mnras, 478, 399, \dodoi{10.1093/mnras/sty850}

\bibitem[{{Bianchi} {et~al.}(2008){Bianchi}, {Chiaberge}, {Piconcelli},
  {Guainazzi}, \& {Matt}}]{Bia08}
{Bianchi}, S., {Chiaberge}, M., {Piconcelli}, E., {Guainazzi}, M., \& {Matt},
  G. 2008, \mnras, 386, 105, \dodoi{10.1111/j.1365-2966.2008.13078.x}

\bibitem[{{Bianchi} \& {Guainazzi}(2007)}]{Bia07}
{Bianchi}, S., \& {Guainazzi}, M. 2007, in American Institute of Physics
  Conference Series, Vol. 924, The Multicolored Landscape of Compact Objects
  and Their Explosive Origins, ed. T.~{di Salvo}, G.~L. {Israel},
  L.~{Piersant}, L.~{Burderi}, G.~{Matt}, A.~{Tornambe}, \& M.~T. {Menna},
  822--829, \dodoi{10.1063/1.2774948}

\bibitem[{{Bianchi} {et~al.}(2001){Bianchi}, {Matt}, \& {Iwasawa}}]{Bia01}
{Bianchi}, S., {Matt}, G., \& {Iwasawa}, K. 2001, \mnras, 322, 669,
  \dodoi{10.1046/j.1365-8711.2001.04156.x}

\bibitem[{{Bigiel} {et~al.}(2015){Bigiel}, {Leroy}, {Blitz}, {Bolatto}, {da
  Cunha}, {Rosolowsky}, {Sandstrom}, \& {Usero}}]{Big15}
{Bigiel}, F., {Leroy}, A.~K., {Blitz}, L., {et~al.} 2015, \apj, 815, 103,
  \dodoi{10.1088/0004-637X/815/2/103}

\bibitem[{{Boller} {et~al.}(2003){Boller}, {Keil}, {Hasinger}, {Costantini},
  {Fujimoto}, {Anabuki}, {Lehmann}, \& {Gallo}}]{Bol03}
{Boller}, T., {Keil}, R., {Hasinger}, G., {et~al.} 2003, \aap, 411, 63,
  \dodoi{10.1051/0004-6361:20031217}

\bibitem[{{Combes} {et~al.}(2019){Combes}, {Garc{\'\i}a-Burillo}, {Audibert},
  {Hunt}, {Eckart}, {Aalto}, {Casasola}, {Boone}, {Krips}, {Viti}, {Sakamoto},
  {Muller}, {Dasyra}, {van der Werf}, \& {Martin}}]{Com19}
{Combes}, F., {Garc{\'\i}a-Burillo}, S., {Audibert}, A., {et~al.} 2019, \aap,
  623, A79, \dodoi{10.1051/0004-6361/201834560}

\bibitem[{{Croton} {et~al.}(2006){Croton}, {Springel}, {White}, {De Lucia},
  {Frenk}, {Gao}, {Jenkins}, {Kauffmann}, {Navarro}, \& {Yoshida}}]{Cro06}
{Croton}, D.~J., {Springel}, V., {White}, S. D.~M., {et~al.} 2006, \mnras, 365,
  11, \dodoi{10.1111/j.1365-2966.2005.09675.x}

\bibitem[{{Di Matteo} {et~al.}(2005){Di Matteo}, {Springel}, \&
  {Hernquist}}]{DiM05}
{Di Matteo}, T., {Springel}, V., \& {Hernquist}, L. 2005, \nat, 433, 604,
  \dodoi{10.1038/nature03335}

\bibitem[{{Elvis} {et~al.}(1994){Elvis}, {Wilkes}, {McDowell}, {Green},
  {Bechtold}, {Willner}, {Oey}, {Polomski}, \& {Cutri}}]{Elv94}
{Elvis}, M., {Wilkes}, B.~J., {McDowell}, J.~C., {et~al.} 1994, \apjs, 95, 1,
  \dodoi{10.1086/192093}

\bibitem[{{Fabbiano} {et~al.}(2017){Fabbiano}, {Elvis}, {Paggi}, {Karovska},
  {Maksym}, {Raymond}, {Risaliti}, \& {Wang}}]{Fab17}
{Fabbiano}, G., {Elvis}, M., {Paggi}, A., {et~al.} 2017, \apjl, 842, L4,
  \dodoi{10.3847/2041-8213/aa7551}

\bibitem[{{Fabbiano} {et~al.}(2019){Fabbiano}, {Paggi}, \& {Elvis}}]{Fab19b}
{Fabbiano}, G., {Paggi}, A., \& {Elvis}, M. 2019, \apjl, 876, L18,
  \dodoi{10.3847/2041-8213/ab1c63}

\bibitem[{{Fabbiano} {et~al.}(2018{\natexlab{a}}){Fabbiano}, {Paggi},
  {Karovska}, {Elvis}, {Maksym}, {Risaliti}, \& {Wang}}]{Fab18a}
{Fabbiano}, G., {Paggi}, A., {Karovska}, M., {et~al.} 2018{\natexlab{a}}, \apj,
  855, 131, \dodoi{10.3847/1538-4357/aab1f4}

\bibitem[{{Fabbiano} {et~al.}(2018{\natexlab{b}}){Fabbiano}, {Paggi},
  {Karovska}, {Elvis}, {Maksym}, \& {Wang}}]{Fab18b}
---. 2018{\natexlab{b}}, \apj, 865, 83, \dodoi{10.3847/1538-4357/aadc5d}

\bibitem[{{Fabbiano} {et~al.}(2020){Fabbiano}, {Paggi}, {Karovska}, {Elvis},
  {Nardini}, \& {Wang}}]{Fab20}
---. 2020, \apj, 902, 49, \dodoi{10.3847/1538-4357/abb5ad}

\bibitem[{{Fabian}(1977)}]{Fab77}
{Fabian}, A.~C. 1977, \nat, 269, 672, \dodoi{10.1038/269672a0}

\bibitem[{{Fabian} {et~al.}(2009){Fabian}, {Vasudevan}, {Mushotzky}, {Winter},
  \& {Reynolds}}]{Fab09}
{Fabian}, A.~C., {Vasudevan}, R.~V., {Mushotzky}, R.~F., {Winter}, L.~M., \&
  {Reynolds}, C.~S. 2009, \mnras, 394, L89,
  \dodoi{10.1111/j.1745-3933.2009.00617.x}

\bibitem[{{Feruglio} {et~al.}(2020){Feruglio}, {Fabbiano}, {Bischetti},
  {Elvis}, {Travascio}, \& {Fiore}}]{Fer20}
{Feruglio}, C., {Fabbiano}, G., {Bischetti}, M., {et~al.} 2020, \apj, 890, 29,
  \dodoi{10.3847/1538-4357/ab67bd}

\bibitem[{{Gao} \& {Solomon}(2004)}]{Gao04}
{Gao}, Y., \& {Solomon}, P.~M. 2004, \apj, 606, 271, \dodoi{10.1086/382999}

\bibitem[{{Garc{\'\i}a-Burillo} {et~al.}(2010){Garc{\'\i}a-Burillo}, {Usero},
  {Fuente}, {Mart{\'\i}n-Pintado}, {Boone}, {Aalto}, {Krips}, {Neri},
  {Schinnerer}, \& {Tacconi}}]{Gar10}
{Garc{\'\i}a-Burillo}, S., {Usero}, A., {Fuente}, A., {et~al.} 2010, \aap, 519,
  A2, \dodoi{10.1051/0004-6361/201014539}

\bibitem[{{Garc{\'\i}a-Burillo} {et~al.}(2014){Garc{\'\i}a-Burillo}, {Combes},
  {Usero}, {Aalto}, {Krips}, {Viti}, {Alonso-Herrero}, {Hunt}, {Schinnerer},
  {Baker}, {Boone}, {Casasola}, {Colina}, {Costagliola}, {Eckart}, {Fuente},
  {Henkel}, {Labiano}, {Mart{\'\i}n}, {M{\'a}rquez}, {Muller}, {Planesas},
  {Ramos Almeida}, {Spaans}, {Tacconi}, \& {van der Werf}}]{Gar14}
{Garc{\'\i}a-Burillo}, S., {Combes}, F., {Usero}, A., {et~al.} 2014, \aap, 567,
  A125, \dodoi{10.1051/0004-6361/201423843}

\bibitem[{{Garcia-Burillo} {et~al.}(2021){Garcia-Burillo}, {Alonso-Herrero},
  {Ramos Almeida}, {Gonzalez-Martin}, {Combes}, {Usero}, {Hoenig}, {Querejeta},
  {Hicks}, {Hunt}, {Rosario}, {Davies}, {Boorman}, {Bunker}, {Burstcher},
  {Colina}, {D{\'\i}az-Santos}, {Gandhi}, {Garcia-Bernete}, {Garcia-Lorenzo},
  {Ichikawa}, {Imanishi}, {Izumi}, {Labiano}, {Levenson}, {Lopez-Rodriguez},
  {Packham}, {Pereira-Santaella}, {Ricci}, {Rigopoulou}, {Rouan}, {Stalevsk},
  {Wada}, \& {Williamson}}]{Gar21}
{Garcia-Burillo}, S., {Alonso-Herrero}, A., {Ramos Almeida}, C., {et~al.} 2021,
  arXiv e-prints, arXiv:2104.10227.
\newblock \doarXiv{2104.10227}

\bibitem[{{Garmire} {et~al.}(2003){Garmire}, {Bautz}, {Ford}, {Nousek}, \&
  {Ricker}}]{Gar03}
{Garmire}, G.~P., {Bautz}, M.~W., {Ford}, P.~G., {Nousek}, J.~A., \& {Ricker},
  George~R., J. 2003, in Society of Photo-Optical Instrumentation Engineers
  (SPIE) Conference Series, Vol. 4851, \procspie, ed. J.~E. {Truemper} \& H.~D.
  {Tananbaum}, 28--44, \dodoi{10.1117/12.461599}

\bibitem[{{Gebhardt} {et~al.}(2000){Gebhardt}, {Bender}, {Bower}, {Dressler},
  {Faber}, {Filippenko}, {Green}, {Grillmair}, {Ho}, {Kormendy}, {Lauer},
  {Magorrian}, {Pinkney}, {Richstone}, \& {Tremaine}}]{Geb00}
{Gebhardt}, K., {Bender}, R., {Bower}, G., {et~al.} 2000, \apjl, 539, L13,
  \dodoi{10.1086/312840}

\bibitem[{{Gehrels}(1986)}]{Geh86}
{Gehrels}, N. 1986, \apj, 303, 336, \dodoi{10.1086/164079}

\bibitem[{{G{\'o}mez-Guijarro} {et~al.}(2017){G{\'o}mez-Guijarro},
  {Gonz{\'a}lez-Mart{\'\i}n}, {Ramos Almeida}, {Rodr{\'\i}guez-Espinosa}, \&
  {Gallego}}]{Gom17}
{G{\'o}mez-Guijarro}, C., {Gonz{\'a}lez-Mart{\'\i}n}, O., {Ramos Almeida}, C.,
  {Rodr{\'\i}guez-Espinosa}, J.~M., \& {Gallego}, J. 2017, \mnras, 469, 2720,
  \dodoi{10.1093/mnras/stx1037}

\bibitem[{{G{\"u}ltekin} {et~al.}(2009){G{\"u}ltekin}, {Richstone}, {Gebhardt},
  {Lauer}, {Tremaine}, {Aller}, {Bender}, {Dressler}, {Faber}, {Filippenko},
  {Green}, {Ho}, {Kormendy}, {Magorrian}, {Pinkney}, \& {Siopis}}]{Gul09}
{G{\"u}ltekin}, K., {Richstone}, D.~O., {Gebhardt}, K., {et~al.} 2009, \apj,
  698, 198, \dodoi{10.1088/0004-637X/698/1/198}

\bibitem[{{Gupta} {et~al.}(2021){Gupta}, {Ricci}, {Tortosa}, {Ueda},
  {Kawamuro}, {Koss}, {Trakhtenbrot}, {Oh}, {Bauer}, {Ricci}, {Privon},
  {Zappacosta}, {Stern}, {Kakkad}, {Piconcelli}, {Veilleux}, {Mushotzky},
  {Caglar}, {Ichikawa}, {Elagali}, {Powell}, {Urry}, \& {Harrison}}]{Gup21}
{Gupta}, K.~K., {Ricci}, C., {Tortosa}, A., {et~al.} 2021, \mnras, 504, 428,
  \dodoi{10.1093/mnras/stab839}

\bibitem[{{Hayashida} {et~al.}(2018){Hayashida}, {Kawabata}, {Hanasaka},
  {Asakura}, {Yoneyama}, {Okazaki}, {Ide}, {Matsumoto}, {Nakajima}, {Awaki}, \&
  {Tsunemi}}]{Hay18}
{Hayashida}, K., {Kawabata}, T., {Hanasaka}, T., {et~al.} 2018, in Society of
  Photo-Optical Instrumentation Engineers (SPIE) Conference Series, Vol. 10699,
  Space Telescopes and Instrumentation 2018: Ultraviolet to Gamma Ray, ed.
  J.-W.~A. {den Herder}, S.~{Nikzad}, \& K.~{Nakazawa}, 106990U,
  \dodoi{10.1117/12.2314181}

\bibitem[{{Henkel} {et~al.}(2018){Henkel}, {M{\"u}hle}, {Bendo}, {J{\'o}zsa},
  {Gong}, {Viti}, {Aalto}, {Combes}, {Garc{\'\i}a-Burillo}, {Hunt}, {Mangum},
  {Mart{\'\i}n}, {Muller}, {Ott}, {van der Werf}, {Malawi}, {Ismail},
  {Alkhuja}, {Asiri}, {Aladro}, {Alves}, {Ao}, {Baan}, {Costagliola}, {Fuller},
  {Greene}, {Impellizzeri}, {Kamali}, {Klessen}, {Mauersberger}, {Tang},
  {Tristram}, {Wang}, \& {Zhang}}]{Hen18}
{Henkel}, C., {M{\"u}hle}, S., {Bendo}, G., {et~al.} 2018, \aap, 615, A155,
  \dodoi{10.1051/0004-6361/201732174}

\bibitem[{{Hickox} \& {Alexander}(2018)}]{Hic18}
{Hickox}, R.~C., \& {Alexander}, D.~M. 2018, \araa, 56, 625,
  \dodoi{10.1146/annurev-astro-081817-051803}

\bibitem[{{Hocuk} \& {Spaans}(2010)}]{Hoc10}
{Hocuk}, S., \& {Spaans}, M. 2010, \aap, 522, A24,
  \dodoi{10.1051/0004-6361/201015055}

\bibitem[{{Hocuk} \& {Spaans}(2011)}]{Hoc11}
---. 2011, \aap, 536, A41, \dodoi{10.1051/0004-6361/201117431}

\bibitem[{{Hori} {et~al.}(2018){Hori}, {Shidatsu}, {Ueda}, {Kawamuro}, {Morii},
  {Nakahira}, {Isobe}, {Kawai}, {Mihara}, {Matsuoka}, {Morita}, {Nakajima},
  {Negoro}, {Oda}, {Sakamoto}, {Serino}, {Sugizaki}, {Tanimoto}, {Tomida},
  {Tsuboi}, {Tsunemi}, {Ueno}, {Yamaoka}, {Yamada}, {Yoshida}, {Iwakiri},
  {Kawakubo}, {Sugawara}, {Sugita}, {Tachibana}, \& {Yoshii}}]{Hor18}
{Hori}, T., {Shidatsu}, M., {Ueda}, Y., {et~al.} 2018, \apjs, 235, 7,
  \dodoi{10.3847/1538-4365/aaa89c}

\bibitem[{{Ichikawa} {et~al.}(2019){Ichikawa}, {Ricci}, {Ueda}, {Bauer},
  {Kawamuro}, {Koss}, {Oh}, {Rosario}, {Shimizu}, {Stalevski}, {Fuller},
  {Packham}, \& {Trakhtenbrot}}]{Ich19}
{Ichikawa}, K., {Ricci}, C., {Ueda}, Y., {et~al.} 2019, \apj, 870, 31,
  \dodoi{10.3847/1538-4357/aaef8f}

\bibitem[{{Imanishi} {et~al.}(2009){Imanishi}, {Nakanishi}, {Tamura}, \&
  {Peng}}]{Ima09}
{Imanishi}, M., {Nakanishi}, K., {Tamura}, Y., \& {Peng}, C.-H. 2009, \aj, 137,
  3581, \dodoi{10.1088/0004-6256/137/3/3581}

\bibitem[{{Inoue} \& {Doi}(2018)}]{Ino18}
{Inoue}, Y., \& {Doi}, A. 2018, \apj, 869, 114,
  \dodoi{10.3847/1538-4357/aaeb95}

\bibitem[{{Isobe} {et~al.}(1990){Isobe}, {Feigelson}, {Akritas}, \&
  {Babu}}]{Iso90}
{Isobe}, T., {Feigelson}, E.~D., {Akritas}, M.~G., \& {Babu}, G.~J. 1990, \apj,
  364, 104, \dodoi{10.1086/169390}

\bibitem[{{Iwasawa} {et~al.}(1997){Iwasawa}, {Fabian}, \& {Matt}}]{Iwa97}
{Iwasawa}, K., {Fabian}, A.~C., \& {Matt}, G. 1997, \mnras, 289, 443,
  \dodoi{10.1093/mnras/289.2.443}

\bibitem[{{Iyomoto} {et~al.}(1996){Iyomoto}, {Makishima}, {Fukazawa},
  {Tashiro}, {Ishisaki}, {Nakai}, \& {Taniguchi}}]{Iyo96}
{Iyomoto}, N., {Makishima}, K., {Fukazawa}, Y., {et~al.} 1996, \pasj, 48, 231,
  \dodoi{10.1093/pasj/48.2.231}

\bibitem[{{Izumi} {et~al.}(2016{\natexlab{a}}){Izumi}, {Kawakatu}, \&
  {Kohno}}]{Izu16b}
{Izumi}, T., {Kawakatu}, N., \& {Kohno}, K. 2016{\natexlab{a}}, \apj, 827, 81,
  \dodoi{10.3847/0004-637X/827/1/81}

\bibitem[{{Izumi} {et~al.}(2018){Izumi}, {Wada}, {Fukushige}, {Hamamura}, \&
  {Kohno}}]{Izu18}
{Izumi}, T., {Wada}, K., {Fukushige}, R., {Hamamura}, S., \& {Kohno}, K. 2018,
  \apj, 867, 48, \dodoi{10.3847/1538-4357/aae20b}

\bibitem[{{Izumi} {et~al.}(2016{\natexlab{b}}){Izumi}, {Kohno}, {Aalto},
  {Espada}, {Fathi}, {Harada}, {Hatsukade}, {Hsieh}, {Imanishi}, {Krips},
  {Mart{\'\i}n}, {Matsushita}, {Meier}, {Nakai}, {Nakanishi}, {Schinnerer},
  {Sheth}, {Terashima}, \& {Turner}}]{Izu16a}
{Izumi}, T., {Kohno}, K., {Aalto}, S., {et~al.} 2016{\natexlab{b}}, \apj, 818,
  42, \dodoi{10.3847/0004-637X/818/1/42}

\bibitem[{{Izumi} {et~al.}(2020){Izumi}, {Nguyen}, {Imanishi}, {Kawamuro},
  {Baba}, {Nakano}, {Kohno}, {Matsushita}, {Meier}, {Turner}, {Michiyama},
  {Harada}, {Mart{\'\i}n}, {Nakanishi}, {Takano}, {Wiklind}, {Nakai}, \&
  {Hsieh}}]{Izu20b}
{Izumi}, T., {Nguyen}, D.~D., {Imanishi}, M., {et~al.} 2020, \apj, 898, 75,
  \dodoi{10.3847/1538-4357/ab9cb1}

\bibitem[{{Kalberla} {et~al.}(2005){Kalberla}, {Burton}, {Hartmann}, {Arnal},
  {Bajaja}, {Morras}, \& {P{\"o}ppel}}]{Kal05}
{Kalberla}, P.~M.~W., {Burton}, W.~B., {Hartmann}, D., {et~al.} 2005, \aap,
  440, 775, \dodoi{10.1051/0004-6361:20041864}

\bibitem[{{Kameno} {et~al.}(2020){Kameno}, {Sawada-Satoh}, {Impellizzeri},
  {Espada}, {Nakai}, {Sugai}, {Terashima}, {Kohno}, {Lee}, \&
  {Mart{\'\i}n}}]{Kam20}
{Kameno}, S., {Sawada-Satoh}, S., {Impellizzeri}, C.~M.~V., {et~al.} 2020,
  \apj, 895, 73, \dodoi{10.3847/1538-4357/ab8bd6}

\bibitem[{{Kawamuro} {et~al.}(2019{\natexlab{a}}){Kawamuro}, {Izumi}, \&
  {Imanishi}}]{Kaw19b}
{Kawamuro}, T., {Izumi}, T., \& {Imanishi}, M. 2019{\natexlab{a}}, \pasj, 71,
  68, \dodoi{10.1093/pasj/psz045}

\bibitem[{{Kawamuro} {et~al.}(2020){Kawamuro}, {Izumi}, {Onishi}, {Imanishi},
  {Nguyen}, \& {Baba}}]{Kaw20}
{Kawamuro}, T., {Izumi}, T., {Onishi}, K., {et~al.} 2020, \apj, 895, 135,
  \dodoi{10.3847/1538-4357/ab8b62}

\bibitem[{{Kawamuro} {et~al.}(2019{\natexlab{b}}){Kawamuro}, {Ueda},
  {Ichikawa}, {Imanishi}, {Izumi}, {Tanimoto}, \& {Matsuoka}}]{Kaw19a}
{Kawamuro}, T., {Ueda}, Y., {Ichikawa}, K., {et~al.} 2019{\natexlab{b}}, \apj,
  881, 48, \dodoi{10.3847/1538-4357/ab2bf6}

\bibitem[{{Kawamuro} {et~al.}(2016{\natexlab{a}}){Kawamuro}, {Ueda}, {Tazaki},
  {Ricci}, \& {Terashima}}]{Kaw16b}
{Kawamuro}, T., {Ueda}, Y., {Tazaki}, F., {Ricci}, C., \& {Terashima}, Y.
  2016{\natexlab{a}}, \apjs, 225, 14, \dodoi{10.3847/0067-0049/225/1/14}

\bibitem[{{Kawamuro} {et~al.}(2013){Kawamuro}, {Ueda}, {Tazaki}, \&
  {Terashima}}]{Kaw13}
{Kawamuro}, T., {Ueda}, Y., {Tazaki}, F., \& {Terashima}, Y. 2013, \apj, 770,
  157, \dodoi{10.1088/0004-637X/770/2/157}

\bibitem[{{Kawamuro} {et~al.}(2016{\natexlab{b}}){Kawamuro}, {Ueda}, {Tazaki},
  {Terashima}, \& {Mushotzky}}]{Kaw16c}
{Kawamuro}, T., {Ueda}, Y., {Tazaki}, F., {Terashima}, Y., \& {Mushotzky}, R.
  2016{\natexlab{b}}, \apj, 831, 37, \dodoi{10.3847/0004-637X/831/1/37}

\bibitem[{{Kawamuro} {et~al.}(2018){Kawamuro}, {Ueda}, {Shidatsu}, {Hori},
  {Morii}, {Nakahira}, {Isobe}, {Kawai}, {Mihara}, {Matsuoka}, {Morita},
  {Nakajima}, {Negoro}, {Oda}, {Sakamoto}, {Serino}, {Sugizaki}, {Tanimoto},
  {Tomida}, {Tsuboi}, {Tsunemi}, {Ueno}, {Yamaoka}, {Yamada}, {Yoshida},
  {Iwakiri}, {Kawakubo}, {Sugawara}, {Sugita}, {Tachibana}, \&
  {Yoshii}}]{Kaw18}
{Kawamuro}, T., {Ueda}, Y., {Shidatsu}, M., {et~al.} 2018, \apjs, 238, 32,
  \dodoi{10.3847/1538-4365/aad1ef}

\bibitem[{{Kohno}(2005)}]{Koh05}
{Kohno}, K. 2005, in American Institute of Physics Conference Series, Vol. 783,
  The Evolution of Starbursts, ed. S.~{H{\"u}ttmeister}, E.~{Manthey},
  D.~{Bomans}, \& K.~{Weis}, 203--208, \dodoi{10.1063/1.2034987}

\bibitem[{{Komossa} {et~al.}(2003){Komossa}, {Burwitz}, {Hasinger}, {Predehl},
  {Kaastra}, \& {Ikebe}}]{Kom03}
{Komossa}, S., {Burwitz}, V., {Hasinger}, G., {et~al.} 2003, \apjl, 582, L15,
  \dodoi{10.1086/346145}

\bibitem[{{Kormendy} \& {Ho}(2013)}]{Kor13}
{Kormendy}, J., \& {Ho}, L.~C. 2013, \araa, 51, 511,
  \dodoi{10.1146/annurev-astro-082708-101811}

\bibitem[{{Koss} {et~al.}(2017){Koss}, {Trakhtenbrot}, {Ricci}, {Lamperti},
  {Oh}, {Berney}, {Schawinski}, {Balokovi{\'c}}, {Baronchelli}, {Crenshaw},
  {Fischer}, {Gehrels}, {Harrison}, {Hashimoto}, {Hogg}, {Ichikawa}, {Masetti},
  {Mushotzky}, {Sartori}, {Stern}, {Treister}, {Ueda}, {Veilleux}, \&
  {Winter}}]{Kos17}
{Koss}, M., {Trakhtenbrot}, B., {Ricci}, C., {et~al.} 2017, \apj, 850, 74,
  \dodoi{10.3847/1538-4357/aa8ec9}

\bibitem[{{Koss} {et~al.}(2021){Koss}, {Strittmatter}, {Lamperti}, {Shimizu},
  {Trakhtenbrot}, {Saintonge}, {Treister}, {Cicone}, {Mushotzky}, {Oh},
  {Ricci}, {Stern}, {Ananna}, {Bauer}, {Privon}, {B{\"a}r}, {De Breuck},
  {Harrison}, {Ichikawa}, {Powell}, {Rosario}, {Sanders}, {Schawinski}, {Shao},
  {Megan Urry}, \& {Veilleux}}]{Kos21}
{Koss}, M.~J., {Strittmatter}, B., {Lamperti}, I., {et~al.} 2021, \apjs, 252,
  29, \dodoi{10.3847/1538-4365/abcbfe}

\bibitem[{{Koyama} {et~al.}(1996){Koyama}, {Maeda}, {Sonobe}, {Takeshima},
  {Tanaka}, \& {Yamauchi}}]{Koy96}
{Koyama}, K., {Maeda}, Y., {Sonobe}, T., {et~al.} 1996, \pasj, 48, 249,
  \dodoi{10.1093/pasj/48.2.249}

\bibitem[{{Koyama} {et~al.}(2007){Koyama}, {Inui}, {Hyodo}, {Matsumoto},
  {Tsuru}, {Maeda}, {Murakami}, {Yamauchi}, {Kissel}, {Chan}, \&
  {Soong}}]{Kat07}
{Koyama}, K., {Inui}, T., {Hyodo}, Y., {et~al.} 2007, \pasj, 59, 221,
  \dodoi{10.1093/pasj/59.sp1.S221}

\bibitem[{{Krips} {et~al.}(2008){Krips}, {Neri}, {Garc{\'\i}a-Burillo},
  {Mart{\'\i}n}, {Combes}, {Graci{\'a}-Carpio}, \& {Eckart}}]{Kri08}
{Krips}, M., {Neri}, R., {Garc{\'\i}a-Burillo}, S., {et~al.} 2008, \apj, 677,
  262, \dodoi{10.1086/527367}

\bibitem[{{Krolik} \& {Kallman}(1984)}]{Kro84}
{Krolik}, J.~H., \& {Kallman}, T.~R. 1984, \apj, 286, 366,
  \dodoi{10.1086/162608}

\bibitem[{{Krolik} \& {Kallman}(1987)}]{Kro87}
---. 1987, \apjl, 320, L5, \dodoi{10.1086/184966}

\bibitem[{{Krumholz} \& {Tan}(2007)}]{Kru07}
{Krumholz}, M.~R., \& {Tan}, J.~C. 2007, \apj, 654, 304, \dodoi{10.1086/509101}

\bibitem[{{Laha} {et~al.}(2018){Laha}, {Guainazzi}, {Piconcelli}, {Gandhi},
  {Ricci}, {Ghosh}, {Markowitz}, \& {Bagchi}}]{Lah18}
{Laha}, S., {Guainazzi}, M., {Piconcelli}, E., {et~al.} 2018, \apj, 868, 10,
  \dodoi{10.3847/1538-4357/aae390}

\bibitem[{{LaMassa} {et~al.}(2012){LaMassa}, {Heckman}, \& {Ptak}}]{LaM12}
{LaMassa}, S.~M., {Heckman}, T.~M., \& {Ptak}, A. 2012, \apj, 758, 82,
  \dodoi{10.1088/0004-637X/758/2/82}

\bibitem[{{LaMassa} {et~al.}(2017){LaMassa}, {Yaqoob}, {Levenson}, {Boorman},
  {Heckman}, {Gandhi}, {Rigby}, {Urry}, \& {Ptak}}]{LaM17}
{LaMassa}, S.~M., {Yaqoob}, T., {Levenson}, N.~A., {et~al.} 2017, \apj, 835,
  91, \dodoi{10.3847/1538-4357/835/1/91}

\bibitem[{{Lamperti} {et~al.}(2017){Lamperti}, {Koss}, {Trakhtenbrot},
  {Schawinski}, {Ricci}, {Oh}, {Landt}, {Riffel}, {Rodr{\'\i}guez-Ardila},
  {Gehrels}, {Harrison}, {Masetti}, {Mushotzky}, {Treister}, {Ueda}, \&
  {Veilleux}}]{Lamperti17}
{Lamperti}, I., {Koss}, M., {Trakhtenbrot}, B., {et~al.} 2017, \mnras, 467,
  540, \dodoi{10.1093/mnras/stx055}

\bibitem[{{Ma} {et~al.}(2020){Ma}, {Elvis}, {Fabbiano}, {Balokovi{\'c}},
  {Maksym}, {Jones}, \& {Risaliti}}]{Ma20}
{Ma}, J., {Elvis}, M., {Fabbiano}, G., {et~al.} 2020, \apj, 900, 164,
  \dodoi{10.3847/1538-4357/abacbe}

\bibitem[{{Magdziarz} \& {Zdziarski}(1995)}]{Mag95}
{Magdziarz}, P., \& {Zdziarski}, A.~A. 1995, \mnras, 273, 837,
  \dodoi{10.1093/mnras/273.3.837}

\bibitem[{{Magorrian} {et~al.}(1998){Magorrian}, {Tremaine}, {Richstone},
  {Bender}, {Bower}, {Dressler}, {Faber}, {Gebhardt}, {Green}, {Grillmair},
  {Kormendy}, \& {Lauer}}]{Mag98}
{Magorrian}, J., {Tremaine}, S., {Richstone}, D., {et~al.} 1998, \aj, 115,
  2285, \dodoi{10.1086/300353}

\bibitem[{{Maloney} {et~al.}(1996){Maloney}, {Hollenbach}, \&
  {Tielens}}]{Mal96}
{Maloney}, P.~R., {Hollenbach}, D.~J., \& {Tielens}, A.~G.~G.~M. 1996, \apj,
  466, 561, \dodoi{10.1086/177532}

\bibitem[{{Marconi} \& {Hunt}(2003)}]{Mar03}
{Marconi}, A., \& {Hunt}, L.~K. 2003, \apjl, 589, L21, \dodoi{10.1086/375804}

\bibitem[{{Marconi} {et~al.}(2004){Marconi}, {Risaliti}, {Gilli}, {Hunt},
  {Maiolino}, \& {Salvati}}]{Mar04}
{Marconi}, A., {Risaliti}, G., {Gilli}, R., {et~al.} 2004, \mnras, 351, 169,
  \dodoi{10.1111/j.1365-2966.2004.07765.x}

\bibitem[{{Marinucci} {et~al.}(2017){Marinucci}, {Bianchi}, {Fabbiano}, {Matt},
  {Risaliti}, {Nardini}, \& {Wang}}]{Mar17}
{Marinucci}, A., {Bianchi}, S., {Fabbiano}, G., {et~al.} 2017, \mnras, 470,
  4039, \dodoi{10.1093/mnras/stx1551}

\bibitem[{{Marinucci} {et~al.}(2013){Marinucci}, {Miniutti}, {Bianchi}, {Matt},
  \& {Risaliti}}]{Mar13}
{Marinucci}, A., {Miniutti}, G., {Bianchi}, S., {Matt}, G., \& {Risaliti}, G.
  2013, \mnras, 436, 2500, \dodoi{10.1093/mnras/stt1759}

\bibitem[{{Marinucci} {et~al.}(2012){Marinucci}, {Risaliti}, {Wang}, {Nardini},
  {Elvis}, {Fabbiano}, {Bianchi}, \& {Matt}}]{Mar12}
{Marinucci}, A., {Risaliti}, G., {Wang}, J., {et~al.} 2012, \mnras, 423, L6,
  \dodoi{10.1111/j.1745-3933.2012.01232.x}

\bibitem[{{Masini} {et~al.}(2016){Masini}, {Comastri}, {Balokovi{\'c}}, {Zaw},
  {Puccetti}, {Ballantyne}, {Bauer}, {Boggs}, {Brandt}, {Brightman},
  {Christensen}, {Craig}, {Gandhi}, {Hailey}, {Harrison}, {Koss}, {Madejski},
  {Ricci}, {Rivers}, {Stern}, \& {Zhang}}]{Mas16}
{Masini}, A., {Comastri}, A., {Balokovi{\'c}}, M., {et~al.} 2016, \aap, 589,
  A59, \dodoi{10.1051/0004-6361/201527689}

\bibitem[{{Matt} {et~al.}(2004){Matt}, {Bianchi}, {Guainazzi}, \&
  {Molendi}}]{Mat04}
{Matt}, G., {Bianchi}, S., {Guainazzi}, M., \& {Molendi}, S. 2004, \aap, 414,
  155, \dodoi{10.1051/0004-6361:20031635}

\bibitem[{{McMullin} {et~al.}(2007){McMullin}, {Waters}, {Schiebel}, {Young},
  \& {Golap}}]{McM07}
{McMullin}, J.~P., {Waters}, B., {Schiebel}, D., {Young}, W., \& {Golap}, K.
  2007, Astronomical Society of the Pacific Conference Series, Vol. 376, {CASA
  Architecture and Applications}, ed. R.~A. {Shaw}, F.~{Hill}, \& D.~J. {Bell},
  127

\bibitem[{{Meijerink} \& {Spaans}(2005)}]{Mei05}
{Meijerink}, R., \& {Spaans}, M. 2005, \aap, 436, 397,
  \dodoi{10.1051/0004-6361:20042398}

\bibitem[{{Meijerink} {et~al.}(2007){Meijerink}, {Spaans}, \& {Israel}}]{Mei07}
{Meijerink}, R., {Spaans}, M., \& {Israel}, F.~P. 2007, \aap, 461, 793,
  \dodoi{10.1051/0004-6361:20066130}

\bibitem[{{Mineo} {et~al.}(2012){Mineo}, {Gilfanov}, \& {Sunyaev}}]{Min12a}
{Mineo}, S., {Gilfanov}, M., \& {Sunyaev}, R. 2012, \mnras, 419, 2095,
  \dodoi{10.1111/j.1365-2966.2011.19862.x}

\bibitem[{{Miyamoto} {et~al.}(2018){Miyamoto}, {Seta}, {Nakai}, {Watanabe},
  {Salak}, \& {Ishii}}]{Miy18}
{Miyamoto}, Y., {Seta}, M., {Nakai}, N., {et~al.} 2018, \pasj, 70, L1,
  \dodoi{10.1093/pasj/psy016}

\bibitem[{{Monje} {et~al.}(2011){Monje}, {Blain}, \& {Phillips}}]{Mon11}
{Monje}, R.~R., {Blain}, A.~W., \& {Phillips}, T.~G. 2011, \apjs, 195, 23,
  \dodoi{10.1088/0067-0049/195/2/23}

\bibitem[{{Murakami} {et~al.}(2000){Murakami}, {Koyama}, {Sakano}, {Tsujimoto},
  \& {Maeda}}]{Mur00}
{Murakami}, H., {Koyama}, K., {Sakano}, M., {Tsujimoto}, M., \& {Maeda}, Y.
  2000, \apj, 534, 283, \dodoi{10.1086/308717}

\bibitem[{{Murphy} \& {Yaqoob}(2009)}]{Mur09}
{Murphy}, K.~D., \& {Yaqoob}, T. 2009, \mnras, 397, 1549,
  \dodoi{10.1111/j.1365-2966.2009.15025.x}

\bibitem[{{Mushotzky} {et~al.}(2019){Mushotzky}, {Aird}, {Barger},
  {Cappelluti}, {Chartas}, {Corrales}, {Eufrasio}, {Fabian}, {Falcone},
  {Gallo}, {Gilli}, {Grant}, {Hardcastle}, {Hodges-Kluck}, {Kara}, {Koss},
  {Li}, {Lisse}, {Loewenstein}, {Markevitch}, {Meyer}, {Miller}, {Mulchaey},
  {Petre}, {Ptak}, {Reynolds}, {Russell}, {Safi-Harb}, {Smith}, {Snios},
  {Tombesi}, {Valencic}, {Walker}, {Williams}, {Winter}, {Yamaguchi}, {Zhang},
  {Arenberg}, {Brandt}, {Burrows}, {Georganopoulos}, {Miller}, {Norman}, \&
  {Rosati}}]{Mus19}
{Mushotzky}, R., {Aird}, J., {Barger}, A.~J., {et~al.} 2019, in Bulletin of the
  American Astronomical Society, Vol.~51, 107.
\newblock \doarXiv{1903.04083}

\bibitem[{{Mushotzky} {et~al.}(2014){Mushotzky}, {Shimizu}, {Mel{\'e}ndez}, \&
  {Koss}}]{Mus14}
{Mushotzky}, R.~F., {Shimizu}, T.~T., {Mel{\'e}ndez}, M., \& {Koss}, M. 2014,
  \apjl, 781, L34, \dodoi{10.1088/2041-8205/781/2/L34}

\bibitem[{{Nakata} {et~al.}(2021){Nakata}, {Hayashida}, {Noda}, {Yoneyama},
  {Matsumoto}, \& {Imanishi}}]{Nak21}
{Nakata}, R., {Hayashida}, K., {Noda}, H., {et~al.} 2021, \pasj,
  \dodoi{10.1093/pasj/psab001}

\bibitem[{{Nemmen} {et~al.}(2006){Nemmen}, {Storchi-Bergmann}, {Yuan},
  {Eracleous}, {Terashima}, \& {Wilson}}]{Nem06}
{Nemmen}, R.~S., {Storchi-Bergmann}, T., {Yuan}, F., {et~al.} 2006, \apj, 643,
  652, \dodoi{10.1086/500571}

\bibitem[{{Nobukawa} {et~al.}(2010){Nobukawa}, {Koyama}, {Tsuru}, {Ryu}, \&
  {Tatischeff}}]{Nob10}
{Nobukawa}, M., {Koyama}, K., {Tsuru}, T.~G., {Ryu}, S.~G., \& {Tatischeff}, V.
  2010, \pasj, 62, 423, \dodoi{10.1093/pasj/62.2.423}

\bibitem[{{Noguchi} {et~al.}(2010){Noguchi}, {Terashima}, {Ishino},
  {Hashimoto}, {Koss}, {Ueda}, \& {Awaki}}]{Nog10}
{Noguchi}, K., {Terashima}, Y., {Ishino}, Y., {et~al.} 2010, \apj, 711, 144,
  \dodoi{10.1088/0004-637X/711/1/144}

\bibitem[{{Ogawa} {et~al.}(2021){Ogawa}, {Ueda}, {Tanimoto}, \&
  {Yamada}}]{Ogw21}
{Ogawa}, S., {Ueda}, Y., {Tanimoto}, A., \& {Yamada}, S. 2021, \apj, 906, 84,
  \dodoi{10.3847/1538-4357/abccce}

\bibitem[{{Ogle} {et~al.}(2003){Ogle}, {Brookings}, {Canizares}, {Lee}, \&
  {Marshall}}]{Ogl03}
{Ogle}, P.~M., {Brookings}, T., {Canizares}, C.~R., {Lee}, J.~C., \&
  {Marshall}, H.~L. 2003, \aap, 402, 849, \dodoi{10.1051/0004-6361:20021647}

\bibitem[{{Ponti} {et~al.}(2010){Ponti}, {Terrier}, {Goldwurm}, {Belanger}, \&
  {Trap}}]{Pon10}
{Ponti}, G., {Terrier}, R., {Goldwurm}, A., {Belanger}, G., \& {Trap}, G. 2010,
  \apj, 714, 732, \dodoi{10.1088/0004-637X/714/1/732}

\bibitem[{{Predehl} \& {Schmitt}(1995)}]{Pre95}
{Predehl}, P., \& {Schmitt}, J.~H.~M.~M. 1995, \aap, 500, 459

\bibitem[{{Puccetti} {et~al.}(2016){Puccetti}, {Comastri}, {Bauer}, {Brandt},
  {Fiore}, {Harrison}, {Luo}, {Stern}, {Urry}, {Alexander}, {Annuar},
  {Ar{\'e}valo}, {Balokovi{\'c}}, {Boggs}, {Brightman}, {Christensen}, {Craig},
  {Gandhi}, {Hailey}, {Koss}, {La Massa}, {Marinucci}, {Ricci}, {Walton},
  {Zappacosta}, \& {Zhang}}]{Pac16}
{Puccetti}, S., {Comastri}, A., {Bauer}, F.~E., {et~al.} 2016, \aap, 585, A157,
  \dodoi{10.1051/0004-6361/201527189}

\bibitem[{{Ramakrishnan} {et~al.}(2019){Ramakrishnan}, {Nagar}, {Finlez},
  {Storchi-Bergmann}, {Slater}, {Schnorr-M{\"u}ller}, {Riffel}, {Mundell}, \&
  {Robinson}}]{Ram19}
{Ramakrishnan}, V., {Nagar}, N.~M., {Finlez}, C., {et~al.} 2019, \mnras, 487,
  444, \dodoi{10.1093/mnras/stz1244}

\bibitem[{{Ricci} {et~al.}(2014){Ricci}, {Ueda}, {Paltani}, {Ichikawa}, {Gand
  hi}, \& {Awaki}}]{Ric14b}
{Ricci}, C., {Ueda}, Y., {Paltani}, S., {et~al.} 2014, \mnras, 441, 3622,
  \dodoi{10.1093/mnras/stu735}

\bibitem[{{Ricci} {et~al.}(2017{\natexlab{a}}){Ricci}, {Trakhtenbrot}, {Koss},
  {Ueda}, {Del Vecchio}, {Treister}, {Schawinski}, {Paltani}, {Oh}, {Lamperti},
  {Berney}, {Gandhi}, {Ichikawa}, {Bauer}, {Ho}, {Asmus}, {Beckmann}, {Soldi},
  {Balokovi{\'c}}, {Gehrels}, \& {Markwardt}}]{Ric17c}
{Ricci}, C., {Trakhtenbrot}, B., {Koss}, M.~J., {et~al.} 2017{\natexlab{a}},
  \apjs, 233, 17, \dodoi{10.3847/1538-4365/aa96ad}

\bibitem[{{Ricci} {et~al.}(2017{\natexlab{b}}){Ricci}, {Trakhtenbrot}, {Koss},
  {Ueda}, {Schawinski}, {Oh}, {Lamperti}, {Mushotzky}, {Treister}, {Ho},
  {Weigel}, {Bauer}, {Paltani}, {Fabian}, {Xie}, \& {Gehrels}}]{Ric17nat}
---. 2017{\natexlab{b}}, \nat, 549, 488, \dodoi{10.1038/nature23906}

\bibitem[{{Rosario} {et~al.}(2019){Rosario}, {Togi}, {Burtscher}, {Davies},
  {Shimizu}, \& {Lutz}}]{Ros19}
{Rosario}, D.~J., {Togi}, A., {Burtscher}, L., {et~al.} 2019, \apjl, 875, L8,
  \dodoi{10.3847/2041-8213/ab1262}

\bibitem[{{Rose} {et~al.}(2019){Rose}, {Edge}, {Combes}, {Gaspari}, {Hamer},
  {Nesvadba}, {Russell}, {Tremblay}, {Baum}, {O'Dea}, {Peck}, {Sarazin},
  {Vantyghem}, {Bremer}, {Donahue}, {Fabian}, {Ferland}, {McNamara}, {Mittal},
  {Oonk}, {Salom{\'e}}, {Swinbank}, \& {Voit}}]{Rose19}
{Rose}, T., {Edge}, A.~C., {Combes}, F., {et~al.} 2019, \mnras, 485, 229,
  \dodoi{10.1093/mnras/stz406}

\bibitem[{{Sandstrom} {et~al.}(2013){Sandstrom}, {Leroy}, {Walter}, {Bolatto},
  {Croxall}, {Draine}, {Wilson}, {Wolfire}, {Calzetti}, {Kennicutt}, {Aniano},
  {Donovan Meyer}, {Usero}, {Bigiel}, {Brinks}, {de Blok}, {Crocker}, {Dale},
  {Engelbracht}, {Galametz}, {Groves}, {Hunt}, {Koda}, {Kreckel}, {Linz},
  {Meidt}, {Pellegrini}, {Rix}, {Roussel}, {Schinnerer}, {Schruba}, {Schuster},
  {Skibba}, {van der Laan}, {Appleton}, {Armus}, {Brandl}, {Gordon}, {Hinz},
  {Krause}, {Montiel}, {Sauvage}, {Schmiedeke}, {Smith}, \& {Vigroux}}]{San13}
{Sandstrom}, K.~M., {Leroy}, A.~K., {Walter}, F., {et~al.} 2013, \apj, 777, 5,
  \dodoi{10.1088/0004-637X/777/1/5}

\bibitem[{{Shangguan} {et~al.}(2020){Shangguan}, {Ho}, {Bauer}, {Wang}, \&
  {Treister}}]{Sha20}
{Shangguan}, J., {Ho}, L.~C., {Bauer}, F.~E., {Wang}, R., \& {Treister}, E.
  2020, \apj, 899, 112, \dodoi{10.3847/1538-4357/aba8a1}

\bibitem[{{Shimizu} {et~al.}(2017){Shimizu}, {Mushotzky}, {Mel{\'e}ndez},
  {Koss}, {Barger}, \& {Cowie}}]{Shi17}
{Shimizu}, T.~T., {Mushotzky}, R.~F., {Mel{\'e}ndez}, M., {et~al.} 2017,
  \mnras, 466, 3161, \dodoi{10.1093/mnras/stw3268}

\bibitem[{{Smith} {et~al.}(2020){Smith}, {Mushotzky}, {Koss}, {Trakhtenbrot},
  {Ricci}, {Wong}, {Bauer}, {Ricci}, {Vogel}, {Stern}, {Powell}, {Urry},
  {Harrison}, {Mejia-Restrepo}, {Oh}, {Baek}, \& {Chung}}]{Smi20}
{Smith}, K.~L., {Mushotzky}, R.~F., {Koss}, M., {et~al.} 2020, \mnras, 492,
  4216, \dodoi{10.1093/mnras/stz3608}

\bibitem[{{Solomon} \& {Vanden Bout}(2005)}]{Sol05}
{Solomon}, P.~M., \& {Vanden Bout}, P.~A. 2005, \araa, 43, 677,
  \dodoi{10.1146/annurev.astro.43.051804.102221}

\bibitem[{{Soltan}(1982)}]{Sol82}
{Soltan}, A. 1982, \mnras, 200, 115, \dodoi{10.1093/mnras/200.1.115}

\bibitem[{{Tanimoto} {et~al.}(2016){Tanimoto}, {Ueda}, {Kawamuro}, \&
  {Ricci}}]{Tan16}
{Tanimoto}, A., {Ueda}, Y., {Kawamuro}, T., \& {Ricci}, C. 2016, \pasj, 68,
  S26, \dodoi{10.1093/pasj/psw008}

\bibitem[{{Tanimoto} {et~al.}(2018){Tanimoto}, {Ueda}, {Kawamuro}, {Ricci},
  {Awaki}, \& {Terashima}}]{Tan18}
{Tanimoto}, A., {Ueda}, Y., {Kawamuro}, T., {et~al.} 2018, \apj, 853, 146,
  \dodoi{10.3847/1538-4357/aaa47c}

\bibitem[{{Tatischeff} {et~al.}(2012){Tatischeff}, {Decourchelle}, \&
  {Maurin}}]{Tat12}
{Tatischeff}, V., {Decourchelle}, A., \& {Maurin}, G. 2012, \aap, 546, A88,
  \dodoi{10.1051/0004-6361/201219016}

\bibitem[{{The Lynx Team}(2018)}]{Lyn18}
{The Lynx Team}. 2018, arXiv e-prints, arXiv:1809.09642.
\newblock \doarXiv{1809.09642}

\bibitem[{{Torrej{\'o}n} {et~al.}(2010){Torrej{\'o}n}, {Schulz}, {Nowak}, \&
  {Kallman}}]{Tor10}
{Torrej{\'o}n}, J.~M., {Schulz}, N.~S., {Nowak}, M.~A., \& {Kallman}, T.~R.
  2010, \apj, 715, 947, \dodoi{10.1088/0004-637X/715/2/947}

\bibitem[{{Treister} {et~al.}(2018){Treister}, {Privon}, {Sartori}, {Nagar},
  {Bauer}, {Schawinski}, {Messias}, {Ricci}, {U}, {Casey}, {Comerford},
  {Muller-Sanchez}, {Evans}, {Finlez}, {Koss}, {Sanders}, \& {Urry}}]{Tre18}
{Treister}, E., {Privon}, G.~C., {Sartori}, L.~F., {et~al.} 2018, \apj, 854,
  83, \dodoi{10.3847/1538-4357/aaa963}

\bibitem[{{Treister} {et~al.}(2020){Treister}, {Messias}, {Privon}, {Nagar},
  {Medling}, {U}, {Bauer}, {Cicone}, {Mu{\~n}oz}, {Evans}, {Muller-Sanchez},
  {Comerford}, {Armus}, {Chang}, {Koss}, {Venturi}, {Schawinski}, {Casey},
  {Urry}, {Sanders}, {Scoville}, \& {Sheth}}]{Tre20}
{Treister}, E., {Messias}, H., {Privon}, G.~C., {et~al.} 2020, \apj, 890, 149,
  \dodoi{10.3847/1538-4357/ab6b28}

\bibitem[{{Ueda} {et~al.}(2014){Ueda}, {Akiyama}, {Hasinger}, {Miyaji}, \&
  {Watson}}]{Ued14}
{Ueda}, Y., {Akiyama}, M., {Hasinger}, G., {Miyaji}, T., \& {Watson}, M.~G.
  2014, \apj, 786, 104, \dodoi{10.1088/0004-637X/786/2/104}

\bibitem[{{Ursini} {et~al.}(2018){Ursini}, {Bassani}, {Panessa}, {Bazzano},
  {Bird}, {Malizia}, \& {Ubertini}}]{Urs18}
{Ursini}, F., {Bassani}, L., {Panessa}, F., {et~al.} 2018, \mnras, 474, 5684,
  \dodoi{10.1093/mnras/stx3159}

\bibitem[{{Valinia} {et~al.}(2000){Valinia}, {Tatischeff}, {Arnaud}, {Ebisawa},
  \& {Ramaty}}]{Val00}
{Valinia}, A., {Tatischeff}, V., {Arnaud}, K., {Ebisawa}, K., \& {Ramaty}, R.
  2000, \apj, 543, 733, \dodoi{10.1086/317133}

\bibitem[{{Venturi} {et~al.}(2017){Venturi}, {Marconi}, {Mingozzi}, {Carniani},
  {Cresci}, {Risaliti}, \& {Mannucci}}]{Ven17}
{Venturi}, G., {Marconi}, A., {Mingozzi}, M., {et~al.} 2017, Frontiers in
  Astronomy and Space Sciences, 4, 46, \dodoi{10.3389/fspas.2017.00046}

\bibitem[{{Viti} {et~al.}(2014){Viti}, {Garc{\'\i}a-Burillo}, {Fuente}, {Hunt},
  {Usero}, {Henkel}, {Eckart}, {Martin}, {Spaans}, {Muller}, {Combes}, {Krips},
  {Schinnerer}, {Casasola}, {Costagliola}, {Marquez}, {Planesas}, {van der
  Werf}, {Aalto}, {Baker}, {Boone}, \& {Tacconi}}]{Vit14}
{Viti}, S., {Garc{\'\i}a-Burillo}, S., {Fuente}, A., {et~al.} 2014, \aap, 570,
  A28, \dodoi{10.1051/0004-6361/201424116}

\bibitem[{{Wang} {et~al.}(2009){Wang}, {Fabbiano}, {Elvis}, {Risaliti},
  {Mazzarella}, {Howell}, \& {Lord}}]{Wan09}
{Wang}, J., {Fabbiano}, G., {Elvis}, M., {et~al.} 2009, \apj, 694, 718,
  \dodoi{10.1088/0004-637X/694/2/718}

\bibitem[{{Wilms} {et~al.}(2000){Wilms}, {Allen}, \& {McCray}}]{Wil00}
{Wilms}, J., {Allen}, A., \& {McCray}, R. 2000, \apj, 542, 914,
  \dodoi{10.1086/317016}

\bibitem[{{Xu} {et~al.}(2016){Xu}, {Liu}, {Gou}, \& {Liu}}]{Xu16}
{Xu}, W., {Liu}, Z., {Gou}, L., \& {Liu}, J. 2016, \mnras, 455, L26,
  \dodoi{10.1093/mnrasl/slv148}

\bibitem[{{Yamada} {et~al.}(2018){Yamada}, {Ueda}, {Oda}, {Tanimoto},
  {Imanishi}, {Terashima}, \& {Ricci}}]{Yam18}
{Yamada}, S., {Ueda}, Y., {Oda}, S., {et~al.} 2018, \apj, 858, 106,
  \dodoi{10.3847/1538-4357/aabacb}

\bibitem[{{Yamada} {et~al.}(2020){Yamada}, {Ueda}, {Tanimoto}, {Oda},
  {Imanishi}, {Toba}, \& {Ricci}}]{Yam20}
{Yamada}, S., {Ueda}, Y., {Tanimoto}, A., {et~al.} 2020, \apj, 897, 107,
  \dodoi{10.3847/1538-4357/ab94b1}

\bibitem[{{Yamada}(1994)}]{Yam94}
{Yamada}, T. 1994, \apjl, 423, L27, \dodoi{10.1086/187227}

\bibitem[{{Yan} {et~al.}(2021){Yan}, {Hickox}, {Chen}, {Ricci}, {Masini},
  {Bauer}, \& {Alexander}}]{Yan21}
{Yan}, W., {Hickox}, R.~C., {Chen}, C.-T.~J., {et~al.} 2021, arXiv e-prints,
  arXiv:2104.10702.
\newblock \doarXiv{2104.10702}

\bibitem[{{Younes} {et~al.}(2019){Younes}, {Ptak}, {Ho}, {Xie}, {Terasima},
  {Yuan}, {Huppenkothen}, \& {Yukita}}]{You19}
{Younes}, G., {Ptak}, A., {Ho}, L.~C., {et~al.} 2019, \apj, 870, 73,
  \dodoi{10.3847/1538-4357/aaf38b}

\bibitem[{{Young} {et~al.}(2001){Young}, {Wilson}, \& {Shopbell}}]{You01}
{Young}, A.~J., {Wilson}, A.~S., \& {Shopbell}, P.~L. 2001, \apj, 556, 6,
  \dodoi{10.1086/321561}

\bibitem[{{Yusef-Zadeh} {et~al.}(2002){Yusef-Zadeh}, {Law}, {Wardle}, {Wang},
  {Fruscione}, {Lang}, \& {Cotera}}]{Yus02}
{Yusef-Zadeh}, F., {Law}, C., {Wardle}, M., {et~al.} 2002, \apj, 570, 665,
  \dodoi{10.1086/340058}

\bibitem[{{Yusef-Zadeh} {et~al.}(2007){Yusef-Zadeh}, {Muno}, {Wardle}, \&
  {Lis}}]{Yus07}
{Yusef-Zadeh}, F., {Muno}, M., {Wardle}, M., \& {Lis}, D.~C. 2007, \apj, 656,
  847, \dodoi{10.1086/510663}

\end{thebibliography}

\end{document}